\documentclass[a4paper,oneside,english,british]{scrbook}
\usepackage[LGR,T1]{fontenc}
\usepackage[utf8x]{inputenc}
\usepackage{xcolor}
\usepackage{babel}
\usepackage{array}
\usepackage{longtable}
\usepackage{float}
\usepackage{textcomp}
\usepackage{multirow}
\usepackage{dsfont}
\usepackage{amsmath}
\usepackage{amssymb}
\usepackage{cancel}
\usepackage{stackrel}
\usepackage{makeidx}
\makeindex
\usepackage{graphicx}
\usepackage{setspace}
\usepackage{wasysym}
\PassOptionsToPackage{normalem}{ulem}
\usepackage{ulem}
\usepackage{nomencl}
\providecommand{\printnomenclature}{\printglossary}
\providecommand{\makenomenclature}{\makeglossary}
\makenomenclature
\usepackage{microtype}
\usepackage[unicode=true,
 bookmarks=true,bookmarksnumbered=true,bookmarksopen=true,bookmarksopenlevel=1,
 breaklinks=false,pdfborder={0 0 1},backref=false,colorlinks=true]
 {hyperref}
\hypersetup{pdftitle={The Measurement and Modelling of Cosmic Ray Muons at KM3NeT Detectors},
 pdfauthor={Piotr Kalaczyński},
 pdfsubject={Muon studies with KM3NeT detectors},
 pdfkeywords={phd, thesis, KM3NeT, astroparticle, neutrino, physics, muon},
 linkcolor=brown,citecolor=green,urlcolor=blue}

\makeatletter

\ProvideTextCommand{\~}{LGR}[1]{\char126#1}

\newcommand{\lyxmathsym}[1]{\ifmmode\begingroup\def\b@ld{bold}
  \text{\ifx\math@version\b@ld\bfseries\fi#1}\endgroup\else#1\fi}

\providecommand{\tabularnewline}{\\}
\newcommand{\lyxdot}{.}

\numberwithin{equation}{section}
\numberwithin{figure}{section}
\numberwithin{table}{section}

\usepackage{ifthen}
\usepackage[a4paper,top=20mm,bottom=30mm,inner=20mm,outer=20mm,headheight=0mm, twoside,bindingoffset=6mm]{geometry}

\usepackage{fancyhdr}
\usepackage[T1]{fontenc}
\usepackage{lmodern}

\usepackage{xcolor} 
\definecolor{airforceblue}{rgb}{0.36, 0.54, 0.66}
\definecolor{electriclime}{rgb}{0.8, 1.0, 0.0}
\definecolor{darkgreen}{rgb}{0.0, 0.5, 0.0}

\usepackage{wallpaper}

\usepackage{textgreek}
\usepackage{upgreek}

\usepackage{siunitx}

\usepackage{caption}
\usepackage[all]{nowidow}

\@ifundefined{showcaptionsetup}{}{%
 \PassOptionsToPackage{caption=false}{subfig}}
\usepackage{subfig}
\makeatother

\usepackage[style=trad-plain,backend=bibtex8,sorting=none,giveninits=true,maxbibnames=2]{biblatex}

\addto\captionsbritish{%
}
\addto\captionsenglish{%
}

\begin{document}
\subject{\selectlanguage{english}%
DOCTORAL THESIS}
\title{The Measurement and Modelling of Cosmic Ray Muons at KM3NeT Detectors}
\author{Author: Piotr Kalaczyński}
\date{Warsaw, September 2023}
\publishers{\selectlanguage{english}%
\includegraphics[width=8cm]{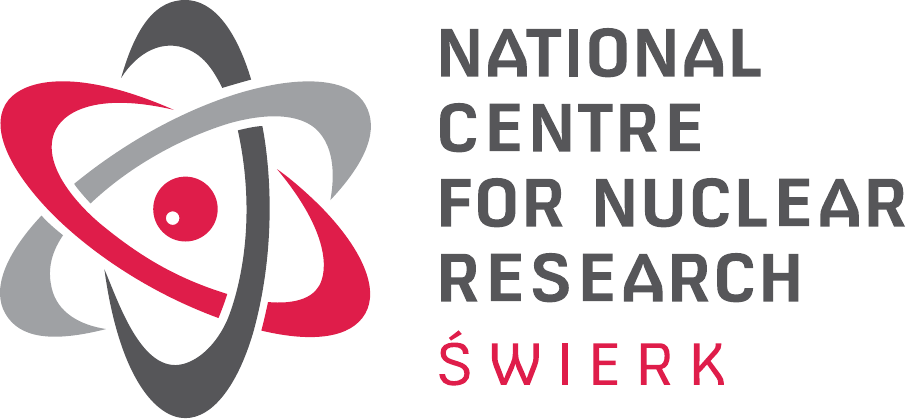}\vspace{\baselineskip}\\
\foreignlanguage{british}{A thesis submitted in fulfilment of the
requirements }\\
\foreignlanguage{british}{for the degree of Doctor of Philosophy}\\
\foreignlanguage{british}{ in the}\\
\foreignlanguage{british}{ High Energy Physics Division}\\
\foreignlanguage{british}{ Department of Fundamental Research}\\
\foreignlanguage{british}{National Centre for Nuclear Research}\ThisURCornerWallPaper{1.0}{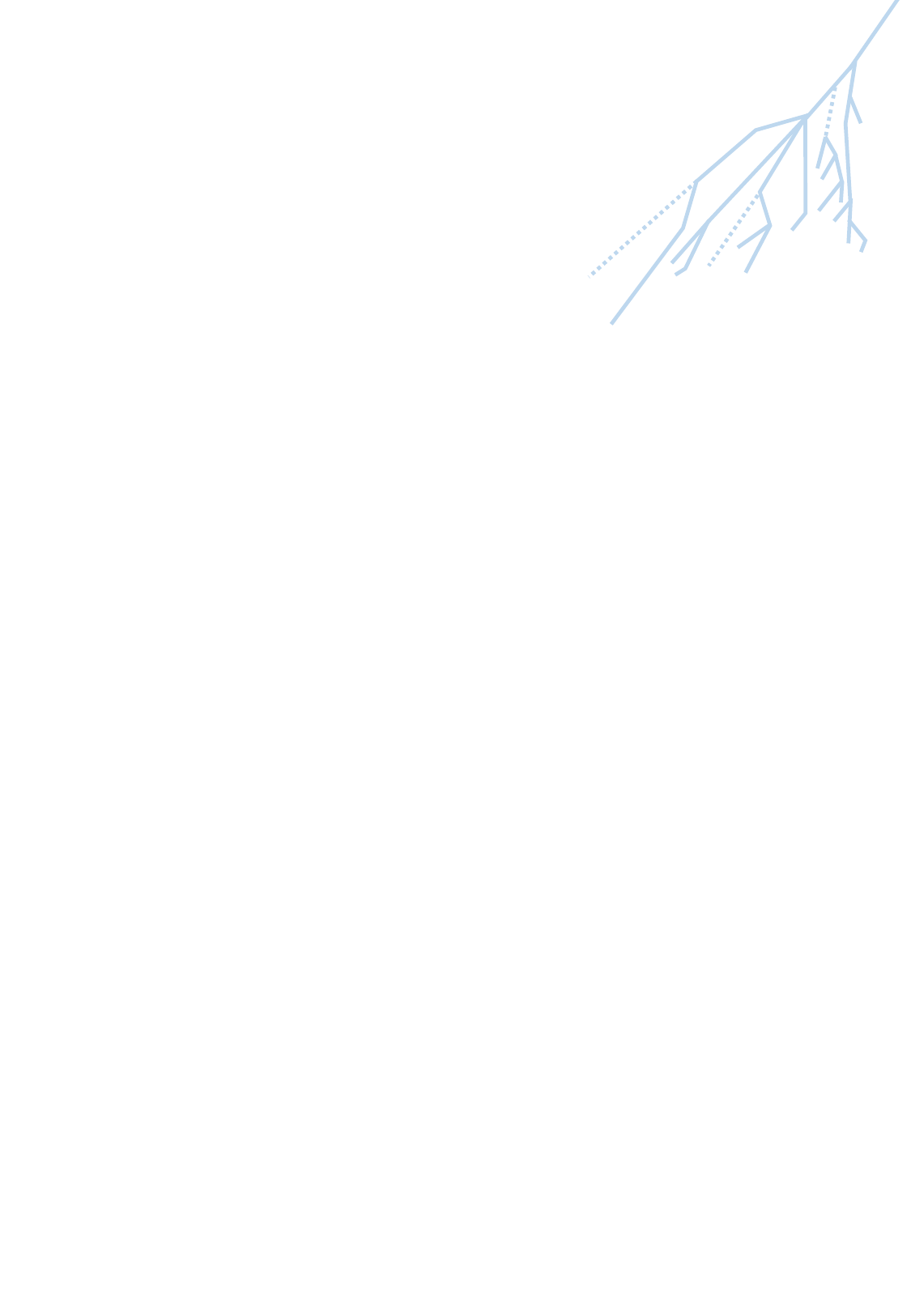}}
\lowertitleback{\textbf{Supervisor of the dissertation}\foreignlanguage{english}{\smallskip{}
}\\
Prof. dr hab. Ewa Rondio\foreignlanguage{english}{\bigskip{}
}\\
\textbf{Co-supervisor of the dissertation}\foreignlanguage{english}{\smallskip{}
}\\
Dr Piotr Mijakowski\foreignlanguage{english}{\bigskip{}
}\\
}
\dedication{I wish to dedicate my work to my deer\thanks{pun intended} ~wife
Daniela\\
who makes me a better person every day and who put up with me devoting
a large portion of the last few years to this work.}

\maketitle
\textbf{\textcolor{white}{d}}

Quotes not necessarily related to the thesis that the author just
fancied to have here:
\begin{quotation}
\vspace{3cm}
`By discovering nature, you discover yourself`

— Maxime Lagacé

\vspace{2cm}
`I am the Super-Kamiokande`

— Piotr Mijakowski

\vspace{2cm}

`Contra vim mortis non crescit salvia in hortis`

(No herb grows in the gardens against the power of death)

— Sigismund III Vasa on his deathbed

\vspace{2cm}

`Woof woof woof$\textrm{!}$`

— Mecenas \& Fiona (our dogs, expressing their irritation at me for
writing this thesis instead of playing with them)\vspace{2cm}

`Indeed, we know what to do with space but do not know what to do
about time, except to make it subservient to space. Most of us seem
to labor for the sake of things of space. As a result we suffer from
a deeply rooted dread of time and stand aghast when compelled to look
into its face. Time to us is sarcasm, a slick treaclierous monster
with a jaw like a furnace incinerating every moment of our lives.
Shrinking, therefore, from facing time, we escape for shelter to things
of space.`

— Abraham Joshua Heschel: `The Sabbath: its meaning for modern man'
\end{quotation}
\textbf{\textcolor{white}{d}}

\chapter*{Declaration of Authorship}

I, Piotr Kalaczyński, declare that this thesis, titled: 

“The Measurement and Modelling of Cosmic Ray Muons at KM3NeT Detectors”

and the work presented in it is my own. 

I confirm that:
\begin{itemize}
\item This work was done wholly or mainly while in candidature for a research
degree at the National Centre for Nuclear Research.
\item Where any part of this thesis has previously been submitted for a
degree or any other qualification at the National Centre for Nuclear
Research or any other institution, this has been clearly stated.
\item Where I have consulted the published work of others, this is always
clearly attributed.
\item Where I have quoted from the work of others, the source is always
given. Except for such quotations, this dissertation is entirely my
work.
\item I have acknowledged all main sources of help.
\item Where the thesis is based on work done by myself jointly with others,
I have made clear exactly what was done by others and what I have
contributed myself.
\end{itemize}
~

Signed: \_\_\_\_\_\_\_\_\_\_\_\_\_\_\_\_\_\_\_\_\_\_\_\_\_\_\_\_\_\_\_\_\_\_\_\_\_\_

Date: ~~ \_\_\_\_\_\_\_\_\_\_\_\_\_\_\_\_\_\_\_\_\_\_\_\_\_\_\_\_\_\_\_\_\_\_\_\_\_\_

\chapter*{Abstract}

\addcontentsline{toc}{chapter}{Abstract} 

Atmospheric muons are the most frequently observed form of cosmic
radiation. Despite this, the existence of the muon flux component
produced in decays of short-lived parent particles, called prompt
muon flux still awaits experimental confirmation. This contribution
to the muon flux is expected to start dominating at high energies,
around PeV, since many of the prompt parent particles are heavy hadrons,
containing charm and strange quarks. The aim of this thesis was two-fold:
to evaluate the possibility of observing the prompt muon flux and
to validate the performance of the KM3NeT detectors.

The KM3NeT experiment is a research network of underwater Cherenkov
neutrino telescopes, currently under construction at two different
locations in the Mediterranean Sea. The ARCA detector, dedicated to
high-energy neutrino astronomy, is located near the coast of Italy
at Portopalo di Capo Passero. The second one is ORCA, and it focuses
on the low-energy atmospheric neutrino physics and is built offshore
Toulon, in France. Even in their intermediate configurations, the
KM3NeT detectors collect vast amounts of muon data. This work makes
use of the muon data to evaluate the current operation of KM3NeT/ARCA
and KM3NeT/ORCA and estimate their future potential.

To study the performance of KM3NeT detectors, an extensive Monte Carlo
(MC) simulation of the muon events has been produced using the CORSIKA
code, coupled with a chain of KM3NeT processing software. It was possible
through significant improvements in the open-source KM3NeT application,
responsible for the transport of simulated particles to the detector.

The MC simulation was put to use in the reconstruction of several
observables: muon bundle energy, total primary energy and muon multiplicity.
The regression of energy and multiplicity was carried out utilising
machine learning tools. The bundle energy reconstruction beat the
standard KM3NeT energy reconstruction by a large margin. Obtaining
predictions of total primary energy and muon multiplicity allowed
for the first measurements of such quantities by an underwater neutrino
telescope.

The developed reconstruction was directly applied in the main physics
analysis, which investigated the potential of KM3NeT detectors to
observe the prompt atmospheric muon flux. The complete KM3NeT detectors
should be able to confirm the prompt signal within 4–6 years of operation,
possibly even sooner if their measurements are combined. Currently,
the sensitivity is strongly limited by systematic uncertainties, which
may be reduced by the time ARCA and ORCA are completed.

\chapter*{Streszczenie}

\addcontentsline{toc}{chapter}{Streszczenie} 

Miony atmosferyczne to najczęściej obserwowana forma promieniowania
kosmicznego. Pomimo tego, istnienie komponentu strumienia mionów powstającego
w rozpadach krótko żyjących cząstek-rodziców, nazywanego natychmiastowym
strumieniem mionów, wciąż pozostaje niepotwierdzone eksperymentalnie.
Oczekuje się, że ten wkład do strumienia mionów powinien dominować
przy wysokich energiach, około PeV, jako że wiele z cząstek-rodziców
natychmiastowych mionów to ciężkie hadrony, zawierające kwarki powabne
i dziwne. Cel tej dysertacji był dwojaki: oszacowanie możliwości obserwacji
natychmiastowego strumienia mionów oraz weryfikacja wydajności detektorów
KM3NeT'u.

Eksperyment KM3NeT jest siecią badawczą, składającą się z podwodnych
czerenkowskich teleskopów neutrinowych, obecnie w fazie konstrukcji
w dwóch różnych lokalizacjach w Morzu Śródziemnym. ARCA to detektor
dedykowany astronomii neutrinowej wysokich energii, zlokalizowany
niedaleko włoskiego wybrzeża przy Portopalo di Capo Passero. Drugi
detektor to ORCA: skupia się on na fizyce niskoenergetycznych neutrin
atmosferycznych i powstaje w pobliżu Tulonu, we Francji. Nawet w ich
pośrednich konfiguracjach, każdy z detektorów KM3NeT zbiera ogromne
ilości danych mionowych. Ta praca korzysta z danych mionowych, aby
przyjrzeć się obecnej efektywności KM3NeT/ARCA i KM3NeT/ORCA, oraz
aby ocenić ich przyszły potencjał.

Aby ocenić wydajność detektorów KM3NeT, została wygenerowana rozległa
symulacja Monte Carlo (MC) dla przypadków mionowych, przy użyciu programu
CORSIKA oraz łańcucha oprogramowania symulacyjnego KM3NeT. Było to
możliwe dzięki wprowadzeniu znacznych usprawnień w publicznie dostępnej
aplikacji KM3NeT-u, transportującej wysymulowane cząstki do detektora.

Symulacja MC została użyta w rekonstrukcji kilku zmiennych: energii
pęku mionów, całkowitej energii cząstki pierwotnej oraz krotności
mionów. Wartości energii i krotności zostały estymowane przy użyciu
narzędzi uczenia maszynowego. Rekonstrukcja energii pęku okazała się
znacząco lepsza od standardowej rekonstrukcji energii używanej przez
KM3NeT. Uzyskanie rekonstrukcji energii cząstki pierwotnej oraz krotności
mionów umożliwiły ponadto pierwsze pomiary takich obserwabli przez
podwodny teleskop neutrinowy.

Przygotowana rekonstrukcja została bezpośrednio zastosowana w głównej
analizie fizycznej, badającej potencjał KM3NeT'u do obserwacji natychmiastowego
strumienia mionów. Ukończone detektory KM3NeT powinny być w stanie
potwierdzić sygnał od natychmiastowego strumienia w przeciągu 4-6
lat działania. Być może nawet wcześniej, jeżeli wyniki ich pomiarów
zostaną połączone. Czułość jest obecnie mocno ograniczona przez niepewności
systematyczne, z których część powinna zostać zredukowana do czasu
ukończenia detektorów ARCA i ORCA.

\tableofcontents{}

\chapter{Introduction}

While it is certainly true that neutrinos are the main focus of the
KM3NeT experiment and still hold many exciting mysteries (see Sec.
\ref{subsec:Neutrinos}), they are not the only particles studied
by neutrino telescopes like KM3NeT. From the perspective of such experiments,
muons are primarily a background, since they are produced both in
interactions of neutrinos, and decays of other particles. It is easy
to fall prey to a deception that they have no secrets left, since
they have been known to physicists and studied by them for much longer
than neutrinos. Muons are a very convenient tool to indirectly study
the physics of cosmic rays and put constraints on their flux. They
offer valuable insight into the development of air showers, primary
cosmic ray flux composition, production of particles associated with
the charmed, strange, or even heavier quarks in the atmosphere and
much more. Comparing the muon data against the Monte Carlo simulation
allows testing and improves both the data acquisition and simulation
of muon production. This thesis investigates the properties of the
muon flux created in interactions of cosmic rays with the air molecules
in the upper atmosphere with a particular emphasis on the prompt muon
flux. The performance of the KM3NeT detectors is validated using the
first collected data and a dedicated event reconstruction.

This work consists of 8 chapters in total. Below, their contents are
briefly summarised:
\begin{enumerate}
\item Gives a preview of the matter of this dissertation, information on
the original contribution of the author, and the motivation for undertaking
this work.
\item Provides an overview of the relevant physics: muons, neutrinos, and
cosmic rays. They are then tied together in the description of the
current understanding of the observed atmospheric flux, with an emphasis
on the muon flux, which is the subject of this thesis. The currently
unconfirmed prompt component of the muon flux, originating from decays
of heavy hadrons and light vector mesons, is of particular interest
with Chap. \ref{chap:prompt_ana} in mind.
\item Introduces the concept of the underwater neutrino telescopes in general,
including a description of the Cherenkov radiation, which they observe.
Furthermore, the KM3NeT experiment is described in much more detail,
both in terms of hardware and it's physics goal \cite{KM3NeT,KM3NeT-LoI-2.0}.
The author of this dissertation is a member of the KM3NeT Collaboration
and the results presented in further chapters were produced entirely
in the KM3NeT context.
\item Describes how muons produced as a result of cosmic ray interactions
in the atmosphere are simulated for the KM3NeT detectors. This includes
a description of the necessary simulation software chain. The author
of this thesis was heavily involved in foundational work in KM3NeT
muon simulations. His contributions include:
\begin{enumerate}
\item Maintenance and development of the Corant software (see Sec. \ref{sec:sea}),
taking care of computation of event weights and conversion of CORSIKA
output files to the ASCII format. This code was used for the earliest
results presented in Sec. \ref{sec:data_MC_Results} and its functionality
was later directly incorporated into gSeaGen (see next point).
\item Maintenance and development of the gSeaGen software (see Sec. \ref{sec:can}
and \ref{sec:gSeaGen_supplementary_material}), originally created
by Carla Distefano to generate neutrino events and transport them
to the boundary of the detector sensitive volume \cite{gSeaGen_2016,gSeaGen-2020}.
The first important contribution of the author of this dissertation
was the adaptation of gSeaGen to be able to propagate muons simulated
by CORSIKA from the sea level to the detector sensitive volume. The
initial approach was to first convert the CORSIKA output (sequential
unformatted $\mathtt{\mathtt{FORTRAN}}$ files \cite{CORSIKA-Userguide})
with Corant and then read and process it further with gSeaGen. At
this stage, the author of this thesis collaborated closely with Alfonso
Andres Garcia Soto. Soon it was realised that, despite its convenience,
conversion to text format done with Corant is in fact inefficient
and limits the entire simulation chain. This inspired the author of
this dissertation to pursue an alternative approach, i.e. directly
read the CORSIKA output files with gSeaGen. This effort built upon
the pre-existing basic readout scripts, which are distributed together
with each CORSIKA release. The first results of these efforts were
teased in the proceedings of the VLVnT2021 conference \cite{gSeaGen_Alfonso_VLVNT2021}.
The author of this thesis expanded the amount of information extracted
from CORSIKA to foster potential physics analyses. \\
Further endeavours moved towards a complete redesign of the muon propagation
routine within gSeaGen, targeting a number of issues: the efficiency
of the code in terms of memory management and runtime, the accuracy
of propagation, and the treatment of the curvature of the Earth and
overall simulation geometry. The resulting enhancement of the code
was particularly noticeable in the case of high-energy simulations,
where the average time needed to simulate 100 events dropped from
12 days to 40 minutes, making the large-scale simulation feasible.
Supplementary material describing some of those developments is available
in Sec. \ref{sec:gSeaGen_supplementary_material}. A publication on
those specific improvements in gSeaGen is currently under internal
review in the KM3NeT Collaboration, with the author of this thesis
as the main author of the paper. In addition, the author of this dissertation
has contributed to adapting gSeaGen to meet the criteria of the open
software policy of KM3NeT, which was described in the proceedings
of the VLVnT2021 conference \cite{gSeaGen_Jutta_VLVnT2021}.
\item Development of official KM3NeT simulation steering scripts, which
control the running of the entire KM3NeT software chain. The author
of this thesis has in particular worked on adapting the code to work
in multiple computing environments and to integrate the CORSIKA processing
into the scripts. This endeavour was necessary to proceed with the
production of the large-scale CORSIKA simulation, described in the
Chap. \ref{chap:Cosmic-Ray-Sim-chain-KM3NeT} and in Sec. \ref{sec:Additional-material-related-to-CORSIKA}.
During first three years of his work within KM3NeT, the author of
this dissertation was the sole person responsible for the CORSIKA
simulations. After that, he was joined by Andrey Romanov, who assisted
him in the work, which included immense amounts of bookkeeping and
monitoring. Suffice it to say that the actual running of the final
simulations themselves took more than a year of continuous running
on two computing clusters: Świerk Computing Centre (CIŚ) in Otwock,
Poland and IN2P3’s Computing Centre in Lyon, France. During the production
of CORSIKA simulation, several issues were identified and reported
to the authors of the code, who addressed them in the release of version
7.7410 \cite{CORSIKA_v77410}. 
\end{enumerate}
Completion of the CORSIKA simulation work was a crucial impulse for
the creation of the cosmic ray physics working group in KM3NeT, led
by Ronald Brujin. Almost all the current analyses in this group rely
on those simulations, with some results already published in \cite{StefanReckThesis,Andrey_and_me_ICRC2023,my-ICRC2021,my-VLVnT2021}.
Additional details on CORSIKA production and improvements in gSeaGen
are given in the appendix in Sec. \ref{sec:Additional-material-related-to-CORSIKA}–\ref{sec:gSeaGen_supplementary_material}.
\item Is dedicated to development of new reconstruction tools for KM3NeT
by the author of this dissertation. The reconstructed quantities are
the number of muons (multiplicity) and energy. To this end, the author
made use of the machine learning tools and the aforementioned CORSIKA
production. The chapter evaluates the performance of trained machine
learning models and compares the results with existing KM3NeT reconstruction:
JMuon, described in Chap. \ref{chap:Cosmic-Ray-Sim-chain-KM3NeT}.
\item Presents the comparisons between the simulations and experimental
data, performed on the first datasets from the KM3NeT detectors. The
results were published by the author of this thesis in \cite{my-ICRC2019,my-ICHEP2020,my-ICRC2021,my-VLVnT2021,Andrey_and_me_ICRC2023}.
For early detector configurations only the distributions of reconstructed
direction and energy were investigated. However, for the most recent
ones the reconstruction of muon multiplicity and primary cosmic ray
energy were included. The experimental datasets and the corresponding
MUPAGE (the other muon simulation software used in KM3NeT, see Sec.
\ref{sec:MUPAGE}) simulations were provided the KM3NeT data processing
and simulation working groups respectively. 
\item Is devoted to the main physics analysis of this work, carried out
by the author of this dissertation within the cosmic ray working group.
The author has developed a procedure to evaluate the contribution
of the prompt component of the atmospheric muon flux or to set an
upper limit on it by interpreting the KM3NeT data. The study made
predictions of the expected sensitivity of KM3NeT detectors to measure
the prompt muon flux in smaller configurations and with complete building
blocks of ARCA and ORCA. The preliminary results of the analysis were
published in \cite{my-VLVnT2021,my-ICRC2021}, and the publication
of the final outcome is planned after completion of this thesis and
submission of the gSeaGen paper.
\item Recapitulates all the obtained results and discusses possible improvements
in the methodology and potential further studies.
\end{enumerate}
The existence of the prompt component in the atmospheric lepton flux
has not been experimentally confirmed yet, although it is widely accepted
that this component must be present. So far, only the BUST experiment
has claimed a significant measurement \cite{BUST-Baksan-prompt-measurement}
and the other ones only set experimental upper bounds \cite{IceCube-prompt-muon-measurement,IC-diffuse-neutrino-flux,LVD_prompt_upper-limit}.
Furthermore, so far no neutrino experiment has performed muon bundle
multiplicity (number of muons in a bundle) measurement. Currently,
the KM3NeT detectors (see Chap. \ref{chap:KM3NeT}) are under construction.
However, already with the muon data collected with the early detector
configurations, it is possible to obtain first results and to validate
the methodology for the full detectors. In particular, the reconstruction
of multiplicity should already be possible with the current detector
configurations and may be compared between the simulations and experimental
data. All these remaining open questions constitute an exciting challenge,
which was taken up in this thesis.

\addcontentsline{toc}{chapter}{Motivation}

\chapter{Muons and neutrinos produced by cosmic rays\label{chap:Atm_mu_nu}}

The main background for observing neutrinos originating directly from
distant astrophysical objects are muons and neutrinos created by interactions
of particles coming from the outer space (cosmic rays, see Sec. \ref{sec:Cosmic-Rays})
with the Earth's atmosphere. This chapter discusses the origin and
properties of those particles and characteristics of the atmospheric
muon and neutrino flux.

\begin{figure}[h]
\centering{}\includegraphics[width=10cm]{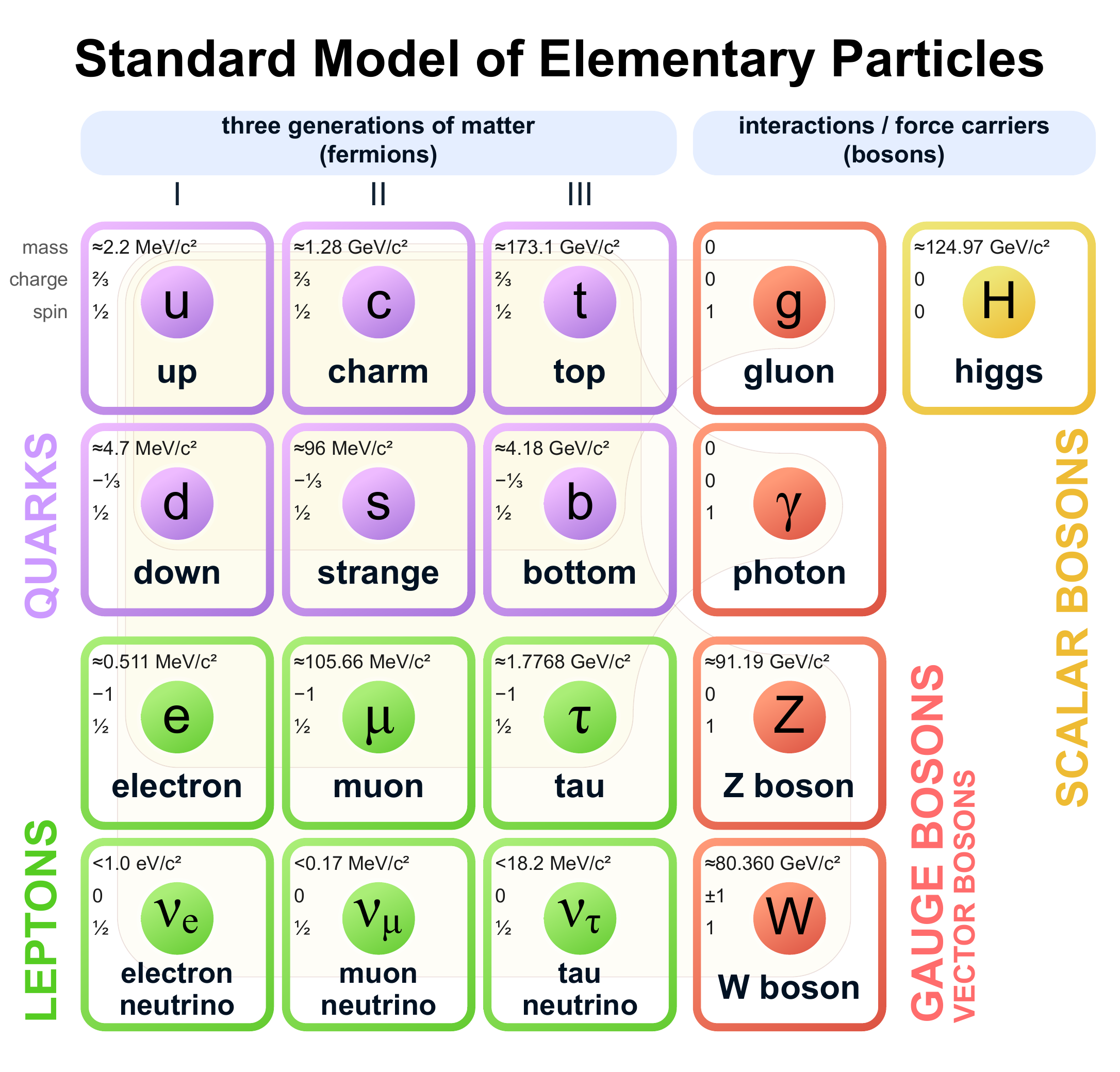}\caption{The Standard Model of particle physics (SM\nomenclature{SM}{standard model of elementary particles}),
with all three neutrinos and their respective charged lepton counterparts
\cite{StandardModelWikipedia}. \label{fig:Standard-Model}}
\end{figure}

\section{Muons\label{subsec:Muons}}

Muons are charged leptons of intermediate mass (see Fig. \ref{fig:Standard-Model}).
They are unstable and decay to electrons and neutrinos: $\mu^{-}\rightarrow e^{-}+\bar{\nu}_{e}+\nu_{\mu}$
(in $\approx100\%$ cases) with a mean lifetime of $\tau_{\mu}=\left(2.1969811\pm0.000022\right)\,$μs
\cite{PDG2022}. Their existence was first confirmed in a cloud chamber
experiment conducted in 1937 by J. C. Street and E. C. Stevenson \cite{MuonDiscovery}
and since then, muon properties have been a subject of extensive studies
\cite{MuonMagneticMomentMeasurementAtFNAL,ANTARES_seasonal,IceCube-prompt-muon-measurement}.

Due to the time dilation effect, muons produced in the atmosphere
with sufficiently high energies may travel tens of kilometres and
reach the surface of the Earth (or even a few km underground or underwater!)
\cite{TimeDilation}. For example, a muon with total energy $E=100\,$TeV
will live on average $\tau=\frac{E}{m_{\mu}c^{2}}\tau_{\mu}\approx2\,$s,
where $c=299\,792\,458\,\frac{\mathrm{m}}{\mathrm{s}}$ is the speed
of light in vacuum and $m_{\mu}=\left(105.6583755\pm0.0000023\right)\,\frac{\mathrm{MeV}}{c^{2}}$
is the muon mass \cite{PDG2022}. This implies that if energy losses
were neglected, the $\mu$ could travel $6\cdot10^{5}\,$km in air!
The real muon ranges are naturally shorter, since energy losses along
such long paths are certainly not negligible. This holds true even
in the air, let alone in more dense media, like water or rock. More
details on the processes through which muons loose energy are given
in the following subsection.

\subsection{Interactions with matter \label{subsec:Interactions-with-matter}}

\begin{figure}[H]
\centering{}\includegraphics[width=12cm]{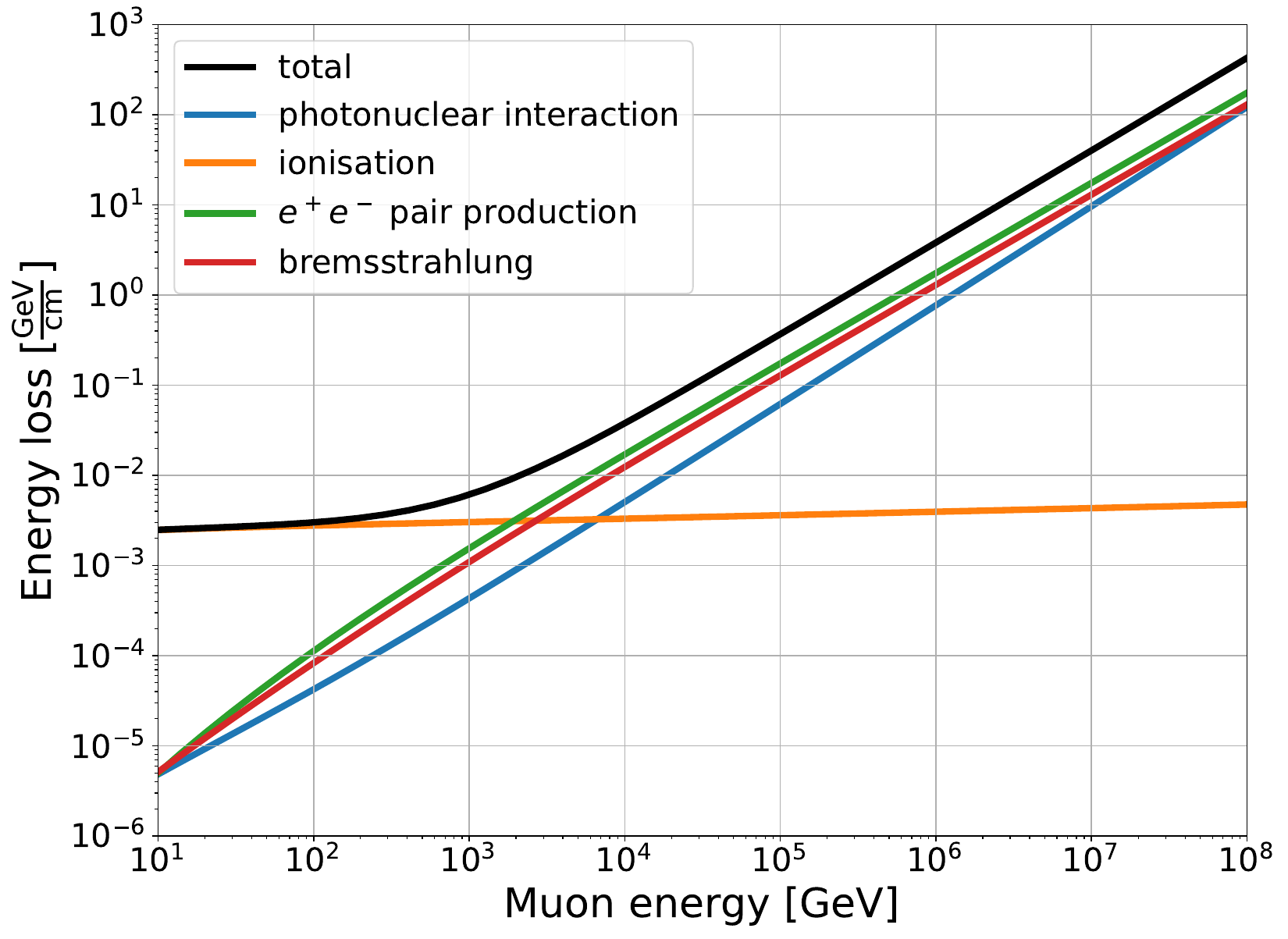}\caption{Continuous muon energy losses in seawater as function of the total
muon energy. The energy losses were computed using default parametrisations
implemented in the PROPOSAL software \cite{PROPOSAL}, which was used
to propagate muons in the gSeaGen code (see Sec. \ref{sec:can} and
\ref{sec:gSeaGen_supplementary_material}). \label{fig:Muon-eloss}}
\end{figure}

Muons carry electric charge and therefore undergo a number of processes
that can lead to a decrease of their energy:
\begin{enumerate}
\item Ionisation – a portion of the electromagnetic field energy of the
muon is transferred to electrons in nearby atoms, freeing some of
them. This results in creation of net positive charge on the atoms
and freely floating negative ions (electrons), hence the name of the
process. Ionisation is dominant at low muon energies, as opposed to
the other three processes (see Fig. \ref{fig:Muon-eloss}).
\item Pair production – creation of electron-positron pairs by exchange
of a photon with an atomic nucleus: $\left(N+\right)\gamma\longrightarrow\left(N+\right)e^{+}+e^{-}$.
The process cannot take place in vacuum.
\item Bremsstrahlung – emission of photons by the muon when deflected by
an electric charge, e.g. of an atomic nucleus.
\item Photonuclear interaction – inelastic scattering of the muon off an
atomic nucleus through virtual photon exchange.
\end{enumerate}

An example with different contributions to muon energy loss in water
is shown in Fig. \ref{fig:Muon-eloss}. The muon ranges in seawater
can be found in Fig. \ref{fig:Muon-range}.

\subsection{Sources of muons\label{subsec:Sources-of-muons}}

Muons are mostly produced in extensive air showers, caused by primary
cosmic ray interactions in the atmosphere (see Sec. \ref{sec:Cosmic-Rays}).
Human-made muon beams can be achieved by colliding hadrons to produce
a beam containing charged pions and kaons. Those mesons subsequently
decay to muons:
\[
\begin{array}{c}
\pi^{+}\longrightarrow\mu^{+}+\nu_{\mu},\\
\pi^{-}\longrightarrow\mu^{-}+\bar{\nu}_{\mu},\\
K^{+}\longrightarrow\mu^{+}+\nu_{\mu},\\
K^{-}\longrightarrow\mu^{-}+\bar{\nu}_{\mu}.
\end{array}
\]

\textcolor{black}{Such muons can be separated from the other collision
products using a shielding (they penetrate much deeper than hadrons)
and focused with a magnetic field. Examples of artificial muon production
are $\mu$ beams created for the measurements of the anomalous magnetic
dipole moment of the muon }\cite{MuonMagneticMomentAtBNL,MuonMagneticMomentMeasurementAtFNAL}
and nucleon structure \cite{COMPASS_experiment,COMPASS_results,Fermilab_muon_beam_nucleon_structure,DIS_muon_nucleus_Fermilab}.
An important application of artificial muon beams will be the planned
muon colliders \cite{Accelerator_roadmap_muon_colliders,muon_colliders_nature,muon_colliders_prd}.
They will take advantage of the properties of muons, such as:
\begin{itemize}
\item much lower energy losses through synchrotron radiation\footnote{Emission of $\gamma$ by relativistic charged particles, when experiencing
acceleration in the direction perpendicular to the direction of motion.}, compared to electrons,
\item being elementary particles, unlike protons, which makes the entire
kinetic energy available in the collision.
\end{itemize}

\section{Neutrinos\label{subsec:Neutrinos}}

Neutrinos are arguably the most elusive particles that have been ever
detected. They only interact via weak and gravitational forces. The
latter was experimentally confirmed in 1998 through the discovery
of neutrino oscillations, which implied that neutrinos must have non-zero
masses (see Sec. \ref{sec:Neutrino-oscillations}). This discovery
was awarded with a shared Nobel Prize in Physics for Takaaki Kajita\footnote{Super-Kamiokande (SK\nomenclature{SK}{Super-Kamiokande}) Collaboration}
and Arthur B. McDonald\footnote{Sudbury Neutrino Observatory (SNO\nomenclature{SNO}{Sudbury neutrino observatory})
Collaboration} in 2015 \cite{Nobel-2015}.

\subsection{Neutrino sources}

All neutrinos are created in weak interactions. Depending on the environment,
in which they are produced, one can distinguish between a number of
neutrino source categories: 
\begin{itemize}
\item Human-made:
\begin{itemize}
\item \textcolor{black}{Nuclear fission reactors}: $\bar{\nu}_{e}$ are
a standard byproduct of nuclear reactor operation. They are created
via the $\beta$ decay: $n\rightarrow p+e^{-}+\bar{\nu}_{e}$. There
was a number of measurements of such neutrinos by KamLAND, Chooz,
Double Chooz, Daya Bay, RENO, DANSS, NEOS, STEREO, SoLid, and CHANDLER
experiments \cite{KamLAND,Chooz,DoubleChooz,DayaBay,RENO,DANSS,NEOS,STEREO,SoLid,CHANDLER}.
The last five are still taking data.
\item \textcolor{black}{Particle accelerators: neutrinos are produced mostly
together with muons as products of pion and kaon decays (see Sec.
}\ref{subsec:Sources-of-muons}\textcolor{black}{). These mesons are
typically produced by colliding accelerated protons on a fixed target.
The most abundant products are muon (anti-)neutrinos, however there
is also a contribution of electron (anti-)neutrinos, which come e.g.
from muon decays or less probable pion or kaon decay channels. A neutrino
beam is produced e.g. at the Japan Proton Accelerator Research Complex
(J-PARC\nomenclature{J-PARC}{Japan proton accelerator research complex})
for the Tokai to Kamioka (T2K\nomenclature{T2K}{Tokai to Kamioka experiment})
experiment \cite{T2K}.}
\end{itemize}
\item \textcolor{black}{Solar: $\nu_{e}$ are created in nuclear reactions
occurring in our star. The leading contribution comes from the proton-proton
chain: }$p+p\rightarrow_{1}^{2}d+e^{+}+\nu_{e}$\textcolor{black}{.
Solar neutrinos have been extensively studied, first by the Homestake
experiment, and then by a number of follow-up projects, including
SAGE, GALLEX, Borexino, SNO, Kamiokande, and SK \cite{SAGE,GALLEX,Borexino_pp_chain_neutrinos,SNO,SuperKamiokande,Homestake,SAGE_2}.}
\item \textcolor{black}{Atmospheric: all three flavours of neutrinos (see
Fig. }\ref{fig:Standard-Model}\textcolor{black}{) are created in
}particle cascades produced by the cosmic rays colliding with atomic
nuclei in the upper atmosphere. Measurements of the atmospheric neutrinos
have been performed by IMB, Kamiokande, SK, ANTARES, IceCube, KM3NeT,
and other experiments \cite{IceCube_seasonal_neutrinos,neutrino-spectra_IC_ANTARES,KM3NeT-LoI-2.0,IMB_and_Kamiokande_atmospheric,SuperKamiokande,Sinopoulou2021_atm_neutrinos}.
\item Cosmic: neutrinos of all flavours directly originating from astrophysical
objects, such as active galactic nuclei (AGNs\nomenclature{AGN}{active galactic nucleus})
and pulsars, or transient events: gamma ray bursts (GRBs\nomenclature{GRB}{gamma ray burst})
and supernovae (SNe\nomenclature{SN}{supernova}). Additionally, $\nu$
coming from the Greitsen-Zatsepin-Kuzmin (GZK) cutoff and relic neutrinos
from the Big Bang nucleosynthesis are counted here. Experiments trying
to observe such sources are KM3NeT, IceCube, SK and Baikal GVD \cite{KM3NeT-LoI-2.0,IC-diffuse-neutrino-flux,Baikal-GVD,SK_diffuse_SN},
with first successful observation of a flaring blazar by IceCube:
\cite{IceCube_blazar,IceCube_blazar_multimessenger}.
\item Geological: neutrinos produced in decays of radioactive isotopes naturally
present in the Earth. Most commonly, these are $\bar{\nu}_{e}$ stemming
from $\beta^{-}$ decay branches of $^{40}\mathrm{K}$, $^{232}\mathrm{Th}$
and $^{238}\mathrm{U}$. Only the last two are detectable by inverse
beta decay (IBD\nomenclature{IBD}{inverse beta decay}): $\bar{\nu}_{e}+p\rightarrow n+e^{+}$,
since $^{40}\mathrm{K}$ decays produce neutrinos below the IBD energy
threshold of $1.8\,$MeV. Measurements of geoneutrinos were performed
by the Borexino and KamLAND experiments \cite{GeoneutrinosBorexino,KamLAND_geoneutrino}.
\end{itemize}

\subsection{Neutrino interactions}

\begin{figure}[H]
\centering{}\includegraphics[width=12cm]{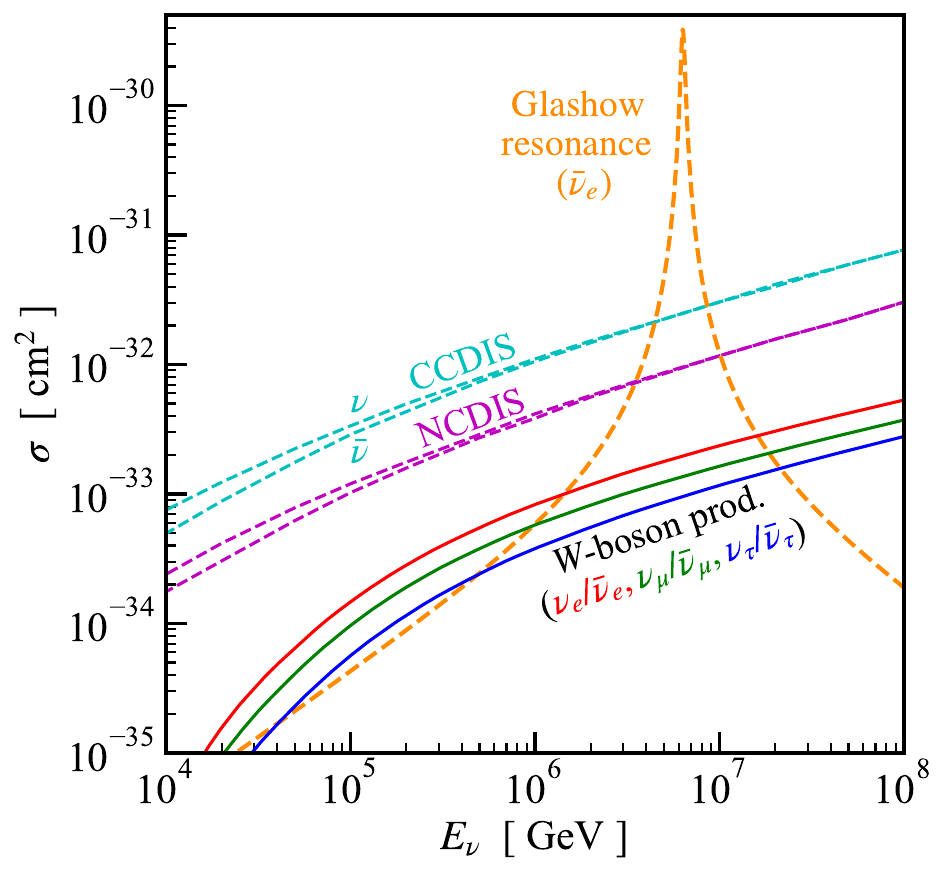}\caption{Cross-sections for $W^{-}$ boson production in (anti-)neutrino scattering
on $^{16}\mathsf{O}$, compared to those for CC and NC deep inelastic
scattering (DIS\nomenclature{DIS}{deep inelastic scattering}), and
the predicted Glashow resonance at $E_{\nu}=6.3\,$PeV \cite{Glashow1960,Glashow_plot}.
The cross-sections are given as a function of the neutrino energy.
The plot was taken from \cite{Glashow_plot}. \label{fig:Cross-sections-with-Glashow}}
\end{figure}

As already mentioned, neutrinos interact only gravitationally and
via weak force. However, gravitational effects are too faint to be
detected. Weak interactions occur through exchange of $W^{\pm}$ or
$Z^{0}$ gauge boson and can be divided into three classes: charged
current (CC\nomenclature{CC}{charged-current weak interaction}),
neutral current (NC\nomenclature{NC}{neutral-current weak interaction}),
and elastic scattering (ES\nomenclature{ES}{elastic scattering})
interactions:

\begin{equation}
\begin{array}{cc}
\mathsf{CC:} & \nu_{l}+N\xrightarrow{W^{\pm}}l+X\\
\mathsf{NC:} & \nu_{l}+N\xrightarrow{Z^{0}}\nu_{l}+X\\
\mathsf{ES:} & \nu_{l}+l\xrightarrow{Z^{0}/W^{\pm}}\nu_{l}+l,
\end{array},
\end{equation}

where $l$ is one of the lepton flavours: $e$, $\mu$ or $\tau$,
$N$ is the target nucleus and $X$ is the resulting hadronic cascade
\cite{neutrino-x-sec}. Interactions of neutrinos with leptons have
lower cross-section (are less probable) than for interactions with
nuclei. However, as can be seen in Fig. \ref{fig:Cross-sections-with-Glashow},
theory predicts that they can become dominant in case of resonant
production of a $W^{-}$ boson, called Glashow resonance:

\begin{equation}
\bar{\nu}_{e}+e^{-}\longrightarrow W^{-},
\end{equation}

at the energy of $E_{\mathsf{Glashow}}\approx\frac{m_{W^{-}}^{2}c^{2}}{2m_{e}}\approx6.3\,$PeV,
where $m_{W^{-}}$ is the $W^{-}$ boson mass and $m_{e}$ is the
mass of an electron \cite{PDG2022,neutrino-x-sec,Glashow_plot}.

Charged particles are produced in all the processes described in this
subsection. They can emit Cherenkov radiation (see Sec. \ref{subsec:Cherenkov-radiation}),
however the event signatures heavily depend on the incident neutrino
flavour and interaction class (CC or NC). Event signatures and reconstruction
for different categories of neutrino-induced events are described
in Sec. \ref{fig:event-topologies} and \ref{sec:reco}.

\subsection{Neutrino oscillations\label{sec:Neutrino-oscillations}}

Neutrino oscillations arise from the interference between the mass
and flavour eigenstates of an operator $\hat{U}_{\alpha n}$:

\begin{equation}
\left|\nu_{\alpha}\right\rangle =\underset{n}{\sum}\hat{U}_{\alpha n}^{\ast}\left|\nu_{n}\right\rangle ,
\end{equation}

\begin{equation}
\left|\nu_{n}\right\rangle =\underset{\alpha}{\sum}\hat{U}_{\alpha n}\left|\nu_{\alpha}\right\rangle ,
\end{equation}

where {*} denotes the complex conjugate, $\left|\nu_{\alpha}\right\rangle $
is an $\alpha=e$, $\mu$, or $\tau$ flavour eigenstate, and $\left|\nu_{n}\right\rangle $
is a mass eigenstate with eigenvalue $m_{n}$ ($n=1$, $2$, $3$).
The operator $\hat{U}_{\alpha n}$ can be represented by the mixing
matrix $U_{\alpha n}$, which is called Pontecorvo-Maki-Nakagawa-Sakata
(PMNS\nomenclature{PMNS matrix}{Pontecorvo-Maki-Nakagawa-Sakata matrix})
matrix. The PMNS matrix is typically parametrized in the following
way:

{\small{}
\begin{equation}
U_{\alpha n}=\begin{bmatrix}U_{e1} & U_{e2} & U_{e3}\\
U_{\mu1} & U_{\mu2} & U_{\mu3}\\
U_{\tau1} & U_{\tau2} & U_{\tau3}
\end{bmatrix}=\underset{\mbox{and\,accelerator}}{\underset{\mbox{atmospheric}}{\underbrace{\begin{bmatrix}1 & 0 & 0\\
0 & c_{23} & s_{23}\\
0 & -s_{23} & c_{23}
\end{bmatrix}}}}\underset{\mbox{and\,reactor}}{\underset{\mbox{accelerator}}{\underbrace{\begin{bmatrix}c_{13} & 0 & s_{13}e^{-i\delta_{\mathsf{CP}}}\\
0 & 1 & 0\\
-s_{13}e^{i\delta_{\mathsf{CP}}} & 0 & c_{13}
\end{bmatrix}}}}\underset{\mbox{reactor}}{\underset{\mbox{solar\,and}}{\underbrace{\begin{bmatrix}c_{12} & s_{12} & 0\\
-s_{12} & c_{12} & 0\\
0 & 0 & 1
\end{bmatrix}}}}\underset{\mbox{are\,Majorana}}{\underset{\mbox{Only\,if\,\ensuremath{\nu}'s}}{\underbrace{\begin{bmatrix}e^{\frac{i\alpha_{1}}{2}} & 0 & 0\\
0 & e^{\frac{i\alpha_{2}}{2}} & 0\\
0 & 0 & 1
\end{bmatrix}}}},
\end{equation}
}{\small\par}

where $s_{kl}\equiv\sin$$\mbox{\ensuremath{\theta}}_{kl}$, $c_{kl}\equiv\cos$$\mbox{\ensuremath{\theta}}_{kl}$,
$\delta_{\mathsf{CP}}$ – CP-violating\nomenclature{CP}{charge conjugation and parity transformation}
phase (charge-parity), and $\alpha_{1}$, $\alpha_{2}$ – Majorana
phases \cite{Majorana,NuFit-4.1-paper}. The latter are expected not
to affect oscillations significantly \cite{Majorana_no_effect_on_osci}.
Atmospheric, accelerator, solar, and reactor refer to the neutrino
sources, described earlier in Sec. \ref{subsec:Neutrinos}, for which
the particular mixing of states is most relevant. Currently, the mixing
parameter values in $U_{\alpha i}$ are still not measured with sufficient
precision to exclude non-unitarity of this $3\times3$ matrix , i.e.
$\left|U_{\alpha i}\right|=U_{\alpha i}^{\dagger}U_{\alpha i}\neq\mathds{1}$
\cite{NuFit_5.2}. Non-unitarity would imply a violation of the postulate
of Quantum Mechanics, stating that only unitary operators can describe
the dynamics of quantum systems \cite{Griffiths2016-Intro-to-QM}.
To preserve unitarity of the $\hat{U}_{\alpha i}$ operator in such
a case, the $3\times3$ matrix would have to be extended by adding
at least one row and column (become $4\times4$ or larger). In other
words, there would have to exist at least one more additional species
of neutrinos, referred to as sterile neutrinos. The term sterile neutrinos
indicates that they do not undergo weak interactions (but do mix with
non-sterile flavours through oscillations).

\selectlanguage{english}%
The probability of conversion from the flavour state $\alpha$ to
flavour state $\beta$ during the propagation is equal to

\begin{equation}
P_{\alpha\rightarrow\beta}\left(t\right)=\left|\left\langle \nu_{\beta}\right.\left|\nu_{\alpha}\left(t\right)\right\rangle \right|^{2},\label{eq:osc_probability-2nu}
\end{equation}

where $t$ is the time and $\left|\nu_{\alpha}\left(t\right)\right\rangle $
is the time-evolved flavour state. In the simplified scenario of 2-flavour
$\nu$ oscillations in vacuum, Eq. \foreignlanguage{british}{\ref{eq:osc_probability-2nu}
can be approximately solved as}

\begin{equation}
P_{\alpha\rightarrow\beta}\left(L\right)=\delta_{\alpha\beta}-\left(2\delta_{\alpha\beta}-1\right)\sin^{2}\left(2\theta_{kl}\right)\sin^{2}\left(\frac{c^{3}}{\hbar}\cdot\frac{\Delta m_{kl}^{2}L}{4E_{\nu}}\right),\label{eq:osc_probability-2nu-final}
\end{equation}

where $\theta_{kl}$ is the mixing angle for the 2-flavour scenario,
$\delta_{\alpha\beta}$ is the Kronecker delta, $\hbar=\frac{h}{2\pi}=1.054571817...\times10^{-34}\,\mathrm{J\cdot\mathrm{s}}$
is the reduced Planck constant, $\Delta m_{kl}^{2}=m_{k}^{2}-m_{l}^{2}$
is the squared mass difference, $E_{\nu}$ is the neutrino energy,
and $L=t\cdot c$ is the distance travelled by the neutrino \cite{PDG2022}.
What is accesible in neutrino experiments is the number of observed
events, which is proportional to $P_{\alpha\rightarrow\beta}\left(L\right)$.
Hence, one may infer from Eq. \foreignlanguage{british}{\ref{eq:osc_probability-2nu-final}
that the accessible neutrino mixing observables are the mixing angles
$\theta_{kl}$ and squared mass differences $\Delta m_{kl}^{2}$,
and that both must be non-zero for oscillations to occur. Their values
have been measured to be: }

\selectlanguage{british}%
\[
\theta_{12}=\left(33.41_{-0.72}^{+0.75}\right)\lyxmathsym{\textdegree},
\]

\[
\Delta m_{21}^{2}\equiv\Delta m_{\mathsf{solar}}^{2}=\left(7.41_{-0.20}^{+0.21}\right)\cdot10^{-5}\,\mathsf{eV^{2}},
\]

and

\[
\theta_{23}=\left(42.2_{-0.9}^{+1.1}\right)\lyxmathsym{\textdegree},
\]

\[
\theta_{13}=\left(8.58\pm0.11\right)\lyxmathsym{\textdegree},
\]

\[
\Delta m_{31}^{2}\approx\Delta m_{32}^{2}\equiv\Delta m_{\mathsf{atmospheric}}^{2}=\left(2.507_{-0.027}^{+0.026}\right)\cdot10^{-3}\,\mathsf{eV^{2}}.
\]
under assumption that $m_{1}<m_{2}<m_{3}$, referred to as normal
neutrino mass ordering (favoured by the global fit) \cite{PDG2022,NuFit_5.2}.
The other possibility is inverted $\nu$ mass ordering: $m_{3}<m_{1}<m_{2}$,
which would result in slightly different fit values and negative sign
of $\Delta m_{\mathsf{atmospheric}}^{2}$. The sign of $\Delta m_{\mathsf{solar}}^{2}$
is already determined to be positive, as a result of the measurements
of the solar $\nu$ spectra from the $^{8}\mathrm{B}$ branch of the
proton-proton chain \cite{Borexino_solar_MSW}.

Among other open questions in the neutrino sector, there are the uncertain
value of the CP-violating phase $\delta_{\mathsf{CP}}$ \cite{T2K_oscillation_parameter_measurement_CP_phase,NuFit-4.1-paper},
the mechanism by which the neutrinos obtain their masses \cite{PDG2022},
whether neutrinos are Majorana\footnote{If neutrinos were Majorana particles, they could act as their own
antiparticles. } or Dirac particles \cite{Majorana,Neutrinoless-double-beta-decay}.

\section{Cosmic rays\label{sec:Cosmic-Rays}}

Cosmic rays (CRs\nomenclature{CR}{cosmic ray}) are charged particles
originating from outer space, with energies sufficient to penetrate
Earth's magnetic field. For the discovery of the CRs during his free
balloon flight measurements, Victor Franz Hess was awarded the Nobel
Prize in Physics in 1936 \cite{Victor-Hess-Nobel-Prize-for-CRs}.
Protons and helium nuclei constitute the majority (99~\%) of the
cosmic rays. There is also a 1~\% contribution to the total flux
from heavier nuclei like lithium, beryllium, boron, carbon, nitrogen,
oxygen, and so on up to iron and from electrons \cite{AMS-02-CR-nuclei-measurement}.
One may also observe CR positrons and antiprotons, however in even
smaller numbers \cite{AMS-02_positron,AMS-02_antiproton,PDG2022}.

\begin{figure}[H]
\centering{}\includegraphics[width=15cm]{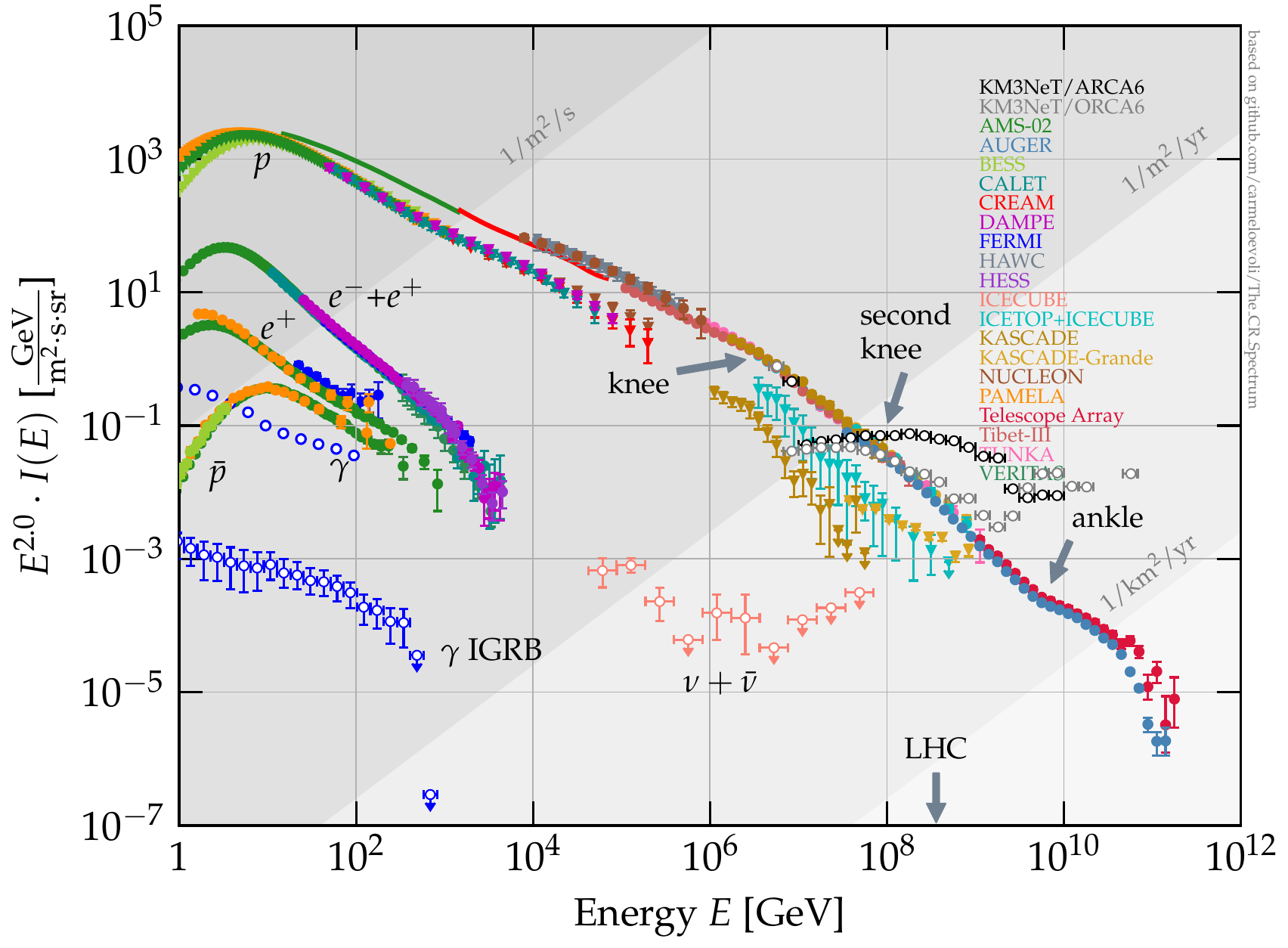}\caption{The cosmic ray all-particle spectrum with data from several experiments
as a function of the primary energy (per nucleus in case of nuclei)
$E$. The figure was adapted from \cite{The_CR_spectrum}. The particle
intensity $I(E)$ is weighted by $E^{2.0}$ to improve the visibility
of spectral features: the \textquoteleft knee\textquoteright{} at $\sim3\cdot10^{6}\,$GeV,
the \textquoteleft second knee\textquoteright{} at $\sim10^{8}\,$GeV,
and the \textquoteleft ankle\textquoteright{} at $\sim5\cdot10^{9}\,$GeV,
separating regions that can be approximated by power laws. The fixed-target
equivalent of the maximal collision energy at the large hadron collider
(LHC\nomenclature{LHC}{large hadron collider}) is marked on the $x$
axis with a grey arrow for comparison. The shaded grey regions indicate
the thresholds for the flux: 1 particle per $\mathrm{m}^{2}\cdot\mathrm{s}$,
per $\mathrm{m}^{2}\cdot\mathrm{yr}$, and $\mathrm{km}^{2}\cdot\mathrm{yr}$.
\cite{PDG2022,HE-CR-with-IceTop,The_CR_spectrum}. \label{fig:The-cosmic-ray-spectrum} }
\end{figure}

Theory predicts an upper limit $E_{\mathsf{GZK}}\simeq50$~EeV on
the energy of the CRs reaching the Earth (see Fig. \ref{fig:The-cosmic-ray-spectrum}),
called Greisen-Zatsepin-Kuzmin cutoff (GZK cutoff\nomenclature{GZK cutoff}{Greisen-Zatsepin-Kuzmin cutoff})
\cite{Greisen1966,Zatsepin-Kuzmin,CR-neutrons-and-GZK}. The main
cause for the GZK cutoff is the strong attenuation of proton energies
in photopion production interactions with the cosmic microwave background
(CMB\nomenclature{CMB}{cosmic microwave background}) photons $\gamma_{\mathsf{CMB}}$.
The inelastic collisions result in a delta resonance, which subsequently
decays producing pions:

\begin{equation}
\begin{array}{c}
p+\gamma_{\mathsf{CMB}}\rightarrow\triangle^{+}\rightarrow n+\pi^{+}\\
p+\gamma_{\mathsf{CMB}}\rightarrow\triangle^{+}\rightarrow p+\pi^{0}
\end{array}.\label{eq:GZK-limit}
\end{equation}

Such process on average leads to 20~\% $E$ loss per collision. This
deceleration mechanism limits the maximal detectable energy for distant
CR sources (further than about 150 million ly) \cite{Zatsepin-Kuzmin,Greisen1966}.
Alternative explanations of the falling slope at the end of the spectrum
in Fig. \ref{fig:The-cosmic-ray-spectrum} include intrinsic CR source
limitations, neutrons being the dominant ultra-high-energy (UHE\nomenclature{UHE}{ultra-high-energy};
with $E>10^{18}\,$eV) CRs instead of protons, Z burst model, top-down
models, violation of Lorentz invariance etc., although these are not
as well-established as GZK cutoff \cite{Z-burst-and-neutrino-clustering-CR,Z-burst-CR,top-down-model-CR,neutrinos-associated-with-top-down-CR,bottom-up-vs-top-down,CR-Lorentz-violation,Lorentz-invariance-violation-CR,CR-and-search-for-Lorentz-Invariance-Violation,UHE-CR-facts-and-myths,UHE-CR-flux-by-Auger,CR-neutrons-and-GZK}.
A popular explanation for the origin of the ankle feature right before
the GZK cutoff (see Fig. \ref{fig:The-cosmic-ray-spectrum}) is the
dominance of the extragalactic over the galactic CR flux at high energies
\cite{CR_spectrum_extragalactic_ankle}.

\subsection{Acceleration mechanisms}

\begin{figure}[H]
\centering{}\subfloat[First order acceleration. \label{fig:First-order-acceleration.}]{\centering{}\includegraphics[width=7cm]{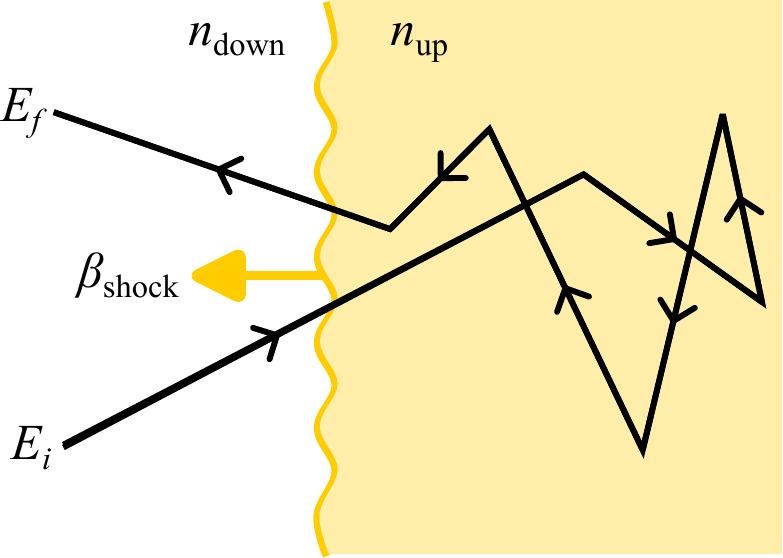}}\hspace*{0.5cm}\subfloat[Second order acceleration. \label{fig:Second-order-acceleration.}]{\centering{}\includegraphics[width=7cm]{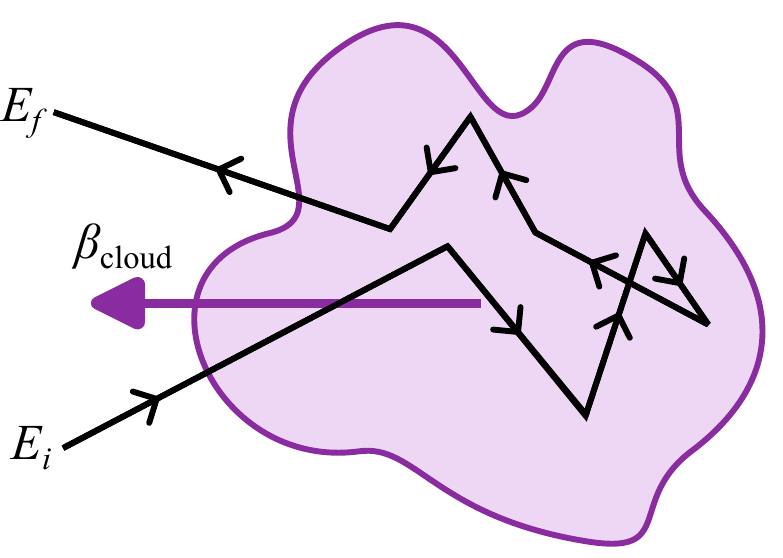}}\caption{Sketches of two Fermi acceleration mechanisms. $E_{i}$ is the initial
and $E_{f}$ is final energy of the accelerated particle. $\beta_{\mathsf{shock}}$and
$\beta_{\mathsf{cloud}}$ are velocities (relative to the speed of
light in vacuum) of the plasma shock front and magnetic cloud respectively.
The particle densities are denoted by $n_{\mathsf{down}}$ for the
downstream and $n_{\mathsf{up}}$ for the upstream plasma \cite{Fermi_acceleration_sketches}.
\label{fig:Fermi-acceleration}}
\end{figure}

It remains unclear how CRs acquire such high energies as in Fig. \ref{fig:The-cosmic-ray-spectrum}.
There are two categories of models that try to explain it: 
\begin{enumerate}
\item Top-down models assume existence of extremely heavy remnant particles
from the Big Bang, that deposit most of their energy into kinetic
energy when decaying into SM particles. Such particles would need
to have lifetimes on the order of the age of the Universe. Those models
are disfavoured by cosmic photon flux measurements by experiments
like Pierre Auger Observatory \cite{UHE-CR-flux-by-Auger,Auger-photons-top-down-CR,top-down-constraints-UHECR-Auger}.
\item Bottom-up theories predict processes taking place in the present Universe
that could accelerate particles to extreme energies. A popular example
of such a process is Fermi acceleration, named after Enrico Fermi
who proposed it in 1949 \cite{Fermi-acceleration}. There are in fact
two kinds: first and second order Fermi acceleration. They are both
sketched in Fig. \ref{fig:Fermi-acceleration}.
\end{enumerate}

In Fermi acceleration, the order of the process is associated with
the efficiency of acceleration:

\begin{equation}
\begin{array}{cc}
\mathsf{First\,order:} & \frac{\left\langle \Delta E\right\rangle }{E}\simeq\frac{4}{3}\beta_{\mathsf{shock}}\\
\mathsf{Second\,order:} & \frac{\left\langle \Delta E\right\rangle }{E}\simeq\frac{4}{3}\beta_{\mathsf{cloud}}^{2}
\end{array}.\label{eq:Fermi-acceleration}
\end{equation}

The one originally proposed by Fermi, where relativistic particles
gain energy by collisions with interstellar clouds is the second order
acceleration. It results in a power law spectrum, but fails to explain
the big abundance of UHE CRs (interstellar clouds are too thin) and
the observed spectral index of about $\gamma=-2.7$ \cite{Fermi-acceleration,UHE-CR-facts-and-myths}.
The first order Fermi acceleration occurs through collisions of relativistic
CRs with SN or AGN shockwaves, which is a more efficient mechanism
($\left|\beta\right|<1$; see Eq. \ref{eq:Fermi-acceleration}). When
moving in a strong magnetic field, CR particles are deflected and
cross the shockwave front repeatedly, gaining energy. This allows
to produce a power law behaviour with a spectral index of $\gamma=-2.0$.
When including CR propagation and acceleration inefficiencies, it
is compatible with the experimentally observed spectrum. First order
Fermi acceleration allows to produce energies up to

\begin{equation}
E_{\mathsf{max}}\simeq Z\cdot\frac{B}{\mathsf{\textrm{μ}G}}\cdot\frac{R_{\mathsf{src}}}{\mathsf{kpc}}\cdot10^{9}\,\mathsf{GeV},\label{eq:1st-Fermi-Emax}
\end{equation}

where $Z$ is the atomic number, $B$ – magnetic field amplitude and
$R_{\mathsf{src}}$ – characteristic size of the source \cite{UHE-CR-facts-and-myths}.

\subsection{Sources}

\begin{figure}[H]
\centering{}\includegraphics[width=13cm]{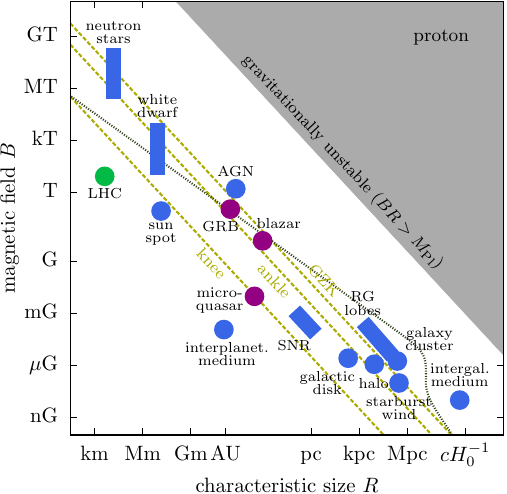}\caption{The so-called Hillas plot. It visualises the energy limitation from
Eq. \ref{eq:1st-Fermi-Emax}. CR source candidates are coloured blue.
Purple circles indicate jet-frame parameters for some object types.
Dashed yellow lines are the lower limits for proton accelerators at
knee, ankle and the GZK cutoff. The dotted grey line is the upper
limit from synchrotron losses and interactions with the CMB. The grey
area corresponds to gravitationally unstable extremely large magnetic
field environments \cite{Fermi-acceleration-sketches,Hillas-plot,Fermi_acceleration_sketches}.
\label{fig:Hillas plot}}
\end{figure}

Since CRs are electrically charged particles, they are deflected by
the magnetic fields they traverse. This makes it nearly impossible
to track a particular CR back to its source. It is only possible for
the most energetic CRs, however at the highest energies insufficient
statistics (see Fig. \ref{fig:The-cosmic-ray-spectrum}) and lack
of knowledge about the CR flux composition become major problems.
For these reasons UHE CR sources are so far unknown. There are however
certain classes of objects, suspected of producing the UHE CRs as
shown in Fig. \ref{fig:Hillas plot}. 

\subsection{Extensive Air Showers\label{sec:Extensive-Air-Showers-EAS}}

\begin{figure}[H]
\centering{}\includegraphics[width=15cm]{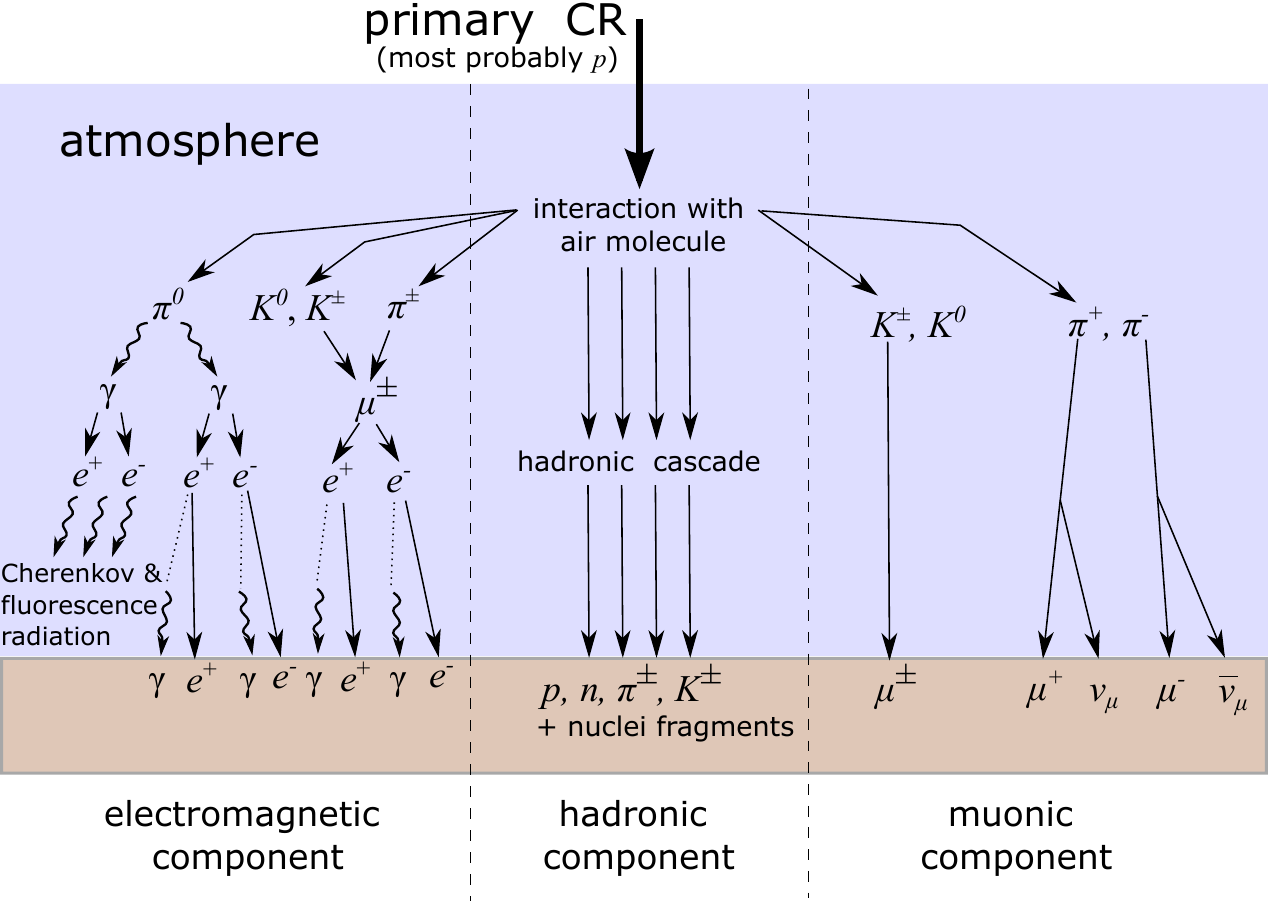}\caption{Structure of an EAS with a separation into the electromagnetic, muonic
and hadronic parts. \label{fig:EAS_schematic}}
\end{figure}

\begin{figure}[H]
\centering{}\subfloat[Second and third hadronic interaction of a vertical 100~GeV proton
shower.]{\begin{centering}
\includegraphics[width=4cm]{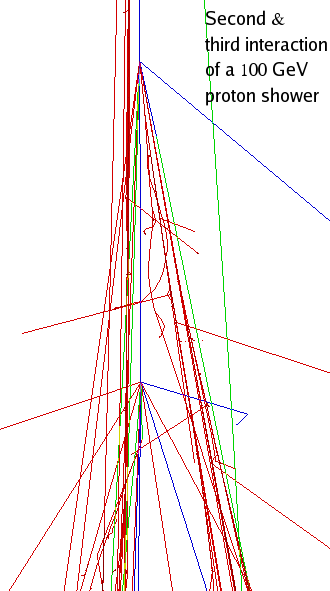}
\par\end{centering}
}~~\subfloat[Compton scattering.]{\begin{centering}
\includegraphics[width=3cm]{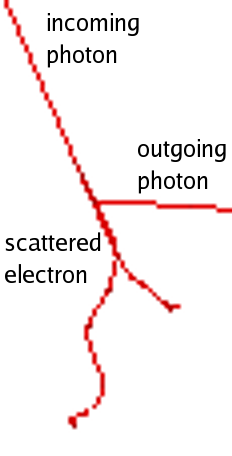}
\par\end{centering}
}~~\subfloat[Deflection in the Earth's magnetic field.]{\begin{centering}
\includegraphics[width=3cm]{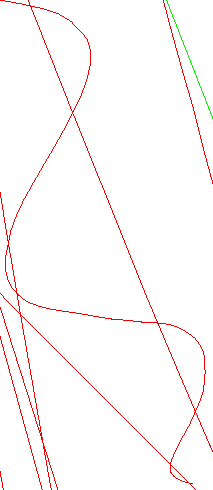}
\par\end{centering}
}~~\subfloat[Bremsstrahlung.]{\begin{centering}
\includegraphics[width=3cm]{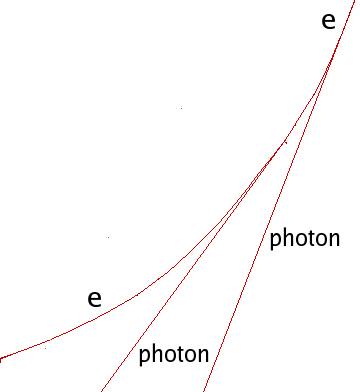}
\par\end{centering}
}\caption{Some processes cut out from the global shower pictures created using
the CORSIKA event generator (see Sec. \ref{sec:CORSIKA}), taken from
\cite{CORSIKA_shower_images}. Electrons and photons are coloured
\textcolor{red}{red}, muons: \textcolor{green}{green} and hadrons:
\textcolor{blue}{blue}. Colour scale is logarithmic and colours are
mixed in case of overlapping tracks (e.g.\textcolor{red}{{} red} + \textcolor{green}{green}
track $\rightarrow$ \textcolor{orange}{yellow} pixel), i.e. dark
color indicates a high track density. \cite{CORSIKA_shower_images}.
\label{fig:selected-processes-in-showers}}
\end{figure}

Extensive air showers (EAS\nomenclature{EAS}{extensive air shower}),
also referred to as simply air showers, are particle cascades induced
by CRs in the upper atmosphere. They can extend through many kilometres,
reaching the Earth's surface and deeper. Fig. \ref{fig:EAS_schematic}
depicts what components a typical EAS consists of. Naturally, air
showers vary a lot in produced secondary particles, depending on the
primary CR type, energy, Earth's magnetic field, atmosphere properties,
first interaction height, and so on. In general, showers caused by
heavier primaries are wider and consist of a larger number of produced
particles. Hence, it may be possible to distinguish between the showers
initiated by the lightest (proton) and heaviest (iron) primary nuclei
\cite{CORSIKA_shower_images}. A bunch of selected processes present
in air showers is displayed in Fig. \ref{fig:selected-processes-in-showers}.

In the case of EAS started by more energetic primaries, the number
of particles produced even in a single shower can be enormous (see
Sec. \ref{sec:Multiplicity}). Despite this, secondary particles tend
to be aligned close to the shower axis (original primary direction),
due to large Lorentz boost in the forward direction. 

In the context of this work, muons are the most important products
of EAS.

Some of the characteristics describing a bundle include:
\begin{enumerate}
\item Muon multiplicity $N_{\mu}$: number of muons in the bundle.
\item Bundle energy $E_{\mathsf{bundle}}$: sum of energies of individual
muons ($E_{\mathsf{bundle}}=\underset{i}{\sum}E_{\mu_{i}}$).
\item Bundle direction $\cos\theta_{\mathsf{zenith}}^{\mu\,\mathsf{bundle}}$:
overall direction of the muon bundle in terms of the cosine of its
zenith angle. Here, different approaches are possible (although all
produce very similar results). One may consider:
\begin{enumerate}
\item arithmetic average of all muon directions: $\cos\theta_{\mathsf{zenith}}^{\mu\,\mathsf{bundle}}=\frac{\underset{i}{\sum}\cos\theta_{\mathsf{zenith}}^{\mu_{i}}}{N_{\mu}}$,
\item direction of the most energetic muon in the bundle,
\item energy-weighted average of all muon directions: $\cos\theta_{\mathsf{bundle}}=\frac{\underset{i}{\sum}E_{\mu_{i}}\cdot\cos\theta_{\mathsf{zenith}}^{\mu_{i}}}{\underset{i}{\sum}E_{\mu_{i}}}=\frac{\underset{i}{\sum}E_{\mu_{i}}\cdot\cos\theta_{\mathsf{zenith}}^{\mu_{i}}}{E_{\mathsf{bundle}}}$.
\end{enumerate}
The option a) was used by default in this work.
\item Bundle width: maximal lateral spread of the bundle.
\item Bundle arrival time spread: difference between the arrival times of
the first and the last muon at the sensitive volume of the detector. 
\item Primary cosmic ray direction $\cos\theta_{\mathsf{prim}}$. For primaries
with high energies it is almost identical to $\cos\theta_{\mathsf{bundle}}$.
\item Primary cosmic ray energy (either per nucleus: $E_{\mathsf{prim}}$
or total: $A\cdot E_{\mathsf{prim}}$, where $A$ is the atomic mass
number).
\end{enumerate}
Out of those listed variables, reconstruction of 1., 2., and 7. is
discussed in Chap. \ref{chap:muon-bundle-reco}, and a reconstruction
of 3. with official KM3NeT software (see Chap. \ref{chap:KM3NeT}
and Sec. \ref{sec:reco}) is used in Chap. \ref{chap:Muon-rate-measurement}.
The multiplicity and bundle energy results are also put to use in
Chap. \ref{chap:prompt_ana}.

\section{Atmospheric flux \label{sec:Atmospheric-flux}}

Secondary particles produced in EAS are usually subdivided into two
categories, based on their lifetimes. The long-lived ($t\apprge10^{-8}\,$s)
ones contribute to the conventional and the short-lived ($t\apprle10^{-13}\,$s)
ones to the prompt atmospheric flux. Both components are described
in this section. From the perspective of KM3NeT detectors (see Chap.
\ref{chap:KM3NeT}), only muons and neutrinos are a relevant signal.
In this work, the focus is on the atmospheric muon flux.

\subsection{Conventional}

Conventional flux is the dominant component of the air showers. It
can be subdivided into the following categories:
\begin{itemize}
\item $\pi^{0}$ producing electromagnetic (EM\nomenclature{EM}{electromagnetic})
cascades through consecutive bremsstrahlung and $e^{-}e^{+}$ pair
creation. This part is in fact not relevant for this work, as EM cascades
do not reach the KM3NeT detectors .
\item Light charged mesons ($\pi^{\pm}$ and $K^{\pm}$), producing hadronic
cascades through scattering processes, and $\nu$ and $\mu$ through
decays.
\end{itemize}
Conventional flux is rather well understood and measured \cite{IceCube-prompt-muon-measurement,IC-diffuse-neutrino-flux,IC-electron-and-tau-neutrino-flux}.

\subsection{Prompt}

\begin{figure}[H]
\begin{centering}
\subfloat[\foreignlanguage{english}{Different components of the atmospheric muon flux and some of the
key contributing mother particles. \label{fig:Different-components-of-atm-mu-flux}}]{\begin{centering}
\includegraphics[width=16cm]{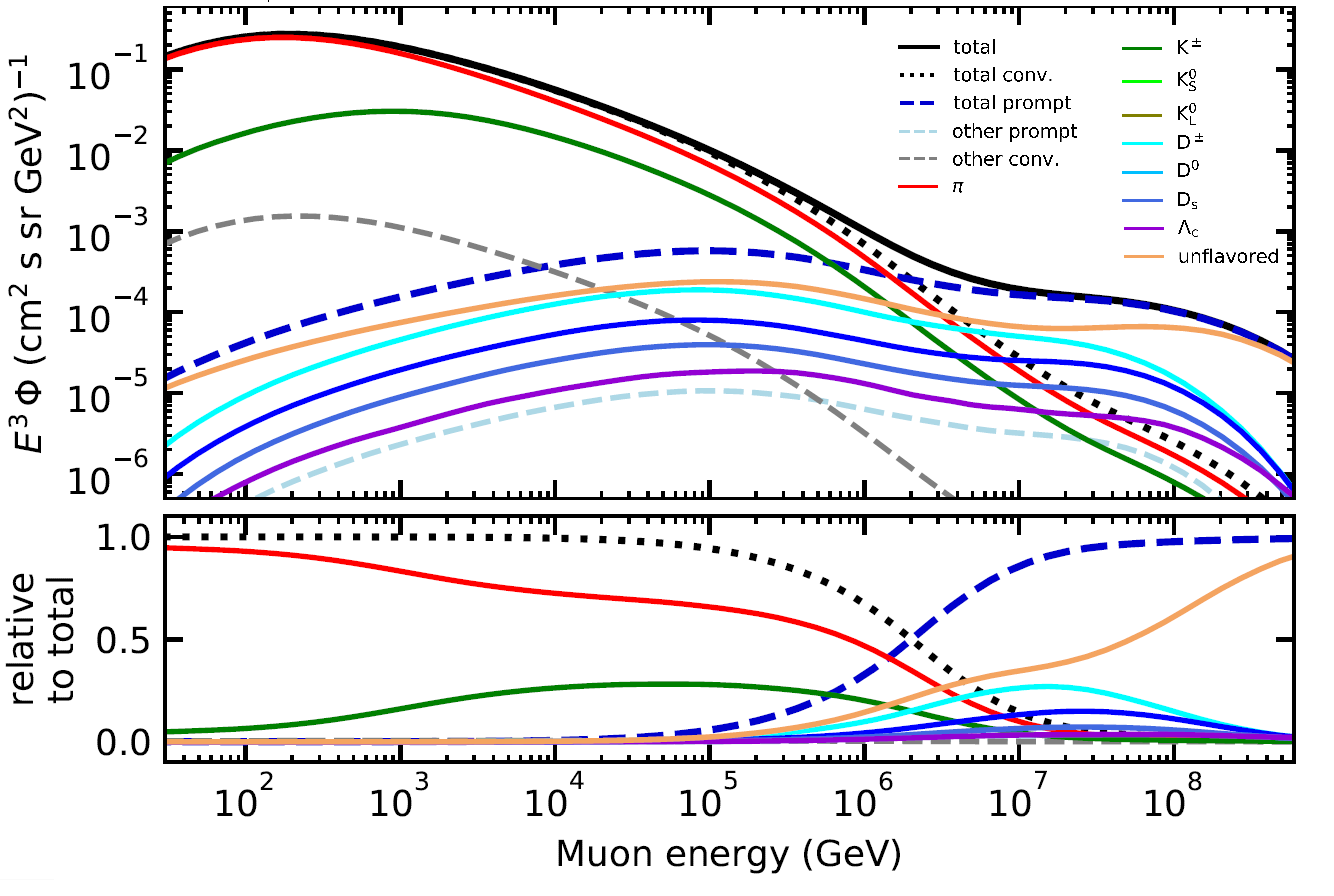}
\par\end{centering}
}
\par\end{centering}
\begin{centering}
\subfloat[Individual contributions to the unflavoured component from Fig. \ref{fig:Different-components-of-atm-mu-flux}.]{\begin{centering}
\includegraphics[width=14cm]{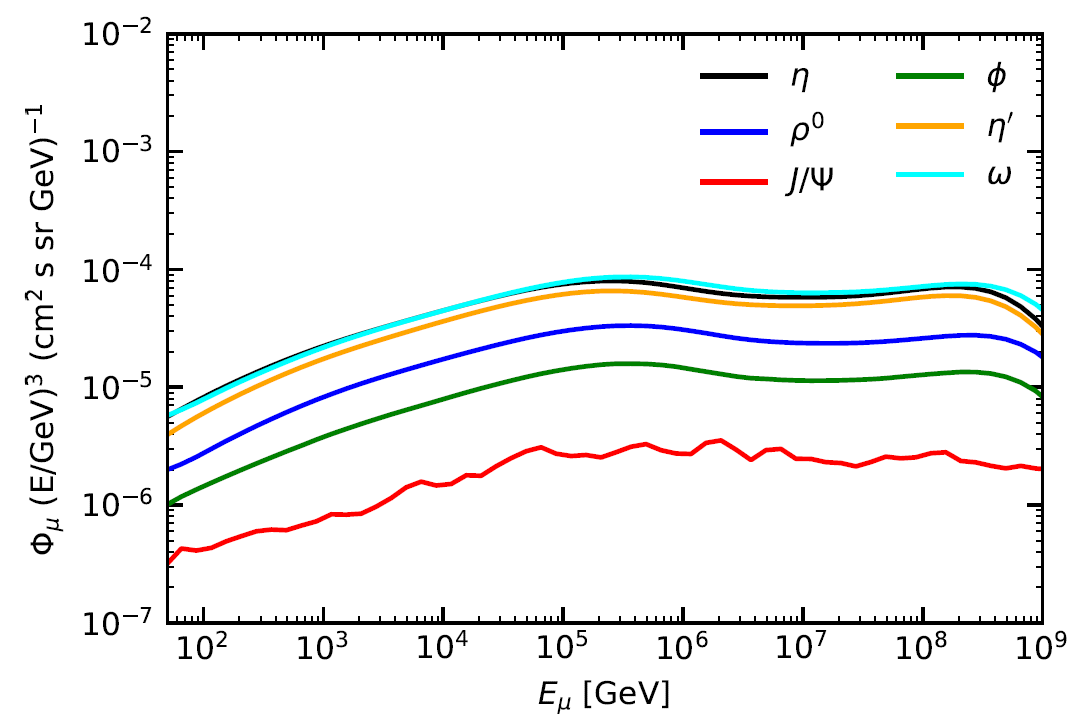}
\par\end{centering}
}
\par\end{centering}
\caption{\foreignlanguage{english}{Contributions to the atmospheric muon flux, taken from \cite{conv-prompt-calculation}.
The plots were made using H3a CR flux model (see Sec. \ref{subsec:Cosmic-Ray-flux-models}),
PYTHIA 8 \cite{Pythia} for decays, and SIBYLL 2.3c as hadronic interaction
model (see Sec. \ref{subsec:Hadronic-interaction-models}). The curves
were plotted for a fixed zenith angle of $\theta_{\mathsf{zenith}}=60\lyxmathsym{\protect\textdegree}$
\cite{SIBYLL}. The \foreignlanguage{british}{unflavoured} component
consists solely of hadrons with up and down valence quarks, as opposed
to the $D$ mesons (containing charm quarks) and kaons (strange quarks).
\label{fig:flux-contributions-SIBYLL}}}
\end{figure}

Prompt flux originates in semileptonic decays of heavy hadrons and
light vector mesons into muons and neutrinos. The contribution of
different parent particles to conventional and prompt components of
muon energy spectrum is shown in Fig. \ref{fig:flux-contributions-SIBYLL}.
The crossing point, where the prompt flux starts to dominate over
the conventional flux, is located around 1~PeV, which is well within
the sensitivity of both KM3NeT detectors. Although the conventional
flux is quite well understood and measured, no neutrino telescope
has yet observed the prompt atmospheric muon or neutrino flux, there
are only upper limits available from IceCube \cite{IceCube-prompt-muon-measurement,IC-diffuse-neutrino-flux,IC-electron-and-tau-neutrino-flux,IC-cascade-data}.
ANTARES, KM3NET, and Baikal-GVD have not tackled the problem so far
\cite{ANTARES,KM3NeT,Baikal-GVD}. In order to predict the potential
of KM3NeT to observe the prompt component, a CORSIKA MC simulation
could be used (see Sec. \ref{sec:CORSIKA}) \cite{CORSIKA}. The ansatz
taken in this work was to test the conventional-only hypothesis against
the combination of prompt and conventional contributions, to evaluate
the expected significance. The measurement in such a scenario would
boil down to substituting the latter with the experimental data. The
analysis is described in detail in Chap. \ref{chap:prompt_ana}.

There are further subtleties concerning the prompt flux modelling,
e.g. the so-called ``intrinsic charm'' \cite{Intrinsic_charm_original_paper_Brodsky1980},
which is a contribution from virtual $c\bar{c}$ pairs inside the
proton. Although virtual, such quark pairs can in principle hadronise
into measurable particles. This effect is included in the SIBYLL 2.3d
model (see Sec. \ref{subsec:Hadronic-interaction-models}), used as
the default high-energy hadronic interaction model for this work \cite{SIBYLL,SIBYLL-2.3d}.

Examples of some relevant leptonic decay channels of short-lived hadrons
in which muons are created directly are listed in Tab. \ref{tab:Selected-leptonic-decays}.

\begin{table}[H]
\caption{Selected leptonic decays of short-lived hadrons resulting in direct
muon production \cite{PDG2022}. \label{tab:Selected-leptonic-decays}}

\centering{}%
\begin{tabular}{|c|c|}
\hline 
Decay channel & Branching fraction $\frac{\Gamma_{i}}{\Gamma}$ {[}\%{]}\tabularnewline
\hline 
\hline 
$J/\psi\rightarrow\mu^{+}+\mu^{-}$ & $5.961\pm0.033$\tabularnewline
\hline 
$D^{0}\rightarrow K^{-}+\mu^{+}+\nu_{\mu}$ & $3.41\pm0.04$\tabularnewline
\hline 
$D_{s}^{+}\rightarrow\mu^{+}+\nu_{\mu}$ & $\left(5.43\pm0.15\right)\cdot10^{-3}$\tabularnewline
\hline 
$D^{+}\rightarrow\mu^{+}+\nu_{\mu}$ & $\left(3.74\pm0.17\right)\cdot10^{-4}$\tabularnewline
\hline 
$\eta\rightarrow\mu^{+}+\mu^{-}+\gamma$ & $\left(3.1\pm0.4\right)\cdot10^{-4}$\tabularnewline
\hline 
\end{tabular}
\end{table}

The full list of mesons and baryons available in the CORSIKA simulation
(see Sec. \ref{sec:CORSIKA}) used in this work, which may contribute
to the prompt muon flux through their decays is compiled in Tab. \ref{sec:Classification-of-parent-particles-by-lifetime}.
The contributions of individual mother particles to the overall flux
can be found in Fig. \ref{fig:parent-contrib-mother}.

Both the conventional and prompt flux are a background for the astrophysical
neutrino flux searches, so far only observed by the IceCube experiment
\cite{IC-diffuse-neutrino-flux,IC-electron-and-tau-neutrino-flux}.
Measuring the prompt flux is of utter importance for the future neutrino
astronomy with KM3NeT/ARCA and KM3NeT/ORCA, and in collaboration with
other experiments. 

\chapter{KM3NeT Experiment \label{chap:KM3NeT}}

KM3NeT is a network of neutrino telescopes located at the bottom of
the Mediterranean Sea. The two sites at which the detectors are currently
installed as well as the involved research institutes are shown in
Fig. \ref{fig:GNN_map}.

\section{Neutrino telescopes in general\label{sec:neutrino-telescopes}}

\begin{figure}[H]
\centering{}\includegraphics[width=16cm]{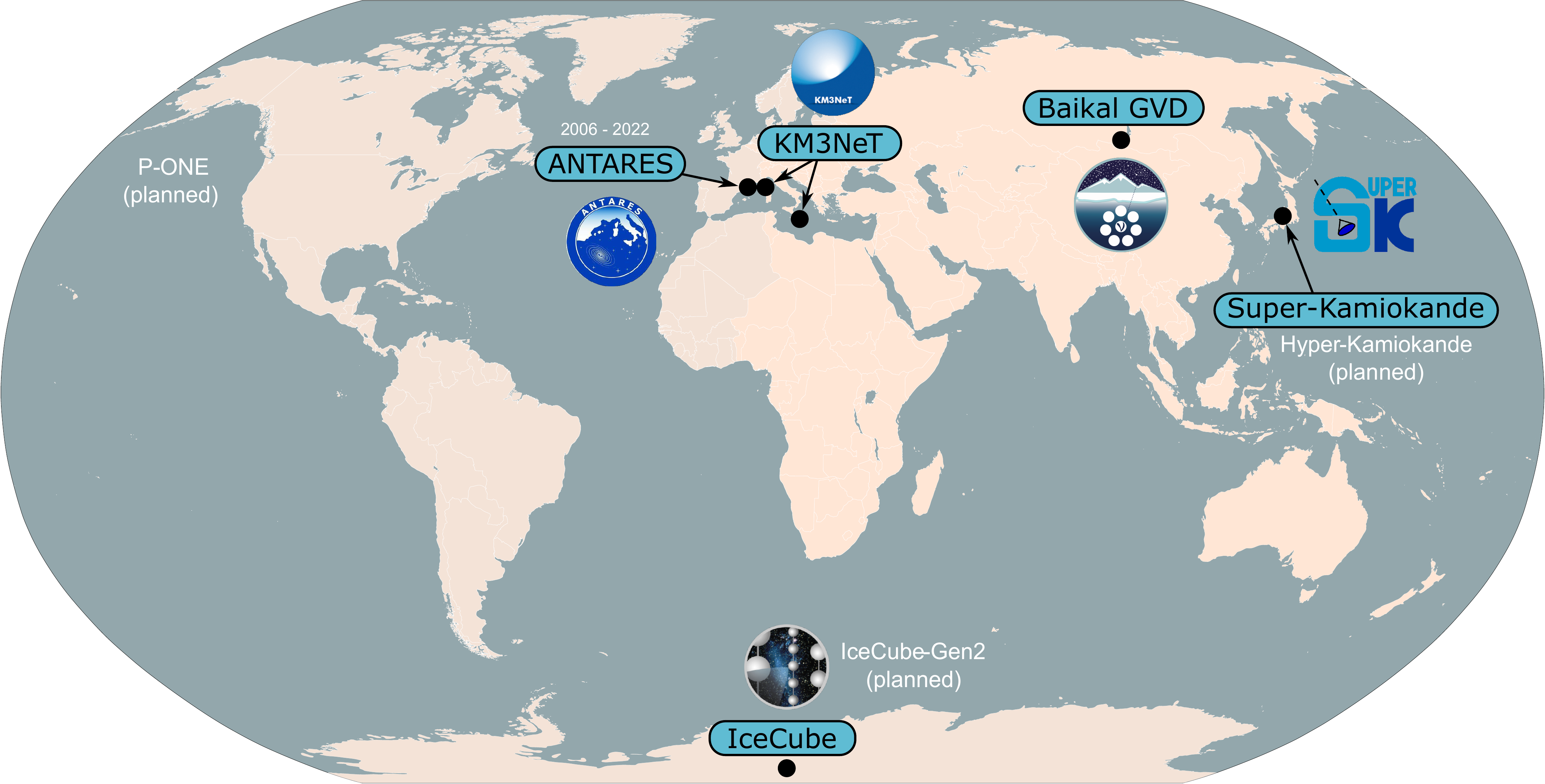}\caption{World map showing the currently operating water- and ice-based neutrino
telescopes. Additionally, the ANTARES detector is mentioned, as it
has finished its duty only very recently in February 2022. In grey,
some planned experiments and upgrades of existing ones are indicated.
\label{fig:GNN_map}}
\end{figure}

KM3NeT is neither the only nor the first water-based neutrino telescope
in existence. In Fig. \ref{fig:GNN_map}, currently operating experiments
are shown, together with recently dismantled ANTARES \cite{ANTARES}.
This does not include the earlier projects, like e.g. AMANDA, the
predecessor of IceCube at the South Pole \cite{AMANDA=000026IC}.
Except for IceCube, which uses Antarctic ice as the detector medium,
all other neutrino telescopes use water in liquid state as the medium.
However, there are still important differences to be noted. Baikal
GVD is located in the freshwater of the lake Baikal in Russia \cite{Baikal-GVD}.
Super Kamiokande (SK) uses artificially purified water, additionally
loaded with $\mathrm{Gd_{2}\left(SO_{4}\right)_{3}\cdot8H_{2}O}$
(gadolinium sulphate octahydrate) in 2020, enclosed inside a stainless
steel tank \cite{SK_water_gd}. The medium for ANTARES and KM3NeT
are the salty waters of the Mediterranean Sea \cite{ANTARES,KM3NeT}.
Apart from the use of solid or liquid water as medium, all discussed
neutrino telescopes share a common detection principle, based on the
Cherenkov light emission in water.

\subsection{Cherenkov radiation \label{subsec:Cherenkov-radiation}}

\begin{figure}[H]
\centering{}\includegraphics[width=8cm]{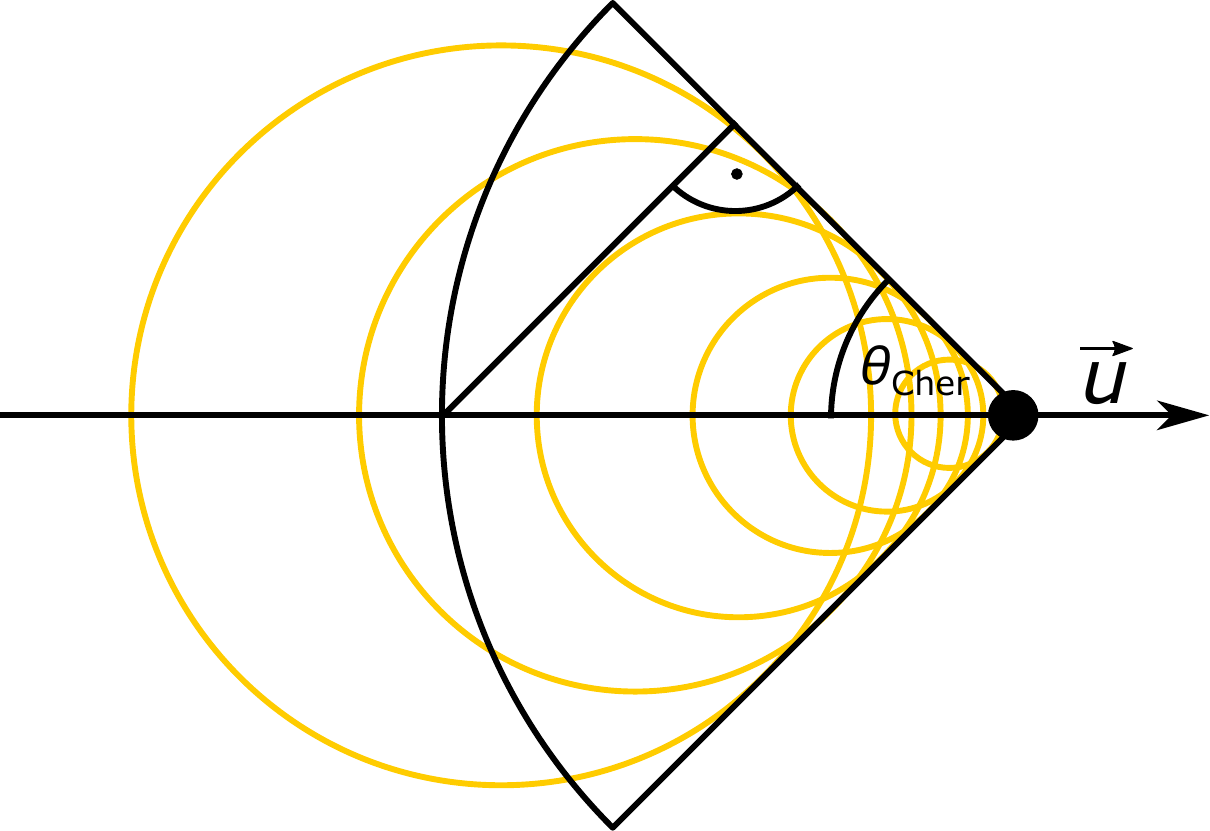}\caption{Schematic geometry of Cherenkov radiation. \textsf{\textit{$\vec{u}$}}
is the velocity of the particle. Emitted light is coloured yellow,
with only a few rings drawn for readability. \label{fig:Cherenkov-cone}}
\end{figure}

Cherenkov radiation is a well-known phenomenon in particle physics.
The effect was named after Pavel Alekseyevich Cherenkov, who shared
the Nobel prize in physics with Ilya Frank and Igor Tamm in 1958 \cite{Nobel_lectures_1942-1962}.
It occurs when an electrically charged particle traverses a medium
with a velocity exceeding the speed of light in that medium. The passage
of the charged particle excites atoms to higher energy levels. Upon
returning to their ground states, the atoms emit photons uniformly
in all directions. However, since the particle is moving faster than
the photons themselves, a conical shock front, analogous to a supersonic
wave behind a jet plane, is formed (see Fig. \ref{fig:Cherenkov-cone}).

The opening angle of the cone $\theta_{\mathsf{Cher}}$ can be computed
by simple trigonometry from the paths the light and the charged particle
will travel in a fixed time:

\begin{equation}
\cos\left(\theta_{\mathsf{Cher}}\right)=\frac{1}{\beta n},\label{eq:cherenkov}
\end{equation}

where $n$ is the refractive index of the medium and $\beta=\frac{u}{c}$
is the velocity of the charged particle, relative to the speed of
light in vacuum $c$ \cite{Cherenkov-radiation}.

The threshold energy above which the effect occurs may be calculated
by inserting Eq. \ref{eq:cherenkov} into the Einstein's energy-mass
relation $E=mc^{2}$ \cite{Griffiths_Intro_to_Electrodynamics_special_relativity}
and setting $\cos\left(\theta_{\mathsf{Cher}}\right)=1$ (to obtain
the extreme case):

\begin{equation}
E_{\mathsf{threshold}}=\frac{m_{0}c^{2}}{\sqrt{1-\frac{1}{n^{2}}}},
\end{equation}

where $m_{0}$ is the rest mass of the particle. For seawater with
a refractive index of about $n_{\mathsf{sea}}\approx1.383$ \cite{ANTARES-water-optical-properties},
the following threshold energies apply:

\begin{table}[H]
\begin{centering}
\caption{Summary of Cherenkov light emission threshold values computed for
different particles in seawater ($n_{\mathsf{sea}}\approx1.383$).
\label{tab:Table-summarizing-Cherenkov}}
\par\end{centering}
\centering{}%
\begin{tabular}{|c|c|c|}
\hline 
Particle & Mass & $E_{\mathsf{threshold}}$\tabularnewline
\hline 
\hline 
\multirow{2}{*}{$e^{\pm}$} & \multirow{2}{*}{511.00~$\frac{\mathsf{keV}}{c^{2}}$} & \multirow{2}{*}{739.74~keV}\tabularnewline
 &  & \tabularnewline
\hline 
\multirow{2}{*}{$\mu^{\pm}$} & \multirow{2}{*}{105.66~$\frac{\mathsf{MeV}}{c^{2}}$} & \multirow{2}{*}{152.96~MeV}\tabularnewline
 &  & \tabularnewline
\hline 
\multirow{2}{*}{$\tau^{\pm}$} & \multirow{2}{*}{1.78~$\frac{\mathsf{GeV}}{c^{2}}$} & \multirow{2}{*}{2.57~GeV}\tabularnewline
 &  & \tabularnewline
\hline 
\multirow{2}{*}{$\pi^{\pm}$} & \multirow{2}{*}{139.57~$\frac{\mathsf{MeV}}{c^{2}}$} & \multirow{2}{*}{202.05~MeV}\tabularnewline
 &  & \tabularnewline
\hline 
\multirow{2}{*}{$K^{\pm}$} & \multirow{2}{*}{493.68~$\frac{\mathsf{MeV}}{c^{2}}$} & \multirow{2}{*}{714.67~MeV}\tabularnewline
 &  & \tabularnewline
\hline 
\multirow{2}{*}{$p$} & \multirow{2}{*}{938.27~$\frac{\mathsf{MeV}}{c^{2}}$} & \multirow{2}{*}{1.35~GeV}\tabularnewline
 &  & \tabularnewline
\hline 
\end{tabular}
\end{table}

As can be seen from Tab. \ref{tab:Table-summarizing-Cherenkov}, for
high-energy neutrino astronomy applications, the Cherenkov threshold
is not an issue. For the low-energy KM3NeT telescope ORCA (see Sec.
\ref{sec:ORCA}), the minimum reconstructable energy will be in the
range 1-5~GeV \cite{KM3NeT-LoI-2.0}. Knowing the refractive index,
one may also compute the typical opening angle $\theta_{\mathsf{Cher}}$
using Eq. \ref{eq:cherenkov}, since $\beta\approx1$ in the relevant
energy range. For seawater this value is approximately equal $43.95\lyxmathsym{\textdegree}$.

The amount of Cherenkov light that will be emitted can be estimated
with the Frank-Tamm formula:

\begin{equation}
\frac{d^{2}N}{d\lambda dx}=\frac{2\pi\alpha z^{2}}{\lambda^{2}}\sin^{2}\left(\theta_{\mathsf{Cher}}\right),\label{eq:Frank-Tamm-formula}
\end{equation}

where $\alpha\approx\frac{1}{137}$ is the fine structure constant,
$z$ is the atomic charge number (for a muon, $z=1$), and $\lambda$
is the wavelength of the Cherenkov radiation \cite{PDG2022}. For
$\text{\ensuremath{\lambda}=}404\,$nm (see the PMT specification
in \cite{KM3NeT-LoI-2.0,KM3NeT-PMT-characterisation}), translating
into $E_{\gamma}=3.07\,$eV, the approximate number of emitted Cherenkov
photons per track length, according to Eq. \ref{eq:Frank-Tamm-formula},
is $\left\langle \frac{dN}{dx}\right\rangle \simeq5.38\cdot10^{4}\,\frac{1}{\mathsf{m}}$.
Assuming a monochromatic emission, this gives the energy loss due
to Cherenkov effect of the order of $\left.\left\langle \frac{dE}{dx}\right\rangle \right|_{\mathsf{Cher}}\simeq165\,\frac{\mathsf{keV}}{\mathsf{m}}$,
which is negligible compared to the ionisation loss of about $\left.\left\langle \frac{dE}{dx}\right\rangle \right|_{\mathrm{ion}}\simeq200\,\frac{\mathsf{MeV}}{\mathsf{m}}$
(see Fig. \ref{fig:Muon-eloss}) \cite{Obodovskiy2019_Radiation_Cherenkov}.

\subsection{Photomultiplier Tubes\label{subsec:Photomultiplier_Tubes}}

\begin{figure}[h]
\centering{}\includegraphics[width=16cm]{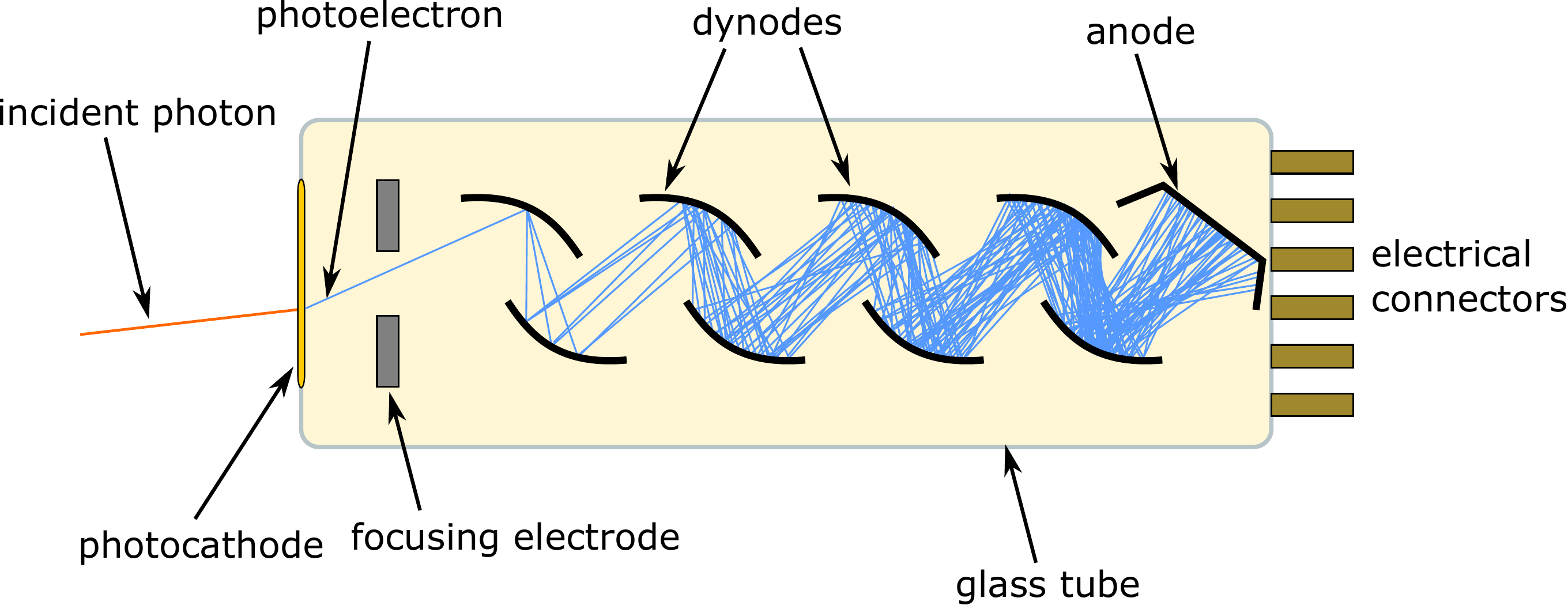}\caption{Sketch of the amplification of the signal from a single photon by
a PMT. \label{fig:PMT-drawing}}
\end{figure}

The water-based neutrino telescopes study charged particles by observing
the Cherenkov light emission they cause. The device, which enables
this is called a photomultiplier tube (PMT\nomenclature{PMT}{photomultiplier tube}).
PMT can convert a faint light signal (potentially even a single photon!)
into a massive cascade of electrons (typically $10^{6}-10^{8}$).
Such a big current flow results in a voltage pulse strong enough to
be recorded by the readout electronics. Fig. \ref{fig:PMT-drawing}
visualizes qualitatively what is happening inside the PMT after an
incident photon hits the photocathode and releases an electron through
photoelectric effect. Such an electron cascade does not happen every
time a $\gamma$ hits the cathode: the probability of it happening
is called quantum efficiency (QE\nomenclature{QE}{quantum efficiency})
and is a property specific to the cathode \cite{Hamamatsu-PMTs}.
It can be calculated as the ratio of the number of emitted photoelectrons
to the number of incident photons. 

PMTs used in KM3NeT experiment are Hamamatsu R12199-02 with a hemispherical
(mushroom) shape (80~mm diameter), 10 dynodes and a standard bi-alkali
photocathode. They have QE of 20\% at photon wavelength of 470~nm
and 28\% at 404~nm, dark count (PMT noise rate) of 1.5~kHz at 15$\,\lyxmathsym{\textdegree}$C,
according to their specification  \cite{KM3NeT-LoI-2.0}. The stated
working temperature matches the actual conditions at the bottom of
the Mediterranean Sea \cite{MediterraneanWaterTemperature,MediterraneanWaterTemperature2}.

\section{KM3NeT detectors and their physics case}

The points, where KM3NeT clearly stands out from the other experiments
are:
\begin{itemize}
\item use of the identical technology under a single collaboration at multiple
sites (which is highly beneficial in terms of systematic effect studies),
\item the deepest underwater location (see Sec. \ref{sec:ARCA}),
\item use of multi-PMT modules (see Sec. \ref{subsec:Digital-Optical-Modules};
much better spatial resolution).
\end{itemize}
KM3NeT detectors are sets of vertically oriented lines with digital
optical modules (DOMs; see Sec. \ref{subsec:Digital-Optical-Modules}),
called detection units (DUs; see Sec. \ref{subsec:Detection-Units}
for more details), arranged in a cylindrical volume. 

\begin{figure}[H]
\begin{centering}
\subfloat[Full KM3NeT detectors shown in comparison with the Eiffel tower and
an imperial star destroyer from Star Wars. Credit: João Coelho. \label{fig:ARCA2x115=000026ORCA115-size}]{\centering{}\includegraphics[width=16cm]{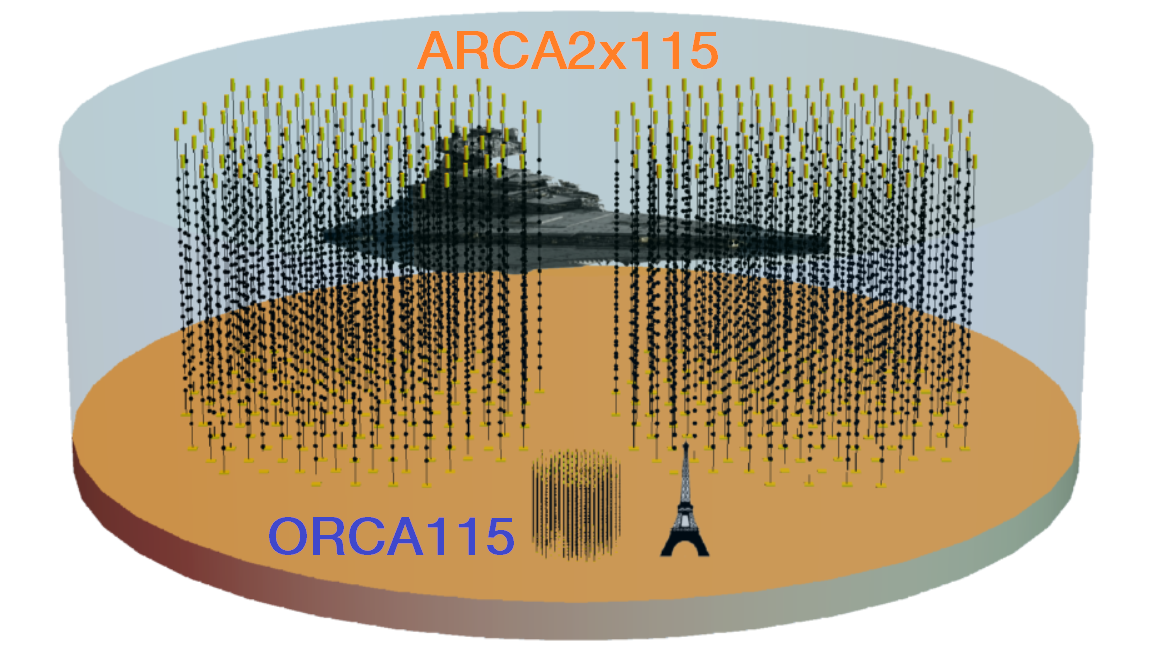}}
\par\end{centering}
\centering{}\subfloat[Sketch of the ARCA6 detector, i.e. ARCA detector with 6 DUs installed.
Data from this detector configuration is used in Chapters \ref{chap:muon-bundle-reco},
\ref{chap:Muon-rate-measurement}, and \ref{chap:prompt_ana}. \label{fig:ARCA6-geometry}]{\centering{}\includegraphics[width=16cm]{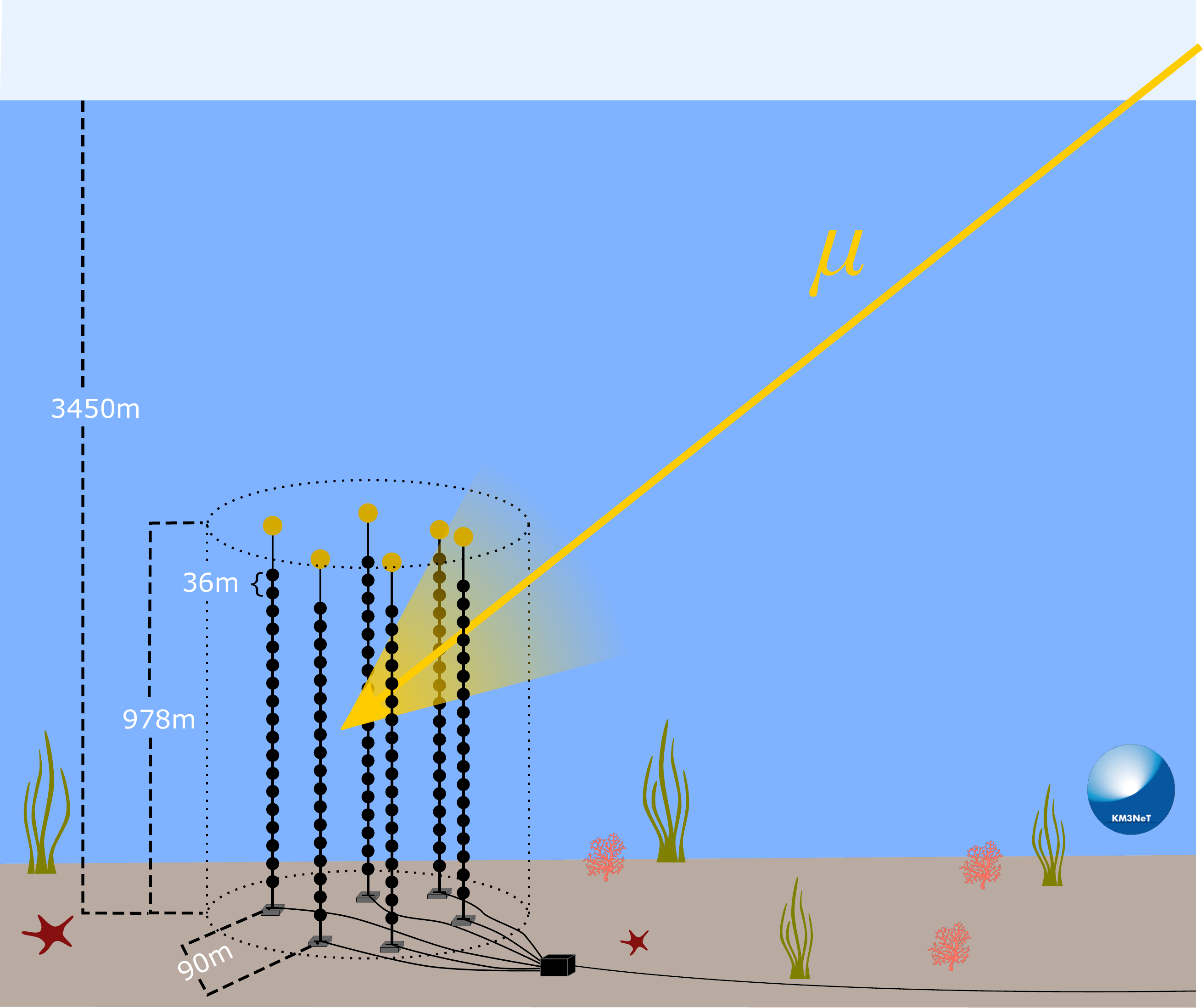}}\caption{Illustrations of the KM3NeT detector geometry. \label{fig:KM3NeT-detector-geometries}}
\end{figure}

In the following subsections, each detector is briefly described.

\subsection{ARCA\label{sec:ARCA}}

ARCA\nomenclature{ARCA}{astroparticle research with cosmics in the abyss}
stands for Astroparticle Research with Cosmics in the Abyss. It is
the bigger of the two KM3NeT detectors, and in its final state it
will consist of 2 building blocks, 115~DUs each (see Sec. \ref{subsec:Detection-Units}
and \ref{subsec:Building-blocks}). The radius of a building block
is approximately 500~m (as in Fig. \ref{fig:Top-view-of-detectors})
and the bottom of the detector will be placed at a depth of 3.5~km
at 36$\lyxmathsym{\textdegree}$16'~N 16$\lyxmathsym{\textdegree}$06'~E,
about 100 km offshore from Portopalo di Capo Passero, Sicily, Italy.
DOMs (see Sec. \ref{subsec:Digital-Optical-Modules}) on a DU are
36~m apart and the distance between DUs is 90m. The dimensions of
the detector are optimized for astrophysical neutrino measurements.
The big detector volume allows observing extremely energetic cosmic
particles. The main objective of ARCA is to confirm the IceCube's
(see \ref{sec:neutrino-telescopes}) measurement of cosmic neutrino
flux \cite{IC-diffuse-neutrino-flux,IC-electron-and-tau-neutrino-flux}
and allow for neutrino astronomy with an unprecedented precision thanks
to a sub-degree angular resolution.

As of August 2023, the completion status of the first ARCA building
block is 19/115 DUs. The construction of KM3NeT/ARCA is expected to
conclude in 2032.

\subsection{ORCA\label{sec:ORCA}}

ORCA\nomenclature{ORCA}{oscillation research with cosmics in the abyss}
(Oscillation Research with Cosmics in the Abyss) is more compact than
ARCA, while maintaining the same number of DUs per building block.
Complete ORCA will comprise a single building block of 115~DUs, distributed
within a 100~m radius at a depth of 2.5~km at 42$\lyxmathsym{\textdegree}$48'~N
06$\lyxmathsym{\textdegree}$02'~E, about 40~km offshore from Toulon,
France. The height of the detector is about 200~m, the spacing between
DOMs on a DU: 9~m and the spacing between DUs: 20~m. As the name
suggests, the detector is designed to study the phenomenon of neutrino
oscillations. A particular emphasis is put on determining the neutrino
mass ordering (see Sec. \ref{subsec:Neutrinos}), since it has never
been measured before and should be well within reach of ORCA. To achieve
this, the few-GeV atmospheric neutrino flux will be studied, as in
this energy range the differences between the expected flux for the
assumption of normal and inverted ordering are very pronounced \cite{ORCA_oscillations}.
The need for a good energy resolution, especially for fainter events
is the reason for ORCA's dense instrumentation. There are also plans
for potential upgrades: making ORCA even more dense (Super-ORCA) \cite{Super-ORCA},
or shooting a neutrino beam from the Protvino accelerator in Russia
to ORCA (Protvino to ORCA: P2O\nomenclature{P2O}{Protvino to ORCA})
\cite{P2O}.

As of August 2023, the ORCA detector consists of 16/115 operational
DUs and should be completed in 2030.

\subsection{Detector design\label{sec:Hardware}}

\begin{figure}[h]
\centering{}\includegraphics[width=16cm]{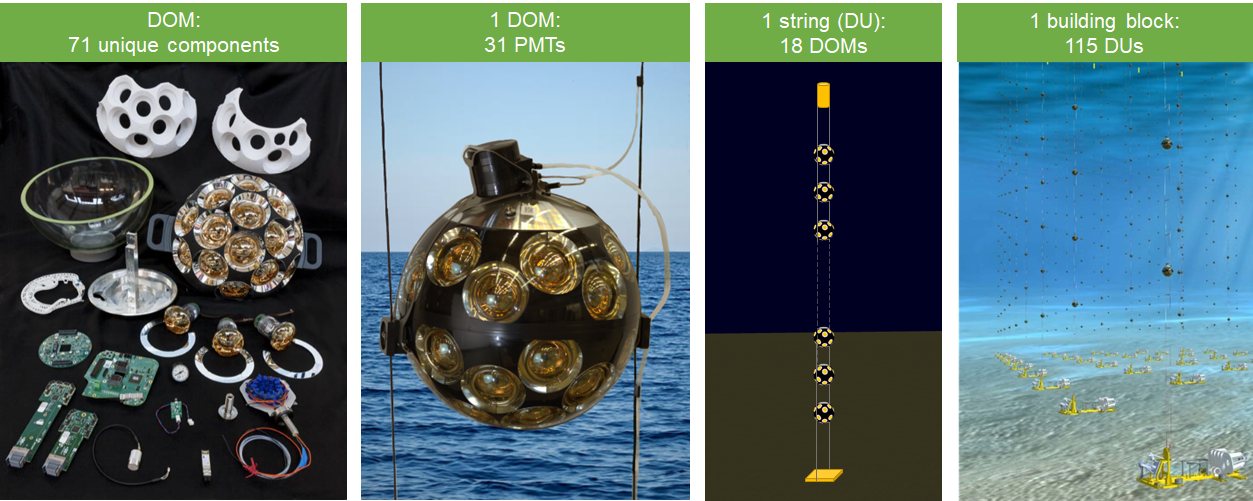}\caption{The summary of the KM3NeT detector design, ordered from tiniest components
to the complete building blocks. More detail on all the components
may be found in \cite{KM3NeT-LoI-2.0,KM3NeT-DOM-paper}. \label{fig:detector_summary}}
\end{figure}

The basic design of both KM3NeT detectors (ARCA, ORCA) is sketched
in Fig. \ref{fig:detector_summary} and their elements are described
in what follows.

\subsubsection{Digital Optical Modules\label{subsec:Digital-Optical-Modules}}

Digital optical module (DOM\nomenclature{DOM}{digital optical module})
is a hermetically enclosed 17'' glass sphere, containing 31 PMTs,
pressure, tilt and acoustic sensors, compass, readout electronics,
and other equipment. A disassembled DOM is presented in Fig. \ref{fig:detector_summary}
on the left. The use of multi-PMT DOMs instead of traditional optical
modules with single large PMT (as e.g. in IceCube or SK) has significant
advantages:
\begin{enumerate}
\item The total photo-cathode area per module is larger: more emitted photons
can be collected.
\item The angular coverage is nearly uniform: light coming from any direction
will be reconstructed equally well.
\item Each of the PMTs inside the DOM has slightly different position and
orientation (see Fig. \ref{fig:detector_summary}). The implication
of this are varying light detection times at different PMTs, which
adds to the directional information.
\item The environmental background (see Sec. \ref{sec:trigger}) may be
more efficiently eliminated by requiring the light signal to be observed
by multiple PMTs on a DOM in coincidence \cite{DepthDependenceMassimiliano}.
\end{enumerate}

\subsubsection{Detection Units\label{subsec:Detection-Units}}

\begin{figure}[H]
\centering{}\includegraphics[height=7cm]{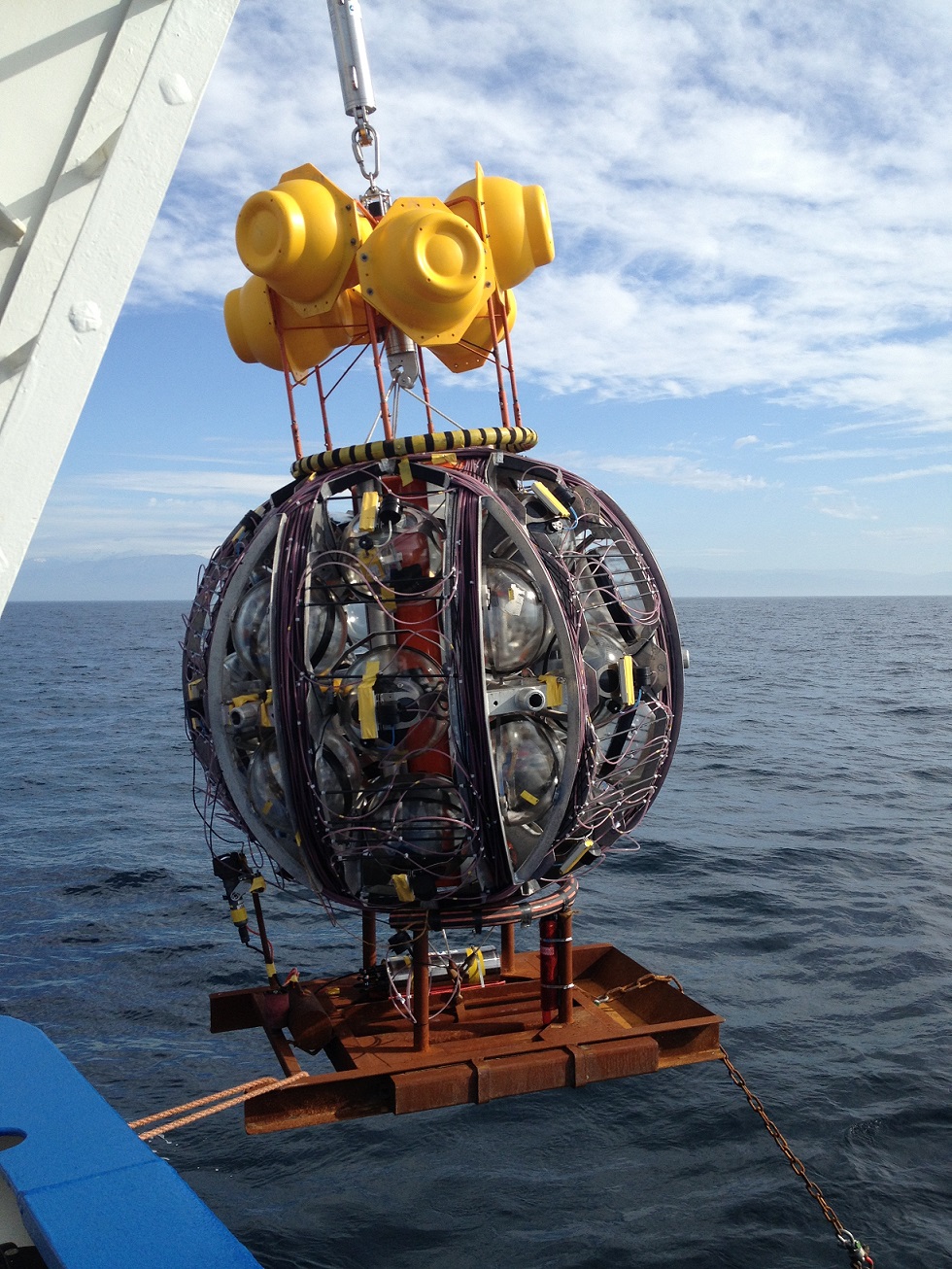}\includegraphics[height=7cm]{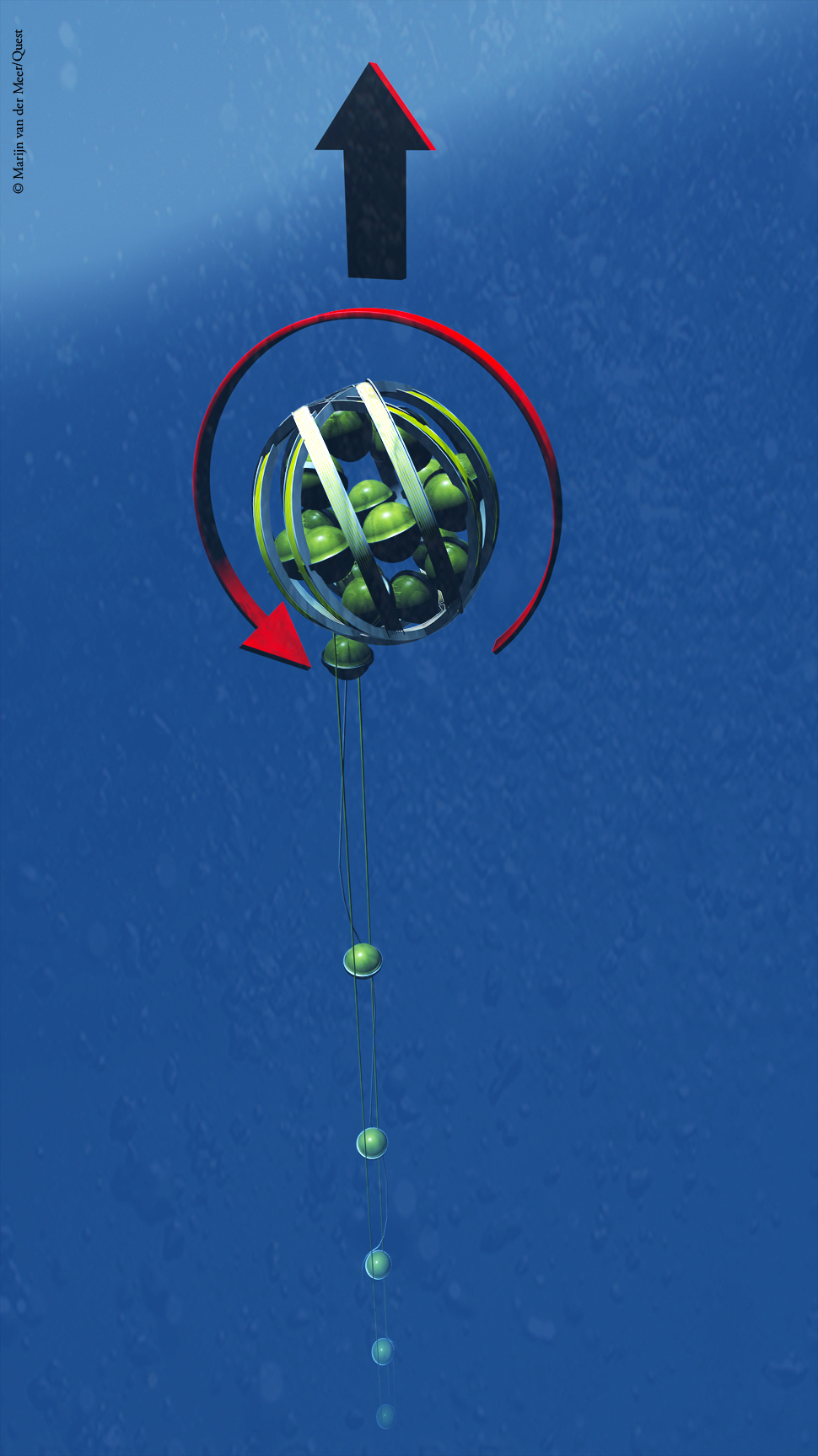}\caption{On the left: a real photo of a DU during the deployment at the French
site. Right: schematic drawing, showing how a LOM unfurls, releasing
the DOMs. \label{fig:DUs}}
\end{figure}

A Detection Unit (DU\nomenclature{DU}{detection unit}) is a set of
18 DOMs attached with titanium collars to two Dyneema$^{\circledR}$
ropes in the shape of a vertical string, mounted at the seabed with
an anchor. DUs are connected to junction boxes at the bottom of the
sea (which in turn are connected to the shore station) through the
electro-optical cable, which contains power lines (400 VDC) and 18
optical fibres to transfer the data. On top of the DU line there is
also a buoy to keep it as vertical as possible (deviations from the
nominal position due to e.g. sea current are tracked, stored and included
when analysing the data). DU deployment at the experimental site is
shown in Fig. \ref{fig:DUs}. Each DU is deployed in a launcher of
optical modules (LOM\nomenclature{LOM}{launcher of optical modules}),
which makes sure that the structure will be properly aligned by guiding
the unfurling process.

\subsubsection{Building blocks \label{subsec:Building-blocks}}

For each detector, a building block  consists of 115 DUs. It is powered
by a set of junction boxes and connected via optical fibres to the
shore station, where the digitized data from the DOMs is sent to and
processed. The adopted approach is `all data to shore', which means
that there is no pre-selection of what will be saved. This implies
a transfer rate up to 25~$\frac{\mathrm{GB}}{\mathrm{s}}$ per building
block (in total 75~$\frac{\mathrm{GB}}{\mathrm{s}}$ for complete
ARCA and ORCA).

\begin{figure}[H]
\centering{}\subfloat[ORCA115.]{\centering{}\includegraphics[width=8cm]{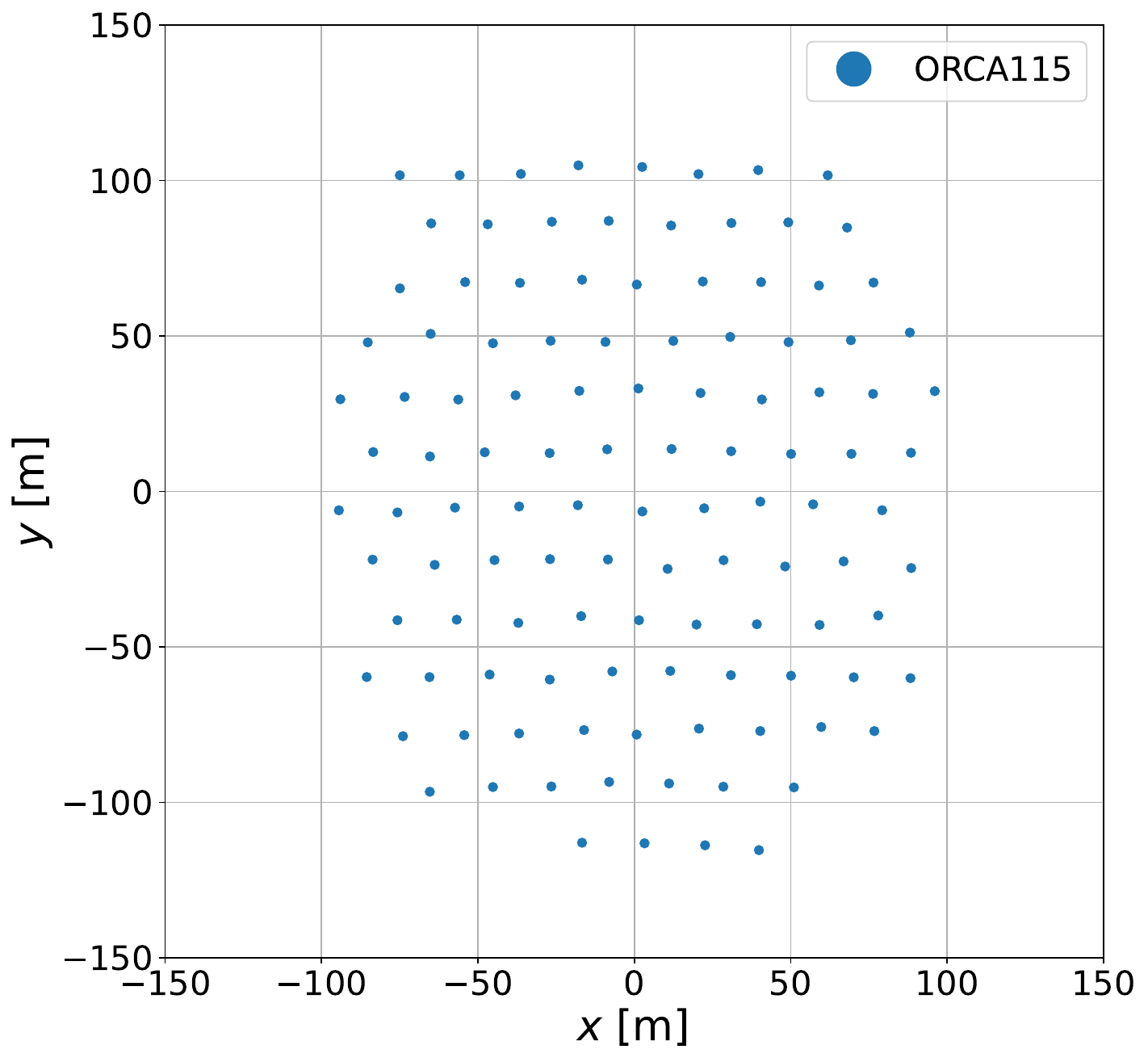}}\subfloat[ARCA115 with ORCA115 for comparison.]{\centering{}\includegraphics[width=8cm]{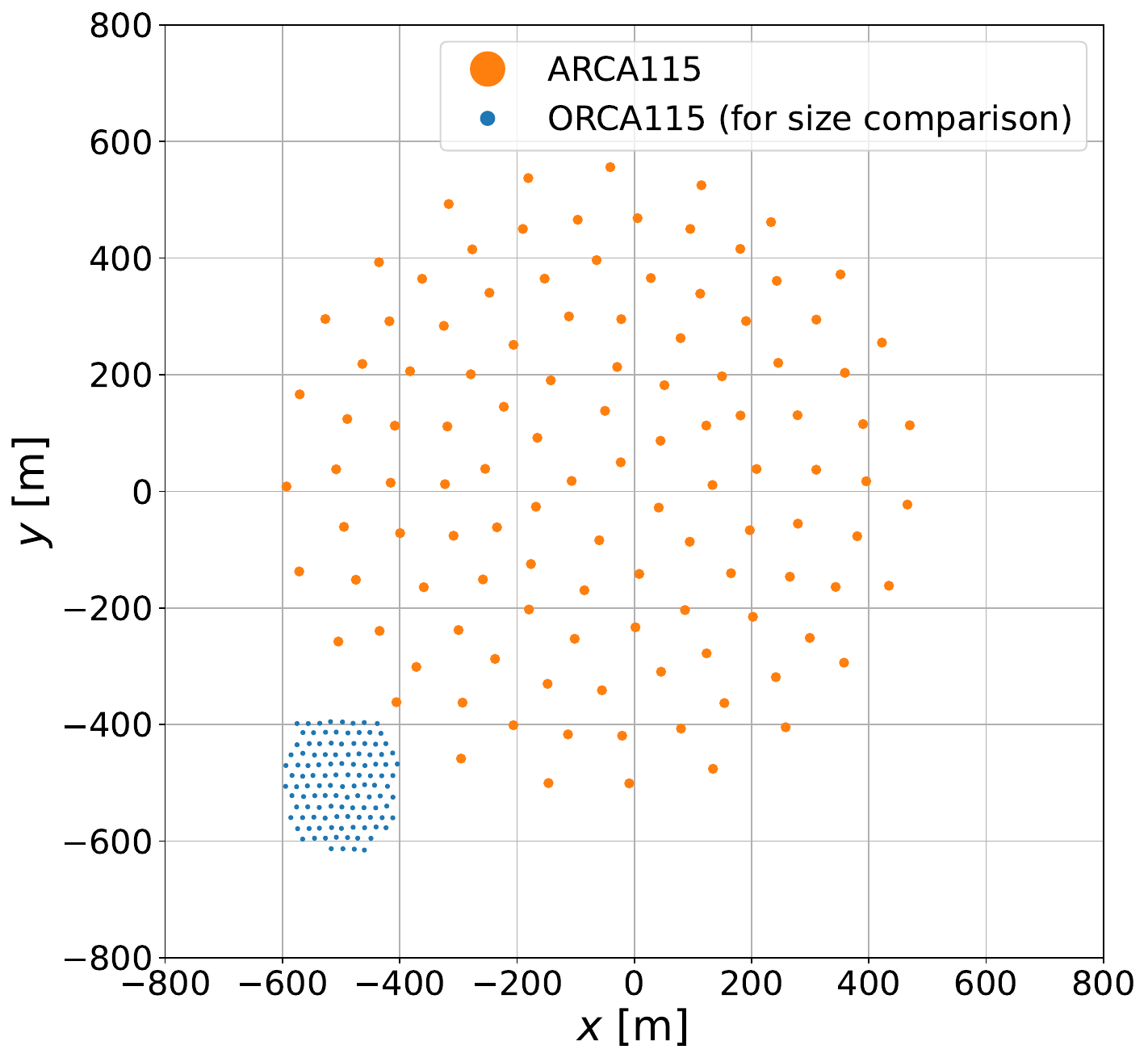}}\caption{KM3NeT detector footprints for complete building block configurations,
as in the CORSIKA simulation used for Chap. \ref{chap:muon-bundle-reco}–\ref{chap:prompt_ana}.
\label{fig:Top-view-of-detectors}}
\end{figure}

The distribution of the DUs in a building block is shown in Fig. \ref{fig:Top-view-of-detectors}.

\subsection{Event topologies}

\begin{figure}[h]
\centering{}\includegraphics[width=16cm]{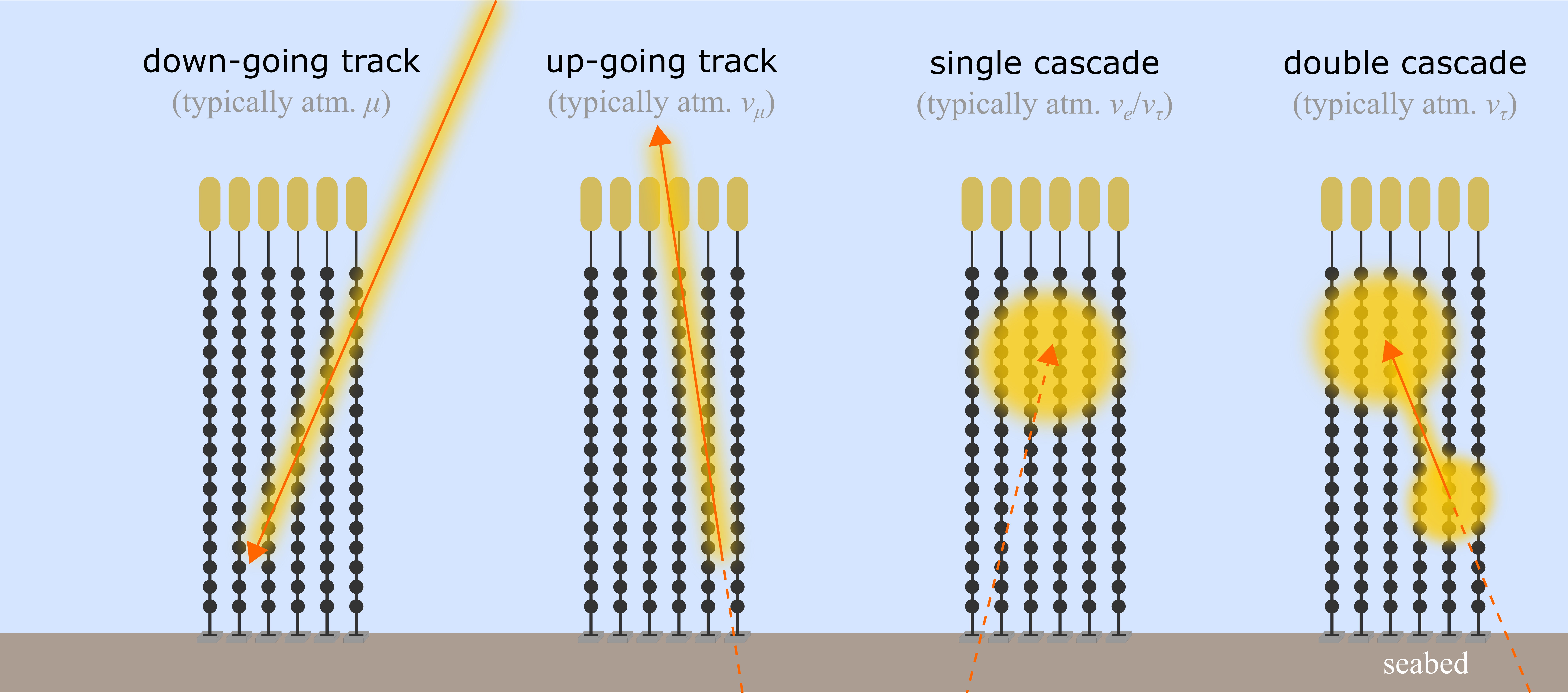}\caption{The summary of different event topologies seen in KM3NeT detectors.
\label{fig:event-topologies}}
\end{figure}

Depending on the particle type, events inside the detector may look
completely different (see Fig. \ref{fig:event-topologies}). In terms
of shape, we distinguish between two categories:
\begin{itemize}
\item track-like events, typically corresponding to muons (coming directly
from the atmosphere or produced in $\nu_{\mu}$ CC interactions),
which do not loose a lot of energy at once (see Sec. \ref{subsec:Muons})
\item cascade-like (often also referred to as shower-like) events, mostly
caused by $\nu_{e}$ or $\nu_{\tau}$ interactions
\end{itemize}

In case of particularly energetic $\nu_{\tau}$ events (which are
rare), two distinct cascades separated by a dim track may be observed:
one from the $\nu_{\tau}$ interaction in which the $\tau$ lepton
is produced and the other from the decay of the $\tau$. The separation
between the cascades depends on tau neutrino energy $E_{\nu_{\tau}}$,
since the $\tau$ decay length is approximately $\left\langle L_{\tau}\right\rangle \sim50\mathrm{m}\cdot\frac{E_{\tau}}{\mathrm{PeV}}$
\cite{KM3NeT-LoI-2.0}, which given the instrumentation density and
energy resolution of ARCA and ORCA should allow to observe such events
with full detectors.

\chapter{Simulation of atmospheric muons in KM3NeT\label{chap:Cosmic-Ray-Sim-chain-KM3NeT}}

This chapter describes the simulation of EAS (see Sec. \ref{sec:Extensive-Air-Showers-EAS})
and their products that can be detected by KM3NeT detectors, with
a particular emphasis on muons.

\section{Event generators\label{sec:generators}}

The two main event generators used in this work: CORSIKA and MUPAGE
\cite{CORSIKA,MUPAGE} are described in the following sections.

\subsection{CORSIKA\label{sec:CORSIKA}}

CORSIKA\nomenclature{CORSIKA}{cosmic ray simulations for kascade}
stands for COsmic Ray SImulations for KAscade and was developed for
KASCADE\nomenclature{KASCADE}{karslruhe shower core and array detector}
(KArslruhe Shower Core and Array DEtector) experiment at Karlsruhe
Institute of Technology, measuring EAS \cite{KASCADE}. It offers
a full event-by-event simulation of air showers with many customization
possibilities. There are various models of low- and high-energy hadronic
interactions available. The simulation can be adapted to almost any
imaginable primary CR flux and composition. The atmosphere density
profile can be taken either from a predefined parametrization or from
a custom fit, supplied by the user. More details on CORSIKA are given
in the following subsections, in Sec. \ref{sec:Additional-material-related-to-CORSIKA},
and in \cite{CORSIKA,CORSIKA-Userguide}. Within this thesis CORSIKA7
was used and unless specified otherwise, CORSIKA always refers to
this major version.

\subsubsection{Implemented processes and particles}

All the processes potentially relevant for EAS are implemented in
CORSIKA and all the secondary particles are explicitly tracked, with
their parameters stored at observation level (in case of KM3NeT simulations,
it is the sea level) \cite{CORSIKA}.

CORSIKA implements the properties of all leptons and most commonly
occurring hadrons. The full list can be found either in \cite{CORSIKA-Userguide},
or in Tab. \ref{tab:List-of-all-particles-in-CORSIKA}.

\subsubsection{Models\label{subsec:Models}}

Here, different classes of physics models are briefly described.

\paragraph{Cosmic ray flux models \label{subsec:Cosmic-Ray-flux-models}}

\begin{figure}[H]
\centering{}\includegraphics[width=16cm]{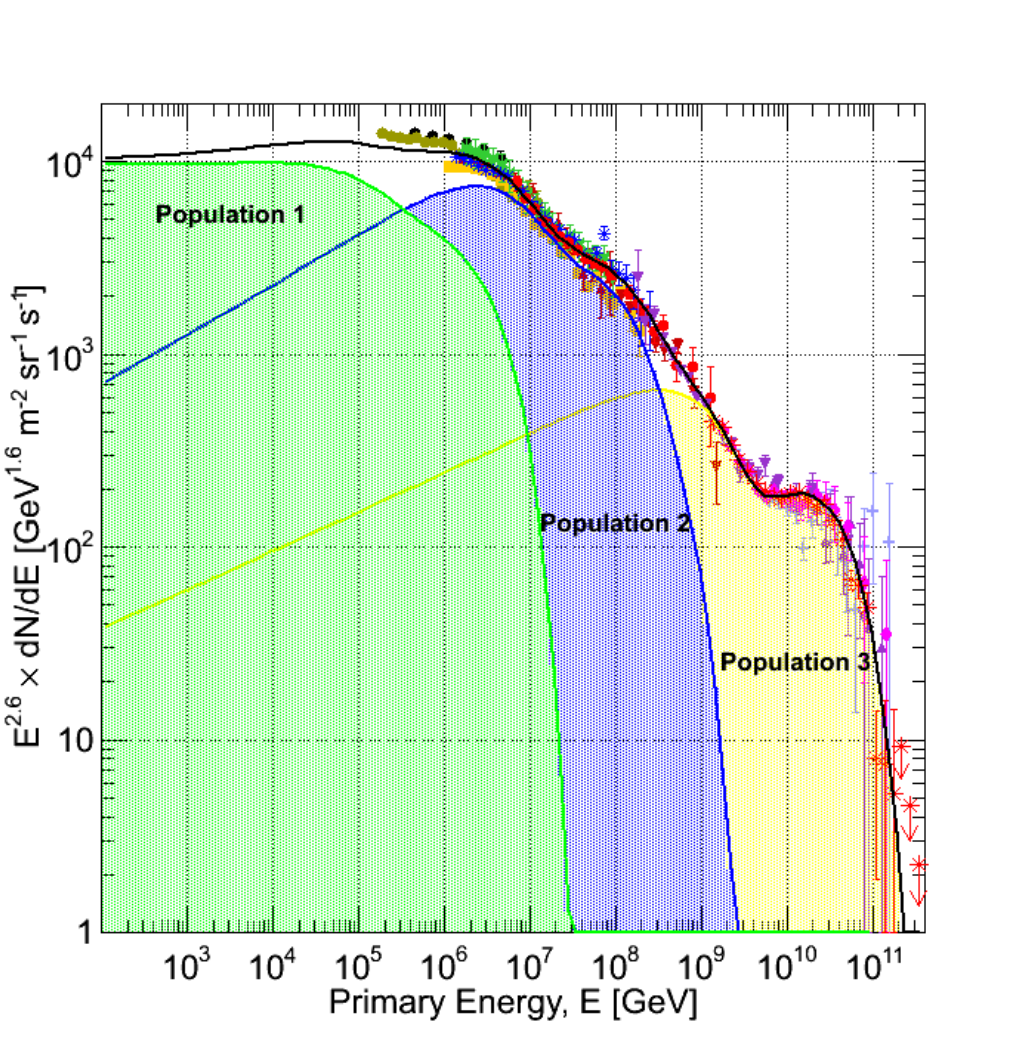}\caption{Fit of the CR flux data to the 3-population scenario from \cite{GST}.
\label{fig:3_populations-1}}
\end{figure}

Cosmic ray flux models describe the total primary CR flux arriving
at the Earth and its composition. The default one used for event weighting
(see Sec. \ref{sec:weights}) in the CORSIKA simulations for KM3NeT
is the Gaisser-Stanev-Tilav model with 3 populations (GST3). It is
a fit to CR data, performed with an assumption of 3 populations of
particle sources:
\begin{enumerate}
\item Supernova remnants (SNR\nomenclature{SNR}{supernova remnant}).
\item Other galactic sources.
\item Extragalactic sources.
\end{enumerate}
Each population has a different magnetic rigidity ($R=\frac{\left|\vec{p}\right|c}{Ze}$)
cutoff, nuclei groups, and spectral indices. The particle spectrum
is approximated with the following formula:

\begin{equation}
\phi_{i}\left(E\right)=\stackrel[j=1]{3}{\sum}a_{i,j}E^{-\gamma_{i,j}}\cdot e^{-\frac{E}{Z_{i}R_{c,j}}},
\end{equation}

where $E$ is the primary cosmic ray energy per nucleus, $i=$(p,
He, C, O, Fe), $a_{i,j}$ are the normalization constants, $\gamma_{i,j}$
are the integral spectral indexes, $Z_{i}$ are the atomic numbers,
and $R_{c,j}$ are the characteristic magnetic rigidities \cite{GST}.
The result of the fit is shown in Fig. \ref{fig:3_populations-1}.

Some of the other available CR flux models can be found in \cite{poly-gonato,ZSP,GaisserHonda,combined_HillasGaisser_and_GaisserHonda_also_crfluxmodels_reference,HillasGaisser_model,GST,Thunman}.
Selected ones were used in Sec. \ref{sec:Systematic-uncertainty-study}
to evaluate the systematic uncertainty related to primary CR flux
modelling.

\paragraph{Hadronic interaction models \label{subsec:Hadronic-interaction-models}}

The models of the hadronic interactions available within CORSIKA are
divided into two categories: low-energy (LE\nomenclature{LE}{low-energy})
and high-energy (HE\nomenclature{HE}{high-energy}), with the threshold
between the two set at $80\,$GeV \cite{CORSIKA-Userguide}.

The choice of LE hadronic interaction models is limited to three options:
FLUKA, GHEISHA, and UrQMD \cite{FLUKA,GHEISHA,UrQMD}. In this work,
the UrQMD model has been used, since it offers a good description
of the physics processes and works stably down to the lowest energies
possible in CORSIKA. It has been verified by the earlier CORSIKA productions,
that GHEISHA implementation tends to fail at the lowest energies.
FLUKA is a state-of-the-art low-energy hadronic interaction model,
however due to limited access to the code and potential copyright
issues, it has not been used. Due to the fact that the KM3NeT detectors
are shielded by more than 2 kilometres of water above them (see Chap.
\ref{chap:KM3NeT}), the adopted LE model is of marginal importance
for muon studies. However, this would no longer hold true for the
simulation of LE neutrino flux arising from the CR-induced showers. 

HE hadronic interaction models are one of the most important choices
made when compiling the CORSIKA code, as they determine the core behaviour
of the program. There is a number of such models available: VENUS,
NEXUS, DPMJET, QGSJET, EPOS, and SIBYLL \cite{VENUS,QGSJET-1,QGSJET-2,QGSJET-II,EPOS-LHC,SIBYLL,SIBYLL-2.1,SIBYLL-2.3c-inclusive-lepton-fluxes,SIBYLL-2.3cFeynmanscaling}.
However, only the latter three are relatively up to date and only
DPMJET and SIBYLL explicitly treat charmed and strange hadron production.
As this is a necessary prerequisite for the prompt muon analysis in
Chap. \ref{chap:prompt_ana}, SIBYLL 2.3d (the most recent available
version at the time) has been used as the default HE hadronic interaction
model \cite{SIBYLL-2.3d}.

\paragraph{Model of the atmosphere\label{par:Model-of-the-atmosphere}}

In CORSIKA, the atmospheric density is modelled by dividing the atmosphere
into 5 distinct layers. The thickness $T\left(h\right)\,\left[\frac{\mathsf{g}}{\mathsf{cm^{2}}}\right]$
of the \foreignlanguage{english}{first 4 layers is parameterised exponentially:}

\selectlanguage{english}%
\begin{equation}
T\left(h\right)=a_{i}+b_{i}\cdot e^{-\frac{h}{c_{i}}},\label{eq:layers_1-4}
\end{equation}

and for the 5th layer linearly:

\begin{equation}
T\left(h\right)=a_{5}-b_{5}\cdot\frac{h}{c_{5}},\label{eq:layer_5}
\end{equation}

where $i$ is the layer number, $a_{i}$, $b_{i}$, $c_{i}$ are the
fitted parameters, and $h$ is the altitude above the sea level. The
upper boundaries of each layer $h_{i}^{\mathrm{upper}}$ can be adjusted
to tune the fit accuracy (see Fig. \foreignlanguage{british}{\ref{fig:atm_fit}}).
CORSIKA comes with a set of predefined density models for different
locations and epochs, and also tools to create custom models: gdastool
(fitting the 5-layer atmosphere to the GDAS database \cite{GDAS})
and bernlohr package \cite{CORSIKA}. In this work, a fit of the NRLMSIS-2.0
\cite{NRLMSIS-2.0} atmosphere model was used. It was fitted to the
5 described by Eq. \ref{eq:layers_1-4} and \ref{eq:layer_5}, using
the fitting utilities of the SciPy package \cite{SciPy}. The details
can be found in \foreignlanguage{british}{Sec. \ref{subsec:Fit-of-the-atmosphere}}.
\selectlanguage{british}%

\subsection{MUPAGE \label{sec:MUPAGE}}

MUPAGE\nomenclature{MUPAGE}{fast atmospheric muon generator for neutrino telescopes based on parametric formulas}
stands for fast atmospheric MUon GEnerator for neutrino telescopes
based on PArametric formulas. As the name suggests, it is not a full
Monte Carlo simulation \cite{MUPAGE}. It samples parametrized muon
bundle distributions to generate muon events on the surface of a cylinder
enclosing the detector (see Sec. \ref{sec:can}). It was designed
for the ANTARES experiment \cite{ANTARES,MUPAGE} and was inherited
by KM3NeT. The main advantage of MUPAGE over CORSIKA is that it is
much faster, however the price for this is a smaller amount of information
that can be extracted from the simulation, e.g. there is completely
no information on particle history. There are ongoing efforts to update
the code by performing new fits of internal MUPAGE parameters, using
both the CORSIKA MC produced within this work \cite{Andrey_Thesis,Andrey_and_me_ICRC2023},
and the currently available data from the KM3NeT detectors \cite{Brian_MUPAGE_tuning}.

\section{Event weighting \label{sec:weights}}

The MC generators simulate a given number of events, however to compare
them against experimental data, a translation to the units of expected
event rate is necessary. This can be achieved by employing event weights
$w_{\mathrm{event}}$. In the following, the weighting schemes for
CORSIKA and MUPAGE are described.

\subsection{CORSIKA\label{subsec:CORSIKA-weighting}}

In CORSIKA muon simulations, there are four types of weights in use:
\begin{itemize}
\item $w_{1}=S$ — the effective surface of the sensitive volume of the
detector (see Sec. \ref{sec:can}) {[}$\mathsf{m^{2}}${]},
\item $w_{2}$ — contain the generation weight {[}$\mathsf{\mathsf{GeV\cdot}m^{2}\cdot sr\cdot\frac{s}{year}}${]}
,
\item $w_{3}$ — so-called ``global weights'' in the unit of rate {[}$\frac{1}{\mathsf{year}}${]},
\item $w_{\mathrm{event}}$ — final weights, used for comparisons between
data and MC.
\end{itemize}
The generation weight $w_{2}$ can be calculated as:

\begin{equation}
w_{2}=\underset{w_{1}}{\underbrace{S}}\cdot I_{\theta}\cdot I_{E_{\mathrm{prim}}}\cdot E_{\mathrm{prim}}^{\gamma}\cdot F,\label{eq:w2_CORSIKA}
\end{equation}

where:

$I_{\theta}=2\pi\left[\cos\left(\theta_{\mathsf{max}}\right)-\cos\left(\theta_{\mathsf{min}}\right)\right]$~{[}sr{]}
— angular phase space factor,

$I_{E_{\mathrm{prim}}}=\left\{ \begin{array}{ccc}
\frac{\left.E_{\mathsf{prim}}^{1-\gamma}\right|_{\mathrm{max}}-\left.E_{\mathsf{prim}}^{1-\gamma}\right|_{\mathrm{min}}}{1-\gamma} & \mathsf{if} & \gamma\neq1\\
\ln\left(\frac{\left.E_{\mathsf{prim}}\right|_{\mathrm{max}}}{\left.E_{\mathsf{prim}}\right|_{\mathrm{min}}}\right) & \mathsf{if} & \gamma=1
\end{array}\right.$ — energy phase space factor,

$E_{\mathrm{prim}}$ — primary energy per nucleus,

$\gamma$ — generation spectral index,

$F=3600\cdot24\cdot365.2422$ — average number of seconds in a year.

The CORSIKA simulation produced in this work was split into four sub-productions,
divided into distinct $E_{\mathrm{prim}}$ ranges, reflected in their
names: TeV\_low, TeV\_high, PeV, and EeV. The simulated primary energy
spectrum $E_{\mathrm{prim}}^{\gamma}$, called generation spectrum,
is different for every CORSIKA sub-production and was tuned to compensate
for the effects of propagation through air and water. It has nothing
to do with the actual CR spectrum observed in nature, it is merely
a way to optimize the statistics of the simulation, similarly to the
number of generated showers for each sub-production. In addition,
the numbers of generated showers (see Eq. \ref{eq:event_weights})
per primary have been adjusted, based on the computation time and
disk space constrains. The details of the CORSIKA production settings
are summarised in Tab. \ref{tab:Simulation-settings-CORSIKA}.

The $w_{3}$ weights are derived from $w_{2}$ by multiplying them
with an assumed CR primary particle flux {\small{}$\phi$ $[\frac{1}{\mathsf{GeV\cdot}\mathsf{m^{2}\cdot sr\cdot s}}]$:}{\small\par}

\begin{equation}
w_{3}=w_{2}\cdot\phi.\label{eq:w_3}
\end{equation}

\begin{figure}[H]
\centering{}\includegraphics[width=12cm]{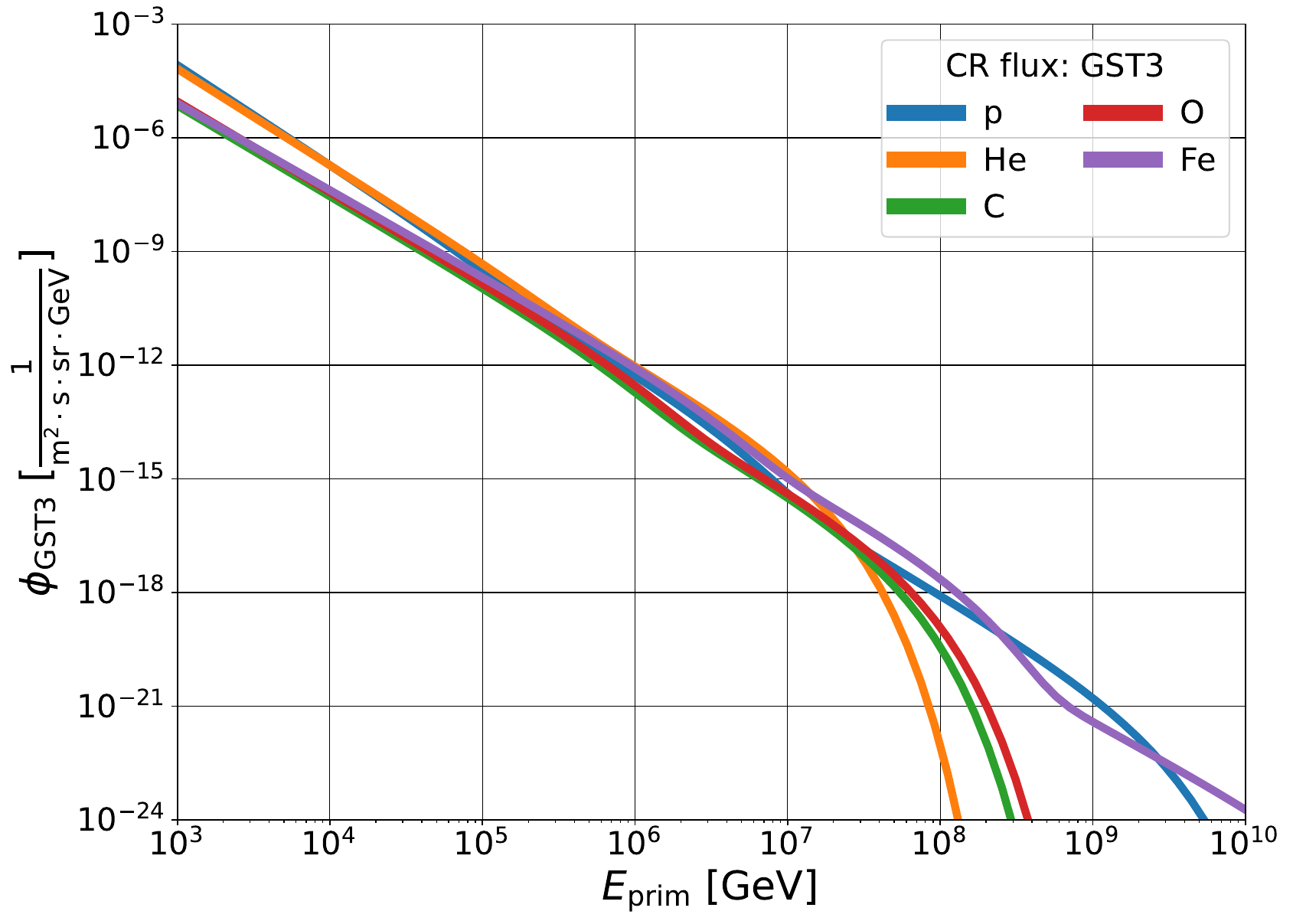}\caption{Flux predicted by the GST3 model as function of the primary energy
for each of the primary particles used in CORSIKA simulation \cite{GST}.
In the flux of iron primaries, the transition between the second and
third source population at around $10^{9}\,$GeV can be seen (see
Sec. \ref{subsec:Cosmic-Ray-flux-models}). \label{fig:CR_flux_plot}}
\end{figure}

As mentioned in Sec. \ref{subsec:Cosmic-Ray-flux-models}, in the
case of CORSIKA simulations the GST3 CR flux model is used \cite{GST}.
The predicted CR flux $\phi$ for the primary energy range relevant
for KM3NeT is shown in Fig. \ref{fig:CR_flux_plot}.

\begin{figure}[H]
\centering{}\includegraphics[width=12cm]{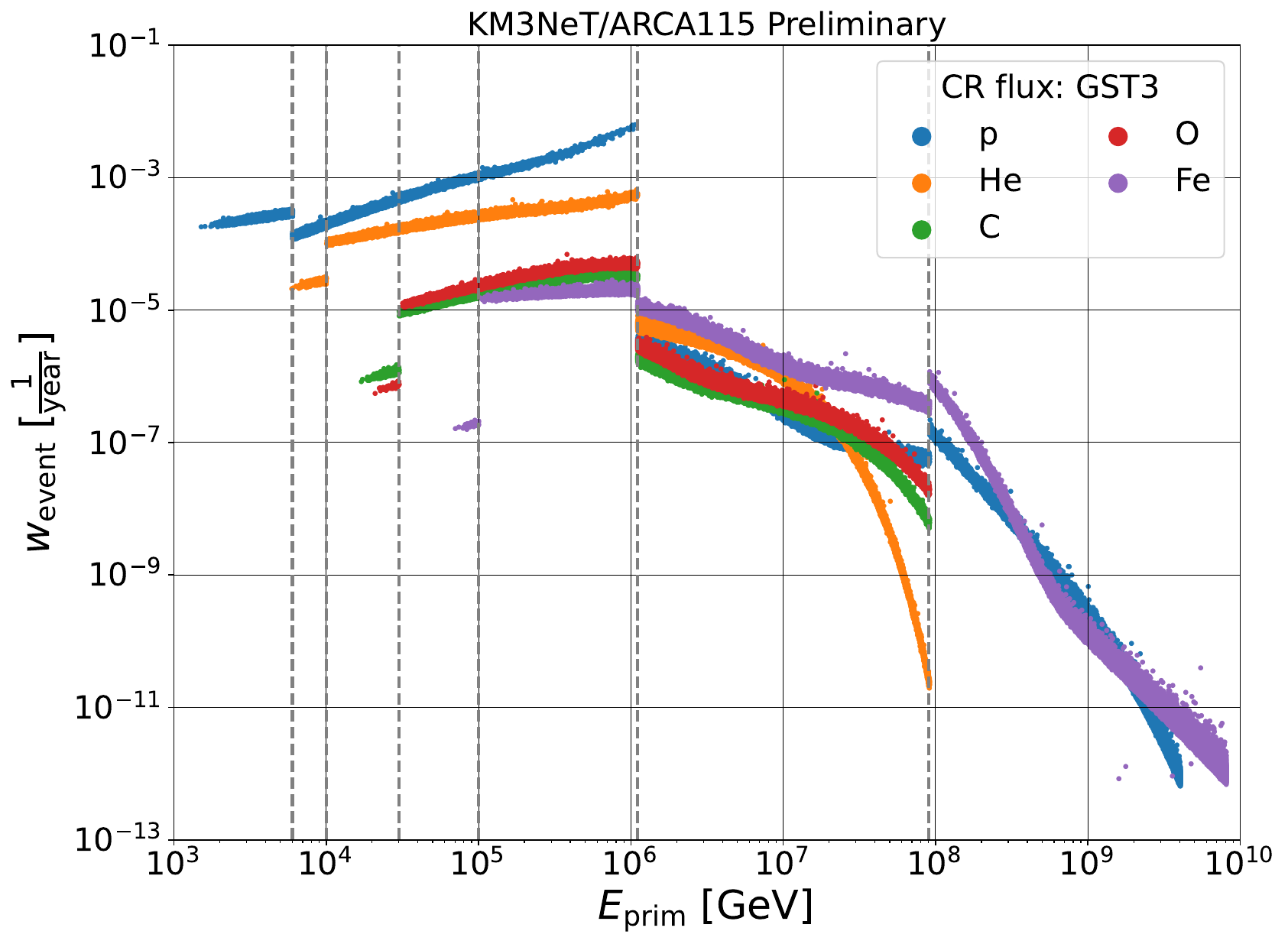}\caption{Event weights as function of the primary energy shown for each of
the primaries separately. The horizontal dashed grey lines are the
boundaries of CORSIKA sub-productions (see Tab. \ref{tab:Simulation-settings-CORSIKA}).
The vertical spread of the data points comes from the angular dependence
of the event weights (see Eq. \ref{eq:w2_CORSIKA}). The plot was
made using the ARCA115 MC. \label{fig:event_weight_plot}}
\end{figure}

To obtain the expected event rates, the final event weights $w_{\mathrm{event}}$
have to be computed:

\begin{equation}
w_{\mathrm{event}}\left(\mathrm{prim},E_{\mathrm{prim}}\right)=\frac{w_{3}}{n_{\mathrm{generated\,showers}}\left(\mathrm{prim},E_{\mathrm{prim}}\right)},\label{eq:event_weights}
\end{equation}

which boils down to dividing $w_{3}$ by the appropriate number of
generated showers. This will result in an event rate in units of $\frac{1}{\mathrm{s}}$.
If one prefers to have it in $\frac{1}{\mathrm{year}}$, the result
has to be multiplied by $F$. It has to be emphasized that $n_{\mathrm{generated\,showers}}$
is in principle a different number for each primary and for each sub-production,
hence the prim and $E_{\mathrm{prim}}$ dependence in Eq. \ref{eq:event_weights}.
If the sub-productions overlapped, the $n_{\mathrm{generated\,showers}}$
in the overlapping regions would be the sum of all the showers from
the overlapping sub-productions. No such overlap was used in the CORSIKA
production, with the intent of avoiding confusion and potential misuse
of the weights by users unaware of such subtlety. The event weight
example for ARCA115 CORSIKA MC is shown in Fig. \ref{fig:event_weight_plot}.

\begin{figure}[H]
\centering{}\includegraphics[width=12cm]{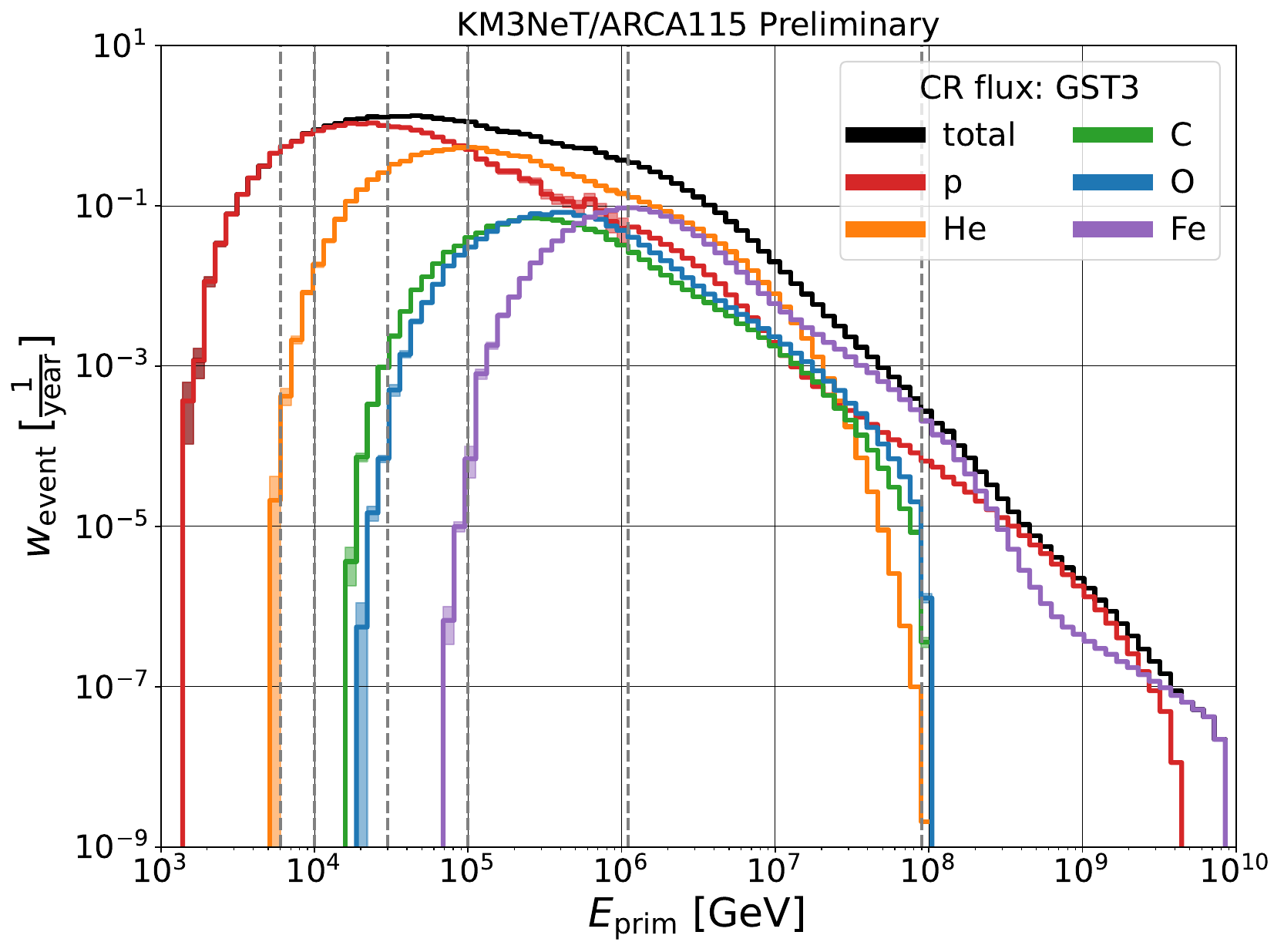}\caption{Weighted primary energy distribution showing the contribution of each
of the primaries. The horizontal dashed grey lines are the boundaries
of the CORSIKA sub-productions (see Tab. \ref{tab:Simulation-settings-CORSIKA}).
The errors are shown by shaded rectangles around the bins and were
computed using Eq. \ref{eq:hist_error}. \label{fig:weighted_Eprim_example}}
\end{figure}

As bizarre as it may seem, such weights as in Fig. \ref{fig:event_weight_plot}
do produce a smooth prediction of the expected rates, as e.g. shown
in Fig. \ref{fig:weighted_Eprim_example}. It is possible because
all the apparent discontinuities and different slopes in Fig. \ref{fig:event_weight_plot}
are exactly accounting for the underlying statistics with which the
CORSIKA MC has been generated.

One more important thing to note is that in the left part of Fig.
\ref{fig:weighted_Eprim_example} there is quite a drastic cutoff
in the spectra, which is not present in the CR flux itself (Fig. \ref{fig:CR_flux_plot}).
This is an effect of the propagation of the shower products (muons).
It is present even directly in the CORSIKA output at the observation
level (sea level in our case), since only the showers reaching the
observation level are saved in CORSIKA. The effect becomes even more
pronounced after propagation through water.

\subsection{MUPAGE}

In the case of MUPAGE, the weighting is rather trivial and the weights
are:

\begin{equation}
w_{\mathrm{MUPAGE}}=\frac{1}{t_{\mathrm{live}}},
\end{equation}

where $t_{\mathrm{live}}$ is the total simulated livetime (active
data taking time), which is analogous to how the experimental data
is weighted for data vs MC comparisons (see Chap. \ref{chap:Muon-rate-measurement}).
The resulting weighted distributions are event rates.

\section{Muon simulation: from sea to reconstruction level\label{sec:Muon-sim-chain}}

\begin{figure}[H]
\centering{}\includegraphics[width=16cm]{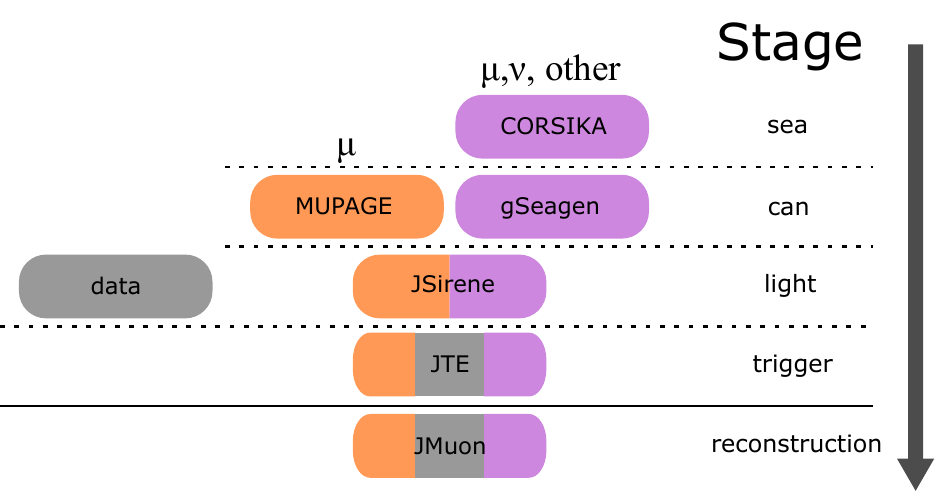}\caption{KM3NeT processing chain. Software used for muon MC simulation (\textcolor{orange}{MUPAGE}
and \textcolor{violet}{CORSIKA}) and \textcolor{gray}{experimental
data} processing is colour-coded respectively. Chain is also subdivided
into stages (levels), starting from the sea level all the way to the
reconstruction level. For MC generators the produced particle types
are indicated. \label{fig:KM3NeT-simulation-chain.}}
\end{figure}

The muon simulations in KM3NeT are typically multi-staged, since after
the muons are generated, they still need to be propagated through
seawater to the detector. Afterwards, the light emission and its subsequent
detection by the PMTs is simulated. A schematic structure of the KM3NeT
muon simulation and data processing software chain is depicted in
Fig. \ref{fig:KM3NeT-simulation-chain.}. All of the following subsections
delve into more detail of each of the processing stages, also called
levels.

\subsection{Sea level\label{sec:sea}}

In context of KM3NeT, the output of CORSIKA simulation is uniquely
at the sea level. The particles are stored at the sea surface, taking
into account the curvature of the Earth, which can be non-negligible
for high-energy horizontal showers \cite{CORSIKA-Userguide}. The
sole reason why the observation level for the KM3NeT CORSIKA simulations
is chosen to be the sea surface is that the propagation through mixed
media (i.e. air + seawater) is currently not supported in CORSIKA7.
However, this will change in the coming release of CORSIKA8 \cite{CORSIKA8}.

The standard CORSIKA output is stored in sequential unformatted $\mathtt{\mathtt{FORTRAN}}$
files \cite{CORSIKA-Userguide}. To further work with such data, it
is convenient to convert it to a more suitable format. 

For the older CORSIKA productions (used e.g. for the ARCA2 and ORCA1
results in Chap. \ref{chap:Muon-rate-measurement}), this was achieved
with the internal KM3NeT software called Corant. It was written in
$\mathtt{C++}$ and performed a conversion to the ASCII (text) format,
preliminarily computed the event weights (the event weights are final
only at the can level; see Sec. \ref{sec:weights} and \ref{sec:can})
and translated the particle tracks to the KM3NeT reference system.
The author of this thesis has contributed a number of developments
to the Corant code and is its current maintainer.

An alternative is to process the file to the can level (see next section)
with gSeaGen using the `-write 2` option, which preserves the complete
CORSIKA sea-level information. The user may pick between ASCII and
ROOT output file formats.

\subsection{Can level\label{sec:can}}

The name can\nomenclature{can}{active volume of the KM3NeT detector, by convention equal to instrumented volume increased by 4 light absorption lengths in water}
level stems from the cylindrical shape of the KM3NeT detectors. If
simulation is at the can level, it means that the secondary particles
(muons and/or neutrinos) are inside the active volume of the detector.
This volume is typically the instrumented volume (minimal cylinder
enclosing the detector) enlarged by 4 light absorption lengths in
seawater\footnote{$4\cdot70\,$m$=280\,$m, this value was also used for the CORSIKA
MC}, to take into account the fact that the light emitted before reaching
the detector may still be observed \cite{Optical_properties_light_absorption_length}.

MUPAGE (described in Sec. \ref{sec:MUPAGE}) directly generates muon
bundles from parametrised zenith, energy, and multiplicity distributions
at the can level. It is possible owing to the fact that it was tuned
on data corresponding roughly to ORCA (and ANTARES) depth \cite{MUPAGE}.
This approach has its drawbacks, for instance MUPAGE does not necessarily
describe the muon flux at other depths (e.g. ARCA) equally well. This
is one of the reasons why a more detailed muon simulation was needed.

CORSIKA is such a comprehensive muon MC. As already mentioned in previous
section, the tool allowing to process CORSIKA output is gSeaGen. gSeaGen
is a code, which can generate neutrino events using GENIE \cite{GENIE}
(NOT used in this work) and propagate $\mu$ and $\nu$. It can be
linked to different libraries that perform the propagation: MUSIC
\cite{MUSIC}, PROPOSAL \cite{PROPOSAL}, Jpp \cite{JSirene}, or
use its own built-in routine — PropaMuon \cite{gSeaGen-2020}. Numerous
contributions to gSeaGen software have been made in scope of this
thesis, and some of them are described in \ref{sec:gSeaGen_supplementary_material}.
gSeaGen is publicly available as an open-source code under \href{https://git.km3net.de/opensource/gseagen}{git.km3net.de/opensource/gseagen}.
The program is maintained and developed by the author of this thesis.

\subsection{Light level\label{sec:light}}

At light level, emission of Cherenkov radiation along the path of
muons and detection of the photons by the PMTs are simulated. An instance
when a PMT records a signal exceeding its voltage threshold (adjusted
during the detector calibration) is called a hit. There are three
codes dedicated to light simulation within KM3NeT, however only one
is used for this work, as indicated in Fig. \ref{fig:KM3NeT-simulation-chain.}.

The default choice for most MC productions, including the CORSIKA
MC, is JSirene. It is a custom application developed for KM3NeT, which
uses multidimensional interpolation tables to speed up the computation
in certain routines \cite{JSirene}. JSirene is a part of the Jpp
framework, consisting of many KM3NeT-specific utility functions, programs,
and services.

The first alternative is KM3Sim. It propagates particles in water,
using the Geant4 software \cite{GEANT4}. It comes with all the advantages
and setbacks of Geant4, which means it offers detailed simulation
of various energy loss processes, however the price to be paid is
a rather high CPU time required for computation. 

Km3 is a $\mathtt{FORTRAN}$-based code for parametric simulation
of light propagation, which was historically developed by the ANTARES
experiment \cite{KM3}. The parametrizations have been obtained from
GEANT3 simulations \cite{GEANT3}. Km3 has been adapted from the ANTARES
version to incorporate the characteristics of KM3NeT DOMs and PMTs. 

\subsection{Trigger level\label{sec:trigger}}

Simulation of the event triggers is performed using the JTriggerEfficiency
(JTE\nomenclature{JTE}{JTriggerEfficiency -- software package for simulation of KM3NeT triggers})
package from the Jpp framework \cite{JSirene}. It mimics the environmental
optical background due to the bioluminescence and radioactive decays
of $^{40}$K (in 89.52\% of the cases a $\beta$ decay producing a
1.31~MeV electron \cite{K40}). On top of adding the simulated background,
JTE emulates the detector response, taking into account the effects
of front-end electronics. The trigger-level simulation may be performed
in a run-by-run (rbr\nomenclature{rbr mode}{run-by-run mode: KM3NeT simulation type,taking into account run-specific  conditions})
mode, i.e. reproducing each data taking run separately (which is typically
done for MUPAGE). The alternative is the non-rbr mode, which is used
by default for the CORSIKA MC, since producing sufficient statistics
for each run with CORSIKA would be too costly. The non-rbr MC provides
a prediction of a mean muon flux, with averaged trigger settings and
PMT parameters used throughout the whole simulation. The trigger algorithms
applied to identify potentially interesting events are exactly the
same ones as used for the real data. There is a number of those, corresponding
to different event classes. For the MC productions considered within
this thesis only two were used: 3DMuon and 3DShower, which are triggers
for muon tracks and for shower-like events respectively. After applying
the trigger algorithms, merging of events overlapping in time is performed,
and the number of overlaying events\footnote{Also referred to as overlays, which was one of the features used in
Chap. \ref{chap:muon-bundle-reco}.} is computed. Here it has to be stressed that the definition of an
event in JTE deviates from what is used throughout the rest of this
thesis (1 event being equivalent to 1 EAS). Here, an event is any
collection of triggered hits within a certain time window (the exact
value depends on the trigger algorithm). Such hits should be more
likely to originate from actual particle interactions, however they
still contain a contribution from the experimental background.

\subsection{Reconstruction level\label{sec:reco}}

Since the main focus of the work is on the muons, the most relevant
event topology are tracks, despite the fact that muons can sometimes
undergo catastrophic energy losses and produce cascade-like signatures
in the detectors \cite{CatastrophicEnergyLosses}. Track events are
reconstructed with help of JMuon algorithm and cascades (showers)
are reconstructed using one of the 2 codes: Aashowerfit (used for
ARCA) and JShowerfit (used for ORCA) \cite{JSirene,ARCA-reconstruction}.
In this work, only JMuon has been used. In the following sections,
reconstruction algorithms for different observables are discussed.

\subsubsection{Direction reconstruction}

The direction of an event is inferred from the timing and charge deposition
information from the PMTs by performing a maximum-likelihood fit.
The assumptions for the fit are that the particle propagates with
the speed of light in vacuum, producing Cherenkov light in the detector,
and that the track is a straight line. The fit takes into account
factors like the optical properties of water and properties of the
DOMs \cite{KM3NeT-LoI-2.0}. The best angular resolution is obtained
for tracks (smaller than $1\lyxmathsym{\textdegree}$, depending on
the energy) and the cascades can be reconstructed with a median angular
resolution of $\sim15\lyxmathsym{\textdegree}$. Thanks to the high
energy and hence also velocity, the muon and muon neutrino directions
are almost parallel. The average angular deviation of the muon from
the neutrino direction can be parametrized as $0.7\lyxmathsym{\textdegree}\cdot\frac{\mathsf{TeV}}{E_{\nu}}$
\cite{KM3NeT-LoI-2.0}.

Cascade direction is harder to reconstruct due to the almost spherical
event topology. For ARCA, the fact that the DOM spacing is bigger
than a typical cascade size of 10~m additionally impairs the spatial
resolution \cite{KM3NeT-LoI-2.0}.

In Chap. \ref{chap:Muon-rate-measurement}, the muon bundle zenith
reconstructed with JMuon is compared between the data and simulations
for a number of intermediate configurations of KM3NeT detectors.

\subsubsection{Energy reconstruction\label{subsec:Energy-reconstruction}}

The energy deposited by the muon is proportional to the amount of
the light seen by the PMTs and allows to approximate its total energy.
The method is likelihood maximization, as for directional reconstruction.
Here it has to be stressed that JMuon always operates on the assumption
of a single muon track, even if the event is in fact a high-multiplicity
muon bundle (see Sec. \ref{sec:Extensive-Air-Showers-EAS}), which
affects the reconstruction quality. 

In case of energy reconstruction, the cascades are easier to reconstruct,
since the number of photons per each deposited GeV of energy is approximately
constant over a broad range of energies \cite{KM3NeT-LoI-2.0,ARCA-reconstruction}.
This is not the case for tracks. There are additional complications,
e.g. the fact that for energies above 100~GeV the tracks are often
uncontained (extending beyond the detector volume).

In Chap. \ref{chap:muon-bundle-reco}, the existing JMuon energy reconstruction
is compared against the dedicated muon bundle energy reconstruction,
developed for this work. In addition, both the standard (JMuon) and
new reconstruction are applied to the data vs MC comparisons in Chap.
\ref{chap:Muon-rate-measurement}.

\subsubsection{Other observables}

Aside from muon direction and energy, JMuon code also reconstructs
the muon vertex (the point in space at which the first seen Cherenkov
photon was emitted), however this information is not used in any way
in this work \cite{KM3NeT-LoI-2.0}.

In Sec. \ref{sec:Extensive-Air-Showers-EAS} it was mentioned that
muon bundles posses certain characteristic properties, among them
the incident primary direction and energy, lateral width, multiplicity,
and bundle energy. None of them has an explicit standard reconstruction
within the KM3NeT simulation framework. The reconstruction of the
primary direction, lateral width, and multiplicity was attempted in
the dissertation of Stefan Reck, in parallel to this work \cite{StefanReckThesis}.
In Chap. \ref{chap:muon-bundle-reco}, reconstruction of the multiplicity
and bundle and primary energies is described. The primary energy reconstruction
is especially important from the point of view of CR composition studies.
Chap. \ref{chap:Muon-rate-measurement} shows the application of all
the developed reconstructions to the data. As shown in Sec. \ref{sec:Phase-space-in-reconstructed},
muon multiplicity and bundle energy are observables that are important
for the prompt muon production.

\chapter{Reconstruction of muon bundle observables \label{chap:muon-bundle-reco}}

This chapter shortly introduces the concepts of machine learning (ML\nomenclature{ML}{machine learning})
and presents the reconstruction of muon bundle properties (see Sec.
\ref{sec:Extensive-Air-Showers-EAS}), performed by the author of
this thesis using ML tools. Prediction of three characteristics was
attempted: bundle energy $E_{\mathsf{bundle}}$, total primary energy
$E_{\mathsf{prim}}\cdot A$ (where $A$ is the number of nucleons
in the nucleus), and muon multiplicity $N_{\mu}$. A number of ML
regression models were tested and the best one was further tuned and
used for all the reconstructions. The $E_{\mathsf{bundle}}$ reconstruction
results were compared against the existing KM3NeT energy reconstruction
JMuon (see Sec. \ref{subsec:Energy-reconstruction}).

\section{Machine learning: a brief introduction \label{sec:Machine-learning-intro}}

\begin{figure}[H]
\centering{}\includegraphics[width=16cm]{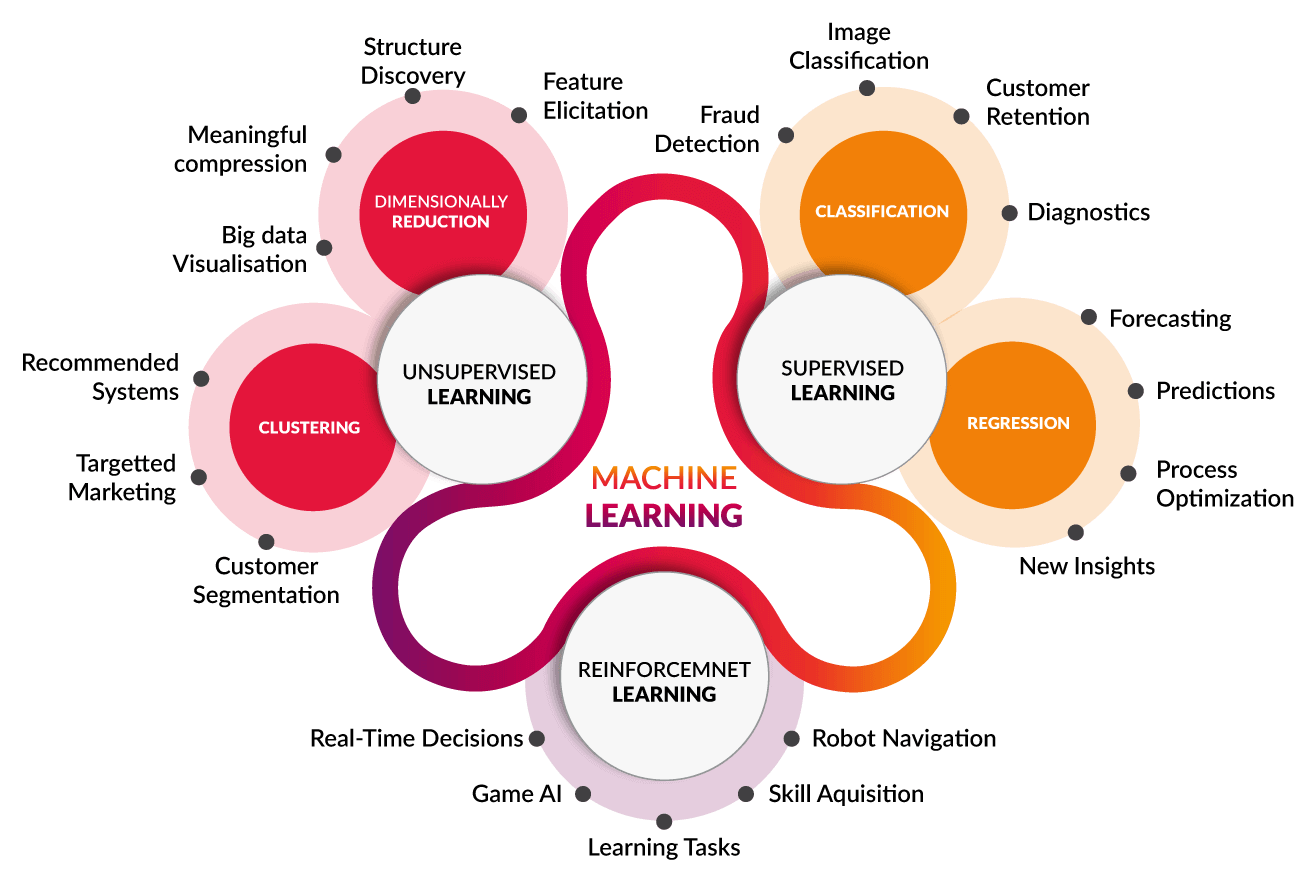}\caption{Diagram depicting main types of machine learning methods and some
of their applications. Taken from \cite{ML_types}. \label{fig:ML-types}}
\end{figure}

Machine learning is a branch of artificial intelligence where the
codes `learn' to make decisions based on data they have been exposed
to. What is explicitly programmed is not the exact response but rather
a method in which the program, often called a model, will learn the
task. The first ML models date back to 1950s. Their architecture attempted
to reproduce the cognitive mechanisms of the human brain by developing
artificial neural networks (NNs\nomenclature{NN}{artificial neural network})
\cite{Term_MachineLearning,FirstMLpaperIBM,Perceptron,MADALINE_PERCEPTRON}.
Since then, the number of available models and their applications
has grown significantly (see Fig. \ref{fig:ML-types}). The types
of ML tasks can be grouped into three main categories:
\begin{enumerate}
\item Supervised learning – predicts the output based on labelled input
data (features).
\begin{enumerate}
\item Classification – predicts categorical labels (e.g. `cat', `dog').
In the case of binary classification (classification with only two
classes), typically one of the labels (classes) is of particular interest
and such a label is referred to as positive label, and analogously
the other one is called negative.
\item Regression – predicts values of a continuous target, e.g. energy of
a particle.
\end{enumerate}
\item Unsupervised learning – learns the structure of the data based on
unlabelled input.
\begin{enumerate}
\item Clustering – looks for groups (clusters) of data points, based on
their similarities and differences.
\item Dimensionality reduction – applies a transformation to the data, preserving
as much information as possible, while shrinking the number of features.
\end{enumerate}
\item Reinforcement learning – puts an agent in an interactive environment,
where it learns by trial and error. The agent receives positive or
negative feedback each time it performs an action. The learning consists
in maximization of the cumulative reward (positive feedback from the
environment) by the agent. The resulting strategy of actions to be
taken in a particular state is called the policy.
\end{enumerate}

The presented categorisation of ML is in fact not the only one in
use. An another way to divide machine learning models is by their
architecture, with the two main types:
\begin{enumerate}
\item Deep learning – the model is a neural network with multiple layers
of neurons. The exact number of layers required for the NN to be considered
deep is as a matter of fact somewhat fluid, however within this work,
the threshold of 2 layers was adopted.
\item Shallow learning – by construction, anything which is not considered
`deep': single-layer NNs and other types of models, e.g. support
vector machines \cite{ML_types,IntroductionToML,ML_encyclopedia}.
\end{enumerate}
In this chapter, application of supervised shallow learning (with
an exception of multi-layer perceptron, see Sec. \ref{subsec:Considered-models})
to regression tasks is presented. In Sec. \ref{subsec:Selection-of-the-best-model-energy}
a single best-performing model was identified and used throughout
the entire chapter to address each of the considered problems.

\subsection{General workflow \label{subsec:General-workflow-ML}}

In all machine learning tasks, there is a common set of steps, which
have to be completed to obtain useful results:
\begin{enumerate}
\item Collecting the data (in the case of this work, from the MC simulation).
\item Data preprocessing:
\begin{enumerate}
\item Handling issues like (partially) missing data points, noise, or inconsistencies
in the data.
\item Scaling the features and weights to make them more convenient to work
with for the models. Too large values or too wide value ranges can
lead to numerical problems in most ML algorithms.
\item Splitting the data into the train, validation, and test sets, typically
reserving the majority of the samples for the training. The difference
between the validation and the test dataset is that the model is tuned
on the validation set, while the test set is only used for the final
results. This means that the test data does not influence the training
of the model in any way and hence, allows for an unbiased evaluation
of the model performance.
\end{enumerate}
\item Finding the optimal model for the task. This may include both researching
available options and directly comparing performance of models.
\item Training and validating the model performance.
\item Tuning the internal parameters of the model, called hyperparameters
(for instance: learning rate). It should be undoubtedly done at the
very end, since it involves repeated re-training, while scanning for
the optimal combination of hyperparameters and is hence computationally
intensive.
\item Evaluation of the results — it is verified, whether the model can
generalise well, i.e. if it can work well also on the data it has
not seen before (the test data).
\end{enumerate}
In practice, many of the steps require re-iteration upon completion
of the other ones. For example, the hyperparameter tuning involves
multiple training and validation steps, and preliminary validation
results can often lead to modifications, e.g. in- or exclusion of
certain models.

\subsection{Considered models\label{subsec:Considered-models}}

In this work, the task was the regression of KM3NeT detector observables,
based on the CORSIKA MC, which is a type of supervised learning. To
this end, a variety of ML algorithms available in the scikit-learn
library \cite{Scikit-learn} and outside it \cite{XGBoost,LightGBM}
were tested. They can be grouped into several categories:
\begin{enumerate}
\item Linear models: they assume that the reconstructed target $y_{\mathrm{pred},i}$
for $i$-th sample is merely a linear combination of $n$ features
$x_{i}=\left(x_{1,i},...,x_{n,i}\right)$ :
\[
y_{\mathrm{pred},i}=c_{0,i}+c_{1,i}\cdot x_{1,i}+...+c_{n,i}\cdot x_{n,i},
\]
where $c_{i}=\left(c_{0,i},...,c_{n,i}\right)$ are constant coefficients,
which must be fitted.
\begin{enumerate}
\item Linear — an implementation of the least squares algorithm. It minimizes
the sum of squared residuals: 
\[
\underset{c}{\min}\left[\stackrel[i=1]{m}{\sum}\left(y_{\mathrm{true},i}-y_{\mathrm{pred},i}\right)^{2}\right],
\]
where $y_{\mathrm{true},i}$ and $y_{\mathrm{pred},i}$ are the true
and predicted (reconstructed) values respectively.
\item Ridge — modifies Linear by adding a penalty term:
\[
\underset{c}{\min}\left[\stackrel[i=1]{m}{\sum}\left(y_{\mathrm{true},i}-y_{\mathrm{pred},i}\right)^{2}+\alpha\cdot\stackrel[i=0]{m}{\sum}\stackrel[j=0]{n}{\sum}c_{j,i}^{2}\right],
\]
 to prevent too large linear coefficients. The regularisation parameter
$\alpha\geq0$ controls the strength of coefficient shrinking.
\item ARD — stands for Automatic Relevance Determination\nomenclature{ARD}{automatic relevance determination}.
It is a type of Bayesian regression, where the regularisation parameters
are not fixed, but tuned to the data presented to the model. In the
particular case of ARD, the priors on $c_{i}$ are drawn from a Gaussian
distribution, with zero mean and standard deviation $\lambda_{i}$:
\[
p\left(\left.c_{i}\right|\lambda_{i}\right)=\mathcal{N}\left(\left.c_{i}\right|0,\lambda_{i}\right),
\]
where $\mathcal{N}\left(\mu,\sigma\right)$ is the Gauss distribution
(often called normal distribution) with a mean $\mu$ and standard
deviation $\sigma$.
\item Lasso (least absolute shrinkage and selection operator) — a linear
model designed to estimate sparse coefficients (i.e. most of them
will be just zero). It does so by minimising:
\[
\underset{c}{\min}\left[\frac{1}{2n_{\mathsf{samples}}}\stackrel[i=1]{m}{\sum}\left(y_{\mathrm{true},i}-y_{\mathrm{pred},i}\right)^{2}+\alpha\cdot\rho\stackrel[i=0]{m}{\sum}\stackrel[j=0]{n}{\sum}c_{j,i}\right].
\]
\item ElasticNet — a combination of Lasso and Ridge, allowing to learn a
sparse model with only few non-zero coefficients, while keeping the
regularisation properties of Ridge. The minimised function is:
\[
\underset{c}{\min}\left[\frac{1}{2m}\stackrel[i=1]{m}{\sum}\left(y_{\mathrm{true},i}-y_{\mathrm{pred},i}\right)^{2}+\alpha\cdot\rho\stackrel[i=0]{m}{\sum}\stackrel[j=0]{n}{\sum}c_{j,i}+\frac{\alpha\cdot\left(1-\rho\right)}{2}\stackrel[i=0]{m}{\sum}\stackrel[j=0]{n}{\sum}c_{j,i}^{2}\right].
\]
\item Lars\nomenclature{LARS}{least-angle regression} (least-angle regression)
— a regression algorithm developed specifically for data with high
dimensionality. It iterates over the following steps:
\begin{enumerate}
\item Initialise all the coefficients to zero: $c_{j,i}=0$.
\item Find the feature $x_{j,i}$ with the strongest correlation to corresponding
$y_{\mathrm{true},i}$.
\item Increase $c_{j,i}$ in the direction of the sign of correlation with
$y_{\mathrm{true},i}$. Stop when some other feature $x_{k,i}$ has
the same correlation with $y_{\mathrm{true},i}-y_{\mathrm{pred},i}$,
as $x_{j,i}$ has.
\item Increase $c_{j,i}$, $c_{k,i}$ in their joint least squares direction,
until some other feature $x_{l}$ has as much correlation with $y_{\mathrm{true},i}-y_{\mathrm{pred},i}$
as they do. 
\item Do the same for $c_{j,i},c_{k,i},c_{l,i}$
\item Carry on until all features are included in the model with non-zero
coefficients.
\end{enumerate}
\item LassoLars v a combination of Lasso and Lars approaches.
\item LassoLarsIC — a version of LassoLars, which uses the Akaike information
criterion (AIC\nomenclature{AIC}{Akaike information criterion}) as
minimisation objective, computed as:
\[
AIC=m\cdot\log\left(2\pi\frac{\stackrel[i=1]{m}{\sum}\left(y_{\mathrm{true},i}-y_{\mathrm{pred},i}\right)^{2}}{m-n}\right)+m+n,
\]
where $n$ is the number of features and $m$ is the number of samples.
\item OrthogonalMatchingPursuit — implements the orthogonal matching pursuit
(OMP\nomenclature{OMP}{orthogonal  matching pursuit}) algorithm \cite{OrthogonalMatchingPursuit_OMP},
designed to work with sparse data. It approximates the optimal solution
for a fixed number of $c_{j,i}\neq0$, by default: $0.1\cdot n$.
\item SGD\nomenclature{SGD}{stochastic gradient descent} (Stochastic Gradient
Descent) — makes use of the stochastic gradient descent \cite{ML_encyclopedia}
method to train the model (minimise the given objective function).
In principle, it can attempt to solve a number of different minimisation
problems. Here, the Ridge regularisation was used.
\item PassiveAggressive — the algorithm starts by initialising all the $c_{j,i}$
coefficients to zero, to then learn them one by one. After each prediction,
a loss is applied instantaneously:
\[
\ell_{\epsilon}\left(y_{\mathrm{true},i}-y_{\mathrm{pred},i}\right)=\left\{ \begin{array}{cc}
0 & \mathrm{if\,\left|y_{\mathrm{true},i}-y_{\mathrm{pred},i}\right|\leq\epsilon}\\
\left|y_{\mathrm{true},i}-y_{\mathrm{pred},i}\right|-\epsilon & \mathrm{otherwise}
\end{array}\right.,
\]
where $\epsilon>0$ controls the accuracy \cite{PassiveAggresive}.
The loss is applied by modifying the coefficients from the current
iteration $t$:
\[
c_{t+1}=c_{t}+\mathrm{sign\left(y_{\mathrm{true}}-y_{\mathrm{pred},t}\right)\frac{\ell_{\epsilon}\left(y_{\mathrm{true},i}-y_{\mathrm{pred},i}\right)}{\stackrel[i=1]{m}{\sum}x_{t,i}^{2}+\frac{1}{2C}}x_{t}},
\]
where $C>0$ is the aggressiveness parameter. This is carried out
until coefficients of all features were set.
\item Robust regressors: try to fit a model on corrupt data (e.g. containing
outliers). Three of such algorithms were used:
\begin{enumerate}
\item RANSAC\nomenclature{RANSAC}{random sample consensus} (RANdom SAmple
Consensus) — iteratively fits a model to random subsets of datapoints
identified as inliers. The inliers are picked by computing the residuals
$y_{\mathrm{true},i}-y_{\mathrm{pred},i}$ and comparing them against
a residual threshold parameter $\epsilon$, requiring $\left|y_{\mathrm{true},i}-y_{\mathrm{pred},i}\right|<\epsilon$
\cite{RANSAC}.
\item TheilSen — it is characterised by robustness against multivariate
outliers, owing to the use of spatial median (a multi-dimensional
generalisation of the median) \cite{TheilSen}.
\item HuberRegressor — it suppresses the outliers by assigning them lower
coefficients. The minimised function is
\[
\underset{c,\lambda}{\min}\left[\stackrel[i=1]{m}{\sum}\left(\lambda+\lambda\cdot H_{\epsilon}\left(\frac{y_{\mathrm{true},i}-y_{\mathrm{pred},i}}{\lambda}\right)\right)+\alpha\stackrel[i=0]{m}{\sum}\stackrel[j=0]{n}{\sum}c_{j,i}^{2}\right],
\]
where 
\[
H_{\epsilon}\left(z\right)=\left\{ \begin{array}{cc}
z^{2} & \mathrm{if\,\left|z\right|<\epsilon}\\
2\epsilon\left|z\right|-\epsilon^{2} & \mathrm{otherwise}
\end{array}\right.,
\]
with a fixed $\epsilon$ parameter, acting as a threshold for classification
as in- or outlier, similarly as for RANSAC \cite{HuberRegressor}.
\end{enumerate}
\end{enumerate}
\item Neighbours-based models: they `memorise' the data they have been
trained on, often in a transformed form. Based on this memory, they
allow assigning new datapoints to already existing clusters. 
\begin{enumerate}
\item KNeighbors — an implementation of the $k$-nearest neighbours algorithm
(kNN\nomenclature{kNN}{k-nearest neighbours algorithm}), which predicts
the target based on the interpolation of the $k$ nearest neighbours
\cite{kNN}. The distance definition is in general adjustable, but
the one used in this work was the standard Euclidean distance.
\end{enumerate}
\item Neural networks: consist of artificial neurons, mimicking the function
of the human brain, as described in Sec. \ref{sec:Machine-learning-intro}.
\begin{enumerate}
\item MLP (multi-layer perceptron\nomenclature{MLP}{multi-layer perceptron })
— one of the most simple NN architectures. It consists of three or
more layers: one input representing the features, one output representing
the target, and at least one non-linear hidden layer. In this thesis,
a MLP with 3 hidden layers with 100, 50 and 25 neurons was used. Each
neuron in the hidden layer transforms the values from the previous
layer as a weighted sum.
\end{enumerate}
\item Decision Trees: models, which predict their target based on a set
of decision rules, learned from the training data. One can think of
a decision tree as a piecewise constant approximation of the problem.
Each split of the decision tree is evaluated in terms of its quality,
i.e. how much it will improve the prediction. 
\item Ensemble methods: they put together predictions of multiple base estimators
to improve the robustness of results. There are two main approaches:
\begin{enumerate}
\item Averaging — base estimators are built independently, sometimes even
in parallel, and their results are averaged. This helps to reduce
the variance of the final result.
\begin{enumerate}
\item RandomForest — creates a 'forest' of decision trees, built from samples
of the training data drawn with replacement. Here, the randomisation
not only reduces variance, but also prevents overfitting (see Sec.
\ref{subsec:Performance-assessment}), which is a common issue of
single decision trees \cite{RandomForests}. 
\item ExtraTrees — extremely randomized trees model goes even further in
randomisation than RandomForest, sacrificing the control over bias
to decrease the variance. Instead of looking for the most discriminative
split thresholds, they are picked randomly for each considered feature
and the best of those becomes the next split \cite{ExtraTrees}.
\item Bagging — a form of meta-estimator, which adds an additional layer
of averaging. It creates several separate instances of a chosen estimator
(in the case of this work, ExtraTrees), and produces a combined prediction
out of those \cite{Bagging}.
\end{enumerate}
\item Boosting: here, the base estimators are built in sequence and they
attempt to reduce the bias of the final result.
\begin{enumerate}
\item AdaBoost — fits a sequence of single estimators (learners), e.g. decision
trees, on different variants of the training data. The data is modified
by applying weights to training samples at each so-called boosting
stages. The initial weights are set to a constant value, but in the
next iterations the weights are reduced for well-predicted and increased
for incorrectly predicted samples. This forces the learners to focus
on the problematic cases, which were missed in the previous stage.
The final prediction is obtained by weighted majority vote or sum
of all estimators \cite{Boosting,Adaboost}.
\item Gradient Boosted Decision Trees (GBDT\nomenclature{GBDT}{gradient boosted decision trees})
— a boosting algorithm able to work with any chosen differentiable
loss function. The name of the model comes from the fact that at each
boosting stage a decision tree is fit on the negative gradient the
loss function. This means that each consequent tree attempts to improve
on the prediction of the previous one. The base estimator for GBDT
is always a decision tree, unlike AdaBoost where it is merely a popular
choice. \cite{GradientBoosting}
\item LightGBM — a highly optimised implementation of a GBDT developed by
Microsoft \cite{LightGBM}. The model is particularly efficient in
working with large datasets, since it works not directly on samples,
but on their histograms. In addition, LightGBM makes use of two novel
techniques. The first one is gradient-based one-side sampling (GOSS\nomenclature{GOSS}{gradient-based one-side sampling})
— it allows to selectively compute the information gain using only
the data with large gradients. The second method is exclusive feature
bundling (EFB\nomenclature{EFB}{exclusive feature bundling}), which
is in essence a type of feature selection. It bundles together mutually
exclusive features (such features will rarely be $\neq0$ at the same
time), allowing to reduce the dimensionality of the problem.
\item XGBoost (Extreme Gradient Boosting) — a competitor of LightGBM, with
a preference for growing deeper trees with more leaves. It also does
not histogram the data. This is reflected in strong performance, but
for the price of inferior memory efficiency and speed (see Fig. \ref{fig:energy_model_comparison}
and \ref{fig:energy_model_comparison-speed}), which can pose problems
for larger datasets \cite{XGBoost}.
\item HistGradientBoosting — is a scikit-learn \cite{Scikit-learn} implementation
of the gradient boosting machine algorithm, inspired by LightGBM.
It has a smaller number of hyperparameters, which makes it potentially
easier to tune.
\end{enumerate}
\end{enumerate}
\end{enumerate}
To limit the number of potential permutations while tuning the performance,
it was decided to pick a single best model early on and apply it to
each of the tasks. The model selection is described in Sec. \ref{subsec:Selection-of-the-best-model-energy}. 

\subsection{Performance assessment \label{subsec:Performance-assessment}}

\begin{figure}[H]
\centering{}\subfloat[Underfitting. \label{fig:underfitting}]{\centering{}\includegraphics[width=5.3cm]{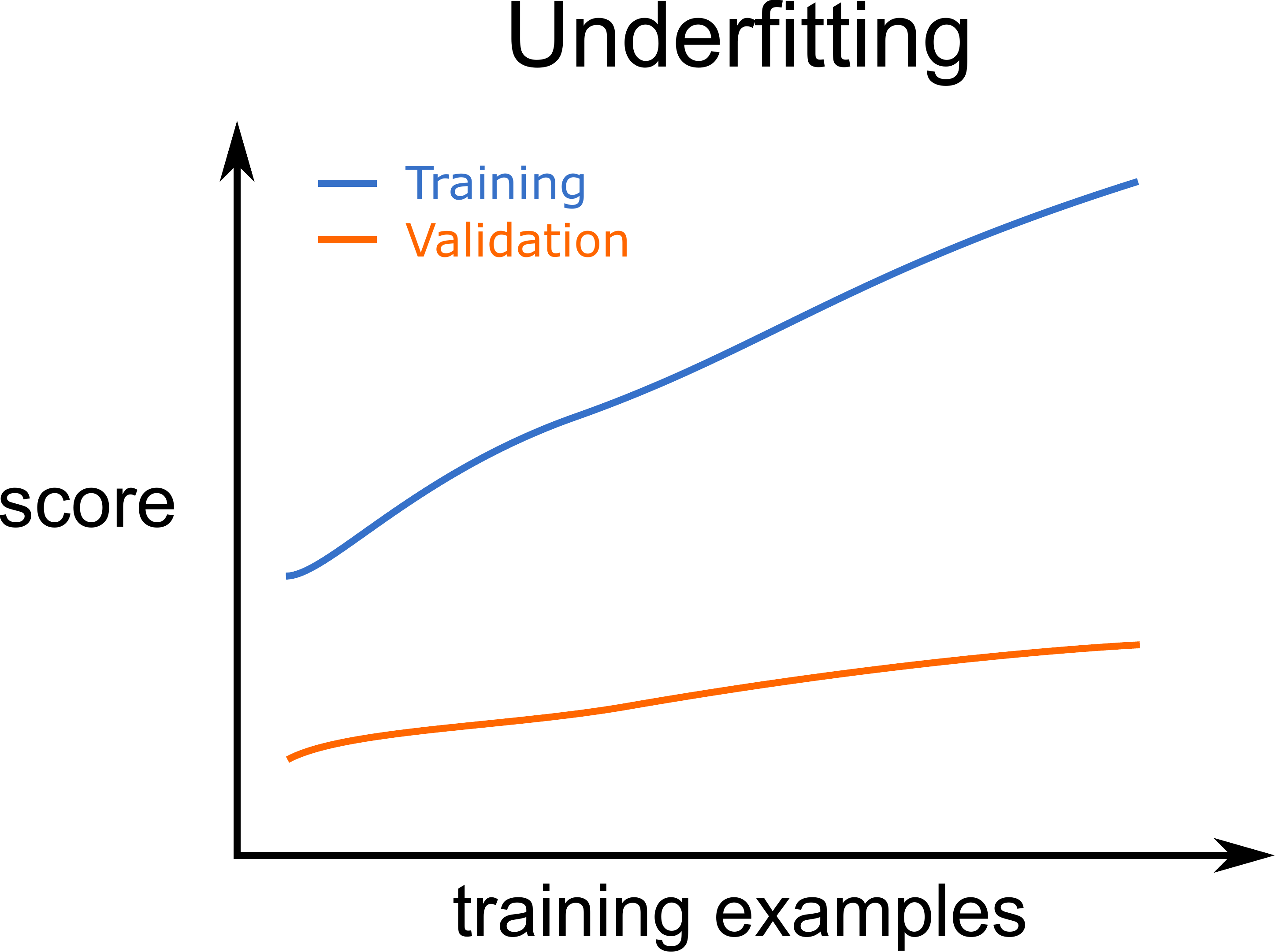}}\subfloat[Good fit. \label{fig:good-fit}]{\centering{}\includegraphics[width=5.3cm]{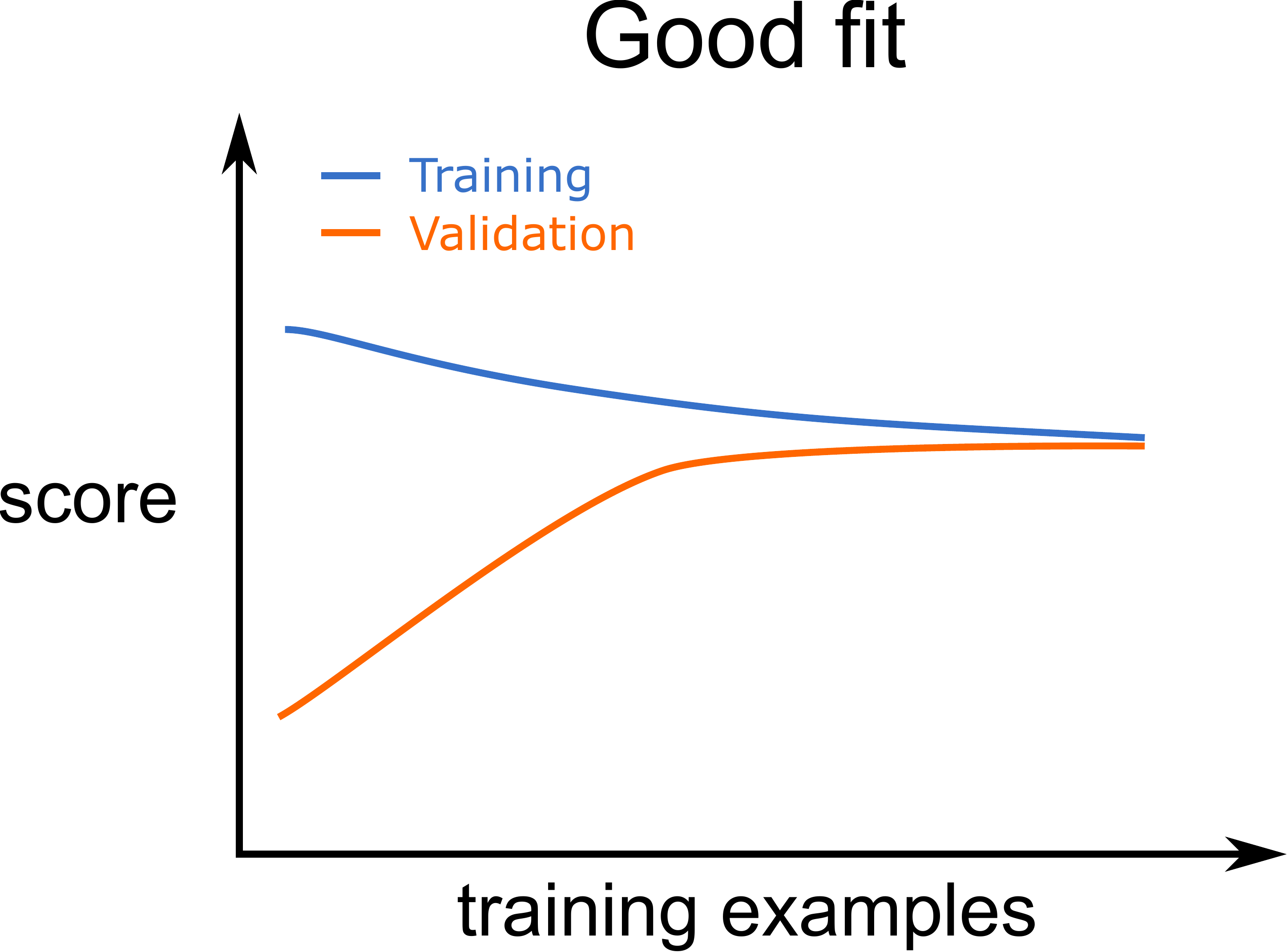}}\subfloat[Overfitting. \label{fig:overfitting}]{\centering{}\includegraphics[width=5.3cm]{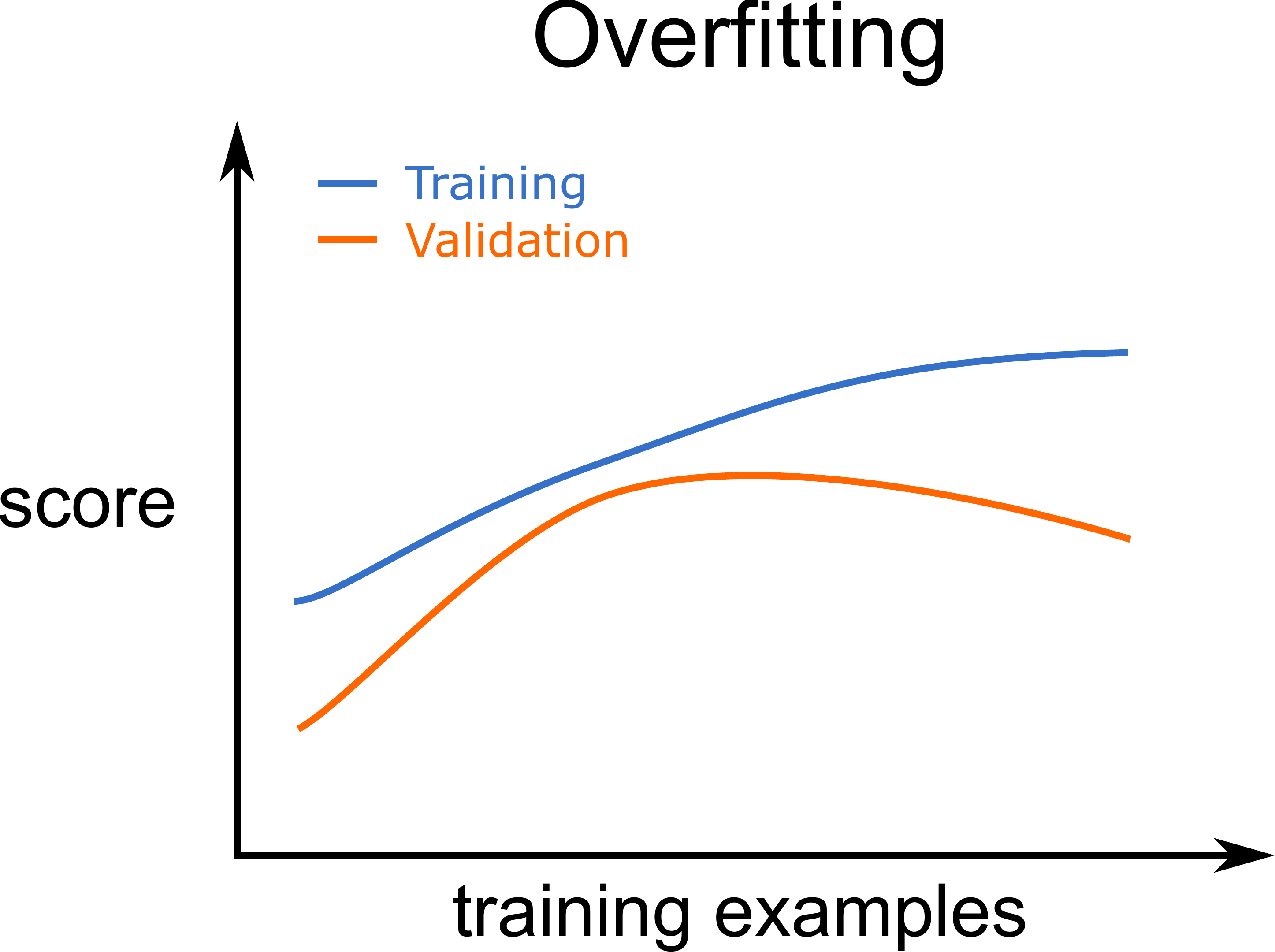}}\caption{Comparison of the three basic behaviours of the learning curves. \label{fig:learning_curves}}
\end{figure}

Inspection of learning curves of the models is a great way to pinpoint
evident issues early on. Learning curves can display either the score
(maximised during training) or the loss function (minimised during
training) versus the number of training examples (events) seen by
the model. Sketches of a few possible cases are shown in Fig. \ref{fig:learning_curves}.
Underfitting means that the model is not able to learn the training
dataset. If the validation score maintains the ascending trend (Fig.
\ref{fig:underfitting}), adding more training examples may be beneficial.
Otherwise, a change or modification of the model to increase its complexity
may be advisable. In the case of a good fit (Fig. \ref{fig:good-fit}),
adding further training data will not yield any significant gain and
may actually lead to overfitting and hence, a loss of generalisability.
The characteristic trait of a learning curve of an overfitted model
(Fig. \ref{fig:overfitting}) is the deflection point in the validation
score, after which it starts to decrease.

To evaluate how well a trained estimator (model) performs, a set of
metrics has to be defined. Here, only the relevant metrics will be
discussed, for a more complete overview see e.g. \cite{Scikit-learn,ML_encyclopedia,Classification_performance_measures}.
To judge if a regression attempt was successful, one should first
look at the correlation plot between the true and predicted values
of a target (see e.g. Fig. \ref{fig:untuned_Ebundle_LightGBM_reco}).
There is a variety of metrics for goodness of a regression \cite{RegressionMetrics,IntroductionToML}.
In this work, two were used:
\begin{enumerate}
\item The (weighted) Pearson correlation coefficient:
\begin{equation}
c\left(y_{\mathrm{true}},y_{\mathrm{pred}}\right)=\frac{\underset{i}{\sum}w_{i}\left(y_{\mathrm{true},i}-\bar{y}_{\mathrm{true}}\right)\left(y_{\mathrm{pred},i}-\bar{y}_{\mathrm{pred}}\right)}{\sqrt{\underset{i}{\sum}w_{i}\left(y_{\mathrm{true},i}-\bar{y}_{\mathrm{true}}\right)^{2}\underset{i}{\sum}w_{i}\left(y_{\mathrm{pred},i}-\bar{y}_{\mathrm{pred}}\right)^{2}}},
\end{equation}
where $y_{\mathrm{true}}$ are the true, and $y_{\mathrm{pred}}$
are the predicted values of a random variable, for which the correlation
is tested. Consequently, $\bar{y}_{\mathrm{true}}$ and $\bar{y}_{\mathrm{pred}}$
are their respective mean values. The weight $w_{i}$ is the event
weight $w_{\mathrm{event}}$ for the $i$-th event. 
\item The (weighted) $R^{2}$-score, called the coefficient of determination:
\begin{equation}
R^{2}\left(y_{\mathrm{true}},y_{\mathrm{pred}}\right)=1-\frac{\underset{i}{\sum}w_{i}\left(y_{\mathrm{true}}-y_{\mathrm{pred}}\right)^{2}}{\underset{i}{\sum}w_{i}\left(y_{\mathrm{true}}-\bar{y}_{\mathrm{true}}\right)^{2}}.\label{eq:R2-score}
\end{equation}
\end{enumerate}
One should note that Eq. \ref{eq:R2-score} does allow negative values,
despite having a square in its symbol. Such values indicate that the
regression is performing worse than just using the mean value ($R^{2}=0$
would imply $y_{\mathrm{pred}}=\bar{y}$). Both the Pearson correlation
coefficient and the coefficient of determination are quoted in all
the correlation plots and 1D distributions in Sec. \ref{sec:Energy}
and \ref{sec:Multiplicity}.

\section{Monte Carlo samples used\label{sec:MC-samples-used}}

\begin{figure}[H]
\centering{}\includegraphics[width=16cm]{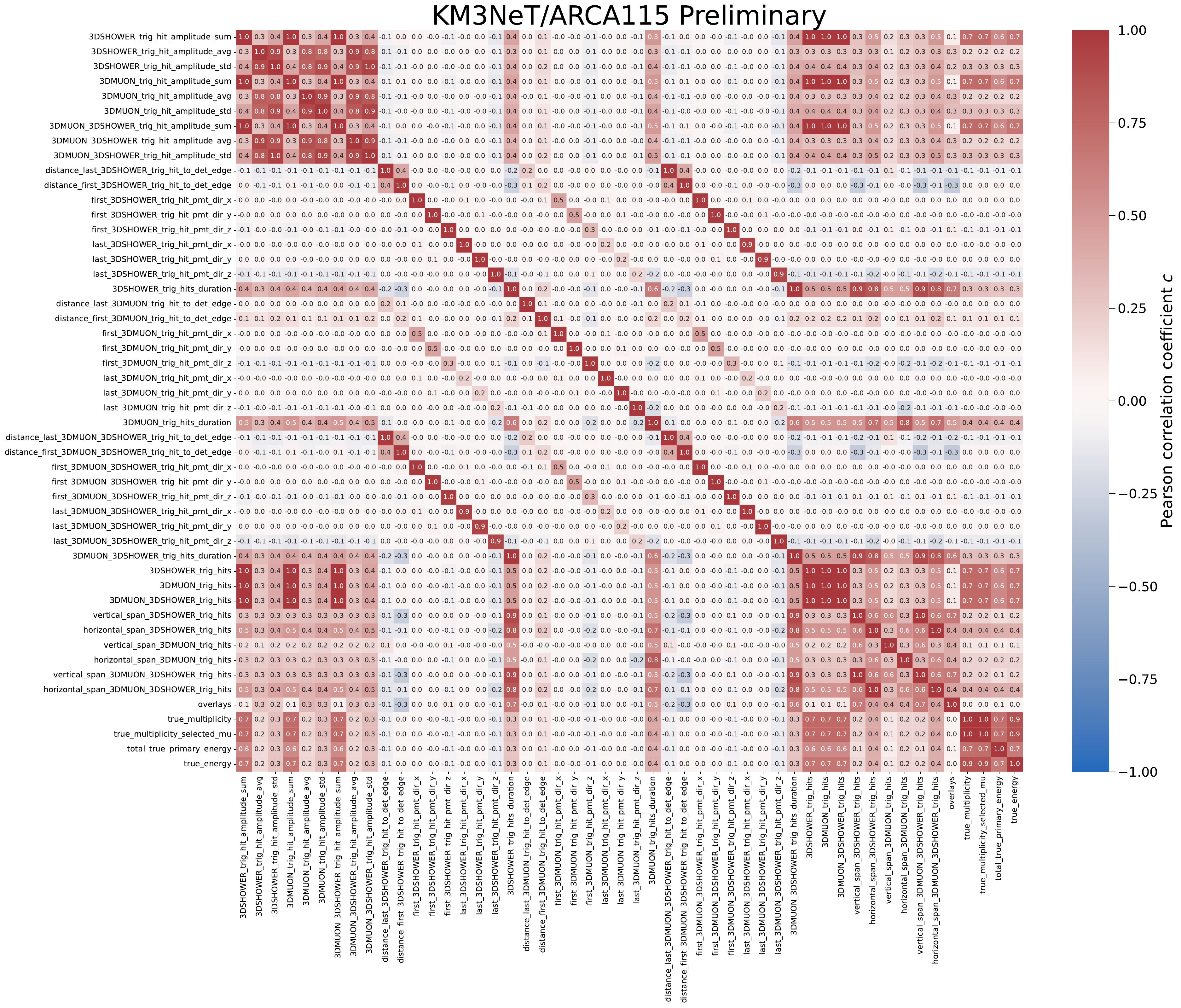}\caption{Correlation matrix of features and targets for ARCA115 MC. Red colour
indicates correlation, blue — anti-correlation. The correlation matrices
for the other three detector configurations may be found in Sec. \ref{subsec:Correlation-matrices-and-dendrograms}.
\label{fig:correlation-matrix}}
\end{figure}

As indicated in Sec. \ref{subsec:General-workflow-ML}, the training
data was obtained from the CORSIKA simulation. The results presented
in the following sections are based on the MC produced for ARCA115,
ARCA6, ORCA115, and ORCA6 (for more details see Sec. \ref{sec:Energy}
and \ref{sec:Additional-material-related-to-CORSIKA}). In each case,
the same set of features, targets, and weights has been extracted.
In fact, the very same features were used both for the multiplicity
and energy reconstruction (Sec. \ref{sec:Multiplicity} and \ref{sec:Energy}).
All the available features, along with the three considered targets
are displayed in Fig. \ref{fig:correlation-matrix} in the form of
a correlation matrix. The features are listed and explained in more
detail in Sec. \ref{subsec:Features-description}. It shows, which
variables (features) may be potentially of greatest importance (strong
correlation with the target) for the reconstruction and how strongly
interconnected they are.

\begin{figure}[H]
\centering{}\includegraphics[width=16cm]{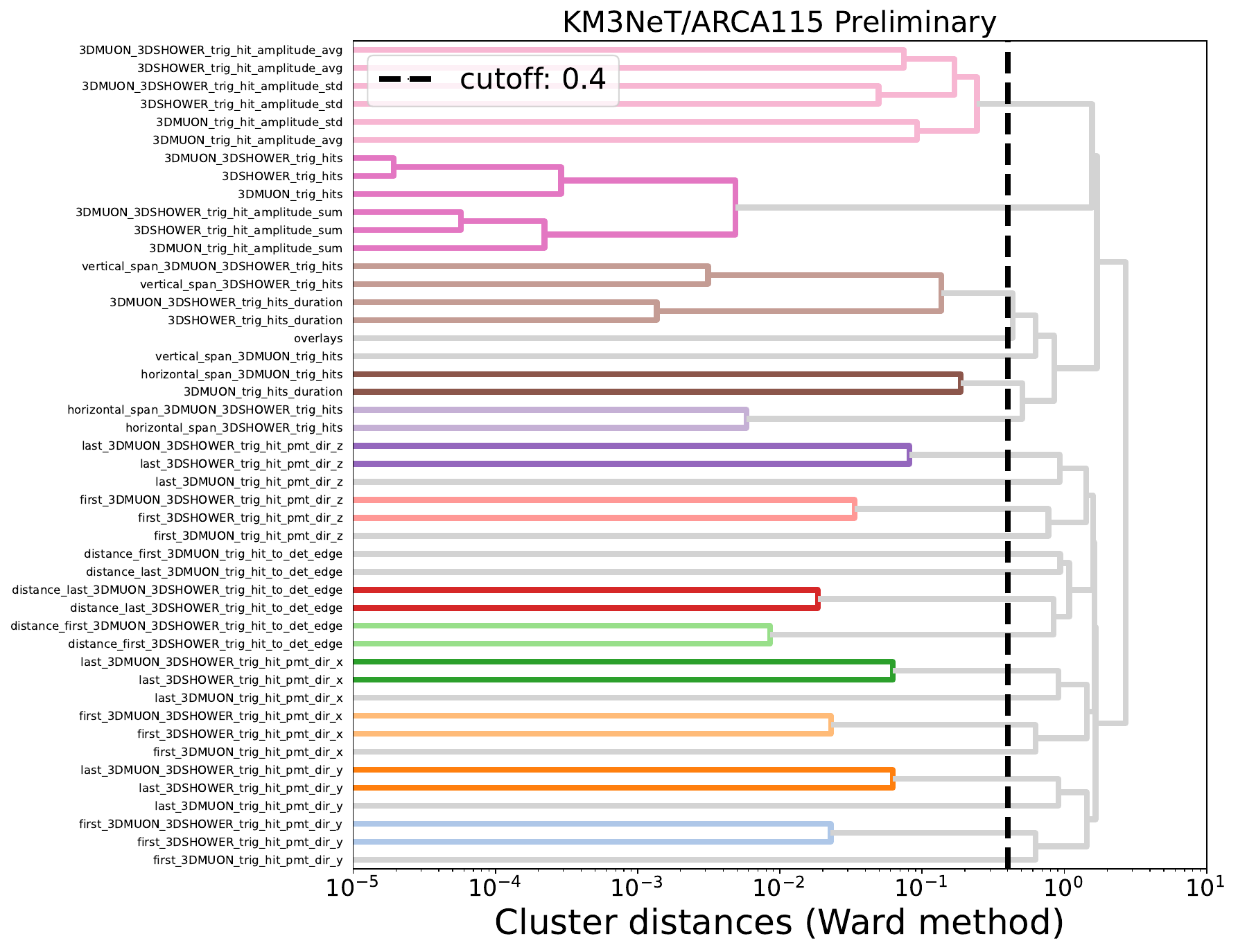}\caption{Visualisation of feature clustering in form of a dendrogram plot,
showing \textit{cluster distances} obtained by computing Ward's linkage
based on the correlation matrix in Fig. \ref{fig:correlation-matrix}
\cite{Ward-linkage-method}. The cutoff for clusters was picked manually
to be 0.4. The image was produced using the ARCA115 CORSIKA MC. The
dendrograms for the other three detector configurations may be found
in Sec. \ref{subsec:Correlation-matrices-and-dendrograms}. \label{fig:dendrogram}}
\end{figure}

To better visualise the intercorrelation of features, it is instructive
to group features into clusters, as shown in Fig. \ref{fig:dendrogram}.
The choice of the \textit{cluster distance} threshold is somewhat
arbitrary. Using the value of 0.4, 23 clusters have been identified.
One could attempt to reduce the number of features by selecting only
the most important ones out of each cluster and putting a cutoff on
the minimal considered feature importance, rejecting any cluster,
landing below that value.

A brief description of all the features may be found in Sec. \ref{subsec:Features-description}.

\subsection{Preprocessing \label{subsec:Preprocessing}}

Before the analysis could be carried out, the datasets for the four
detector configurations were carefully put together. The procedure
was as follows:
\begin{enumerate}
\item The whole dataset was split into a train, validation, and test set
in proportions 64:16:20. Due to the nature of CORSIKA simulation weighting
(see Eq. \ref{eq:event_weights}), the events from a single run could
not be divided. Instead, entire runs were split between the samples.
To avoid accidental mixing of events from different datasets, the
datasets were stored in separate files.
\item Features were transformed to resemble normally distributed data with
a mean of 0 and a variance of 1 (using the StandardScaler utility
implemented in scikit-learn framework\cite{Scikit-learn}). Feature
scaling is a necessity for a number of ML models, which would otherwise
either perform very badly or in some cases run into complete failure.
\item The sample weights (event weights to be used for training, validation,
and testing) were computed in two variants: unscaled $w_{\mathrm{event}}$
(as in Eq. \ref{eq:event_weights}), and scaled according to:
\begin{equation}
w_{\mathsf{scaled}}=\frac{\log_{10}w_{\mathrm{event}}+\max\left(\left|\log_{10}w_{\mathrm{event}}\right|\right)\cdot1.01}{\max\left(\left|\log_{10}w_{\mathrm{event}}\right|\right)},\label{eq:scaled_sample_weights}
\end{equation}
motivated by the fact that their value range spans from roughly $10^{-10}$
up to $10^{14}$, which could cause problems for some models, similarly
to unscaled features with large dynamic ranges. The $\max\left(\left|\log_{10}w_{\mathrm{event}}\right|\right)$
terms ensure that the resulting scaled weights $w_{\mathsf{scaled}}$
have positive values.
\end{enumerate}

\subsection{Summary of datasets \label{subsec:Summary-of-datasets}}

After preprocessing and splitting the CORSIKA MC, the following datasets
were used for the reconstruction tasks:

\begin{table}[H]
\begin{centering}
\caption{Size summary of the datasets used to train the ML models. \label{tab:ML-dataset-summary}}
\par\end{centering}
\centering{}%
\begin{tabular}{|c|c|c|c|}
\hline 
Detector & Training set {[}\# events{]} & Validation set {[}\# events{]} & Test set {[}\# events{]}\tabularnewline
\hline 
\hline 
ARCA115 & 14~374~415 & 3~598~133 & 4~489~177\tabularnewline
\hline 
ARCA6 & 3~894~293 & 974~371 & 1~215~552\tabularnewline
\hline 
ORCA115 & 15~439~946 & 3~857~715 & 4~825~073\tabularnewline
\hline 
ORCA6 & 6~154~895 & 1~539~988 & 1~923~475\tabularnewline
\hline 
\end{tabular}
\end{table}

\section{Reconstruction of energy\label{sec:Energy}}

The energy reconstruction is described first, although the methodology
for multiplicity reconstruction was very similar. Most of the intermediate
results are only presented for ARCA115 for the sake of readability,
however the exactly same procedure was followed for each of the four
detector configurations. 

\subsection{Bundle energy \label{subsec:Bundle-energy}}

\begin{figure}[H]
\centering{}\subfloat[All multiplicities. \label{fig:Jmuon_all}]{\centering{}\includegraphics[width=8cm]{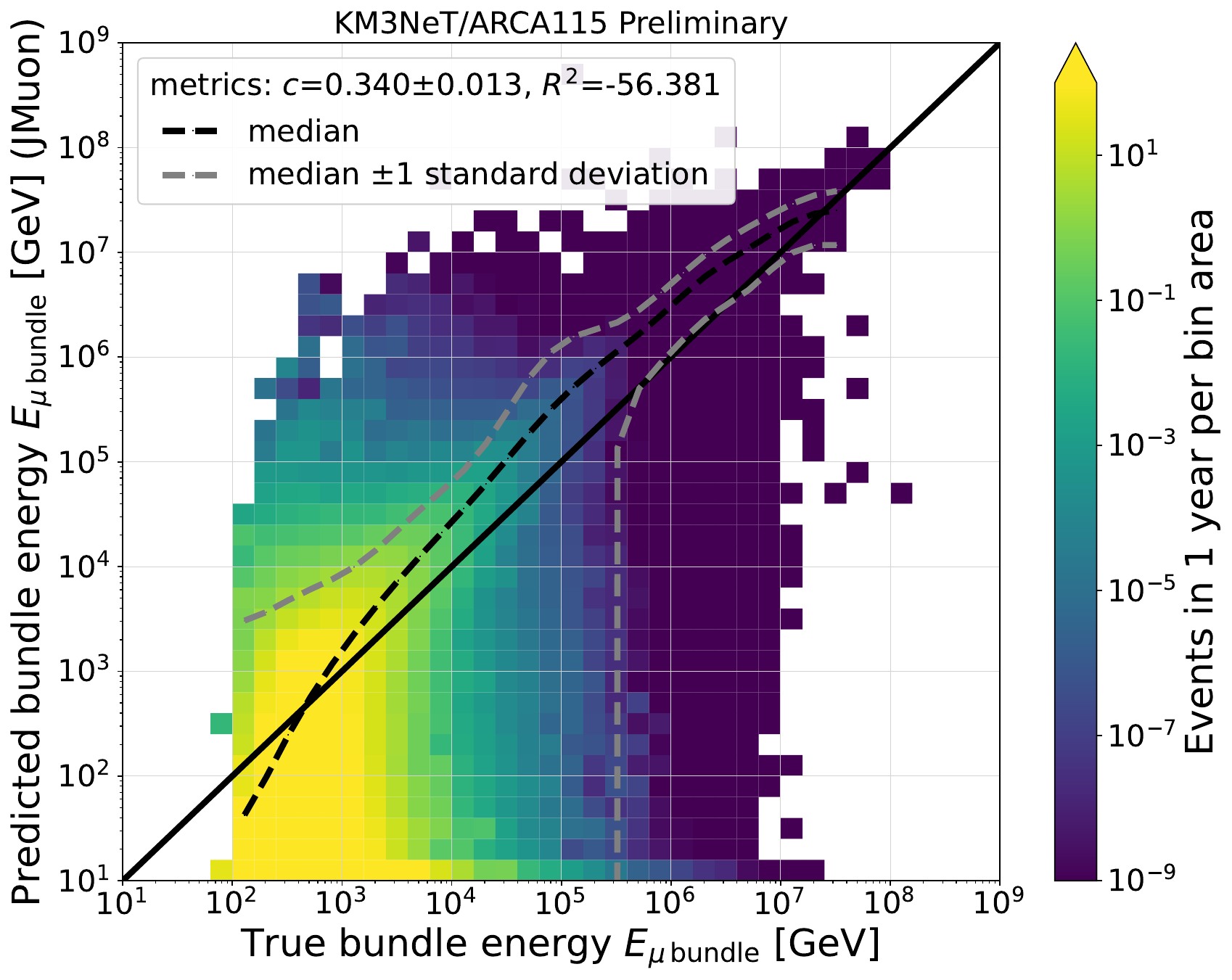}}\subfloat[Single muon events (multiplicity 1). \label{fig:JMuon_single_mu}]{\centering{}\includegraphics[width=8cm]{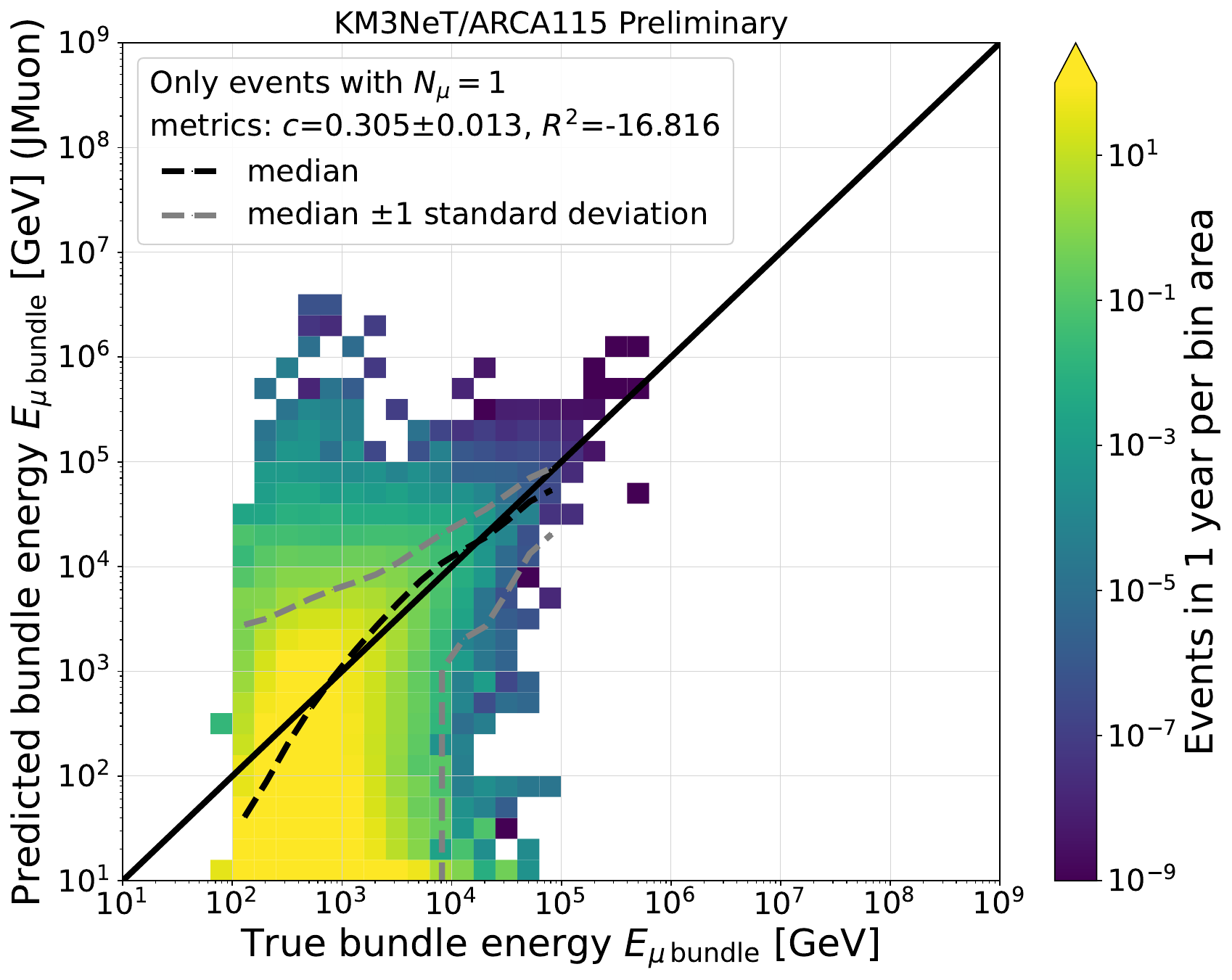}}\caption{Comparison of the energy predicted by the JMuon reconstruction and
true bundle energy. The reconstruction was applied to ARCA115 CORSIKA
MC. Plots for the remaining configurations (ARCA6, ORCA115, ORCA6)
may be found in Sec. \ref{sec:Performance-of-JMuon}. \label{fig:official_E_reco}}
\end{figure}

\begin{figure}[H]
\centering{}\subfloat[All multiplicities. ]{\centering{}\includegraphics[width=8cm]{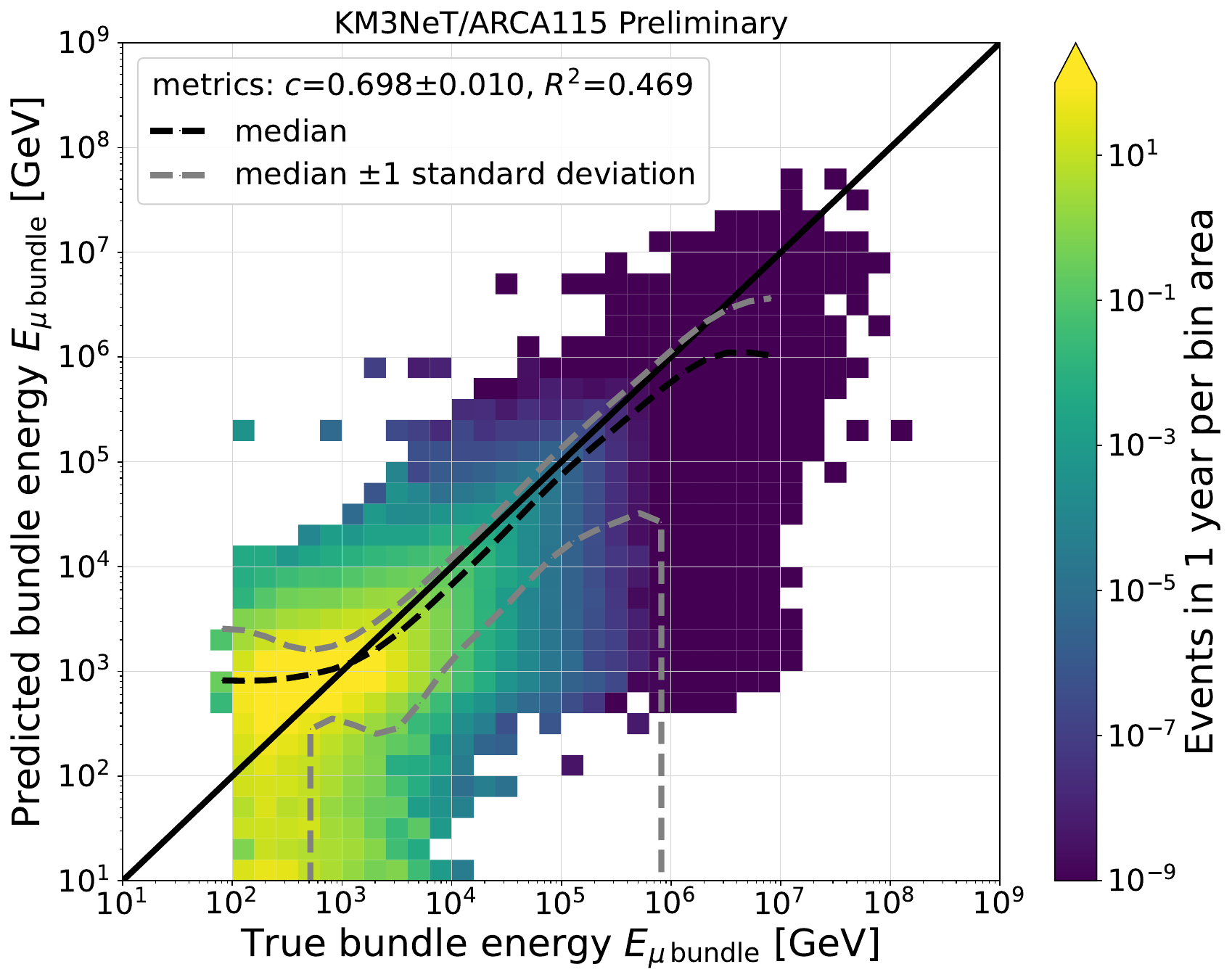}}\subfloat[Single muon events (multiplicity 1).]{\centering{}\includegraphics[width=8cm]{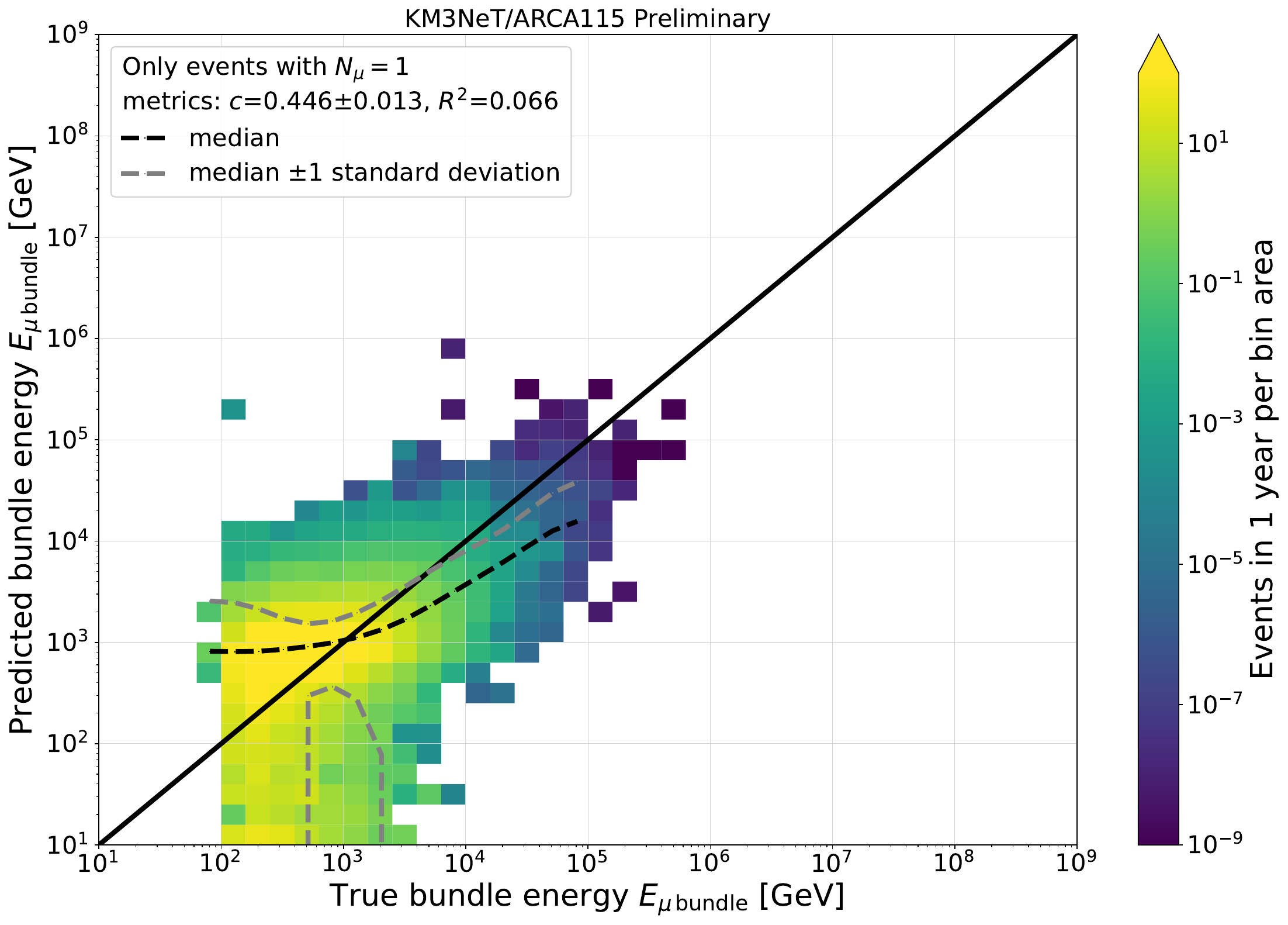}}\caption{Comparison of predicted and true bundle energy for the LightGBM reconstruction.
The reconstruction was applied to ARCA115 CORSIKA MC. \label{fig:untuned_Ebundle_LightGBM_reco}}
\end{figure}

\begin{figure}[H]
\centering{}\subfloat[All multiplicities. \label{fig:Jmuon_all-1}]{\centering{}\includegraphics[width=8cm]{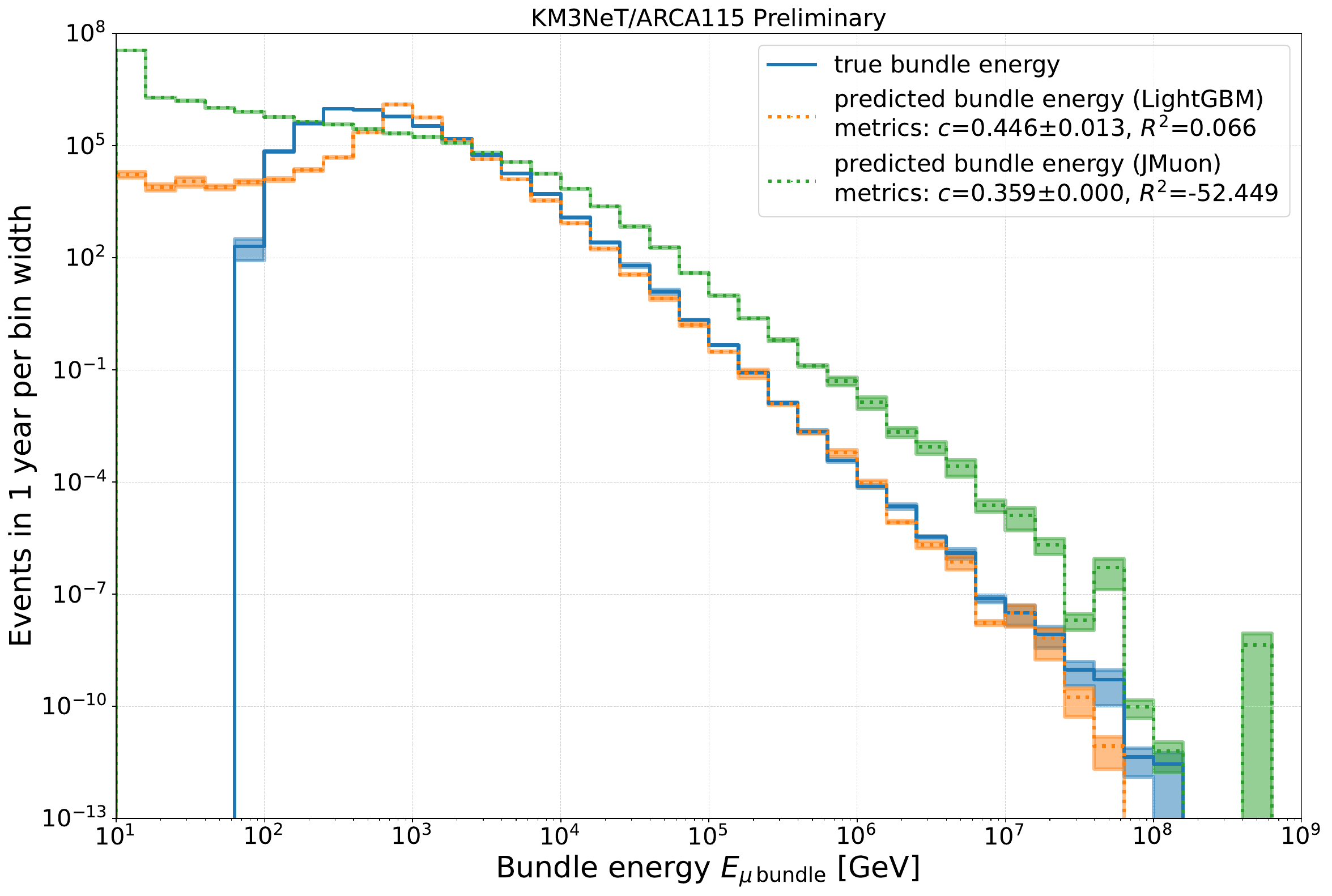}}\subfloat[Only single muon events (multiplicity 1). \label{fig:JMuon_single_mu-1}]{\centering{}\includegraphics[width=8cm]{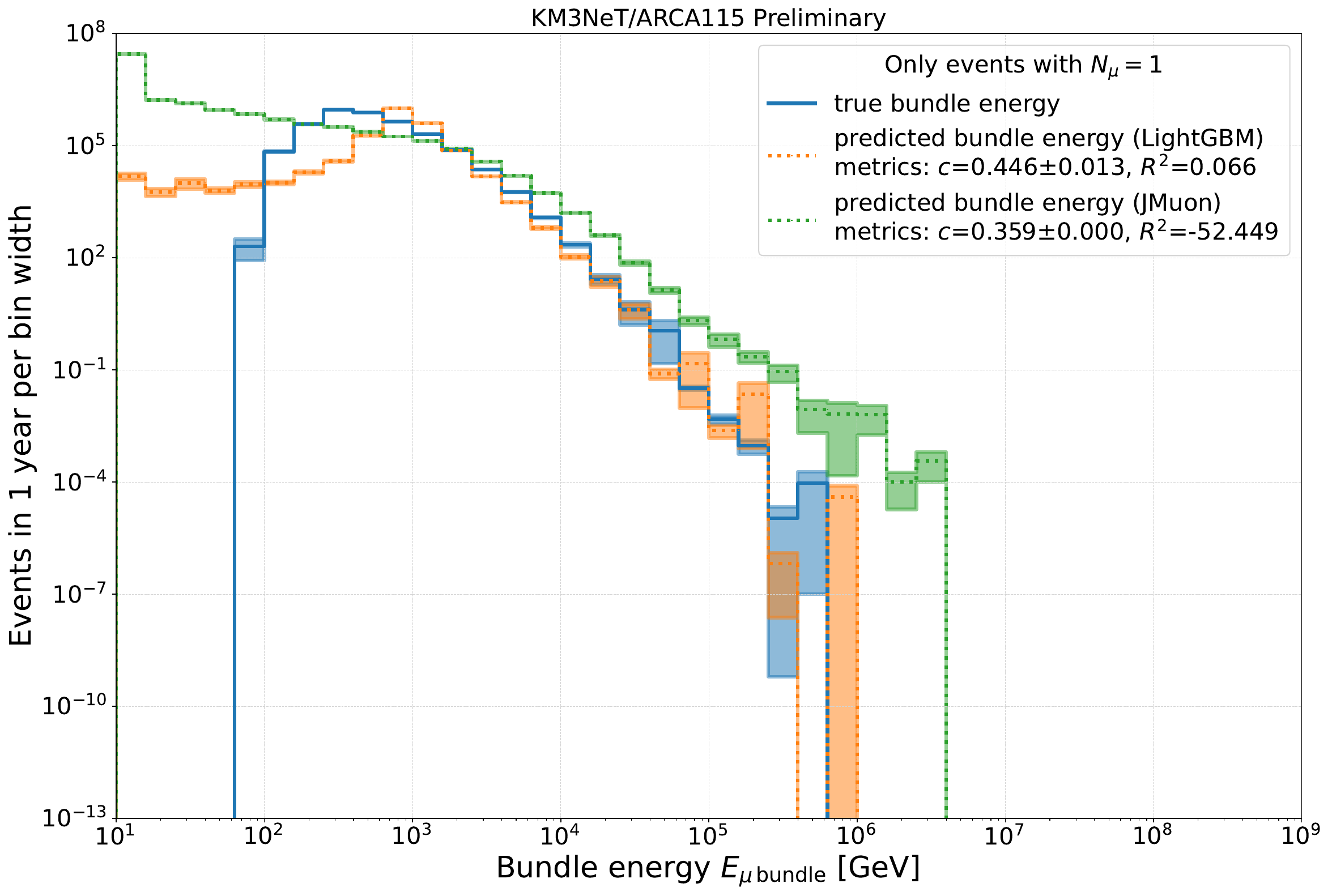}}\caption{Comparison of predicted and true distributions of bundle energy for
the JMuon and LightGBM reconstruction, applied to ARCA115 CORSIKA
MC. Corresponding plots for the remaining configurations (ARCA6, ORCA115,
ORCA6) may be found in Sec. \ref{sec:Performance-of-JMuon}. The error
bands were computed using Eq. \ref{eq:hist_error}. \label{fig:official_E_reco-vs-untuned_LightGBM}}
\end{figure}

The idea to reconstruct the muon bundle energy was inspired by the
available official KM3NeT reconstruction. The performance of the muon
energy reconstruction code JMuon (see Sec. \ref{sec:reco}) is shown
in Fig. \ref{fig:official_E_reco}. Here it has to be noted that JMuon
was designed to reconstruct a single muon track per event. Hence,
an evaluation of its performance in the case of single muon events
was included in Fig. \ref{fig:JMuon_single_mu}. Such events were
selected by requiring the true multiplicity equal to 1. The goal of
this work was to predict the bundle energy, as the JMuon reconstruction
seemed not to deliver satisfactory results. Even an untuned ML result
(Fig. \ref{fig:untuned_Ebundle_LightGBM_reco}) was clearly performing
better than the standard reconstruction (JMuon), which became even
more evident in Fig. \ref{fig:official_E_reco-vs-untuned_LightGBM}.

\subsubsection{Choice of the best model \label{subsec:Selection-of-the-best-model-energy}}

\begin{figure}[H]
\centering{}\includegraphics[width=16cm]{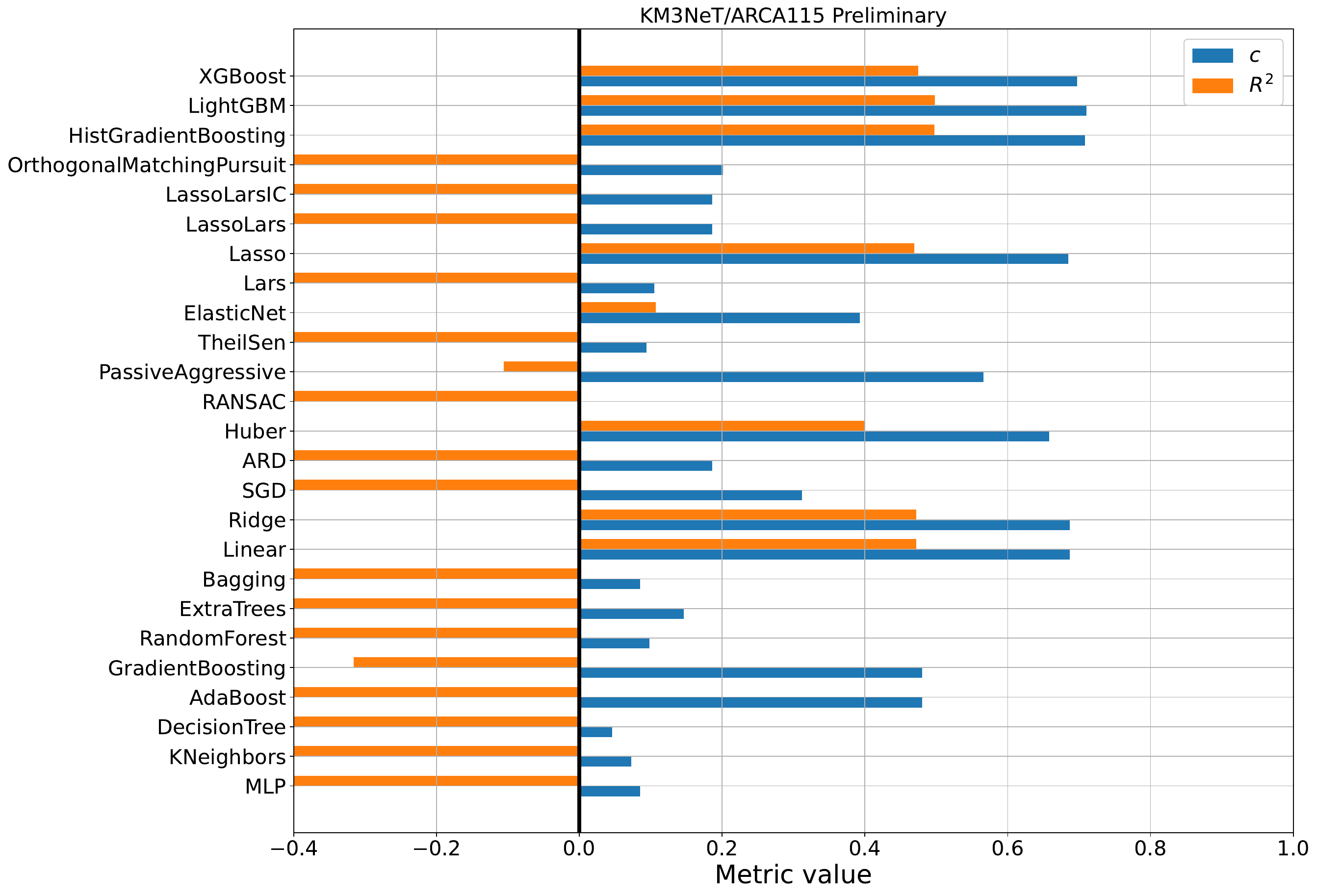}\caption{Performance comparison of selected models in terms of correlation
$c$ and $R^{2}-$score (for both metrics 1.0 is the perfect score).
\label{fig:energy_model_comparison}}
\end{figure}

\begin{figure}[H]
\centering{}\includegraphics[width=16cm]{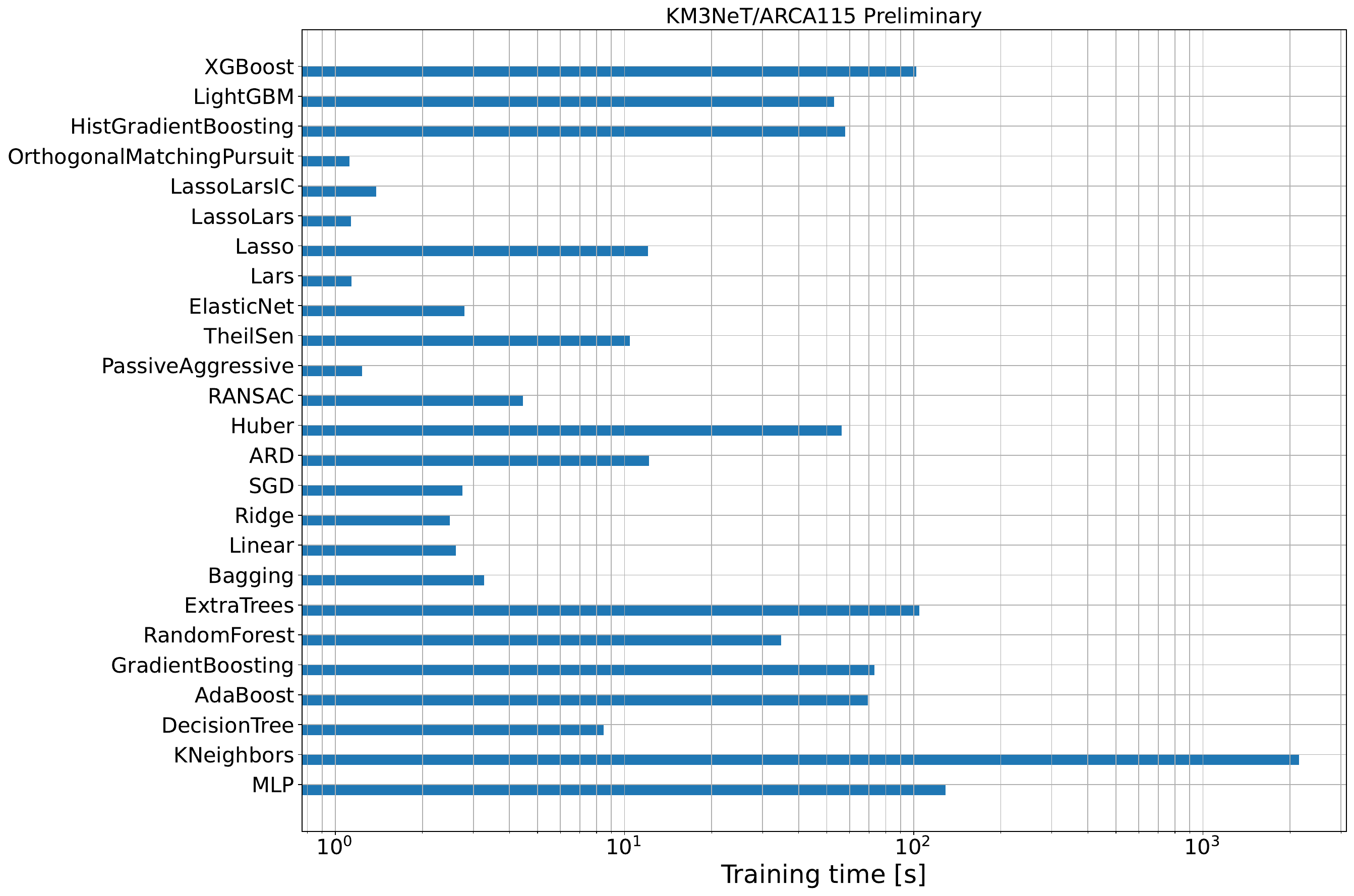}\caption{Performance comparison of selected models in terms of execution speed.
\label{fig:energy_model_comparison-speed}}
\end{figure}

The best model was picked from a wide range of candidates, briefly
introduced in Sec. \ref{subsec:Considered-models} and listed in Fig.
\ref{fig:energy_model_comparison} and \ref{fig:energy_model_comparison-speed}.
All estimators were trained using a fraction of the ARCA115 training
data (50 000 events out of 14 374 415 at hand; see Tab. \ref{tab:ML-dataset-summary})
and evaluated on the full validation set. This was dictated by the
memory and execution time limitations. 

It was presumed that the best model would outperform the others, regardless
of the detector configuration, or available training data. Furthermore,
an assumption was made that the same learner would be the best both
for the muon bundle energy and for the primary energy. This is well-grounded,
as the bundle energy is the primary energy reduced by the interactions
in the shower prior to the arrival of the muons at the can. Moreover,
it was presupposed that the full validation dataset was sufficient
to compare the performance reliably. The models were compared in terms
of the obtained Pearson correlation coefficient $c$ and the coefficient
of determination $R^{2}$ (see Sec. \ref{sec:Machine-learning-intro}).
In addition, their execution times on an Intel i5-2400 CPU were measured.
Since a number of estimators (MLP, KNeighbors, ARD, PassiveAggressive,
TheilSen, Lars, LassoLars, LassoLarsIC, and OrthogonalMatchingPursuit;
see Sec. \ref{subsec:Considered-models}) did not support sample weights
at all, they were trained unweighted (equivalent to training with
all weights equal to 1). In the case of the remaining models, it was
possible that training with scaled or unscaled weights (see Sec. \ref{subsec:Preprocessing})
could yield better results (and neither is intrinsically wrong). Hence,
each estimator supporting weights was trained twice: with scaled and
with unscaled weights, and the better of the two results was used.
The values of the metrics were always computed using the unscaled
flux weights, since they provide an approximation of what may be expected
from the experimental data. 

The results are compiled in Fig. \ref{fig:energy_model_comparison},
and the winner was LightGBM. It achieved the best regression metrics,
while remaining the fastest among the three best models, beating XGBoost
and HistGradientBoosting (Fig. \ref{fig:energy_model_comparison-speed}).
This was possible, among the other factors, thanks to its approach
of internally histogramming the data and working on the histogram
counts, rather than on individual events \cite{LightGBM}. In this
entire chapter, LightGBM was used as the default model for all regression
tasks.

\subsubsection{Inspection of learning}

\begin{figure}[H]
\centering{}\includegraphics[width=14cm]{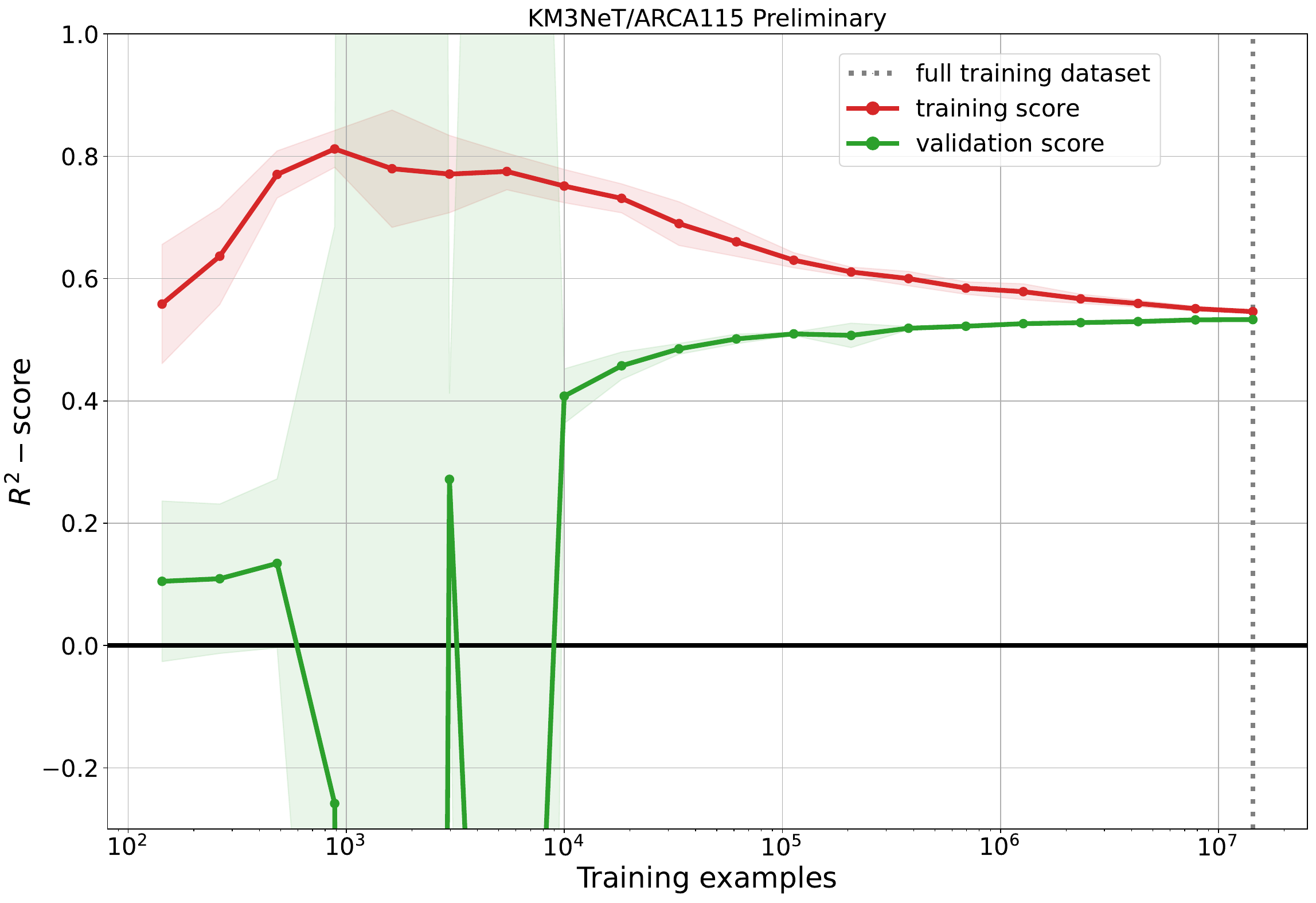}\caption{$R^{2}-$scores for the training and validation sets as function of
the number of seen training examples. The reconstruction was applied
to ARCA115 CORSIKA MC. The $R^{2}$ metric values were computed taking
into account the unscaled event weights. For each point, the training
was performed 7 times on a randomly selected part of the training
dataset, each time with a different random seed. The solid line with
dots represents the median of the score and the surrounding band —
one standard deviation. The validation score was computed using the
entire validation dataset each time. \label{fig:Ebundle-learning-curves}}
\end{figure}

As mentioned in Sec. \ref{subsec:Performance-assessment}, checking
the learning curve of the model is an important step in assuring the
correctness of the results. The learning curves of LightGBM training
for the bundle energy regression task are presented in Fig. \ref{fig:Ebundle-learning-curves}.
As may be seen, the model was still slightly underfitting (see Fig.
\ref{fig:learning_curves}), however the improvement with adding more
training data was rather slow and the cost of simulating more events
would be substantial. When learning on small samples of the training
dataset, one can see very rapid changes in the scores: this is an
expected behaviour. The model was hardly able to generalise (negative
validation score) because it was exposed to too little examples. Relatively
small samples of the dataset had low chance of being representative
for the complete one. As the training sample size grew above $10^{4}$,
the training started to converge to a score of around 0.5. The final
result was further improved by hyperparameter tuning (see Sec. \ref{subsec:Hyperparameter-tuning}).

\subsubsection{Feature importances\label{subsec:Feature-importances-Ebundle}}

This section investigates the profitability of restricting the features
in use to only the most important ones, as hinted in Sec. \ref{sec:MC-samples-used}.
Fig. \ref{fig:Ebundle-feature-importance} visualises how essential
each of the features was. The features, which boost the resulting
$R^{2}$-score the most have been selected by applying a cutoff on
the feature importance, requiring a positive importance for each such
feature. In addition, only one feature with the highest importance
was picked out of each cluster (the colour coding corresponds to Fig.
\ref{fig:dendrogram}). The features selected in this manner are listed
in Tab. \ref{tab:Features-selected-Ebundle}. The importance was dominated
by 3DMUON\_3DSHOWER\_trig\_hits, which is the number of hits triggered
by both the 3DMuon and 3DSHOWER trigger algorithms, followed by 3DSHOWER\_trig\_hits,
3DMUON\_trig\_hits, and overlays (see Sec. \ref{sec:light}).
\begin{center}
\begin{figure}[H]
\centering{}\includegraphics[width=16cm]{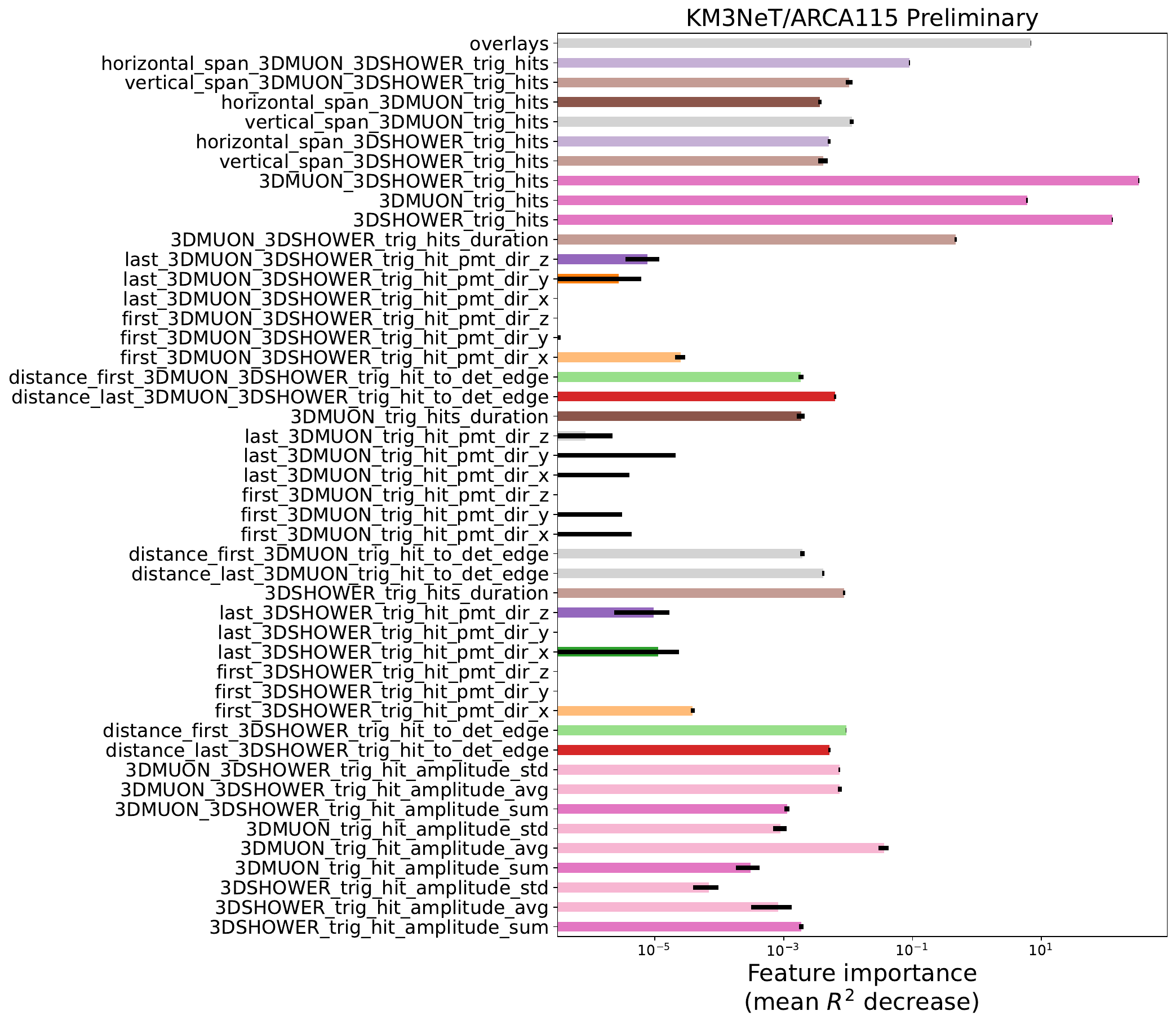}\caption{Feature importance plot, showing how much each feature contributes
to training. The figure has been obtained by performing 10 permutations
for each feature to ensure a stable result. The standard deviation
for each feature is denoted by a black bar. The colour coding of features
is consistent with Fig. \ref{fig:dendrogram} to show the clusters,
and their most important members. Feature importance plots for the
other detector configurations are gathered in Sec. \ref{subsec:Feature-importances}.
\label{fig:Ebundle-feature-importance}}
\end{figure}
\par\end{center}

\begin{table}[H]
\begin{centering}
\caption{Features selected by feature importance and clustering (see Fig. \ref{fig:Ebundle-feature-importance}
and \ref{fig:dendrogram}). For description of individual features,
refer to Sec. \ref{subsec:Features-description}. \label{tab:Features-selected-Ebundle}}
\par\end{centering}
\centering{}{\small{}}%
\begin{tabular}{|c|c|}
\hline 
{\small{}Feature} & {\small{}Importance $\pm$ 1 standard deviation}\tabularnewline
\hline 
\hline 
{\small{}distance\_first\_3DMUON\_trig\_hit\_to\_det\_edge} & {\small{}$0.0020\pm0.0001$}\tabularnewline
\hline 
{\small{}horizontal\_span\_3DMUON\_trig\_hits} & {\small{}$0.0036\pm0.0002$}\tabularnewline
\hline 
{\small{}distance\_last\_3DMUON\_trig\_hit\_to\_det\_edge} & {\small{}$0.0041\pm0.0002$}\tabularnewline
\hline 
{\small{}distance\_last\_3DMUON\_3DSHOWER\_trig\_hit\_to\_det\_edge} & {\small{}$0.0063\pm0.0002$}\tabularnewline
\hline 
{\small{}distance\_first\_3DSHOWER\_trig\_hit\_to\_det\_edge} & {\small{}$0.0093\pm0.0001$}\tabularnewline
\hline 
{\small{}vertical\_span\_3DMUON\_trig\_hits} & {\small{}$0.0114\pm0.0008$}\tabularnewline
\hline 
{\small{}3DMUON\_trig\_hit\_amplitude\_avg} & {\small{}$0.0359\pm0.0065$}\tabularnewline
\hline 
{\small{}horizontal\_span\_3DMUON\_3DSHOWER\_trig\_hits} & {\small{}$0.0900\pm0.0016$}\tabularnewline
\hline 
{\small{}3DMUON\_3DSHOWER\_trig\_hits\_duration} & {\small{}$0.47\pm0.02$}\tabularnewline
\hline 
{\small{}overlays} & {\small{}$6.84\pm0.06$}\tabularnewline
\hline 
{\small{}3DMUON\_3DSHOWER\_trig\_hits} & {\small{}$325\pm8$}\tabularnewline
\hline 
\end{tabular}{\small\par}
\end{table}

\begin{figure}[H]
\begin{centering}
\subfloat[All features.]{\centering{}\includegraphics[width=8cm]{Plots/ML-reco/Ebundle_reco/A115/v6\lyxdot 9_ARCA115_ML_LightGBM_true_energy_all_features_true_vs_pred}}\subfloat[Selected features (importance \& clustering).]{\centering{}\includegraphics[width=8cm]{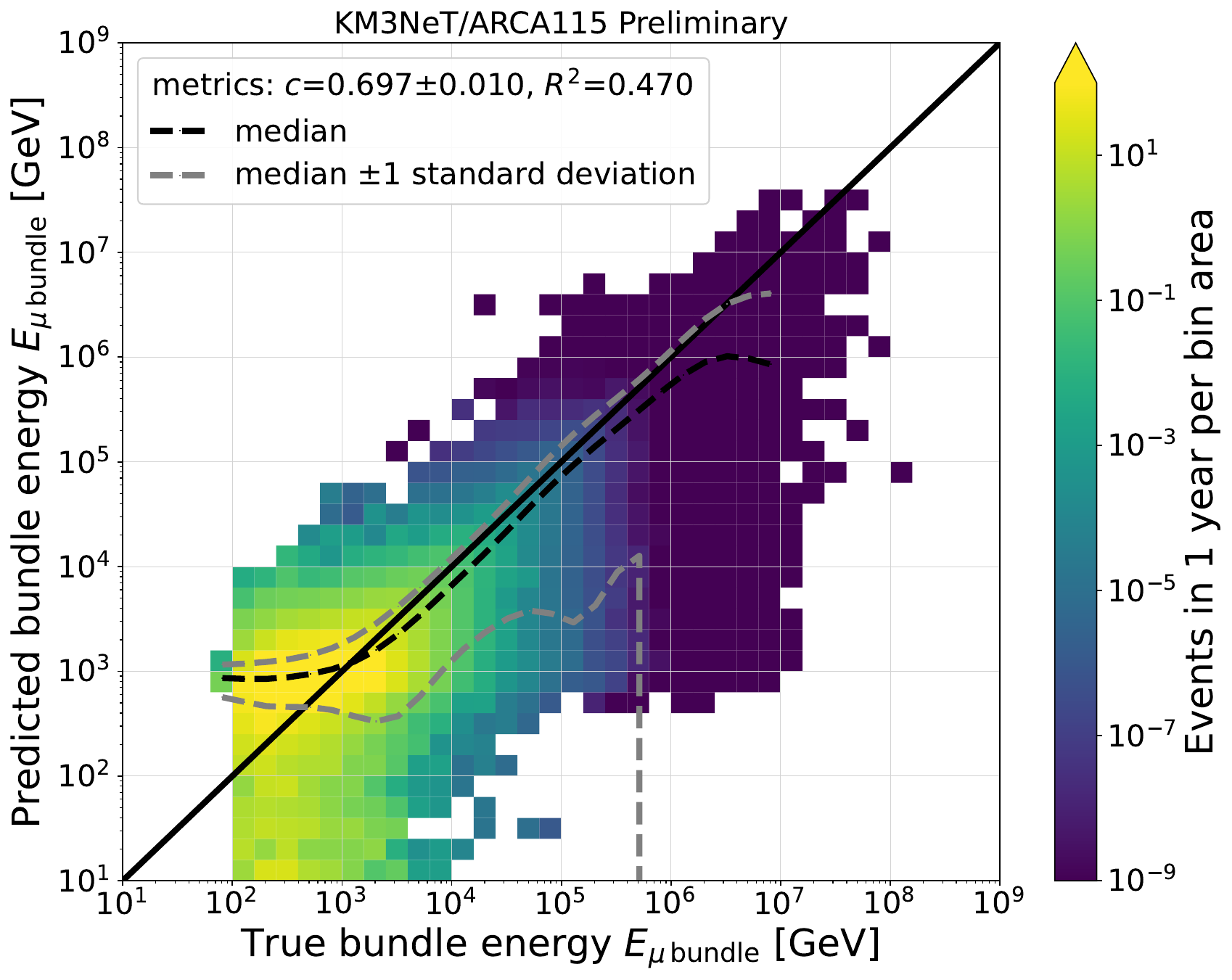}}
\par\end{centering}
\centering{}\subfloat[Only 3DMUON\_3DSHOWER\_trig\_hits. \label{fig:Only-3DMUON_trig_hits.}]{\centering{}\includegraphics[width=8cm]{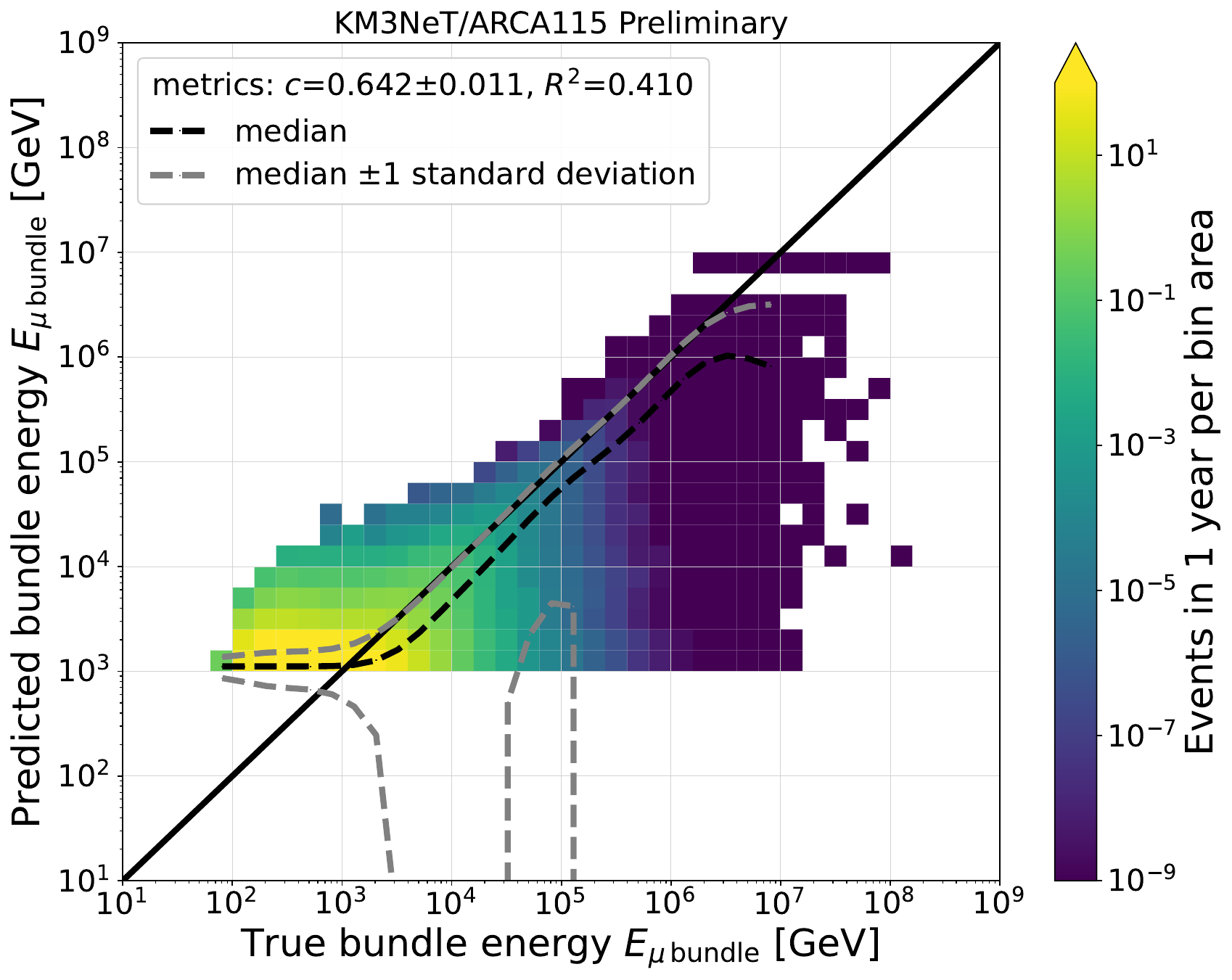}}\subfloat[Selected features (positive importance). \label{fig:Selected-features-positive-importance}]{\centering{}\includegraphics[width=8cm]{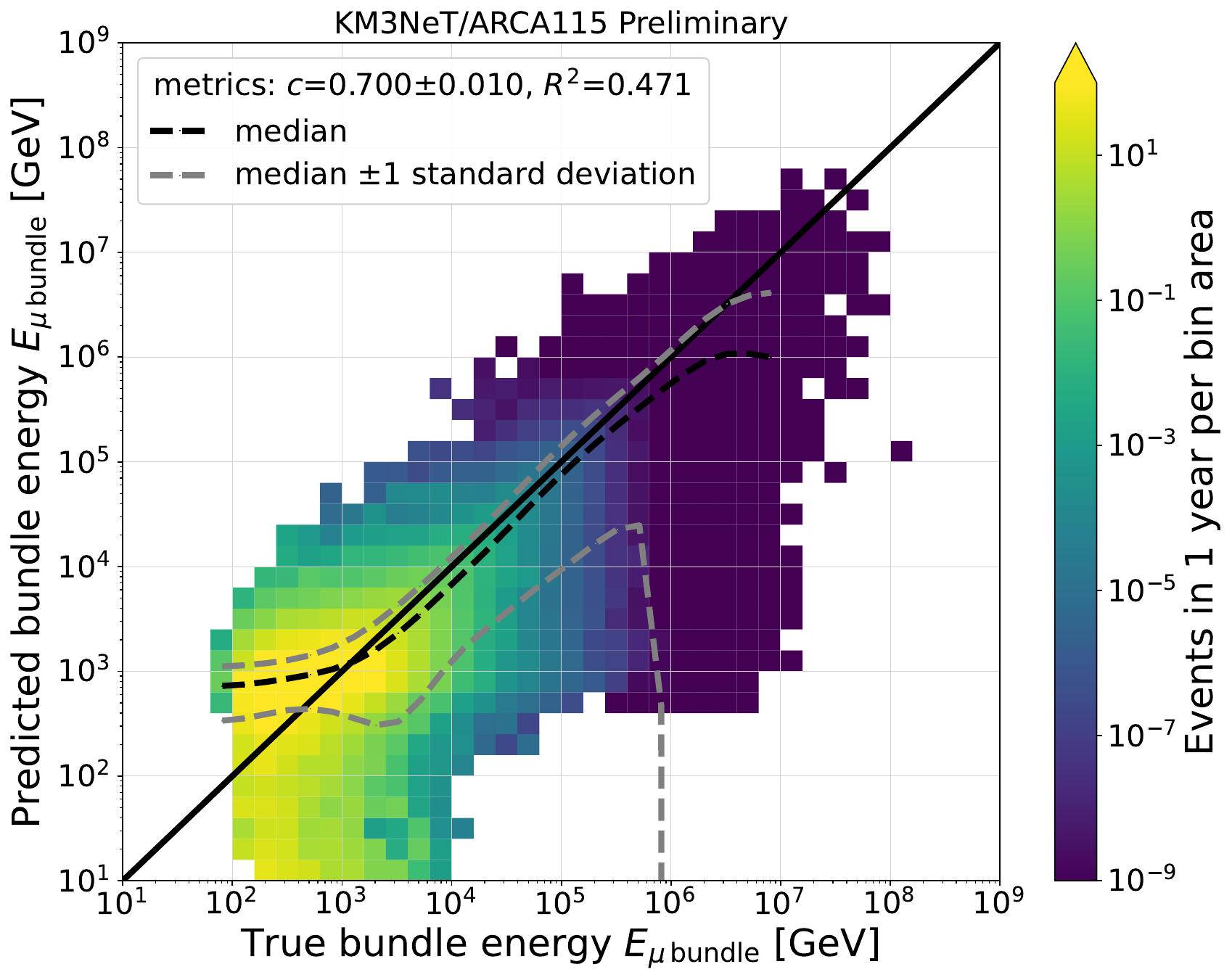}}\caption{Predicted bundle energy plotted against the MC truth with and without
feature selection. Both the correlation $c$ and $R^{2}-$score values
were computed bin-wise. \label{fig:Ebundle-feature-selection-comaprison}}
\end{figure}

A comparison of performance with and without feature selection is
presented in Fig. \ref{fig:Ebundle-feature-selection-comaprison}.
The results do not differ much, however the number of selected features
was only 11, as compared against 46 without the selection. The training
time was reduced from 3~min 19~s down to 38.8~s, and the the $R^{2}-$score
improved by $0.2\,\%$, while the correlation coefficient worsened
only by $0.1\,\%$. Fig. \ref{fig:Only-3DMUON_trig_hits.} shows the
result if only the most important feature (3DMUON\_3DSHOWER\_trig\_hits)
was used. It confirms that the energy measurement is in essence equivalent
to counting the hits caused by the muons. However, both metrics were
noticeably worse, which meant that the additional information provided
by the other features does make a difference. Lastly, Fig. \ref{fig:Selected-features-positive-importance}
shows the outcome if features with negative or null importance (see
Fig. \ref{fig:Ebundle-feature-importance}) were not used. In this
case not only the learning was faster (1~min 59~s), due to bringing
the number of features down to 35, but also both the correlation and
$R^{2}$ improved. Since the ultimate deciding factor here was not
the code speed, but rather the achieved scores, the feature selection
using only the positive-importance features was adopted.

\subsubsection{Results \label{subsec:Results-Ebundle}}

\begin{figure}[H]
\begin{centering}
\subfloat[ARCA115.\label{fig:Ebundle_tuned_result_ARCA115.}]{\centering{}\includegraphics[width=8cm]{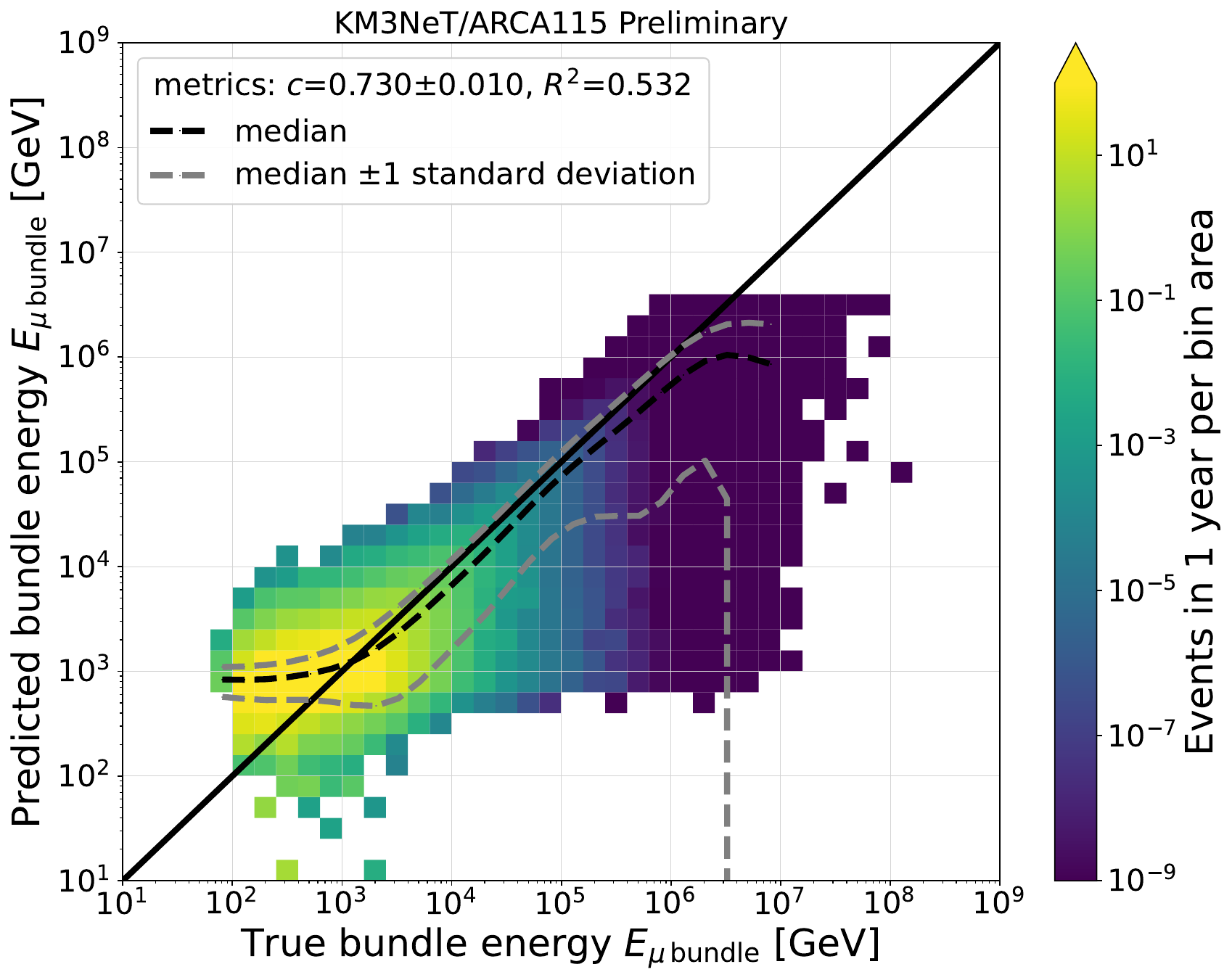}}\subfloat[ARCA6.]{\centering{}\includegraphics[width=8cm]{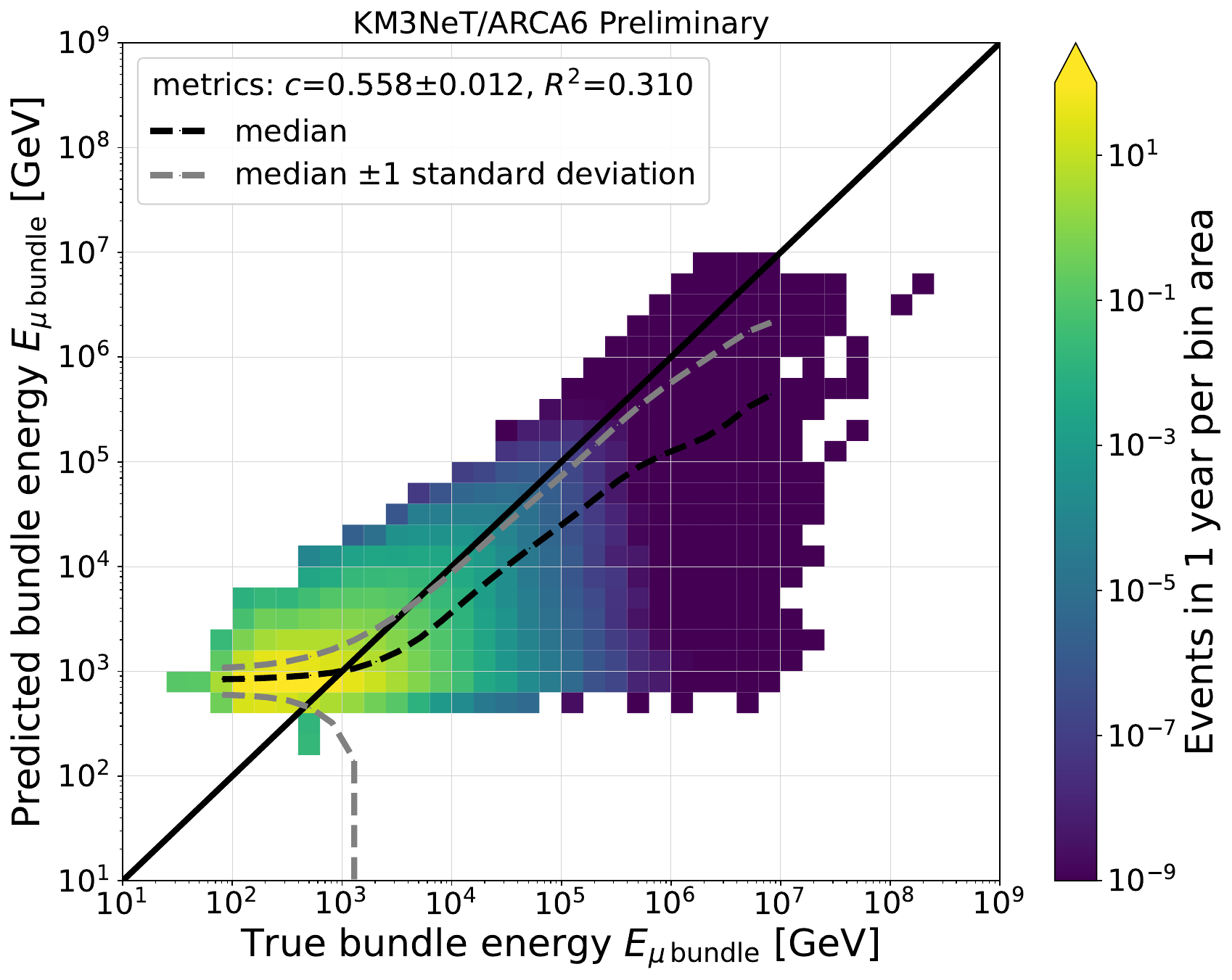}}
\par\end{centering}
\centering{}\subfloat[ORCA115.]{\centering{}\includegraphics[width=8cm]{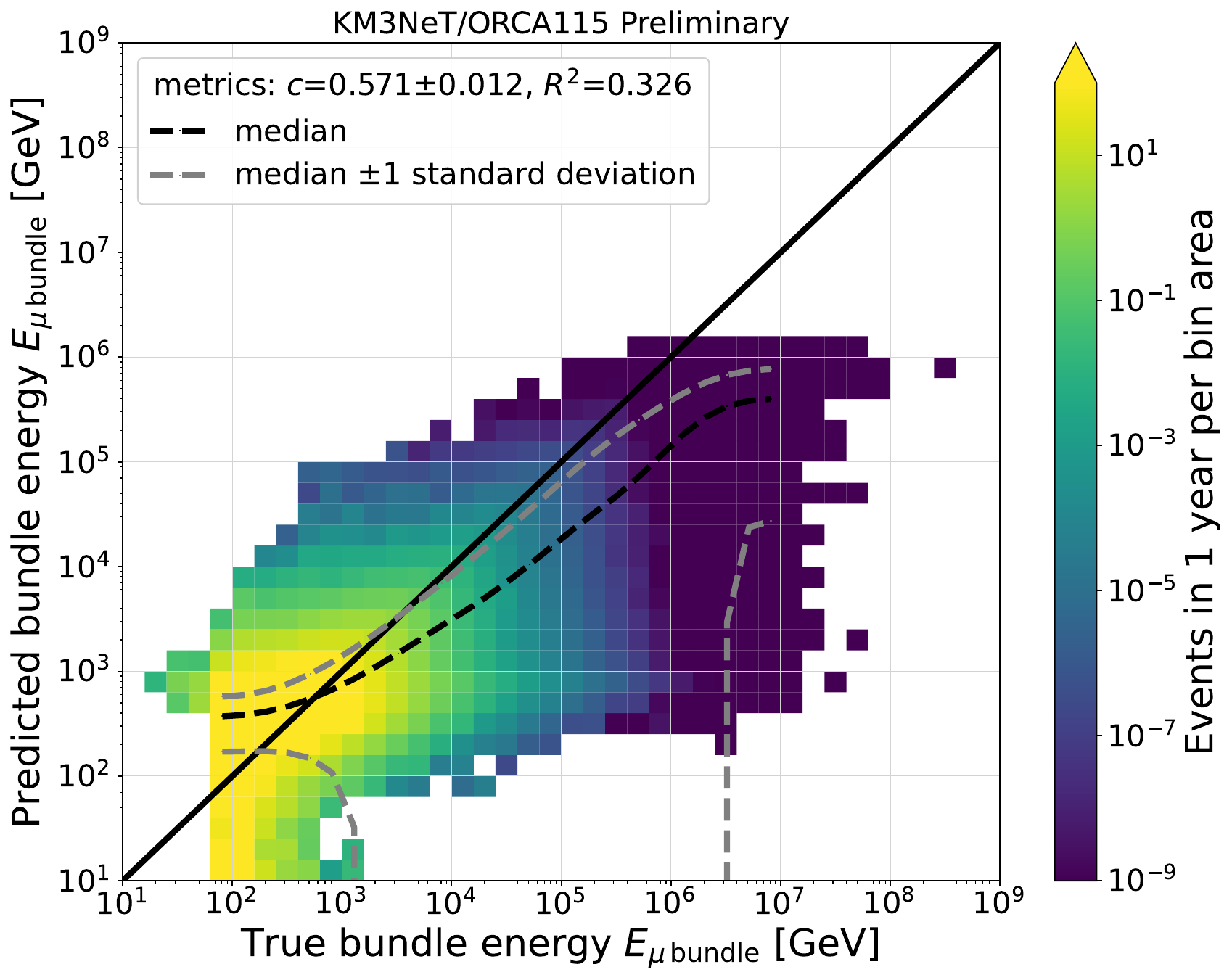}}\subfloat[ORCA6.]{\centering{}\includegraphics[width=8cm]{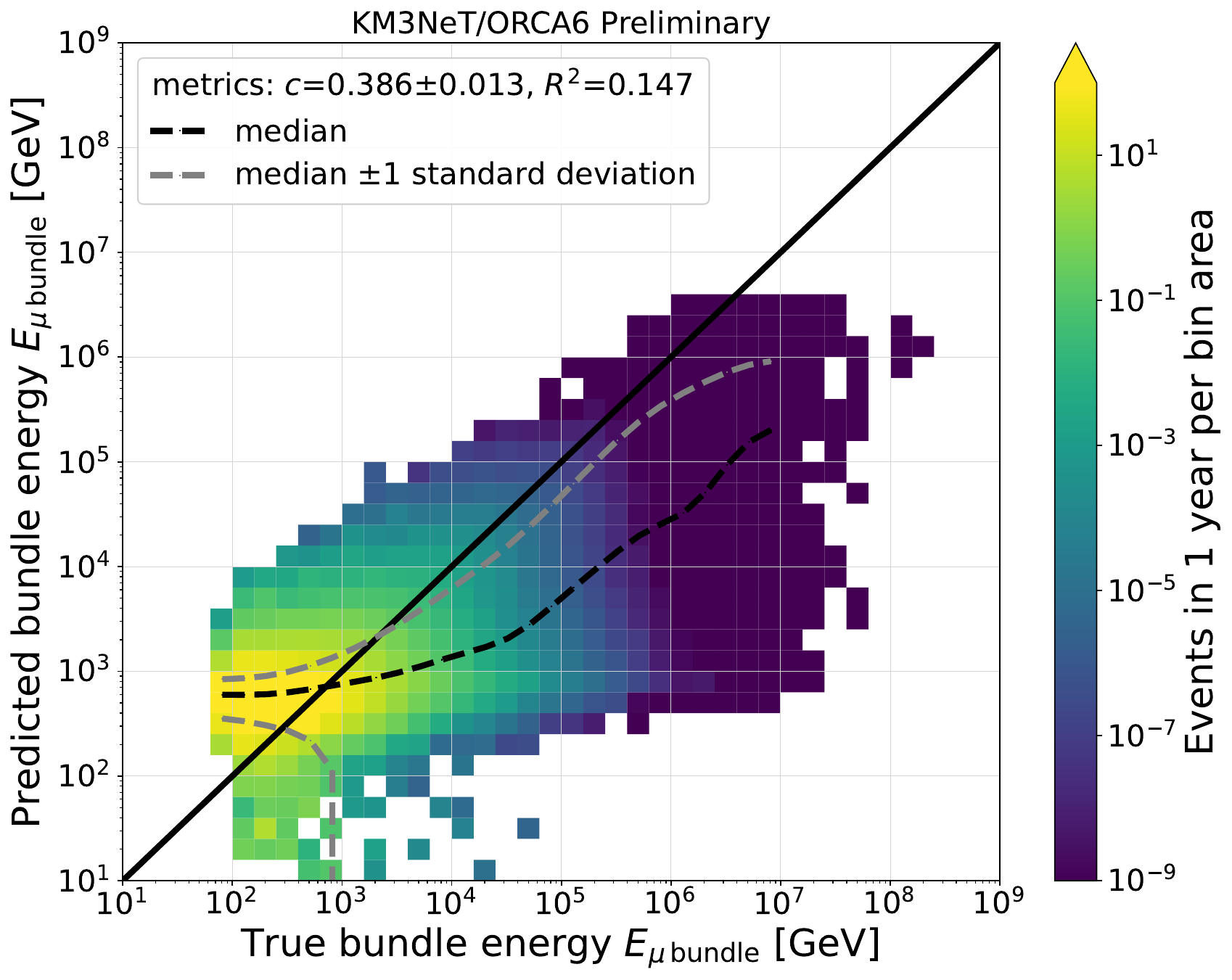}}\caption{Comparison of muon bundle energy reconstructed with LigthGBM against
the true value. Both the correlation $c$ and $R^{2}-$score values
were computed on the bin values, not individual events. \label{fig:Ebundle_reco_results}}
\end{figure}

\begin{figure}[H]
\begin{centering}
\subfloat[ARCA115.]{\centering{}\includegraphics[width=8cm]{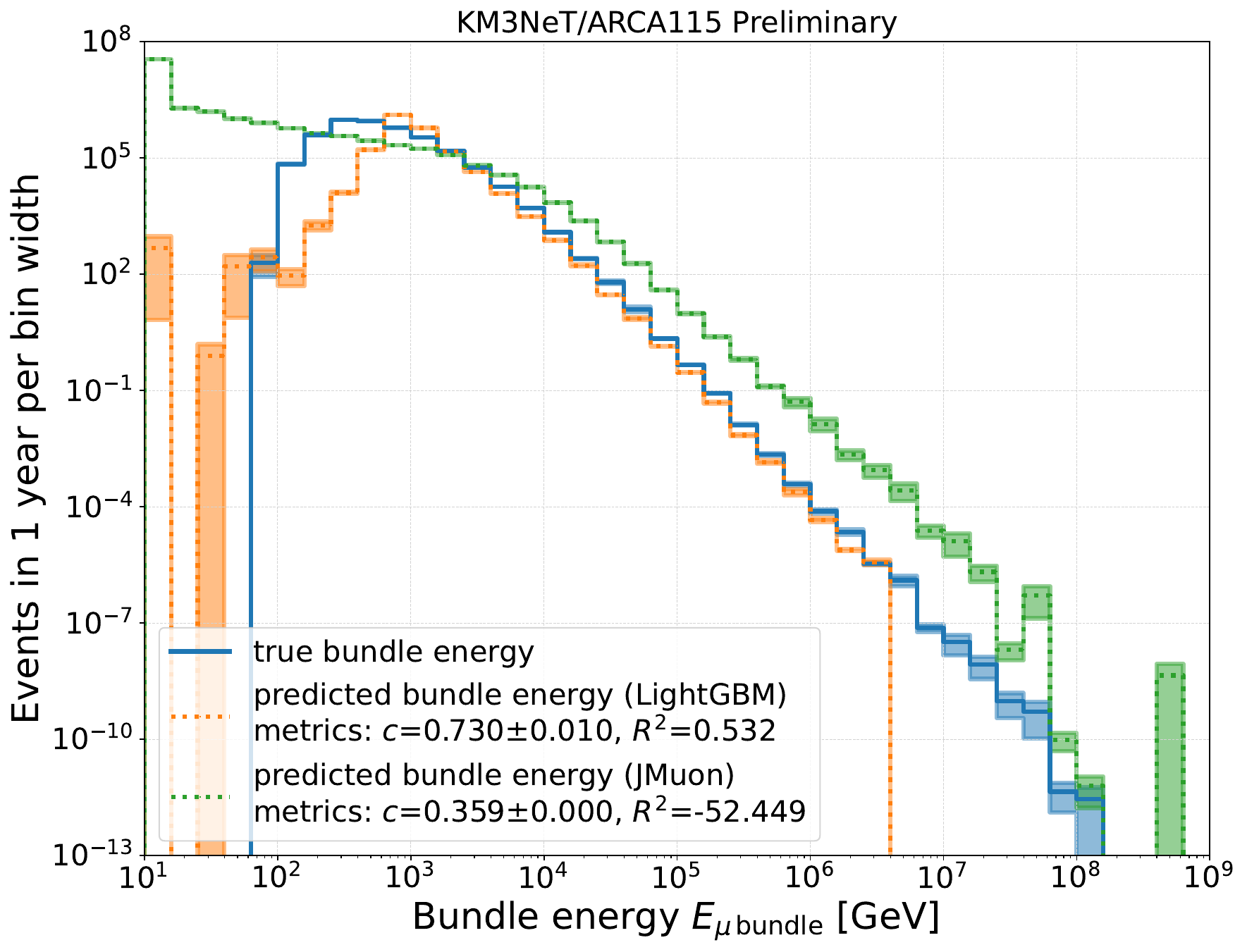}}\subfloat[ARCA6.]{\centering{}\includegraphics[width=8cm]{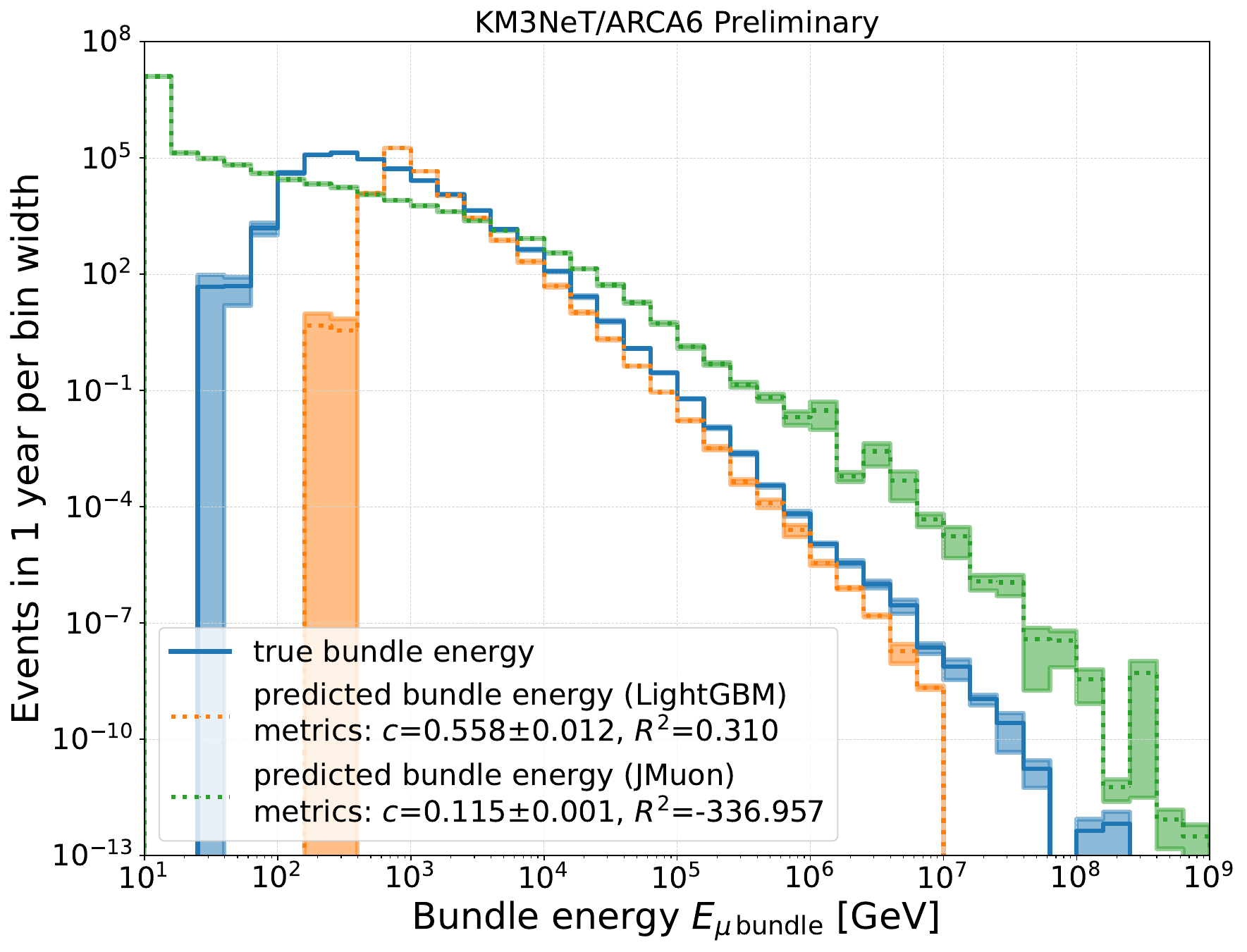}}
\par\end{centering}
\centering{}\subfloat[ORCA115.]{\centering{}\includegraphics[width=8cm]{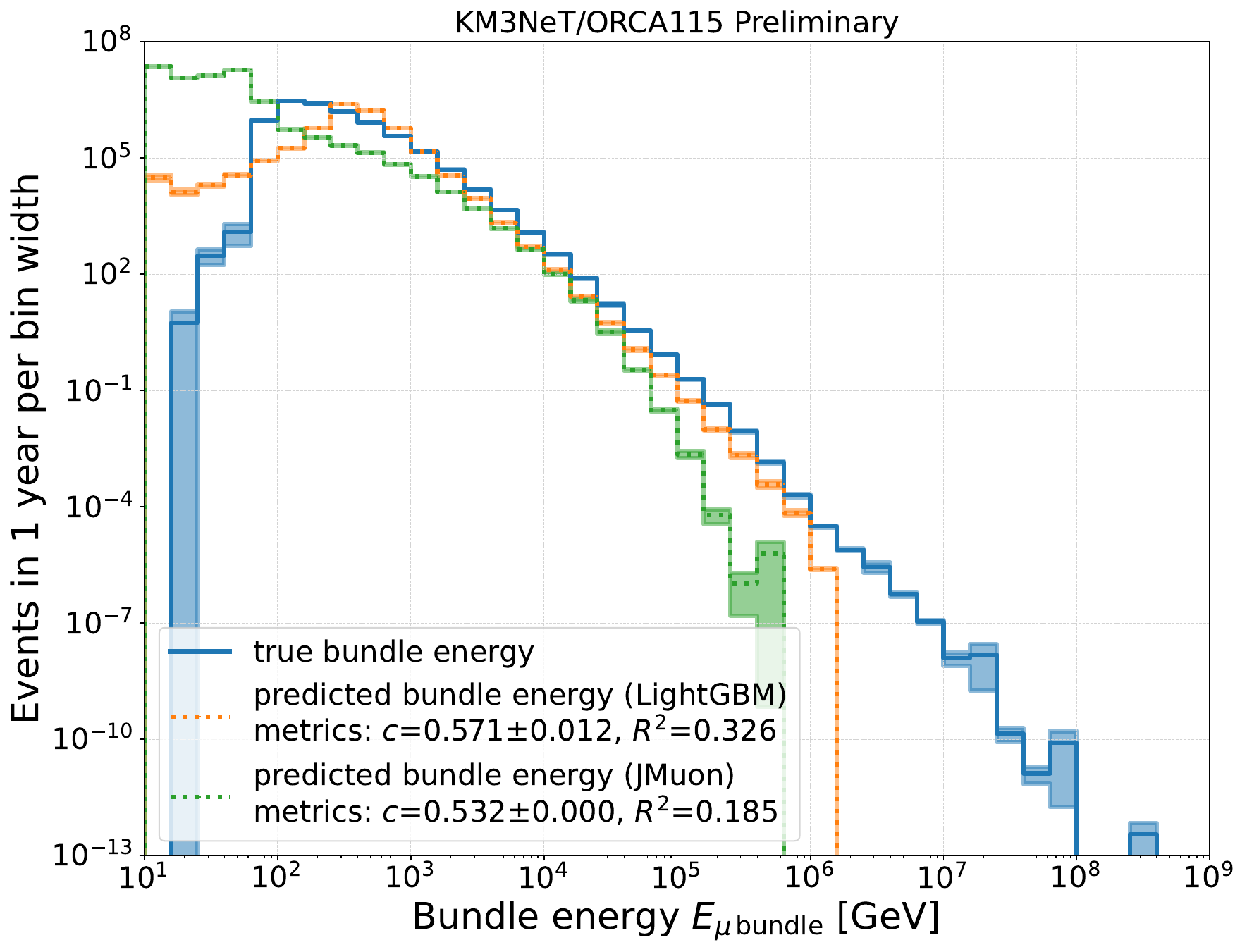}}\subfloat[ORCA6.]{\centering{}\includegraphics[width=8cm]{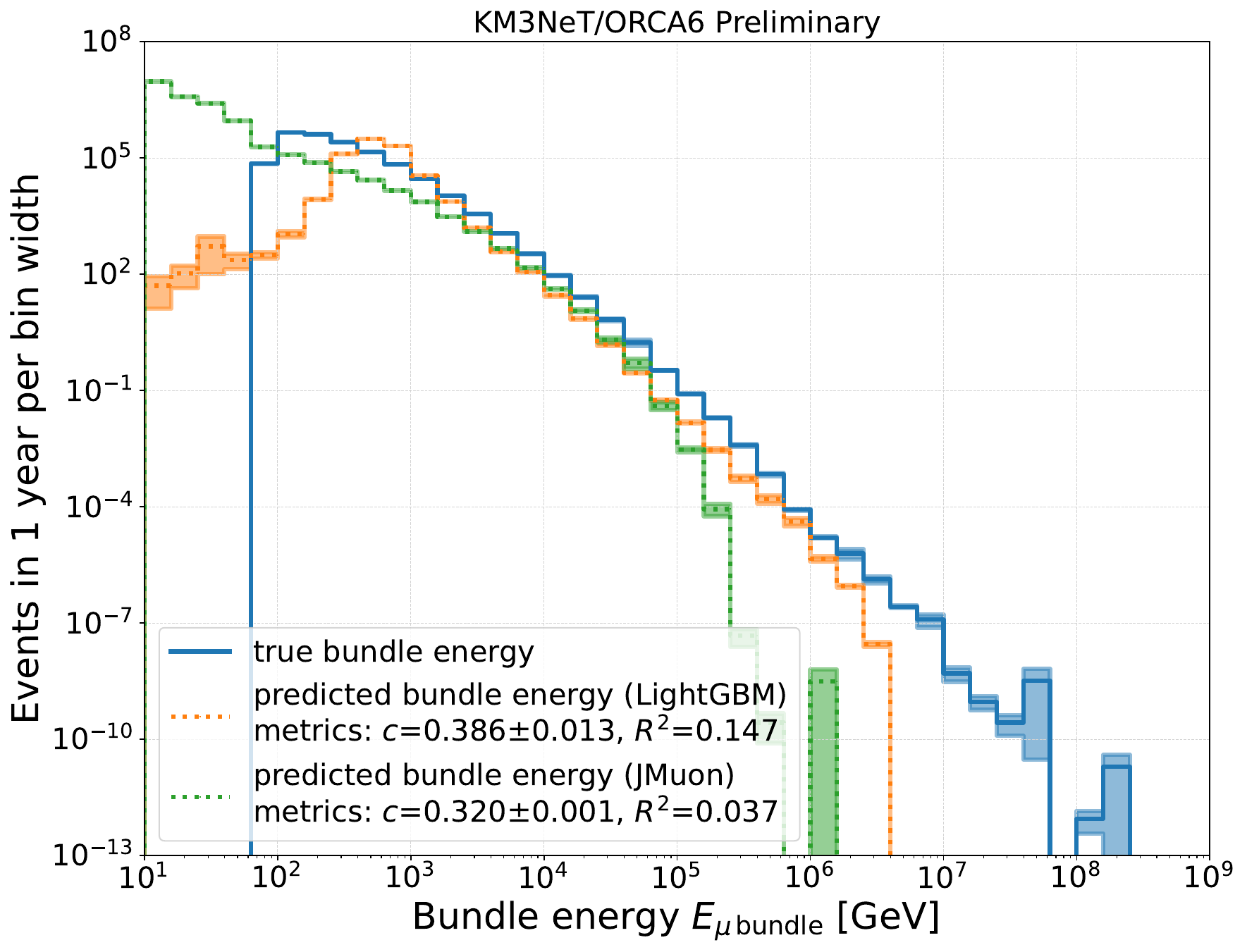}}\caption{Comparison of muon bundle energy reconstructed with LigthGBM against
the true value in 1D. The error bands were computed using Eq. \ref{eq:hist_error}.
\label{fig:Ebundle_reco_results-1D}}
\end{figure}

The final outcome of muon bundle energy reconstruction is shown in
Fig. \ref{fig:Ebundle_reco_results}. For this result, LightGBM with
tuned hyperparameters from Tab. \ref{tab:LightGBM-tuned-hyperparameters-summary}
has been used. In the previous sections, all the steps were only shown
for ARCA115, with the intent of making the description of the procedure
clearer. The same procedure was carried out for each of the detector
configurations. From the point of view of data vs MC comparisons (Chap.
\ref{chap:Muon-rate-measurement}), it is encouraging to see that
the shape of the distribution is retained above $E_{\mathrm{bundle}}\sim1\,$TeV
for each of the detector configurations (see Fig. \ref{fig:Ebundle_reco_results-1D}).
The performance of LightGBM-based regression clearly outperforms JMuon
in the whole range of energies.

\subsection{Total primary energy \label{subsec:Primary-energy}}

The results obtained for bundle energies encouraged an attempt at
reconstructing the energy of the primary nucleus as well. The very
same method has been adopted. In this case, there was no standard
reconstruction to compare against. Given the very close relation between
the bundle energy and the primary energy, neither separate model selection,
nor feature selection, nor hyperparameter tuning was performed, and
the results from Sec. \ref{subsec:Bundle-energy} were directly adopted.

\begin{figure}[H]
\begin{centering}
\subfloat[ARCA115.]{\centering{}\includegraphics[width=8cm]{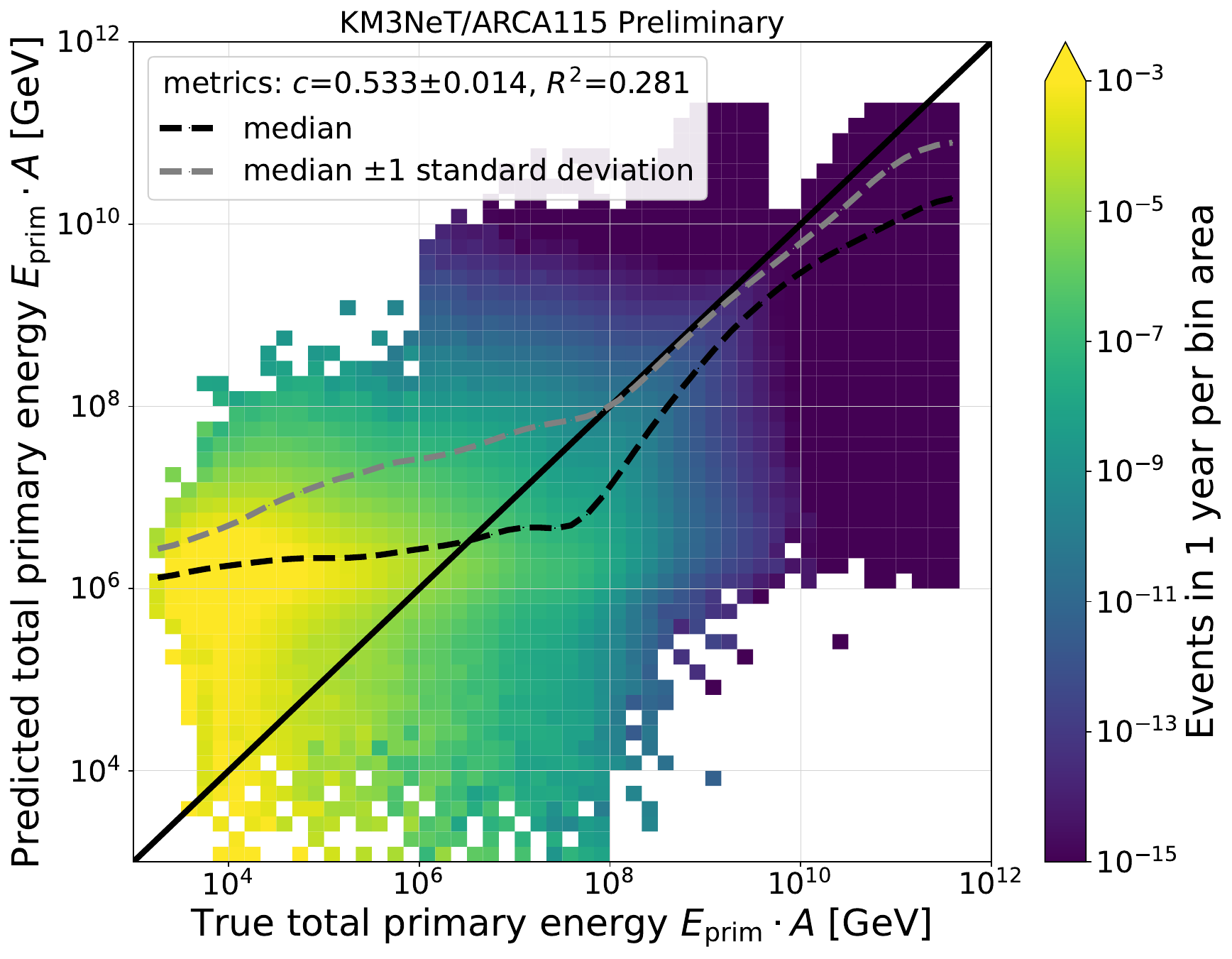}}\subfloat[ARCA6. ]{\centering{}\includegraphics[width=8cm]{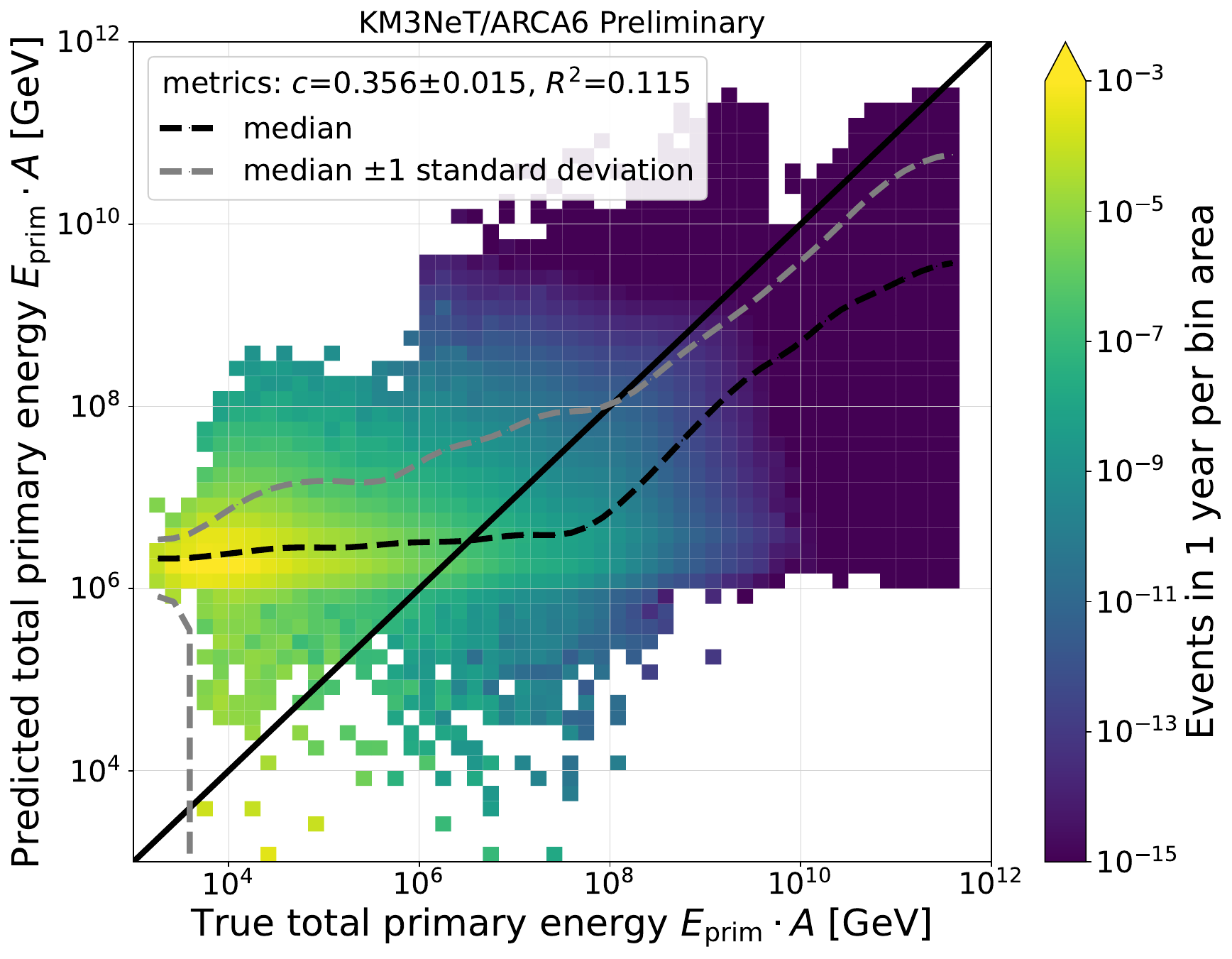}}
\par\end{centering}
\centering{}\subfloat[ORCA115. ]{\centering{}\includegraphics[width=8cm]{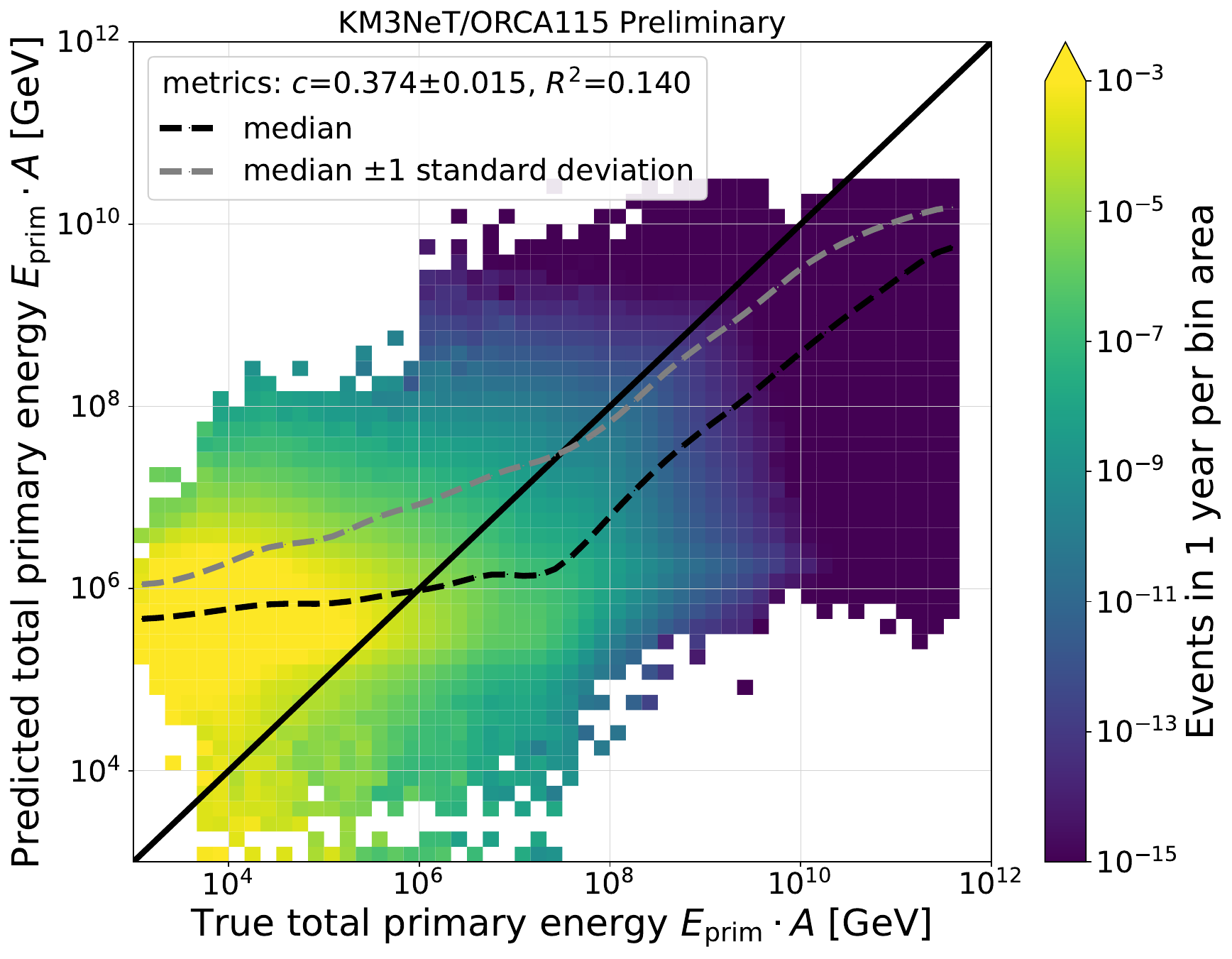}}\subfloat[ORCA6. ]{\centering{}\includegraphics[width=8cm]{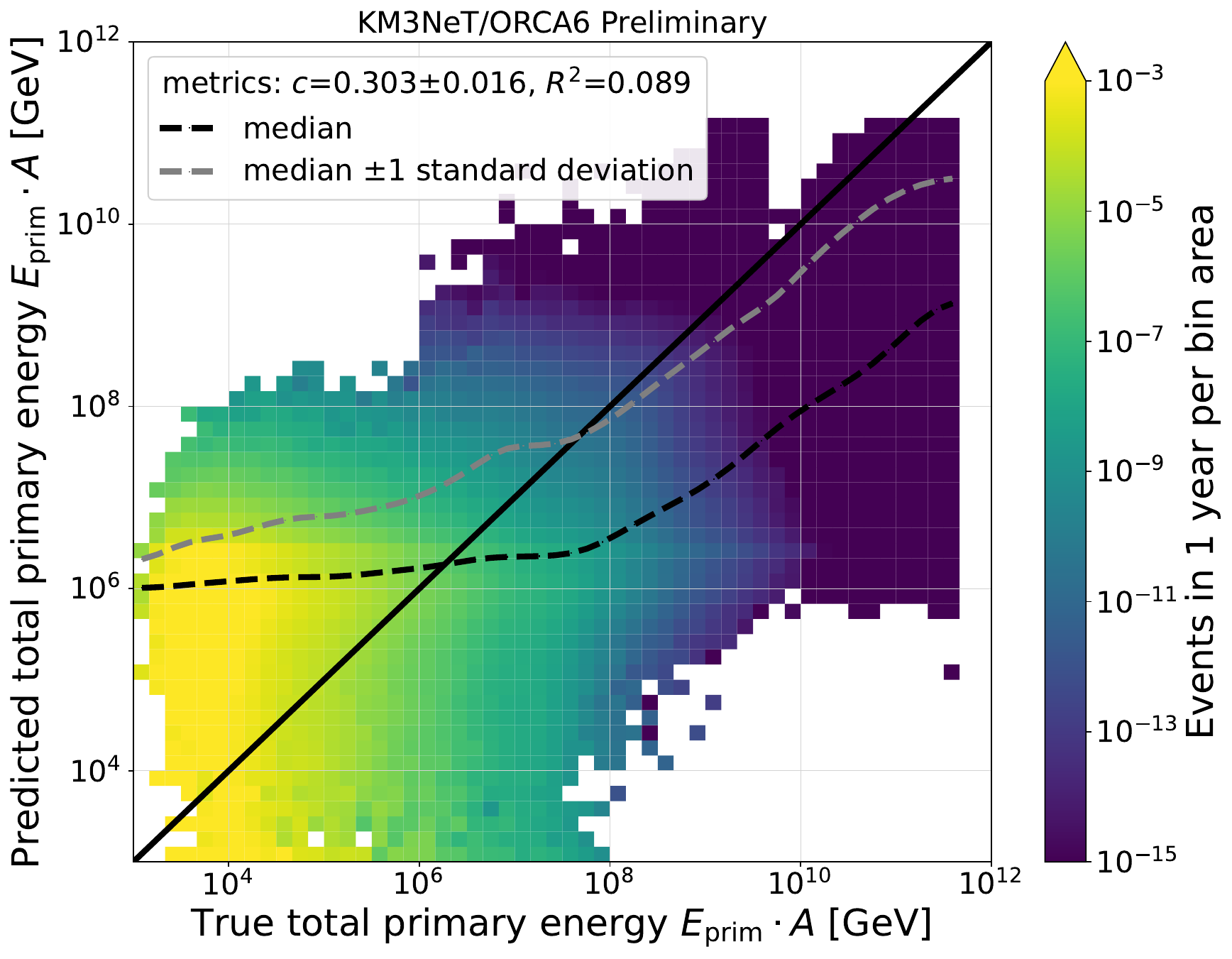}}\caption{Comparison of total primary energy reconstructed with the LigthGBM
model against the true value. The lower grey dashed lines are not
visible, because their values are negative. \label{fig:Eprim_reco_results}}
\end{figure}

\begin{figure}[H]
\begin{centering}
\subfloat[ARCA115.]{\centering{}\includegraphics[width=8cm]{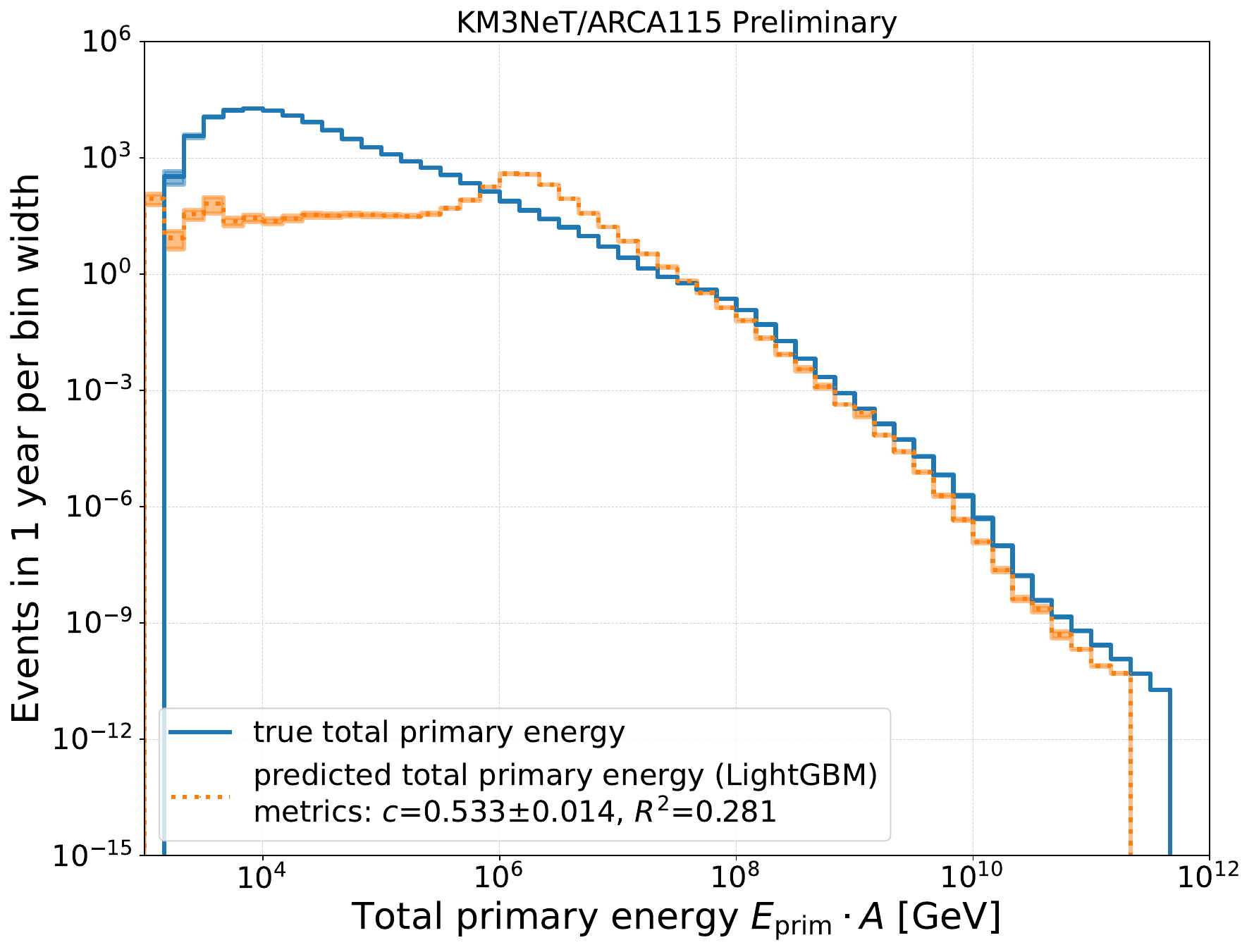}}\subfloat[ARCA6.]{\centering{}\includegraphics[width=8cm]{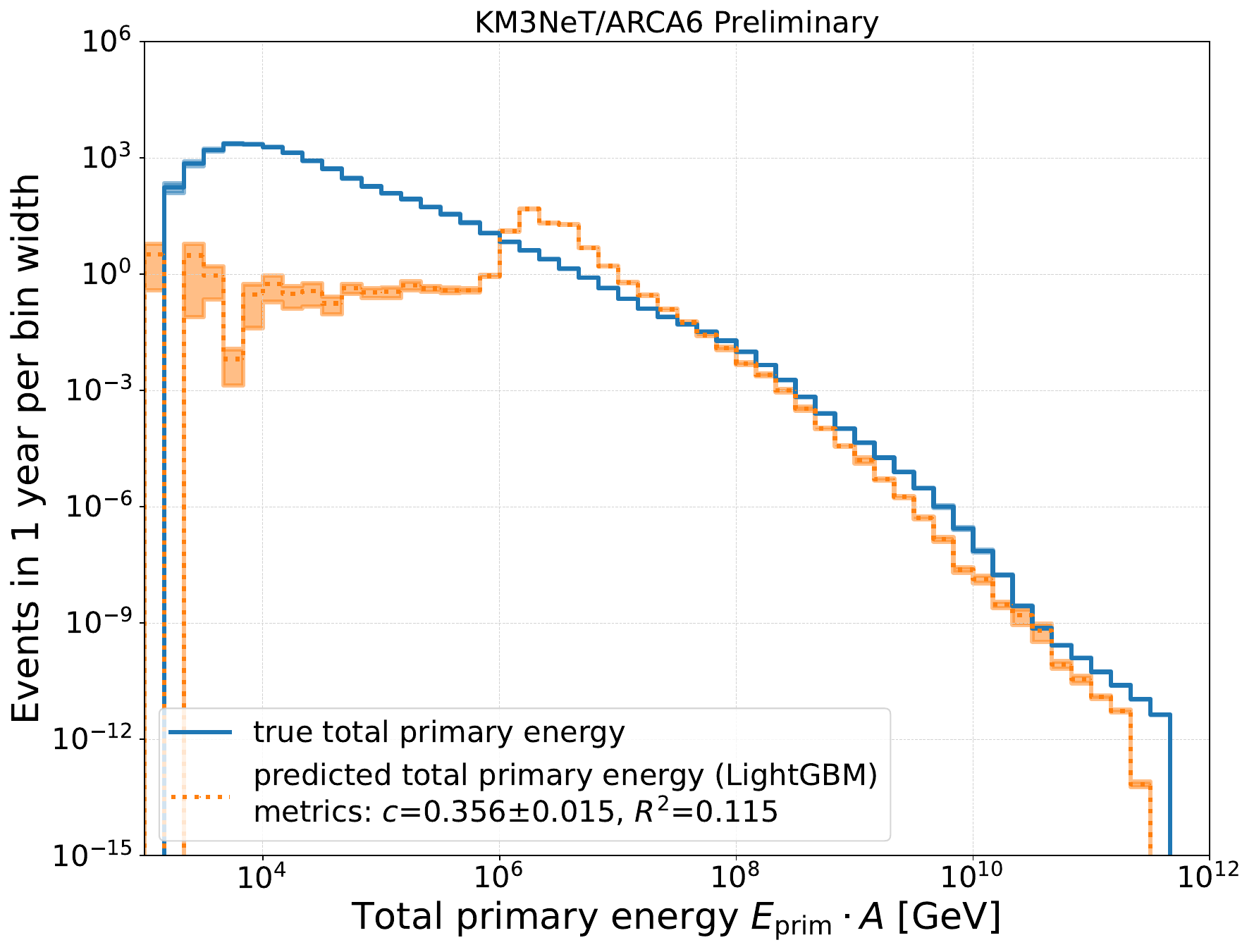}}
\par\end{centering}
\centering{}\subfloat[ORCA115.]{\centering{}\includegraphics[width=8cm]{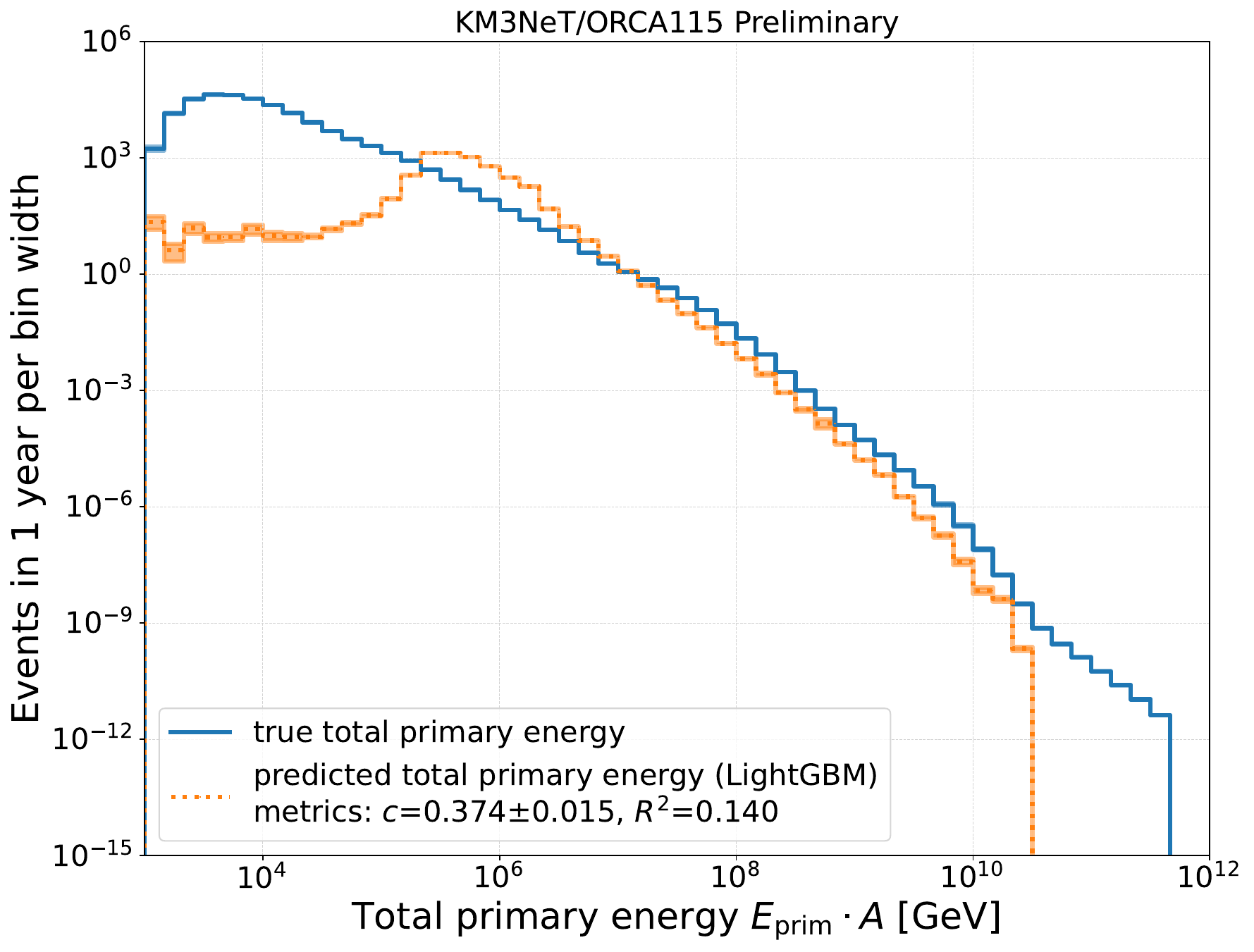}}\subfloat[ORCA6.]{\centering{}\includegraphics[width=8cm]{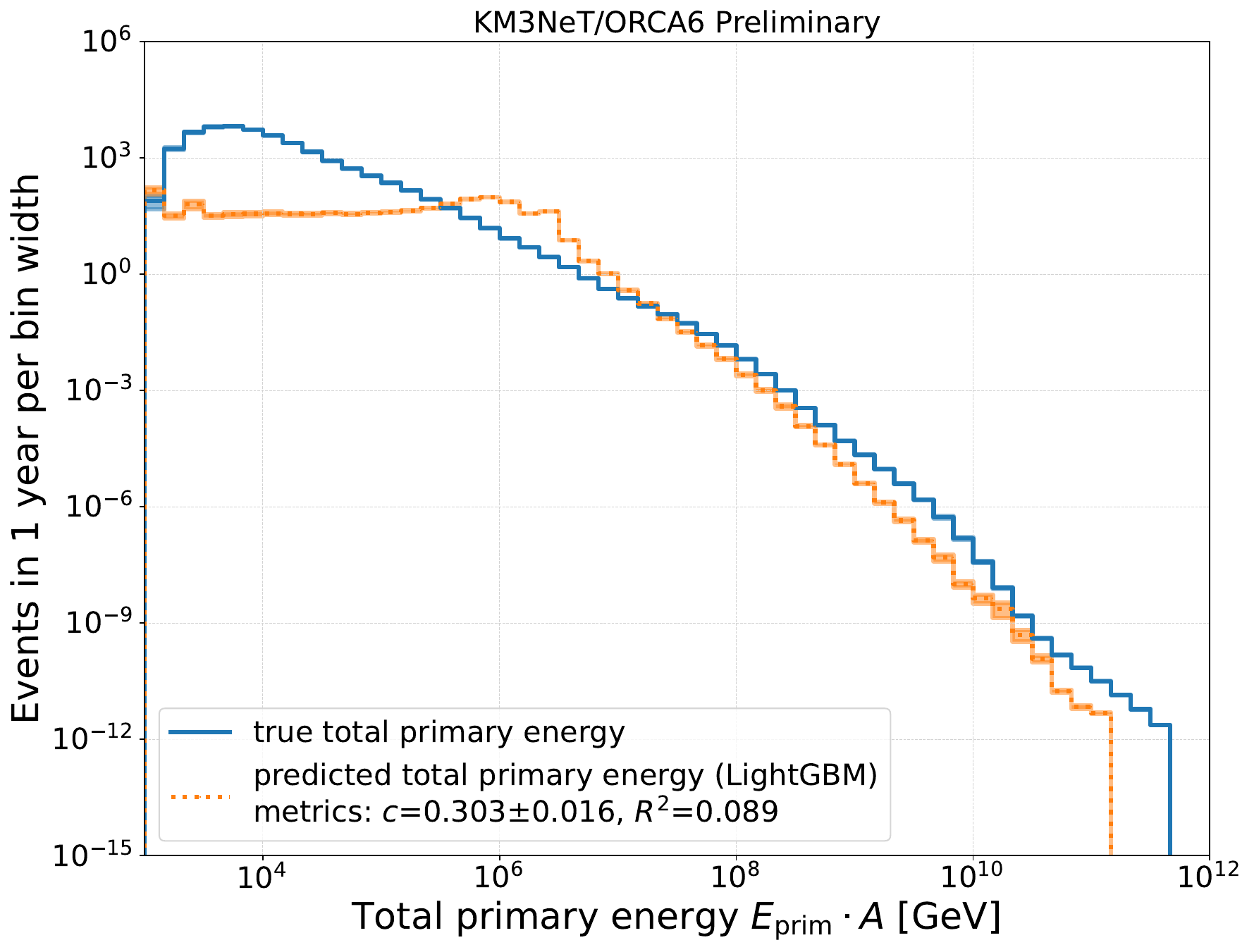}}\caption{Comparison of total primary energy reconstructed with the LigthGBM
model against the true value in 1D. The error bands were computed
using Eq. \ref{eq:hist_error}. \label{fig:Eprim_reco_results-1D}}
\end{figure}

The primary energy reconstruction described within this section, concerns
the total energy of the primary nucleus $E_{\mathrm{prim}}\cdot A$,
which should not be confused with the energy per nucleon $E_{\mathrm{prim}}$,
used as the CORSIKA input (see Sec. \ref{subsec:Simulation-inputs}).
The results are shown in Fig. \ref{fig:Eprim_reco_results} and \ref{fig:Eprim_reco_results-1D}.
The predictions are clearly less accurate than for the bundle energy
— both metric values are much lower compared to Fig. \ref{fig:Ebundle_tuned_result_ARCA115.}.
This was expected, since the reconstruction was forced to predict
the energy of the parts of the shower, which did not make it to the
detector at all. Having this in mind, the obtained reconstruction
is in fact extraordinary. Above the predicted primary energy of few
PeV, with some difference between the detector configurations, the
quality of the reconstruction seems to be good enough to allow for
CR-related studies.

\section{Reconstruction of multiplicity \label{sec:Multiplicity}}

Reconstruction of individual tracks in multi-track events is a complex
task, in the case of extensive air showers often practically impossible
without a very fine-grained detector. However, as is demonstrated
in this section, reconstruction of the number of muon tracks was possible
using machine learning algorithms. The result of the procedure in
Sec. \ref{subsec:Selection-of-the-best-model-energy} was adopted
for the estimator selection, given the similarity of the reconstruction
task and the same set of used features. Hence, in the following, LightGBM
has been used as the default model.

\subsection{Muon selection \label{subsec:Muon-selection}}

\begin{figure}[H]
\centering{}\includegraphics[width=16cm]{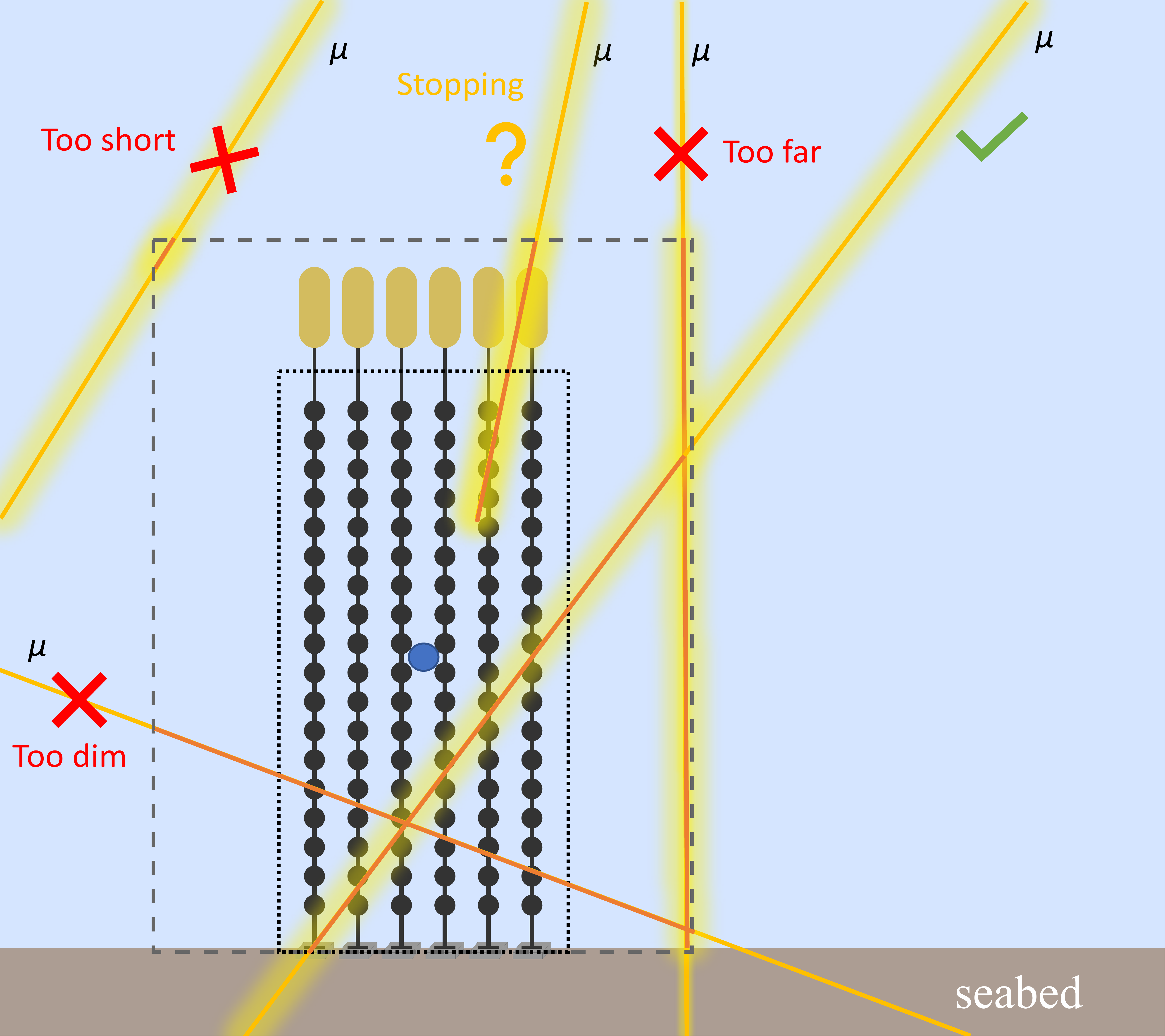}\caption{Sketch of various cases potentially relevant for the muon selection.
\label{fig:muon_selection_cases}}
\end{figure}

Not every muon can be reliably counted. There are certain cases, when
the muon might produce very few, or even no hits in the detector whatsoever.
They are sketched in Fig. \ref{fig:muon_selection_cases} and condense
into two categories: 
\begin{enumerate}
\item Energetic — low-energy muon emits less light and may just stop.
\item Spatial — light from the muon that is too far may not reach the optical
modules.
\end{enumerate}
It has to be noted that the muons created by $\nu_{\mu}$ interactions
are neglected here, as the muon-only MC was used. This is well justified,
since the neutrino-induced muon flux contributes to less than 1\%
of the overall muon flux, even at a depth of ARCA \cite{ARCA6_results}.
In fact, there are other secondaries (e.g. $\Sigma$, $\Lambda$,
$p$, $n$, $\pi^{\pm}$, $K^{\pm}$), which reach the sea level,
and could contribute to the muon flux at the KM3NeT detectors. However,
their contribution to the total flux is of the order of 2.7~\% (evaluated
using the CORSIKA datasets used within this chapter), and was hence
neglected.

\begin{figure}[H]
\begin{centering}
\subfloat[ARCA115.]{\centering{}\includegraphics[width=8cm]{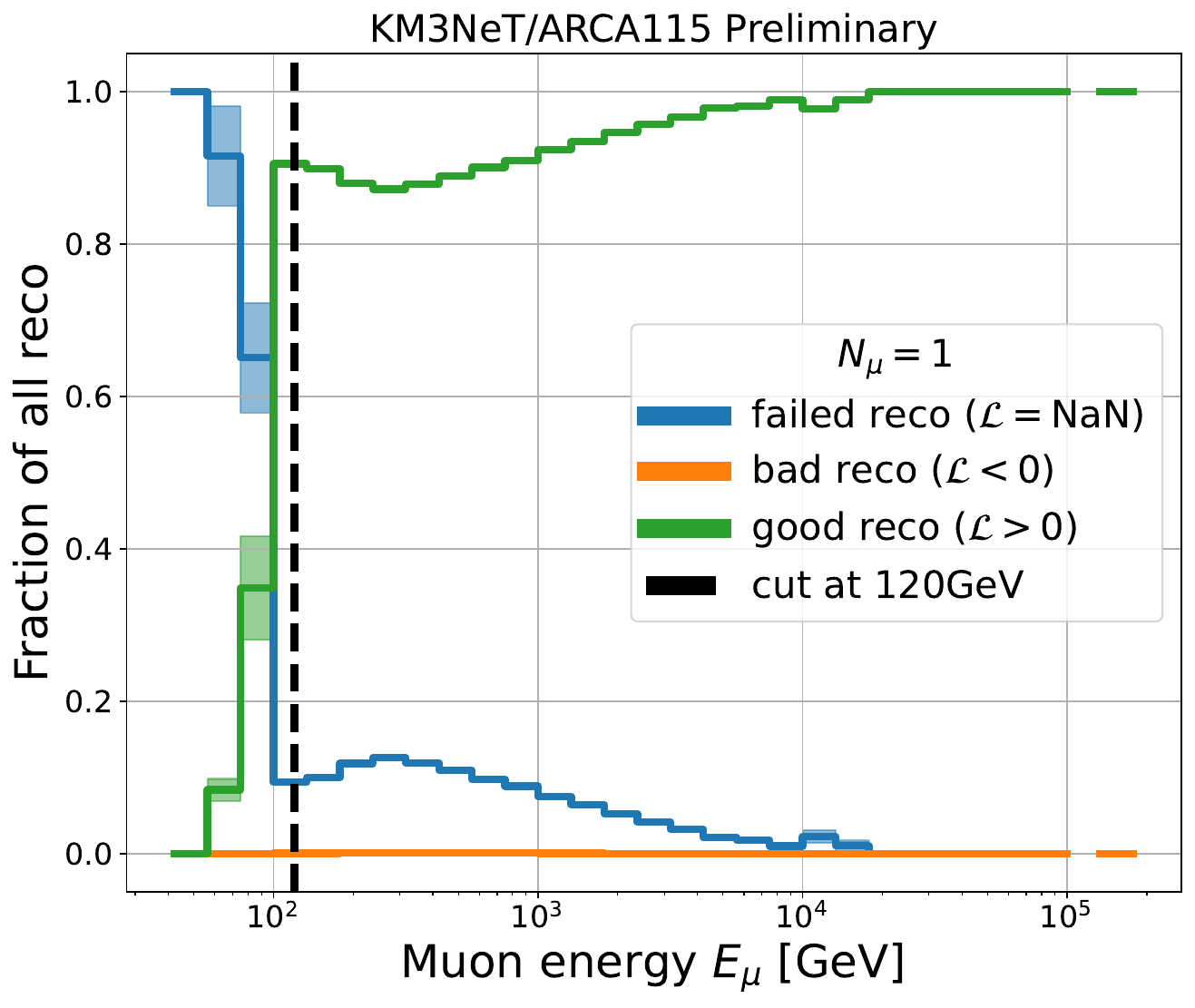}}\subfloat[ARCA6.]{\centering{}\includegraphics[width=8cm]{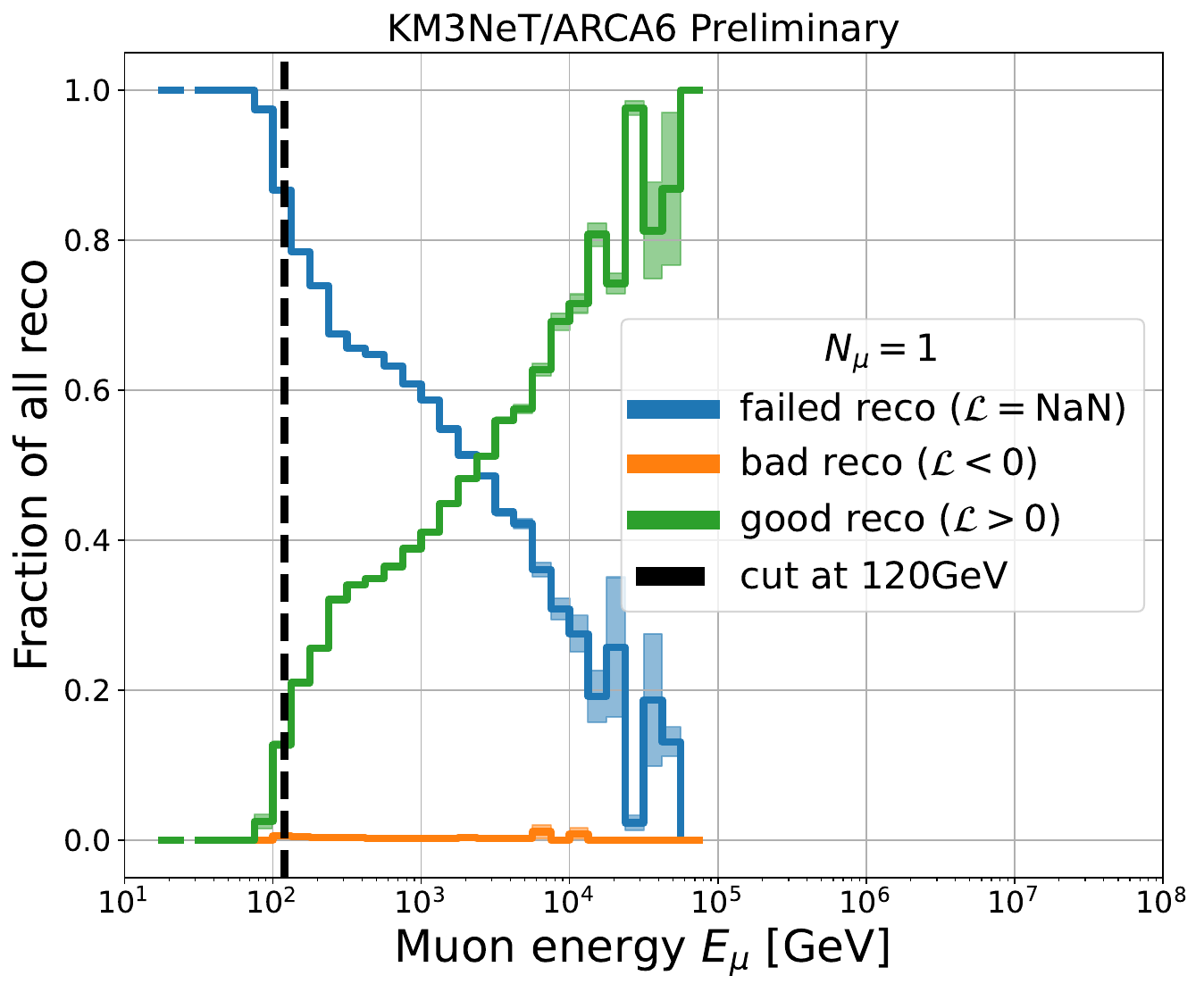}}
\par\end{centering}
\centering{}\subfloat[ORCA115.]{\centering{}\includegraphics[width=8cm]{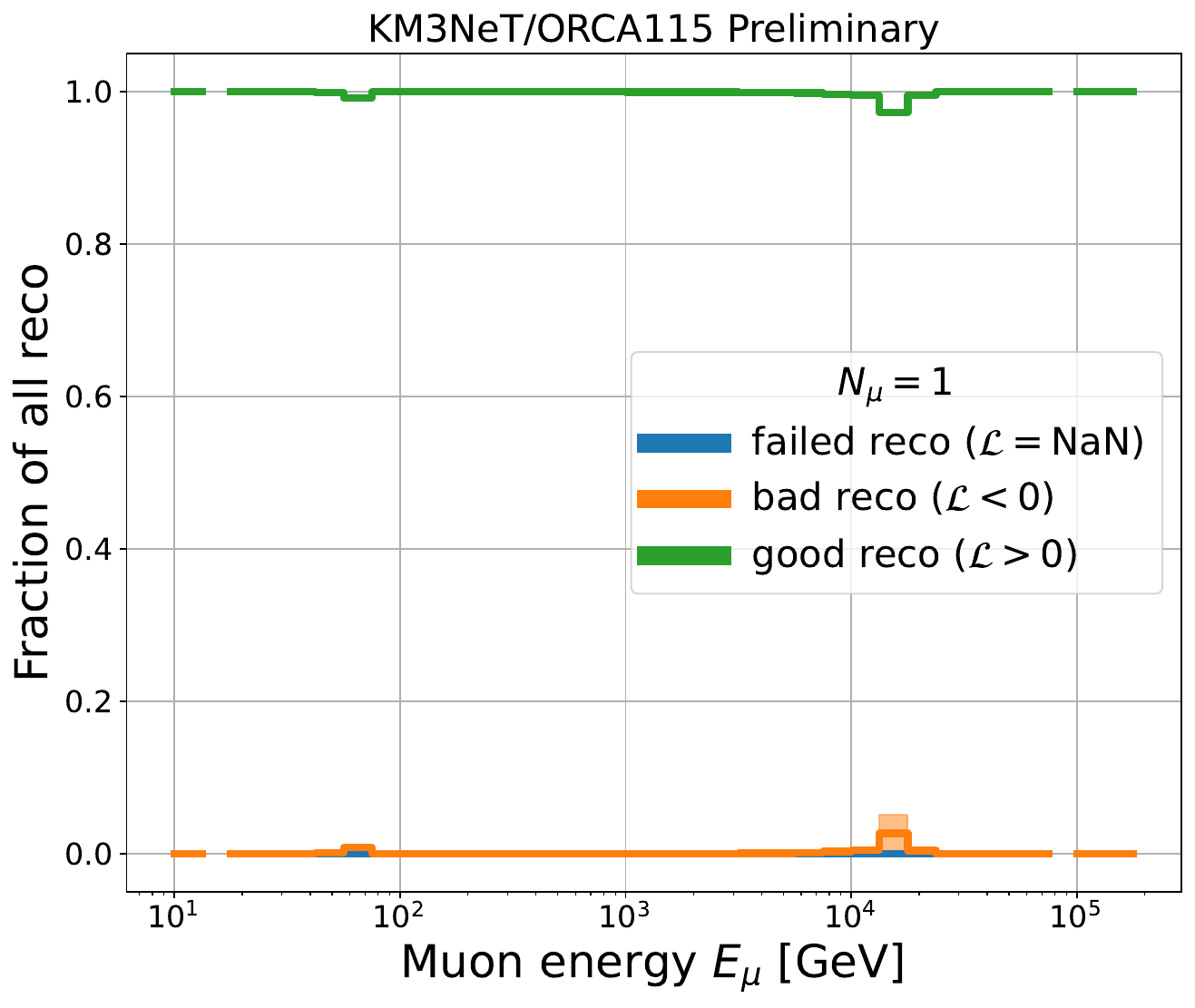}}\subfloat[ORCA6.]{\centering{}\includegraphics[width=8cm]{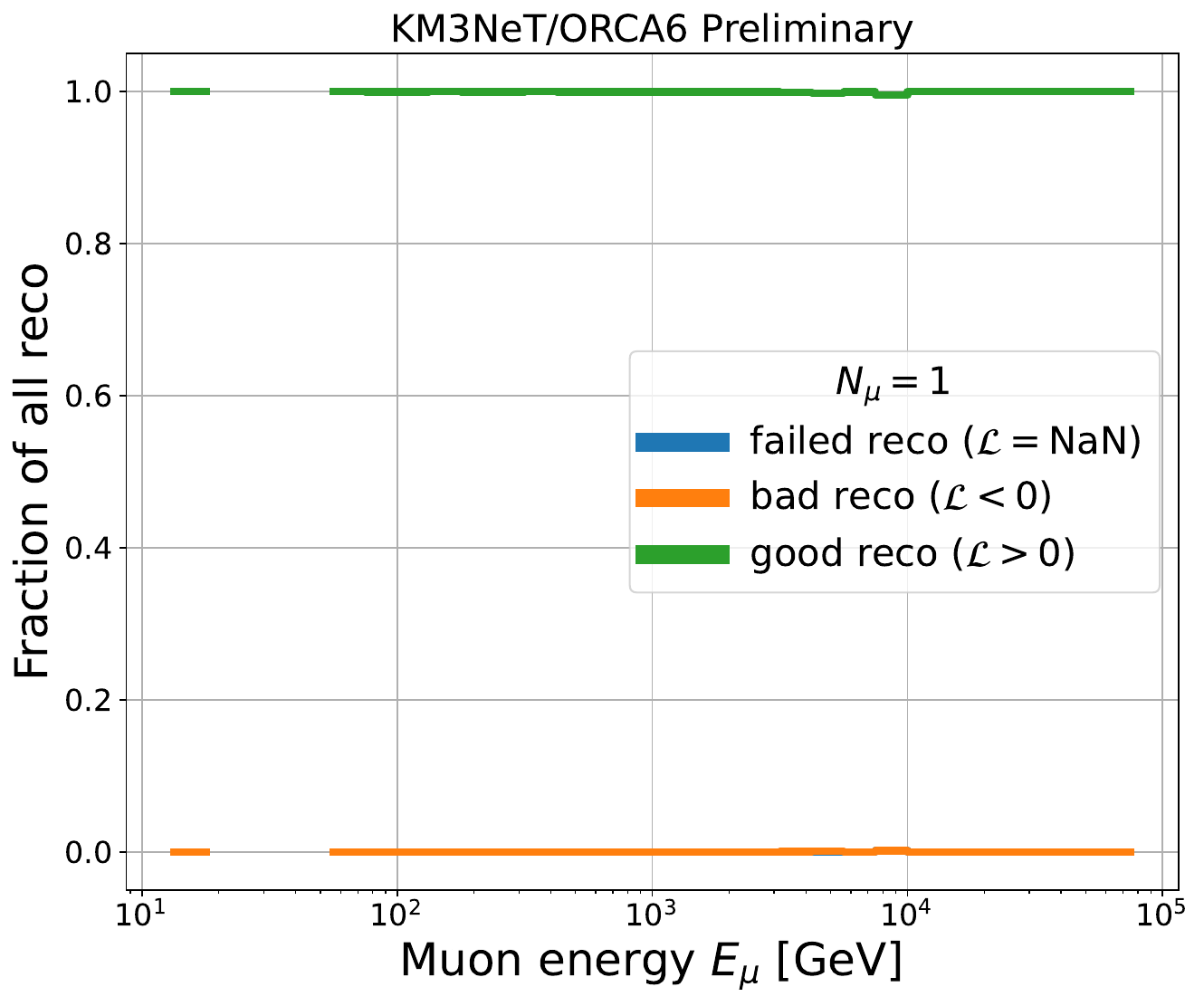}}\caption{Fractions of good ($\mathcal{L}>0$), bad ($\mathcal{L}<0$), and
completely failed ($\mathcal{L}=\mathrm{NaN}$) reconstructions as
function of $E_{\mu}$. Only single muon events (multiplicity $N_{\mu}=1$)
were used, and the reconstruction algorithm was JMuon. Plots are shown
for different detector configurations, and have the energy cut indicated
by a black dashed line (if applicable). \label{fig:reco-fractions-energy}}
\end{figure}

\begin{figure}[H]
\begin{centering}
\subfloat[ARCA115.]{\centering{}\includegraphics[width=8cm]{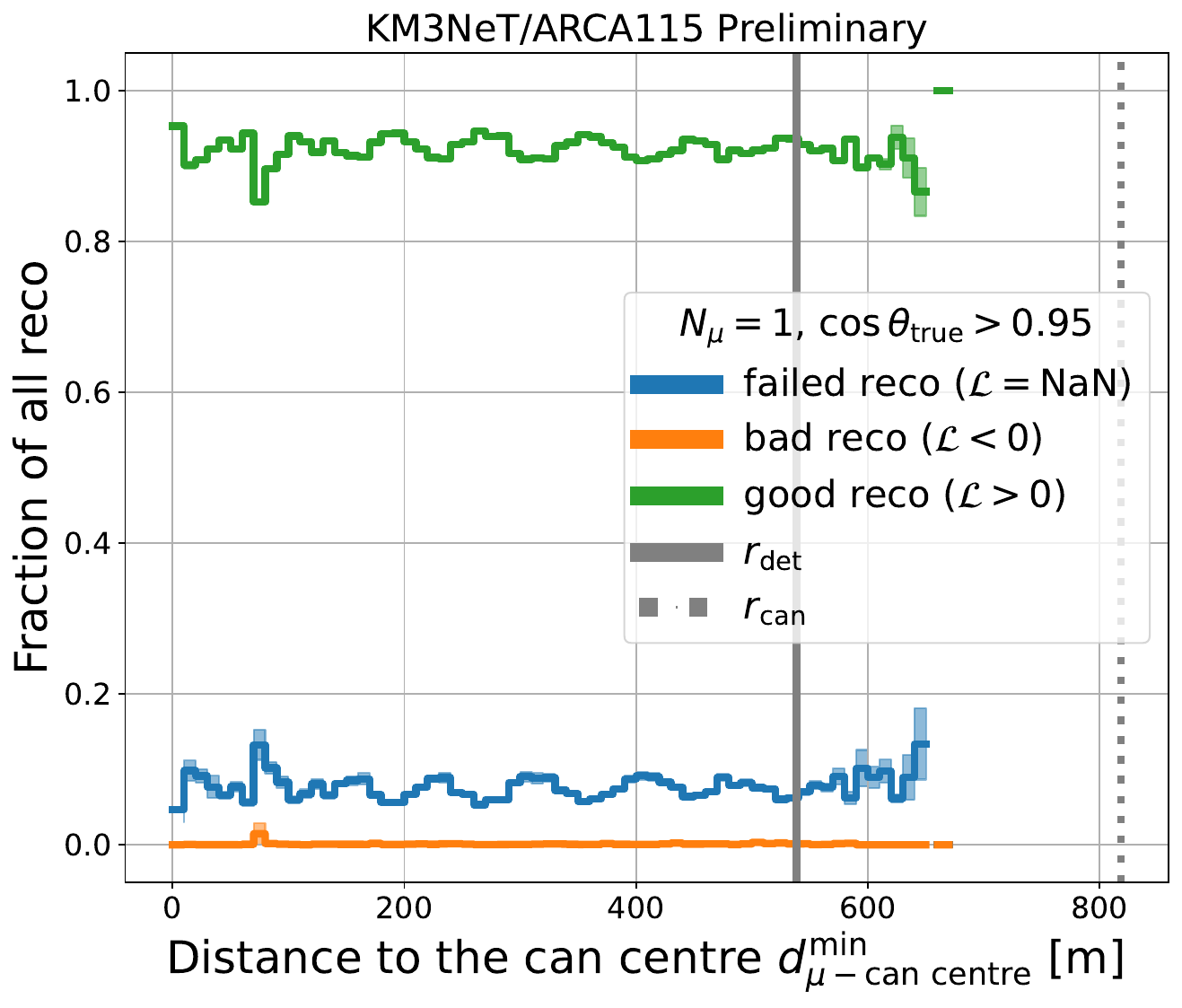}}\subfloat[ARCA6.]{\centering{}\includegraphics[width=8cm]{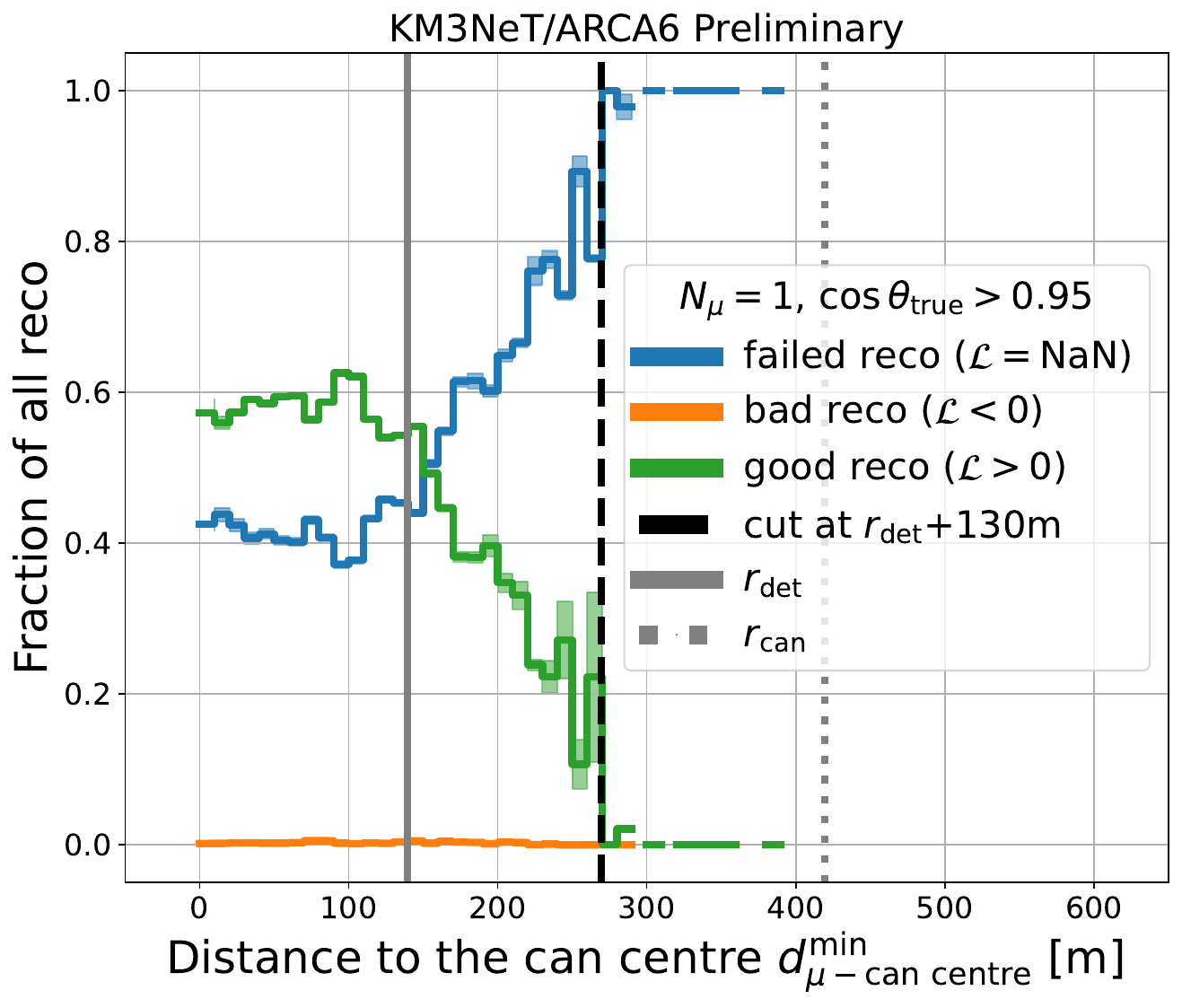}}
\par\end{centering}
\centering{}\subfloat[ORCA115.]{\centering{}\includegraphics[width=8cm]{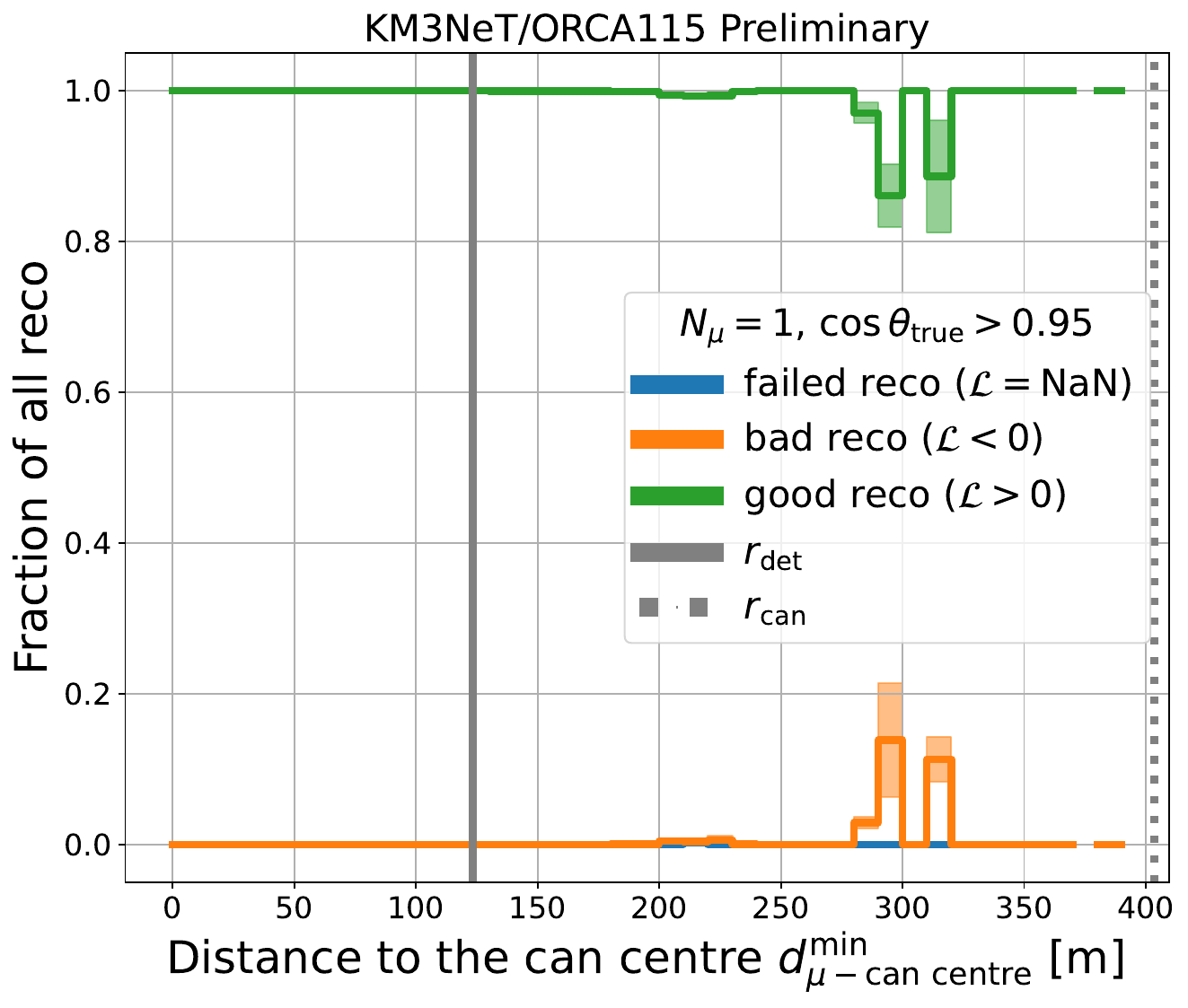}}\subfloat[ORCA6.]{\centering{}\includegraphics[width=8cm]{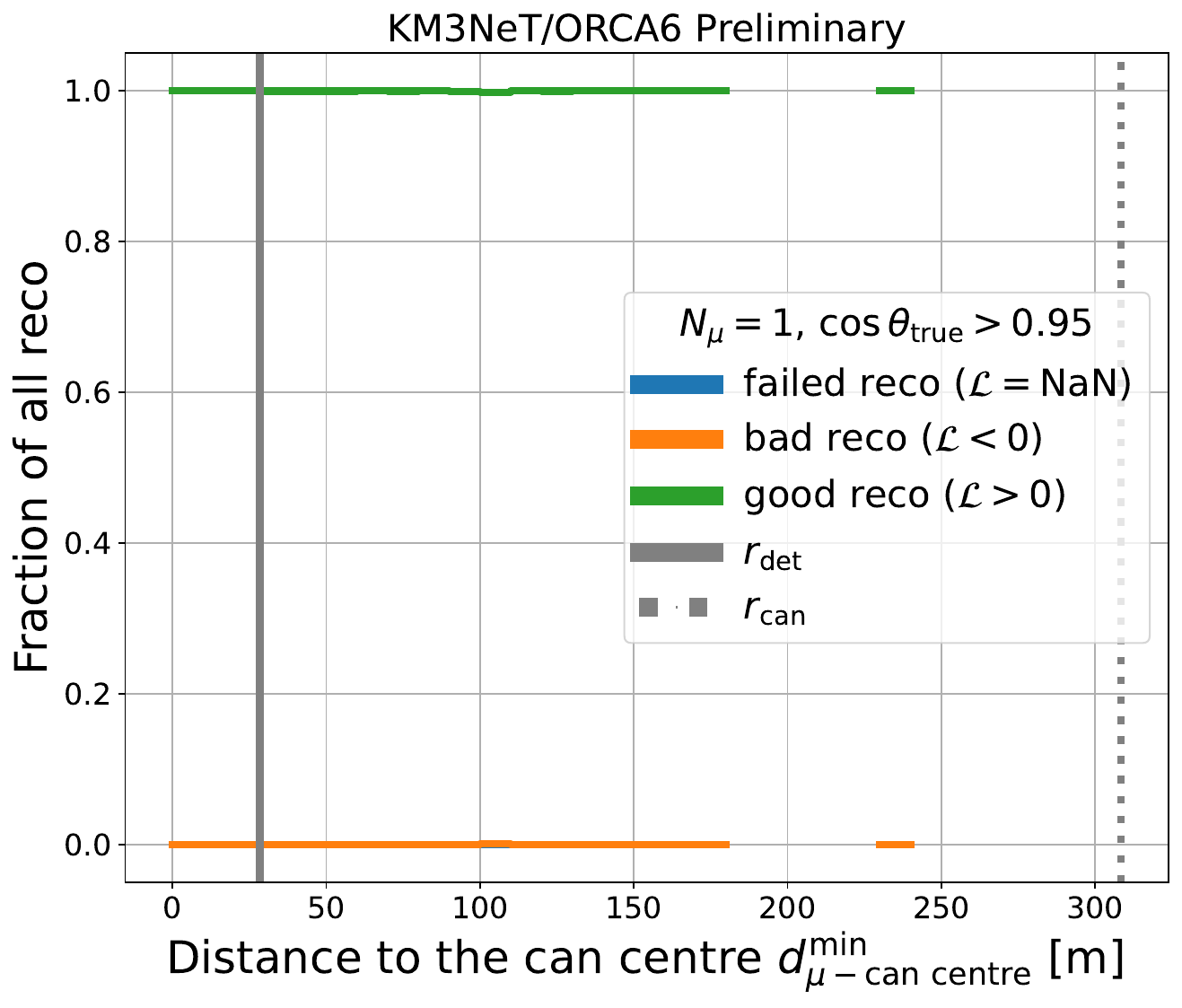}}\caption{Fractions of good ($\mathcal{L}>0$), bad ($\mathcal{L}<0$), and
completely failed ($\mathcal{L}=\mathrm{NaN}$) reconstructions as
function of $d_{\mu-\mathrm{can\,center}}^{\mathrm{min}}$ . Only
vertical ($\cos\theta_{\mathrm{true}}>0.95$) single muon events (multiplicity
$N_{\mu}=1$) were used, and the reconstruction algorithm was JMuon.
Plots are shown for different detector configurations, and have the
distance cut indicated by a black dashed line (if applicable). \label{fig:reco-fractions-distance}}
\end{figure}

\begin{figure}[H]
\begin{centering}
\subfloat[ARCA115.]{\centering{}\includegraphics[width=8cm]{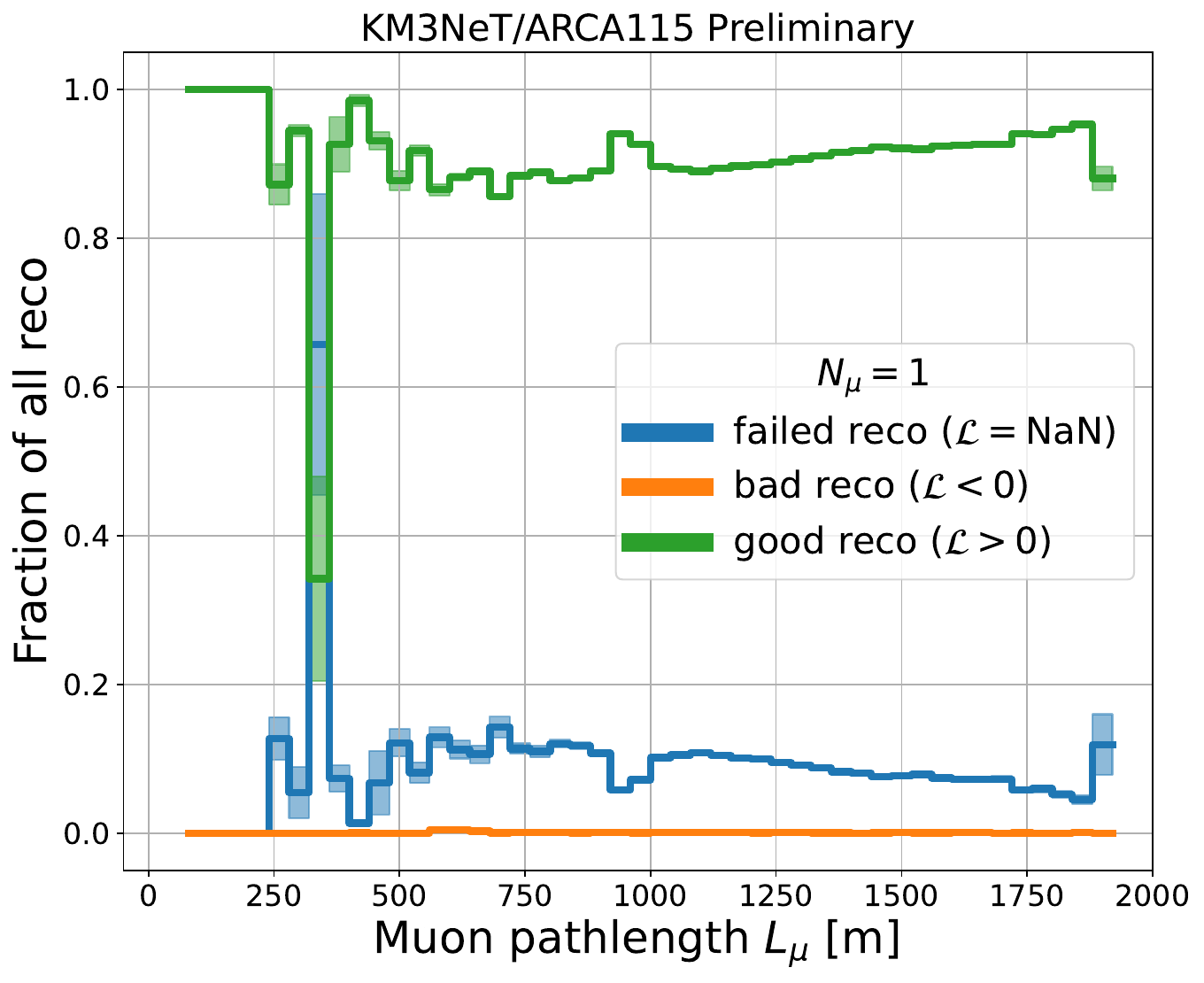}}\subfloat[ARCA6.]{\centering{}\includegraphics[width=8cm]{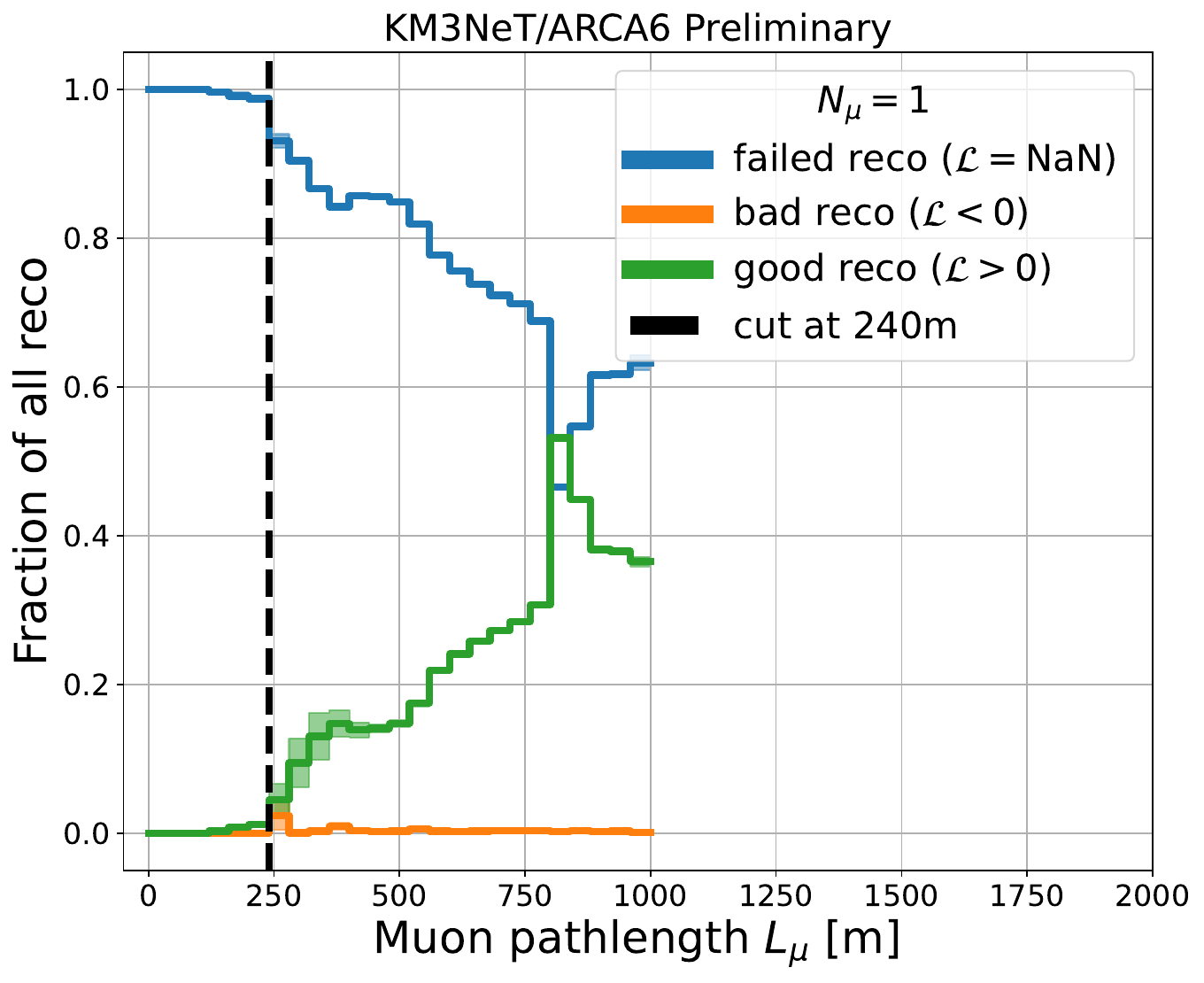}}
\par\end{centering}
\centering{}\subfloat[ORCA115.]{\centering{}\includegraphics[width=8cm]{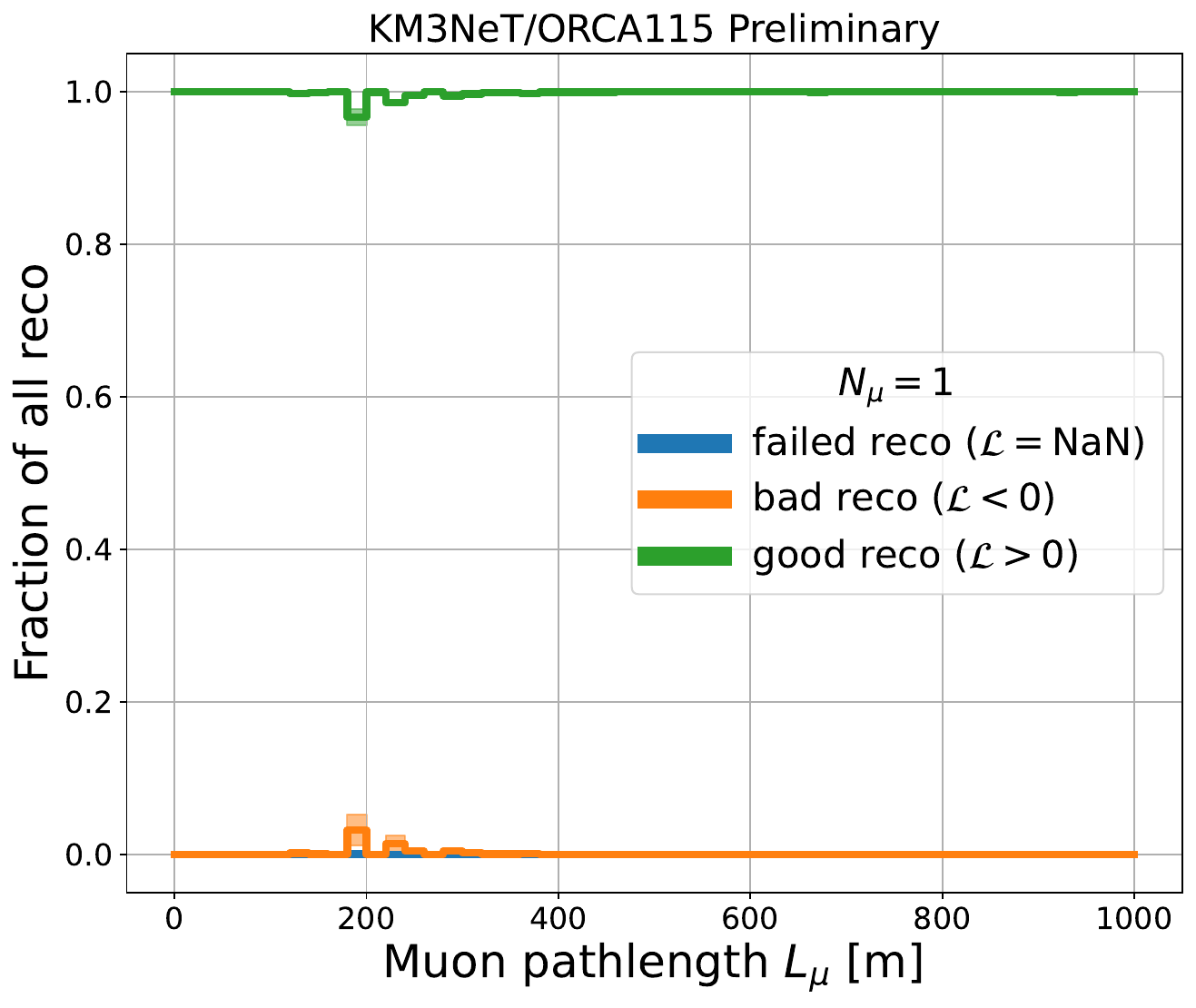}}\subfloat[ORCA6.]{\centering{}\includegraphics[width=8cm]{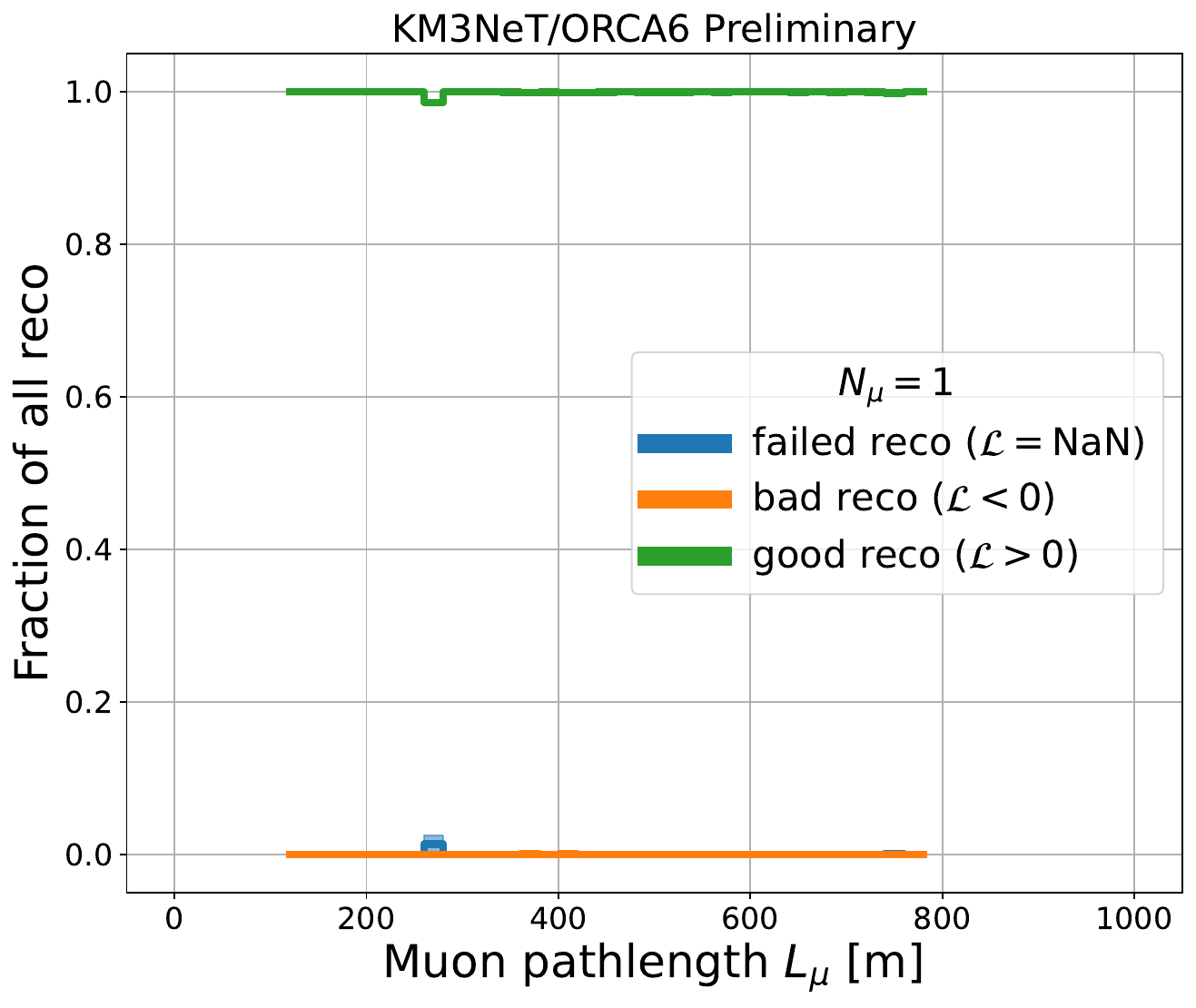}}\caption{Fractions of good ($\mathcal{L}>0$), bad ($\mathcal{L}<0$), and
completely failed ($\mathcal{L}=\mathrm{NaN}$) reconstructions as
function of muon pathlength $L_{\mu}$. Only single muon events (multiplicity
$N_{\mu}=1$) were used, and the reconstruction algorithm was JMuon.
Plots are shown for different detector configurations, and have the
pathlength cut indicated by a black dashed line (if applicable). \label{fig:reco-fractions-pathlength}}
\end{figure}

Based on these considerations and having a standard single muon track
reconstruction (JMuon) at hand, a set of three control plots (see
Fig. \ref{fig:reco-fractions-energy}, \ref{fig:reco-fractions-distance},
and \ref{fig:reco-fractions-pathlength}) was devised. They show the
reconstruction quality of JMuon, which is quantified in terms of the
likelihood parameter $\mathcal{L}$, as a function of:
\begin{itemize}
\item muon energy $E_{\mu}$,
\item minimum distance of the muon track from the can centre $d_{\mu-\mathrm{can\,center}}^{\mathrm{min}}$
(blue dot in Fig. \ref{fig:muon_selection_cases}): only vertical
muons with $\cos\theta>0.95$ ($\theta$ is the true zenith angle
of the muon, and $\cos\theta=1.0$ is perfectly vertical) are considered,
\item muon pathlength $L_{\mu}$ under assumption of a straight line track
(neglecting the scattering), determined after fixing the $d_{\mu-\mathrm{can\,center}}^{\mathrm{min}}$
value, and thus, the volume of interest.
\end{itemize}
For all three listed variables, only events with true muon multiplicity
$N_{\mu}=1$ were used. The motivation was to ensure that each likelihood
value $\mathcal{L}$ would correspond to a single muon. Moreover,
in the case of $d_{\mu-\mathrm{can\,center}}^{\mathrm{min}}$ (Fig.
\ref{fig:reco-fractions-distance}), only vertical ($\cos\theta>0.95$)
muons were investigated. The muon direction was fixed along the vertical
axis of the detector for two reasons. Firstly, the statistics of horizontal
single muon events in the Monte Carlo was insufficient. The second
reason was the intention to simplify the problem and look only for
one value of maximal allowed distance from the optical modules $d_{\mathrm{max}}$,
both in horizontal, and vertical direction. Such an approach resulted
in a cylindrical volume of interest, with radius $r=r_{\mathsf{det}}+d_{\mathrm{max}}$,
and height $h=h_{\mathrm{det}}+d_{\mathrm{max}}$, where $r_{\mathrm{det}}$
and $h_{\mathrm{det}}$ are the radius and height of the smallest
cylinder enclosing the instrumented volume (the detector). The value
of $d_{\mathrm{max}}$ (see Tab. \ref{tab:Selections-of-muons}) was
decided based on the $d_{\mu-\mathrm{can\,centre}}^{\mathrm{min}}$
plot (Fig. \ref{fig:reco-fractions-distance}) for each of the detector
configurations. In the described approach, differences in the inter-DOM
and inter-DU spacings, and due to inclination other than horizontal
or vertical were neglected.

The muon selections for the muon multiplicity reconstruction were
derived from Fig. \ref{fig:reco-fractions-energy}, \ref{fig:reco-fractions-distance},
and \ref{fig:reco-fractions-pathlength}, and were compiled in Tab.
\ref{tab:Selections-of-muons}. Somewhat expected, no cuts seemed
to be needed for ORCA (thanks to much denser instrumentation than
ARCA), therefore the energy cut has been set to 1~GeV, which is the
expected lower threshold for the reconstruction, according to \cite{KM3NeT-LoI-2.0}.
An alternative approach could have been the energy threshold matching
the point, where the LightGBM bundle energy starts to fail (see Sec.
\ref{subsec:Results-Ebundle}) or to set a certain cutoff at the difference
between the true energy and JMuon prediction, however neither of the
two was adopted. The energy cut for ARCA was the same regardless of
the configuration, and for ARCA6, additional spatial cuts were found
to be beneficial. Those selections were applied by default in the
following sections.

\begin{table}[H]
\begin{centering}
\caption{Selection of muons to be counted when computing the multiplicity for
different detector configurations. \label{tab:Selections-of-muons}}
\par\end{centering}
\centering{}%
\begin{tabular}{|c|c|c|c|}
\hline 
Detector & Minimal $E_{\mu}$ {[}GeV{]} & $d_{\mathrm{max}}$ {[}m{]} & Minimal $L_{\mu}$ {[}m{]}\tabularnewline
\hline 
\hline 
ARCA115 & $120$ & $-$ & $-$\tabularnewline
\hline 
ARCA6 & $120$ & $269.4$ & $240$\tabularnewline
\hline 
ORCA115 & $1$ & $-$ & $-$\tabularnewline
\hline 
ORCA6 & $1$ & $-$ & $-$\tabularnewline
\hline 
\end{tabular}
\end{table}

\subsection{Verification of the muon selection}

The effect of introducing the muon selection (see Sec. \ref{subsec:Muon-selection})
was the modification of the target for the reconstruction. Instead
of predicting the true number of all muons in an event, the model
attempted to evaluate the true number of all muons meeting the selection
criteria (Tab. \ref{tab:Selections-of-muons}) in an event. This number
is smaller or equal than the one without selection, and there were
events for which such multiplicity with muon selection was equal to
zero. The impact of the muon selection was demonstrated on ARCA6,
as there, it is expected to be most pronounced. As one may see in
Fig. \ref{fig:Nmu-muon-selection-comparison}, the selection improves
both metric values. After inspection of the bottom-left part of Fig.
\ref{fig:With-muon-selection.}, it is evident that the reconstruction
struggled with $N_{\mu}=0$ events, mostly betting for $N_{\mu}=1$.
True multiplicity 0 cases were relatively rare ($<0.4\%$ of all events),
while not counting barely visible muons clearly boosted the overall
LightGBM's performance. The muon selection was used as the default
in all further sections.
\begin{center}
\begin{figure}[H]
\centering{}\subfloat[Without muon selection.]{\centering{}\includegraphics[width=8cm]{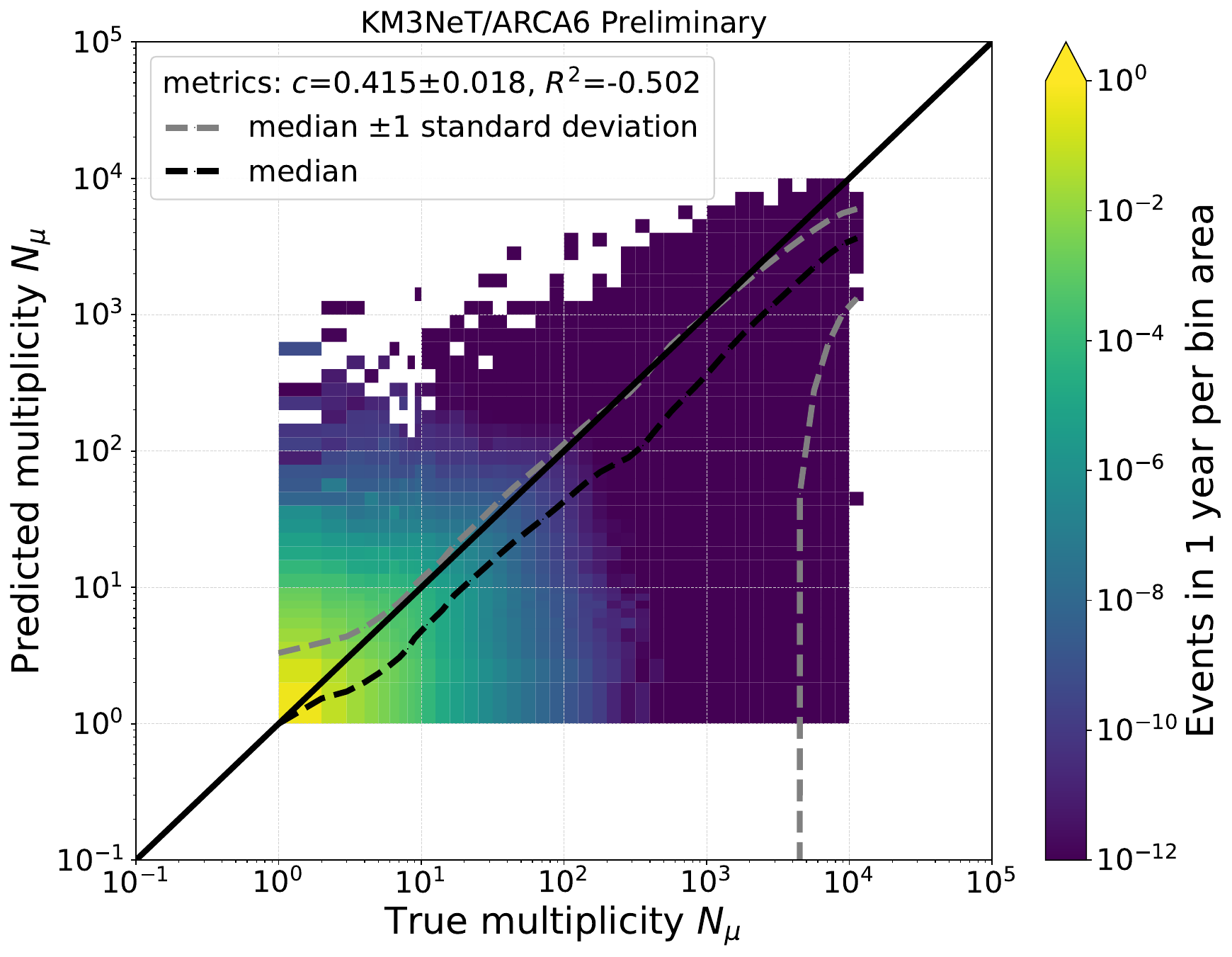}}\subfloat[With muon selection.\label{fig:With-muon-selection.}]{\centering{}\includegraphics[width=8cm]{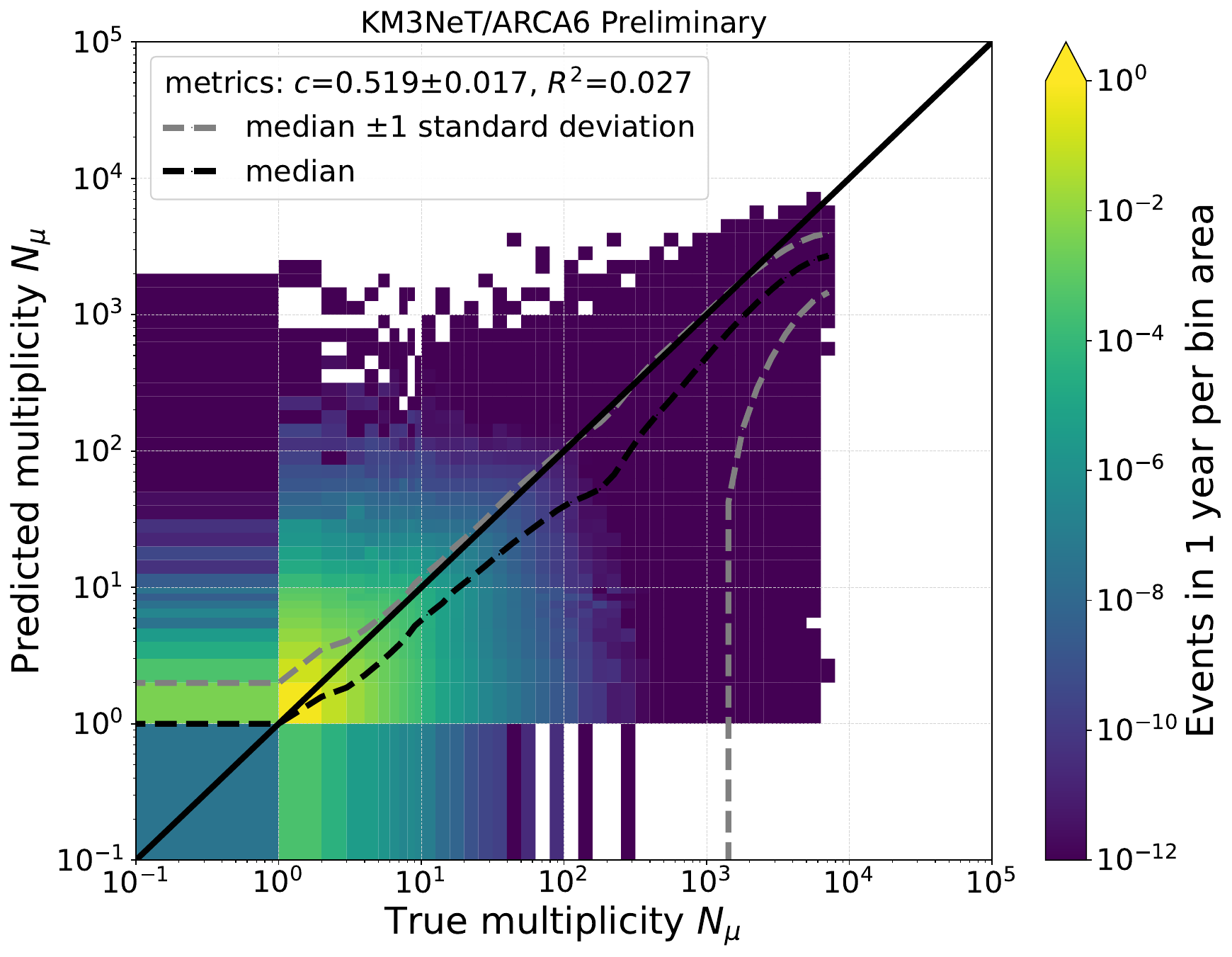}}\caption{Predicted muon multiplicity plotted against the MC truth with and
without muon selection. In each of the cases, the LightGBM model was
trained on features selected according to results of Sec. \ref{subsec:Feature-importances-Nmu}.
Both the correlation $c$ and $R^{2}-$score values were computed
on the bin values, not individual events. \label{fig:Nmu-muon-selection-comparison}}
\end{figure}
\par\end{center}

\subsection{Feature importances\label{subsec:Feature-importances-Nmu}}

Feature importances for the multiplicity reconstruction have been
studied analogously to Sec. \ref{sec:Energy}. The result is shown
in Fig. \ref{fig:Nmu-feature-importance} and \ref{fig:Nmu-feature-selection-comaprison}.
The same behaviour has been observed — the most profitable selection
was the one picking only features with positive importance, hence
the other selections were not included in Fig. \ref{fig:Nmu-feature-selection-comaprison}.
The correlation coefficient $c$ has in fact slightly worsened, however
within the statistical uncertainty, while the $R^{2}$-score improved
by 8~\%. Interestingly, in this particular case the feature selection
also purged the events falsely reconstructed as $N_{\mu}=0$. However,
this is not a general rule, as one can see in Sec. \ref{subsec:Results-simple-regr}.
\begin{center}
\begin{figure}[H]
\centering{}\includegraphics[width=16cm]{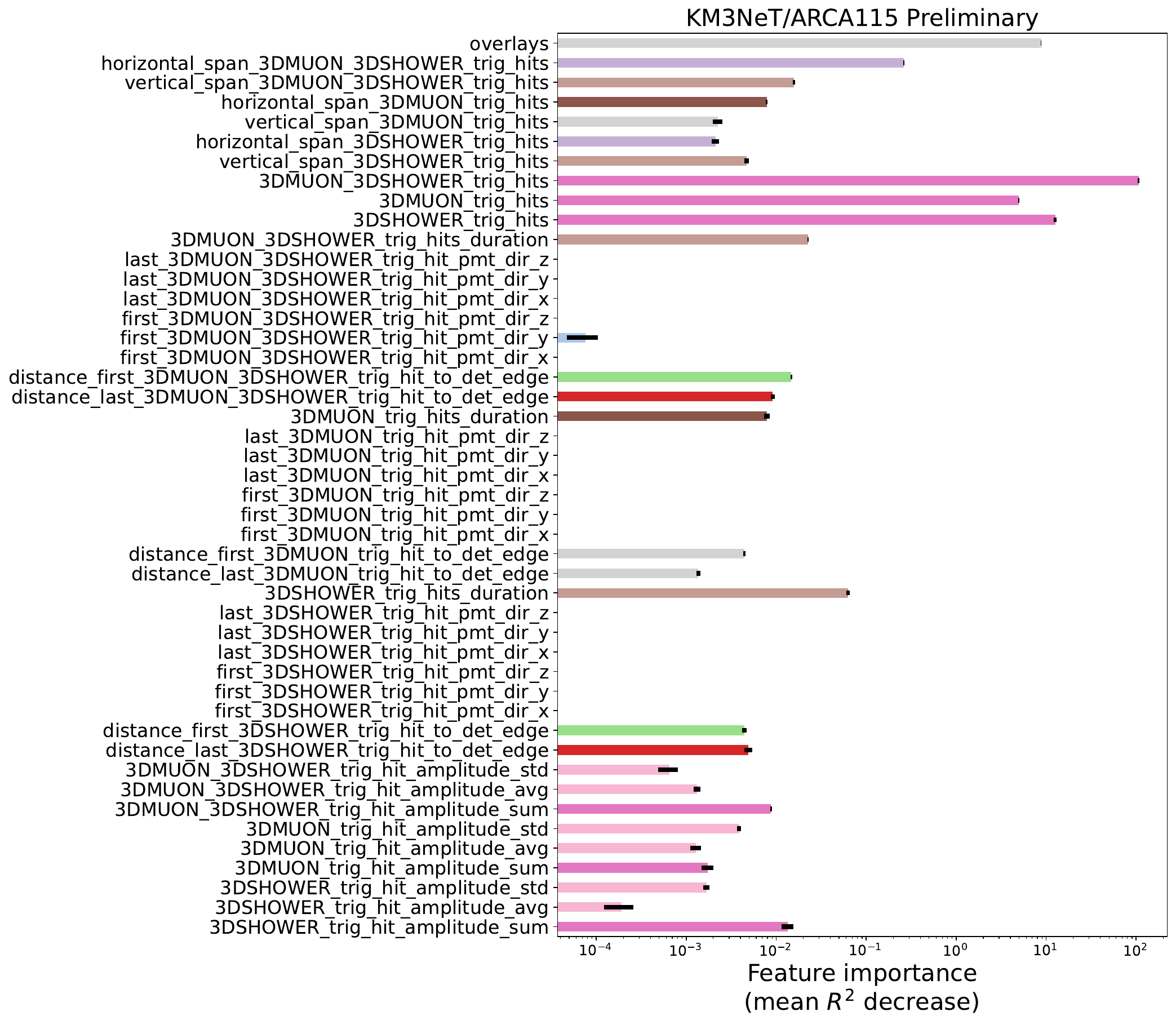}\caption{Feature importance plot, showing how much each feature contributed
to the training. The figure has been obtained by performing 10 permutations
on a trained model to ensure a stable result. The standard deviation
for each feature is denoted by a black bar. The colour coding of features
is consistent with Fig. \ref{fig:dendrogram} to show the feature
clusters and their most important members. Feature importance plots
for the other detector configurations were gathered in Sec. \ref{subsec:Feature-importances}.
\label{fig:Nmu-feature-importance}}
\end{figure}
\par\end{center}

\begin{center}
\begin{figure}[H]
\centering{}\subfloat[All features.]{\centering{}\includegraphics[width=8cm]{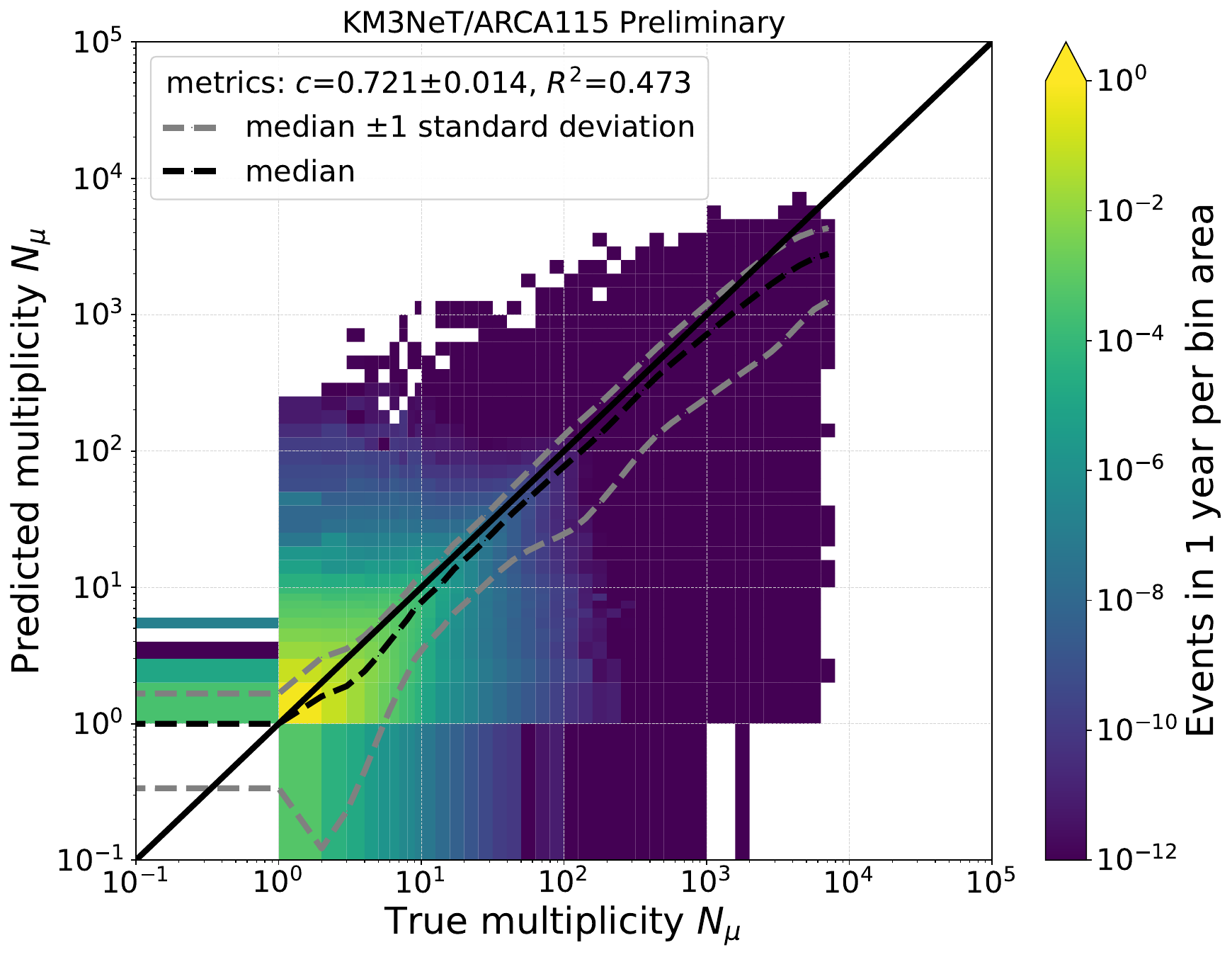}}\subfloat[Selected features (positive importance). \label{fig:Selected-features-positive-importance-Nmu}]{\centering{}\includegraphics[width=8cm]{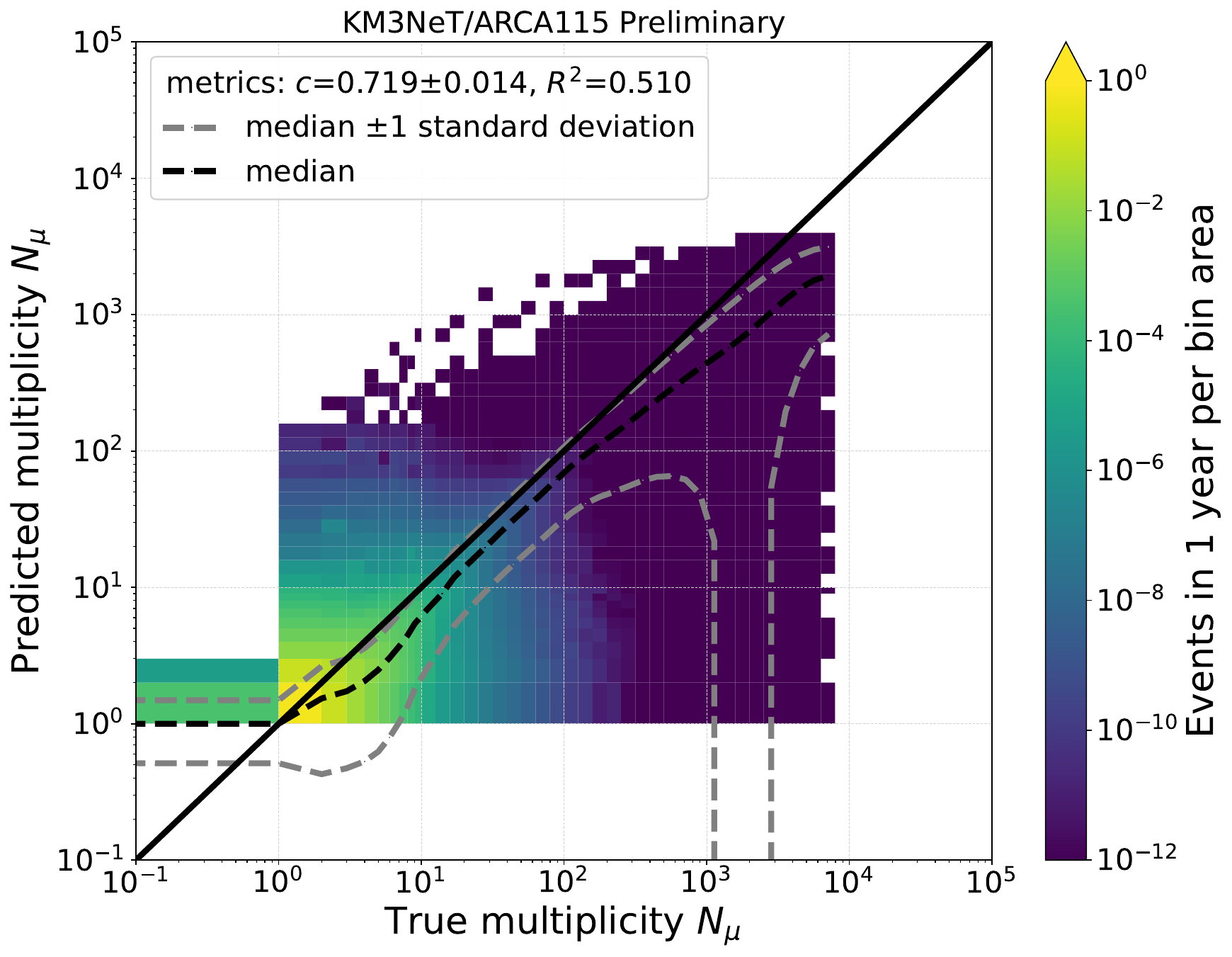}}\caption{Predicted muon multiplicity plotted against the MC truth with and
without feature selection. Both the correlation and $R^{2}-$score
values were computed on the bin values, not individual events. \label{fig:Nmu-feature-selection-comaprison}}
\end{figure}
\par\end{center}

\subsection{Hyperparameter tuning}

The LightGBM hyperparameters have been tuned in the same way as for
the energy reconstruction. The resulting parameter values are documented
in Tab. \ref{tab:LightGBM-tuned-hyperparameters-summary}.

\subsection{Results\label{subsec:Results-simple-regr}}

\begin{figure}[H]
\begin{centering}
\subfloat[ARCA115.]{\centering{}\includegraphics[width=8cm]{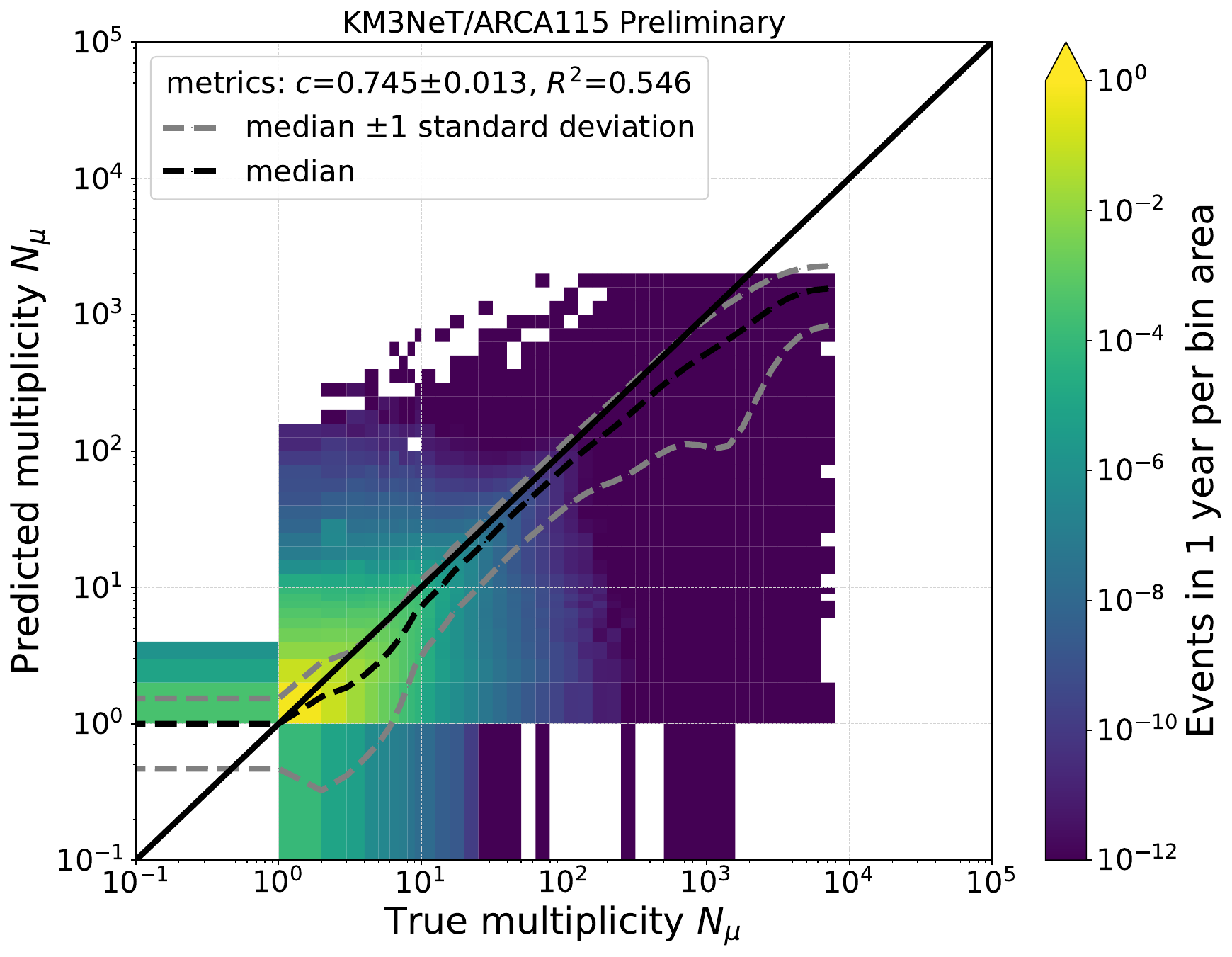}}\subfloat[ARCA6.]{\centering{}\includegraphics[width=8cm]{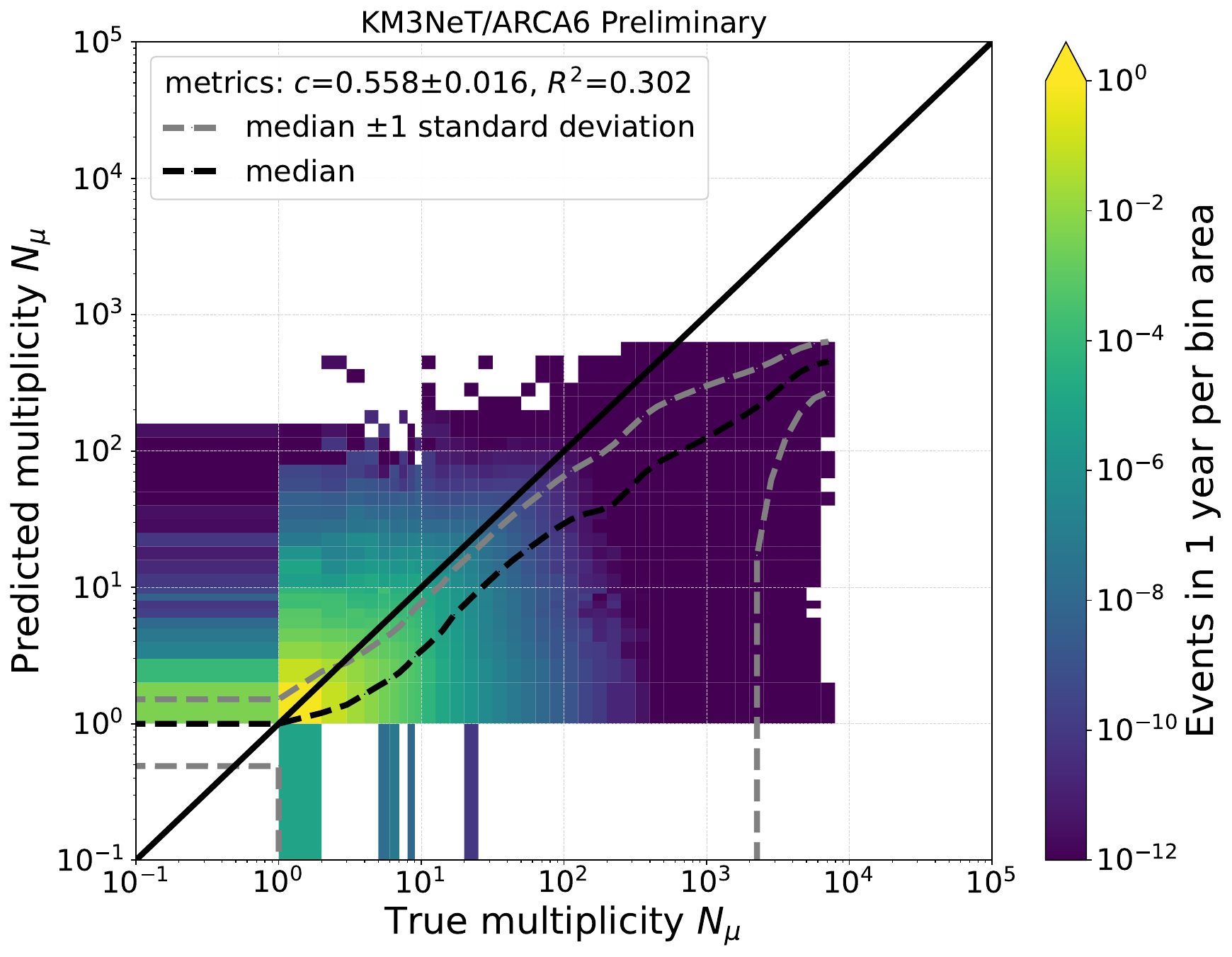}}
\par\end{centering}
\centering{}\subfloat[ORCA115.]{\centering{}\includegraphics[width=8cm]{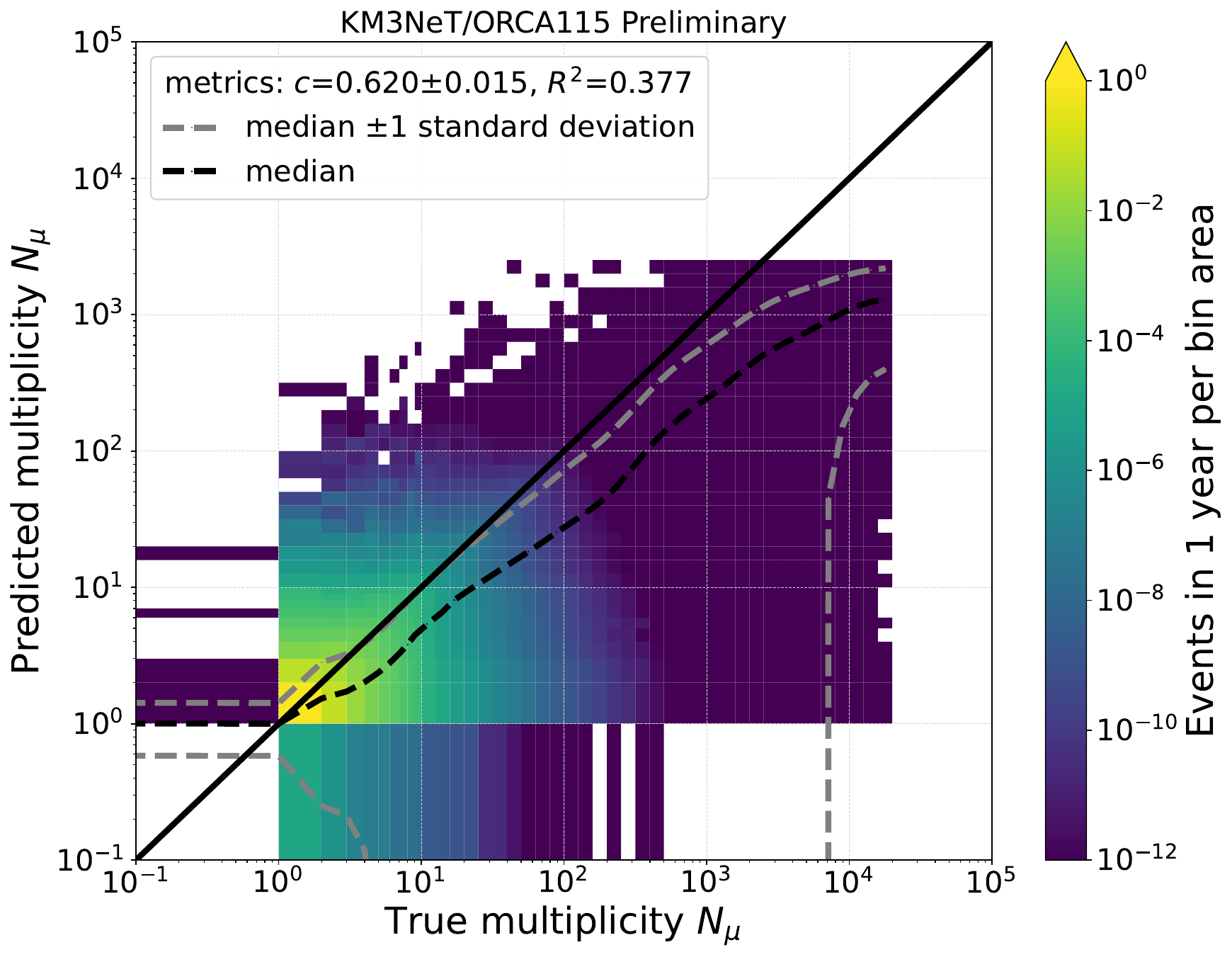}}\subfloat[ORCA6.]{\centering{}\includegraphics[width=8cm]{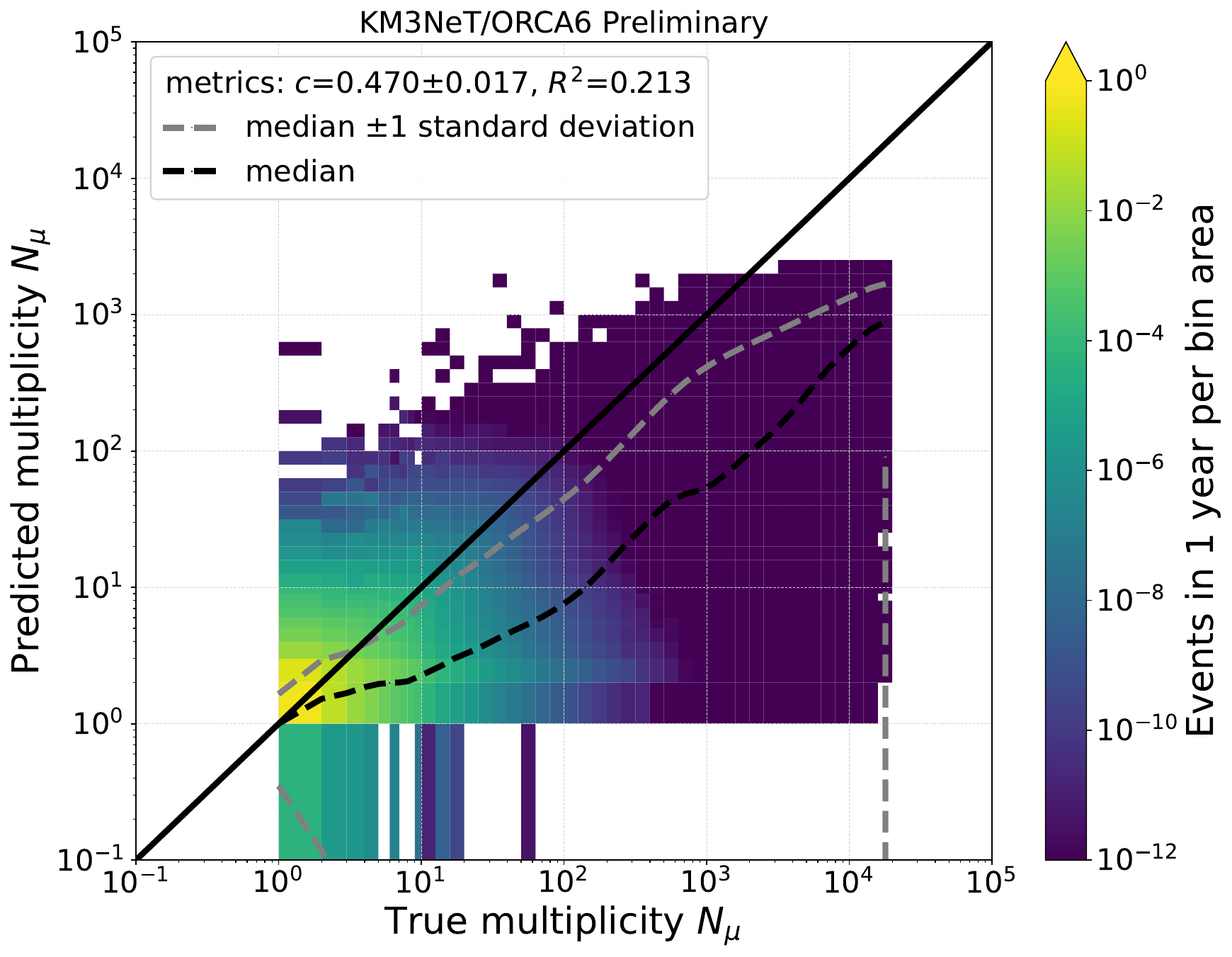}}\caption{Comparison of muon multiplicity reconstructed with the LigthGBM model
against the true value. Both the correlation and $R^{2}-$score values
were computed on the bin values, not individual events. \label{fig:Nmu_simple_reco_results}}
\end{figure}

\begin{figure}[H]
\begin{centering}
\subfloat[ARCA115.]{\centering{}\includegraphics[width=8cm]{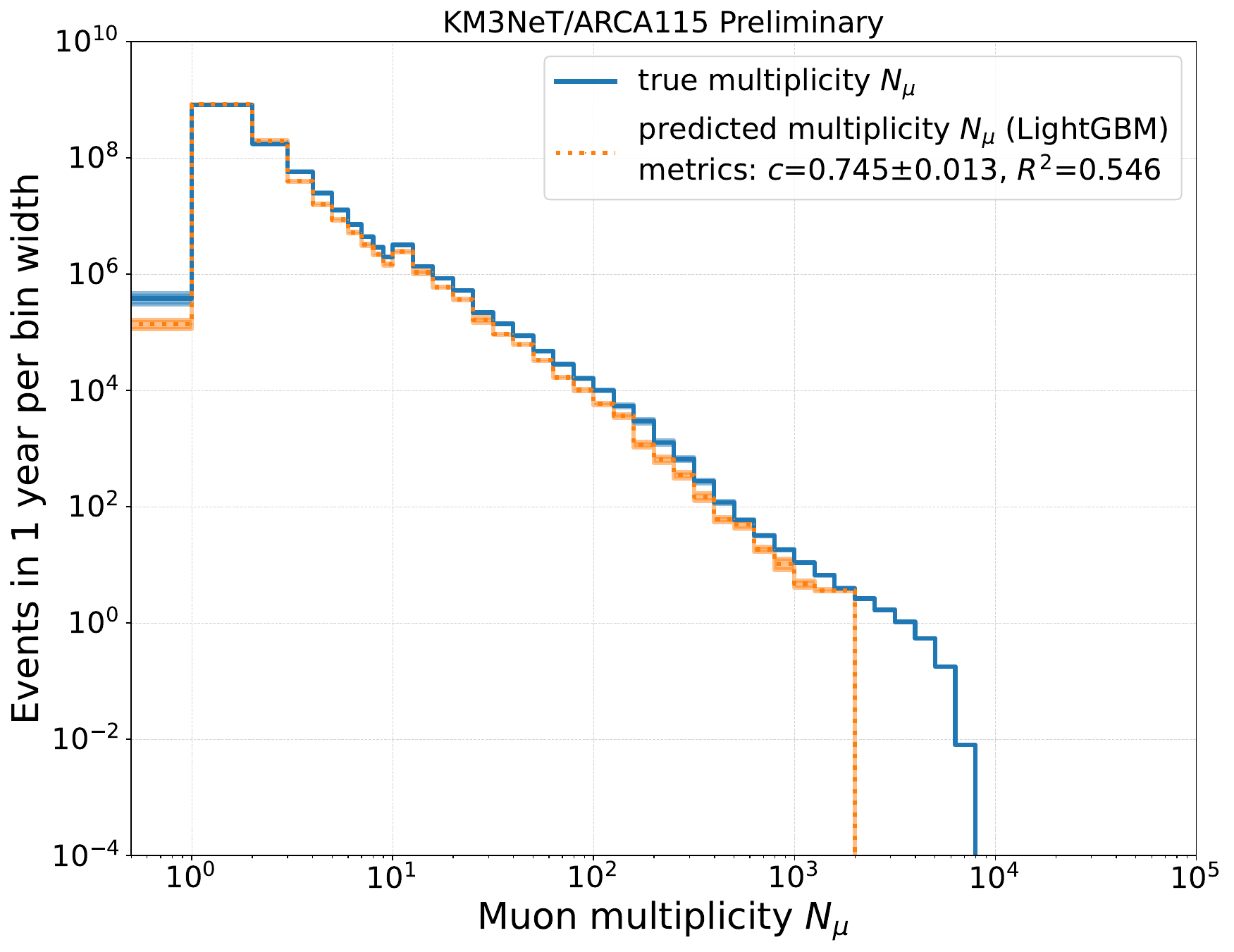}}\subfloat[ARCA6.]{\centering{}\includegraphics[width=8cm]{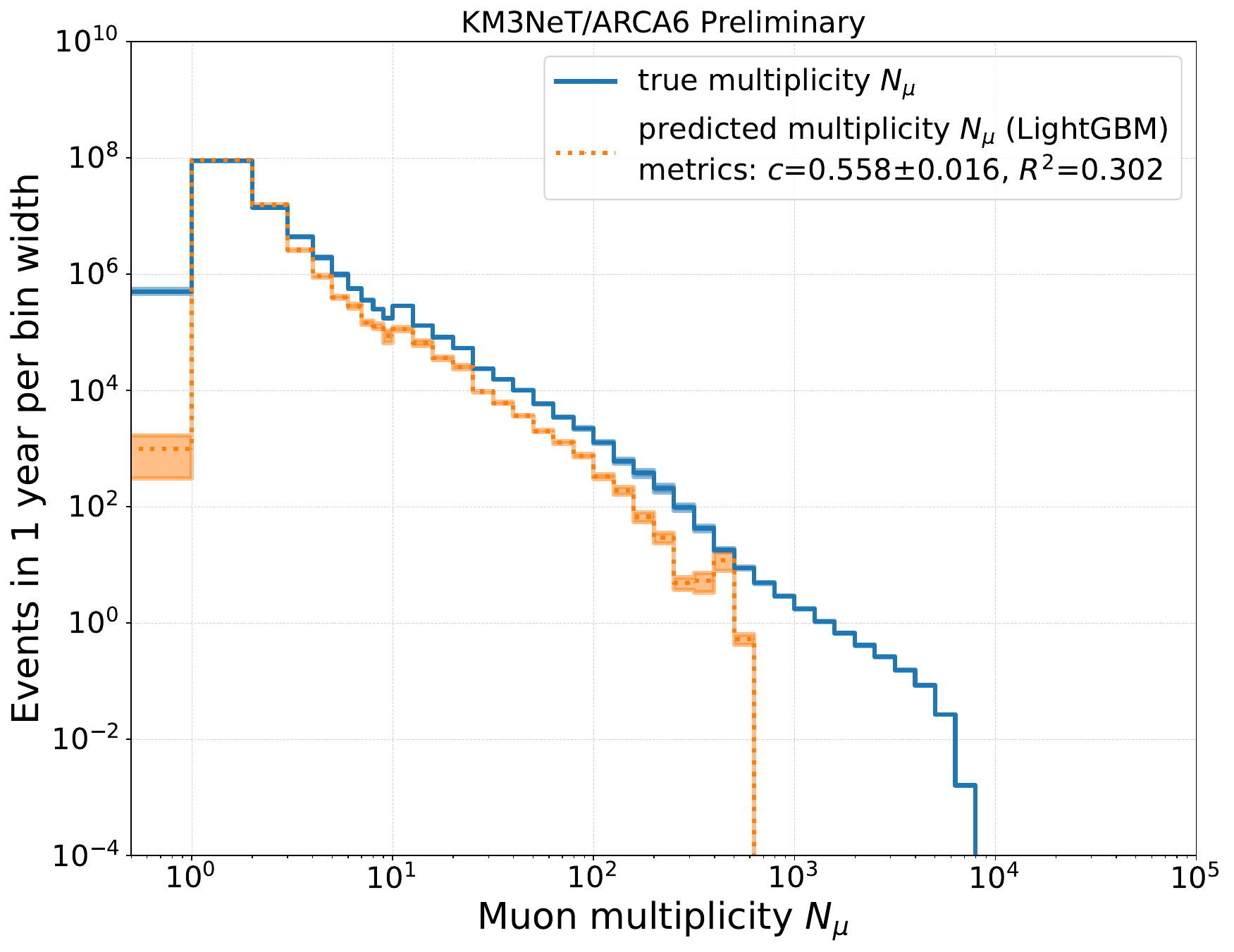}}
\par\end{centering}
\centering{}\subfloat[ORCA115.]{\centering{}\includegraphics[width=8cm]{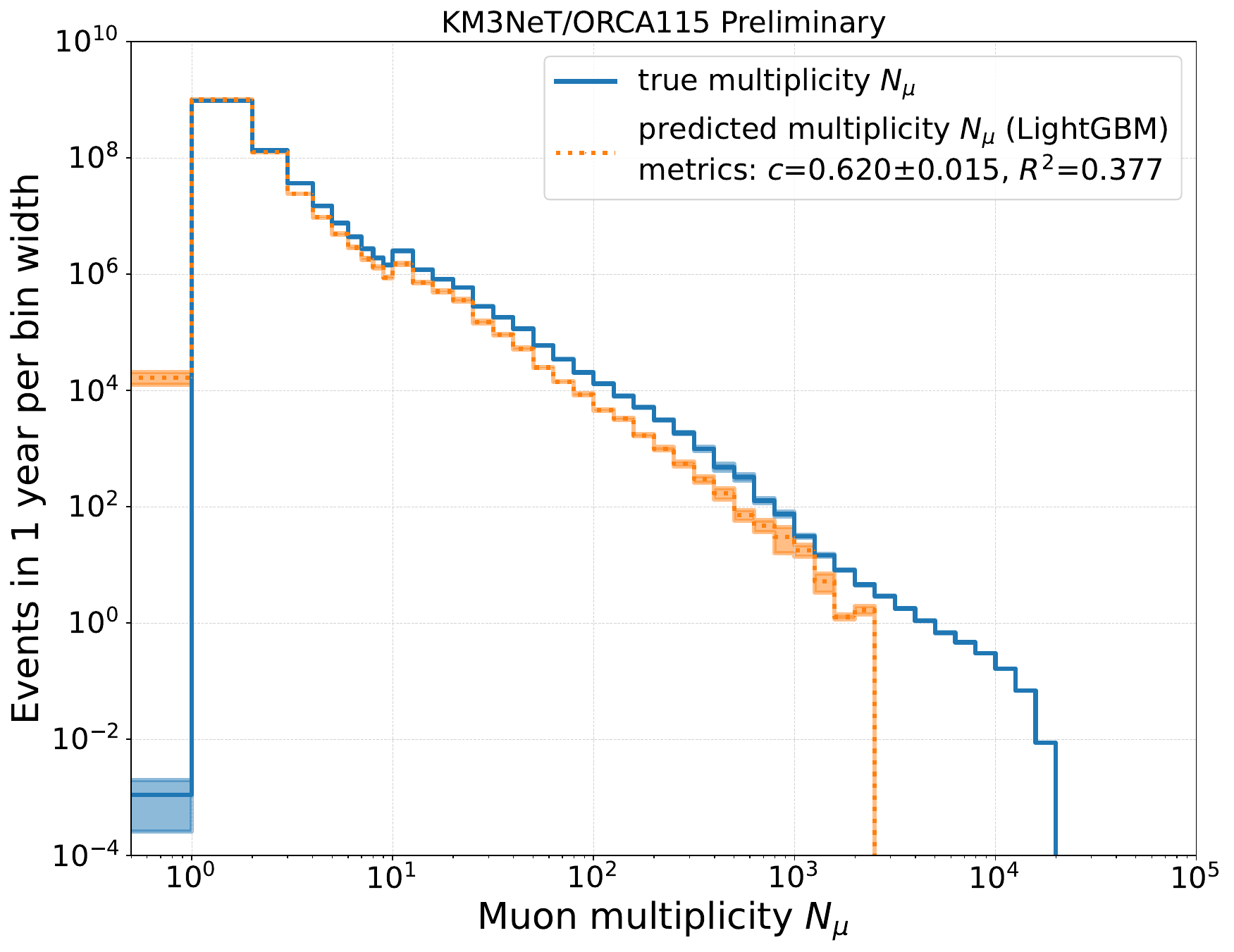}}\subfloat[ORCA6.]{\centering{}\includegraphics[width=8cm]{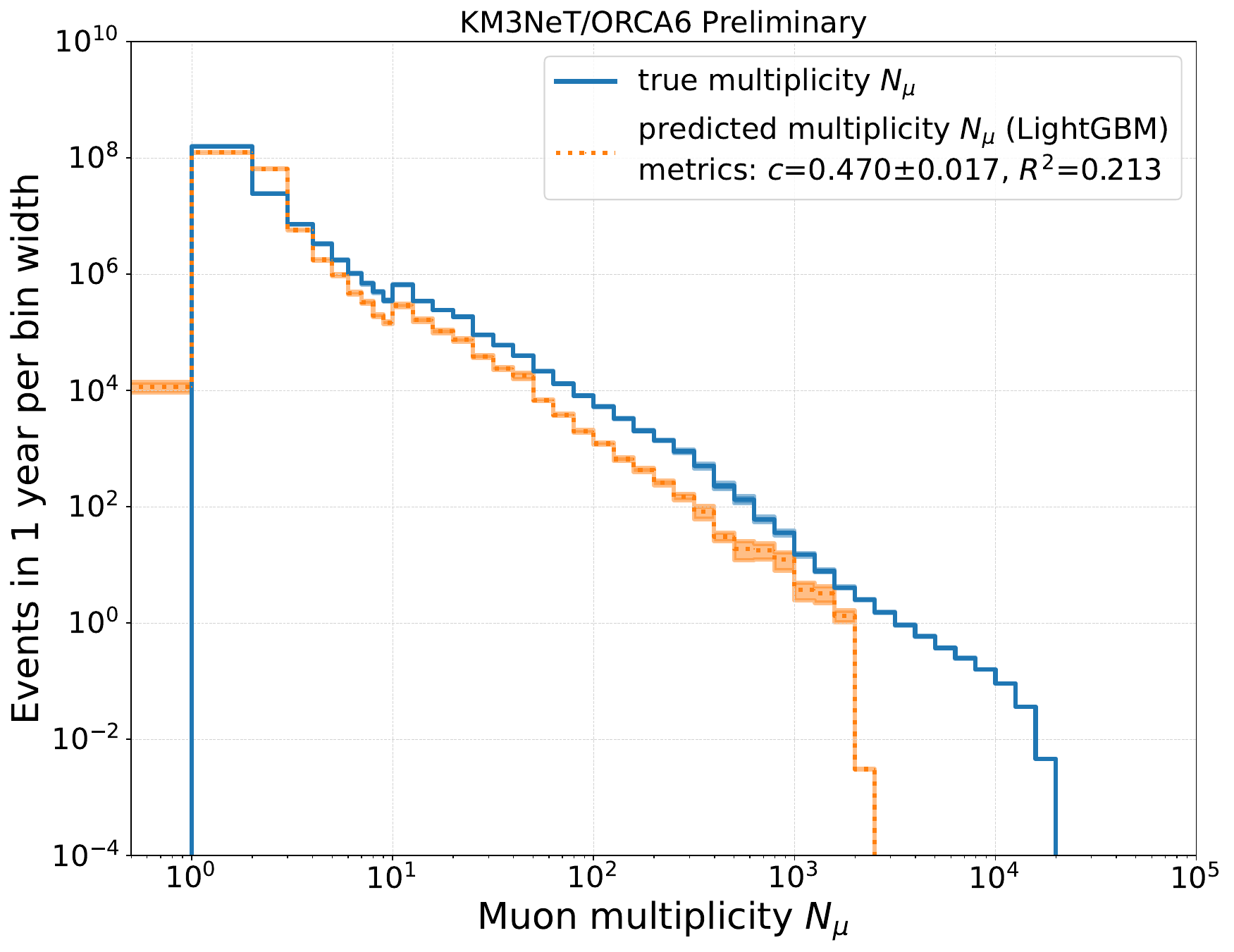}}\caption{Comparison of muon multiplicity reconstructed with the LigthGBM model
against the true value in 1D. Both the correlation and $R^{2}-$score
values were computed on the bin values, not individual events. \label{fig:Nmu_reco_results-1D}}
\end{figure}

Using the parameters from Tab. \ref{tab:LightGBM-tuned-hyperparameters-summary},
the final results were computed and are presented in Fig. \ref{fig:Nmu_simple_reco_results}
and \ref{fig:Nmu_reco_results-1D}. The best reconstruction was achieved
for ARCA115, which would indicate that apparently a detector with
larger instrumented volume has an advantage over the same number of
DOMs packed into a more dense configuration (ORCA). This should not
come as a surprise, when one considers e.g. how the muon trackers
around the ATLAS detector are designed: they are the most sparsely
instrumented part of the detector \cite{ATLAS_muon_tracker}. This
is a direct consequence of the fact that a typical muon signature
in a detector is an elongated track (although muons can and sometimes
do produce particle showers!). The effect of improved performance
upon adding additional DOMs is evident for both ARCA and ORCA. Fig.
\ref{fig:Nmu_reco_results-1D} demonstrates very good reproduction
of the shape of the muon multiplicity distributions in the low and
intermediate $N_{\mu}$ range (except for multiplicity 0), with a
growing underprediction towards higher values of multiplicity. 

\chapter{Muon rate measurement with KM3NeT detectors \label{chap:Muon-rate-measurement}}

This chapter is devoted to the comparison of the simulated MC and
experimental data for a number of KM3NeT detector configurations,
listed in the next section. A comparison of data against the Monte
Carlo simulation is the ultimate test of the reliability of the MC.
The outcomes are shown in terms of weighted distributions of reconstructed
observables. The author has published earlier results of the data
vs MC comparisons in \cite{my-ICRC2019,my-ICRC2021,my-VLVnT2021,my-ICHEP2020}.
For the most recent ones (ARCA6, ORCA6), he has evaluated the systematic
uncertainties on the event weights (Sec. \ref{sec:Systematic-uncertainty-study}),
except for the two incorporated from the work of A. Romanov \cite{Andrey_Thesis},
and prepared an event quality selection (Sec. \ref{sec:Event-quality-selection}).
In addition, the energy and multiplicity reconstruction described
in Chap. \ref{chap:muon-bundle-reco} were applied to ARCA6 and ORCA6
data. The systematic uncertainties were combined with statistical
uncertainty according to Eq. \ref{eq:hist_error_2} for simulated
events from the CORSIKA MC productions prepared by the author of this
dissertation.

\section{Simulated samples and experimental data}

At the time of compiling this thesis, data and MC for several detector
configurations of both ARCA and ORCA were available. The datasets
used in this chapter are listed in Tab. \ref{tab:data_and_MC_datasets}.
From the indicated run ranges, the selected ones corresponding to
particularly stable data taking periods were used.

\begin{table}[H]
\begin{centering}
\caption{Available data and MC simulations. \label{tab:data_and_MC_datasets}}
\par\end{centering}
\centering{}{\scriptsize{}}%
\begin{tabular}{|c|c|c|c|c|c|c|}
\hline 
\multirow{2}{*}{\textbf{\scriptsize{}Detector}} & \multicolumn{3}{c|}{{\scriptsize{}data/MUPAGE}} & \multicolumn{3}{c|}{{\scriptsize{}CORSIKA}}\tabularnewline
\cline{2-7} \cline{3-7} \cline{4-7} \cline{5-7} \cline{6-7} \cline{7-7} 
 & \textbf{\scriptsize{}Runs} & \textbf{\scriptsize{}Dates} & \textbf{\scriptsize{}Livetime {[}d{]}} & {\scriptsize{}Code version} & {\scriptsize{}$n_{\mathrm{generated\,showers}}$} & {\scriptsize{}$E_{\mathrm{prim}}$ range}\tabularnewline
\hline 
\hline 
{\scriptsize{}ARCA2} & {\scriptsize{}5009 - 5408} & {\scriptsize{}23.12.2016 - 08.03.2017} & {\scriptsize{}57.71} & {\scriptsize{}v7.6400} & {\scriptsize{}$2.5\cdot10^{9}$} & {\scriptsize{}1~TeV$-$1~EeV}\tabularnewline
\hline 
{\scriptsize{}ARCA6} & {\scriptsize{}9635 - 10286} & {\scriptsize{}12.05.2021 - 10.09.2021} & {\scriptsize{}113.63} & {\scriptsize{}v7.7410} & {\scriptsize{}$1.44\cdot10^{10}$} & {\scriptsize{}1~TeV$-$8~EeV}\tabularnewline
\hline 
{\scriptsize{}ORCA1} & {\scriptsize{}2867 - 3250} & {\scriptsize{}09.11.2017 - 13.12.2017} & {\scriptsize{}33.77} & {\scriptsize{}v7.6400} & {\scriptsize{}$2.5\cdot10^{9}$} & {\scriptsize{}1~TeV$-$1~EeV}\tabularnewline
\hline 
{\scriptsize{}ORCA4} & {\scriptsize{}5378 - 7219} & {\scriptsize{}01.07.2019 - 25.01.2020} & {\scriptsize{}188.17} & {\scriptsize{}v7.6400} & {\scriptsize{}$2.5\cdot10^{9}$} & {\scriptsize{}1~TeV$-$1~EeV}\tabularnewline
\hline 
{\scriptsize{}ORCA6} & {\scriptsize{}7231 - 9264} & {\scriptsize{}27.01.2020 - 07.01.2021} & {\scriptsize{}328.38} & {\scriptsize{}v7.7410} & {\scriptsize{}$1.44\cdot10^{10}$} & {\scriptsize{}1~TeV$-$8~EeV}\tabularnewline
\hline 
\end{tabular}{\scriptsize\par}
\end{table}

CORSIKA MC each time was generated in non-rbr mode, i.e. with averaged
PMT parameters and trigger settings (see Sec. \ref{sec:trigger}).
This means that the muon flux simulated with CORSIKA is averaged over
time. For MUPAGE, depending on the detector configuration, both simulations
in rbr (ARCA2, ORCA1, ORCA4) and in non-rbr mode (ARCA6, ORCA6) were
used.

.

\section{Systematic uncertainty study \label{sec:Systematic-uncertainty-study}}

The goal of the study described here was to estimate the systematic
uncertainty on the atmospheric muon bundle rate. The systematics were
included event-wise, as in Eq. \ref{eq:hist_error_2}, hence the sought
quantity was the uncertainty on event weight as function of the primary
energy: $\Delta w\left(E_{\mathrm{prim}}\right)$. The following five
sources of systematic uncertainty were considered in this work:
\begin{enumerate}
\item Primary cosmic ray flux models.
\item High-energy hadronic interaction models.
\item Seasonal differences in the atmospheric density profile.
\item Spread of the PMT efficiency.
\item Varying light absorption length.
\end{enumerate}
The methods by which individual uncertainties were evaluated are explained
in the following subsections. The last 3 listed uncertainties were
evaluated separately for KM3NeT/ARCA and KM3NeT/ORCA.

There were a number of sources of systematic uncertainty, which were
neglected in this study:
\begin{enumerate}
\item The Earth's magnetic field strength variations: see Sec. \ref{subsec:Magnetic-field-strength}.
\item Atmospheric density models: due to lack of recent models, satisfying
the requirements of CORSIKA simulations for KM3NeT (see Sec. \ref{subsec:Fit-of-the-atmosphere}).
\item Primary CR flux composition (see Sec. \ref{subsec:Primary-CR-flux-syst}).
\item Low-energy hadronic interaction models: they have marginal influence
on the CORSIKA results, as they are only used below 80~GeV in air
\cite{CORSIKA-Userguide}. In the case of KM3NeT, muons with energies
of 80~GeV and extremely unlikely to reach even the top of ORCA (see
Sec. \ref{sec:CORSIKA-benchmarking-and-opti}).
\item Differences in the run settings, which were averaged out in the non-rbr
MC simulations. Based on MUPAGE studies in \cite{Andrey_Thesis},
the difference between the rbr and non-rbr MC is on the order of $\sim\pm5\,\%$.
\end{enumerate}
The obtained uncertainties are shown collectively in Fig. \ref{fig:combined_uncertainties}.

\subsection{Primary cosmic ray flux models\label{subsec:Primary-CR-flux-syst}}

The uncertainty coming from the differences between the predictions
of the total CR primary flux $\phi_{\mathrm{CR}}$ by the different
models, was estimated directly based on those flux predictions. This
was the most natural way to evaluate it, as the event weights are
computed using the CR flux (see Eq. \ref{eq:w_3}). The total (all-nuclei)
flux was used, since this is precisely what is indirectly observed
by KM3NeT by looking at the atmospheric muon bundles. One could try
to infer the underlying composition of primary CRs, however as demonstrated
in \cite{StefanReckThesis}, it is no trivial task. The uncertainty
related to the unknown underlying primary CR composition was neglected
in this work and the composition as in the default primary CR flux
model: GST3, was assumed. 

The uncertainty on the event weight was estimated using:
\begin{equation}
\ensuremath{\ensuremath{\left.\ensuremath{\Delta w_{\mathrm{cr}}}\right|_{\mathsf{upper}}=\frac{\left.\phi_{\mathrm{CR}}\left(E_{\mathrm{prim}}\right)\right|_{\mathsf{max}}-\left.\phi_{\mathrm{CR}}\left(E_{\mathrm{prim}}\right)\right|_{\mathsf{default}}}{\left.\phi_{\mathrm{CR}}\left(E_{\mathrm{prim}}\right)\right|_{\mathsf{default}}}\cdot100\%}}
\end{equation}

and

\begin{equation}
\ensuremath{\ensuremath{\left.\ensuremath{\Delta w_{\mathrm{cr}}}\right|_{\mathsf{lower}}=\frac{\left.\phi_{\mathrm{CR}}\left(E_{\mathrm{prim}}\right)\right|_{\mathsf{default}}-\left.\phi_{\mathrm{CR}}\left(E_{\mathrm{prim}}\right)\right|_{\mathsf{min}}}{\left.\phi_{\mathrm{CR}}\left(E_{\mathrm{prim}}\right)\right|_{\mathsf{default}}}\cdot100\%.}}
\end{equation}

The considered CR flux models included: GST3 (default), GST4, H3a,
H4a, cH3a, and cH4a \cite{GST,combined_HillasGaisser_and_GaisserHonda_also_crfluxmodels_reference,HillasGaisser_model}.
Models like poly-gonato or ZS \cite{poly-gonato,ZatsepinSokolskaya,Thunman,GaisserHonda}
were excluded from the evaluation, as they are clearly outdated and
do not describe the CR flux above the knee. 

\subsection{High-energy hadronic interaction models\label{subsec:HE-hadronic-interaction-model-syst}}

The uncertainty due to the modelling of the HE hadronic interactions
was computed based on a~set of dedicated CORSIKA mini-productions.
The mini-productions were processed in the same way as the mass production,
used for the development of the event reconstruction in Chap. \ref{chap:muon-bundle-reco}
and analysis in Chap. \ref{chap:prompt_ana}. The only differences
were:
\begin{itemize}
\item zenith angle fixed to $0\lyxmathsym{\textdegree}$ (vertically downgoing
events only),
\item fixed primary energies were sampled: $1,\,2,\,3,\,6,\,20,\,60,\,200\,$TeV
and $2\,$PeV (instead of a continuous~spectrum),
\item the total number of generated showers per primary: $1.11\cdot10^{7}$,
\item each of the mini-productions was done for a different HE hadronic
interaction model: EPOS LHC (v3400), SIBYLL 2.3d, or QGSJET II-04.
\end{itemize}
The first two changes were introduced to minimize the fluctuations
related to the sampling itself. The second and third also aimed at
reducing the required computational resources. Fixing the zenith relies
of the assumption that the uncertainty only weakly depends on the
direction, as compared to the energy. This has been verified for single
muons using the simplified simulation using \cite{MCEq}, with the
result of sub-percent variation of muon flux due to varying the zenith
angle. The uncertainty on the event weight was evaluated in an analogous
manner as in Sec. \ref{subsec:Primary-CR-flux-syst}:
\begin{equation}
\ensuremath{\left.\ensuremath{\Delta w_{\mathrm{had}}}\right|_{\mathsf{upper}}=\frac{\left.f_{\mathrm{surv}}\left(E_{\mathrm{prim}}\right)\right|_{\mathsf{max}}-\left.f_{\mathrm{surv}}\left(E_{\mathrm{prim}}\right)\right|_{\mathsf{default}}}{\left.f_{\mathrm{surv}}\left(E_{\mathrm{prim}}\right)\right|_{\mathsf{default}}}\cdot100\%},
\end{equation}

\begin{equation}
\ensuremath{\ensuremath{\left.\ensuremath{\Delta w_{\mathrm{had}}}\right|_{\mathsf{lower}}=\frac{\left.f_{\mathrm{surv}}\left(E_{\mathrm{prim}}\right)\right|_{\mathsf{default}}-\left.f_{\mathrm{surv}}\left(E_{\mathrm{prim}}\right)\right|_{\mathsf{min}}}{\left.f_{\mathrm{surv}}\left(E_{\mathrm{prim}}\right)\right|_{\mathsf{default}}}\cdot100\%}},
\end{equation}

where $f_{\mathrm{surv}}=\frac{N_{\mathrm{survived}}}{N_{\mathrm{generated}}}$
is the fraction of showers surviving until the sea level and the default
model is SIBYLL 2.3d. At the sea level, the shower content is already
dominated by the muons and neutrinos, with a contribution of only
$\sim2.7\%$ from the hadrons (based on the CORSIKA mass production).
For this reason, the selection of the hadronic interaction model could
be assumed to be irrelevant for the transport of muons from the sea
to the can.

\subsection{Atmospheric density profiles\label{subsec:Atmospheric-density-profiles}}

The values of $\left.\ensuremath{\Delta w_{\mathrm{atm}}}\right|_{\mathsf{upper}}$
and $\left.\ensuremath{\Delta w_{\mathrm{atm}}}\right|_{\mathsf{lower}}$,
related to different atmospheric density profiles were calculated
exactly in the same way, as in Sec. \ref{subsec:HE-hadronic-interaction-model-syst},
but with the HE hadronic interaction model fixed to SIBYLL 2.3d and
for the following atmosphere fits:
\begin{itemize}
\item evaluated at ARCA in summer (June),
\item evaluated at ARCA in winter (December),
\item evaluated at ORCA in summer (June),
\item evaluated at ORCA in winter (December),
\item evaluated for a prediction averaged between ARCA and ORCA and over
a time period of 2 years (default).
\end{itemize}
The first four were taken from \cite{Thomas_Heid_phd_thesis} and
the fifth one is the same as in Sec. \ref{subsec:Fit-of-the-atmosphere}. 

For the ARCA result, only the averaged and the ARCA summer and winter
atmosphere fits were used. Analogously, for ORCA the considered atmosphere
fits were the averaged one and the ones for ORCA summer and winter.

\subsection{Photomultiplier tube efficiency\label{subsec:PMT-efficiency}}

The uncertainty related to the variations in PMT efficiencies was
adapted from the study performed by A. Romanov \cite{Andrey_Thesis}.
There, muon bundle rates were compared using MUPAGE simulations with
PMT efficiencies set to the nominal value or to the nominal value
modified by $\pm10\,\%$ . The result was an uncertainty of $\ensuremath{\Delta w_{\mathrm{pmt}}}\pm5\,\%$
for ORCA6 and $\Delta w_{\mathrm{pmt}}\pm20\,\%$ for ARCA6. The difference
between the detectors stems from the higher density of instrumentation
with PMTs of the ORCA telescope. The effect of the detector size was
neglected and the uncertainty for ORCA6 was assumed to be the same
as for ORCA115. Analogously, the one for ARCA6 was set equal to ARCA115
as well.

\subsection{Light absorption length}

The same study by A. Romanov mentioned in Sec. \ref{subsec:PMT-efficiency}
encompassed the comparison of MUPAGE simulation results with varied
absorption length of light in seawater. In this case, the absorption
length was varied by $\pm10\,\%$, following the results from \cite{Optical_properties_light_absorption_length}.
The obtained uncertainties were $\Delta w_{\mathrm{abs}}=\pm5\,\%$
for ORCA6 and $\Delta w_{\mathrm{abs}}=\pm15\,\%$ for ARCA6, again
with the difference between the two coming from different detector
geometries. In ORCA light has shorter path to travel between the DOMs,
and hence ORCA is less affected by the uncertainty due to the light
absorption. The uncertainty for ORCA115 was taken to be the same as
for ORCA6 and for ARCA115 the same as ARCA6.

\subsection{Combined results \label{subsec:Results-systematics}}

\begin{figure}[H]
\centering{}\subfloat[Uncertainty for ARCA.]{\centering{}\includegraphics[width=8cm]{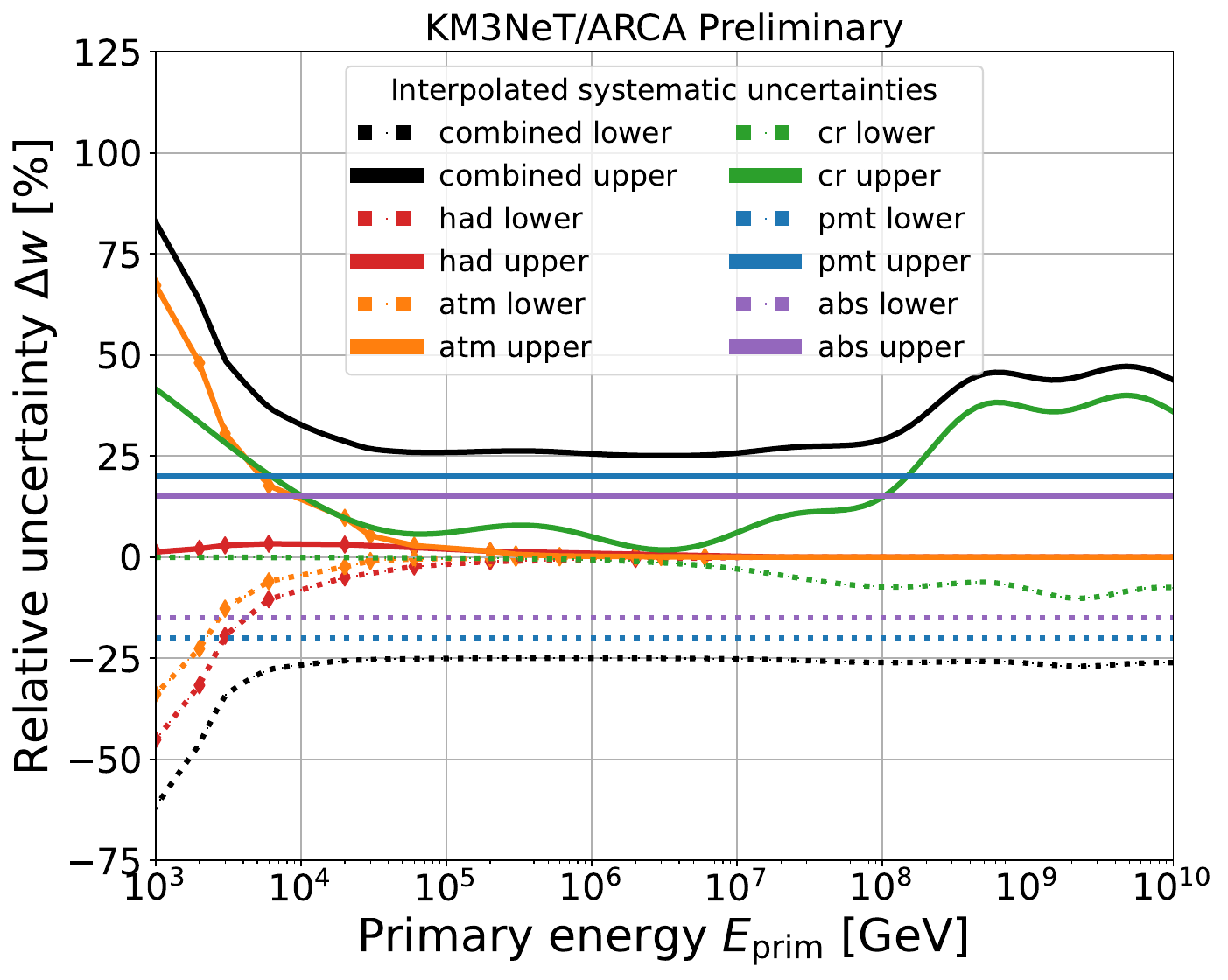}}\subfloat[Uncertainty for ORCA.]{\centering{}\includegraphics[width=8cm]{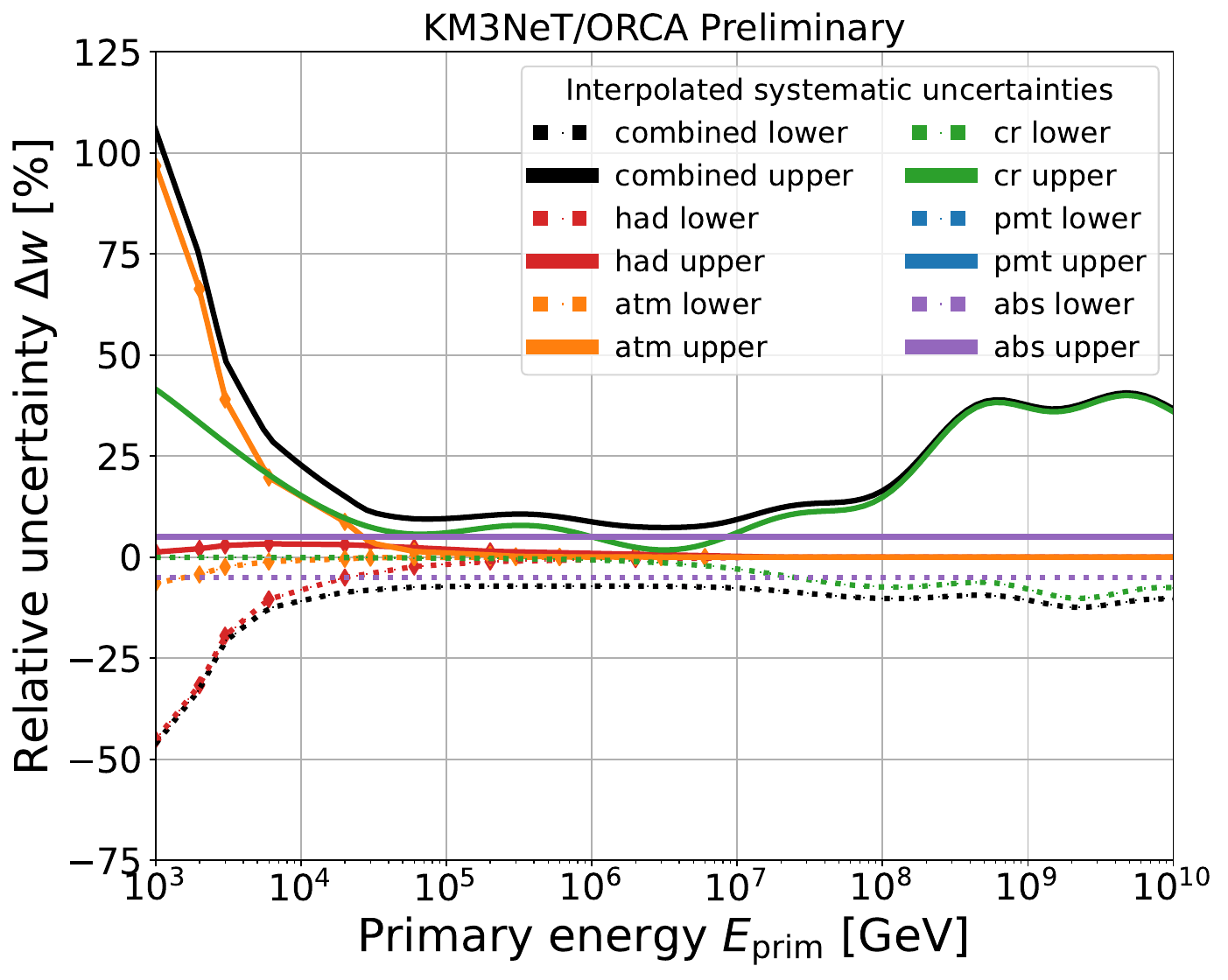}}\caption{Interpolated relative systematic uncertainties for ARCA and ORCA.
Both individual contributions and the combined uncertainties are plotted.
The upper uncertainties are shown as solid lines and the lower uncertainties:
as dashed lines. The square points indicate the sampled energies,
over which the interpolation was done. There was no interpolation
for cr, pmt and abs. \label{fig:combined_uncertainties}}
\end{figure}

The final result on systematic uncertainty was obtained by adding
the five contributions in quadrature:

\begin{equation}
\Delta w_{\mathrm{syst}}=\sqrt{\left(\Delta w_{\mathrm{had}}\right)^{2}+\left(\Delta w_{\mathrm{cr}}\right)^{2}+\left(\Delta w_{\mathrm{atm}}\right)^{2}+\left(\Delta w_{\mathrm{pmt}}\right)^{2}+\left(\Delta w_{\mathrm{abs}}\right)^{2}},
\end{equation}

where had, cr, atm, pmt, and abs stand for uncertainties related to
HE hadronic interaction models, primary CR flux models, atmospheric
seasonal density variations, PMT (see Sec. \ref{subsec:Photomultiplier_Tubes})
efficiency and light absorption length, respectively. The result of
combining the uncertainties is displayed in Fig. \ref{fig:combined_uncertainties},
separately for ARCA and ORCA.

As one may note in Fig. \ref{fig:combined_uncertainties}, there is
a plateau of minimal systematic uncertainty for both KM3NeT/ARCA and
KM3NeT/ORCA in the range of primary energies between 10~TeV and 100~PeV.
Nevertheless, the minimal uncertainties are still large, on the order
of 25~\% for ARCA and 10~\% for ORCA. The low-energy regime is mostly
dominated by the atmospheric uncertainty. This can be intuitively
understood in the following way: the varying atmospheric density profiles
result in different first interaction heights and also different total
atmosphere thicknesses that have to be traversed (see Sec. \ref{subsec:Fit-of-the-atmosphere}).
For a very low-energy shower it may be critical for reaching the sea-level.
Here it must be stressed that what is meant are muons from such a
shower reaching the sea: neutrinos easily travel even longer distances.
It has been verified that below 10~TeV, the mean survival rate of
showers starts to vary considerably between showers starting at different
heights. The difference between events that start above and below
24~km reaches 50~\% at $E_{\mathrm{prim}}=1\,$TeV. The difference
in what and how many particles are produced in the hadronic interactions
can also become crucial for the survival of air showers. For this
reason the hadronic uncertainty also rises at low primary energies.
This only happens for the lower hadronic uncertainty, since SIBYLL
2.3d tends to produce on average more low-energy muons, which are
able to reach the sea.

At the highest energies, the atmospheric and hadronic uncertainties
become irrelevant, since the events are energetic enough to reach
the sea, regardless of the starting point, atmospheric density or
the interactions undergone along the way. However, what does become
important are the uncertainties related to the primary CR flux modelling.
This is a limitation imposed by the insufficient experimental data
in this regime, as the CR flux drops sharply with increasing energy.
Similarly, the cosmic ray flux uncertainty at the lowest energies
results from the limited experimental data at lowest energies at the
time of design of the models. Admittedly, there is one more recent,
data-driven model: Global Spline Fit (GSF) \cite{GSF-Global_Spline_Fit},
which was not considered in this work and could be included in future
extensions. The upper cosmic ray flux uncertainty is larger than the
lower one due to the choice of the GST3 as the default CR flux model. 

The results from Fig. \ref{fig:combined_uncertainties} have been
included in the ARCA6 and ORCA6 plots in Sec. \ref{sec:data_MC_Results}
and in results in Chap. \ref{chap:prompt_ana}.

\section{Event quality selection \label{sec:Event-quality-selection}}

The comparison of experimental data against MC simulations typically
requires purging the datasets of the following two classes of events:
\begin{enumerate}
\item Generally poorly reconstructed, e.g. low-energy muons, skimming the
outer parts of the can. Such $\mu$ may produce little to none detectable
photons (see the discussion in Sec. \ref{subsec:Muon-selection}).
\item Not included in the standard KM3NeT MC productions:
\begin{enumerate}
\item pure noise events: events containing solely environmental background
due to $^{40}$K radioactive decays in seawater (see Sec. \ref{sec:trigger})
and electronics noise \cite{KM3NeT-LoI-2.0},
\item events related to hardware malfunctions, e.g. sparking DOMs (DOMs
with a short-circuit) or afterpulses (additional delayed signal, which
occurs after the main response of the PMT to a detected photon) \cite{KM3NeT-DOM-paper,KM3NeT-PMT-characterisation,Hamamatsu-PMTs}.
\end{enumerate}
\end{enumerate}

This section is dedicated to the quality selection further used in
Sec. \ref{sec:data_MC_Results} for the most recent detector configurations:
ARCA6 and ORCA6. The cut on JMuon likelihood $\mathcal{L}$ was performed
for both detectors, as $\mathcal{L}$ is a generic quantity evaluating
the feasibility of reconstruction of an event. Undoubtedly, it would
be more optimal to have a separate, dedicated measure of the ML reconstruction
uncertainty, similarly to what was done in \cite{StefanReckThesis}.
Nevertheless, the employed cuts improved the data vs MC agreement
for low-level variables, as demonstrated in the following two subsections.

The event quality selection, applied in Sec. \ref{subsec:ARCA6} and
\ref{subsec:ORCA6}, is a combination of the minimum likelihood cut
($\mathcal{L}>50$ for ARCA and $\mathcal{L}>280$ for ORCA, see Sec.
\ref{subsec:ARCA6-quality-cuts} and \ref{subsec:ORCA6-quality-cuts})
and reconstruction reliability cuts, summarized in Tab. \ref{tab:Reliability_selection}.

\begin{table}[H]
\begin{centering}
\caption{Lower prediction reliability bounds for different observables reconstructed
by JMuon and the ML models, based on results from Chap. \ref{chap:muon-bundle-reco}
and Sec. \ref{sec:Performance-of-JMuon}. In the case of ORCA115 JMuon
energy reconstruction, no limit could be set based on the available
CORSIKA datasets, however $1\,$GeV was taken as the expected lowest
achievable energy with ORCA \cite{KM3NeT-LoI-2.0}. \label{tab:Reliability_selection}}
\par\end{centering}
\centering{}%
\begin{tabular}{|c|c|c|c|c|c|}
\hline 
Reconstruction & Observable & ARCA115 & ARCA6 & ORCA115 & ORCA6\tabularnewline
\hline 
\hline 
\multirow{2}{*}{JMuon} & $\cos\left(\theta_{\mathrm{zenith}}\right)$ & - & - & - & -\tabularnewline
\cline{2-6} \cline{3-6} \cline{4-6} \cline{5-6} \cline{6-6} 
 & Energy & $10\,$GeV & $1\,$TeV & $1\,$GeV & $60\,$GeV\tabularnewline
\hline 
\multirow{3}{*}{ML} & Bundle energy $E_{\mathrm{bundle}}$ & $1\,$TeV & $1\,$TeV & $400\,$GeV & $600\,$GeV\tabularnewline
\cline{2-6} \cline{3-6} \cline{4-6} \cline{5-6} \cline{6-6} 
 & Total primary energy $A\cdot E_{\mathrm{prim}}$ & $6\,$PeV & $8\,$PeV & $3\,$PeV & $5\,$PeV\tabularnewline
\cline{2-6} \cline{3-6} \cline{4-6} \cline{5-6} \cline{6-6} 
 & Multiplicity $N_{\mu}$ & 1 & 1 & 1 & 1\tabularnewline
\hline 
\end{tabular}
\end{table}

\subsection{ARCA6 \label{subsec:ARCA6-quality-cuts}}

\begin{figure}[H]
\centering{}\includegraphics[width=8cm]{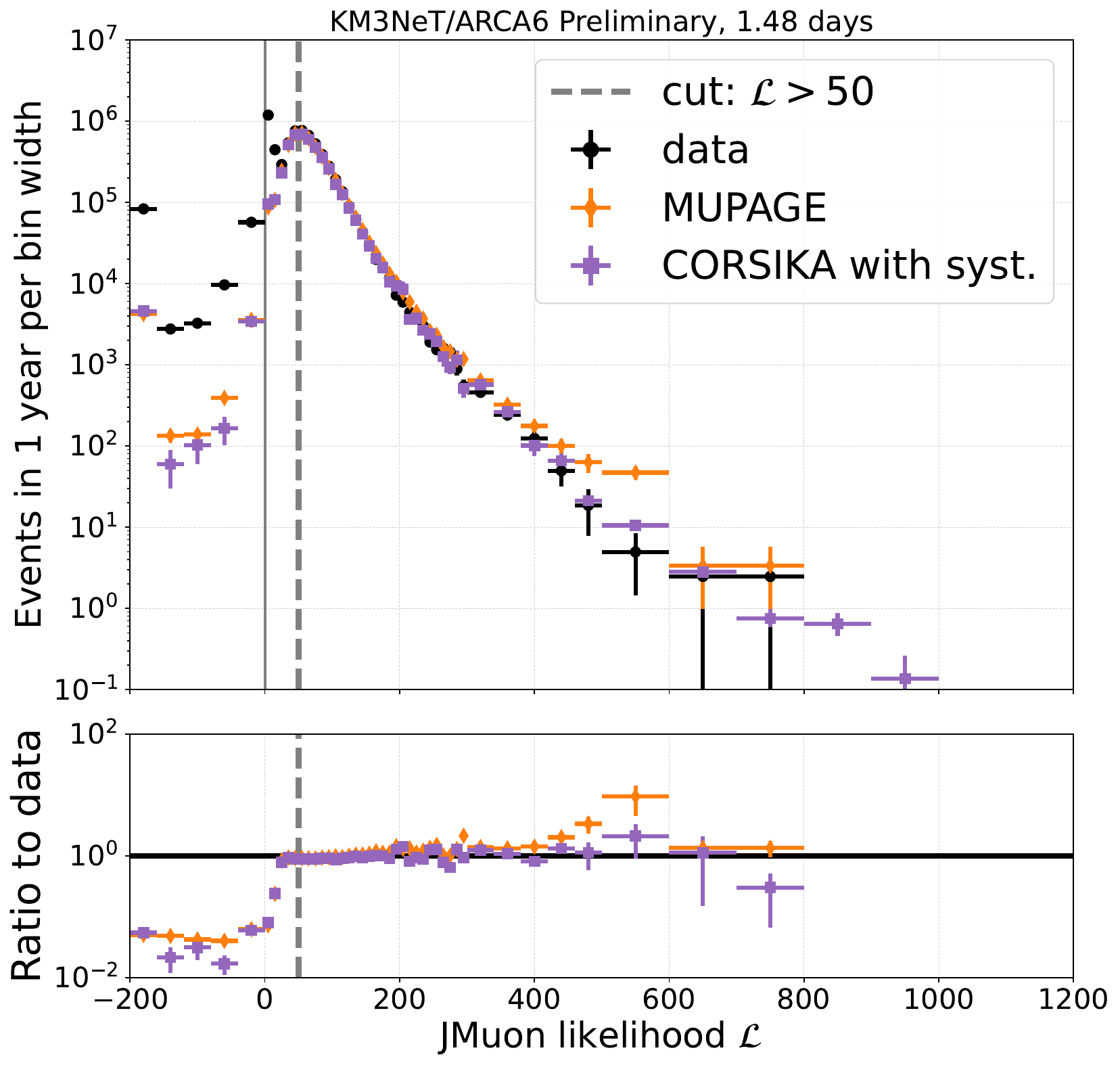}\caption{The comparison of the distributions of the JMuon likelihood $\mathcal{L}$
for ARCA6 data and MC simulations. \label{fig:ARCA6_JMuon_likelihood}}
\end{figure}

In this section, selected tigger-level observables for ARCA6 are shown
to investigate the effect of the cut on the JMuon likelihood $\mathcal{L}$.
The adopted cut value was $\mathcal{L}=50$, which followed the previous
ARCA6 data analysis work \cite{Sinopoulou2021_atm_neutrinos}. It
can be verified by looking at Fig. \ref{fig:ARCA6_JMuon_likelihood},
that such a selection indeed focuses on a region with better agreement
between the data and MC.

\begin{figure}[H]
\centering{}\subfloat[No event quality selection.]{\centering{}\includegraphics[width=8cm]{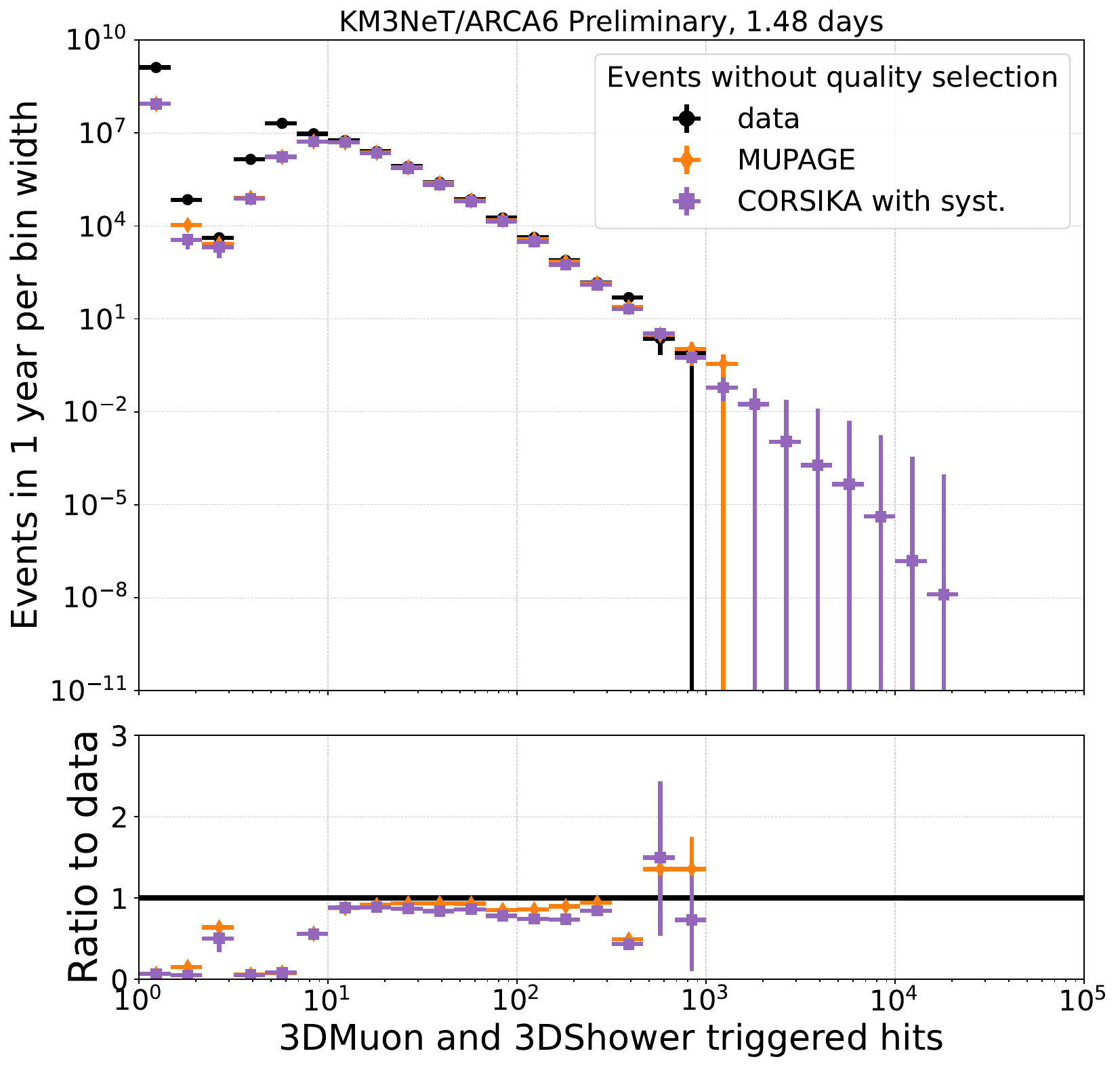}}\subfloat[Only events with $\mathcal{L}>50$.]{\centering{}\includegraphics[width=8cm]{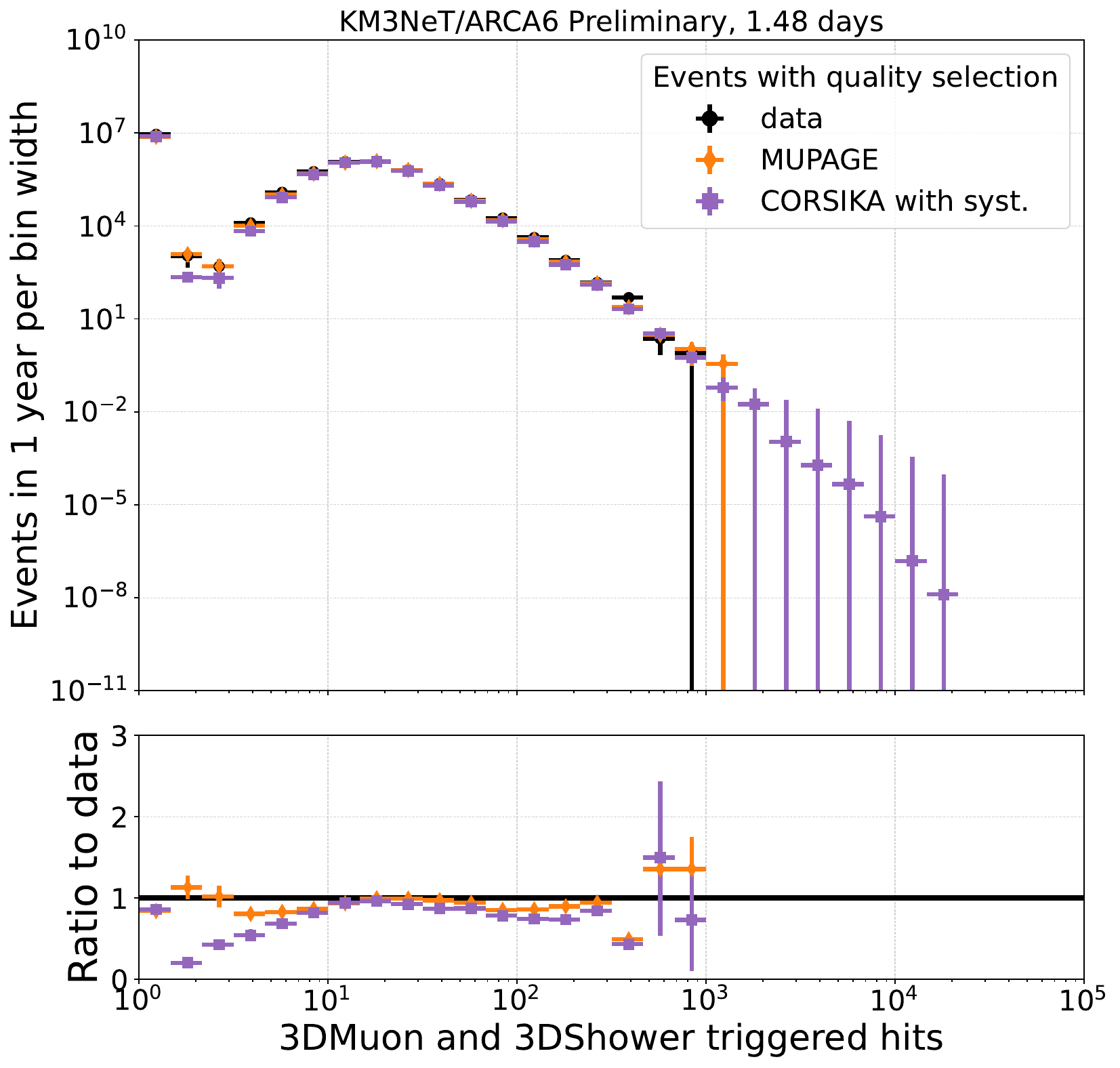}}\caption{The comparison of the distributions of the number of hits triggered
by the 3DMuon and 3DShower triggers for ARCA6 data and MC simulations.
\label{fig:ARCA6_3DMUON_3DSHOWER_trig_hits}}
\end{figure}

\begin{figure}[H]
\centering{}\subfloat[No event quality selection.]{\centering{}\includegraphics[width=8cm]{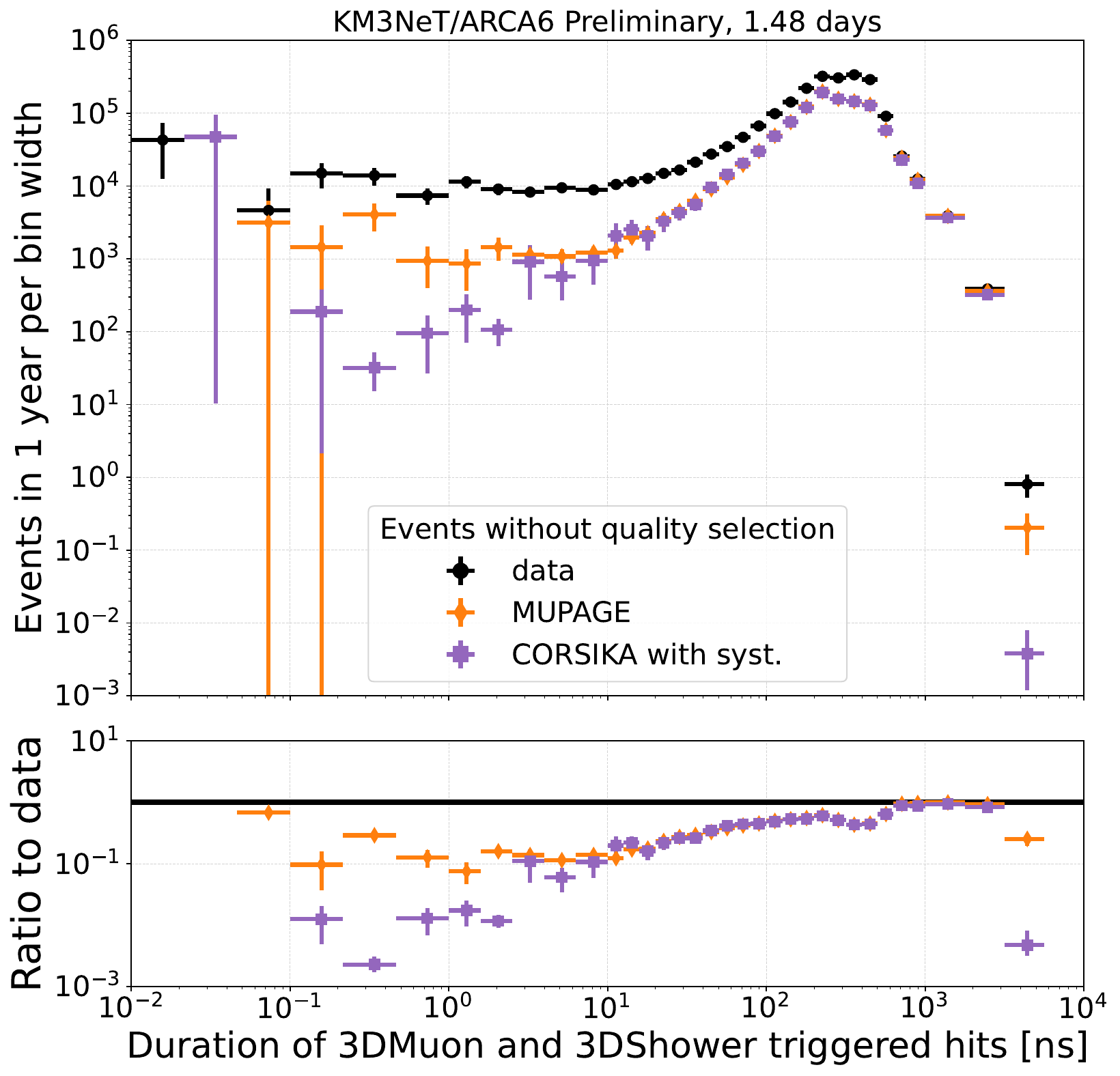}}\subfloat[Only events with $\mathcal{L}>50$.]{\centering{}\includegraphics[width=8cm]{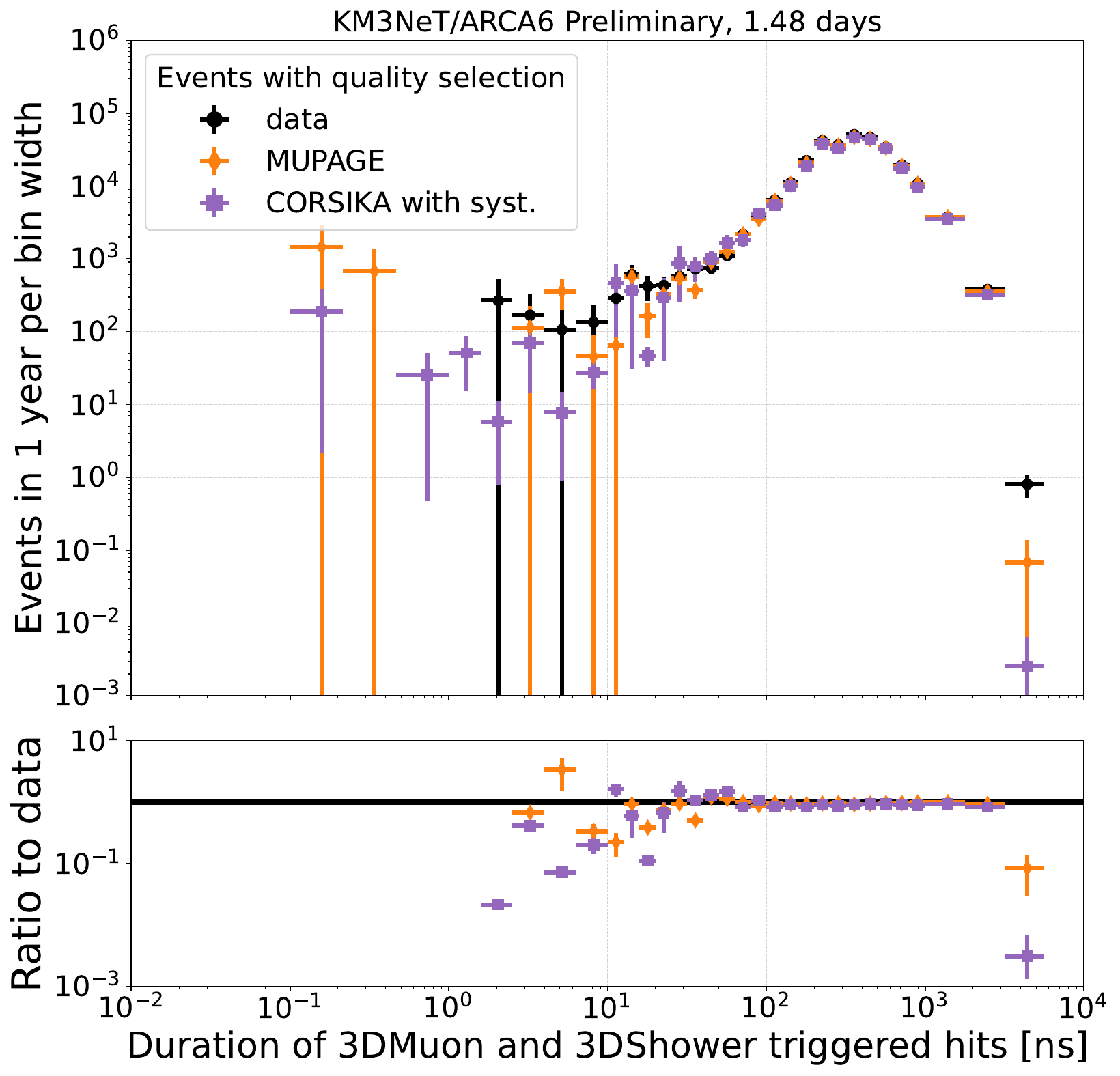}}\caption{The comparison of the distributions of the duration of hits triggered
by the 3DMuon and 3DShower triggers for ARCA6 data and MC simulations.
\label{fig:ARCA6_3DMUON_3DSHOWER_trig_hits-duration}}
\end{figure}

The first investigated trigger-level quantity was the number of hits
passing the requirements of both the 3DMuon and 3DShower trigger,
which was picked as the most important feature for the reconstruction
in Sec. \ref{subsec:Feature-importances-Ebundle}. The comparison
of experimental and simulated data is presented in Fig. \ref{fig:ARCA6_3DMUON_3DSHOWER_trig_hits}.
Both MCs were generated with averaged PMT parameters and trigger settings
(in non-rbr mode, see Sec. \ref{sec:trigger}). In addition, the distributions
of the duration of events with hits triggered by both 3DMuon and 3DShower
were plotted in Fig. \ref{fig:ARCA6_3DMUON_3DSHOWER_trig_hits-duration}.

As demonstrated on examples of Fig. \ref{fig:ARCA6_3DMUON_3DSHOWER_trig_hits}
and \ref{fig:ARCA6_3DMUON_3DSHOWER_trig_hits-duration}, the agreement
of the trigger-level variables clearly improved after applying the
cut on the JMuon likelihood. It has to be noted that such variables
are entirely independent of the JMuon reconstruction, since they are
computed at trigger level (see Sec. \ref{sec:Muon-sim-chain}). This
leads to a conclusion that the likelihood is an effective proxy of
the `reconstructability' of an event. In other words, it measures
the feasibility of reconstruction of an event, with a caveat that
it is somewhat biased towards the JMuon reconstruction. The applied
quality cut mostly rejects the sub-ns events with few triggered hits,
which are indeed expected to be more challenging to reconstruct. Such
short events are dominated by the pure noise, which is not included
in the simulations, hence the surplus of the experimental data over
the MC.

\subsection{ORCA6 \label{subsec:ORCA6-quality-cuts}}

In this section, a set of plots for ORCA6, analogous to the ones from
Sec. \ref{subsec:ARCA6-quality-cuts} is presented. The likelihood
cut for ORCA6 was fixed at $\mathcal{L}=280$.

\begin{figure}[H]
\centering{}\includegraphics[width=8cm]{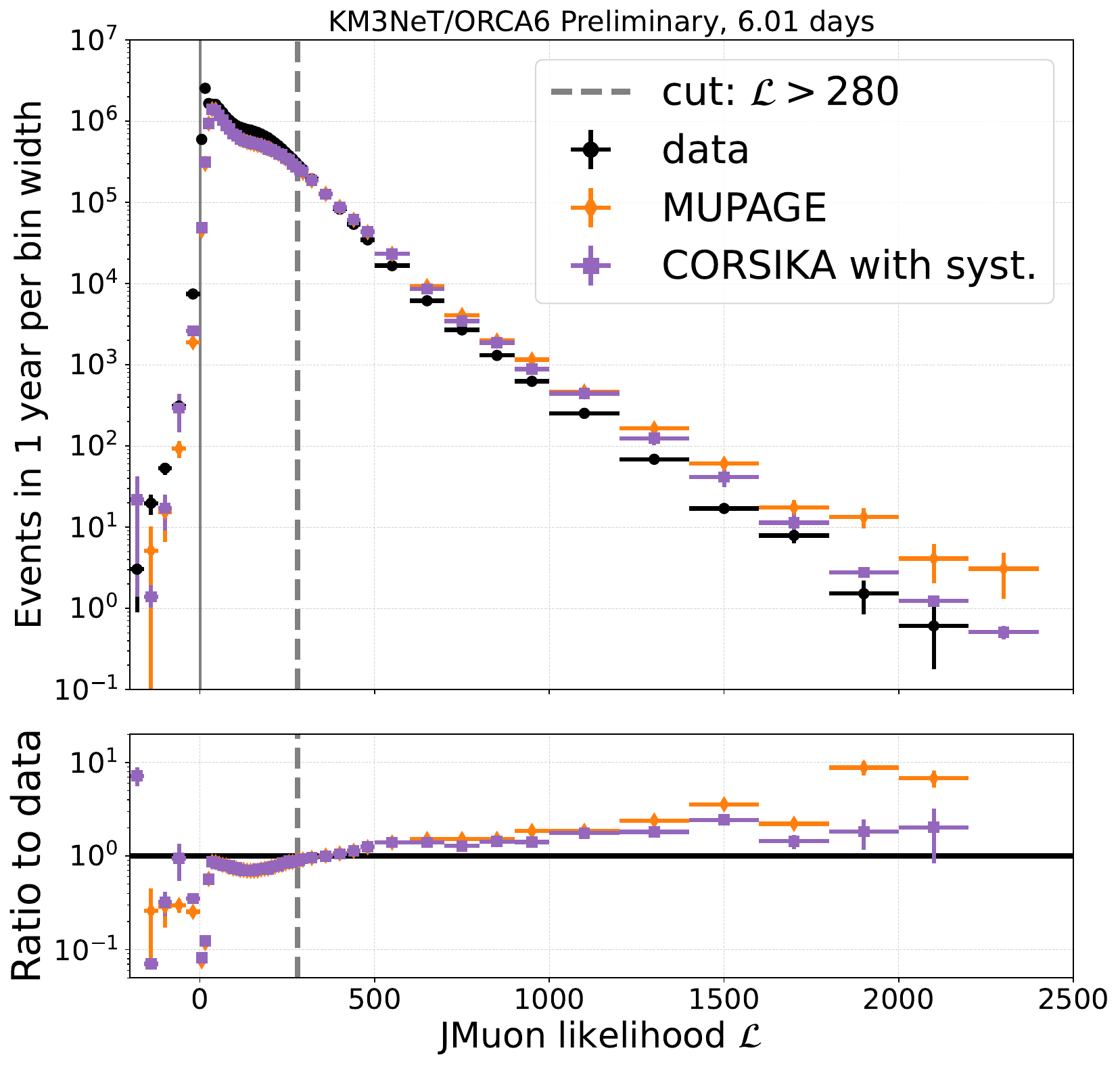}\caption{The comparison of the distributions of the JMuon likelihood $\mathcal{L}$
for ORCA6 data and MC simulations. \label{fig:ORCA6_JMuon_likelihood}}
\end{figure}

\begin{figure}[H]
\centering{}\subfloat[No event quality selection.]{\centering{}\includegraphics[width=8cm]{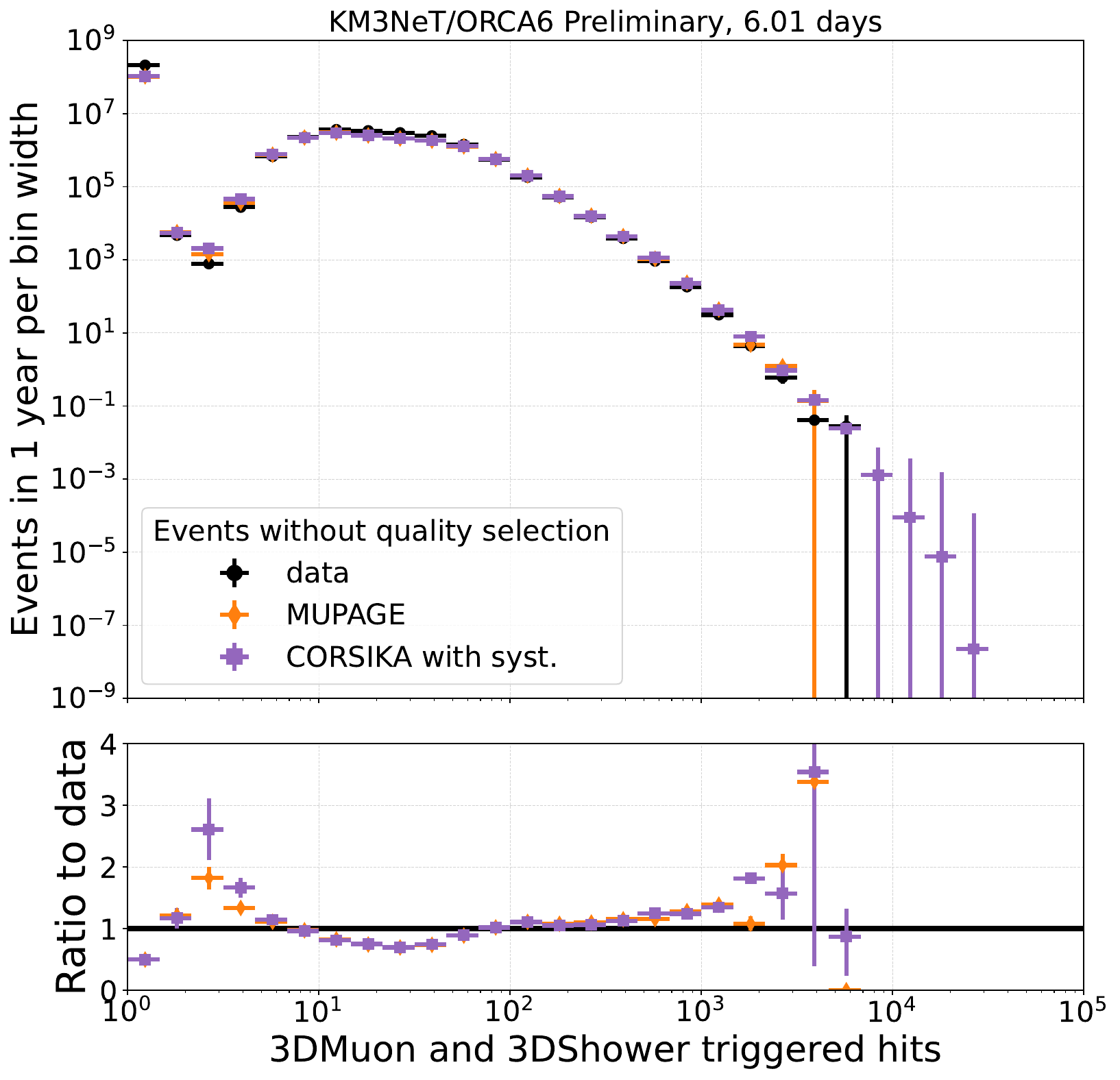}}\subfloat[Only events with $\mathcal{L}>280$.]{\centering{}\includegraphics[width=8cm]{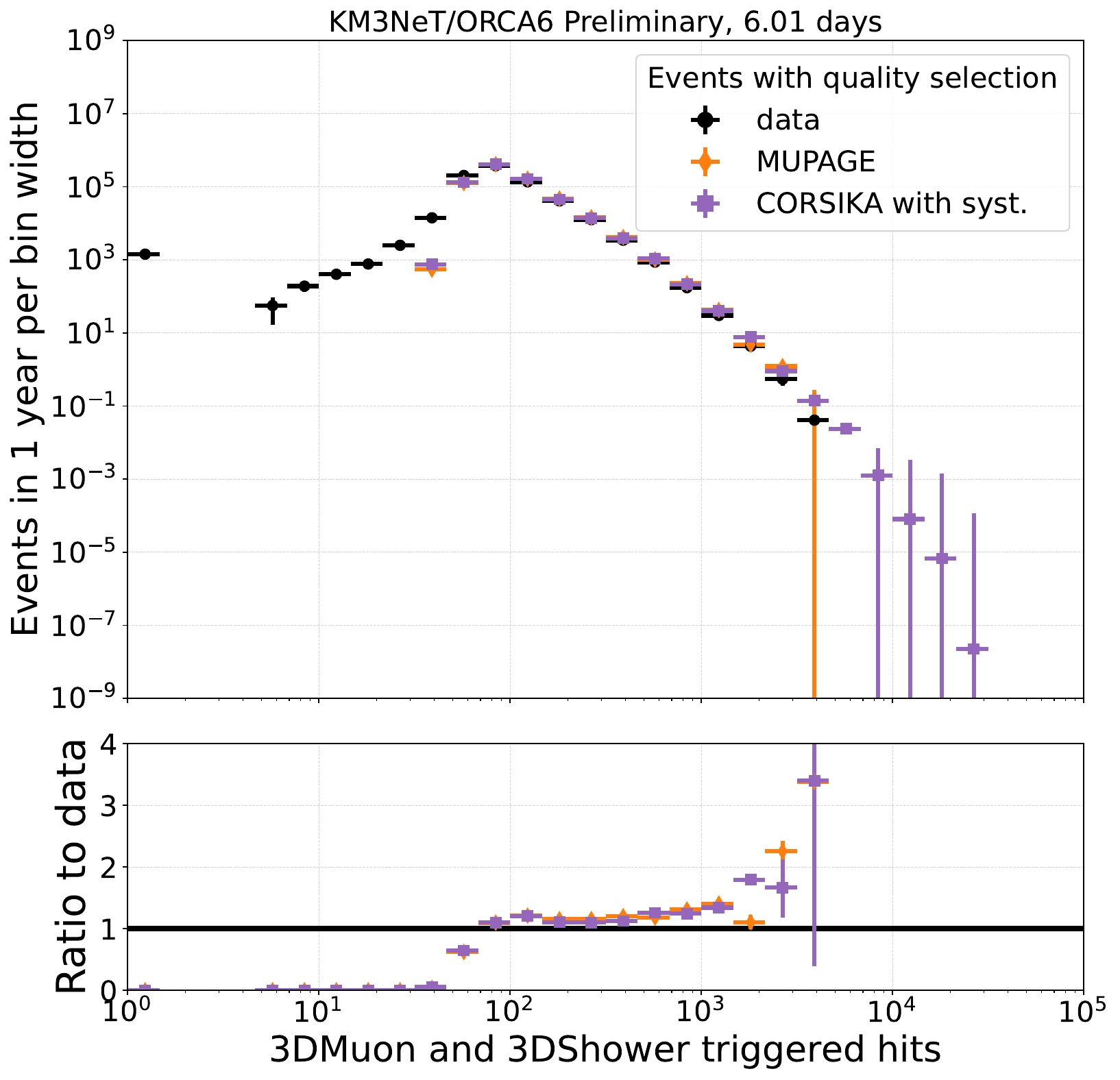}}\caption{The comparison of the distributions of the number of hits triggered
by the 3DMuon and 3DShower triggers for ORCA6 data and MC simulations.
\label{fig:ORCA6_3DMUON_3DSHOWER_trig_hits}}
\end{figure}

\begin{figure}[H]
\centering{}\subfloat[No event quality selection.]{\centering{}\includegraphics[width=8cm]{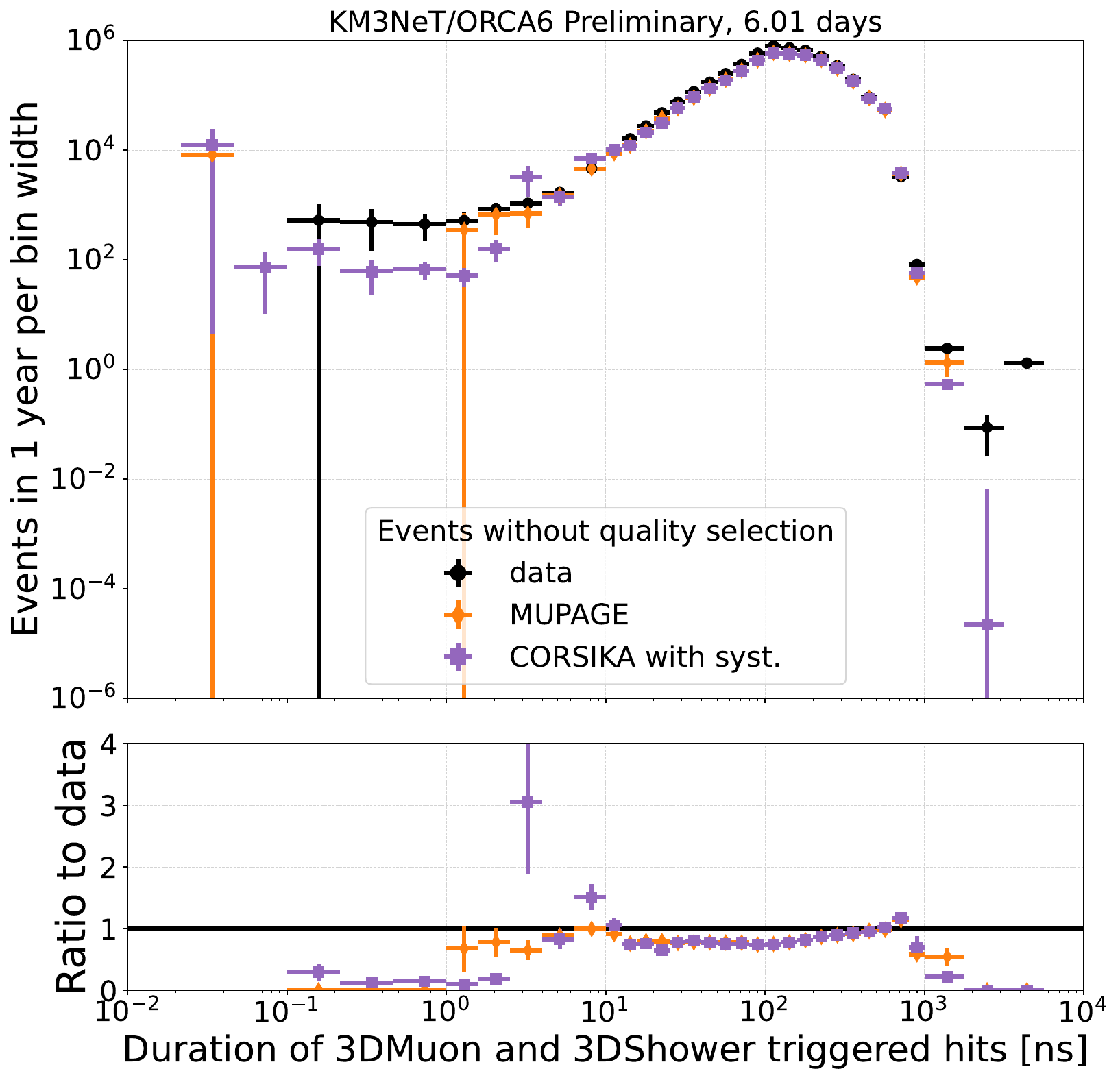}}\subfloat[Only events with $\mathcal{L}>280$.]{\centering{}\includegraphics[width=8cm]{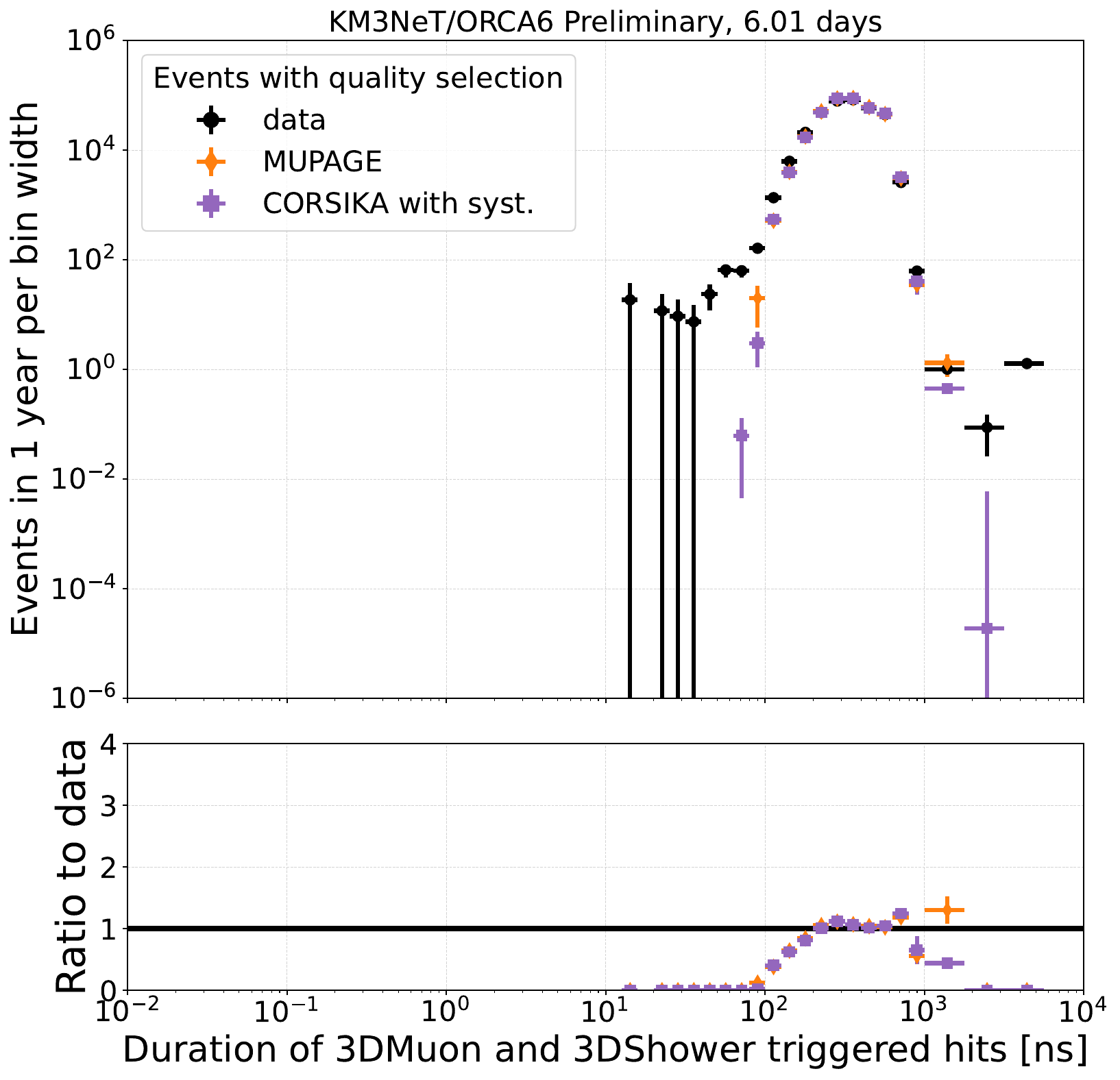}}\caption{The comparison of the distributions of the duration of hits triggered
by the 3DMuon and 3DShower triggers for ORCA6 data and MC simulations.
\label{fig:ORCA6_3DMUON_3DSHOWER_trig_hits-duration}}
\end{figure}

As clear already from Fig. \ref{fig:ORCA6_JMuon_likelihood}, the
agreement between the data and MC simulations for ORCA6 still required
significant improvement. Both MUPAGE and CORSIKA simulations have
shown consistent behaviour, however their ratio to experimental data
did not settle around $1$ at high $\mathcal{L}$ values, as was the
case for ARCA6. One of the known factors causing the difference between
the experimental data and MC simulation were the differences between
the rbr (data) and non-rbr (MC) processing modes, which were larger
than for ARCA6. This is tied to generally less stable data taking
of ORCA6. The much stricter cut at $\mathcal{L}=280$ was introduced
to remedy these problems to some extent. As one can see, most of the
rejected events are again ones with few triggered hits and short duration.

With the quality of results in mind, the ORCA6 data was not used in
Chap. \ref{chap:prompt_ana}. Nevertheless, the comparison of reconstructed
observables is still shown in Sec. \ref{subsec:ORCA6}. The exact
reasons for the observed discrepancies are still under close investigation
within the KM3NeT Collaboration. They were certainly at least partially
connected with hardware issues, which have affected the stability
of ORCA6 data taking. Recovery and replacement of the faulty DUs is
foreseen during the upcoming ORCA sea campaigns (rounds of deployment
of further DUs and other hardware). In parallel, there are ongoing
efforts to improve the entire KM3NeT simulation chain. The work of
this thesis, in particular the developments in the gSeaGen code (see
Sec. \ref{sec:gSeaGen_supplementary_material}), has been a significant
contribution in this area. 

\section{Comparisons of data and Monte Carlo simulations \label{sec:data_MC_Results}}

In the following, selected ARCA and ORCA experimental data sets were
compared against the MUPAGE and CORSIKA muon simulations. To minimize
the influence of the seasonal variations on the experimental data,
the runs covering the full available periods of data taking were used
both for the data and MUPAGE simulation (if using the rbr simulation).
The seasonal variations of the muon flux observed by ANTARES, which
was located near the ORCA site, were on the order of 2~\% \cite{ANTARES_seasonal}.
The systematic uncertainties have been included only in the most recent
results: for ARCA6 and ORCA6. The results for ARCA2, ORCA1, and ORCA4
have been published prior to completion of the systematic uncertainty
study and were hence not modified \cite{my-ICRC2019,my-ICRC2021,my-VLVnT2021,my-ICHEP2020}.
The primary CR flux model assumed for CORSIKA MC weighting was GST3
\cite{GST}, as mentioned in Sec. \ref{sec:CORSIKA}.

\subsection{KM3NeT/ARCA\label{subsec:KM3NeT/ARCA}}

In this section, comparisons of measured and simulated distributions
of different reconstructed muon bundle observables for ARCA2 and ARCA6
were collected. ARCA2 and ARCA6 are the intermediate configurations
of the ARCA detector with 2 and 6 installed DUs respectively.

\subsubsection{ARCA2\label{subsec:ARCA2-data-MC}}

For the ARCA2 configuration, reconstructed muon bundle direction (average
muon $\cos\theta_{\mathsf{zenith}}$) and energy (sum of $E_{\mu}$)
distributions were compared between experimental data, MUPAGE and
CORSIKA, using the datasets indicated in Tab. \ref{tab:data_and_MC_datasets}.
No cuts have been applied.

\begin{figure}[H]
\centering{}\subfloat[Reconstructed cosine of the zenith angle of the bundle. \label{fig:ARCA2_plots-1}]{\centering{}\includegraphics[width=8cm]{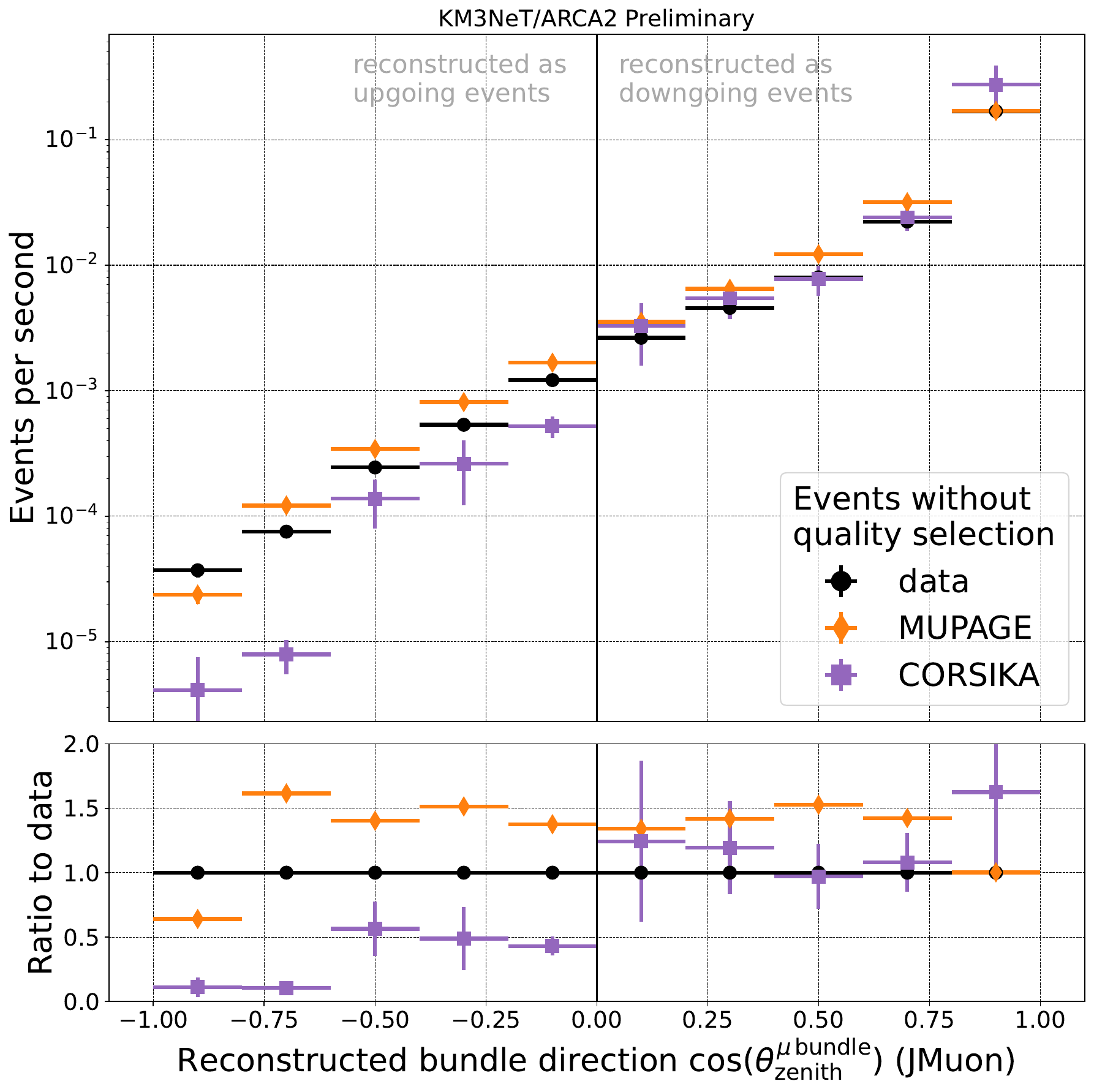}}\subfloat[Reconstructed bundle energy. \label{fig:ARCA2_plots-2}]{\centering{}\includegraphics[width=8cm]{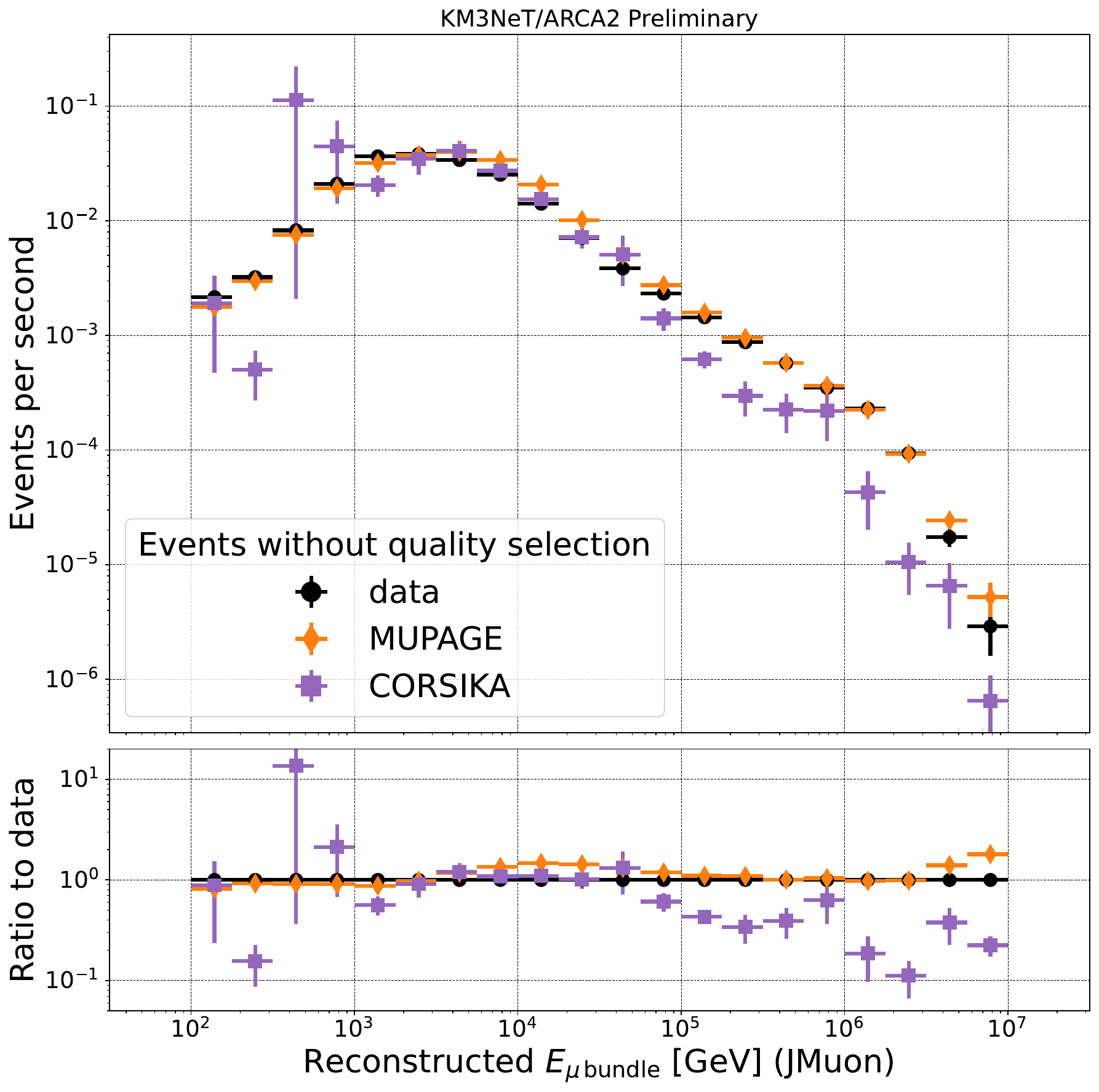}}\caption{The muon bundle rates for the ARCA2 detector. The data and MUPAGE
samples both have the livetimes of about 20~d. The plots have been
published in \cite{my-ICRC2019,my-ICRC2021,my-VLVnT2021}. \label{fig:ARCA2_plots}}
\end{figure}

Results presented in Fig. \ref{fig:ARCA2_plots} are one of the very
first KM3NeT muon measurements and have been an invaluable proof of
concept for the detector operation and the validity of KM3NeT simulations.
At the same time, the observed discrepancies were a clear sign, where
further improvement was needed. In the zenith plot (Fig. \ref{fig:ARCA2_plots-1}),
the events reconstructed as upgoing ($\cos\theta_{\mathsf{zenith}}^{\mu\,\mathrm{bundle}}<0$)
are in fact muons for which the direction was flipped by JMuon (possibly
with a very small contribution of actual upgoing events from neutrino
interactions in the data). A number of possible causes of the clear
underprediction of the data by CORSIKA at higher energies (Fig. \ref{fig:ARCA2_plots-2})
have been identified and were discussed in \cite{my-ICRC2019,my-ICRC2021,my-VLVnT2021}.
Notably, one of the underlying issues was solved by improving the
propagation of muons through water with gSeaGen (some of these developments
are described in Sec. \ref{sec:gSeaGen_supplementary_material}).

\subsubsection{ARCA6 \label{subsec:ARCA6}}

For the ARCA6 results, the selected runs corresponded to a livetime
of 1.46 days. It may seem small at first, however it has to be noted
that this translates to 3~013~982 events, out of which all were
reconstructed by the ML-based reconstruction and 287~553 by JMuon.
The used runs were intentionally picked to be separated by possibly
large periods of time, to ensure at least partial averaging out of
the seasonal variations of the muon flux. For CORSIKA, the ARCA6 test
dataset from Sec. \ref{subsec:Summary-of-datasets} was used. An event
quality selection from Sec. \ref{subsec:ARCA6-quality-cuts} has been
applied to both MCs and to the data. 

\begin{figure}[H]
\centering{}\subfloat[Reconstructed cosine of the zenith angle of the bundle. \label{fig:ARCA6_plots-JMuon-zenith}]{\centering{}\includegraphics[width=7.5cm]{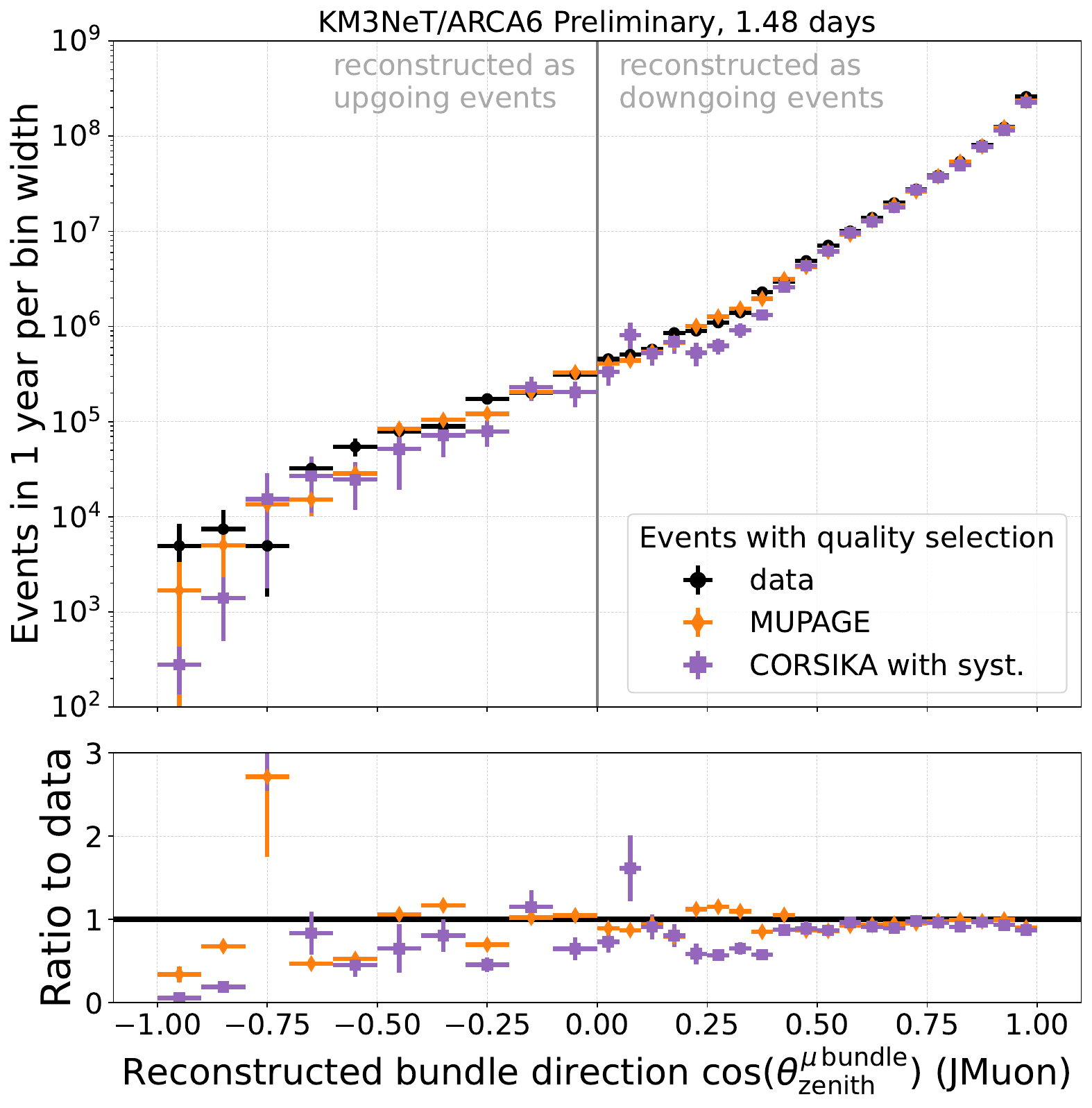}}\subfloat[Reconstructed bundle energy. \label{fig:ARCA6_plots-JMuon-energy}]{\centering{}\includegraphics[width=8cm]{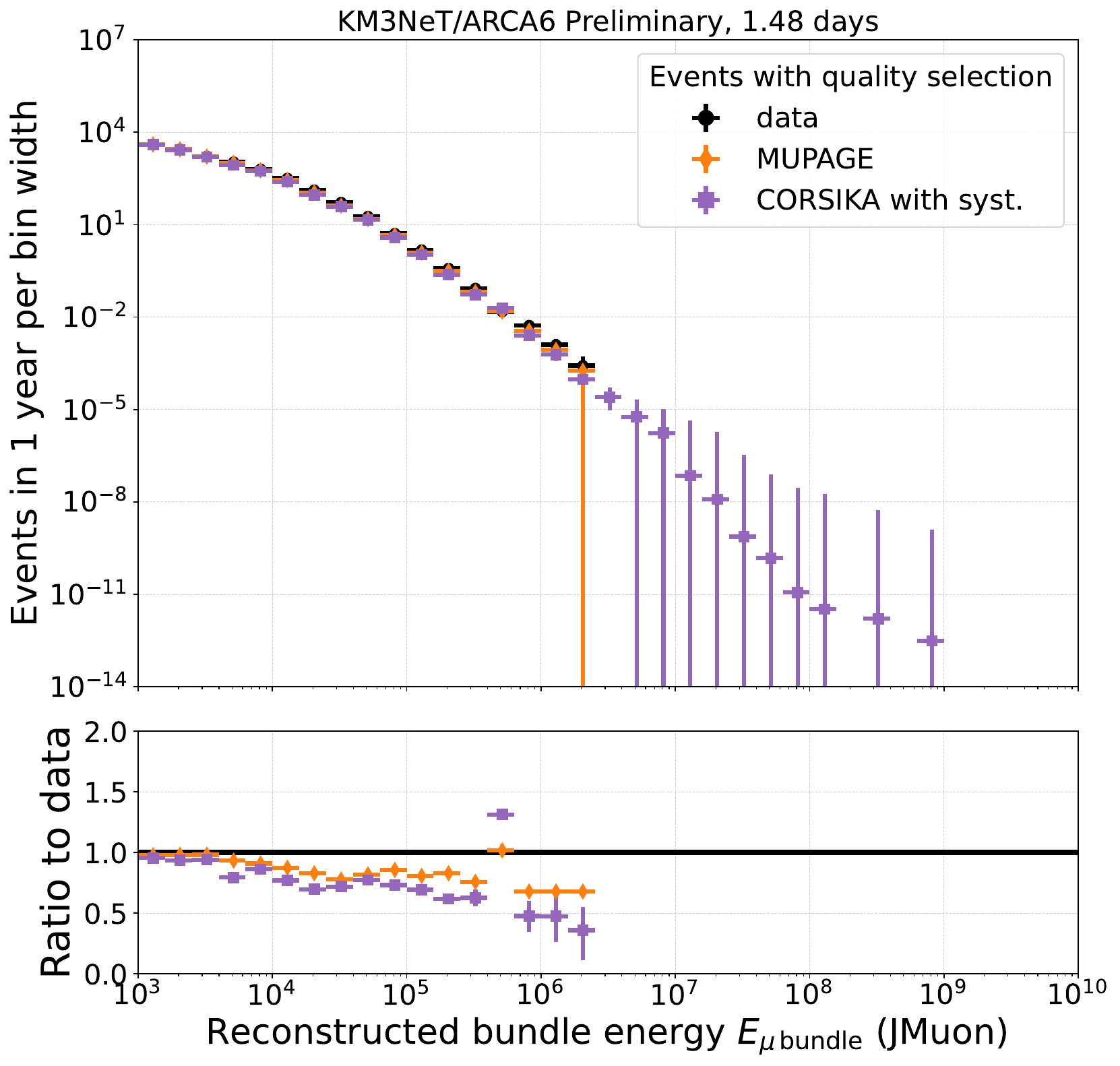}}\caption{The muon bundle rates for the ARCA6 detector as function of the direction
and energy reconstructed by JMuon \label{fig:ARCA6_plots-JMuon}}
\end{figure}

The standard JMuon reconstruction results are presented in Fig. \ref{fig:ARCA6_plots-JMuon}.
The agreement between data and MC in zenith has improved with respect
to ARCA2 results, however most of the upgoing events were still misreconstructed
downgoing ones, as in Sec. \ref{subsec:ARCA2-data-MC}. The reconstructed
energy distributions in Fig. \ref{fig:ARCA6_plots-JMuon-energy} match
well below few TeV. As one can verify in Fig. \ref{fig:official_E_reco-A6_all},
this is in fact a region, where JMuon prediction largely underestimates
the true energy. Above TeV energies, both CORSIKA and MUPAGE underestimate
the data. MUPAGE is closer to the experimental data, since it was
tuned on the KM3NeT data from earlier detector configurations. For
CORSIKA, the behaviour is consistent with results reported in Sec.
\ref{subsec:ARCA2-data-MC}, however the ratio is closer to one than
it was for ARCA2. This is an improvement directly resulting from optimisation
of the gSeaGen code, especially in terms of muon propagation (see
Sec. \ref{sec:gSeaGen_supplementary_material}). MUPAGE, on the other
hand, displays a different behaviour than it did in Fig. \ref{fig:ARCA2_plots-2}.
This stems from the fact that MUPAGE simulation in non-rbr mode was
used for ARCA6, while for ARCA2 it was in rbr mode. This points out
that there is a need for run-by-run simulations, especially in the
case of intermediate KM3NeT detector configurations, which typically
do not collect data for much longer than a year before next detection
units are deployed. The caveat here is that more detailed simulations
with CORSIKA in rbr mode are simply not feasible, due to enormous
computational cost. This was addressed in \cite{Andrey_Thesis}, where
the MUPAGE parametrization was tuned to the CORSIKA simulation produced
within this work. 

\begin{figure}[H]
\centering{}\subfloat[Reconstructed muon bundle energy. \label{fig:ARCA6_plots-LightGBM-Ebundle}]{\centering{}\includegraphics[width=8cm]{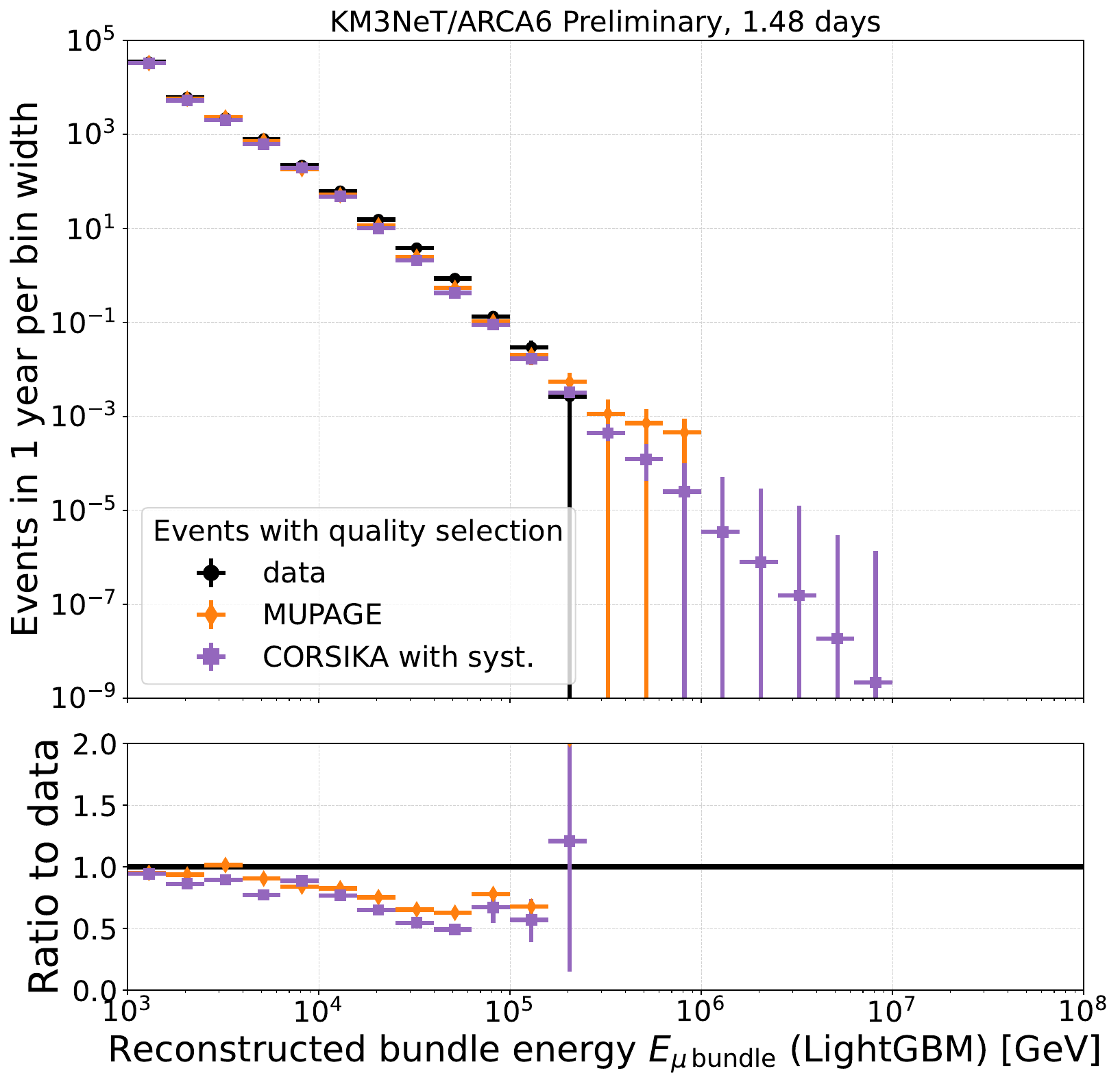}}\subfloat[Reconstructed total primary energy. \label{fig:ARCA6_plots-LightGBM-Eprim}]{\centering{}\includegraphics[width=8cm]{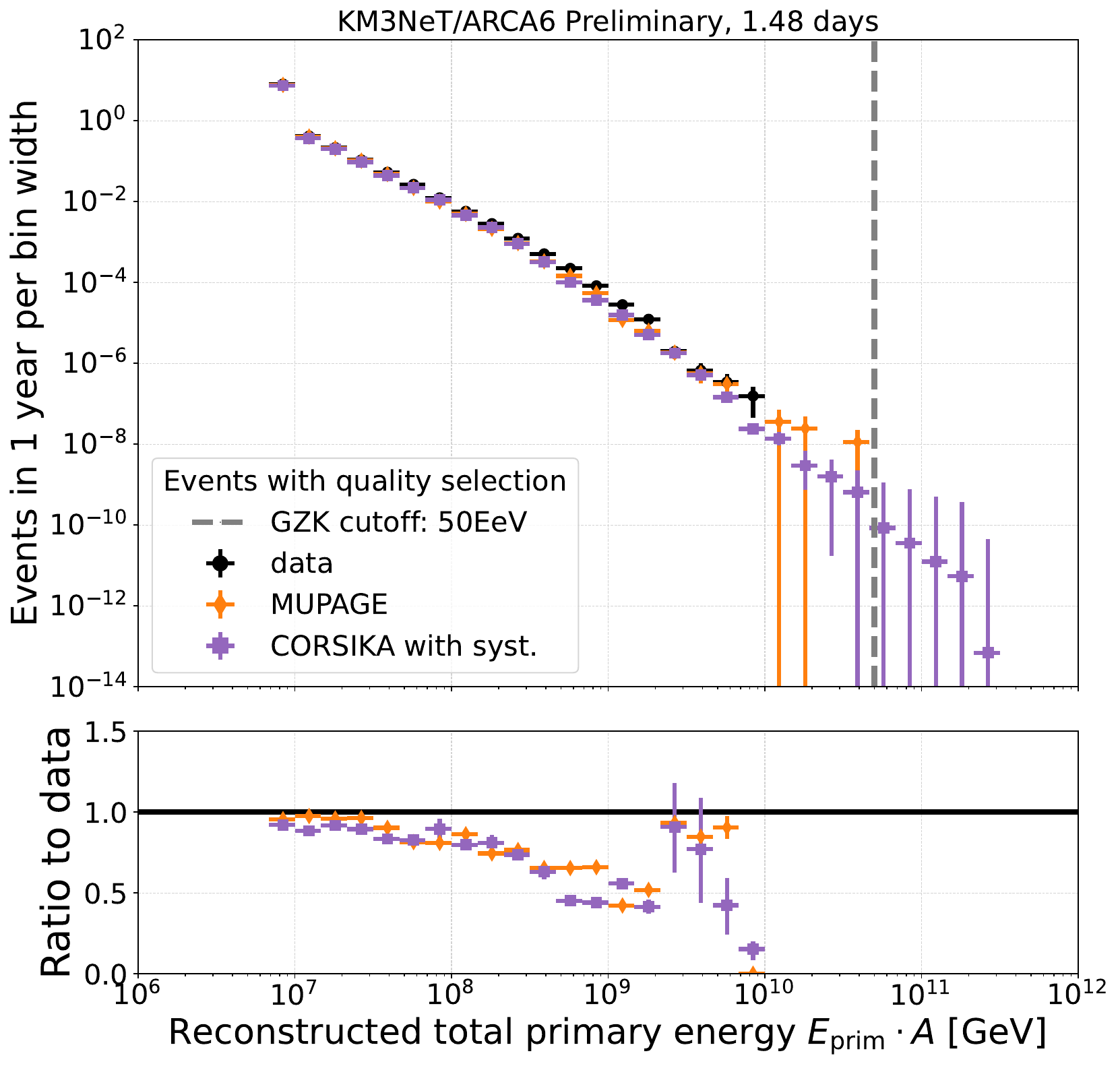}}\caption{ML-based energy reconstruction results for the ARCA6 detector. \label{fig:ARCA6_plots-LightGBM-energy}}
\end{figure}

Another open question is, whether the JMuon energy reconstruction,
which is developed under a~hypothesis of a single muon track is a
suitable tool for reconstruction of high-energy muon bundles. On one
hand, muons in EAS are highly collinear due to very forward production,
on the other hand it is highly unlikely to encounter an EeV shower
producing a bundle with multiplicity equal to one. An alternative
was provided by the energy reconstruction developed within this work
(see Chap. \ref{chap:muon-bundle-reco}). The results of the LightGBM-based
reconstruction applied to the experimental data, MUPAGE, and CORSIKA
MC is shown in Fig. \ref{fig:ARCA6_plots-LightGBM-energy}. The reconstructed
bundle energy in Fig. \ref{fig:ARCA6_plots-LightGBM-Ebundle} does
not reach as high values as the JMuon energy reconstruction in Fig.
\ref{fig:ARCA6_plots-JMuon-energy}. This is expected, since LightGBM
reconstruction tends to underpredict the true energy, especially at
higher energies (see Sec. \ref{subsec:Bundle-energy}). In Fig. \ref{fig:ARCA6_plots-LightGBM-Ebundle},
the energy prediction can be trusted above few TeV (see Fig. \ref{fig:Ebundle_reco_results}).
The underestimation of the data by the MC simulations is consistent
between Fig. \ref{fig:ARCA6_plots-LightGBM-Ebundle} and \ref{fig:ARCA6_plots-JMuon-energy},
hence such a behaviour cannot be attributed to the reconstruction.
It implies that there is a need for more accurate modelling of CR
air showers. 

The total primary energy reconstruction behaves in a similar manner
to the bundle energy reconstruction, as the two quantities are tightly
related. According to Fig. \ref{fig:Eprim_reco_results}, the primary
energy prediction offers sensible (under-)predictions above 10~PeV.
The GZK cutoff value of 50~EeV has been marked in Fig. \ref{fig:ARCA6_plots-LightGBM-Eprim}
to guide the eye \cite{Zatsepin-Kuzmin}. As one may note, the data
does not reach near that point, hence it is not possible to draw any
conclusion yet. The CORSIKA simulation extends beyond 50~EeV, however
the GZK cutoff is not implemented in any way in the CORSIKA MC (nor
in MUPAGE).

\begin{figure}[H]
\centering{}\includegraphics[width=8cm]{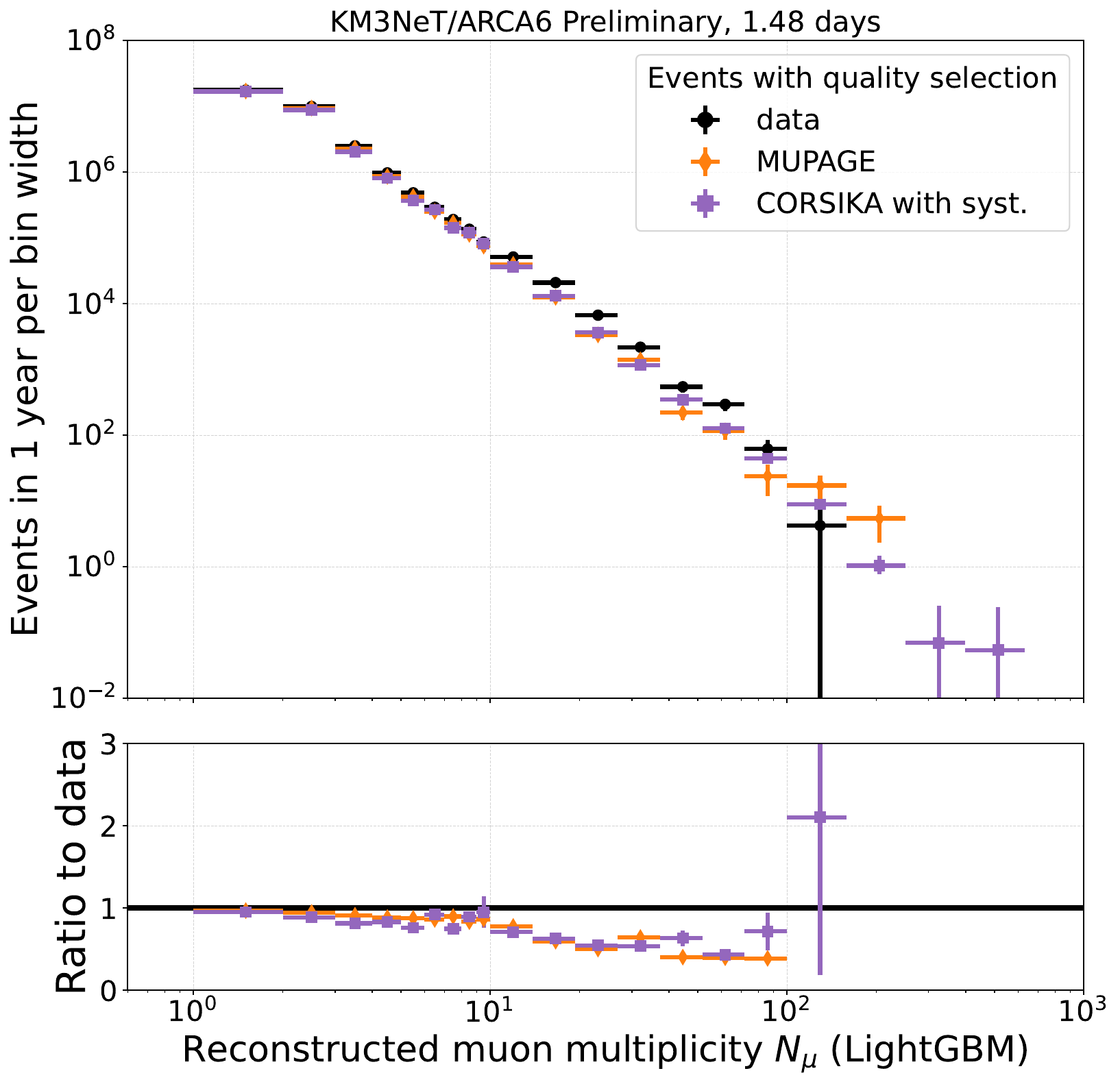}\caption{ML-based multiplicity reconstruction result the ARCA6 detector. \label{fig:ARCA6_plots-LightGBM-multiplicity}}
\end{figure}

The multiplicity reconstructed in Fig. \ref{fig:ARCA6_plots-LightGBM-multiplicity}
is the one with muon selection, as in Sec. \ref{subsec:Results-simple-regr}.
The MUPAGE simulation generally tends to better match the data, which
is expected as it was tuned on ORCA4 data. Both simulations struggle
to accurately describe the data for bundles with more than 10 muons.
This is again an indication that further work in CR shower simulations
is required.

\subsection{KM3NeT/ORCA}

Here, the results of data vs MC comparisons for ORCA1, ORCA4, and
ORCA6 were gathered. Analogous to Sec. \ref{subsec:KM3NeT/ARCA},
ORCA1, ORCA4, and ORCA6 are the intermediate ORCA configurations with
1, 4, and 6 installed DUs respectively.

\subsubsection{ORCA1}

\begin{figure}[H]
\centering{}\includegraphics[width=8cm]{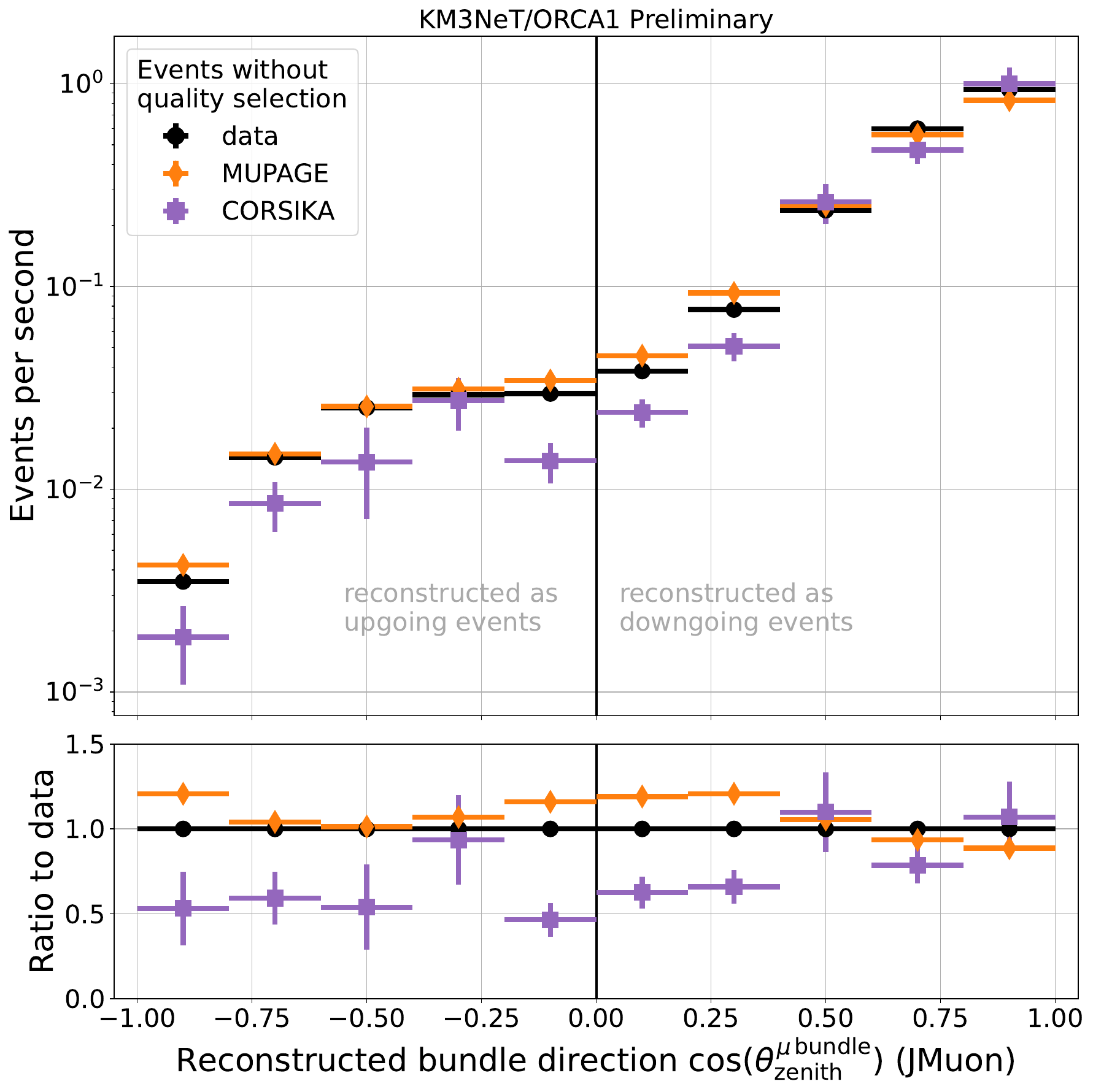}\caption{The muon bundle rates as a function of the reconstructed cosine of
the zenith angle of the bundle for the ORCA1 detector. The data and
MUPAGE samples both have the livetimes of about 24 days. The plot
has been published in \cite{my-ICRC2019}. \label{fig:ORCA1_zenith}}
\end{figure}

In the case of the ORCA1, no energy reconstruction has been performed,
as this was one of the earliest detector configurations. The reconstructed
zenith angle is shown in Fig. \ref{fig:ORCA1_zenith}. A very similar
behaviour was observed as for ARCA2 (Fig.\ref{fig:ARCA2_plots-1}).
In this case no cuts have been applied.

\subsubsection{ORCA4\label{subsec:ORCA4-data-MC}}

\begin{figure}[H]
\centering{}\subfloat[Reconstructed cosine of the zenith angle of the bundle. \label{fig:ORCA4_plots-1}]{\centering{}\includegraphics[width=8cm]{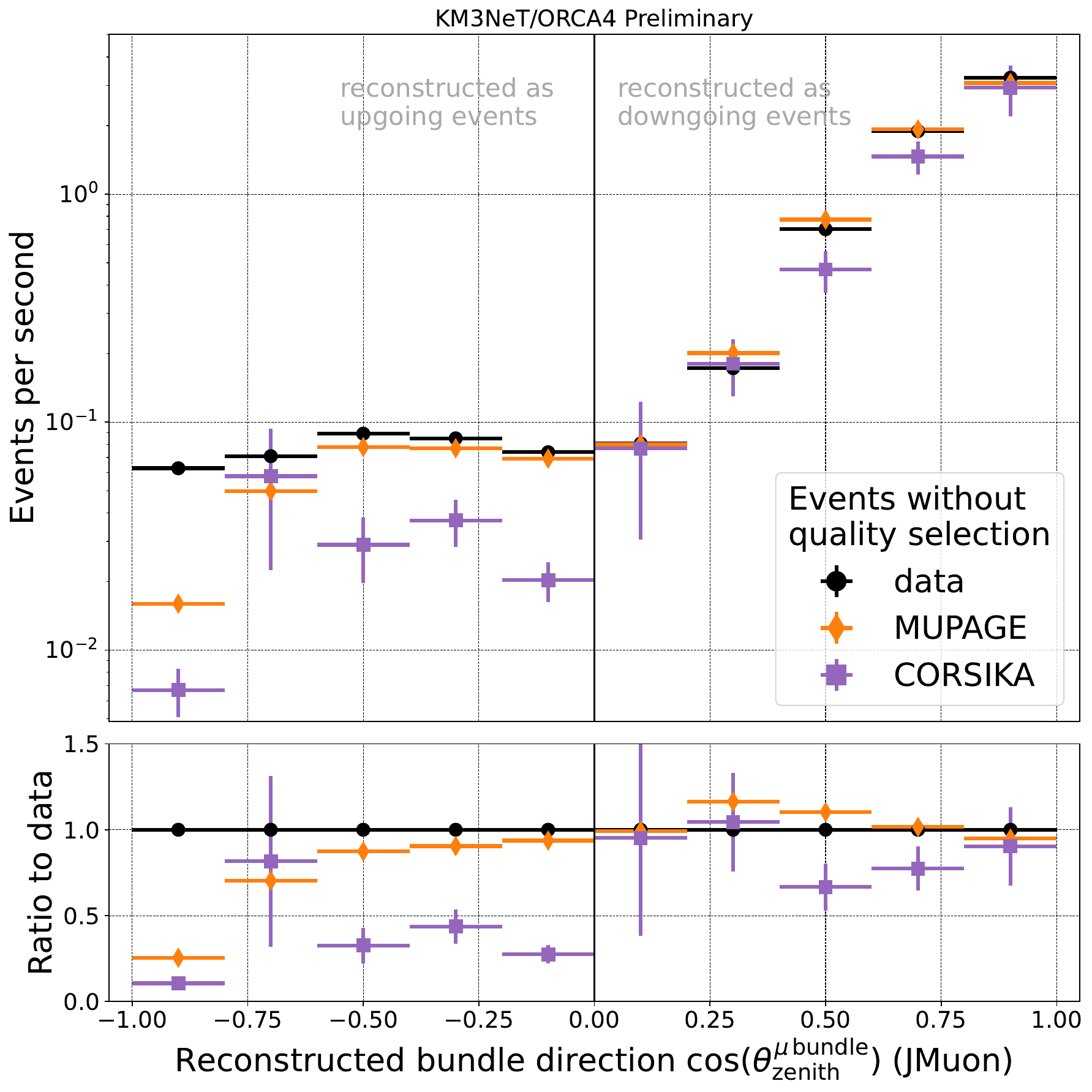}}\subfloat[Reconstructed bundle energy. \label{fig:ORCA4_plots-2}]{\centering{}\includegraphics[width=8cm]{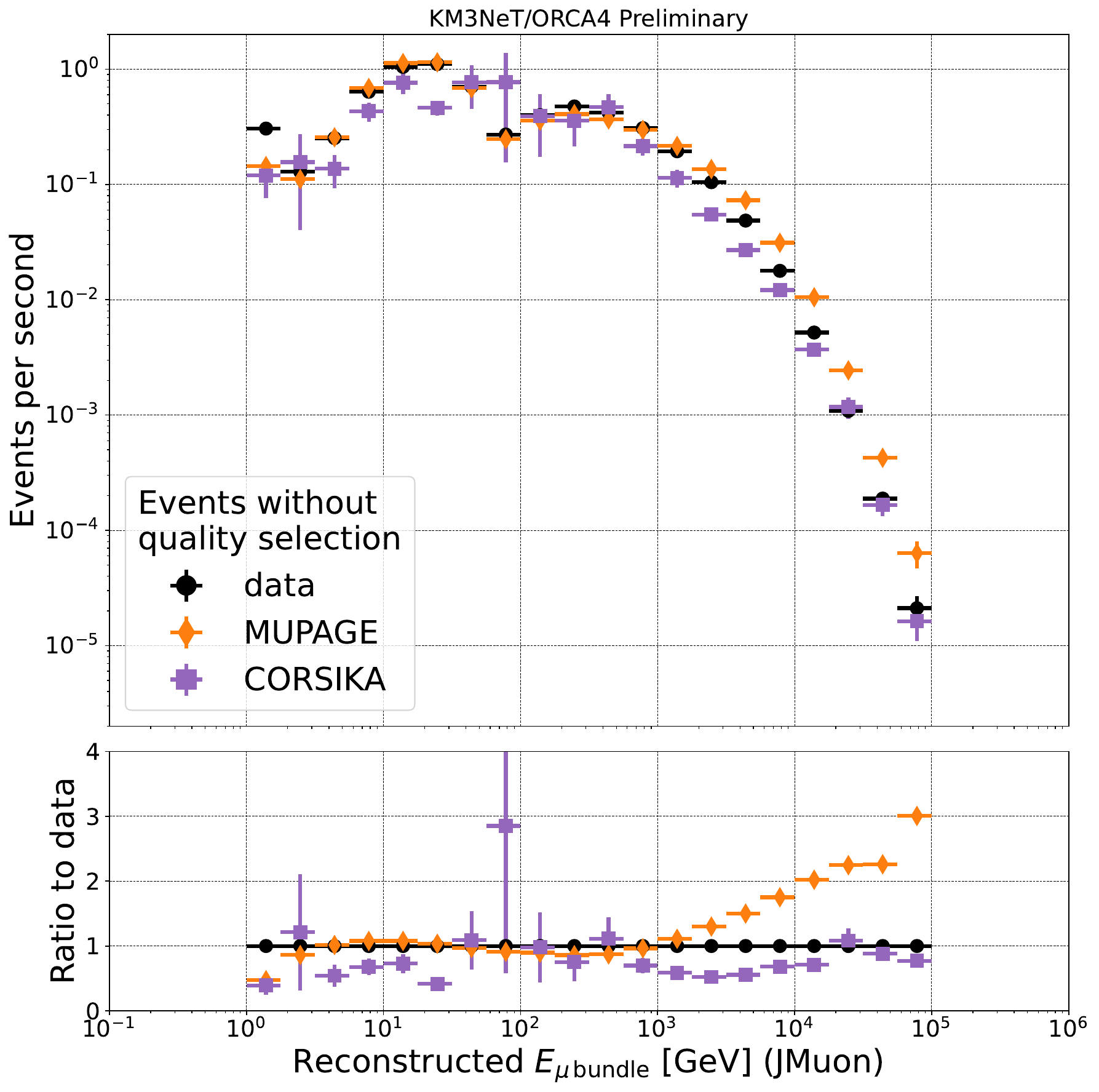}}\caption{The muon bundle rates for the ORCA4 detector. The data and MUPAGE
samples both have the livetimes of about 10 days. The plots have been
published in \cite{my-ICRC2021,my-VLVnT2021,my-ICHEP2020}. \label{fig:ORCA4_plots}}
\end{figure}

The JMuon reconstruction of zenith and energy for ORCA4 with no cuts
applied is presented in Fig.\ref{fig:ORCA4_plots}. The agreement
between the data and both Monte Carlos for the downgoing muons has
significantly improved with respect to the ORCA1 result, which can
be attributed to advances in both simulation and reconstruction software.
Interestingly, here MUPAGE seems to overestimate the high-energy event
rate (see Fig. \ref{fig:ORCA4_plots-2}). The excess of upgoing events
in data is most probably due to pure noise events, as e.g. in Fig.
\ref{fig:ARCA6_plots-JMuon-zenith}.

\subsubsection{ORCA6 \label{subsec:ORCA6}}

The ORCA6 results were derived using a subsample of the runs from
Tab. \ref{tab:data_and_MC_datasets} with a livetime of 6.01 days.
The runs were selected sparsely, analogously to what was done for
ARCA6 in Sec. \ref{subsec:ARCA6}. This was equivalent to 4~574~887
events reconstructed with LightGBM and 4~571~428 with JMuon). Here,
the reconstruction failure rate of JMuon is noticeably lower, compared
with ARCA6 (Sec. \ref{subsec:ARCA6}), owing to more dense instrumentation
of ORCA. Regarding CORSIKA, the ORCA6 test dataset from Sec. \ref{subsec:Summary-of-datasets}
was used. The event quality selection from Sec. \ref{subsec:ORCA6-quality-cuts}
has been applied to both MCs and to the data.

\begin{figure}[H]
\centering{}\subfloat[Reconstructed cosine of the zenith angle of the bundle. \label{fig:ORCA6_plots-JMuon-zenith}]{\centering{}\includegraphics[width=7.5cm]{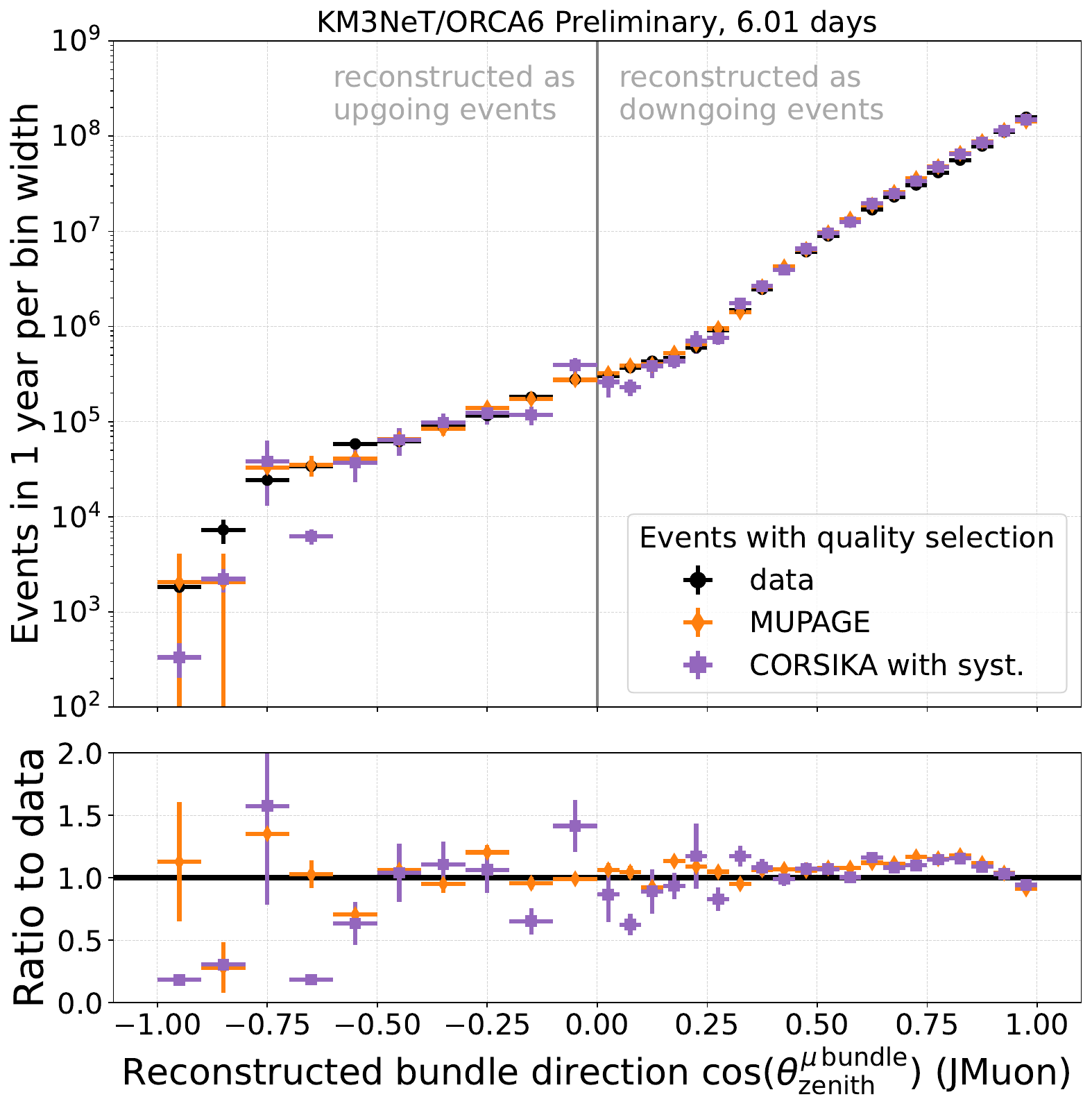}}\subfloat[Reconstructed bundle energy. \label{fig:ORCA6_plots-JMuon-energy}]{\centering{}\includegraphics[width=8cm]{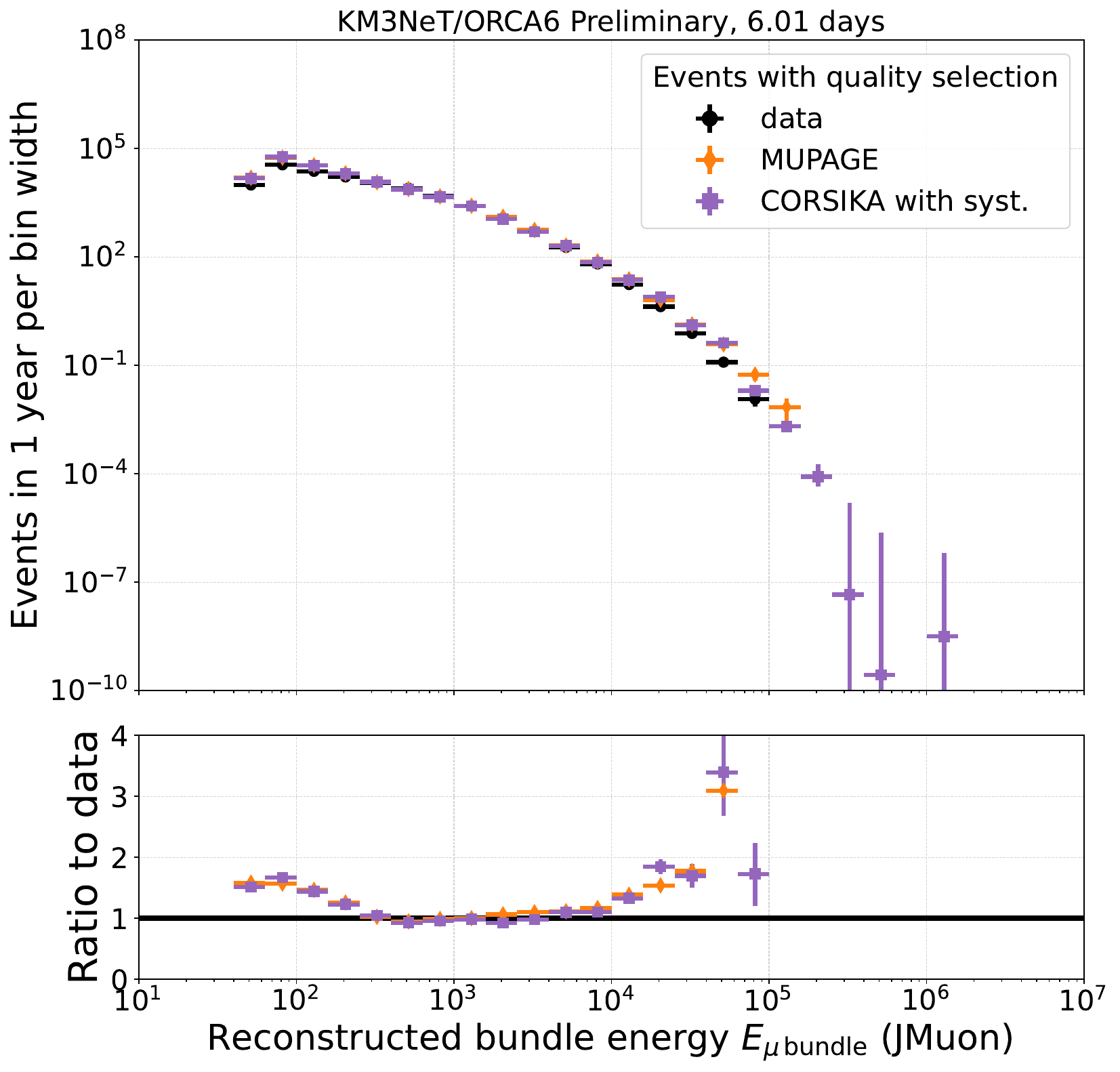}}\caption{The muon bundle rates for the ORCA6 detector as function of the direction
and energy reconstructed by JMuon. \label{fig:ORCA6_plots-JMuon}}
\end{figure}

Distributions of zenith and energy for ORCA6 as reconstructed by JMuon
can be found in Fig. \ref{fig:ORCA6_plots-JMuon}. The stricter cut
on likelihood seems to have eliminated the upgoing pure noise events.
Given the challenges noted in Sec. \ref{subsec:ORCA6-quality-cuts},
the agreement between the simulated and measured direction of showers
is remarkably good. It has also improved with respect to results for
ORCA1 and ORCA4.

\begin{figure}[H]
\centering{}\subfloat[Reconstructed muon bundle energy. \label{fig:ORCA6_plots-LightGBM-Ebundle}]{\centering{}\includegraphics[width=8cm]{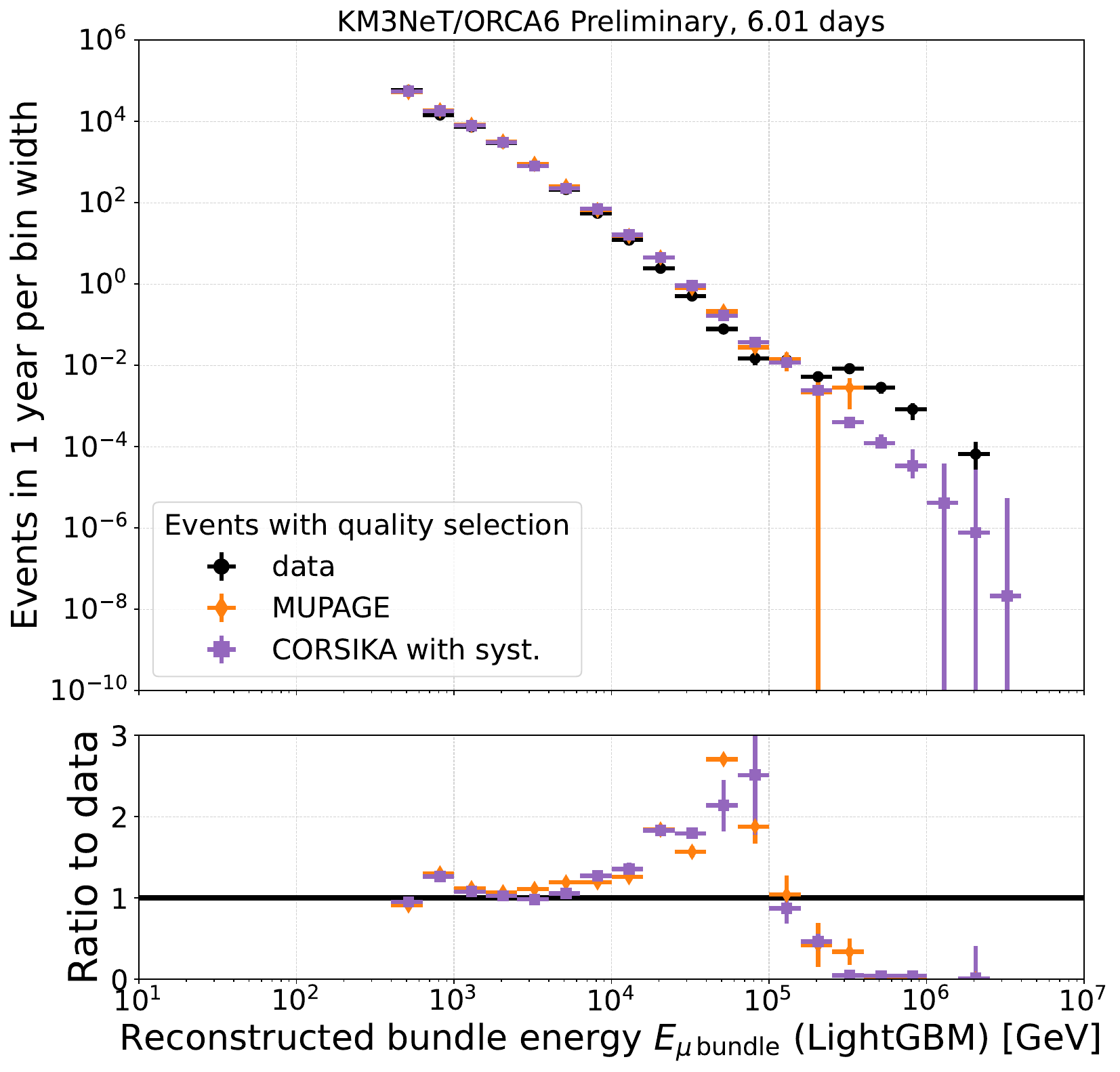}}\subfloat[Reconstructed primary energy. \label{fig:ORCA6_plots-LightGBM-Eprim}]{\centering{}\includegraphics[width=8cm]{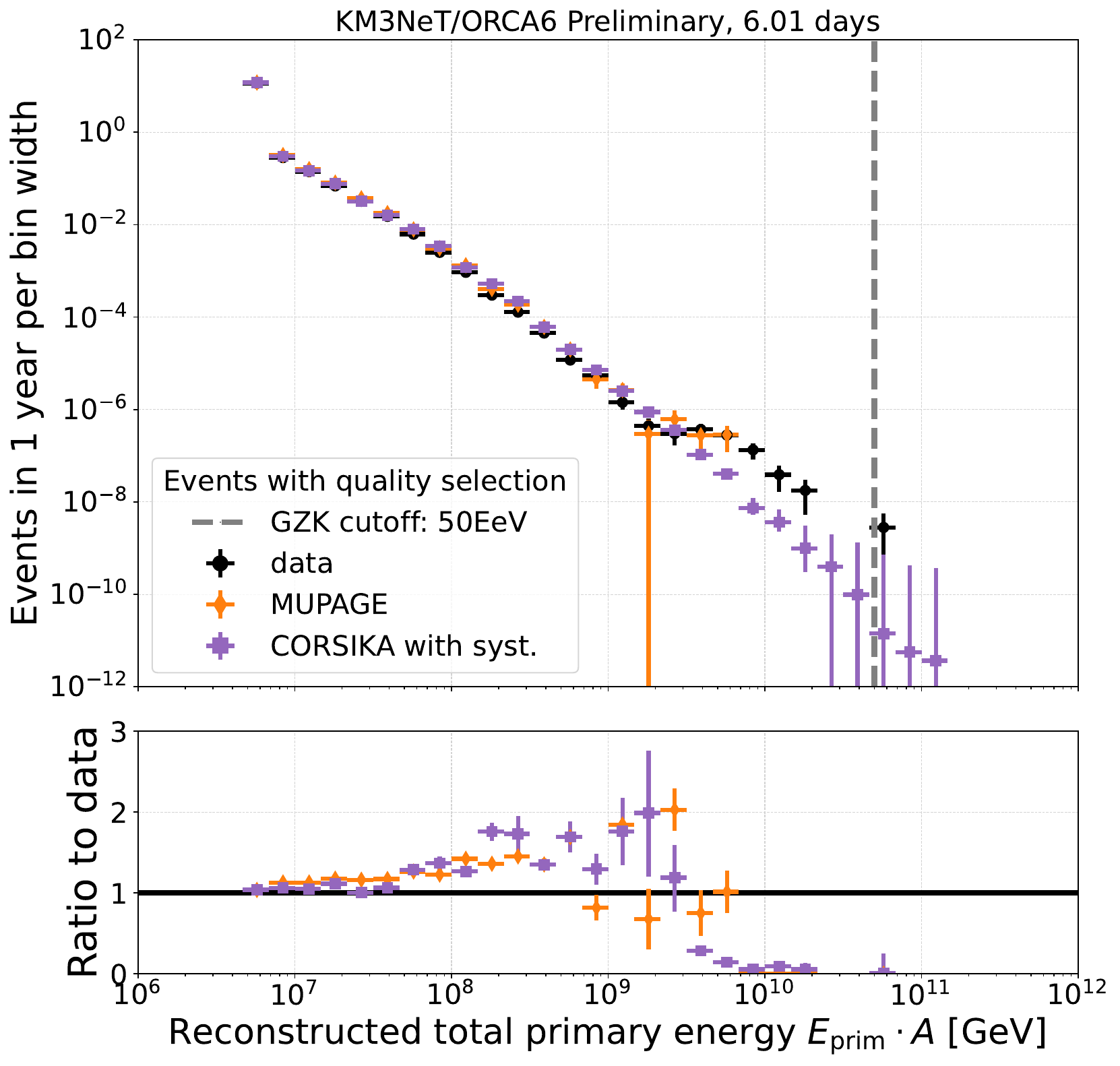}}\caption{ML-based energy reconstruction results the ORCA6 detector. \label{fig:ORCA6_plots-LightGBM-energy}}
\end{figure}

It might come as a surprise that in Fig. \ref{fig:ORCA6_plots-JMuon-energy}
both simulations seem to overpredict the experimental data, which
was only the case for MUPAGE in earlier results. However, upon inspection
of Fig. \ref{fig:official_E_reco-O6_all} one has to conclude that
for ORCA6 the JMuon energy reconstruction results are simply not reliable
above 3~TeV. Such a problem does not occur for LightGBM-based reconstruction
(see Fig. \ref{fig:Ebundle_reco_results}), hence one may claim that
it is related to the inaccuracy of JMuon in reconstruction of muon
bundle energies. In Fig. \ref{fig:ORCA6_plots-LightGBM-Ebundle},
the experimental data and simulations match well up to about 20~TeV
and there is a large excess of data over the CORSIKA MC above 200~TeV,
however confirming it with a larger collected (and simulated for MUPAGE)
livetime would be preferable. The observed excess seems to be much
larger than statistical fluctuations, even including potential unidentified
sources of systematic uncertainty. The region, where it occurs, roughly
matches where the contribution from the prompt muon component is expected,
however at this stage it would be premature to draw any conclusions,
given the remaining challenges in the simulation of atmospheric muons.

The total primary energy reconstruction in Fig. \ref{fig:ORCA6_plots-LightGBM-Eprim}
yielded results consistent to the ones for ARCA6 in Fig. \ref{fig:ARCA6_plots-LightGBM-Eprim},
also demonstrating an excess at high energies, as one would expect
after seeing Fig. \ref{fig:ORCA6_plots-LightGBM-Ebundle}. For ORCA6
data, there is a single data point at the GZK cutoff value, unlike
for ARCA6. It is the first such a measurement with KM3NeT detectors,
and it still remains to be verified with a more conclusive dataset.
As was already noted when discussing the ARCA6 results, neither CORSIKA
nor MUPAGE implements the GZK cutoff directly.

\begin{figure}[H]
\centering{}\includegraphics[width=8cm]{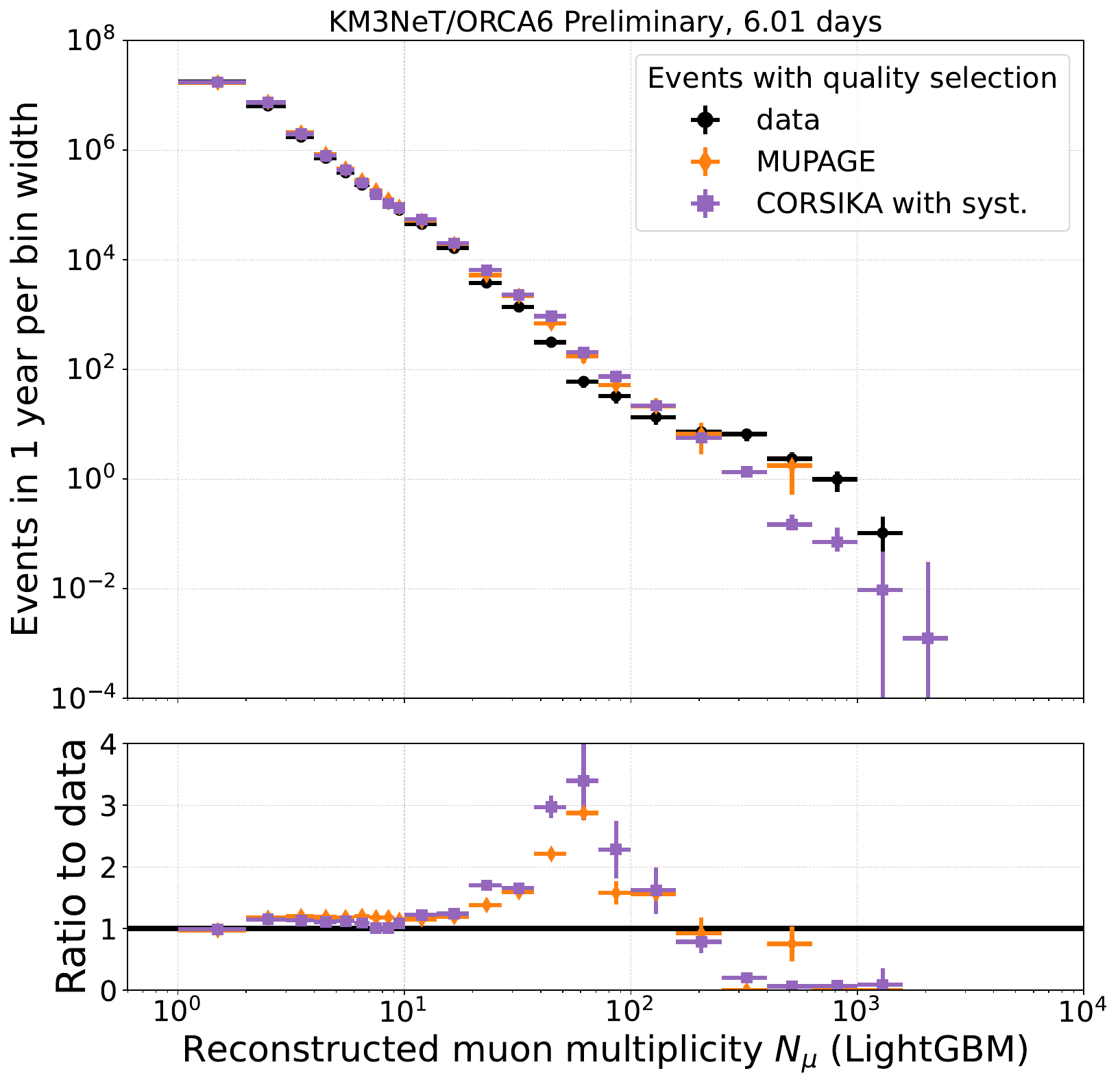}\caption{ML-based multiplicity reconstruction result the ORCA6 detector. \label{fig:ORCA6_plots-LightGBM-multiplicity}}
\end{figure}

The excess of the data over the simulations observed for the energy
reconstructions is present in the reconstructed multiplicity as well
(Fig. \ref{fig:ORCA6_plots-LightGBM-multiplicity}). This is expected,
as bundle energy and muon multiplicity are strongly correlated observables,
especially at high energies: the value of Pearson correlation coefficient
is around 0.7 for true and 0.9 for reconstructed variables. Despite
the poor agreement above multiplicity 10, both MUPAGE and CORSIKA
seem to be reproducing the experimental data quite well at low multiplicities,
which could be owed to the dense instrumentation of ORCA.

\section{Summary}

A clear progress has been made in quality of the KM3NeT atmospheric
muon simulations. Still, further improvements in the simulation software
are necessary for a reliable measurement of the prompt muon flux (see
Chap. \ref{chap:prompt_ana}) or to determine, whether the KM3NeT
indeed observe the GZK cutoff \cite{Greisen1966}. It may be anticipated
that by the time of completion of KM3NeT/ARCA and KM3NeT/ORCA construction,
such measurements will be possible.

The underprediction of the high-energy muon rates by the Monte Carlo
simulations is not unique to KM3NeT. It was seen by a number of experiments
and a term `Muon Puzzle' was coined for it \cite{MuonPuzzle}. One
of the possible explanations for this can be a larger than currently
predicted prompt muon flux (see Chap. \ref{chap:prompt_ana}). What
is novel here is also the fact that the excess is also visible in
muon multiplicity (admittedly, not a completely independent channel).
The excess is more pronounced in the ORCA6 results as compared to
ARCA6 mainly due to 6 times larger livetime. There are also the factors
of shallower depth at which ORCA is located (additional boost of collected
statistics) and its smaller size, which allows to collect less data
in the same time, however their impact is not as large.

The measurement of the all-nuclei primary CR flux was demonstrated
to be possible, even with intermediate configurations of the KM3NeT
detectors. The total primary energy reconstruction has yielded sensible
results, although burdened with large, partially irreducible, inaccuracy
related to the information loss as the shower propagates and looses
energy.

\chapter{Sensitivity of KM3NeT detectors to the prompt muon flux \label{chap:prompt_ana}}

\begin{figure}[H]
\centering{}\includegraphics[width=16cm]{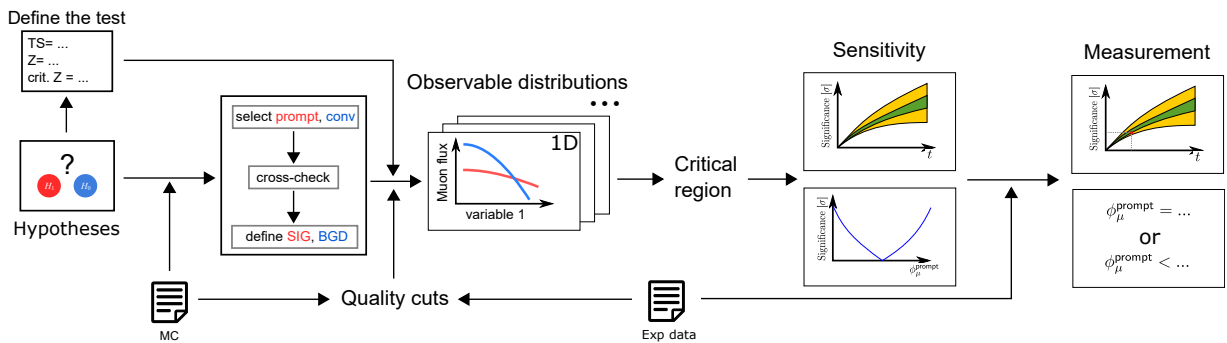}\caption{The outline of the analysis. \label{fig:prompt-ana-scheme}}
\end{figure}

This chapter presents the study investigating the potential of KM3NeT
detectors to observe the \textcolor{red}{prompt} muon flux. The general
scheme of the analysis is presented in Fig. \ref{fig:prompt-ana-scheme}.
The analysis starts with definitions of the hypotheses and how they
should be tested. Consequently, signal \textcolor{red}{SIG} and background
\textcolor{blue}{BGD} are defined, based on studying the properties
of available CORSIKA MC simulation. Afterwards, distributions of reconstructed
observables, split into containing either pure \textcolor{blue}{BGD}
or \textcolor{blue}{BGD} with \textcolor{red}{SIG} are investigated.
By computing the test statistic from them, a set of cuts that will
maximize the sensitivity is procured (defining the so-called `critical
region'). Next, expected significances as function of the data taking
time and \textcolor{red}{prompt} flux normalisation are computed using
the selected region. After inserting the experimental data, the \textcolor{red}{prompt}
muon rate fit or a limit on this rate is obtained.

The hypotheses of this analysis were:
\begin{itemize}
\item null hypothesis $H_{0}$: the atmospheric muon flux consists solely
of \textcolor{blue}{conventional} muons.
\item alternate hypothesis $H_{1}$: the atmospheric muon flux is a combination
of \textcolor{red}{prompt} and \textcolor{blue}{conventional} contributions.
\end{itemize}
The following study evaluated the capability of KM3NeT/ARCA and KM3NeT/ORCA
to reject $H_{0}$ by performing a significance test. The discovery
level of significance was set at $Z_{\mathrm{discovery}}=5\sigma$.
The exact formulation of the test may be found in Sec. \ref{sec:Significance-test},
after the MC-based definitions of signal \textcolor{red}{SIG} and
background \textcolor{blue}{BGD}. The sensitivity results in this
chapter were derived for ARCA115 and ORCA115 (complete building blocks)
and ARCA6 and ORCA6 (already available data, see Tab. \ref{tab:ML-dataset-summary}).
Since the agreement between the ARCA6 and ORCA6 data and simulations
in Chap. \ref{chap:Muon-rate-measurement} was not yet sufficient
to claim good description of the non-sensitive region, no measurement
was performed.

\section{Separation of conventional and prompt muons\label{sec:Selection-of-conventional-and-prompt}}

As pointed out in Sec. \ref{sec:Atmospheric-flux}, \textcolor{blue}{conventional}
and \textcolor{red}{prompt} are two distinct muon categories with
their characteristic properties. These differences can be utilized
to separate the two contributions to the atmospheric muon flux in
the CORSIKA MC. In practice, one can either look at the parent particles
or at where in the shower the muon was produced. Here, the first is
used for the definition and the latter helps to gain a better understanding
of the sample and to cross-validate the results (see Sec. \ref{sec:Control-plots-for}).

\subsection{Parent particles}

Each secondary particle in CORSIKA has two parent particles assigned:
a mother and a grandmother. This is the case regardless of the actual
number of parent particles of the muon. It is possible to distinguish
between the \textcolor{blue}{conventional} and \textcolor{red}{prompt}
muons by setting a threshold on lifetimes of the parent particles.
In the case of this analysis, the lifetime of $K_{\mathsf{S}}^{0}$
was selected as a reference point. Every parent particle, which decays
on average faster than $K_{\mathsf{S}}^{0}$ is declared \textcolor{red}{prompt}
(see Sec. \ref{sec:Classification-of-parent-particles-by-lifetime}),
following the approach in \cite{conv-prompt-calculation}. The simulated
contributions of muons with particular parent particles are shown
in Fig. \ref{fig:parent-contribution}. As expected, the muon rates
are dominated by \textcolor{blue}{conventional} muons produced by
$\pi^{\pm}$ and $K^{\pm}$ mothers (see Sec.\ref{sec:Atmospheric-flux}).
It has to be stressed that even if a grandmother particle is short-lived
(\textcolor{red}{prompt}), it can still lead to a production of a
\textcolor{blue}{conventional} muon, if the mother particle is \textcolor{blue}{conventional}
or if there are other interactions in between. To keep track of the
latter, CORSIKA introduces interaction counters (see Sec. \ref{subsec:Parent-particle-generations}).

By further inspection of Fig. \ref{fig:parent-contrib-grandmother},
it is not uncommon that a nucleus is saved as the grandmother particle.
This may happen in two cases:
\begin{enumerate}
\item When there is no other parent particle between the primary and the
mother particle of the muon. This implies that either the primary
nucleus will be saved as the grandmother particle, or the muon will
be saved as its own mother particle, if it underwent scattering on
its way (see Fig. \ref{fig:parent-contrib-mother}).
\item If the incident nucleus (or an air particle) is fragmented and a new
nucleus is formed (in the dedicated CORSIKA simulation produced for
this analysis, only $p$, $He$, $C$, $O$ and $Fe$ nuclei are used
as primaries, see Sec. \ref{subsec:Cosmic-Ray-flux-models} and Tab.
\ref{tab:Simulation-settings-CORSIKA}). 
\end{enumerate}
In the second scenario, the new nucleus may produce a particle cascade,
leading to the production of a muon. However, production of such intermediate
nuclei implies that only a fraction of the original primary energy
is available, meaning that the produced muons may have worse chance
to reach the detector. 
\begin{center}
\begin{figure}[H]
\centering{}\subfloat[Contribution of different mother particles. \label{fig:parent-contrib-mother}]{\centering{}\includegraphics[width=8cm]{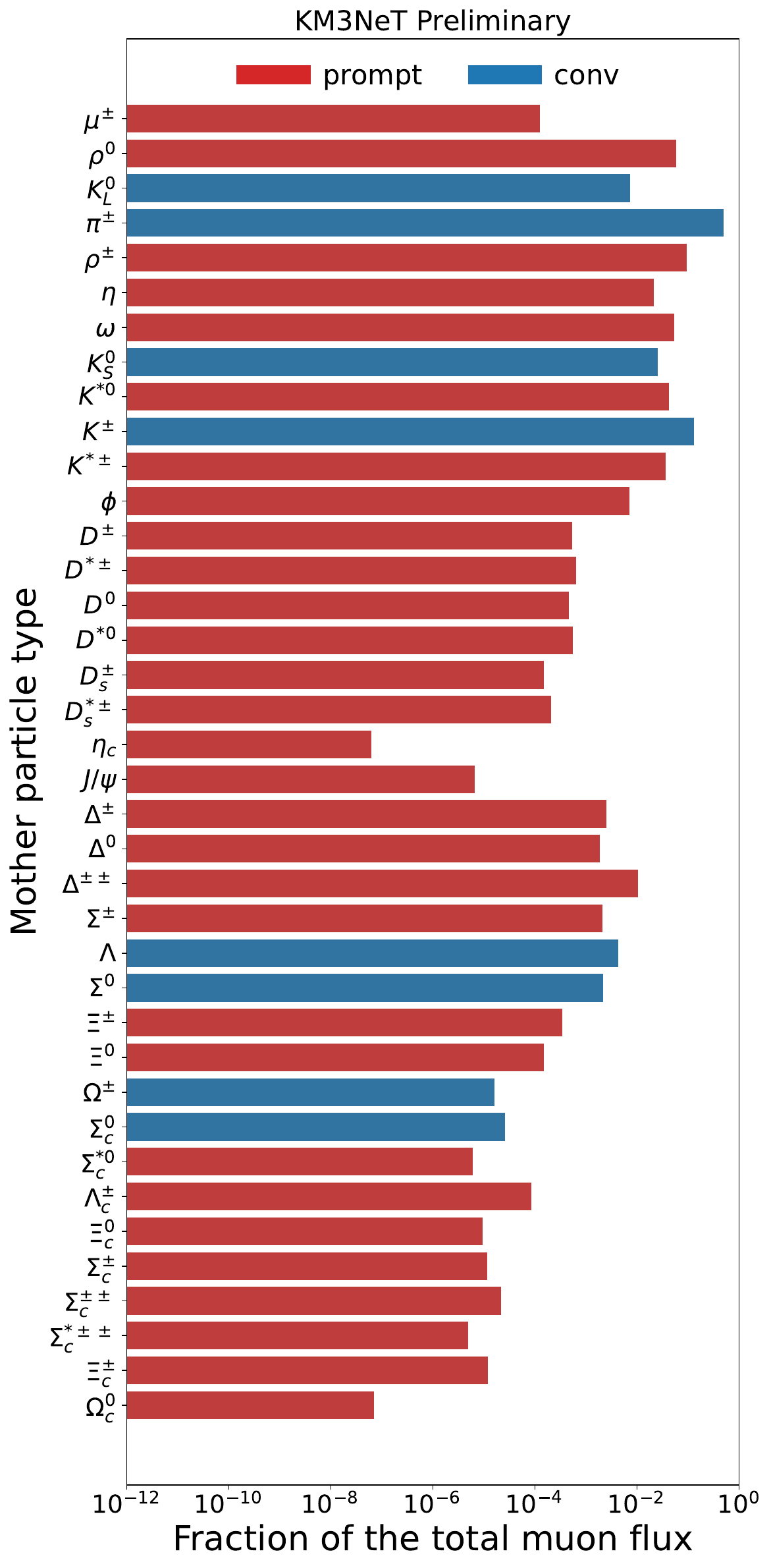}}\subfloat[Contribution of different grandmother particles. \label{fig:parent-contrib-grandmother}]{\centering{}\includegraphics[width=8cm]{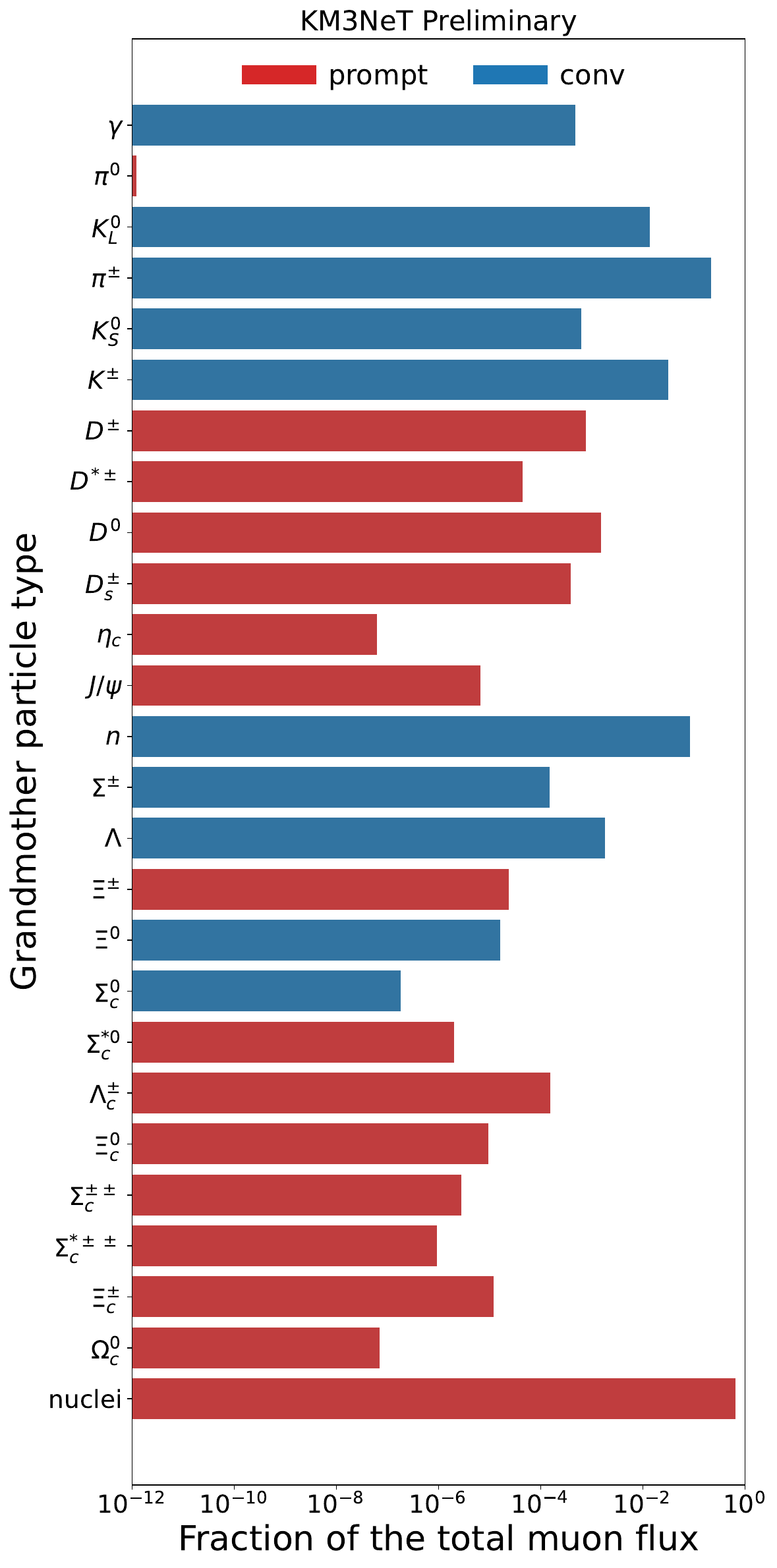}}\caption{Contributions of different parent particles in the CORSIKA simulation
at sea level. The particles and anti-particles were counted together.
The parent particles that may potentially result in a \textcolor{red}{prompt}
muon are coloured red and the ones that will only produce \textcolor{blue}{conventional}
muons are coloured blue. \label{fig:parent-contribution}}
\end{figure}
\par\end{center}

\subsection{Parent particle generations\label{subsec:Parent-particle-generations}}

A muon may have more than just two parent particles, however only
the two most recent ones are stored in the CORSIKA7 output. In the
case of EeV showers, the average number of muon parents is $\sim8$
and can be even $15$ \cite{EAS_genealogy_muon}. Saving all of them
could be highly inefficient, given that a single EeV primary CR may
often produce thousands of muons. Not to mention that the number of
secondaries from EM cascades, simulated by CORSIKA as well is even
larger (by several orders of magnitude). This is the main motivation
for the introduction of \foreignlanguage{english}{interaction} counters
in CORSIKA. From the point of view of this analysis, the relevant
ones are:
\selectlanguage{english}%
\begin{itemize}
\item hadronic interaction counter $h$, encoding the number of hadronic
interactions and/or decays,
\item electromagnetic (EM) interaction counter $e$, summarizing the number
of $e^{\pm}$ interactions (including emission of low-$E$ particles),
\end{itemize}
where the only desired value of the latter is 0 (see Tab. \foreignlanguage{british}{\ref{tab:Prompt-muon-possible-combinations}}
and Sec. \foreignlanguage{british}{\ref{subsec:Cross-check-of-the-prompt-def}}).
\foreignlanguage{british}{The CORSIKA output stores EM and hadronic
interaction counters of the muon ($e_{\mu},$$h_{\mu}$) and its mother
particle ($e_{\mathrm{mother}},h_{\mathrm{mother}}$).} More details
on the CORSIKA counters can be found in \cite{CORSIKA-Userguide}.
The hadronic interaction counter $h$ is typically incremented by
1 after each interaction or decay, however there are exceptions to
this rule \cite{CORSIKA-Userguide,CORSIKA-EHISTORY}. The incrementation
scheme is presented below, with the channels that contribute to the
\foreignlanguage{british}{\textcolor{red}{prompt}} muon flux marked
with ${\color{red}\blacksquare}$:
\begin{itemize}
\item +0:
\begin{itemize}
\item $\Delta\rightarrow p+\pi$ 
\item $\Delta\rightarrow n+\pi$ 
\item $K^{*}\rightarrow K+\pi$ 
\item $\rho^{0}\rightarrow\pi^{+}+\pi^{-}$
\item $\rho^{\pm}\rightarrow\pi^{\pm}+\pi^{0}$ ${\color{red}\blacksquare}$
(through $\pi^{0}$)
\item $\omega\rightarrow\pi^{+}+\pi^{-}\,(+\pi^{0})$ ${\color{red}\blacksquare}$
(through $\pi^{0}$)
\item $\omega\rightarrow\pi^{0}+\gamma$ ${\color{red}\blacksquare}$ (through
$\pi^{0}$)
\item other short-lived resonances ${\color{red}\blacksquare}$
\end{itemize}
\item +1:
\begin{itemize}
\item hadron + air $\rightarrow$ hadron + $X$ ${\color{red}\blacksquare}$
\item $\gamma$ + air $\rightarrow$ hadron + $X$
\item $\gamma$ + air $\rightarrow$ $\rho^{0}$ (+ recoil nucleus)
\item $\gamma$ + air $\rightarrow$ $\omega$ (+ recoil nucleus)
\item $\gamma$ (+ air ) $\rightarrow$ $\mu^{+}\mu^{-}$
\item all hadronic decays except $\pi^{\pm}$:
\begin{itemize}
\item $\pi^{0}$, $\eta$ ${\color{red}\blacksquare}$
\item $K^{\pm}$, $K_{\mathsf{S}}^{0}$, $K_{\mathsf{L}}^{0}$
\item all strange baryons ${\color{red}\blacksquare}$
\item all charmed hadrons ${\color{red}\blacksquare}$
\end{itemize}
\item $\pi^{\pm}$ if its hadronic interaction counter $h_{\pi^{\pm}}\geq49$
\end{itemize}
\item +31: (but clipped at 99)
\begin{itemize}
\item $D$ $\rightarrow$ $\mu$ + $X$ ${\color{red}\blacksquare}$
\item other charmed mesons ${\color{red}\blacksquare}$
\end{itemize}
\item +51:
\begin{itemize}
\item $\pi^{\pm}\rightarrow\mu^{\pm}+\nu_{\mu}$ (if $h_{\pi^{\pm}}<49$,
otherwise +1)
\end{itemize}
\end{itemize}
\newpage{}
\begin{itemize}
\item +500:
\begin{itemize}
\item $\mu$ pair production
\item photonuclear interaction in $\mu$ history
\end{itemize}
\end{itemize}
\selectlanguage{british}%

\subsection{Definition of a prompt muon}

The definition of what would be considered a \textcolor{red}{prompt}
muon in the CORSIKA simulation was constructed with the intent to
retain as much control over the particle history as possible. For
this reason, the muons with more than two parent particles were excluded,
making use of the hadronic and electromagnetic generation counters
(see Sec. \ref{subsec:Parent-particle-generations}). The criteria
for a muon to be considered \textcolor{red}{prompt} are summarized
in Tab. \ref{tab:Prompt-muon-possible-combinations}. Every muon,
which does not meet those, is by construction declared \textcolor{blue}{conventional}.
The muons with \textcolor{red}{prompt} grandmother and muon mother
were rejected, motivated by the fact that in the MC they are produced
only far away from the shower start (see Fig. \ref{fig:d_mu_prod_prompt_mu}).

\begin{table}[H]
\begin{centering}
\caption{Parent particle and generation counter combinations for \textcolor{red}{prompt}
muons. \label{tab:Prompt-muon-possible-combinations}}
\par\end{centering}
\centering{}%
\begin{tabular}{|c|c|c|c|c|}
\hline 
Grandmother & Mother & $h_{\mathrm{mother}}$ & $h_{\mu}$ & EM counters ($e_{\mu}$ and $e_{\mathrm{mother}}$)\tabularnewline
\hline 
\hline 
primary & muon & $1$ & $1$ & \multirow{4}{*}{0}\tabularnewline
\cline{1-4} \cline{2-4} \cline{3-4} \cline{4-4} 
primary & \textcolor{red}{prompt} & $\leq2$ & $\leq2$ & \tabularnewline
\cline{1-4} \cline{2-4} \cline{3-4} \cline{4-4} 
\multirow{2}{*}{\textcolor{red}{prompt}} & \multirow{2}{*}{\textcolor{red}{prompt}} & $2$ & $32$ & \tabularnewline
\cline{3-4} \cline{4-4} 
 &  & $3$ & $33$ & \tabularnewline
\hline 
\end{tabular}
\end{table}

\section{Control plots for the selection \label{sec:Control-plots-for}}

To ensure that the selection of \textcolor{blue}{conventional} and
\textcolor{red}{prompt} muons proposed in Tab. \ref{tab:Prompt-muon-possible-combinations}
is reasonable, a number of sanity checks were performed. They are
discussed in this section.

\subsection{Muon arrival time\label{subsec:Muon-arrival-time}}

\begin{figure}[H]
\begin{centering}
\includegraphics[width=8cm]{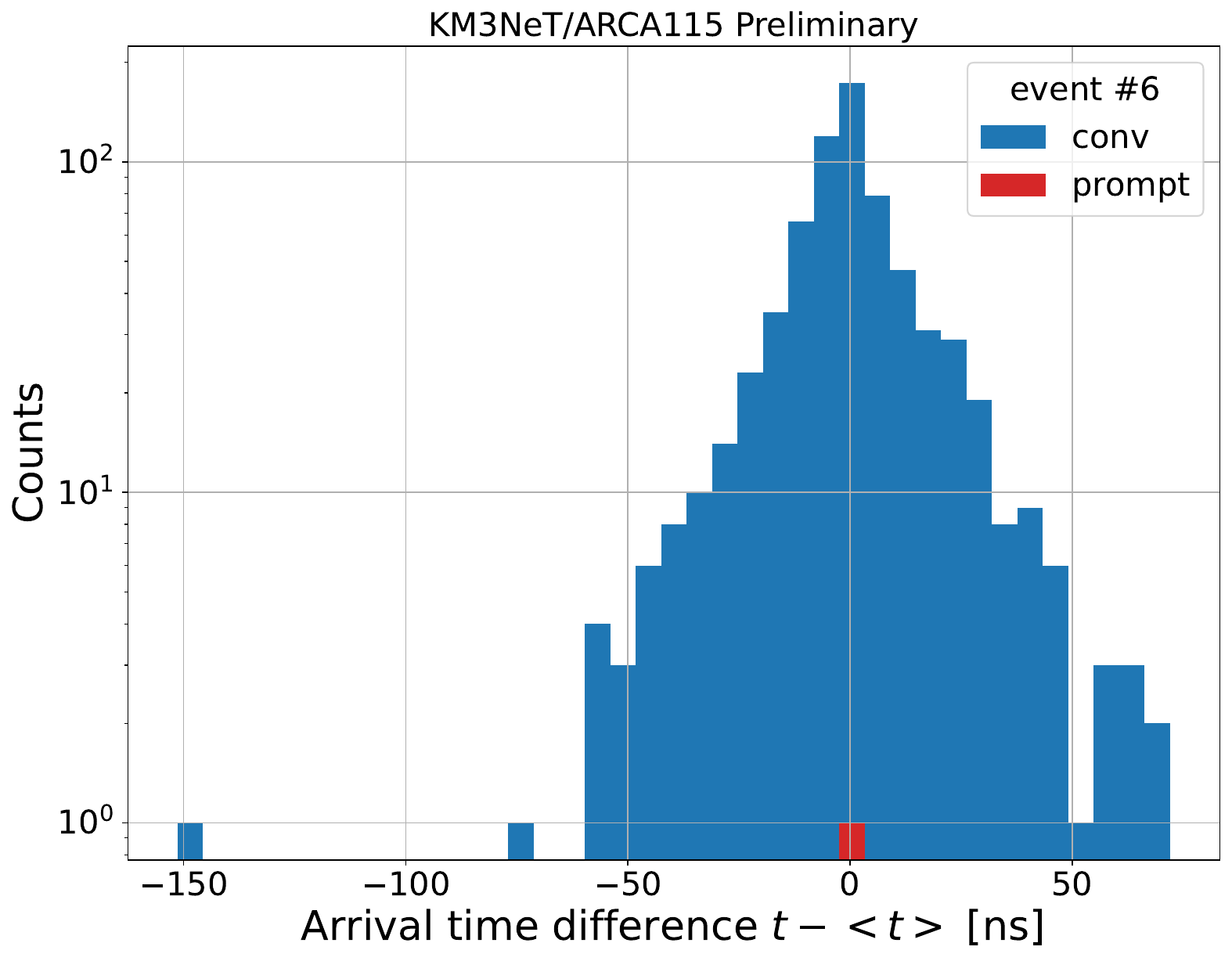}\includegraphics[width=8cm]{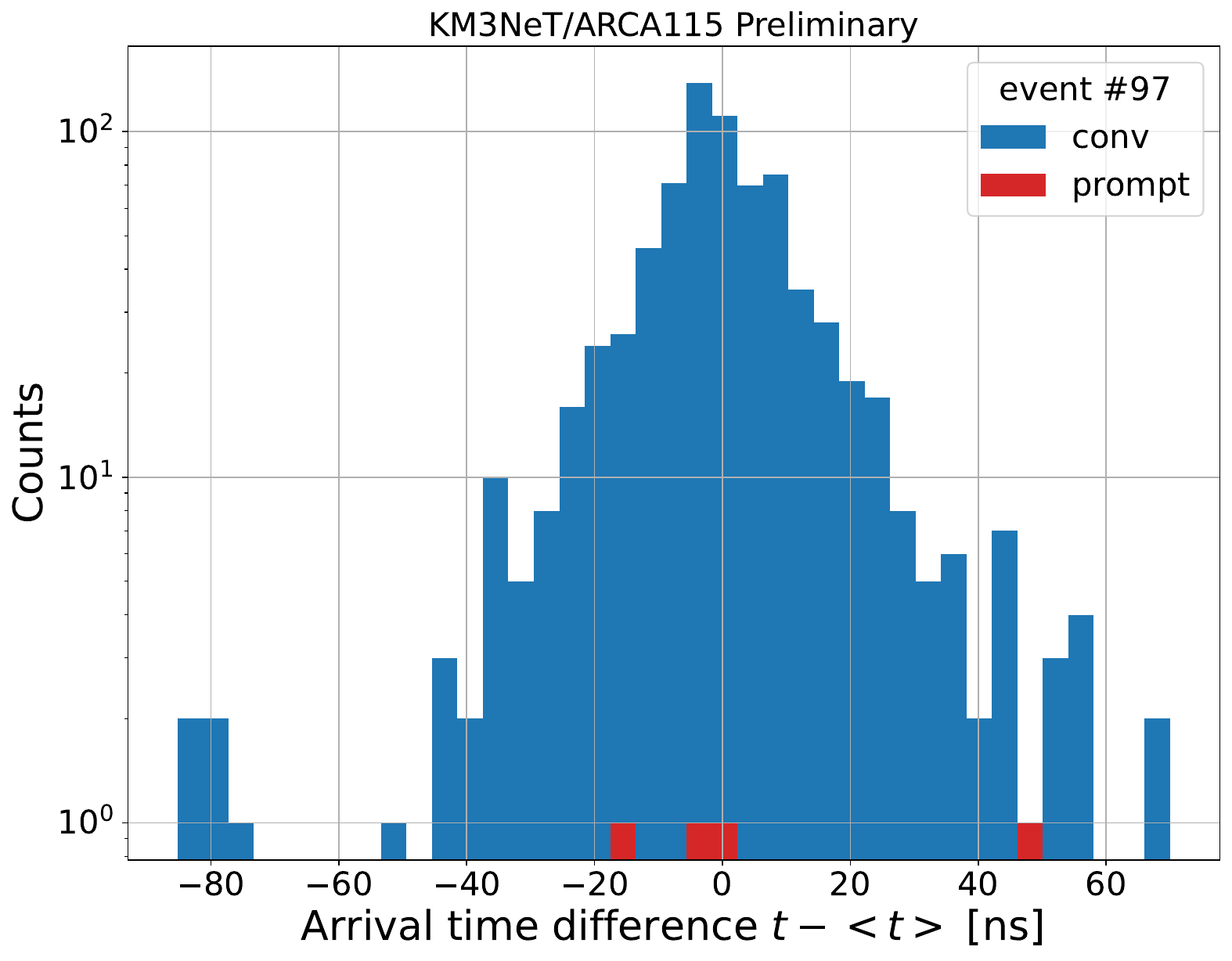}
\par\end{centering}
\begin{centering}
\includegraphics[width=8cm]{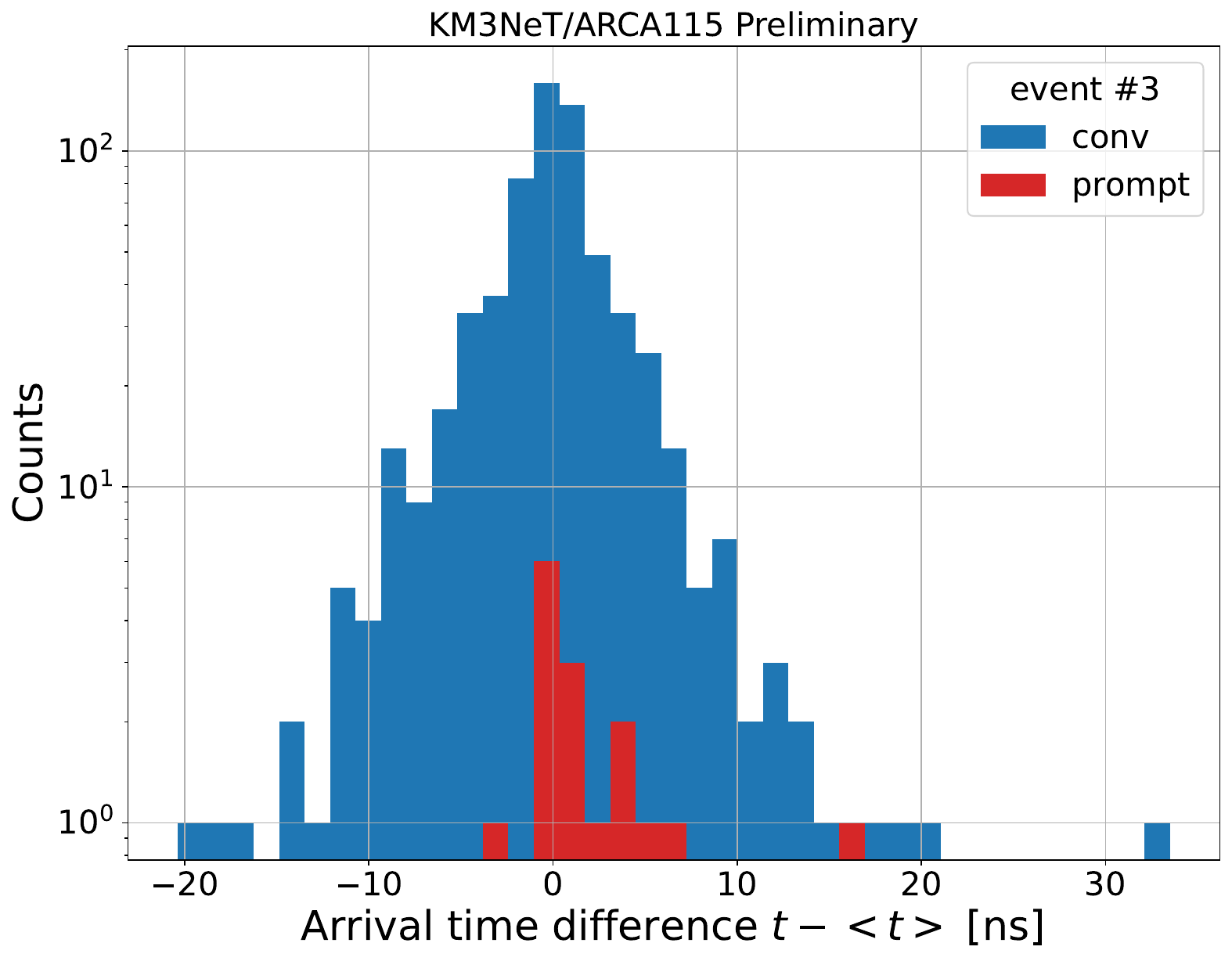}\includegraphics[width=8cm]{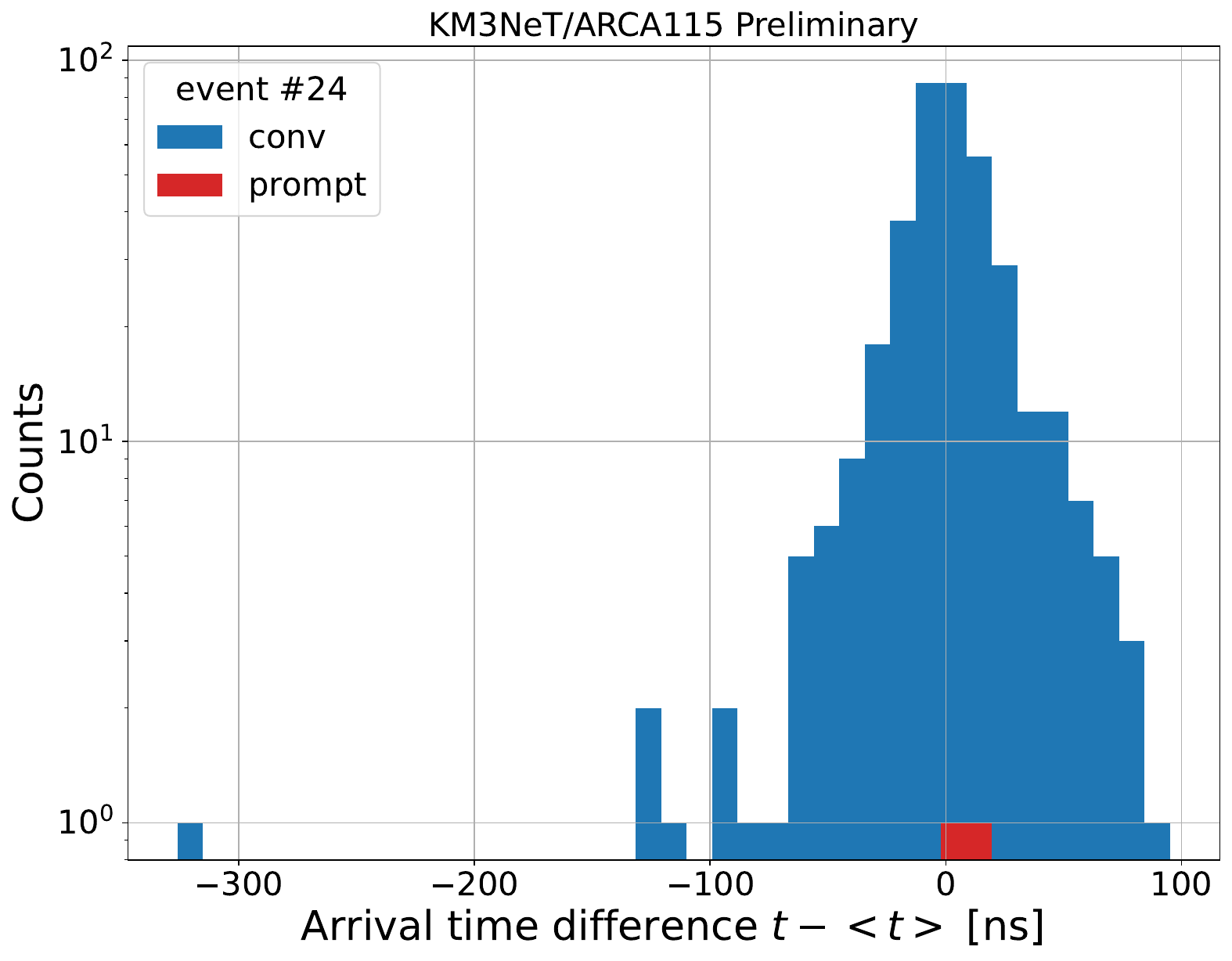}
\par\end{centering}
\centering{}\caption{Distributions of the arrival time difference, where for each muon
the average arrival time is subtracted. The arrival time is defined
as the time since the first interaction of the primary to the arrival
of a muon at the ARCA115 can. In the top row there are two examples
of events started by proton primaries and the showers shown in the
bottom row were caused by iron primaries. Each plot is an unweighted
histogram of a single muon bundle event at the can level. The events
from the EeV sub-production were intentionally picked here, to increase
the chance of observing \textcolor{red}{prompt} muons. \label{fig:prompt-times}}
\end{figure}

The time distribution could potentially reveal a distinct feature
coming from the \textcolor{red}{prompt} contribution. If so, it would
allow for an event-by-event detection of the showers containing \textcolor{red}{prompt}
muons. However, after inspecting a number of events, examples of which
are shown in Fig. \ref{fig:prompt-times}, it became clear that it
is not feasible. \textcolor{red}{Prompt} muons are typically not really
the leading muons in the bundle, understood as being the first to
arrive. They arrive at around the same time as most of the \textcolor{blue}{conventional}
ones do. In some cases there are a few muons arriving earlier than
the rest, but they are \textcolor{blue}{conventional}. This can be
understood by considering the fact that muons, which are produced
early on, but e.g. in decays of $\pi^{\pm}$ or $K^{\pm}$, are still
considered \textcolor{blue}{conventional}. The lifetime of a particle
is only the average time after which the particle is expected to decay.
The low probability of a sooner decay can be compensated by the huge
amount of produced \textcolor{blue}{conventional} muons, exceeding
the numbers of \textcolor{red}{prompt} muons by orders of magnitude.

A cross-check of the range of observed values can be done by estimating
the time it takes a muon, travelling at nearly the speed of light
in vacuum $c$, to reach the detector. The minimal and maximal observed
first interaction heights are $\sim3\,$km and $\sim500\,$km (see
\foreignlanguage{english}{Fig. \ref{fig:1st_interaction_height}})
above the sea level. From the sea surface, there are $\sim2\,$km
or $\sim3\,$km left to the ORCA and ARCA detector respectively. By
dividing those distances by $c$, one arrives at travel time estimates
in the range 0.01-2~ms. The absolute values of the arrival times
used to compute the time differences in Fig. \ref{fig:prompt-times}
were between 0.39 and 2.38~ms, which is in good agreement with the
estimated time.

\subsection{Muon energy share\label{subsec:Muon-energy-share}}

\begin{figure}[H]
\centering{}\includegraphics[width=16cm]{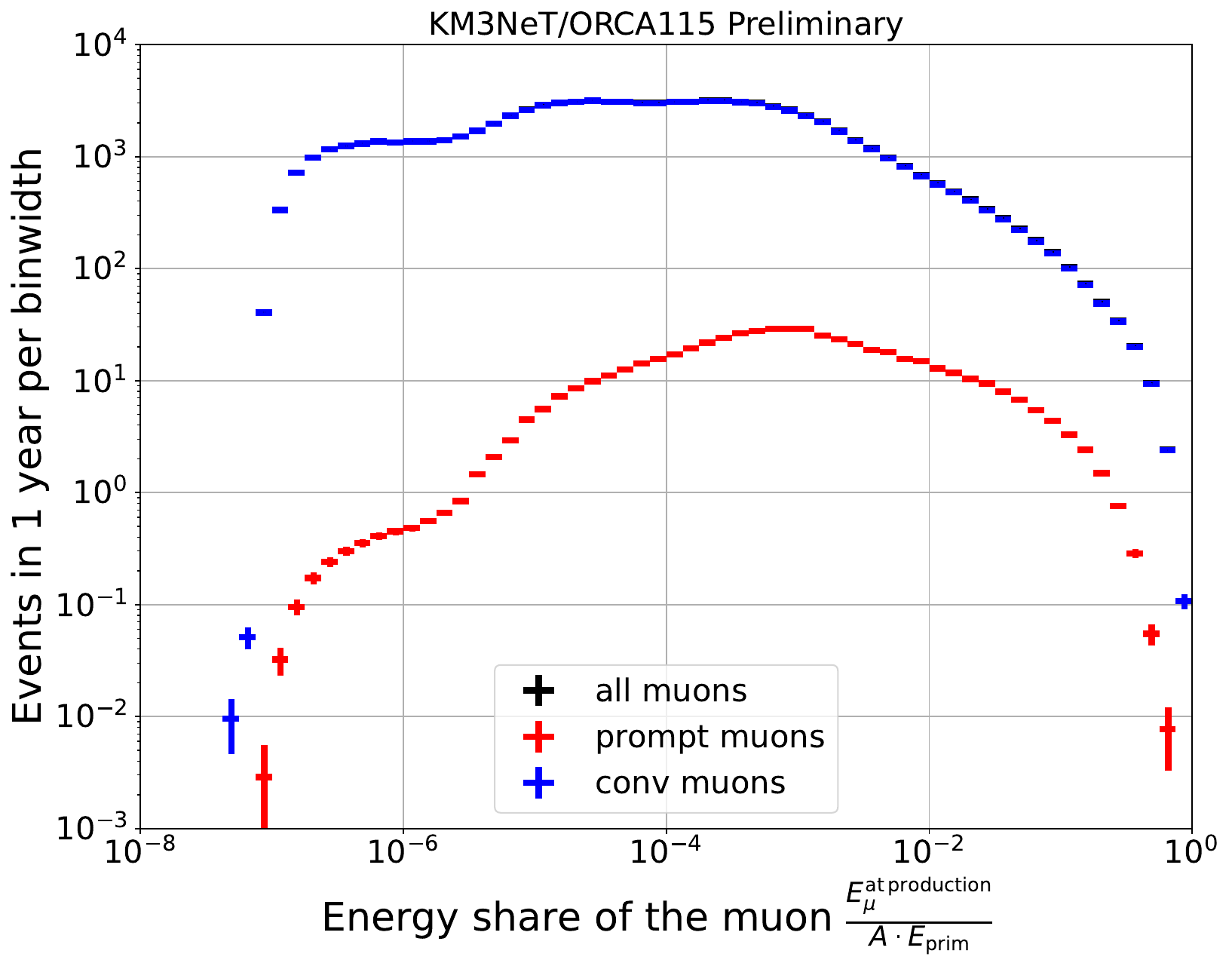}\caption{Distribution of the fraction of the original total primary energy
held by the muon when it was created. $A$ is the atomic mass number
(the number of nucleons) and $E_{\mathrm{prim}}$ is the primary energy
per nucleon. The plot was produced for muons reaching the ORCA115
detector. The error bands were computed using Eq. \ref{eq:hist_error}
and marked with shaded rectangles (mostly very small, but there is
e.g. a larger one on the left edge of the \textcolor{blue}{conv} distribution).
\label{fig:prompt-energy-share}}
\end{figure}

Since \textcolor{red}{prompt} muons are expected to have systematically
higher energies, it means that they should typically take up a bigger
portion of the initial available energy ($E_{\mathsf{prim}}$). This
was tested and the outcome is shown in Fig. \ref{fig:prompt-energy-share}.
As one may see, even though the \textcolor{red}{prompt} curve does
not exceed the \textcolor{blue}{conventional} one, it does indeed
have a flatter slope towards higher energy fractions. The multi-peak
structure of the distribution is a result of superposition of contributions
from different primary nuclei, with the biggest energy share belonging
to hydrogen (proton) primaries.

\subsection{Muon production point}

The average distance travelled by the parent particle before decaying
$\left\langle d\right\rangle $ may be estimated knowing the particle
type and its energy. The basic formula is:

\begin{equation}
\left\langle d\right\rangle =v\cdot\tau,\label{eq:dist}
\end{equation}

where $v$ is the particle's velocity and $\tau$ – its lifetime in
the lab reference frame. Since EAS produce highly energetic particles,
travelling at speeds close to the speed of light in vacuum, the time
dilation effect has to be taken into account:

\begin{equation}
\tau=\gamma\tau_{0},
\end{equation}

where $\gamma=\frac{1}{\sqrt{1-(\frac{v}{c})^{2}}}=\frac{E}{m_{0}c^{2}}$,
$m_{0}$ is the rest mass, and $\tau_{0}$ is the lifetime in the
rest frame. Inserting into Eq. \ref{eq:dist} and rearranging gives
the final formula:

\begin{equation}
\left\langle d\right\rangle =\frac{\sqrt{E^{2}-m_{0}^{2}c^{4}}}{m_{0}c^{2}}\cdot c\tau_{0}.
\end{equation}

In Tab. \ref{tab:Examples-of-average-distances}, a few examples were
calculated to get a feeling for the possible range of values. As one
may note, they are spread over many orders of magnitude.

\begin{table}[H]
\caption{Examples of average distances travelled by 4 different parent particles
before decaying. The energies are roughly the lower and upper bound
for what can be expected. The particles with $E<10^{3}\,$GeV will
seldom produce a $\mu$ that can reach the detector and particles
with $E>10^{8}\,$GeV are produced extremely rarely. \label{tab:Examples-of-average-distances}}

\centering{}%
\begin{tabular}{|c|c|c|c|c|}
\cline{2-5} \cline{3-5} \cline{4-5} \cline{5-5} 
\multicolumn{1}{c|}{} & \multirow{2}{*}{$E$~{[}GeV{]}} & \multirow{2}{*}{$m_{0}\,${[}$\frac{\mathsf{MeV}}{c^{2}}${]}} & \multirow{2}{*}{$\tau_{0}\,${[}s{]}} & \multirow{2}{*}{$\left\langle d\right\rangle $~{[}m{]}}\tabularnewline
\multicolumn{1}{c|}{} &  &  &  & \tabularnewline
\hline 
\multirow{4}{*}{$\pi^{\pm}$} & \multirow{2}{*}{$10^{3}$} & \multirow{4}{*}{139.57} & \multirow{4}{*}{$2.60\cdot10^{-8}$} & \multirow{2}{*}{$5.58\cdot10^{4}$}\tabularnewline
 &  &  &  & \tabularnewline
\cline{2-2} \cline{5-5} 
 & \multirow{2}{*}{$10^{8}$} &  &  & \multirow{2}{*}{$2.15\cdot10^{9}$}\tabularnewline
 &  &  &  & \tabularnewline
\hline 
\multirow{4}{*}{$D_{s}^{+}$} & \multirow{2}{*}{$10^{3}$} & \multirow{4}{*}{1968.34} & \multirow{4}{*}{$5.04\cdot10^{-13}$} & \multirow{2}{*}{$0.08$}\tabularnewline
 &  &  &  & \tabularnewline
\cline{2-2} \cline{5-5} 
 & \multirow{2}{*}{$10^{8}$} &  &  & \multirow{2}{*}{$7.60\cdot10^{3}$}\tabularnewline
 &  &  &  & \tabularnewline
\hline 
\multirow{4}{*}{$\eta_{c}$} & \multirow{2}{*}{$10^{3}$} & \multirow{4}{*}{2983.90} & \multirow{4}{*}{$2.06\cdot10^{-23}$} & \multirow{2}{*}{$2.07\cdot10^{-12}$}\tabularnewline
 &  &  &  & \tabularnewline
\cline{2-2} \cline{5-5} 
 & \multirow{2}{*}{$10^{8}$} &  &  & \multirow{2}{*}{$2.07\cdot10^{-7}$}\tabularnewline
 &  &  &  & \tabularnewline
\hline 
\multirow{4}{*}{$K_{\mathsf{S}}^{0}$} & \multirow{2}{*}{$10^{3}$} & \multirow{4}{*}{497.61} & \multirow{4}{*}{$8.95\cdot10^{-11}$} & \multirow{2}{*}{$53.92$}\tabularnewline
 &  &  &  & \tabularnewline
\cline{2-2} \cline{5-5} 
 & \multirow{2}{*}{$10^{8}$} &  &  & \multirow{2}{*}{$5.39\cdot10^{6}$}\tabularnewline
 &  &  &  & \tabularnewline
\hline 
\end{tabular}
\end{table}

\begin{figure}[H]
\begin{centering}
\includegraphics[width=16cm]{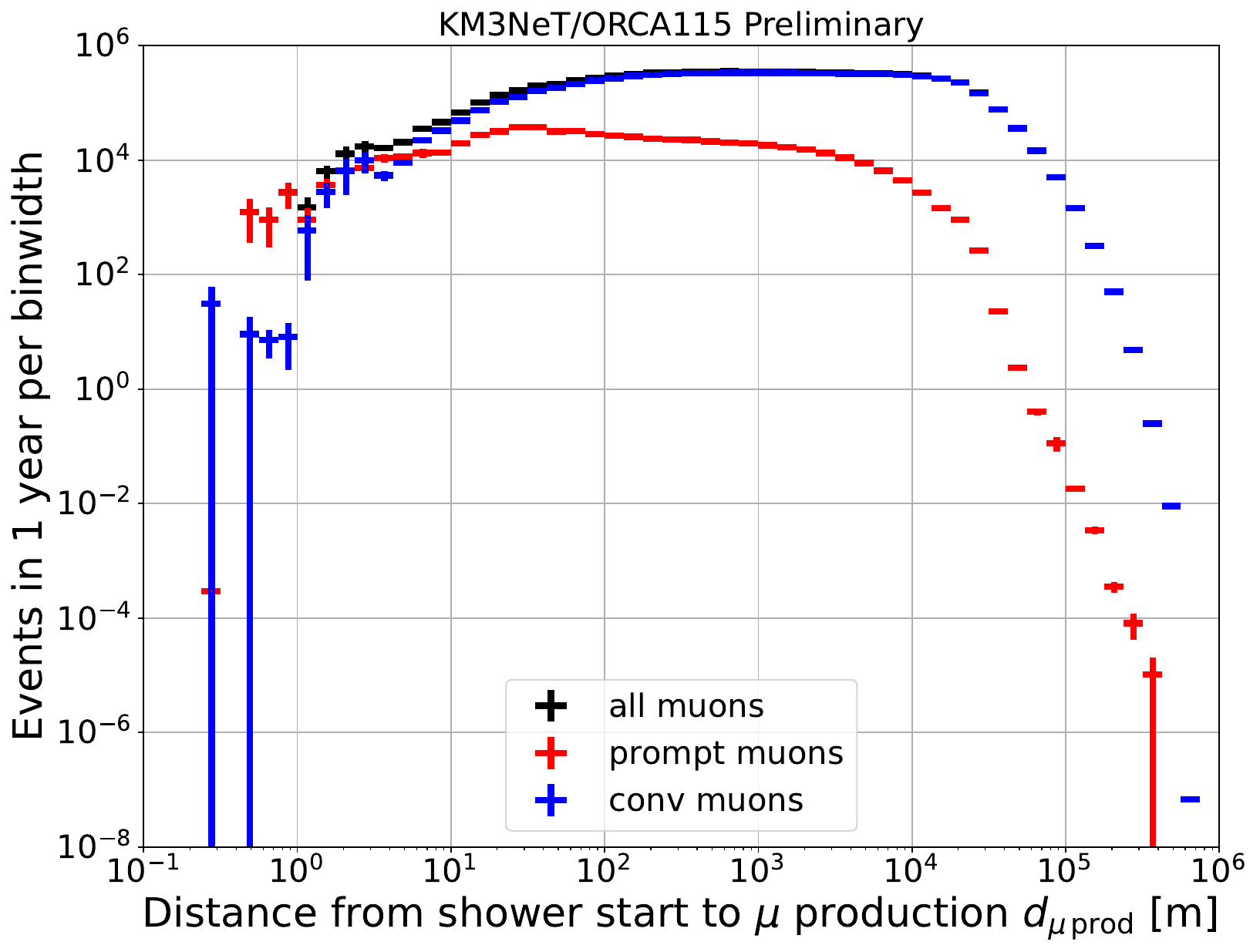}
\par\end{centering}
\centering{}\caption{Distributions of distance from the first interaction point of a primary
to the point at which a muon was created $d_{\mu\,\mathrm{prod}}$.
The plot was produced for muons at sea level, which were part of showers
reaching the ORCA115 detector. This means that muons stopping before
reaching ORCA115 were included. The error bars were computed using
Eq. \ref{eq:hist_error}. \label{fig:d_mu_prod}}
\end{figure}

The distributions of the distance between the first interaction of
the primary and the creation of the muon $d_{\mu\,\mathrm{prod}}$
are depicted in Fig. \ref{fig:d_mu_prod}. Even though the \textcolor{red}{prompt}
muons do not necessarily arrive in the detectors first (see Sec. \ref{subsec:Muon-arrival-time}),
they are less likely to be produced far from the first interaction
of the primary nucleus. This is a consequence of defining the \textcolor{red}{prompt}
muons as the ones created in decays of parent particles with particularly
short lifetimes (see Sec. \ref{sec:Selection-of-conventional-and-prompt}).
The upper limit of $d_{\mu\,\mathrm{prod}}\approx300\,$km is below
the expected maximal distances for $\pi^{\pm}$ and $K_{\mathrm{S}}^{0}$
from Tab. \ref{tab:Examples-of-average-distances}. However, it is
fully consistent both with Fig. \ref{fig:1st_interaction_height}
and with the bounds of the simulation geometry: for maximally horizontal
showers (at 87$\lyxmathsym{\textdegree}$ from the zenith in the CORSIKA
simulation used in this work), the distance could be at most 2155.31~km.
Even if not for this limitation, Tab. \ref{tab:Examples-of-average-distances}
gives only rough estimates, not taking into account energy losses
and deflection of the trajectory during particle propagation.

In Sec. \ref{subsec:Cross-check-of-the-prompt-def}, a complete set
of $d_{\mu\,\mathrm{prod}}$ plots, cross-checking the \textcolor{red}{prompt}
muon definition in Tab. \ref{tab:Prompt-muon-possible-combinations}
may be found.

\section{Definition of signal and background \label{sec:Definition-of-signal-and-background}}

Up to this point, a distinction was made between \textcolor{red}{prompt}
and \textcolor{blue}{conventional} muons. However, as was very clearly
demonstrated in Chap. \ref{chap:muon-bundle-reco} and \ref{chap:Muon-rate-measurement},
the observed events consist only partially of single muons. Therefore,
the signal and background in this analysis must be defined in terms
of entire showers, not individual muons. Consequently, three categories
of events (showers observed at the detectors) were introduced:
\begin{itemize}
\item TOTAL – all showers,
\item \textcolor{red}{SIG} – showers with at least 1 \textcolor{red}{prompt}
muon (not directly used for the hypothesis test),
\item \textcolor{blue}{BGD} – showers with 0 \textcolor{red}{prompt} muons
(100\% \textcolor{blue}{conventional}).
\end{itemize}
Since \textcolor{red}{prompt} and \textcolor{blue}{conventional} components
were simulated together in CORSIKA MC, they share a common phase-space.
When selecting shower for \textcolor{blue}{BGD}, a part of the simulated
muon rate is discarded. To compensate for this, \textcolor{blue}{BGD}
had to be reweighted (for TOTAL it was not necessary). The motivation
for this was that for the null hypothesis $H_{0}$, \textcolor{blue}{BGD}
constitutes the entirety of the muon flux and the MC prediction should
be adapted accordingly. The details of reweighting are discussed in
the following subsection.

\subsection{Reweighting of background}

\begin{figure}[H]
\centering{}\includegraphics[width=12cm]{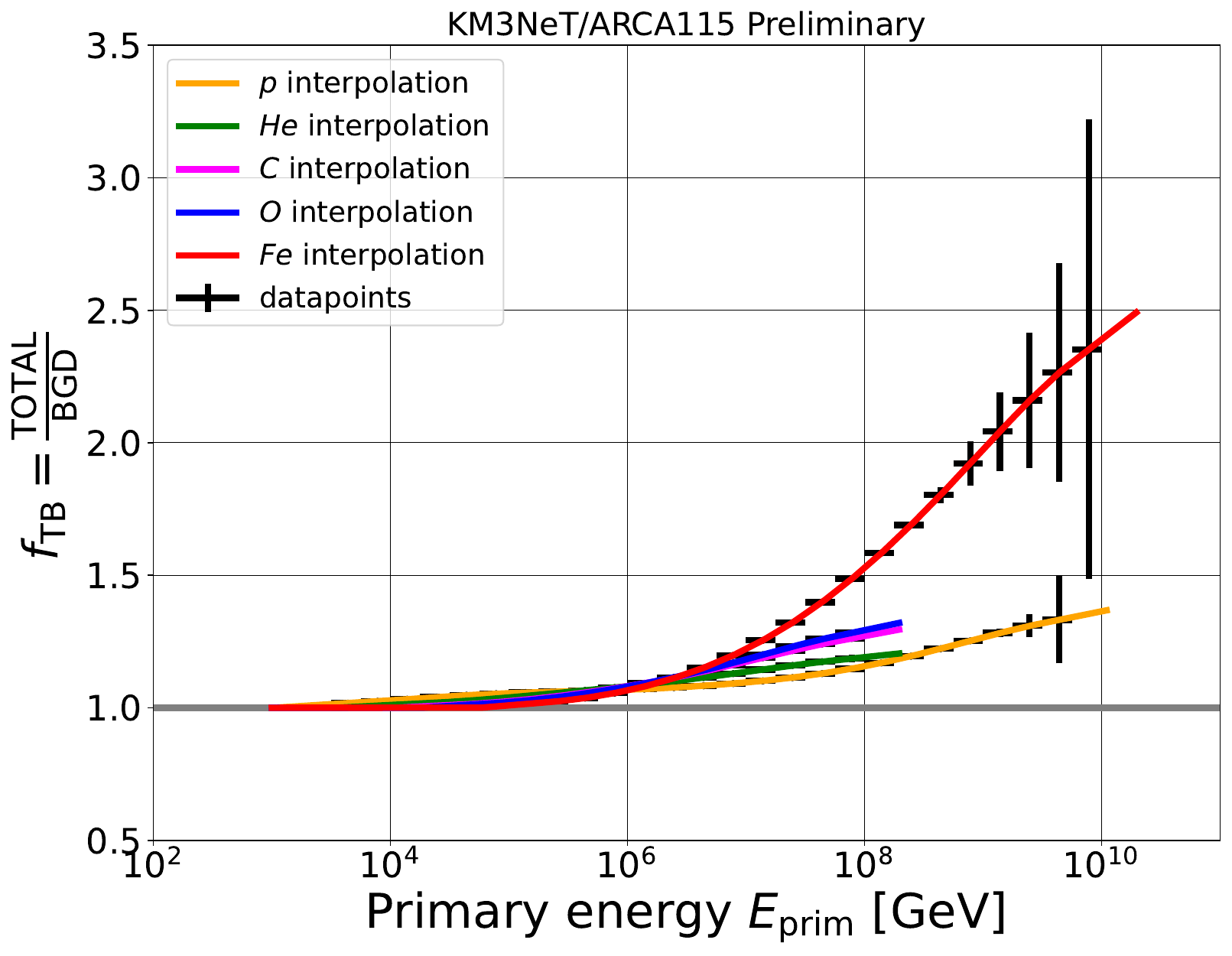}\caption{Interpolated fraction of all generated events to the number of\textbf{\textcolor{blue}{{}
}}\textcolor{blue}{BGD} events as a function of the primary type and
energy. The values were computed for ARCA115 at reconstruction level.
\label{fig:f_TB}}
\end{figure}

The \textcolor{red}{SIG} events were discarded and thus, not all generated
showers were used, which had to be reflected in the weights. In principle,
the goal was to obtain such a distribution of \textcolor{blue}{BGD},
as if it was simulated alone. The number of discarded events (`taken
away' by \textcolor{red}{SIG}) is a function of the primary type
and energy $E_{\mathsf{prim}}$, as are the event weights themselves.
A ratio between the number of events falling into the TOTAL category
divided by the number of \textcolor{blue}{BGD} events $f_{\mathsf{TB}}\left(\mathsf{prim},E_{\mathsf{prim}}\right)$
is plotted in Fig. \ref{fig:f_TB}. The final weight for \textcolor{blue}{BGD}
was computed as:

\begin{equation}
w_{\mathsf{{\color{blue}BGD}}}=w_{\mathrm{event}}\left(\mathrm{prim},E_{\mathrm{prim}}\right)\cdot f_{\mathsf{TB}}\left(\mathrm{prim},E_{\mathrm{prim}}\right),
\end{equation}

where $w_{\mathrm{event}}\left(\mathrm{prim},E_{\mathrm{prim}}\right)$
are the event weights (see Eq. \ref{eq:event_weights}) that would
be used for TOTAL. The $f_{\mathsf{TB}}$ in Fig. \ref{fig:f_TB}
was evaluated from unweighted TOTAL and \textcolor{blue}{BGD} distributions
at reconstruction level under assumption that the ratio does not change
significantly from the generation level (which is not directly accessible).
An ideal solution would be a separate MC simulation without the \textcolor{red}{prompt}
contribution, however it was not possible with CORSIKA v7.7410 for
technical reasons. For every detector configuration, a common $f_{\mathsf{TB}}$
evaluated using the ARCA115 dataset was used, as it had the smallest
uncertainty on the values at high energies. It was verified that the
$f_{\mathsf{TB}}$ interpolations obtained using the other datasets
produce consistent results.

\section{Significance test\label{sec:Significance-test}}

The test of the background-only hypothesis $H_{0}$ (defined in the
beginning of this chapter) performed in this analysis followed the
procedure proposed in \cite{Poisson_significance_Li1983,Poisson_significanceCousins2008,Poisson_significance_Basso,Poisson_significance_Cowan2011}.
It was based on the profile likelihood ratio for Poisson-distributed
measurements. The assumption that the measurements followed Poisson
statistics was justified, since the measured quantity was the number
of observed muon bundles, which is an example of a counting experiment.
The function $q_{0}$, used to maximize the power of the null hypothesis
test of an analysis, is often referred to as the test statistic (TS\nomenclature{TS}{test statistic}).
The TS used in this analysis was:

{\footnotesize{}
\begin{equation}
q_{0}=\left\{ \begin{array}{cc}
2\cdot\left[N_{\mathsf{TOTAL}}\cdot\ln\left(\frac{N_{\mathsf{TOTAL}}\cdot\left(N_{\mathsf{{\color{blue}BGD}}}+\sigma_{\mathsf{{\color{blue}BGD}}}^{2}\right)}{N_{\mathsf{{\color{blue}BGD}}}^{2}+N_{\mathsf{TOTAL}}\cdot\sigma_{\mathsf{{\color{blue}BGD}}}^{2}}\right)-\frac{N_{\mathsf{TOTAL}}^{2}}{\sigma_{\mathsf{{\color{blue}BGD}}}^{2}}\cdot\ln\left(1+\frac{\sigma_{\mathsf{{\color{blue}BGD}}}^{2}\left(N_{\mathsf{TOTAL}}-N_{\mathsf{{\color{blue}BGD}}}\right)}{N_{\mathsf{{\color{blue}BGD}}}\cdot\left(N_{\mathsf{{\color{blue}BGD}}}+\sigma_{\mathsf{{\color{blue}BGD}}}^{2}\right)}\right)\right] & \mathrm{for\,N_{\mathsf{TOTAL}}\geq N_{\mathsf{{\color{blue}BGD}}}}\\
0 & \mathrm{for\,N_{\mathsf{TOTAL}}<N_{{\color{blue}BGD}}}
\end{array}\right.,\label{eq:TestStatistic}
\end{equation}
}{\footnotesize\par}

where $N_{\mathsf{TOTAL}}$ and $N_{\mathsf{{\color{blue}BGD}}}$
are the expected numbers of TOTAL and \textcolor{blue}{BGD} events
respectively and $\sigma_{\mathsf{{\color{blue}BGD}}}$ is the uncertainty
on \textcolor{blue}{BGD}. In the case of a measurement on experimental
data, $N_{\mathsf{TOTAL}}$ is equal to the number of measured events.
In the sensitivity study, the CORSIKA MC was used as the Asimov dataset
\cite{Poisson_significance_Cowan2011}. The value of $\sigma_{\mathsf{{\color{blue}BGD}}}$
was computed in quadrature from the statistical and systematic uncertainty:

\begin{equation}
\sigma_{\mathsf{{\color{blue}BGD}}}=\sqrt{\left(\sigma_{\mathsf{{\color{blue}BGD}}}^{\mathrm{stat}}\right)^{2}+\left(\sigma_{\mathsf{{\color{blue}BGD}}}^{\mathrm{syst}}\right)^{2}}.
\end{equation}

The systematic uncertainty $\sigma_{\mathsf{{\color{blue}BGD}}}^{\mathrm{syst}}$
was evaluated using the method described in Sec. \ref{sec:Systematic-uncertainty-study},
i.e. it included the uncertainty due to different HE hadronic interaction
models, CR flux models, seasonal variations of the atmosphere, PMT
efficiency, and light absorption in seawater. Since the histograms
were weighted, the error $\sigma_{\mathsf{{\color{blue}BGD}}}^{\mathrm{stat}}$
is \uline{not} a square root of $N_{\mathsf{{\color{blue}BGD}}}$
(see Sec. \ref{sec:Derivation-of-the-errors-weighted-histo}). The
derivation of Eq. \ref{eq:TestStatistic} may be found in \cite{Poisson_significance_Li1983,Poisson_significanceCousins2008}.
Following those references, the median significance may be found by
using the asymptotic approximation (Wilks’ theorem \cite{WilksTheorem}):

\begin{equation}
Z=\sqrt{q_{0}}.\label{eq:Significance}
\end{equation}

\subsection{Scaling with time and prompt normalisation}

The TS was always computed for a certain time period and \textcolor{red}{prompt}
muon flux normalisation $\phi_{\mathsf{{\color{red}prompt\,\text{\ensuremath{\mu}}}}}$.
The event rates themselves were treated as constant in time (averaged
over seasonal variations), which was a necessary approximation when
using very computationally intensive CORSIKA simulations. The yearly
variations were neglected (see Sec. \ref{subsec:Fit-of-the-atmosphere}).
These simplifying assumptions allowed to scale $q_{0}$ obtained from
the MC. Based on Sec. \ref{sec:Derivation-of-the-errors-weighted-histo},
the values of each bin and their errors could be scaled with time
$t$ as follows:

\[
\begin{array}{c}
N\left(t\right)=\left(\frac{t}{t_{\mathrm{nom}}}\right)\cdot\underset{i}{\sum}w_{i}=\left(\frac{t}{t_{\mathrm{nom}}}\right)\cdot N\left(t_{\mathrm{nom}}\right)\\
\Delta N\left(t\right)=\sqrt{\left(\frac{t}{t_{\mathrm{nom}}}\right)\underset{i}{\sum}w_{i}^{2}}=\sqrt{\frac{t}{t_{\mathrm{nom}}}}\cdot\Delta N\left(t_{\mathrm{nom}}\right)
\end{array},
\]

where $t_{\mathrm{nom}}=1\,$year is the nominal time. This, after
inserting into Eq. \ref{eq:TestStatistic}, leads to: 
\begin{equation}
q_{0}\left(t\right)=\left(\frac{t}{t_{\mathrm{nom}}}\right)\cdot q_{0}\left(t_{\mathrm{nom}}\right).\label{eq:time_scaling_of_TS}
\end{equation}

Scaling by the \textcolor{red}{prompt} muon flux normalisation is
effectively a scaling of \textcolor{red}{SIG} and thus only affects
TOTAL, not \textcolor{blue}{BGD}:

\[
N_{\mathsf{{\color{red}SIG}}}\left(\phi_{\mathsf{{\color{red}prompt\,\text{\ensuremath{\mu}}}}}\right)=\left(\frac{\phi_{\mathsf{{\color{red}prompt\,\text{\ensuremath{\mu}}}}}}{\phi_{\mathsf{{\color{red}prompt\,\mu}}}^{\mathrm{nom}}}\right)\cdot\underset{i}{\sum}w_{i}=\left(\frac{\phi_{\mathsf{{\color{red}prompt\,\text{\ensuremath{\mu}}}}}}{\phi_{\mathsf{{\color{red}prompt\,\mu}}}^{\mathrm{nom}}}\right)\cdot N_{\mathsf{{\color{red}SIG}}}\left(\phi_{\mathsf{{\color{red}prompt\,\mu}}}^{\mathrm{nom}}\right),
\]

where $\phi_{\mathsf{{\color{red}prompt\,\mu}}}^{\mathrm{nom}}$is
the nominal \textcolor{red}{prompt} muon flux normalisation and is
equal to the flux predicted by the CORSIKA MC. This time, Eq. \ref{eq:TestStatistic}
did not simplify to a compact formula as for the time scaling. One
can see it by substituting $N_{\mathsf{TOTAL}}=N_{\mathsf{{\color{red}SIG}}}+N_{\mathsf{{\color{blue}BGD}}}$
in the case of $N_{\mathsf{TOTAL}}\geq N_{\mathsf{{\color{blue}BGD}}}$:

{\footnotesize{}
\[
q_{0}=2\cdot\left[(N_{\mathsf{{\color{red}SIG}}}+N_{\mathsf{{\color{blue}BGD}}})\cdot\ln\left(\frac{(N_{\mathsf{{\color{red}SIG}}}+N_{\mathsf{{\color{blue}BGD}}})\cdot\left(N_{\mathsf{{\color{blue}BGD}}}+\sigma_{\mathsf{{\color{blue}BGD}}}^{2}\right)}{N_{\mathsf{{\color{blue}BGD}}}^{2}+(N_{\mathsf{{\color{red}SIG}}}+N_{\mathsf{{\color{blue}BGD}}})\cdot\sigma_{\mathsf{{\color{blue}BGD}}}^{2}}\right)-\frac{(N_{\mathsf{{\color{red}SIG}}}+N_{\mathsf{{\color{blue}BGD}}})^{2}}{\sigma_{\mathsf{{\color{blue}BGD}}}^{2}}\cdot\ln\left(1+\frac{\sigma_{\mathsf{{\color{blue}BGD}}}^{2}\cdot N_{\mathsf{{\color{red}SIG}}}}{N_{\mathsf{{\color{blue}BGD}}}\cdot\left(N_{\mathsf{{\color{blue}BGD}}}+\sigma_{\mathsf{{\color{blue}BGD}}}^{2}\right)}\right)\right].
\]
}{\footnotesize\par}

There is no possibility for the flux scaling factors to cancel out,
since only $N_{\mathsf{{\color{red}SIG}}}$ undergoes scaling. Nevertheless,
this posed merely an inconvenience, not a real technical problem for
the computation.

\section{Selection of the critical region \label{sec:Phase-space-in-reconstructed}}

The critical region is a part of the considered phase-space, in which
the sensitivity of the hypothesis test is the greatest. This section
demonstrates the procedure for choosing the optimal critical region.

To perform the same selection on the MC simulation and on experimental
data, it was necessary to restrict to the observables reconstructed
in the detector. The considered phase-space was 3-dimensional: zenith,
energy, and multiplicity. The distributions of each of the variables,
together with corresponding TS values for each bin are plotted in
Fig. \ref{fig:Distributions-of-reconstructed-zenith}, \ref{fig:Distributions-of-reconstructed-energy},
and \ref{fig:Distributions-of-reconstructed-multiplicity}. In those
figures, only events meeting the quality criteria outlined in Sec.
\ref{sec:Event-quality-selection} were included. Following the convention
from Chap. \ref{chap:muon-bundle-reco}, only the plots for ARCA115
are shown. The zenith distribution has been rejected as not particularly
useful, due to negligible difference between TOTAL and \textcolor{blue}{BGD},
resulting TS values consistent with zero in every bin (see bottom
panel of Fig. \ref{fig:Distributions-of-reconstructed-zenith}).
\begin{center}
\begin{figure}[H]
\centering{}\includegraphics[width=10cm]{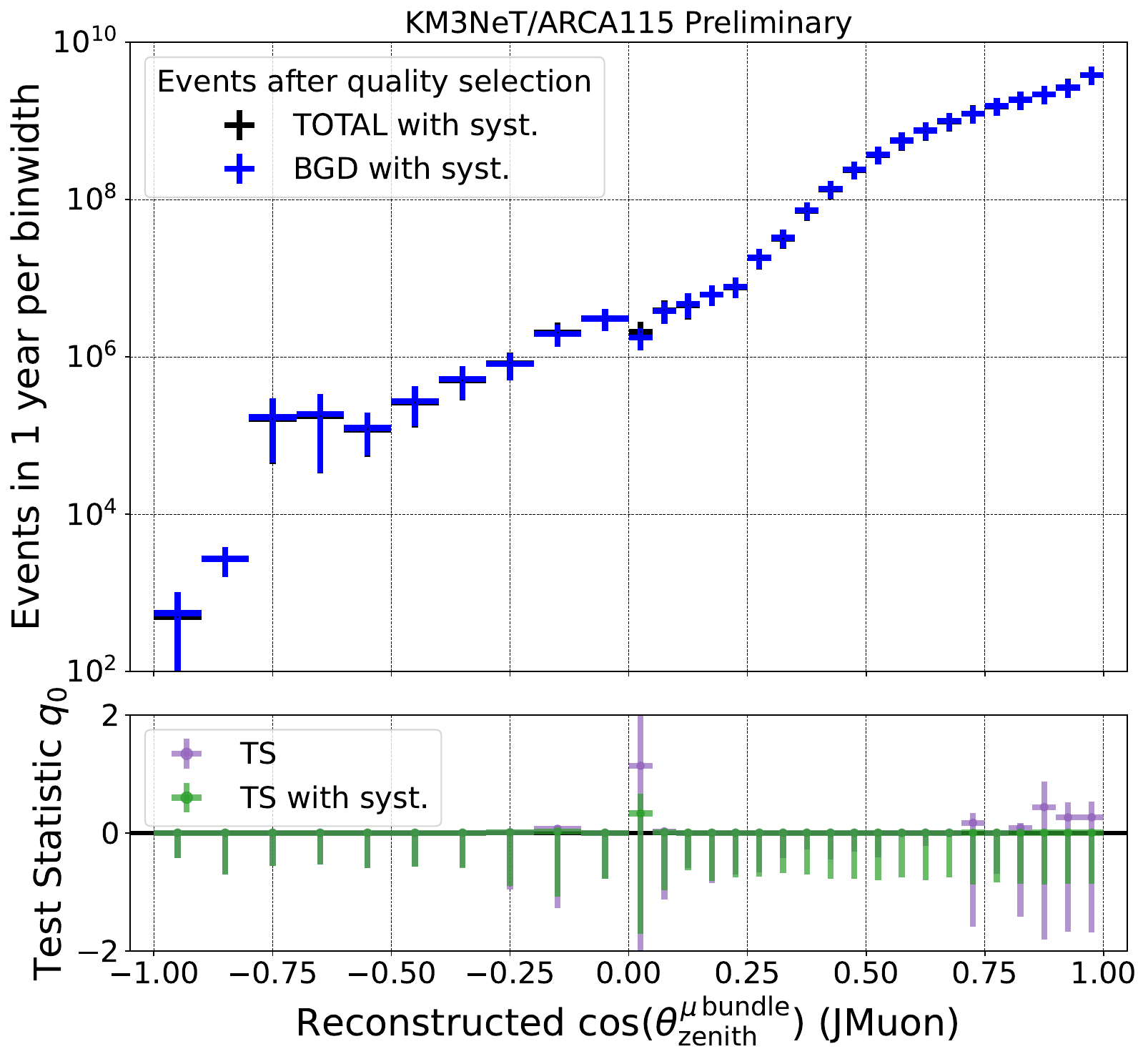}\caption{Distributions of reconstructed zenith for ARCA115 events passing the
quality selection from Sec. \ref{sec:Event-quality-selection}. The
TS values were computed according to Eq. \ref{eq:TestStatistic} for
each bin. The histograms and TS values are shown computed with and
without the systematic uncertainties. \label{fig:Distributions-of-reconstructed-zenith}}
\end{figure}
\par\end{center}

\begin{center}
\begin{figure}[H]
\centering{}\includegraphics[width=10cm]{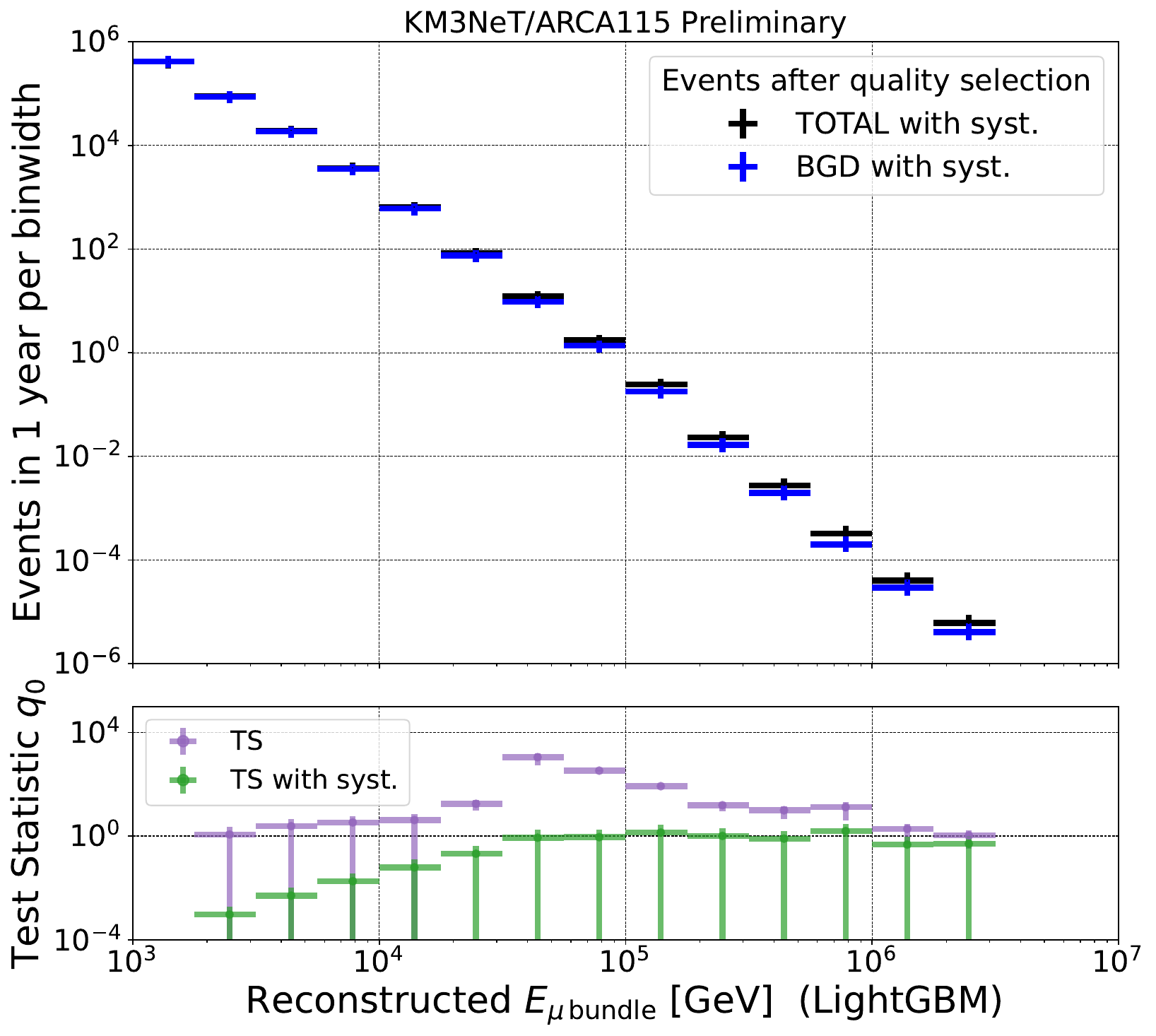}\caption{Distributions of reconstructed muon bundle energy for ARCA115 events
passing the quality selection from Sec. \ref{sec:Event-quality-selection}.
The TS values were computed according to Eq. \ref{eq:TestStatistic}
for each bin. The histograms and TS values are shown computed with
and without the systematic uncertainties. \label{fig:Distributions-of-reconstructed-energy}}
\end{figure}
\par\end{center}

\begin{center}
\begin{figure}[H]
\centering{}\includegraphics[width=10cm]{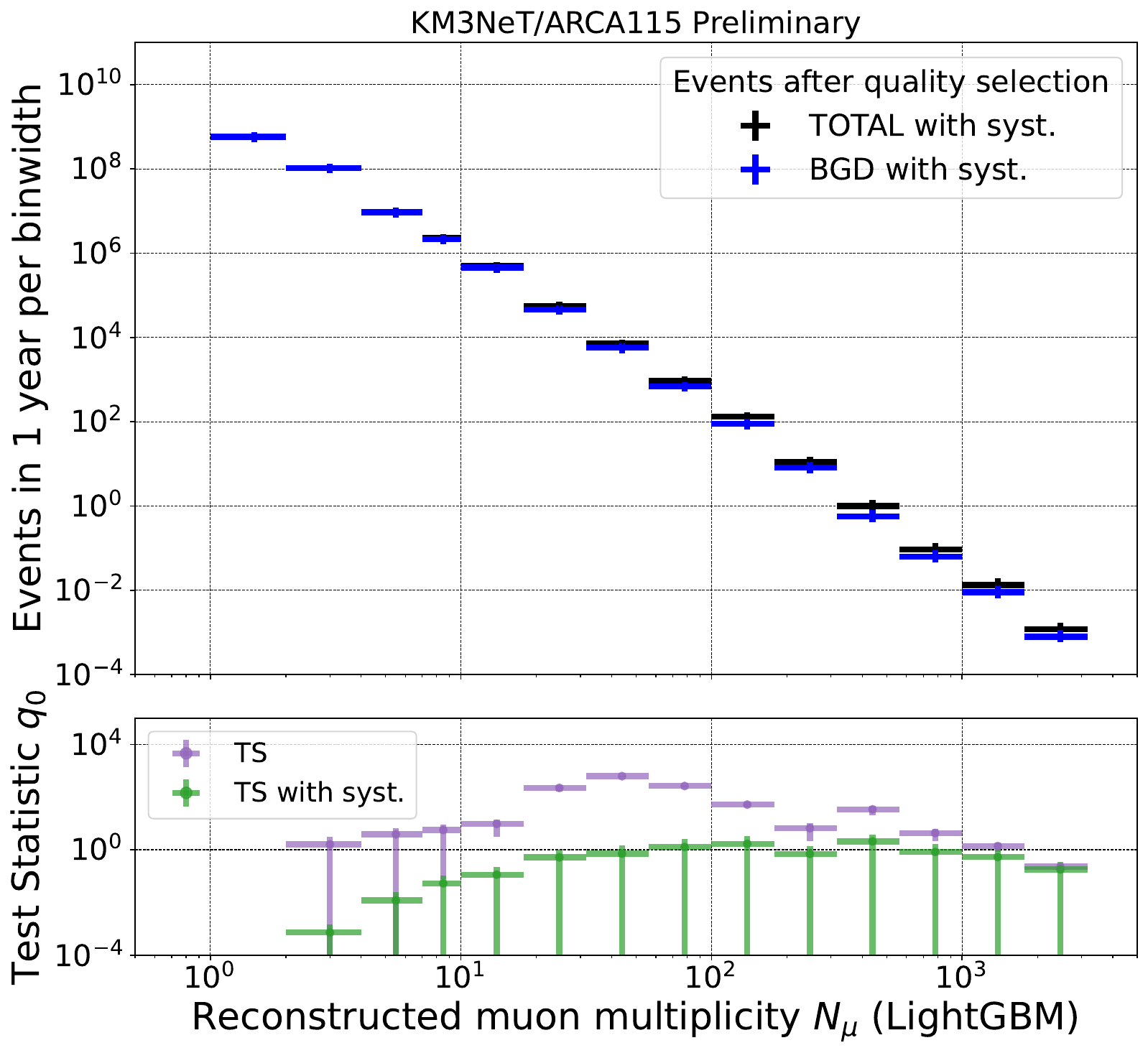}\caption{Distributions of reconstructed multiplicity for ARCA115 events passing
the quality selection from Sec. \ref{sec:Event-quality-selection}.
The TS values were computed according to Eq. \ref{eq:TestStatistic}
for each bin. The histograms and TS values are shown computed with
and without the systematic uncertainties. \label{fig:Distributions-of-reconstructed-multiplicity}}
\end{figure}
\par\end{center}

\begin{figure}[H]
\centering{}\subfloat[Scan for ARCA115.]{\centering{}\includegraphics[width=8cm]{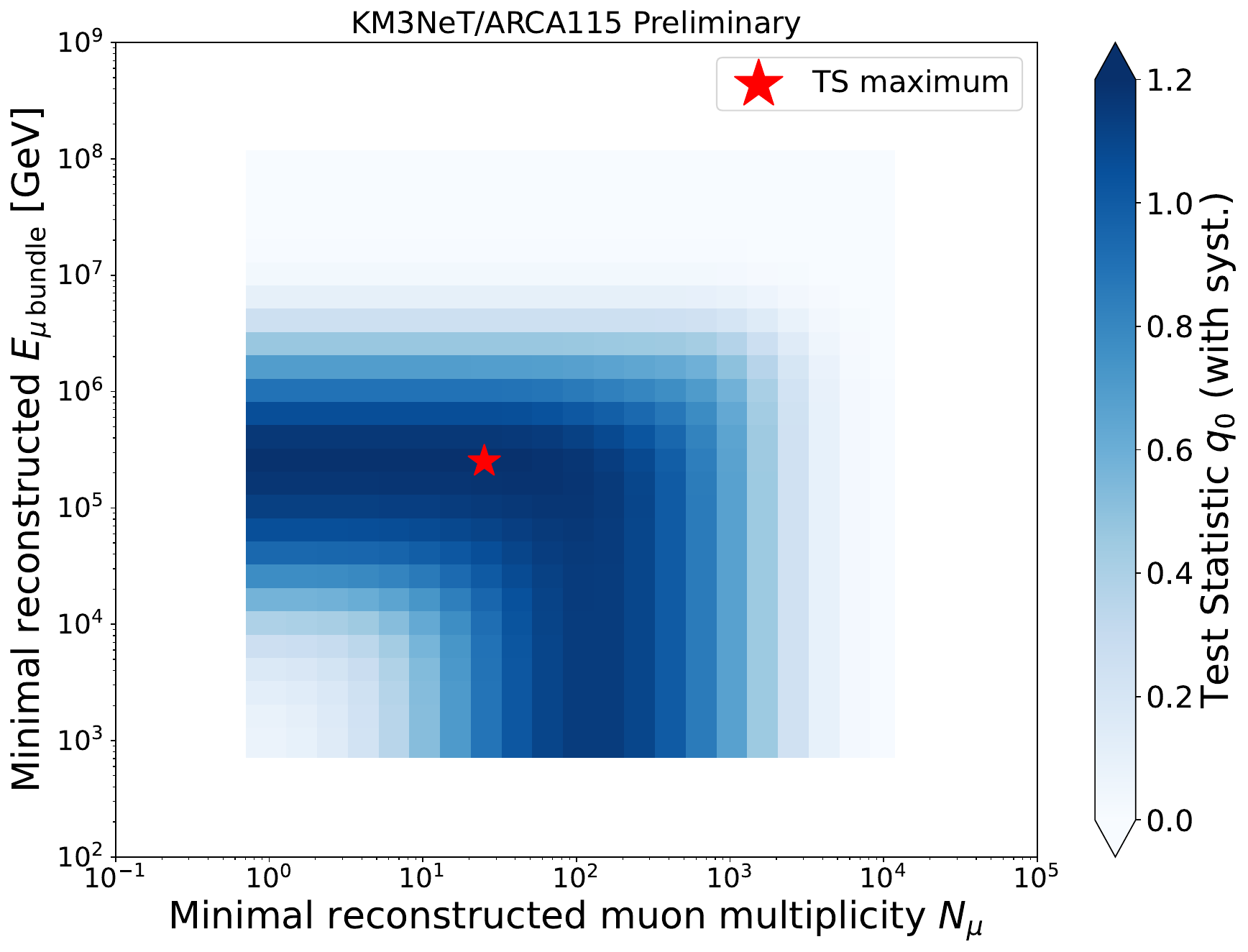}}\subfloat[Scan for ORCA115.]{\centering{}\includegraphics[width=8cm]{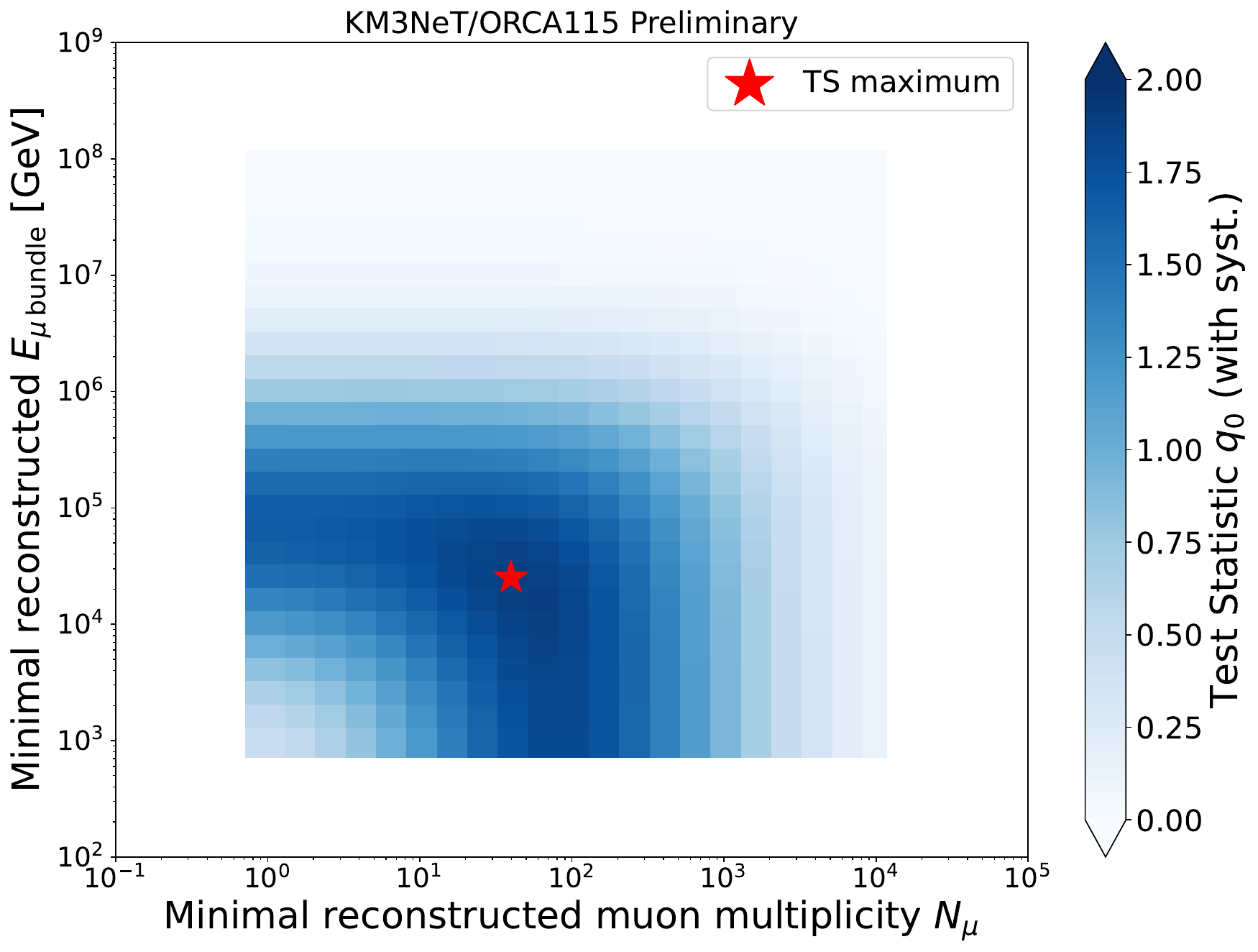}}\caption{Scans over minimal energy and multiplicity, identifying the optimal
selection of the critical region. The significance computed with the
systematic uncertainties was maximized and the maximum is denoted
with the red star. The obtained data was smoothed by convolving it
with a Gaussian kernel to minimize the impact of statistical fluctuations
on the selection of the critical region. \label{fig:crit_region_optimisation-2D}}
\end{figure}

The plots in Fig. \ref{fig:Distributions-of-reconstructed-zenith},
\ref{fig:Distributions-of-reconstructed-energy}, and \ref{fig:Distributions-of-reconstructed-multiplicity}
already roughly indicate, what could be a reasonable critical region.
However, a more direct optimization approach has been adopted to determine
its boundary. A range of minimum reconstructed bundle energy $E_{\mathrm{bundle}}^{\mathrm{min}}$
and minimum reconstructed muon multiplicity $N_{\mu}^{\mathrm{min}}$
values has been scanned, investigating the impact on the resulting
significance. Such scans for ARCA115 and ORCA115 are shown in Fig.
\ref{fig:crit_region_optimisation-2D}, and for the remaining detector
configurations (ARCA6 and ORCA6 respectively), the same values were
used, as summarized in Tab. \ref{tab:Bounds-of-the-critical-region}.
Applying identical critical regions, regardless of the detector configuration,
aimed at ensuring the comparability of results. No upper bounds on
energy and multiplicity were set, as it was found that they do not
boost the power of the hypothesis test. The difference between KM3NeT/ARCA
and KM3NeT/ORCA in Tab. \ref{tab:Bounds-of-the-critical-region} is
compliant with the expectations coming from the detector geometries.
ARCA as the larger detector, dedicated to study of higher energies,
requires a higher value of $E_{\mathrm{bundle}}^{\mathrm{min}}$ ,
whereas the more densely instrumented ORCA is sensitive to lower energies.
ORCA is located shallower than ARCA and hence observes events with
systematically larger multiplicities (see Fig. \ref{fig:Nmu_reco_results-1D}),
which is reflected in the larger value of $N_{\mu}^{\mathrm{min}}$.

\begin{table}[H]
\begin{centering}
\caption{Lower bounds of the critical regions for each of the considered detector
configurations. \label{tab:Bounds-of-the-critical-region}}
\par\end{centering}
\centering{}%
\begin{tabular}{|c|c|c|}
\hline 
Detector & $E_{\mathrm{bundle}}^{\mathrm{min}}$ & $N_{\mu}^{\mathrm{min}}$\tabularnewline
\hline 
\hline 
ARCA115 & $251$~TeV & $25$\tabularnewline
\hline 
ARCA6 & $251$~TeV & $25$\tabularnewline
\hline 
ORCA115 & $25$~TeV & $40$\tabularnewline
\hline 
ORCA6 & $25$~TeV & $40$\tabularnewline
\hline 
\end{tabular}
\end{table}

A potential refinement of the analysis presented in this chapter could
be a non-rectangular critical region, introduced either by specifying
a two-dimensional function of $E_{\mathrm{bundle}}$ and $N_{\mu}$
or by computing the TS on a grid of very fine bins. Both approaches
could lead to a boost of the power of the test, however they bear
a risk of producing over-tuned results, especially with overly dense
binning. For this reason, a robust single-bin method was used here.
A further alternative would be direct classification of events as
\textcolor{red}{SIG} and \textcolor{blue}{BGD}. This has been attempted,
however the developed classifier strongly underestimated \textcolor{blue}{BGD},
which could compromise the entire analysis. For this reason, such
an approach was disqualified.

\section{Sensitivity results\label{sec:Sensitivity-results}}

\begin{figure}[H]
\centering{}\subfloat[Sensitivity as function of the data taking time. \label{fig:sensitivity_full_vs_time}]{\centering{}\includegraphics[width=8cm]{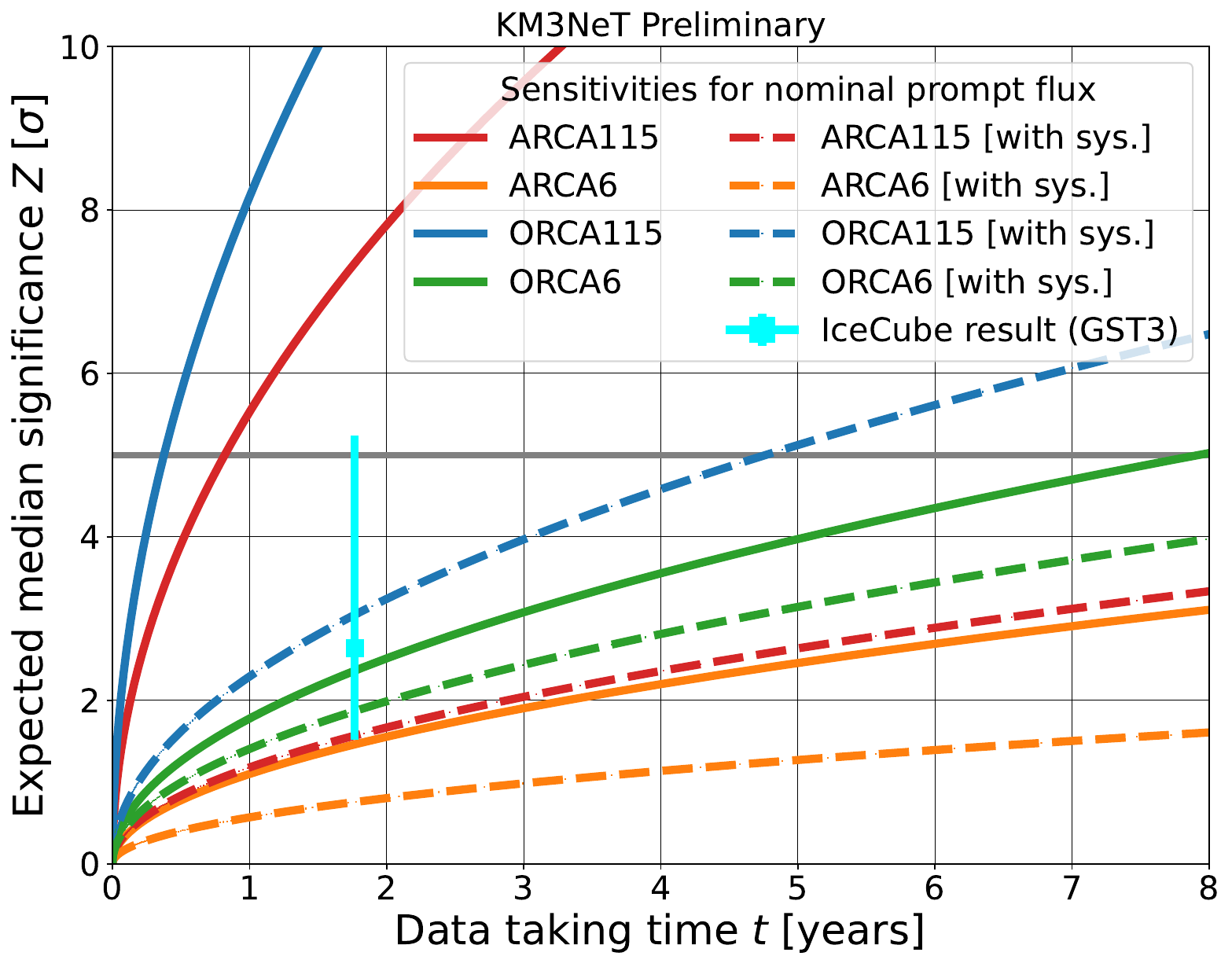}}\subfloat[Sensitivity as function of the \textcolor{red}{prompt} muon bundle
flux normalisation. \label{fig:sensitivity_full_vs_prompt_norma}]{\centering{}\includegraphics[width=8cm]{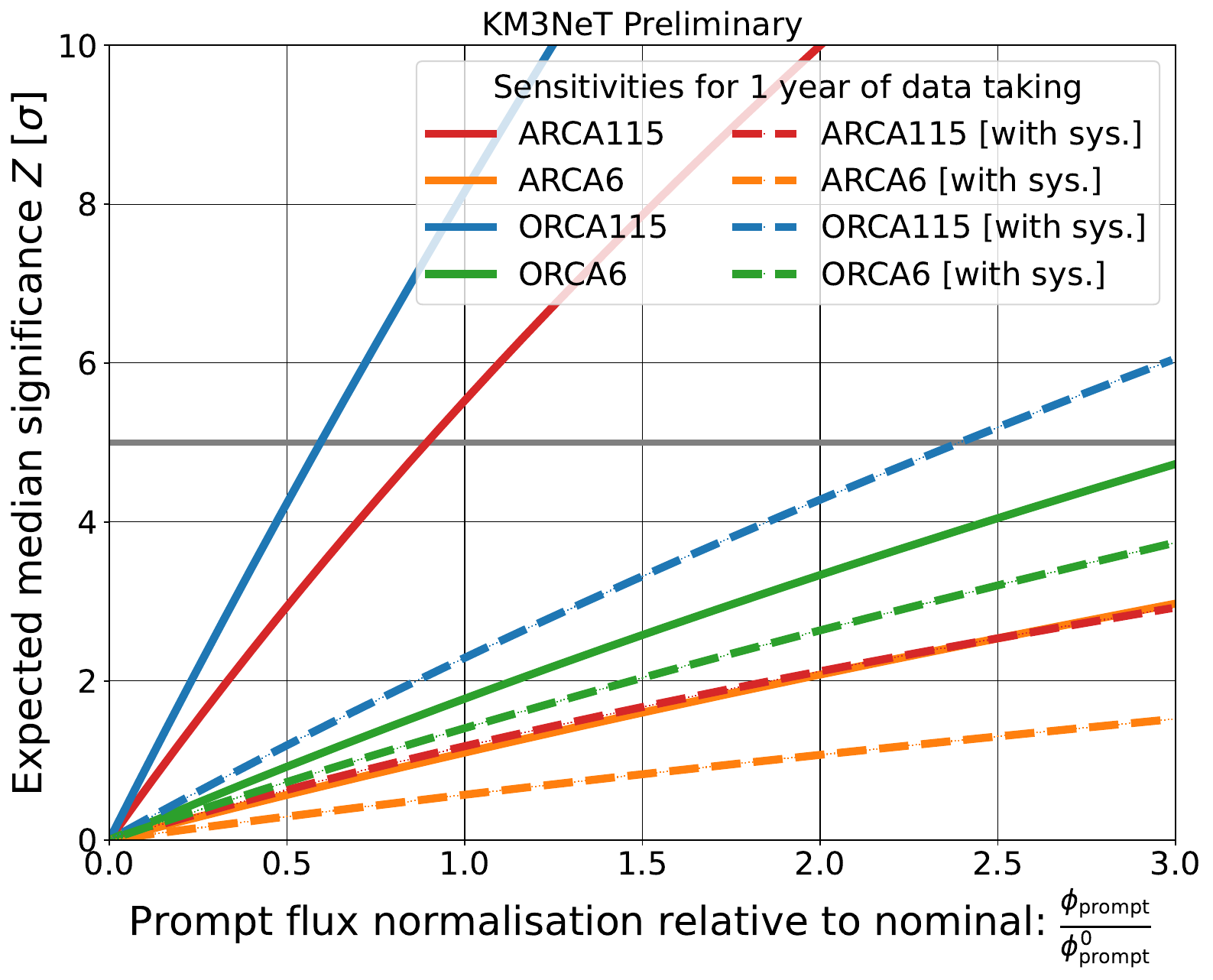}}\caption{Sensitivities of different detector configurations to the \textcolor{red}{prompt}
muon bundle signal (as defined in Sec. \ref{sec:Definition-of-signal-and-background}).
The solid lines were computed only with the statistical uncertainty
and the dashed ones: with the total uncertainty, including the systematics
from Sec. \ref{sec:Systematic-uncertainty-study}. The IceCube result
was taken from \cite{IceCube-prompt-paper-characterisation-of-the-atm-mu-flux}
and was obtained for the GST3 CR flux model and livetime of 645.4~d,
while the error bar shows the variation for different CR flux models.
\label{fig:sensitivity_full_det}}
\end{figure}

In this section, the final results of the analysis are presented.
Sensitivity to exclude the null hypothesis $H_{0}$ was expressed
in terms of expected median significance, computed from the TS using
Eq. \ref{eq:Significance}. The sensitivities of ARCA115, ARCA6, ORCA115,
and ORCA6 detectors are shown in Fig. \ref{fig:sensitivity_full_det}.
Naturally, the complete building blocks are more sensitive, since
they collect more data and can reconstruct the events with better
precision. A comparison of the results of this analysis against the
results obtained by IceCube using data from approximately 2 years
of operation was included as well \cite{IceCube-prompt-paper-characterisation-of-the-atm-mu-flux}
. However, it has to be taken with a grain of salt, given the vast
differences in the analysis methods and in the experimental setups,
to name a few:
\begin{enumerate}
\item The definition of the \textcolor{red}{prompt} component:
\begin{enumerate}
\item IceCube (IC) analysis: \textcolor{red}{prompt} flux prediction obtained
by reweighting the CORSIKA MC, assuming it to be purely \textcolor{blue}{conventional}
(which is a crude approximation). The reweighting followed the, somewhat
outdated, ERS model \cite{ERS-prompt-flux-model}. 
\item This analysis: direct \textcolor{red}{prompt} muon flux prediction
from CORSIKA.
\end{enumerate}
\item The HE hadronic interaction model:
\begin{enumerate}
\item IC analysis: SIBYLL 2.1.
\item This analysis: SIBYLL 2.3d.
\end{enumerate}
\item Observables used:
\begin{enumerate}
\item IC analysis: reconstructed energy and direction of the most energetic
muon in the bundle.
\item This analysis: reconstructed muon bundle energy and reconstructed
number of visible muons.
\end{enumerate}
\item IceCube is located shallower than ARCA and ORCA, which means that
it measures a larger muon flux, than the KM3NeT detectors. 
\item The DOM and string (equivalent of a DU in KM3NeT) spacing in IceCube
is different than in KM3NeT.
\item The total photo-cathode areas of the PMTs for IceCube is roughly $261\,\mathrm{m^{2}}$,
while for one KM3NeT building block it is $292.5\,\mathrm{m^{2}}$.
\item IceCube did not consider the CR flux models as a source of the systematic
uncertainty. Instead, they have provided separate significance values
for the assumptions of different models – this is what the error bars
in Fig. \ref{fig:sensitivity_full_vs_time} were derived from.
\end{enumerate}

\begin{figure}[H]
\centering{}\includegraphics[width=8cm]{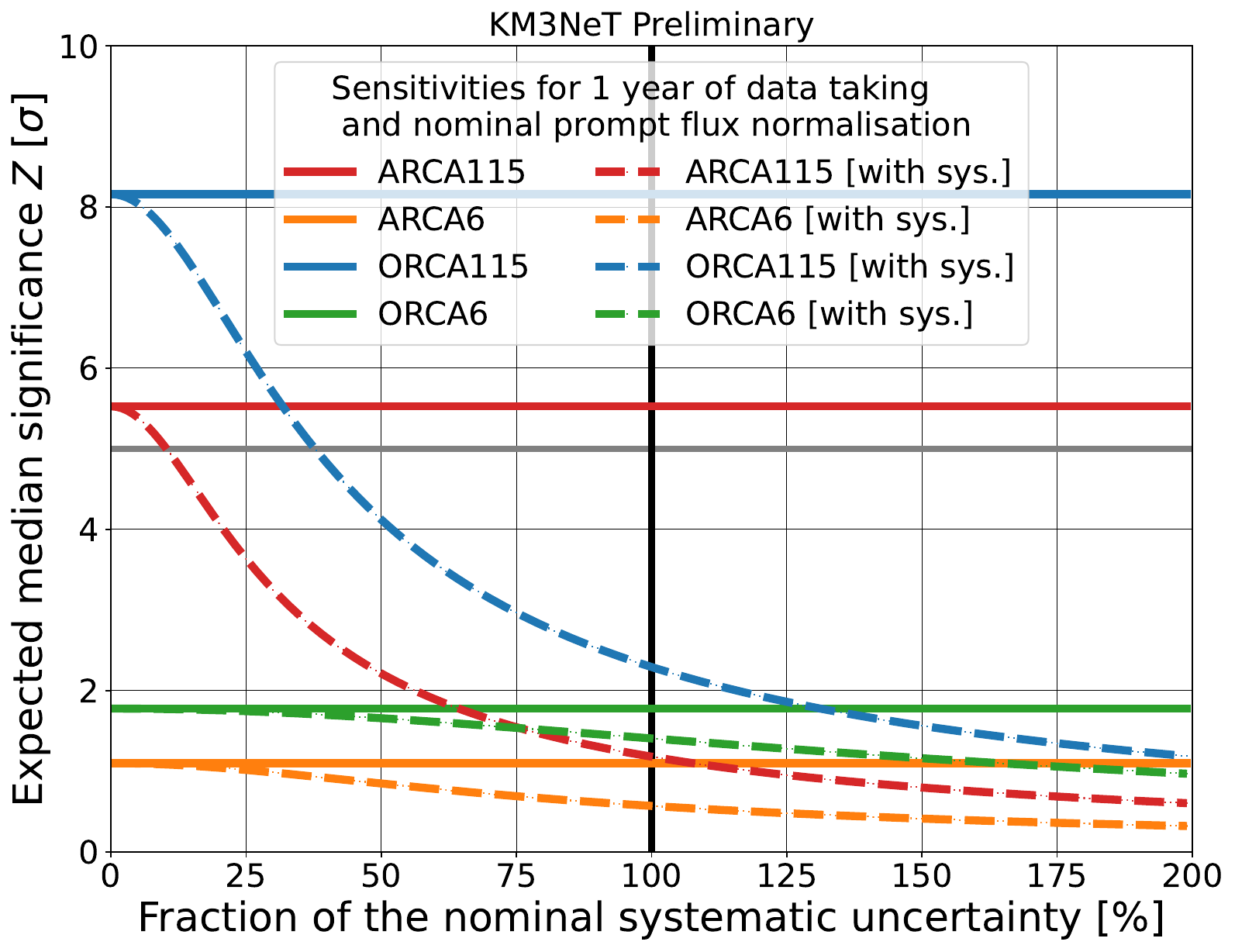}\caption{Sensitivities of different detector configurations to the \textcolor{red}{prompt}
muon bundle signal (as defined in Sec. \ref{sec:Definition-of-signal-and-background}).
The solid lines were computed only with the statistical uncertainty
and the dashed ones: with the total uncertainty, including the systematics
from Sec. \ref{sec:Systematic-uncertainty-study}. The systematic
uncertainty was then varied as function of the fraction of the nominal
value. \label{fig:sensitivity_vs_syst}}
\end{figure}

The predictions of Fig. \ref{fig:sensitivity_full_det} look rather
optimistic for KM3NeT, despite the evident strong limitation by systematic
uncertainties. It is expected that by the time of completion of ARCA
and ORCA the uncertainties related to the HE hadronic interactions
will be reduced, thanks to advances in theoretical models \cite{EPOS3,QGSJET3,SIBYLL-future}.
The scale of improvement is hard to predict and will largely depend
on the availability of complementary measurements. The most anticipated
data are the oxygen runs at LHC \cite{proton-oxygen-for-air-shower-physics-at-LHC,LHCf-proton-oxygen-collisions,ICRC2023-LHC-oxygen-collisions}.
Accumulation of new data from experiments measuring the cosmic rays
directly might shrink the uncertainty related to the CR flux modelling
as well \cite{Direct-Measurements-CR,direct-measurements-CR-calet}.
In fact, already the incorporation of the currently available experimental
data into newer models should provide a significant improvement in
this regard. Upon inspection of Fig. \ref{fig:sensitivity_full_vs_time},
even in the worst case: with unchanged uncertainties, KM3NeT/ORCA
should be able to exclude the \textcolor{blue}{BGD}-only scenario
at $5\sigma$ in about 5 years. As demonstrated in Fig. \ref{fig:sensitivity_vs_syst},
the same may be achieved in just 1 year if the systematic uncertainty
can be shrunk to 37\% of the current level. 

The most crucial systematic uncertainties for this analysis are the
ones related to: CR flux models, PMT efficiencies, and light absorption
length (see Sec. \ref{subsec:Results-systematics}). The first relies
on primary CR measurements by other experiments and corresponding
theoretical advancements \cite{AMS-02-CR-nuclei-measurement,HE-CR-with-IceTop,ICRC2023-Auger-data-CR-modelling}.
The current efforts focus on improving the understanding of the primary
CR spectrum at the ankle (see Fig. \ref{fig:The-cosmic-ray-spectrum})
and beyond, which is highly relevant for this work. To make a rough
prediction of the capability of the full KM3NeT detectors, let us
assume that by the time of completion of ARCA and ORCA, the CR flux
model uncertainty would be reduced by 50~\%. The uncertainties due
to PMT efficiencies and light absorption length were evaluated by
A. Romanov in \cite{Andrey_Thesis} as function of the zenith angle,
leading to constant value estimates in Sec. \ref{subsec:Results-systematics}.
The zenith distributions are dominated by the low-energy events, hence
the same uncertainties in the energy range most relevant for this
work may in fact be lower. To investigate this, an updated study with
larger high-energy statistics would be needed. This is feasible, as
the study can be done with MUPAGE, as was the original one. Let us
suppose that the uncertainties would be at $\sim70\,\%$ of the original
values and that the resulting total uncertainty (together with the
50\% reduction of the CR flux model uncertainty) would be $\sim60\,\%$
of the nominal one. This would imply a discovery at $5\sigma$ in
less than 2 years with ORCA115 and in about 6 years with ARCA115.
For ORCA6 and ARCA6 the same could be achieved in approximately 10
and 51 years respectively, which clearly illustrates the benefit of
having larger detector configurations. Such a long timescale is certainly
not the fate of KM3NeT, since as of August 2023, the currently installed
configurations are already ARCA19 and ORCA16.

One should note that the sensitivity for ARCA is indicated for 1 building
block (ARCA115), yet the complete detector will consist of two such
blocks, doubling the amount of collected data. From Eq. \ref{eq:Significance}
and \ref{eq:time_scaling_of_TS} it follows that the corresponding
significance will be $\sqrt{2}$ times larger. Furthermore, the results
of ARCA and ORCA may in principle be combined and the data collected
with intermediate configurations of both detectors could be put to
use as well. Producing such a result during the construction phase
of KM3NeT will surely present an exciting technical challenge.

\chapter{Conclusions \label{chap:Conclusions}}

This chapter contains a brief summary of this thesis and some thoughts
on possible improvements and extensions.

\section{Summary}

The work of this thesis revolved around few main points: CORSIKA simulation
(Chap. \ref{chap:Cosmic-Ray-Sim-chain-KM3NeT}, Sec.\ref{sec:CORSIKA}
and \ref{sec:gSeaGen_supplementary_material}), ML-based muon bundle
observable reconstruction (Chap. \ref{chap:muon-bundle-reco}, Sec.
\ref{sec:Muon-bundle-reconstruction-supplement}), comparisons of
experimental data against the Monte Carlo simulations (Chap. \ref{chap:Muon-rate-measurement}),
and the \textcolor{red}{prompt} muon analysis (Chap. \ref{chap:prompt_ana}
and Sec. \ref{sec:Prompt-muon-analysis-supplement}). They are all
briefly recapitulated in the following sections.

\subsection{CORSIKA Monte Carlo production}

The procurement of the CORSIKA simulation for KM3NeT detectors has
proven to be a complex and challenging task for a number of reasons.
Obtaining reasonable results and covering the entire phase space of
interest required careful selection of inputs. The simulation settings
had to be iteratively adjusted after generating the test productions,
e.g. to evaluate the optimal energy thresholds. Some inputs, for instance
the Earth's magnetic field strength or the atmosphere layer parameters
had to be first estimated from or fitted to a model, which should
meet certain specifications (see Sec. \ref{par:Model-of-the-atmosphere}).
Even after finalising all the settings, running the CORSIKA MC still
posed a challenge in terms of the required computing power, storage,
and bookkeeping effort. The process was partially automated, yet still
manual scrutiny was, and will most likely remain a necessity.

The CORSIKA simulation produced for this work has already found use
in a number of analyses within the KM3NeT collaboration and is expected
to continue to contribute to the scientific success of KM3NeT. The
applications of the CORSIKA MC outside of this work included reconstruction
of muon bundle properties, measurement of the primary CR flux composition,
tuning of the MUPAGE code, classification of muons stopping inside
the KM3NeT detectors, and more \cite{StefanReckThesis,Andrey_Thesis,Andrey_and_me_ICRC2023,Stopping-muons-Louis-ICRC2023}.

\subsection{Muon bundle reconstruction}

As was shown in Chapters \ref{chap:muon-bundle-reco}–\ref{chap:prompt_ana},
the ML-based muon bundle property reconstruction has yielded valuable
results. Even the regression of the total primary energy was successful,
despite the intrinsic loss of large portion of the original information
about the shower. This opens a window to interesting analysis possibilities
from the point of view of the CR physics. After improving the agreement
between the data and MC, the KM3NeT datapoints could be included in
the primary CR spectrum (Fig. \ref{fig:The-cosmic-ray-spectrum}).
The ML bundle energy reconstruction has proven to be a much better
description of reality, than the traditional JMuon reconstruction
designed under single muon hypothesis. The reconstruction of muon
multiplicity was the first such reconstruction performed for an underwater
neutrino telescope. It allowed to boost the power of the hypothesis
test in Chap. \ref{chap:prompt_ana} by rejecting the less sensitive
phase space. 

\subsection{Data vs simulation comparisons}

Chap. \ref{chap:Muon-rate-measurement} has demonstrated that even
very early KM3NeT detector configurations, like ORCA1 and ARCA2, were
not merely a proof of concept. They were already capable of delivering
physically sound results. In addition, thanks to the development of
dedicated reconstruction, the first measurements of muon multiplicity
and total primary energy with an underwater neutrino telescope were
performed on ARCA6 and ORCA6 data. By comparing the outcomes for different
detector configurations, one can note clear progress in the description
of the data. It was achieved by both advancing the detector construction,
and significant improvements in simulation and reconstruction software.
A consistent picture emerges from looking at the data: the high-energy
and high-multiplicity muon bundle rates are severely underestimated
by the simulations. This is consistent with observations of other
experiments and indicates that there is a need for further work in
modelling of hadronic interaction models \cite{MuonPuzzle}.

\subsection{Prompt muon analysis}

The sensitivities obtained for ARCA115 and ORCA115 (Sec. \ref{sec:Sensitivity-results})
predict a bright future for KM3NeT. Even the results for ARCA6 and
ORCA6 look encouraging, although the detectors in these configurations
have been in operation for too short (each for less than a year).
To provide a definite confirmation, whether the \textcolor{red}{prompt}
muon flux component is present in the atmospheric muon flux or not,
more data will be needed. Currently, the strongest limitation to the
sensitivity comes from large systematic uncertainties (Sec. \ref{subsec:Results-systematics}).
The comparison against experimental data for ARCA6 and ORCA6 suggests
that there is indeed an excess of high-energy and high-multiplicity
events. However, a reliable measurement will require further work
on improving the simulations to accurately describe the data in the
insensitive region of the phase space.

The comparison with the IceCube results suggests that it also has
a great potential to confirm the existence of the \textcolor{red}{prompt}
muon flux component. In fact, with more than 10 years of accumulated
livetime, IceCube is most probably already able to make a significant
measurement, or at least set a strong upper limit on the \textcolor{red}{prompt}
muon flux. The only hindrance may be the need for a dedicated simulation.

\section{Outlook}

There is a number of routes that have been followed in this thesis,
yet there are even more that have been not. Some of those are listed
and grouped topically in the following sections.

\subsection{CORSIKA Monte Carlo production\label{subsec:CORSIKA-Monte-Carlo-outlook}}

In the case of the CORSIKA simulation, one could think of a number
of improvements. There are CR flux models, which require other CR
primaries than the ones simulated for this work ($p$, $He$, $C$,
$O$, $Fe$) \cite{HillasGaisser_model,combined_HillasGaisser_and_GaisserHonda_also_crfluxmodels_reference,poly-gonato,GSF-Global_Spline_Fit}.
Such a model could in principle also be used as the default one for
the analysis. However, adding more primaries would considerably increase
the computational cost of the CORSIKA MC, since one must produce sufficient
statistics for each of the primaries.

Producing a rbr CORSIKA simulation would certainly be extremely valuable
for a number of analyses. However, it would be a costly endeavour,
perhaps not even feasible. Besides the need to generate sufficient
statistics for each run, the atmosphere fit (see Sec. \ref{subsec:Fit-of-the-atmosphere})
should ideally be redone for each run,or at least each month, since
the atmospheric conditions vary seasonally.

The propagation of muon bundles simulated with CORSIKA has been fully
integrated into gSeaGen. However, those muons often come mixed with
neutrinos, and they remain to be included in the implementation. Notably,
such a combined simulation of $\mu$ and $\nu$ would be necessary
to carry out a self-veto study for KM3NeT \cite{Self-veto}.

Already with the current CORSIKA7 simulation, di-muon production from
the Drell-Yan processes \cite{Drell-Yan} could be studied closer
by finding muons with common parent particles. 

A MC production with CORSIKA8 \cite{CORSIKA8}, once it is officially
released, would be invaluable for cross-checking the CORSIKA7 results.
In particular, the new feature allowing to simulate showers across
different media would enable a direct validation of the gSeaGen code.
Furthermore, CORSIKA8 is expected to provide much more detailed particle
history information and a more flexible atmosphere modelling.

\subsection{Muon bundle reconstruction}

The reconstruction of muon bundle observables developed in Sec. \ref{chap:muon-bundle-reco}
is in principle extensible to other detector configurations beyond
ARCA6 and ORCA6. It requires processing the existing CORSIKA MC for
the desired detector configuration to be able to train the estimator.
For instance, processing a MC for ORCA4 could allow a direct comparison
with the deep learning muon multiplicity reconstruction developed
in parallel to this work by S. Reck \cite{StefanReckThesis}.

Besides extending to other detector configurations, reconstruction
of further muon bundle observables listed in Sec. \ref{sec:Extensive-Air-Showers-EAS}
could certainly be attempted. To this end, it might be beneficial
to consider including additional information in the training. In this
work, each feature was a single number per event, however the spatial
and temporal distributions of hits are in principle accessible. Furthermore,
complementary data could be extracted from the other sensors besides
the PMTs, which are already installed in the DOMs \cite{KM3NeT-DOM-paper}.
The technical challenge would be how to deal with such irregular and
highly multidimensional data in an efficient manner. In a broader
perspective, a step towards diversifying the collected information
could be an extension of the KM3NeT detectors, e.g. by installing
a (sea-)surface detector array above ARCA or ORCA (or both), similarly
to IceTop array above IceCube \cite{IceTop}.

The muon selection, described in Sec. \ref{subsec:Muon-selection}
could be further refined, perhaps without relying on the likelihood
from the classical reconstruction (JMuon). Instead, a direct measure
of reconstruction confidence, similar to what was done by S. Reck
in \cite{StefanReckThesis} could be used. The approximation of cylindrical
symmetry of the detector in the geometric part of the selection could
have compromised the results to some extent, especially for ORCA (see
Fig. \ref{fig:Top-view-of-detectors}). Hence, the performance could
profit from developing a more robust approach the performance.

\subsection{Data vs simulation comparisons}

An obvious extension of the comparisons of experimental data and simulations
would be to include datasets from newer detector configurations. This
could allow to include more runs, resulting in a longer total livetime,
preferably sampled from a period of 1 year or longer. This is crucial
from the point of view of studying the seasonal variations in the
muon rates (or averaging them out accurately). In addition, the agreement
between the data and MC could be improved by applying a stricter event
selection, utilising more of the available low-level information. 

From the MC side, there are still ongoing efforts to improve the light
simulation, which are most relevant for ORCA and should be included
in the CORSIKA production when possible. As mentioned in Chap. \ref{chap:Muon-rate-measurement},
pure noise is currently not included in the official simulations,
it was evaluated only in standalone studies. Adding it could foster
the event selection and improve the understanding of the observed
discrepancies when comparing MC against the data.

\subsection{Systematic uncertainties}

Most of the systematic uncertainties evaluated in Sec. \ref{sec:Systematic-uncertainty-study}
could be either reduced by improvement of modelling, or refined through
more detailed studies. 

The first category includes the total CR flux uncertainty due to differences
between the models, which can be expected to shrink as a result of
new direct CR measurements and theoretical advancements. Similarly,
the next generation of HE hadronic interaction models is in the making.
They will benefit from oxygen runs at LHC \cite{proton-oxygen-for-air-shower-physics-at-LHC,LHCf-proton-oxygen-collisions,ICRC2023-LHC-oxygen-collisions},
which are crucial for better understanding of interactions in the
atmosphere. The uncertainty related to the density variations of the
atmosphere itself may be reduced by generating separate CORSIKA productions
with dedicated atmosphere fits. This could be done both for the ARCA
and ORCA locations, and for different months.

The uncertainty evaluation related to the light absorption length
in water and the PMT efficiency would surely benefit from larger statistics,
allowing to express the uncertainty as the function of the energy.
What is more, the PMT efficiency study should be preferably done on
a simulation in rbr mode (see Sec. \ref{sec:trigger}), since the
PMT parameters can vary from run to run.

Along with reducing the leading uncertainties, the neglected ones
mentioned in Sec. \ref{sec:Systematic-uncertainty-study} should ideally
be included.

\subsection{Prompt muon analysis}

The \textcolor{red}{prompt} muon analysis could be refined in a number
of ways. The sensitivity could unquestionably benefit from a combination
of ARCA and ORCA results. Moreover, the results from intermediate
KM3NeT detector configurations could in principle be consolidated.
Another way to boost the sensitivity would be including more observables,
for instance the muon bundle diameter, or time spread of the event. 

Selection of the critical region is an aspect of the analysis, which
could use further investigation. In the earlier phases of this work,
direct classification into \textcolor{red}{SIG} and \textcolor{blue}{BGD}
was attempted, although with unsatisfactory results. The contribution
of \textcolor{blue}{BGD} was largely underestimated in this approach,
which could lead to a false discovery. To avoid it, this approach
was abandoned. An alternative would be more complex critical region
boundary, either defined by a function, or by binning in energy and
multiplicity. The latter might require additional caution, not to
tune the selection of the binning to statistical fluctuations. This
is a real risk due to lower statistics available for high-energy (and
multiplicity) events.

By default, MUPAGE does not simulate the $\mu$ flux consistent with
\textcolor{blue}{BGD}. However, there may be a way around this. Following
the success of the MUPAGE tuning on CORSIKA in A. Romanov's work \cite{Andrey_Thesis},
one could attempt a specialized tuning of MUPAGE on the \textcolor{blue}{BGD}-only
CORSIKA. This would not only allow for easier generation of considerable
statistics of pure \textcolor{blue}{BGD} simulation, but also to produce
a rbr MC with the tuned MUPAGE.

The discrimination between \textcolor{red}{prompt} and \textcolor{blue}{conventional}
muons could be made more precise with CORSIKA8 \cite{CORSIKA8}. It
will come with more complete particle history, including exact numbers
of interactions, which will remove the ambiguity in the case of longer
particle genealogies.

A very natural extension of the analysis would be to study the sensitivity
of KM3NeT to the \textcolor{red}{prompt} neutrinos, however this will
require implementing propagation of $\nu$ simulated with CORSIKA,
as mentioned in Sec. \ref{subsec:CORSIKA-Monte-Carlo-outlook}. An
another possible variation of the analysis could investigate the sensitivity
to the presence of heavy quarks ($c$, $s$, $t$, $b$) in the atmospheric
flux. Such a heavy-quark component is in fact partially overlapping
with the \textcolor{red}{prompt} definition. It is more inclusive,
since secondaries with one heavy and other light parent particles
would still be counted as contributing to the heavy flux. For \textcolor{red}{prompt}
muons, a single \textcolor{blue}{conventional} parent particle was
already excluding such a secondary $\mu$ (or $\nu$, if we consider
the proposed extension). It was verified, that excluding the absence
of charmed and heavier hadrons in observed EAS is indeed easier than
excluding the \textcolor{blue}{conventional}-only scenario. This alternative
approach was not pursued further, since currently the main hindrance
to any measurement is the accuracy of simulations (see Chap. \ref{chap:Muon-rate-measurement}),
which was thoroughly addressed in this work.

\appendix

\chapter{Appendix}

In this chapter, all the supplementary materials for the thesis have
been gathered.

\section{CORSIKA \label{sec:Additional-material-related-to-CORSIKA}}

This section contains miscellaneous information concerning the CORSIKA
MC production.

\subsection{Simulation inputs\label{subsec:Simulation-inputs}}

The CORSIKA production, on which all the results of Chap. \ref{chap:muon-bundle-reco},
the ARCA6 and ORCA6 results in Chap. \ref{chap:Muon-rate-measurement},
and the entire analysis in Chap. \ref{chap:prompt_ana} are based,
was ran with CORSIKA v7.7410 \cite{CORSIKA_v77410}. Tab. \ref{tab:Compilation-settings-CORSIKA}
summarises the compilation options, which were common for all sub-productions:
TeV\_low, TeV\_high, PeV, and EeV. The sub-production naming refers
roughly to simulated $E_{\mathrm{prim}}$ ranges, as can be verified
in Tab. \ref{tab:Compilation-settings-CORSIKA}, which details all
the relevant CORSIKA settings used in the simulation.

\begin{table}[H]
\begin{centering}
\caption{Compilation settings used for CORSIKA. The complete list of possible
options is documented in \cite{CORSIKA-Userguide}. \label{tab:Compilation-settings-CORSIKA}}
\par\end{centering}
\centering{}%
\begin{tabular}{|c|c|c|}
\cline{2-3} \cline{3-3} 
\multicolumn{1}{c|}{} & \textbf{\scriptsize{}Value} & {\scriptsize{}Description}\tabularnewline
\hline 
\multirow{3}{*}{{\scriptsize{}main options}} & {\scriptsize{}SIBYLL-2.3d \cite{SIBYLL-2.3d}} & {\scriptsize{}HE hadronic interaction model}\tabularnewline
\cline{2-3} \cline{3-3} 
 & {\scriptsize{}URQMD-1.3cr \cite{UrQMD}} & {\scriptsize{}LE hadronic interaction model}\tabularnewline
\cline{2-3} \cline{3-3} 
 & {\scriptsize{}VOLUMEDET} & {\scriptsize{}non-flat (volume) detector geometry}\tabularnewline
\hline 
\multirow{7}{*}{{\scriptsize{}additional options}} & {\scriptsize{}CURVED} & {\scriptsize{}Earth's curvature is accounted for}\tabularnewline
\cline{2-3} \cline{3-3} 
 & {\scriptsize{}PYTHIADIR} & {\scriptsize{}particle decays are handled by PYTHIA \cite{Pythia}}\tabularnewline
\cline{2-3} \cline{3-3} 
 & {\scriptsize{}EHISTORY} & {\scriptsize{}parent particle information for muons is saved in the
output}\tabularnewline
\cline{2-3} \cline{3-3} 
 & {\scriptsize{}NEUTRINO} & {\scriptsize{}$\nu$'s from $\pi$, $K$, and $\mu$ decays are tracked
explicitly (but without interaction)}\tabularnewline
\cline{2-3} \cline{3-3} 
 & {\scriptsize{}CHARM} & {\scriptsize{}charmed particle and $\tau$ lepton production is enabled}\tabularnewline
\cline{2-3} \cline{3-3} 
 & {\scriptsize{}UPWARD} & {\scriptsize{}upward (or nearly horizontal) particles are not discarded}\tabularnewline
\cline{2-3} \cline{3-3} 
 & {\scriptsize{}LPM} & {\scriptsize{}LPM effect \cite{LPM} is simulated}\tabularnewline
\hline 
\end{tabular}
\end{table}

\newpage{}
\begin{table}[H]
\centering{}\caption{Simulation settings used for CORSIKA. \label{tab:Simulation-settings-CORSIKA}}
\end{table}

\begin{center}
\begin{longtable}[c]{|c|c|c|c|}
\hline 
\textbf{\scriptsize{}Prod} & \textbf{\scriptsize{}Setting} & \textbf{\scriptsize{}Value} & \textbf{\scriptsize{}Param name}\tabularnewline
\hline 
\hline 
{\scriptsize{}TeV\_low} & \multirow{4}{*}{{\scriptsize{}Primaries used}} & {\scriptsize{}p, He, C, O, Fe} & \multirow{4}{*}{{\scriptsize{}-}}\tabularnewline
\cline{1-1} \cline{3-3} 
{\scriptsize{}TeV\_high} &  & {\scriptsize{}p, He, C, O, Fe} & \tabularnewline
\cline{1-1} \cline{3-3} 
{\scriptsize{}PeV} &  & {\scriptsize{}p, He, C, O, Fe} & \tabularnewline
\cline{1-1} \cline{3-3} 
{\scriptsize{}EeV} &  & {\scriptsize{}p, Fe} & \tabularnewline
\hline 
{\scriptsize{}all} & {\scriptsize{}Primary codes} & {\scriptsize{}14, 402, 1206, 1608, 5626 (p, He, C, O, Fe)} & {\scriptsize{}PRMPAR}\tabularnewline
\hline 
{\scriptsize{}TeV\_low} & \multirow{4}{*}{{\scriptsize{}Generated showers per file}} & {\scriptsize{}$\begin{array}{c}
1.5\cdot10^{7}\,(\mathsf{p})\\
10^{7}\,(\mathsf{He})\\
5\cdot10^{6}\,(\mathsf{C})\\
5\cdot10^{6}\,(\mathsf{O})\\
5\cdot10^{6}\,(\mathsf{Fe})
\end{array}$} & \multirow{4}{*}{{\scriptsize{}NSHOW}}\tabularnewline
\cline{1-1} \cline{3-3} 
{\scriptsize{}TeV\_high} &  & {\scriptsize{}$\begin{array}{c}
6\cdot10^{6}\,(\mathsf{p})\\
4\cdot10^{6}\,(\mathsf{He})\\
1.5\cdot10^{6}\,(\mathsf{C})\\
1.5\cdot10^{6}\,(\mathsf{O})\\
4\cdot10^{5}\,(\mathsf{Fe})
\end{array}$} & \tabularnewline
\cline{1-1} \cline{3-3} 
{\scriptsize{}PeV} &  & {\scriptsize{}$\begin{array}{c}
4\cdot10^{4}\,(\mathsf{p})\\
4\cdot10^{4}\,(\mathsf{He})\\
2.5\cdot10^{4}\,(\mathsf{C})\\
2.5\cdot10^{4}\,(\mathsf{O})\\
2\cdot10^{4}\,(\mathsf{Fe})
\end{array}$} & \tabularnewline
\cline{1-1} \cline{3-3} 
{\scriptsize{}EeV} &  & {\scriptsize{}100} & \tabularnewline
\hline 
{\scriptsize{}TeV\_low} & \multirow{4}{*}{{\scriptsize{}$E_{\mathsf{prim}}$ range}} & {\scriptsize{}$\begin{array}{c}
400<\frac{E_{\mathsf{prim}}}{\mathsf{GeV}}<2\cdot10^{4}\,(\mathsf{p})\\
1.6\cdot10^{3}<\frac{E_{\mathsf{prim}}}{\mathsf{GeV}}<2\cdot10^{4}\,(\mathsf{He})\\
4.8\cdot10^{3}<\frac{E_{\mathsf{prim}}}{\mathsf{GeV}}<6\cdot10^{4}\,(\mathsf{C})\\
6.4\cdot10^{3}<\frac{E_{\mathsf{prim}}}{\mathsf{GeV}}<8\cdot10^{4}\,(\mathsf{O})\\
2.24\cdot10^{4}<\frac{E_{\mathsf{prim}}}{\mathsf{GeV}}<3\cdot10^{5}\,(\mathsf{Fe})
\end{array}$} & \multirow{4}{*}{{\scriptsize{}ERANGE}}\tabularnewline
\cline{1-1} \cline{3-3} 
{\scriptsize{}TeV\_high} &  & {\scriptsize{}$\begin{array}{c}
6\cdot10^{3}\,(\mathsf{p})\\
1\cdot10^{4}\,(\mathsf{He})\\
3\cdot10^{4}\,(\mathsf{C})\\
3\cdot10^{4}\,(\mathsf{O})\\
1\cdot10^{5}\,(\mathsf{Fe})
\end{array}<\frac{E_{\mathsf{prim}}}{\mathsf{GeV}}<1.1\cdot10^{6}$} & \tabularnewline
\cline{1-1} \cline{3-3} 
{\scriptsize{}PeV} &  & {\scriptsize{}$1.1\cdot10^{6}<\frac{E_{\mathsf{prim}}}{\mathsf{GeV}}<0.9\cdot10^{8}$} & \tabularnewline
\cline{1-1} \cline{3-3} 
{\scriptsize{}EeV} &  & {\scriptsize{}$0.9\cdot10^{8}<\frac{E_{\mathsf{prim}}}{\mathsf{GeV}}<\begin{array}{c}
4.0\cdot10^{9}\,(\mathsf{p})\\
8.0\cdot10^{9}\,(\mathsf{Fe})
\end{array}$} & \tabularnewline
\hline 
{\scriptsize{}TeV\_low} & \multirow{4}{*}{{\scriptsize{}$E_{\mathsf{prim}}$ spectrum slope}} & {\scriptsize{}-3.0} & \multirow{4}{*}{{\scriptsize{}ESLOPE}}\tabularnewline
\cline{1-1} \cline{3-3} 
{\scriptsize{}TeV\_high} &  & {\scriptsize{}-3.5, -3.0, -3.0, -3.0, -2.5 (p, He, C, O, Fe)} & \tabularnewline
\cline{1-1} \cline{3-3} 
{\scriptsize{}PeV} &  & {\scriptsize{}-2.0} & \tabularnewline
\cline{1-1} \cline{3-3} 
{\scriptsize{}EeV} &  & {\scriptsize{}0.0} & \tabularnewline
\hline 
{\scriptsize{}TeV\_low} & \multirow{4}{*}{{\scriptsize{}Primary zenith angle range}} & {\scriptsize{}$0<\frac{\theta}{\mathsf{deg}}<\begin{array}{c}
75\,(\mathsf{p})\\
75\,(\mathsf{He})\\
60\,(\mathsf{C})\\
60\,(\mathsf{O})\\
65\,(\mathsf{Fe})
\end{array}$} & \multirow{4}{*}{{\scriptsize{}THETAP}}\tabularnewline
\cline{1-1} \cline{3-3} 
{\scriptsize{}TeV\_high} &  & {\scriptsize{}$0<\frac{\theta}{\mathsf{deg}}<87$} & \tabularnewline
\cline{1-1} \cline{3-3} 
{\scriptsize{}PeV} &  & {\scriptsize{}$0<\frac{\theta}{\mathsf{deg}}<87$} & \tabularnewline
\cline{1-1} \cline{3-3} 
{\scriptsize{}EeV} &  & {\scriptsize{}$0<\frac{\theta}{\mathsf{deg}}<87$} & \tabularnewline
\hline 
{\scriptsize{}all} & {\scriptsize{}Azimuth angle range} & {\scriptsize{}$180<\frac{\phi}{\mathsf{deg}}<180$} & {\scriptsize{}PHIP}\tabularnewline
\hline 
{\scriptsize{}all} & {\scriptsize{}Observation level} & {\scriptsize{}0 cm (sea level)} & {\scriptsize{}OBSLEV}\tabularnewline
\hline 
{\scriptsize{}all} & {\scriptsize{}Starting altitude} & {\scriptsize{}$0~\frac{\mathsf{g}}{\mathsf{cm^{2}}}$} & {\scriptsize{}FIXCHI}\tabularnewline
\hline 
{\scriptsize{}all} & {\scriptsize{}Magnetic field} & {\scriptsize{}$B_{x}=25.2179$~μT, $B_{z}=38.4848\,$μT} & {\scriptsize{}MAGNET}\tabularnewline
\hline 
\multirow{2}{*}{{\scriptsize{}all}} & {\scriptsize{}Flags for hadronic} & \multirow{2}{*}{{\scriptsize{}0 0 0 0 0 2}} & \multirow{2}{*}{{\scriptsize{}HADFLAG}}\tabularnewline
 & {\scriptsize{}interactions and fragmentation} &  & \tabularnewline
\hline 
\multirow{2}{*}{{\scriptsize{}all}} & {\scriptsize{}Energy cuts} & \multirow{2}{*}{{\scriptsize{}$(300,300,10^{3},10^{3})$~GeV}} & \multirow{2}{*}{{\scriptsize{}ECUTS}}\tabularnewline
 & {\scriptsize{}(hadrons/nuclei, $\mu$, $e$, $\gamma$)} &  & \tabularnewline
\hline 
\multirow{11}{*}{{\scriptsize{}all}} & \multirow{11}{*}{{\scriptsize{}Atmosphere fit}} & {\scriptsize{}0} & {\scriptsize{}ATMOD}\tabularnewline
\cline{3-4} \cline{4-4} 
 &  & {\scriptsize{}(1750000, 4500000, 7300000, 10130000, 12500000)~cm} & {\scriptsize{}ATMLAY}\tabularnewline
\cline{3-4} \cline{4-4} 
 &  & {\scriptsize{}-58.44897651964439, 0.5019792446185176,} & \multirow{3}{*}{{\scriptsize{}ATMA}}\tabularnewline
 &  & {\scriptsize{}-0.014658740225644349, -4.286764182377298e-05,} & \tabularnewline
 &  & {\scriptsize{}0.0030020492001918493} & \tabularnewline
\cline{3-4} \cline{4-4} 
 &  & {\scriptsize{}1071.309922924368, 1380.8205315315772,} & \multirow{3}{*}{{\scriptsize{}ATMB}}\tabularnewline
 &  & {\scriptsize{}450.1250293064572, 5714.833843117572,} & \tabularnewline
 &  & {\scriptsize{}-38.86175122739144} & \tabularnewline
\cline{3-4} \cline{4-4} 
 &  & {\scriptsize{}865209.9870960491, 622615.8339434739,} & \multirow{3}{*}{{\scriptsize{}ATMC}}\tabularnewline
 &  & {\scriptsize{}790046.9277382238, 599218.1788640362,} & \tabularnewline
 &  & {\scriptsize{}-141350551294.98242} & \tabularnewline
\hline 
{\scriptsize{}all} & {\scriptsize{}Additional information on $\nu$} & {\scriptsize{}T} & {\scriptsize{}NUADDI}\tabularnewline
\hline 
{\scriptsize{}all} & {\scriptsize{}Additional information on $\mu$} & {\scriptsize{}T} & {\scriptsize{}MUADDI}\tabularnewline
\hline 
{\scriptsize{}all} & {\scriptsize{}$\mu$ multiple scattering angle} & {\scriptsize{}T} & {\scriptsize{}MUMULT}\tabularnewline
\hline 
{\scriptsize{}all} & {\scriptsize{}EM interaction flags} & {\scriptsize{}F T} & {\scriptsize{}ELMFLG}\tabularnewline
\hline 
\multirow{2}{*}{{\scriptsize{}all}} & {\scriptsize{}Multiple scattering} & \multirow{2}{*}{{\scriptsize{}1}} & \multirow{2}{*}{{\scriptsize{}STEPFC}}\tabularnewline
 & {\scriptsize{}step length factor} &  & \tabularnewline
\hline 
\multirow{2}{*}{{\scriptsize{}all}} & {\scriptsize{}Outer radius for NKG} & \multirow{2}{*}{{\scriptsize{}200.E2}} & \multirow{2}{*}{{\scriptsize{}RADNKG}}\tabularnewline
 & {\scriptsize{}lateral density distribution} &  & \tabularnewline
\hline 
\multirow{2}{*}{{\scriptsize{}all}} & {\scriptsize{}Longitudinal distribution,} & \multirow{2}{*}{{\scriptsize{}F 10 F F}} & \multirow{2}{*}{{\scriptsize{}LONGI}}\tabularnewline
 & {\scriptsize{}step size, fit, out} &  & \tabularnewline
\hline 
{\scriptsize{}all} & {\scriptsize{}Cut on $\gamma$ factor for printout} & {\scriptsize{}$10^{4}$} & {\scriptsize{}ECTMAP}\tabularnewline
\hline 
{\scriptsize{}all} & {\scriptsize{}Max number of printed events} & {\scriptsize{}1} & {\scriptsize{}MAXPRT}\tabularnewline
\hline 
\end{longtable}
\par\end{center}

\smallskip{}

\begin{figure}[H]
\centering{}\includegraphics[width=12cm]{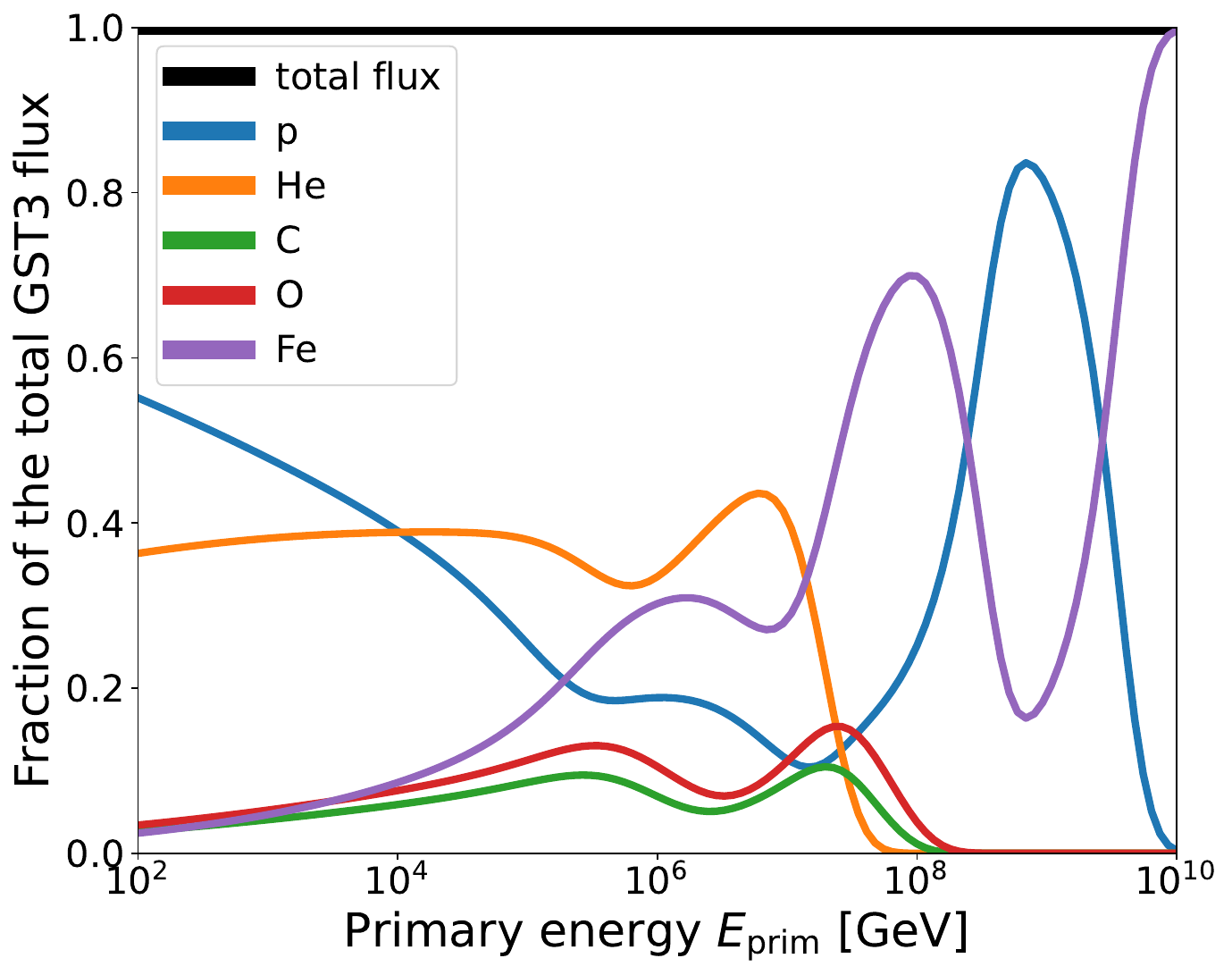}\caption{Primary CR composition assumed in the GST3 CR flux model \cite{GST}.
\label{fig:gst3-composition}}
\end{figure}

The parameters in Tab. \ref{tab:Simulation-settings-CORSIKA} were
optimised for a simulation of atmospheric muon fluxes at ARCA and
ORCA detectors. The atmosphere fit (see Sec. \ref{subsec:Fit-of-the-atmosphere})
and magnetic field model were averaged between the two detector locations
and also over the time (seasonal and yearly variations). The EeV sub-production
was only ran for proton and iron primaries, since only those have
a non-negligible contribution to the all-nuclei CR flux in the GST3
model, which assumes the composition, as shown in Fig. \ref{fig:gst3-composition}.
For more details, see the following sections.

\subsection{Fit of the atmosphere\label{subsec:Fit-of-the-atmosphere}}

\begin{figure}[H]
\centering{}\includegraphics[width=12cm]{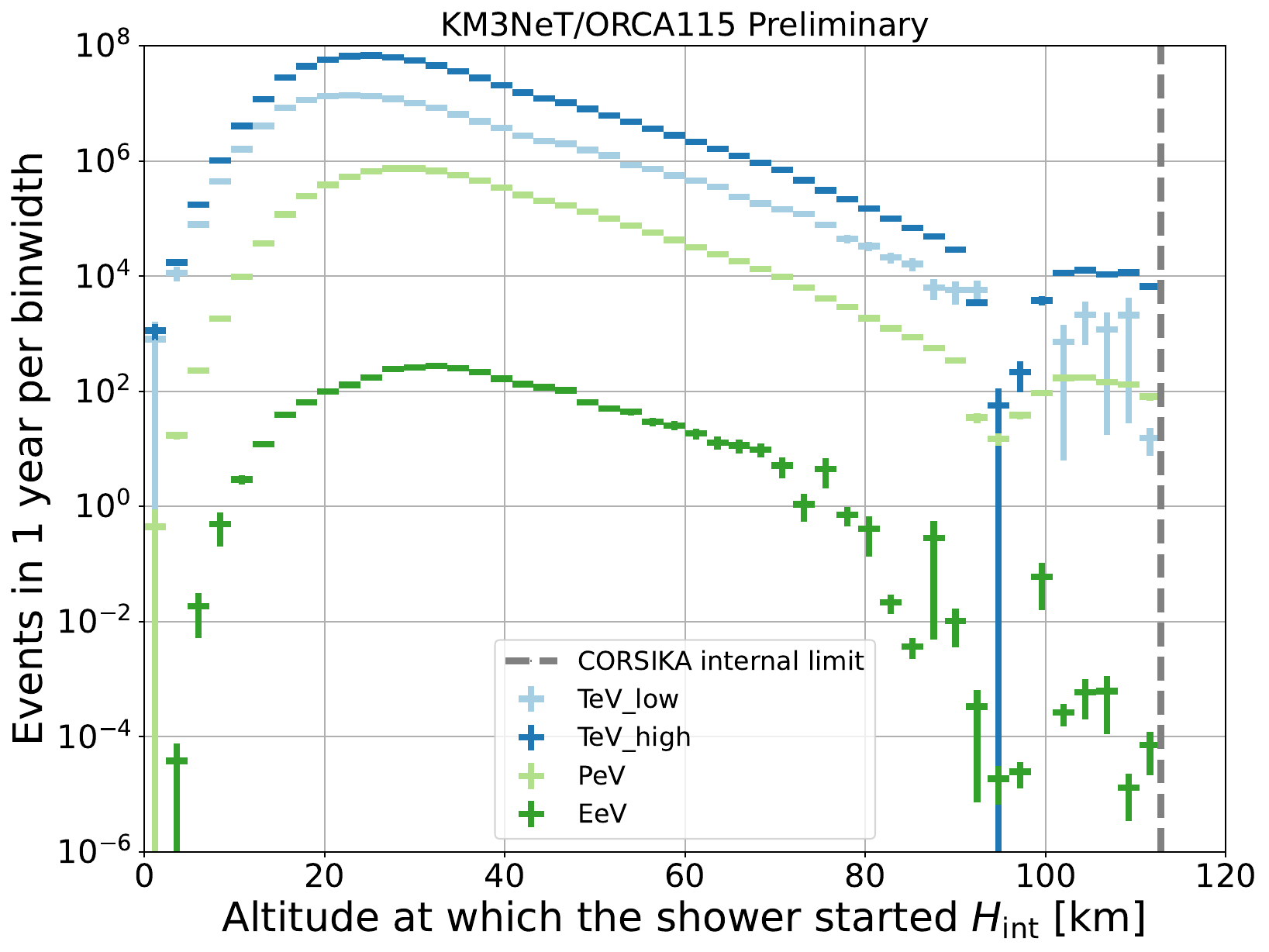}\caption{Distributions of the first interaction height $H_{\mathsf{int}}$
of CORSIKA showers for different sub-productions (see Tab. \ref{tab:Simulation-settings-CORSIKA}).
Using the weighted simulation (see Sec. \ref{sec:weights}), the expected
numbers of events in ORCA115 after 1 year of data taking were estimated.
The dip and peak structures above 90~km stem from the fact that the
simulation already used the fit from Fig. \ref{fig:atm_fit}. The
fit overpredicts the density at highest altitudes, leading to premature
start of some showers. \label{fig:1st_interaction_height}}
\end{figure}

The range of the atmospheric heights (altitudes) $h$ relevant for
the fit has been determined from the first interaction height $H_{\mathsf{int}}$
distribution (Fig. \ref{fig:1st_interaction_height}). It was concluded
that a fit should reproduce the model of the atmosphere at least till
$\sim70\,$km, which was the limit of validity (within $\pm5\%$)
of the previously used fit \cite{Thomas_Heid_phd_thesis}. The fit
had to cover the lower altitudes as well, to ensure CORSIKA propagates
the showers down to the sea level ($h=0$~km) through a realistic
approximation of the atmosphere. The maximum extent of the atmosphere
in CORSIKA is $112.8$~km and above, the density is clipped to zero
by the code, regardless of the actual fit range \cite{CORSIKA-Userguide}.
The NRLMSIS-2.0 model was well suited here, since it covers the heights
up to thousands of km, unlike e.g. the GDAS model, supplied together
with CORSIKA, which only goes up to $h=25\,$km \cite{GDAS,CORSIKA-Userguide}.
A further advantage of NRLMSIS-2.0 was the flexibility of choice of
any geographical location of the detector, which was fixed in the
predefined models within CORSIKA.

\begin{figure}[H]
\centering{}\includegraphics[width=12cm]{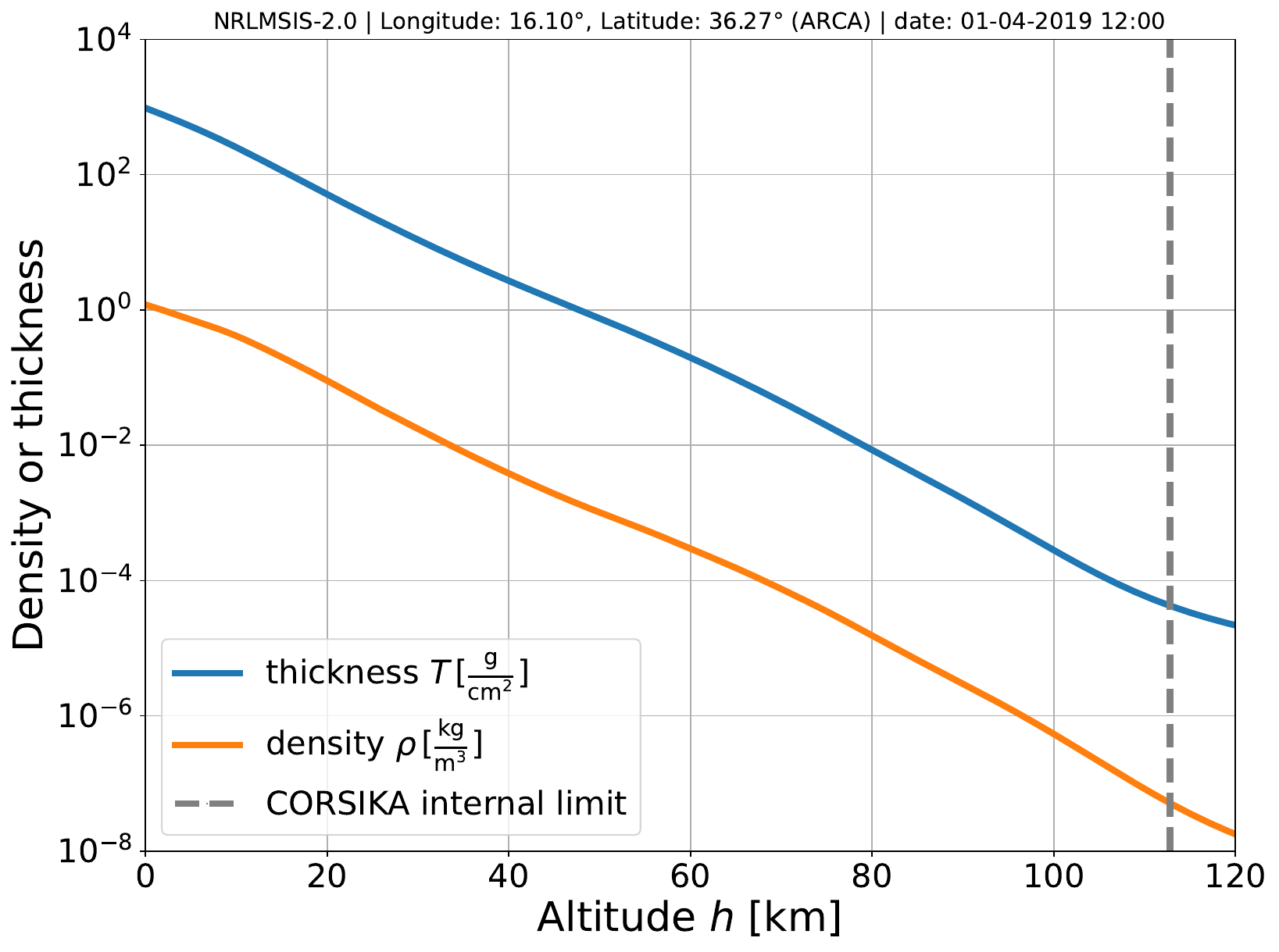}\caption{Atmospheric density and thickness as a function of height above the
ARCA site, according to \foreignlanguage{english}{NRLMSIS-2.0}. \label{fig:atm_density_into_thickness}}
\end{figure}

\begin{figure}[H]
\centering{}\subfloat[Thickness $T$. \label{fig:different_hmax-1}]{\centering{}\includegraphics[width=8cm]{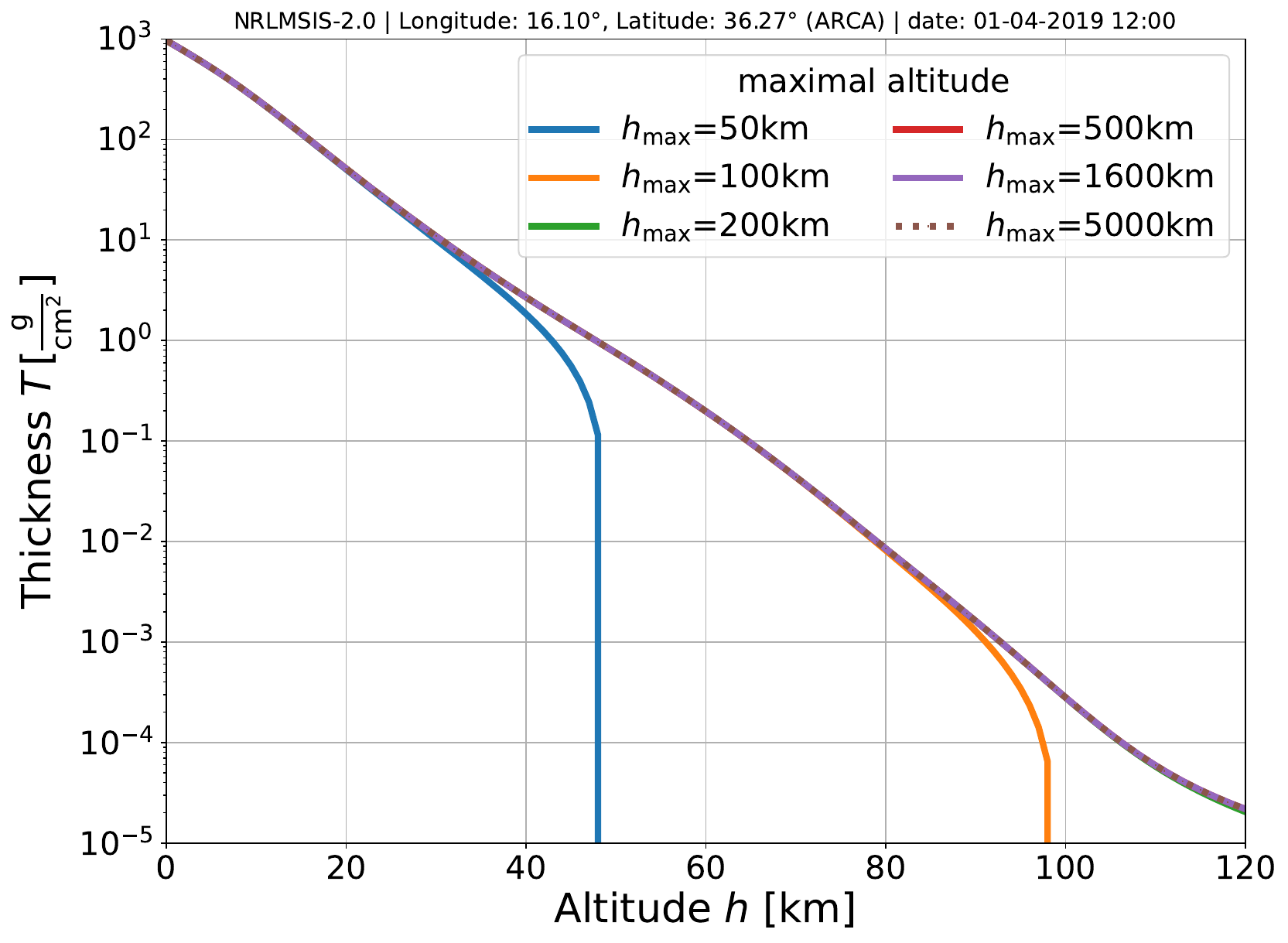}}\subfloat[Relative difference with respect to $T_{h_{\mathsf{max}}=5000\,\mathrm{km}}$.
\label{fig:different_hmax-2}]{\centering{}\includegraphics[width=8cm]{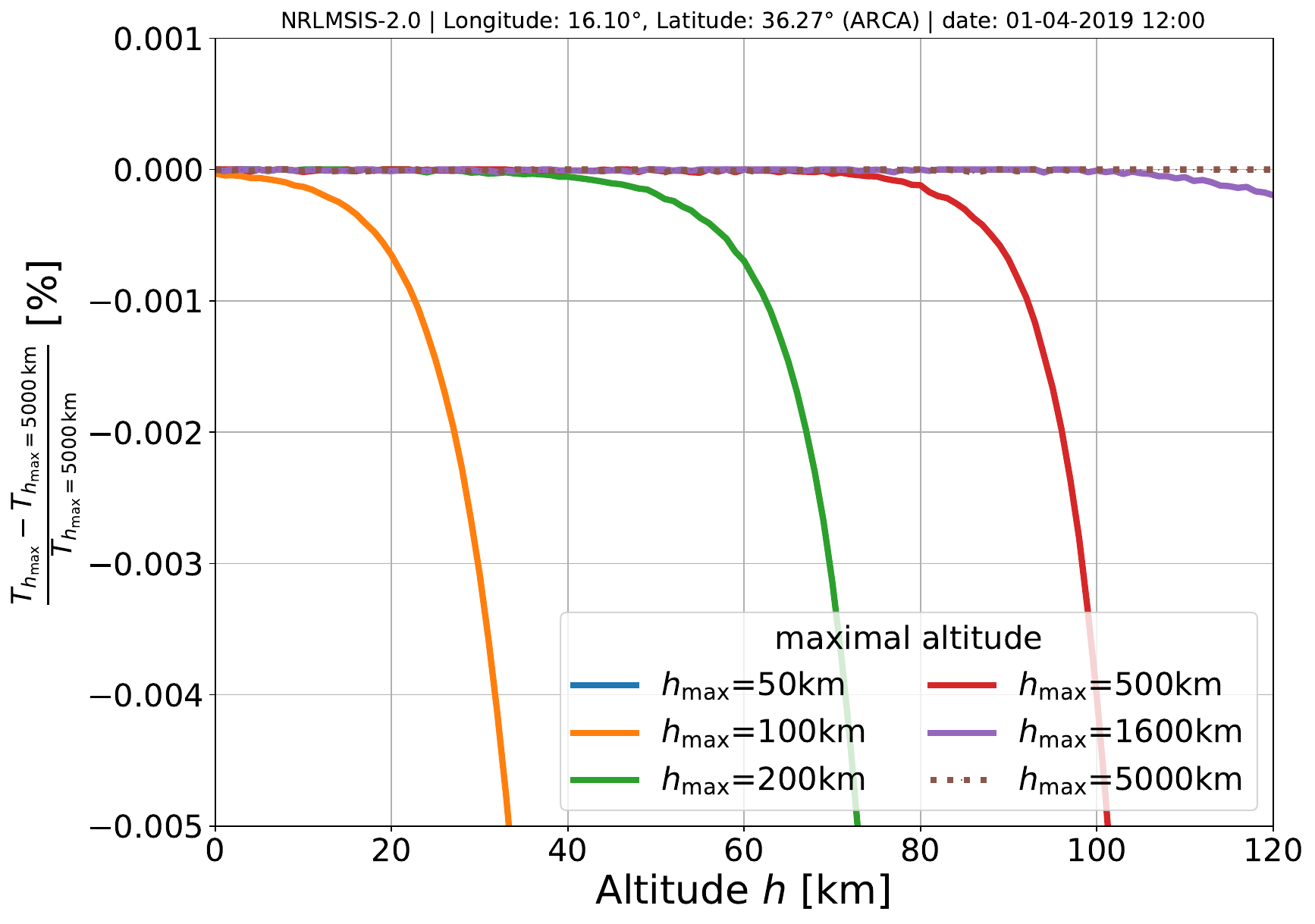}}\caption{Atmospheric thickness as function of height computed for different
$h_{\mathsf{max}}$ values. The location is at the ARCA site. \label{fig:different_hmax}}
\end{figure}

The NRLMSIS-2.0 model does not directly predict the atmospheric thickness
$T\left(h_{0}\right)$ at a certain height, which is the input required
by CORSIKA. Instead, it provides an estimate of the density of the
atmosphere at a fixed location and time, which can be recalculated
into atmospheric thickness (see Fig. \ref{fig:atm_density_into_thickness})
by integrating over the atmosphere above the height of interest $h_{0}$
:

\begin{equation}
T\left(h_{0}\right)=\stackrel[h_{0}]{h_{\mathsf{max}}}{\int}\rho dh,
\end{equation}

where $h_{\mathsf{max}}$ is the height at which the atmosphere is
assumed to end (have negligible density). When computed numerically,
the integral becomes a sum over a histogram with very fine steps (in
this case 1~m) and the atmosphere has to be clipped at a certain
point. The range of altitudes interesting from the point of view of
CORSIKA simulations for KM3NeT extends up to $\sim113\,$km, however
it is most important to cover the low altitudes around 20-40~km (see
Fig. \ref{fig:1st_interaction_height}). Fig. \ref{fig:different_hmax}
demonstrates the effect of varying the $h_{\mathsf{max}}$ parameter.
The clipping point was picked to be $h_{\mathsf{max}}=1600$~km,
which ensures that up to $112.8$~km the model prediction is within
$0.0005\%$ from the unclipped case, in Fig. \ref{fig:different_hmax}
approximated by $h_{\mathsf{max}}=5000\,$km.

\begin{figure}[H]
\centering{}\includegraphics[width=12cm]{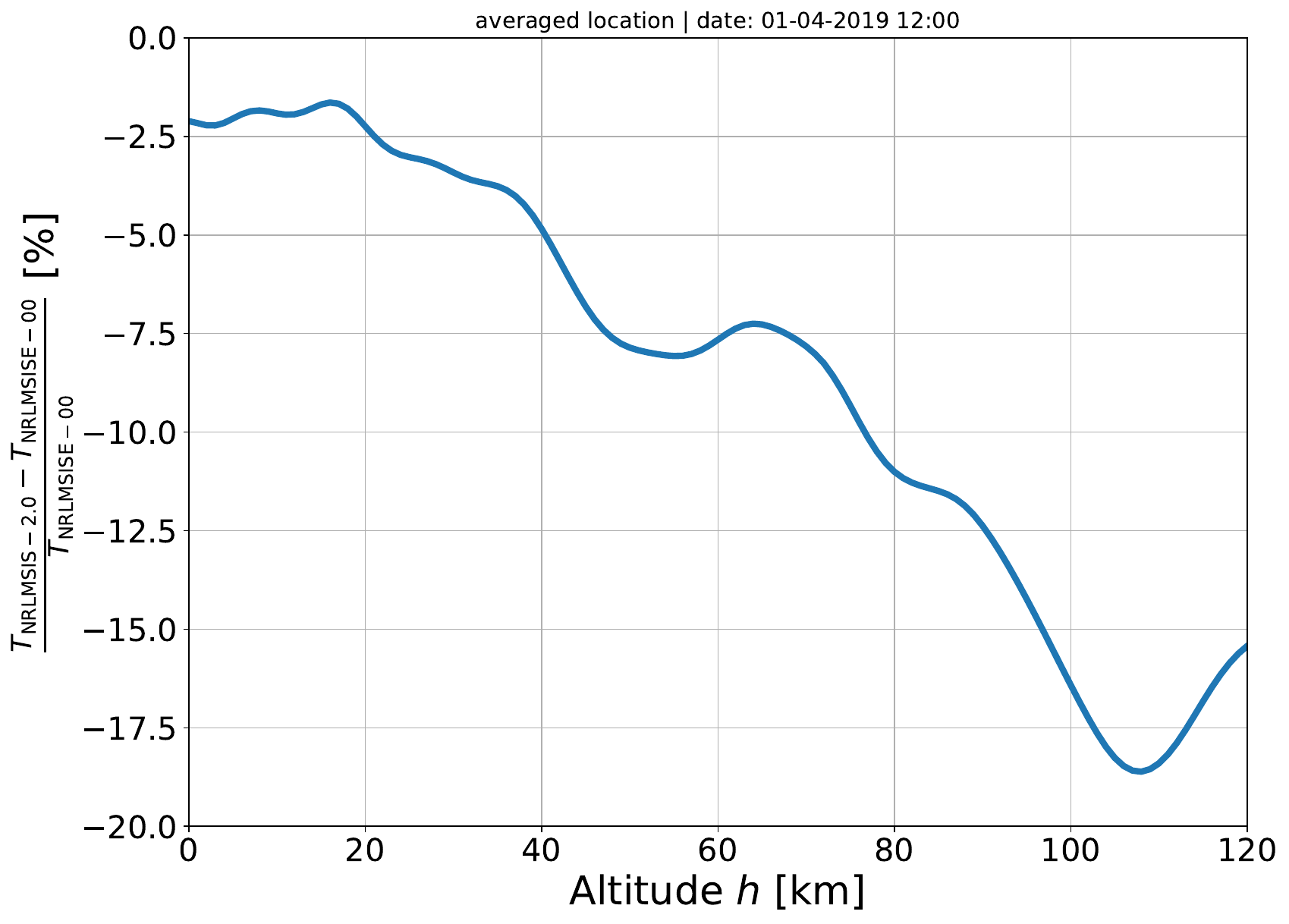}\caption{Ratio of atmospheric thickness of \foreignlanguage{english}{NRLMSIS-2.0
and }NRLMSISE-00 to their average, at ARCA site\foreignlanguage{english}{.}
\label{fig:NRLMSIS-00-comparison}}
\end{figure}

There is in fact an older atmospheric density model: NRLMSISE-00 \cite{NRLMSISE-00},
which was a predecessor of NRLMSIS-2.0. NRLMSISE-00 was used in the
KM3NeT CORSIKA test simulations, prior to this work. The relative
difference between the two is between 1\% and 18~\%, depending on
the altitude (see Fig. \ref{fig:NRLMSIS-00-comparison}). The NRLMSIS-2.0
was a significant upgrade of NRLMSISE-00, addressing a number of issues
\cite{NRLMSIS-2.0}:
\begin{enumerate}
\item Thermospheric densities were fully coupled to the entire temperature
profile from the ground to the exosphere, while in NRLMSISE-00 they
were treated independently from the lower layers of the atmosphere.
\item Upper thermosphere in the model was tuned to better match new satellite
mass density measurements.
\item Many new temperature measurements in the mesosphere, stratosphere,
and troposphere have been included.
\item A number of new atomic oxygen and hydrogen measurements in the mesosphere
were added. 
\end{enumerate}
Not all changes are listed, however the changes in NRLMSIS-2.0 are
mostly focused on altitudes below 100 km \cite{NRLMSIS-2.0}, which
are the most relevant ones for the CORSIKA simulations in this thesis.

\begin{figure}[H]
\centering{}\includegraphics[width=12cm]{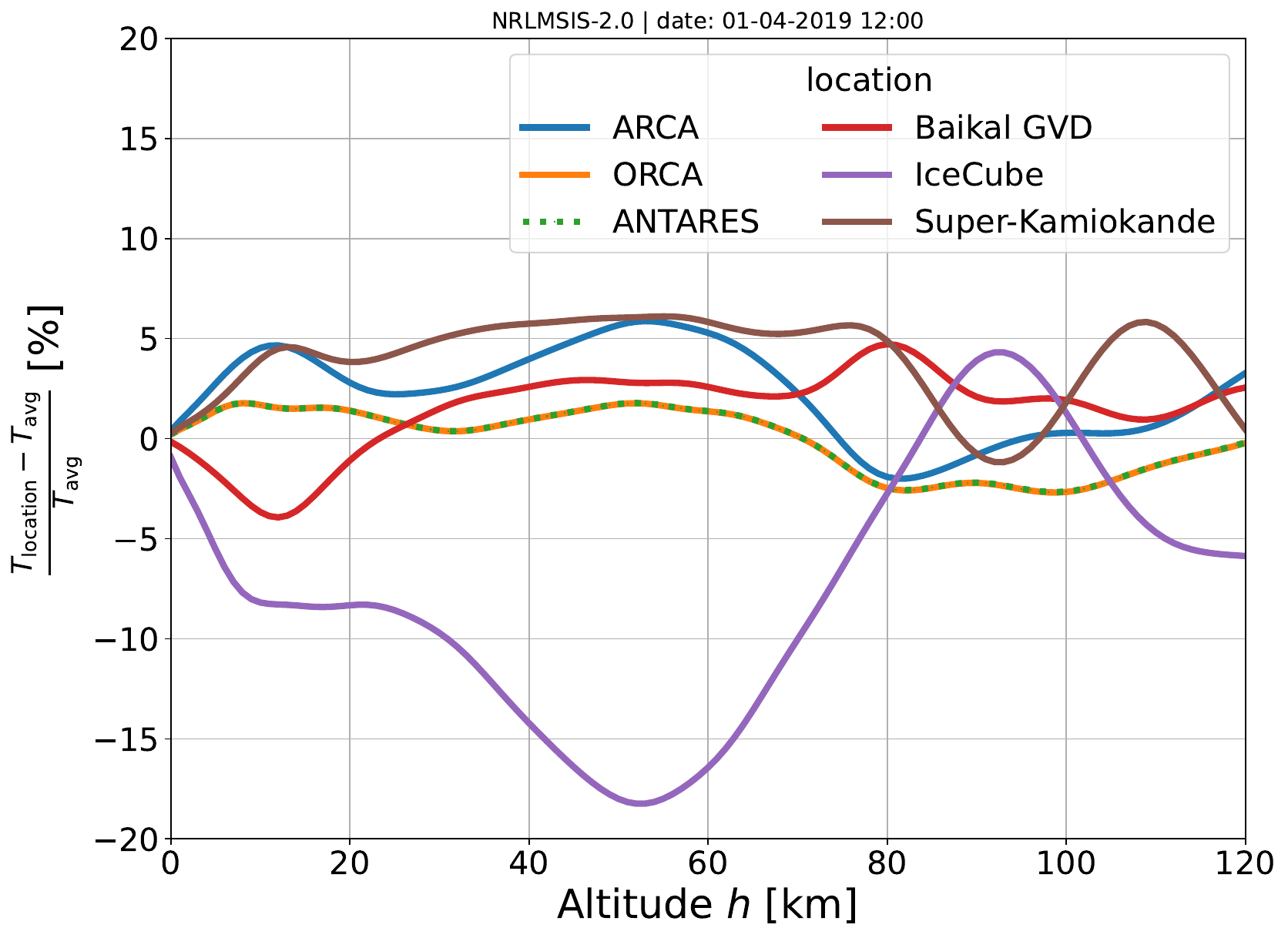}\caption{Ratio of atmospheric thickness computed at ARCA, ORCA, and other experimental
sites from Fig. \ref{fig:GNN_map} to their average, plotted as function
of the height\foreignlanguage{english}{. IceCube atmosphere clearly
stands out from the other ones at low altitudes and it is a real effect
of the different atmospheric conditions at the geographic South Pole
\cite{IceCube_seasonal_muons,IceCube_seasonal_neutrinos}.} \label{fig:atm_location}}
\end{figure}

An important issue to investigate were the variations in model predictions
depending on the chosen location and time. The comparison of the atmosphere
at ARCA and ORCA sites, along with ANTARES x, Baikal GVD, IceCube
and Super-Kamiokande (SK) sites added for comparison \cite{ANTARES_seasonal,Baikal-GVD,IceCube_seasonal_muons,AMANDA=000026IC,SuperKamiokande},
is shown in Fig. \ref{fig:atm_location}. An interesting observation
from Fig. \ref{fig:atm_location} is that the differences between
the atmospheric properties of both KM3NeT sites and Baikal and SK
seem to be rather small at the considered altitudes. This indicates
that these two experiments could potentially share a CORSIKA MC production
with KM3NeT if there are not other factors preventing it (e.g. the
magnetic field strength). The same conclusion is certainly true for
ANTARES, however its case is more trivial, as it was (the detector
was dismantled on 12.02.2022) located very close to ORCA. IceCube
on the other hand, due to its unique location at the geographic South
Pole, has noticeably different atmospheric thickness, which seems
to be prone to larger variations with height.

\begin{figure}[H]
\centering{}\includegraphics[width=12cm]{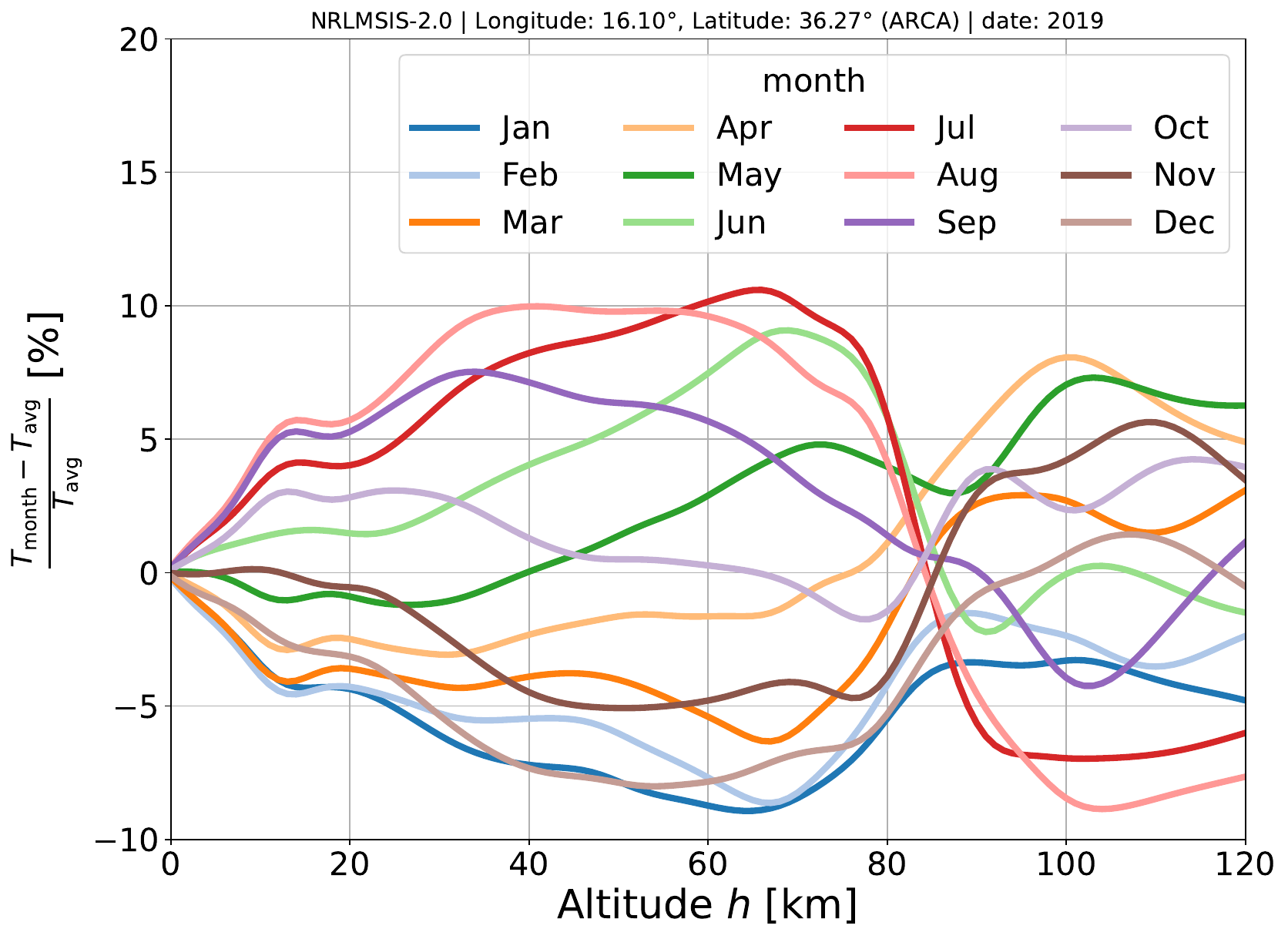}\caption{Ratio of atmospheric thickness for different months in 2019 to their
average\foreignlanguage{english}{. The choice of the year was made
only such that it overlaps with the actual data taking by KM3NeT detectors.
The result is shown for the ARCA location.} \label{fig:atm_month}}
\end{figure}

\begin{figure}[H]
\centering{}\includegraphics[width=12cm]{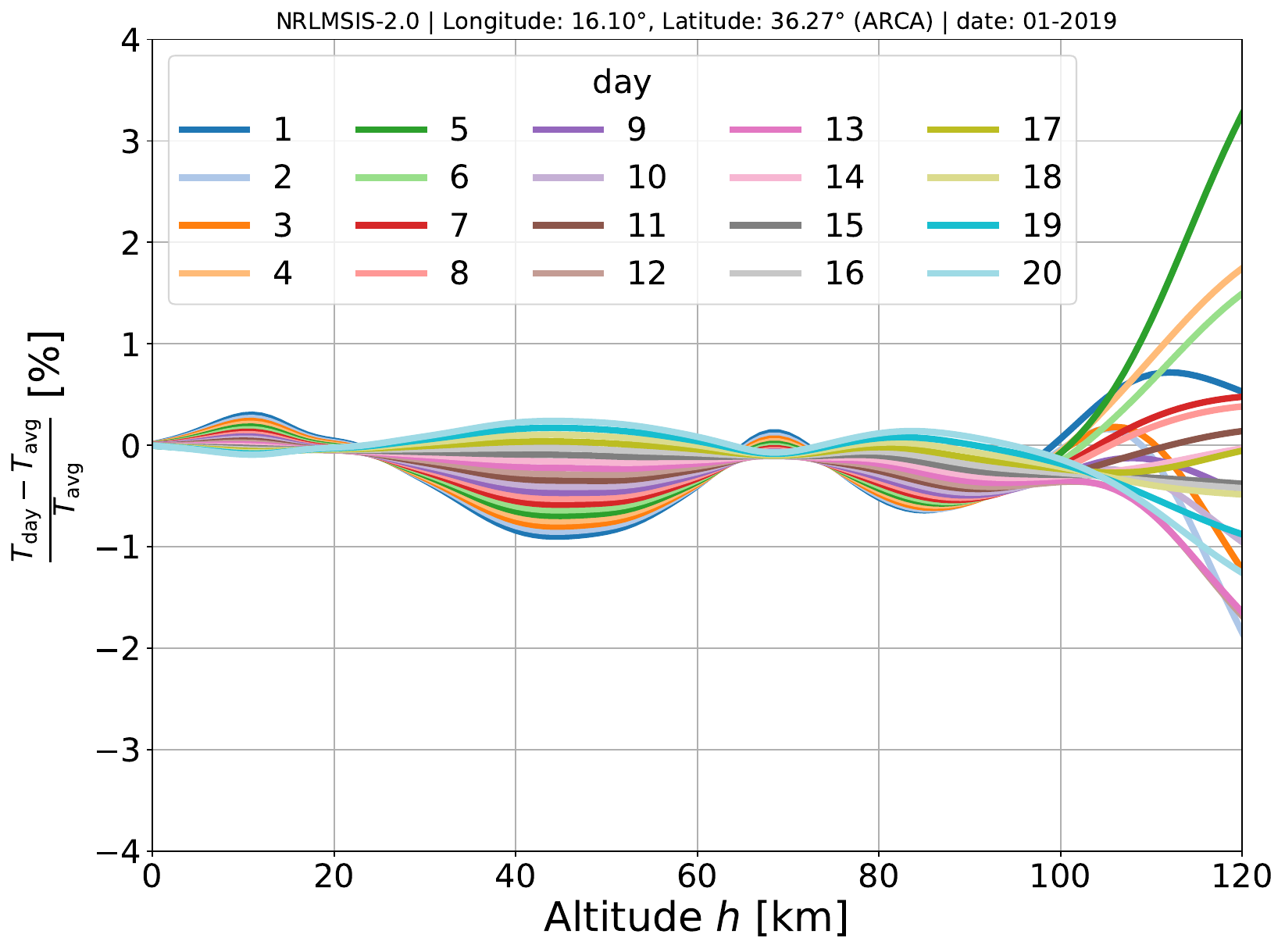}\caption{Ratio of atmospheric thickness for individual days of January 2019
to their average\foreignlanguage{english}{. The result is shown for
the ARCA location. The day-to-day changes in the atmospheric density
are rather negligible (however, they accumulate into the monthly variations,
which are not).} \label{fig:atm_day}}
\end{figure}

\begin{figure}[H]
\centering{}\includegraphics[width=12cm]{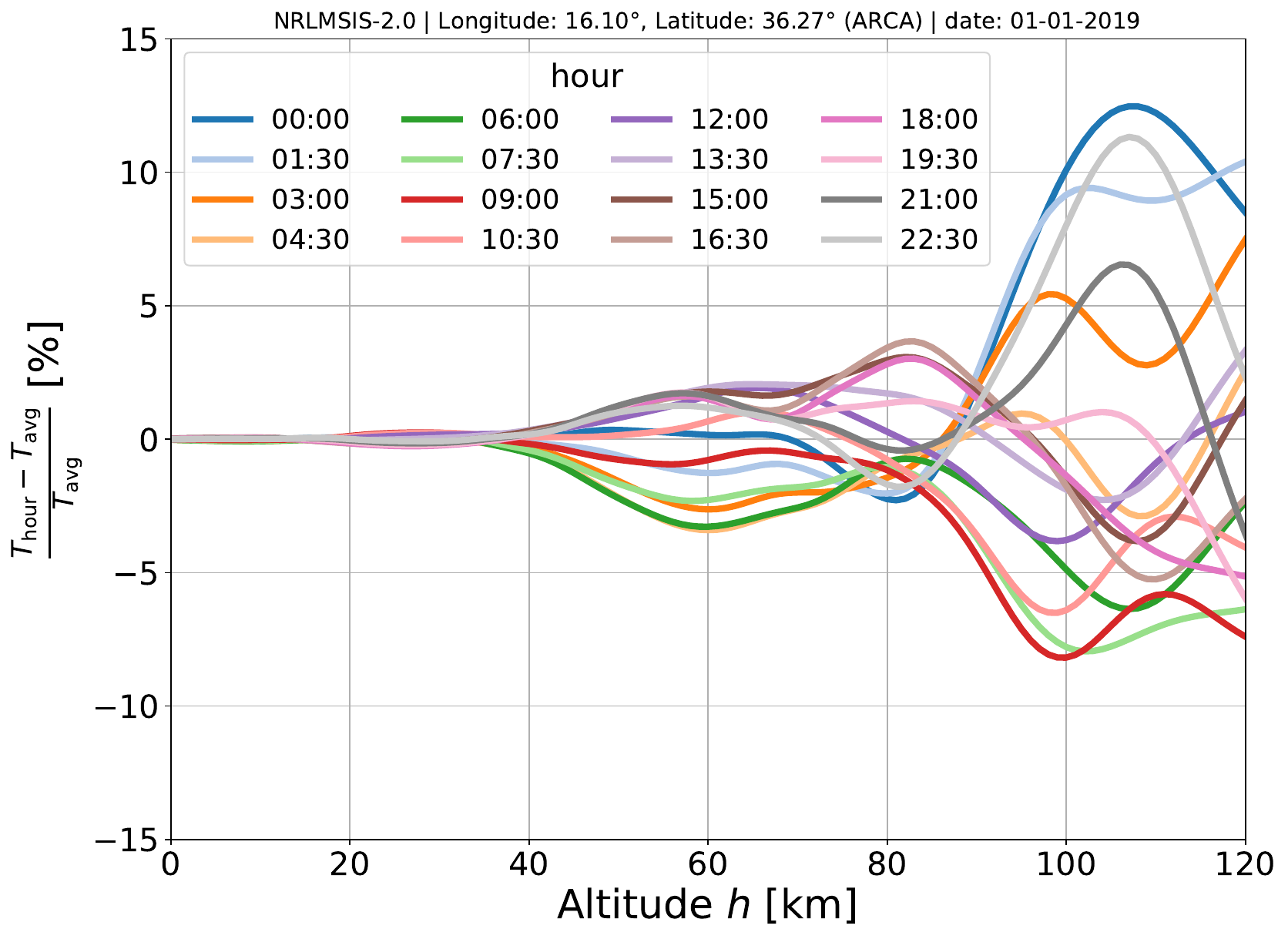}\caption{Ratio of atmospheric thickness for individual hours to their average\foreignlanguage{english}{
on 01.01.2019. The result is shown for the ARCA location. The hourly
changes in the atmosphere, especially at higher altitudes are much
stronger than the daily variations (Fig. }\ref{fig:atm_day}). This
is connected with the day-night cycle and the corresponding heating
and cooling of the atmosphere, which affects the density. \label{fig:atm_hour}}
\end{figure}

Seasonal variations in muon flux have already been measured by the
IceCube and ANTARES experiments to be about 10~\% and 2~\% respectively
\cite{IceCube_seasonal_neutrinos,IceCube_seasonal_muons,ANTARES_seasonal}.
Monthly atmosphere variability predicted by NRLMSIS-2.0 is shown in
Fig. \ref{fig:atm_month} and the thickness varies up to $\sim\pm10\,\%$
at around $70\,$km above the sea level. However, one should refrain
from a conclusion that the expected variation in the muon flux will
also be on the order of $10\,$\%. As was shown in Fig. \ref{fig:1st_interaction_height},
most of the showers typically start at $\sim30$~km, where the thickness
varies by about 6~\%. The seasonal changes may in fact shift the
peak of the $H_{\mathsf{int}}$ distribution, since there will be
a bigger or smaller slant depth before reaching a certain height.
Generally, the lower the first interaction happens, the better is
the chance for the shower to reach the sea surface, and hence the
higher muon flux can be expected. Neutrino flux should be less affected,
since neutrinos do not loose energy on their way. There is typically
much less difference between the days of a single month. In Fig. \ref{fig:atm_day},
one can see that the variations are below 1~\% in the range of interest.
Stronger effects may be seen in hourly changes, associated with the
day-night cycle, as shown in Fig. \ref{fig:atm_hour}. The effect
is the strongest at very high altitudes, higher than the typical first
interaction height (see Fig. \ref{fig:1st_interaction_height}). For
this reason, the effect on the total atmospheric muon flux may be
hard to measure. The yearly differences were not plotted, however
they are very small, of the order of 0.01~\%.

\begin{figure}[H]
\centering{}\subfloat[Atmospheric thickness fit.]{\centering{}\includegraphics[width=8cm]{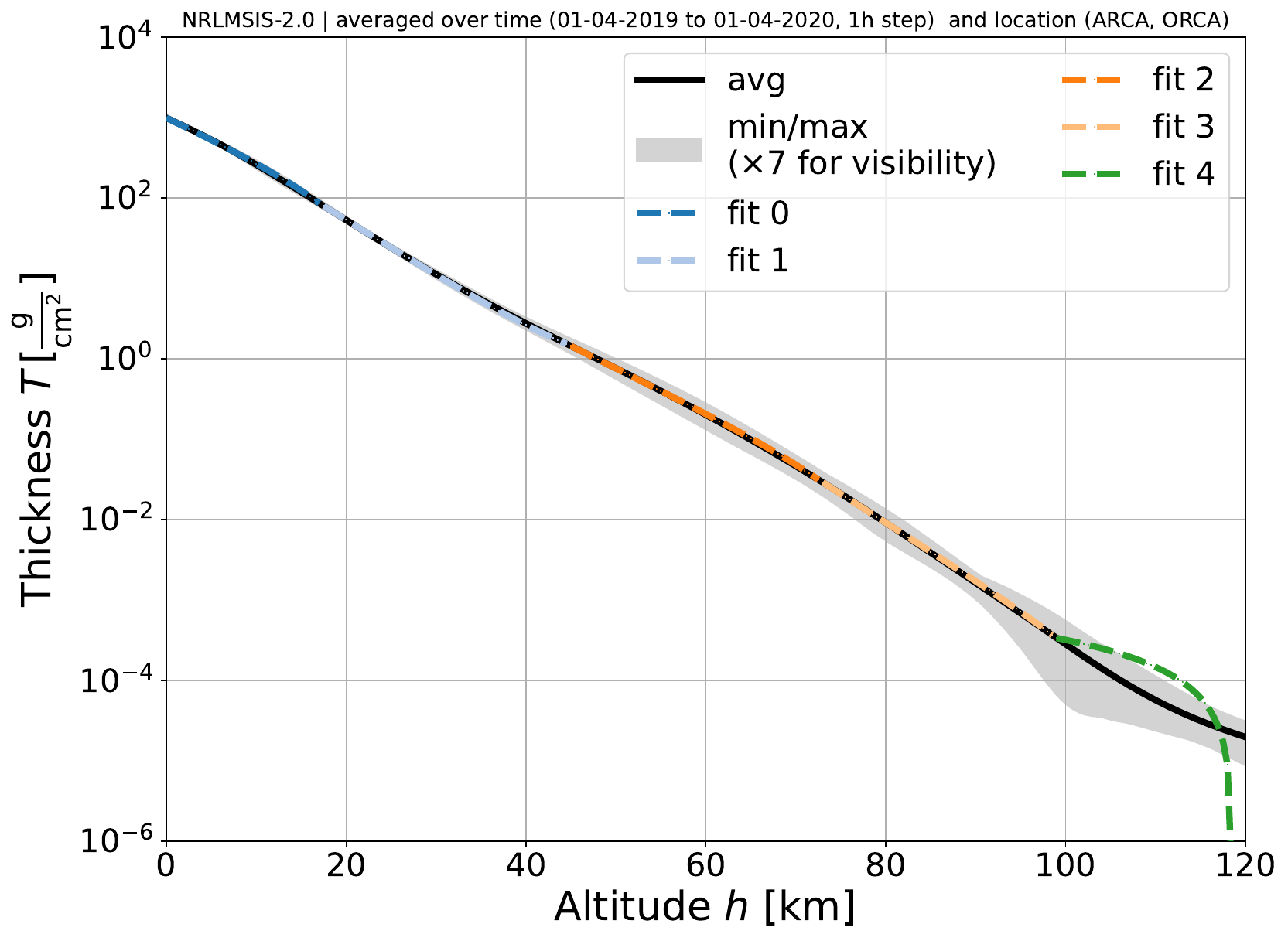}}\subfloat[Ratio of atmospheric thickness to the average thickness from the model.]{\centering{}\includegraphics[width=8cm]{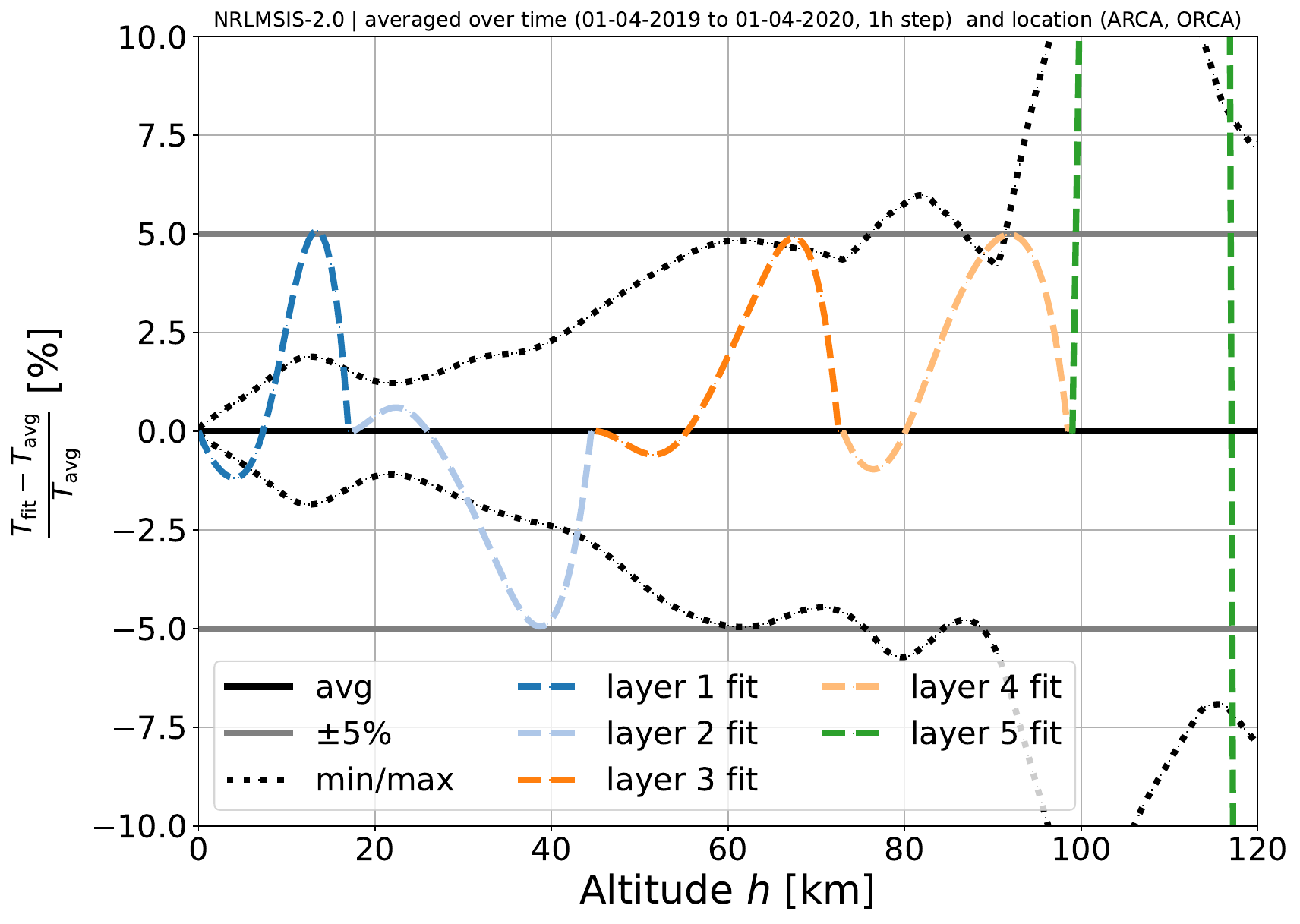}}\caption{Atmospheric thickness \foreignlanguage{english}{fit performed for
5 CORSIKA layers (}colour-coded\foreignlanguage{english}{). The avg
stands for the average thickness, as predicted by }NRLMSIS-2.0 and
min/max indicates the corresponding possible extremal values when
varying the location and time. The horizontal grey bands indicate
the $\pm5\,\%$ validity interval. \label{fig:atm_fit}}
\end{figure}

The fit of the NRLMSIS-2.0 model has been performed on the prediction
averaged over the time (period from 01.04.2019 to 01.04.2020, with
a step of 1 hour) and location (ARCA and ORCA sites). Each of the
5 CORSIKA atmosphere layers (see Eq. \ref{eq:layers_1-4} and \ref{eq:layer_5})
was fitted separately. Each of the fits was bound at least from one
side by the requirement that the layers should connect exactly. The
resulting fit, presented in Fig. \ref{fig:atm_fit}, is stable and
within $\pm5\,\%$ around the average thickness from the model up
to 99~km. This is a significant improvement with respect to the previously
used fit, which was stable within $\pm5\,\%$ only up to 70~km \cite{Thomas_Heid_phd_thesis}.
The fit parameters are reported in Tab. \ref{tab:Simulation-settings-CORSIKA}
as `Atmosphere fit'.

\subsection{Magnetic field strength \label{subsec:Magnetic-field-strength}}

The mean Earth's magnetic field strength used for CORSIKA simulations
was estimated using the International Geomagnetic Reference Field
(IGRF) \cite{IGRF}. CORSIKA requires the north and east components
of the magnetic field ($B_{x}$ and $B_{z}$ respectively) as input.
The values $B_{x}=25.2179\,$μT and $B_{z}=38.4848\,$μT were computed
for the location in between the ARCA and ORCA sites: (40.825214$\lyxmathsym{\textdegree}$
N, 10.442726$\lyxmathsym{\textdegree}$ E), and averaged over the
year 2019. It was also checked that, according to IGRF model, the
yearly variations in the magnetic field components are at most of
the order of $\pm6\,\%$, which allows for rather stable results,
regardless of the date used.

\subsection{Benchmarking and energy thresholds\label{sec:CORSIKA-benchmarking-and-opti}}

Here, a number of tests performed on CORSIKA MC are highlighted. Their
main purpose was to optimise the settings in Tab. \ref{tab:Simulation-settings-CORSIKA},
and assess the feasibility of CORSIKA mass production with the computational
resources available within The KM3NeT Collaboration. 

The impact of the minimal electron and photon energies $E_{e}^{\mathsf{min}}$
and $E_{\gamma}^{\mathsf{min}}$ was checked to be of little importance
for CORSIKA muon simulations and set to 1~TeV. Varying each by two
orders of magnitude only results in at most 5~\% difference in expected
muon rate at sea level, and at can level the effect was even weaker.

\begin{figure}[H]
\centering{}\includegraphics[width=12cm]{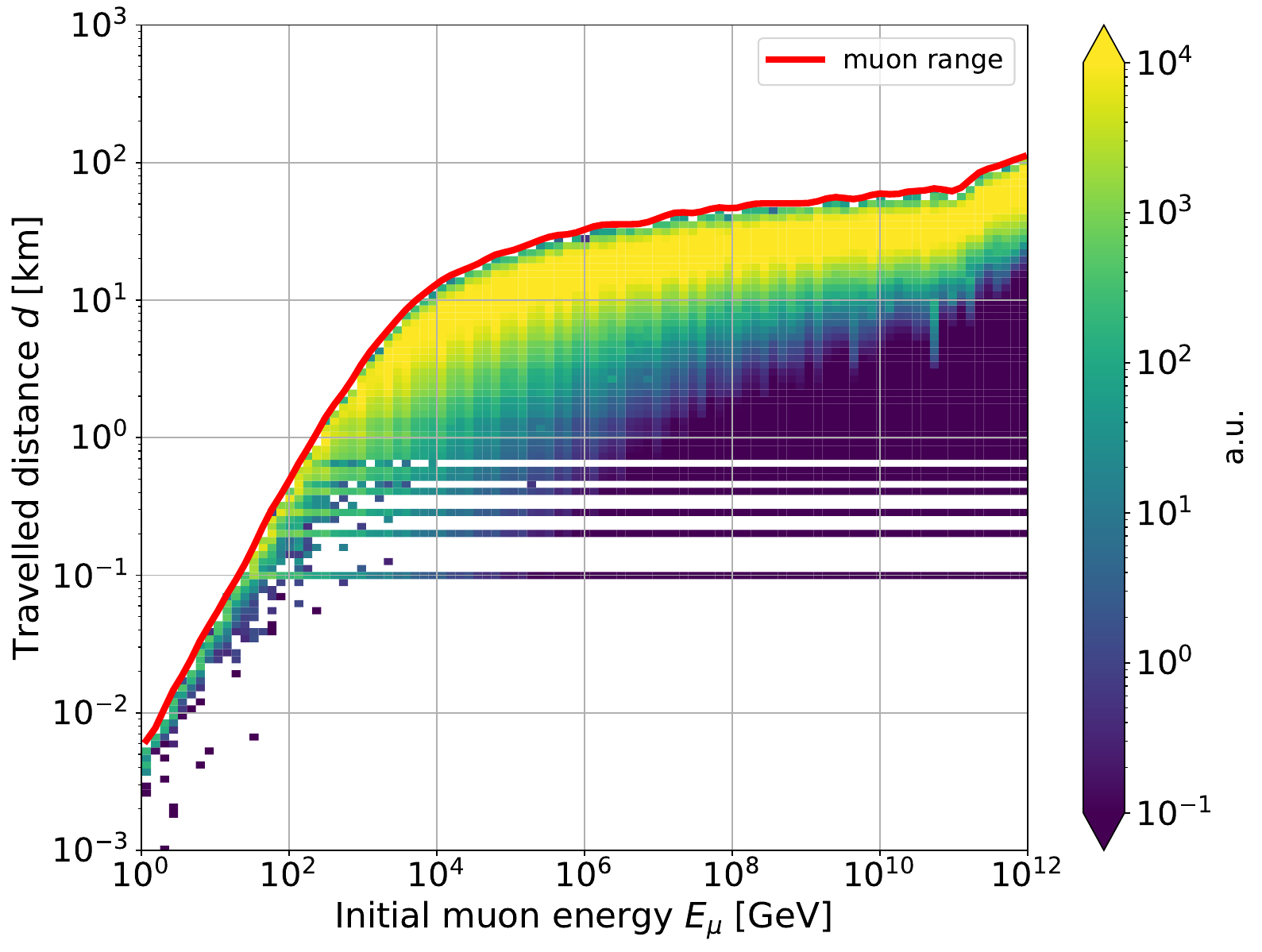}\caption{Distance $d$ travelled by muons in seawater as a function of their
initial energy $E_{\mu}$. The red line is the maximum of the histogram
increased by half of the standard deviation and is an estimate of
the muon range in seawater. The horizontal lines in the histogram
are an artefact of the fact that the muon positions were recorded
every 100~m (or after stopping, but this did not contribute to the
effect; see Sec. \ref{sec:muon_lateral_deflection}). \label{fig:Muon-range}}
\end{figure}

The muon and hadron/nuclei energy thresholds play a much more important
role in KM3NeT CORSIKA simulations. The threshold for hadrons/nuclei
in practice has turned out to be equivalent to a threshold on neutrino
energy, hence in the following it is referred to as $E_{\nu}^{\mathsf{min}}$.
The choice of the minimal muon energy $E_{\mu}^{\mathsf{min}}$ was
dictated by the muon range in seawater and the shortest possible distance
to the can, which is the vertical distance from the sea surface to
the top of the can of ORCA: $d_{\mathsf{min}}=2440\mathsf{m}-476.5\mathsf{m}=1963.5\mathsf{m}$.
Muon range estimation, produced along with the other results from
Sec. \ref{sec:muon_lateral_deflection}, is shown in Fig. \ref{fig:Muon-range}.
Based on this result, the value of $E_{\mu}^{\mathsf{min}}=300\,$GeV
was adopted. Since muon neutrinos can produce only muons with smaller
energies than their own, the same cut was used for $E_{\nu}^{\mathsf{min}}$.

\subsubsection{Surviving shower fraction}

\begin{figure}[H]
\subfloat[Surviving fraction for different $E_{\mu}^{\mathsf{min}}$ values.]{\centering{}\includegraphics[width=8cm]{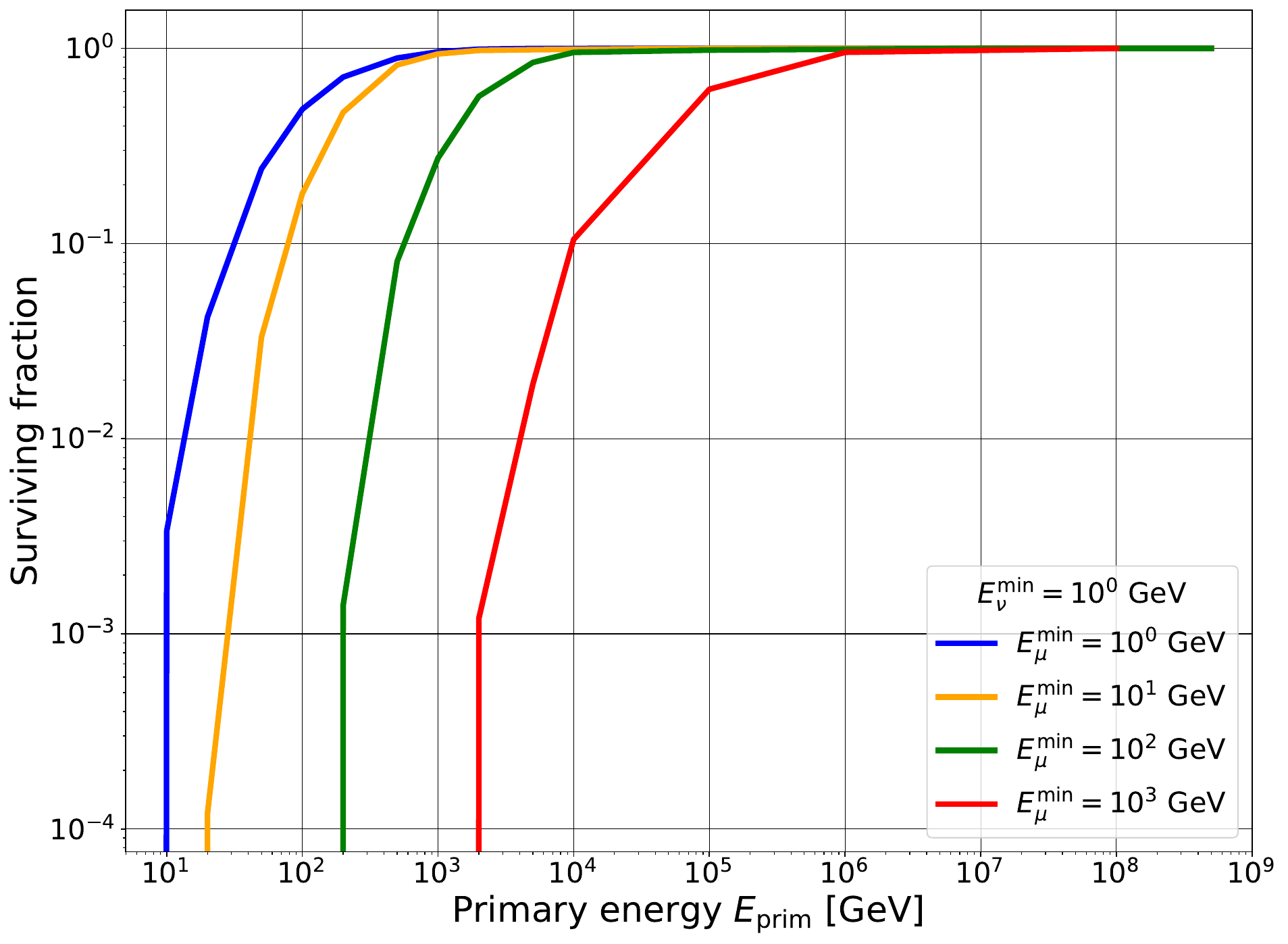}}\subfloat[Surviving fraction for different $E_{\nu}^{\mathsf{min}}$ values.]{\centering{}\includegraphics[width=8cm]{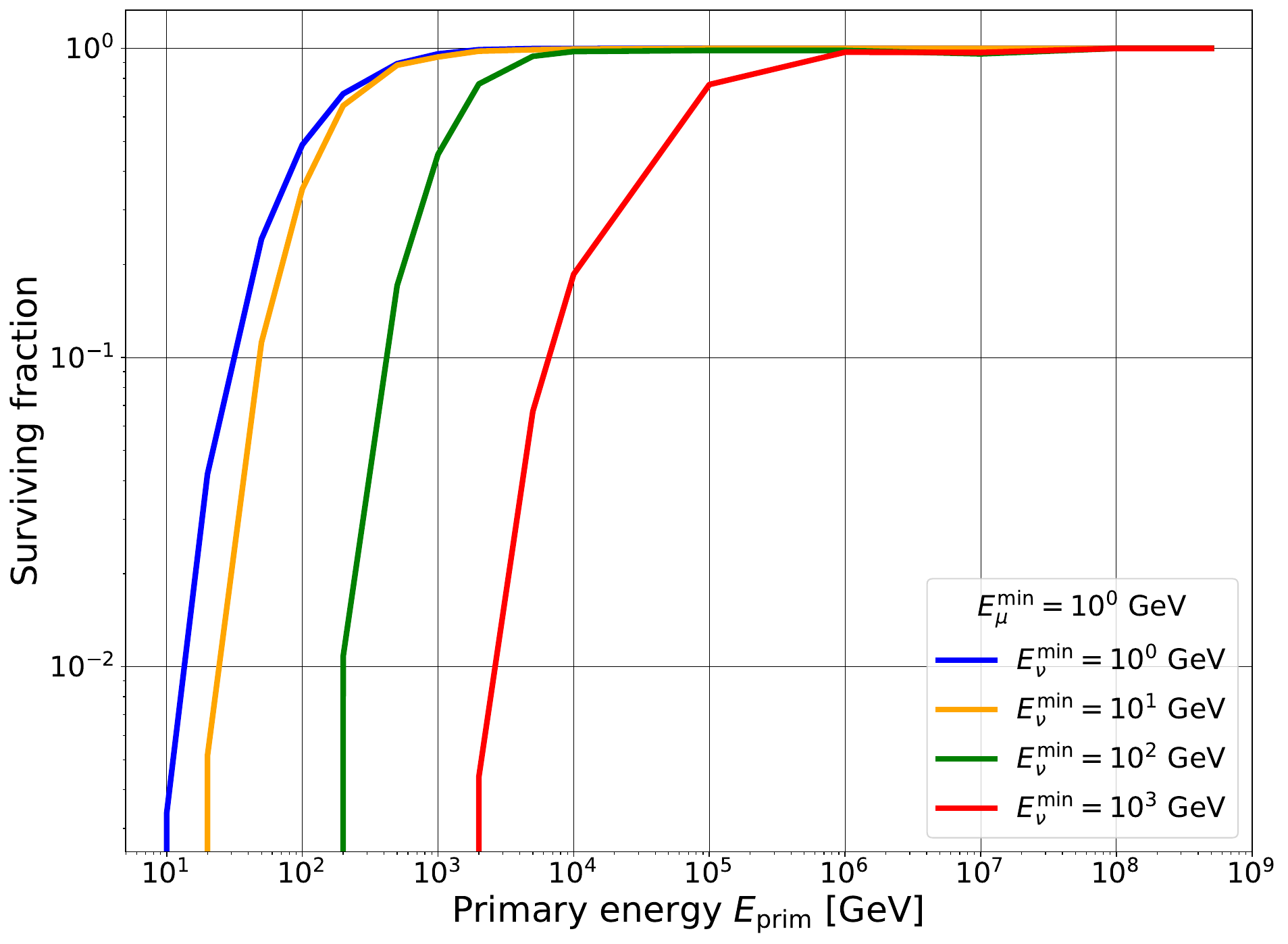}}\caption{Fraction of the generated showers, producing particles reaching the
sea level. \label{fig:Surviving_fraction}}
\end{figure}

The first investigated issue was the success rate of showers as a
function of the primary energy $E_{\mathsf{prim}}$. The results shown
in Fig. \ref{fig:Surviving_fraction} indicate that the $E_{\nu}^{\mathsf{min}}$
is slightly more constraining than $E_{\mu}^{\mathsf{min}}$, which
is related to the much smaller probability of interaction for neutrinos,
compared against muons (in fact, in CORSIKA simulation $\nu$ are
not interacted at all, only tracked). Thus, even very low-energy neutrinos
make it to the sea level, which is not true for muons, loosing energy
on their way. One may also note that showers with $E_{\mathsf{prim}}\gtrsim1\,$PeV
almost always survive.

\subsubsection{Storage and CPU time}

\begin{figure}[H]
\subfloat[Required disk space for different $E_{\mu}^{\mathsf{min}}$ values..]{\centering{}\includegraphics[width=8cm]{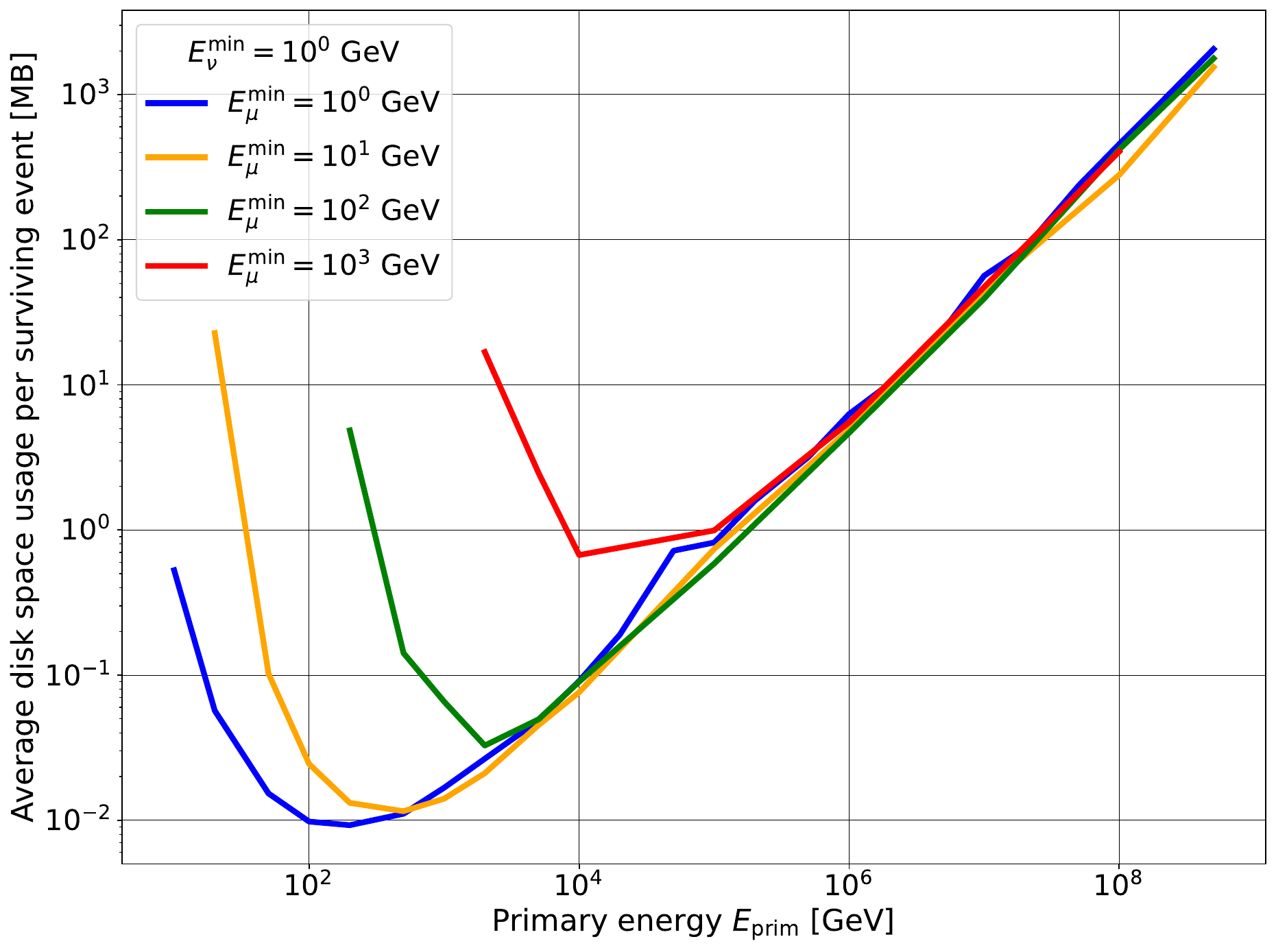}}\subfloat[Required disk space for different $E_{\nu}^{\mathsf{min}}$ values.]{\centering{}\includegraphics[width=8cm]{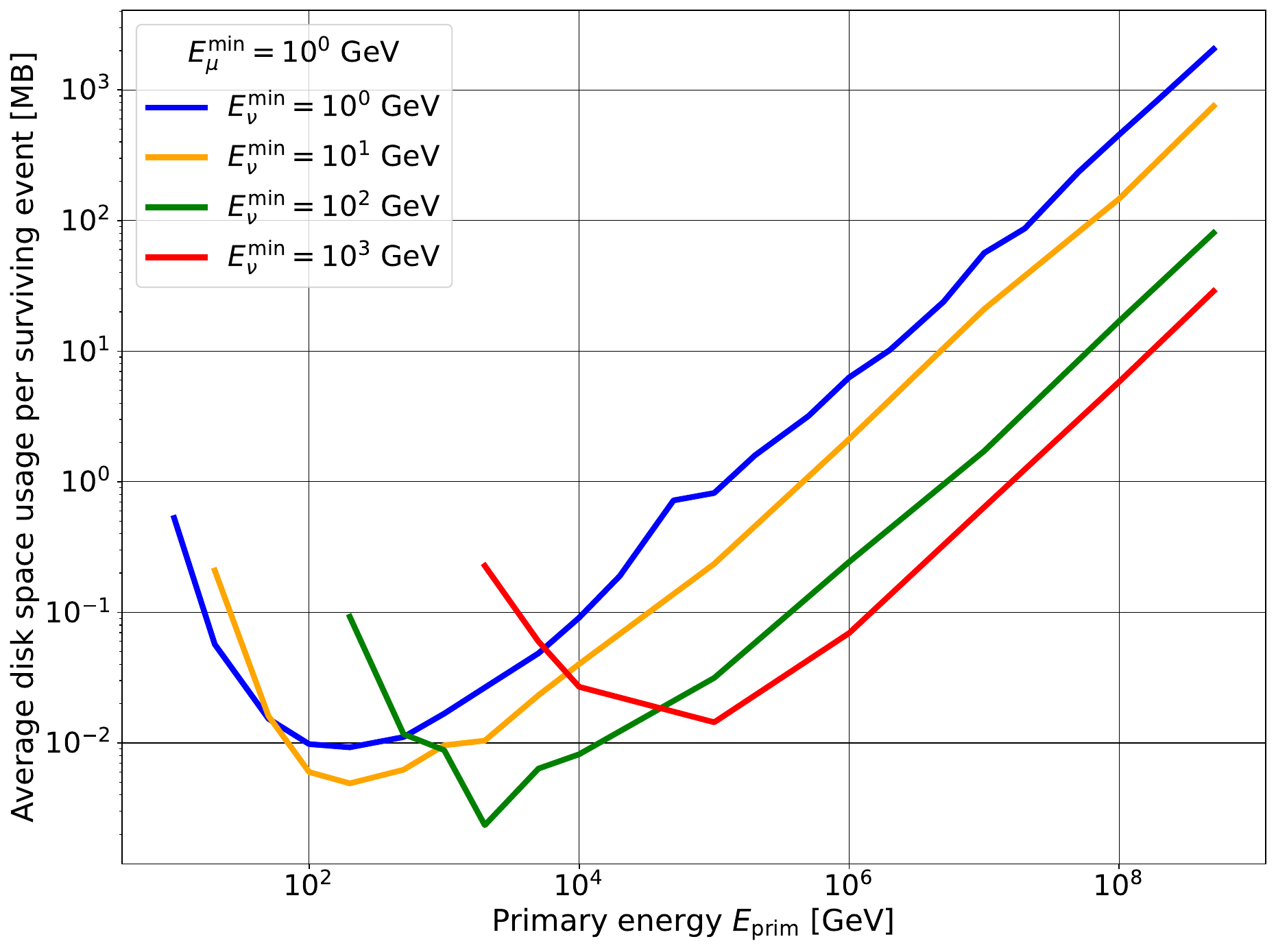}}\caption{Storage per successful event that is required to store CORSIKA  files
as a function of the primary energy. \label{fig:storage}}
\end{figure}

\begin{figure}[H]
\subfloat[Required CPU time for different $E_{\mu}^{\mathsf{min}}$ values..]{\centering{}\includegraphics[width=8cm]{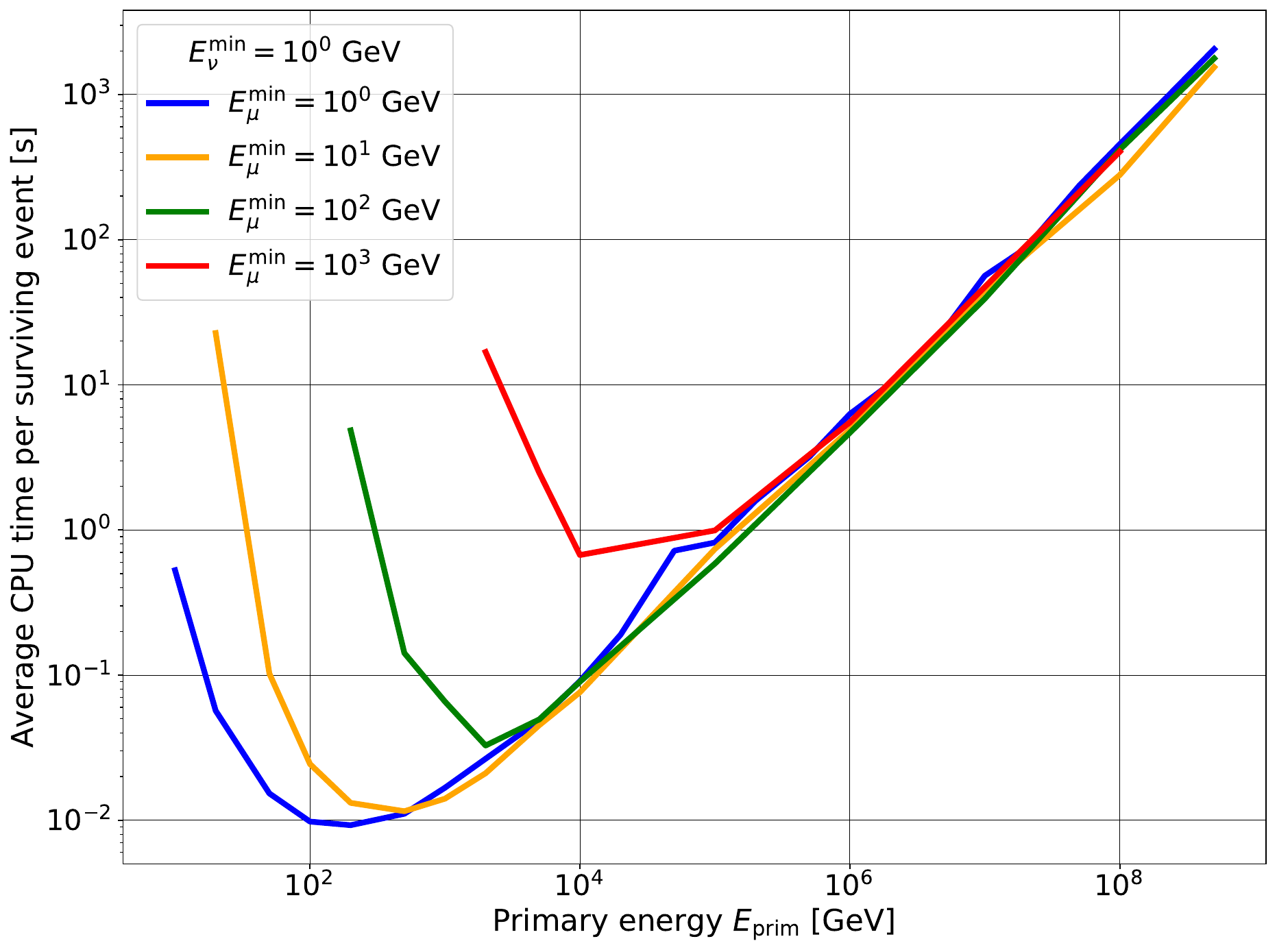}}\subfloat[Required CPU time for different $E_{\nu}^{\mathsf{min}}$ values.]{\centering{}\includegraphics[width=8cm]{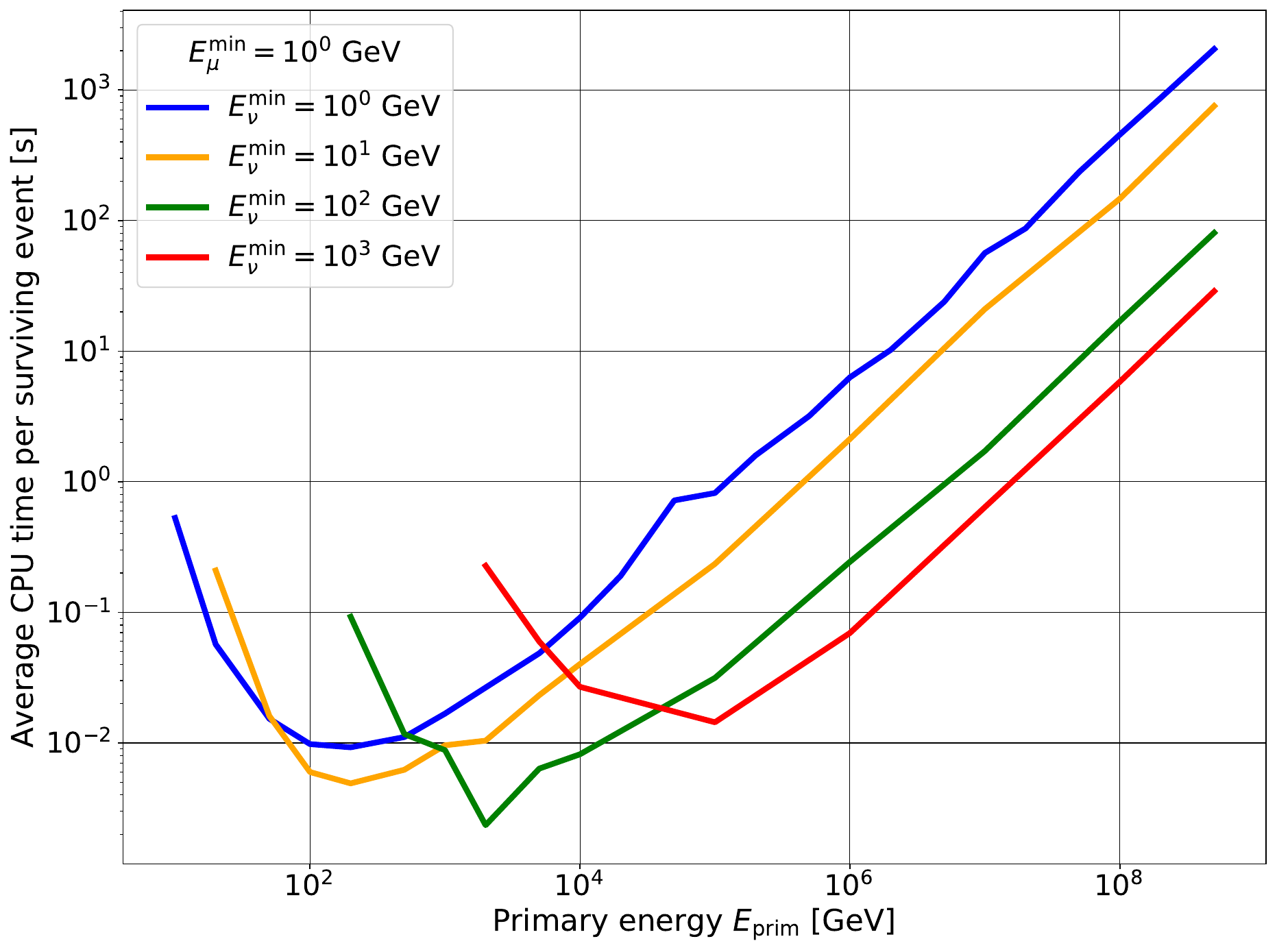}}\caption{CPU time per successful event that is required to process CORSIKA
files as a function of the primary energy. \label{fig:time}}
\end{figure}

Next, the storage and the time needed to produce a certain number
of events at sea have been investigated. All the performance checks
were performed on Intel(R) Xeon(R) CPU E5-2680 v2 @ 2.80GHz with 126GB
RAM. As visible in Fig. \ref{fig:storage} and \ref{fig:time}, both
required CPU time and disk space seem to have one global minimum (naturally,
shifting with the energy thresholds). Beyond a certain $E_{\mathsf{prim}}$
range around this minimum, the simulation seems if not completely
impossible, then at least very challenging. The increase in required
resources is particularly dramatic towards lower energies, as there,
the chance to produce secondaries that can possibly reach the sea
diminishes rapidly. On the other hand, the atmospheric muon and neutrino
fluxes peak at low energies, which means it is extremely important
to properly simulate that region. This has been one of the motivations
for splitting up the originally defined TeV sub-production into two,
more fine-tuned ones: TeV\_low and TeV\_high (see Tab. \ref{tab:Simulation-settings-CORSIKA}).

\section{Formulae}

This section gathers the derivations of some of the equations used
throughout this thesis.

\subsection{Errors for the weighted histograms\label{sec:Derivation-of-the-errors-weighted-histo}}

The value of each bin in the weighted histogram is given by

\[
N_{i}=\underset{j}{\sum}w_{ij},
\]

where $w_{ij}$ is the weight of the $j$-th event, falling into the
$i$-th bin. The error of this quantity is related to its variance:

\begin{equation}
\Delta N_{i}=\sqrt{\mathsf{Var}\left(\underset{j}{\sum}w_{ij}\right)}=\sqrt{\underset{j}{\sum}\mathsf{Var}\left(w_{ij}\right)}=\sqrt{\underset{j}{\sum}w_{ij}^{2}}.\label{eq:hist_error}
\end{equation}

The systematic uncertainties can be included by modifying the event
weights: 

\begin{equation}
\Delta N_{i}=\sqrt{\underset{j}{\sum}\left[w'_{ij}\right]^{2}}=\sqrt{\underset{j}{\sum}\left[w_{ij}\cdot\left(1+\Delta w_{\mathrm{syst},\,j}\right)\right]^{2}},\label{eq:hist_error_2}
\end{equation}

where $\Delta w_{\mathrm{syst},\,j}$ is evaluated for the $j$-th
event according to the results from Sec. \ref{subsec:Results-systematics}.

\subsection{The shortest distance between a point and a line\label{sec:The-shortest-distance}}

The result of this derivation was applied to a number of tasks within
this thesis, notably during the implementation of a more accurate
calculation of event weights in gSeaGen (see Sec. \ref{sec:Computation-of-DistaMax}).
The general vector formula for the point-line distance is:

\[
d=\frac{\left|\left(\vec{x}_{0}-\vec{x}_{1}\right)\times\left(\vec{x}_{0}-\vec{x}_{2}\right)\right|}{\left|\vec{x}_{2}-\vec{x}_{1}\right|},
\]

where $\vec{x}_{0}$ is the point and $\vec{x}_{1}$ and $\vec{x}_{2}$
lie on the line \cite{Point-Line-dist-WolframAlpha}. By expanding
into versor notation:

{\tiny{}
\[
d=\frac{\left|\left[\left(x_{0}-x_{1}\right)\hat{x}+\left(y_{0}-y_{1}\right)\hat{y}+\left(z_{0}-z_{1}\right)\hat{z}\right]\times\left[\left(x_{0}-x_{2}\right)\hat{x}+\left(y_{0}-y_{2}\right)\hat{y}+\left(z_{0}-z_{2}\right)\hat{z}\right]\right|}{\sqrt{\left(x_{2}-x_{1}\right)^{2}+\left(y_{2}-y_{1}\right)^{2}+\left(z_{2}-z_{1}\right)^{2}}}=
\]
}{\tiny\par}

{\tiny{}
\begin{multline*}
=\frac{\left|\left[\left(y_{0}-y_{1}\right)\left(z_{0}-z_{2}\right)-\left(z_{0}-z_{1}\right)\left(y_{0}-y_{2}\right)\right]\hat{x}+\left[\left(z_{0}-z_{1}\right)\left(x_{0}-x_{2}\right)-\left(x_{0}-x_{1}\right)\left(z_{0}-z_{2}\right)\right]\hat{y}+\left[\left(x_{0}-x_{1}\right)\left(y_{0}-y_{2}\right)-\left(y_{0}-y_{1}\right)\left(x_{0}-x_{2}\right)\right]\hat{z}\right|}{\sqrt{\left(x_{2}-x_{1}\right)^{2}+\left(y_{2}-y_{1}\right)^{2}+\left(z_{2}-z_{1}\right)^{2}}}=
\end{multline*}
}{\tiny\par}

{\tiny{}
\begin{multline*}
=\frac{\left|\left[\cancel{y_{0}z_{0}}-y_{0}z_{2}-y_{1}z_{0}+y_{1}z_{2}\cancel{-z_{0}y_{0}}+z_{0}y_{2}+z_{1}y_{0}-z_{1}y_{2}\right]\hat{x}+\left[\cancel{z_{0}x_{0}}-z_{0}x_{2}-z_{1}x_{0}+z_{1}x_{2}\cancel{-x_{0}z_{0}}+x_{0}z_{2}+x_{1}z_{0}-x_{1}z_{2}\right]\hat{y}+..\right.}{\sqrt{\left(x_{2}-x_{1}\right)^{2}+\left(y_{2}-y_{1}\right)^{2}+\left(z_{2}-z_{1}\right)^{2}}}
\end{multline*}
}{\tiny\par}

{\tiny{}
\begin{multline*}
\frac{\left...+\left[\cancel{x_{0}y_{0}}-x_{0}y_{2}-x_{1}y_{0}+x_{1}y_{2}\cancel{-y_{0}x_{0}}+y_{0}x_{2}+y_{1}x_{0}-y_{1}x_{2}\right]\hat{z}\right|}{\sqrt{\left(x_{2}-x_{1}\right)^{2}+\left(y_{2}-y_{1}\right)^{2}+\left(z_{2}-z_{1}\right)^{2}}}=
\end{multline*}
}{\tiny\par}

{\tiny{}
\[
=\frac{\sqrt{\left[y_{0}\left(z_{1}-z_{2}\right)+z_{0}\left(y_{2}-y_{1}\right)+y_{1}z_{2}-z_{1}y_{2}\right]^{2}+\left[z_{0}\left(x_{1}-x_{2}\right)+x_{0}\left(z_{2}-z_{1}\right)+z_{1}x_{2}-x_{1}z_{2}\right]^{2}+\left[x_{0}\left(y_{1}-y_{2}\right)+y_{0}\left(x_{2}-x_{1}\right)+x_{1}y_{2}-y_{1}x_{2}\right]^{2}}}{\sqrt{\left(x_{2}-x_{1}\right)^{2}+\left(y_{2}-y_{1}\right)^{2}+\left(z_{2}-z_{1}\right)^{2}}}
\]
}{\tiny\par}

The fact that one point on the trajectory line and the directional
cosines $\cos\theta_{x,y,z}$ are known allowed to write down the
following relations:

\begin{equation}
\left\{ \begin{array}{c}
x_{2}=x_{1}+\cos\theta_{x}\cdot\left|\vec{v}\right|\cdot t\\
y_{2}=y_{1}+\cos\theta_{y}\cdot\left|\vec{v}\right|\cdot t\\
z_{2}=z_{1}+\cos\theta_{z}\cdot\left|\vec{v}\right|\cdot t
\end{array}\right.,
\end{equation}

where $\vec{v}$ is the velocity vector of a particle, trajectory
of which is approximated by the line, and $t$ is the time duration
of its travel. In this case, the choice of $\left|\vec{v}\right|\cdot t$
was completely arbitrary and, in fact, this factor would cancel out
in the further simplification. Thus, for convenience, it was picked
to be $\left|\vec{v}\right|\cdot t=1$, simplifying the above equation
to:

\begin{equation}
\left\{ \begin{array}{c}
x_{2}=x_{1}+\cos\theta_{x}\\
y_{2}=y_{1}+\cos\theta_{y}\\
z_{2}=z_{1}+\cos\theta_{z}
\end{array}\right..\label{eq:directional_cosines}
\end{equation}

Using Equation \ref{eq:directional_cosines} and shorthand notation:
$\cos\theta_{i}=c_{i}$ ($i=x,y,z$), one could simplify the expression
for $d$ as:

{\tiny{}
\[
d=\sqrt{\frac{\left[-y_{0}c_{z}+z_{0}c_{y}+y_{1}\left(\cancel{z_{1}}+c_{z}\right)-z_{1}\left(\cancel{y_{1}}+c_{y}\right)\right]^{2}+\left[-z_{0}c_{x}+x_{0}c_{z}+z_{1}\left(\cancel{x_{1}}+c_{x}\right)-x_{1}\left(\cancel{z_{1}}+c_{z}\right)\right]^{2}+\left[-x_{0}c_{y}+y_{0}c_{x}+x_{1}\left(\cancel{y_{1}}+c_{y}\right)-y_{1}\left(\cancel{x_{1}}+c_{x}\right)\right]^{2}}{c_{x}^{2}+c_{y}^{2}+c_{z}^{2}}}=
\]
}{\tiny\par}

{\tiny{}
\[
=\sqrt{\frac{\left[-y_{0}c_{z}+z_{0}c_{y}+y_{1}c_{z}-z_{1}c_{y}\right]^{2}+\left[-z_{0}c_{x}+x_{0}c_{z}+z_{1}c_{x}-x_{1}c_{z}\right]^{2}+\left[-x_{0}c_{y}+y_{0}c_{x}+x_{1}c_{y}-y_{1}c_{x}\right]^{2}}{c_{x}^{2}+c_{y}^{2}+c_{z}^{2}}}=
\]
}{\tiny\par}

{\tiny{}
\[
=\sqrt{\frac{\left[\left(y_{1}-y_{0}\right)c_{z}+\left(z_{0}-z_{1}\right)c_{y}\right]^{2}+\left[\left(z_{1}-z_{0}\right)c_{x}+\left(x_{0}-x_{1}\right)c_{z}\right]^{2}+\left[\left(x_{1}-x_{0}\right)c_{y}+\left(y_{0}-y_{1}\right)c_{x}\right]^{2}}{c_{x}^{2}+c_{y}^{2}+c_{z}^{2}}}=
\]
}{\tiny\par}

and finally, using the basic property of directional cosines, i.e.
$c_{x}^{2}+c_{y}^{2}+c_{z}^{2}=1$, the final formula for the shortest
point-line distance reads:

\begin{equation}
d=\sqrt{\left[\left(y_{1}-y_{0}\right)c_{z}+\left(z_{0}-z_{1}\right)c_{y}\right]^{2}+\left[\left(z_{1}-z_{0}\right)c_{x}+\left(x_{0}-x_{1}\right)c_{z}\right]^{2}+\left[\left(x_{1}-x_{0}\right)c_{y}+\left(y_{0}-y_{1}\right)c_{x}\right]^{2}}.\label{eq:shortest-point-line-distance}
\end{equation}

\subsection{Intersection of a point and a sphere \label{sec:line-sphere-intersection}}

Following the derivation in \cite{Line-sphere-Paul-Borke}, the line-sphere
intersection can be found by solving the following quadratic equation
for $d_{\mathsf{intersect}}$:
\[
0=a\cdot d_{\mathsf{intersect}}^{2}+b\cdot d_{\mathsf{intersect}}+c.
\]
The coefficients are: 
\[
a=\left(x_{2}-x_{1}\right)^{2}+\left(y_{2}-y_{1}\right)^{2}+\left(z_{2}-z_{1}\right)^{2},
\]
\[
b=2\cdot\left[\left(x_{2}-x_{1}\right)\cdot\left(x_{1}-x_{\mathsf{s}}\right)+\left(y_{2}-y_{1}\right)\cdot\left(y_{1}-y_{\mathsf{s}}\right)+\left(z_{2}-z_{1}\right)\cdot\left(z_{1}-z_{\mathsf{s}}\right)\right],
\]
\[
c=x_{\mathsf{s}}^{2}+y_{\mathsf{s}}^{2}+z_{\mathsf{s}}^{2}+x_{1}^{2}+y_{1}^{2}+z_{1}^{2}-2\cdot\left[x_{\mathsf{s}}x_{1}+y_{\mathsf{s}}y_{1}+z_{\mathsf{s}}z_{1}\right]-r^{2},
\]
where $\left(x_{1},y_{1},z_{1}\right)$ and $\left(x_{2},y_{2},z_{2}\right)$
are two arbitrary points along the line, $\left(x_{\mathsf{s}},y_{\mathsf{s}},z_{\mathsf{s}}\right)$
is the centre of the sphere, and $r$ is the radius of the sphere.
Switching to the usual KM3NeT convention, using the directional cosines
(see Sec. \ref{sec:The-shortest-distance}), the second point can
be picked to be: $\left(x_{2},y_{2},z_{2}\right)=\left(x_{1}+c_{x},y_{1}+c_{y},z_{1}+c_{z}\right)$.
Such a choice, together with locating the centre sphere at the origin
of the coordinate system: $\left(x_{\mathsf{s}},y_{\mathsf{s}},z_{\mathsf{s}}\right)=\left(0,0,0\right)$,
allows to simplify the coefficients to:

\[
a=\left(\cancel{x_{1}}+c_{x}\cancel{-x_{1}}\right)^{2}+\left(\cancel{y_{1}}+c_{y}\cancel{-y_{1}}\right)^{2}+\left(\cancel{z_{1}}+c_{z}\cancel{-z_{1}}\right)^{2}=c_{x}^{2}+c_{y}^{2}+c_{z}^{2}=1,
\]
\[
b=2\cdot\left[\left(\cancel{x_{1}}+c_{x}\cancel{-x_{1}}\right)\cdot\left(x_{1}-0\right)+\left(\cancel{y_{1}}+c_{y}\cancel{-y_{1}}\right)\cdot\left(y_{1}-0\right)+\left(\cancel{z_{1}}+c_{z}\cancel{-z_{1}}\right)\cdot\left(z_{1}-0\right)\right]=2\cdot\left[c_{x}x_{1}+c_{y}y_{1}+c_{z}z_{1}\right],
\]

\[
c=0+0+0+x_{1}^{2}+y_{1}^{2}+z_{1}^{2}\cancel{-2\cdot\left[0\cdot x_{1}+0\cdot y_{1}+0\cdot z_{1}\right]}-r^{2}=x_{1}^{2}+y_{1}^{2}+z_{1}^{2}-r^{2}.
\]

The first point $\left(x_{1},y_{1},z_{1}\right)$ can be then any
known point lying on the line.

\section{gSeaGen \label{sec:gSeaGen_supplementary_material}}

This section contains a more detailed description of selected developments
in the gSeaGen code performed for this work.

\subsection{Muon lateral deflection in water \label{sec:muon_lateral_deflection}}

\begin{figure}[H]
\centering{}\includegraphics[width=12cm]{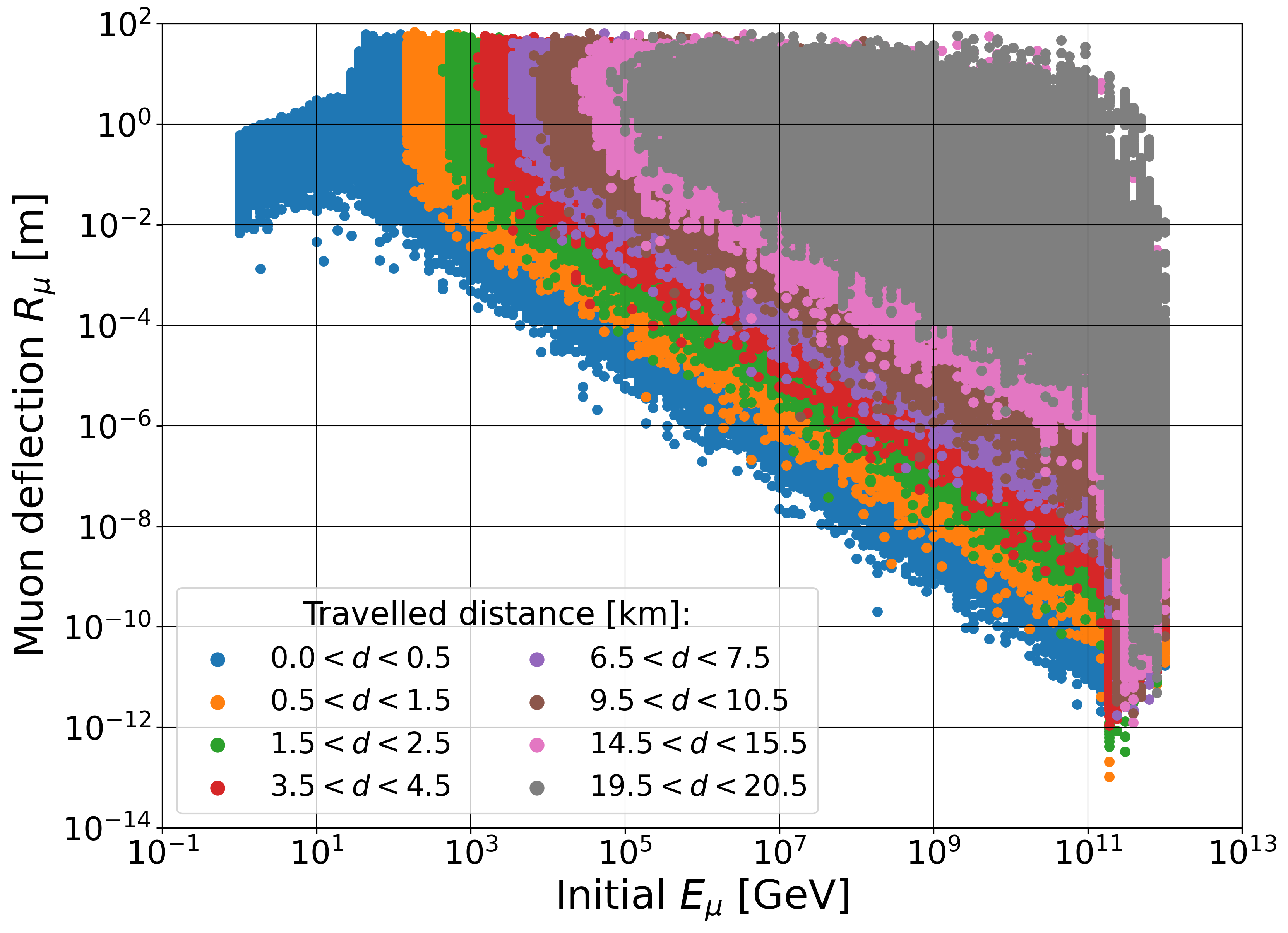}\caption{Lateral deflection of muons $R_{\mu}$ as a function of their initial
energy $E_{\mu}$ and distance $d$ travelled in seawater. \label{fig:Lateral-deflection}}
\end{figure}

While investigating possibilities to improve the gSeaGen code, lateral
deflection of muons in seawater was studied. The result in Fig. \ref{fig:Lateral-deflection}
was obtained by using gSeaGen with PROPOSAL as muon propagator and
shooting $\mu$ vertically downwards through 10000~km of seawater
until they stopped. Such a geometry allowed to conveniently compute
the lateral deflection as $R_{\mu}=\sqrt{x^{2}+y^{2}}$ and ensured
that all the muons would have enough space to stop. Muon energy range
between $1\,$GeV and $1\,$ZeV ($10^{12}\,$GeV) was scanned, with
1000 muons generated per sampled energy. The $\mu$ deflection $R_{\mu}$
was saved every 100~m, or after the muon stopped. For the largest
muon energies, the longest distances travelled were slightly more
than 100~km. The geometrical limit for the travelled distance in
CORSIKA simulation for KM3NeT is about 66~km and can be derived using
the maximum depth (ARCA site depth: $d_{\mathsf{max}}=3450\mathsf{m}$)
and the maximum allowable zenith angle of $87\text{°}$ (see Tab.
\ref{tab:Simulation-settings-CORSIKA}).

Fig. \ref{fig:Lateral-deflection} is very rich in physics. The vertical
left edges for different travelled distances demonstrate that a certain
minimal energy is required for a muon to be able to travel a particular
distance. The upper limit on the lateral deflection seems to be roughly
100~m, almost regardless of initial muon energy. An exception from
this is the initial $E_{\mu}$ region 1-50~GeV, where the dominating
energy loss source is ionisation. At higher energies, radiative energy
losses start to dominate and generally, $\mu$ energy losses increase
sharply. As initial $E_{\mu}$ increases, smaller radial (lateral)
deflections are possible for fixed travelled distances, which is a
result of muons having stronger Lorentz boosts and the fact that high-momentum
particles are just harder to deflect. It is reflected in the downward
slope in Fig. \ref{fig:Lateral-deflection}. The falling upper edge
of the distribution for $19.5<\frac{d}{\mathsf{km}}<20.5$ is not
special, the ones for other distributions are simply buried underneath.
The reason for such falling edges is the very same as for the log-log-linear
slope on the bottom of the distributions: for high-energy muons big
lateral deflections are just less likely.

\begin{figure}[H]
\centering{}\includegraphics[width=12cm]{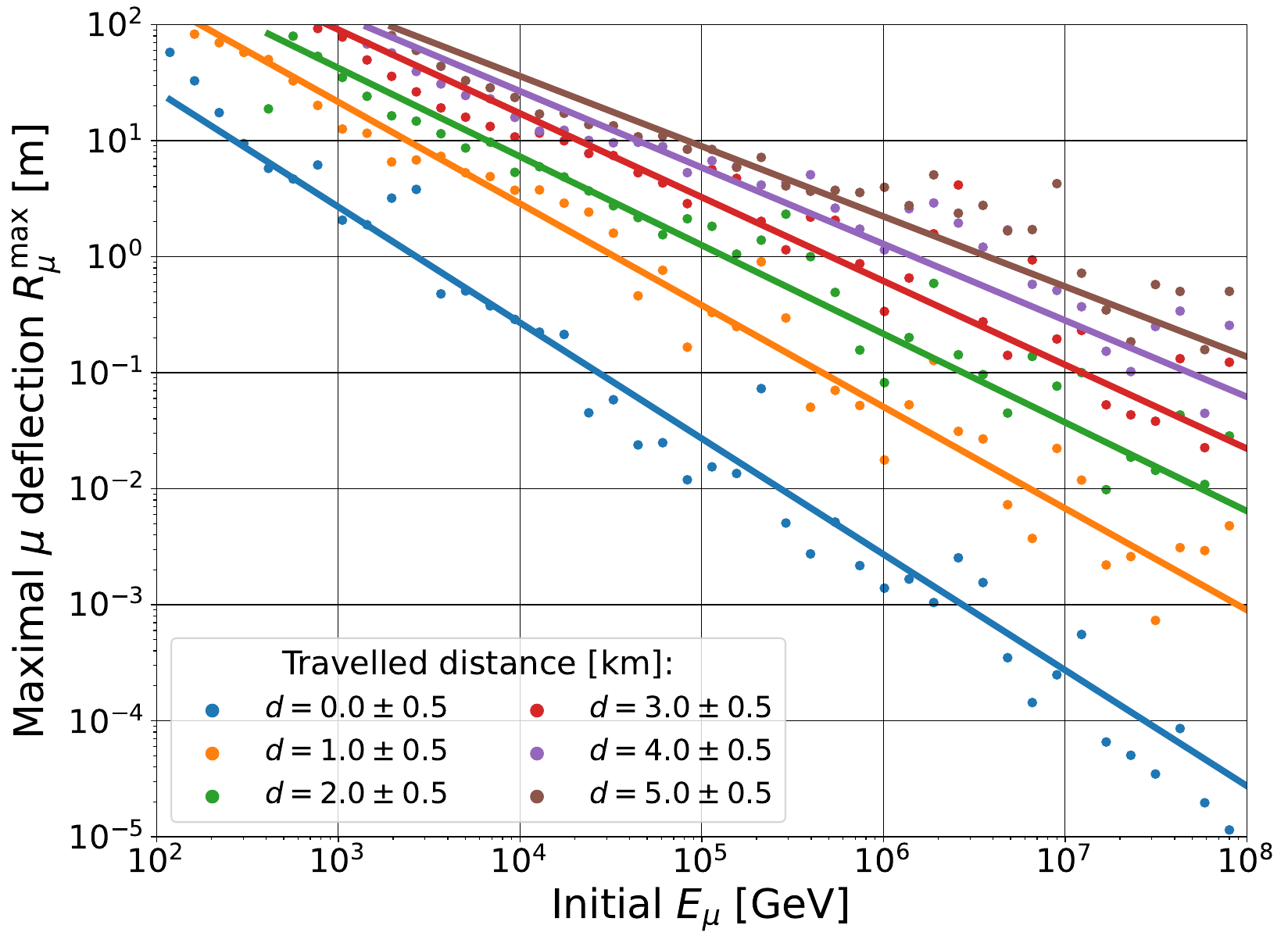}\caption{Maximal lateral muon deflections fitted as a function of $E_{\mu}$.
The fits were based on Eq. \ref{eq:fitting_function}. \label{fig:fit_examples}}
\end{figure}

\begin{figure}[H]
\centering{}\subfloat[Dependence of the fit parameter $a_{E_{\mu}}$ on the distance. \label{fig:fitting_the_fit_1}]{\centering{}\includegraphics[width=8cm]{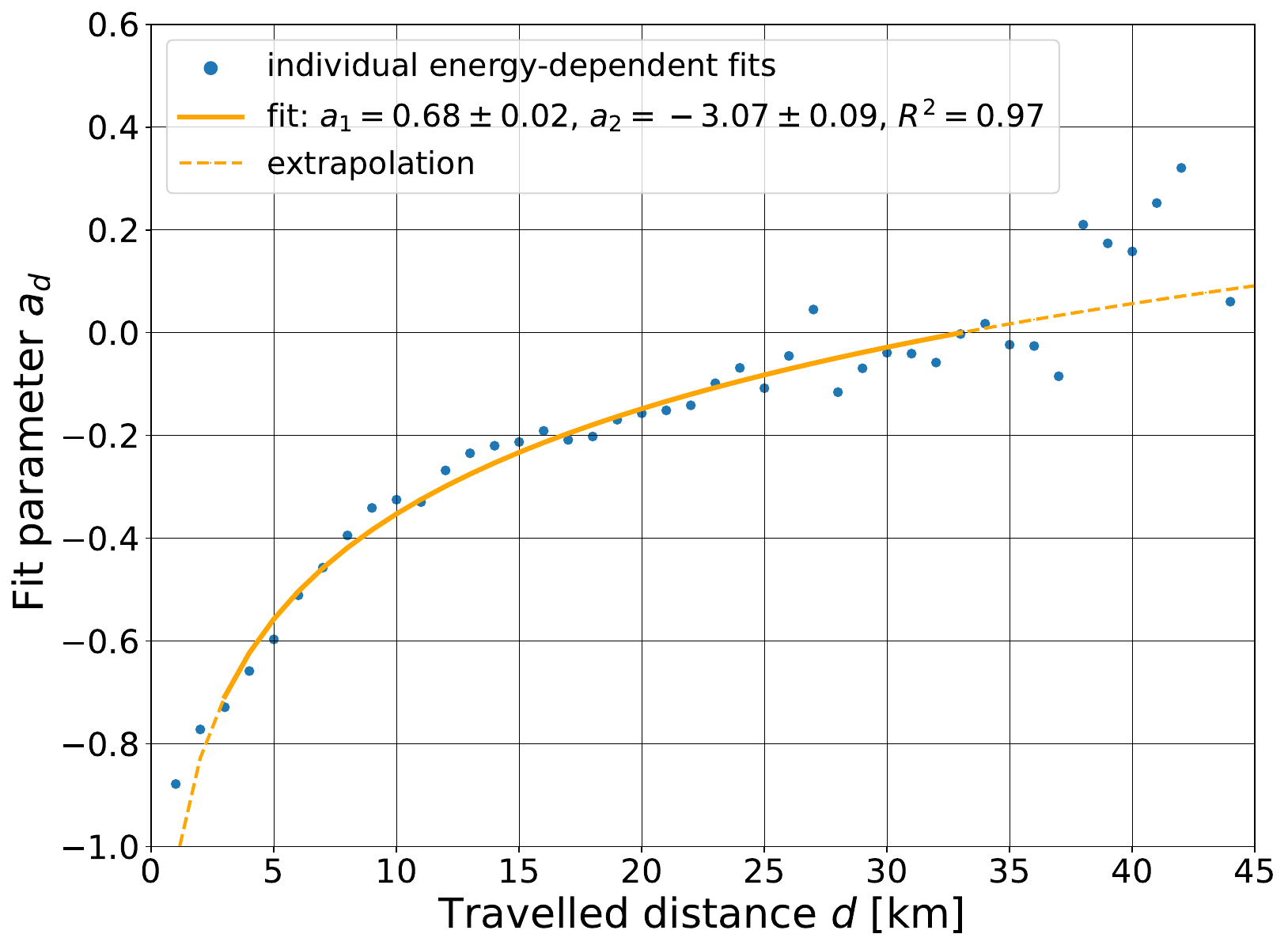}}\subfloat[Dependence of the fit parameter $b_{E_{\mu}}$ on the distance. \label{fig:fitting_the_fit_2}]{\centering{}\includegraphics[width=8cm]{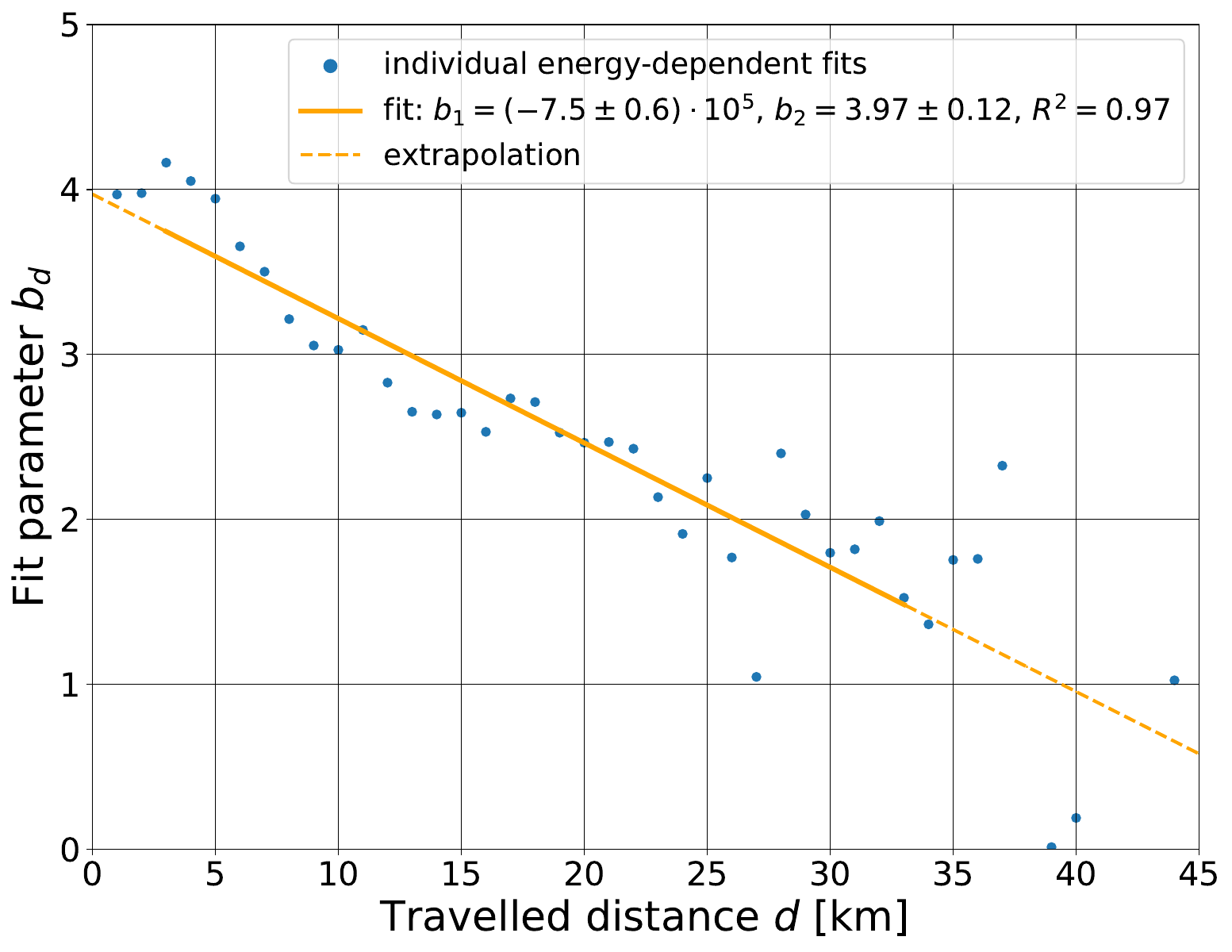}}\caption{Energy-dependent fit parameters of the maximal lateral muon deflection
fitted as functions of travelled distance. \label{fig:fitting_the_fit}}
\end{figure}

It was possible to fit the maximal deflections linearly in log-log
scale as function of $E_{\mu}$ in the range from $10^{2}\,$GeV to
$10^{8}\,$GeV, which is the relevant range for gSeaGen. The fitting
function was:

\begin{equation}
R_{\mu}^{\mathsf{max}}(E_{\mu})=10^{a_{E_{\mu}}\cdot\log_{10}(E_{\mu})+b_{E_{\mu}}}\,\mathsf{m},\label{eq:fitting_function}
\end{equation}

where $a_{E_{\mu}}$ and $b_{E_{\mu}}$ are the fit parameters. Instead
of the maximum value of the energy bin, the mean value incremented
by 10 standard deviations was used as the maximal deflection: $R_{\mu}^{\mathsf{max}}=R_{\mu}^{\mathsf{avg}}+10\cdot R_{\mu}^{\mathsf{std}}$.
The reasoning behind this was that the maximal values tend to scatter
more and the standard deviation gives a more smooth result. In addition,
the $R_{\mu}^{\mathsf{max}}$ was set up to slightly exceed the sampled
datapoints, in order to compensate for the finite sample size. Examples
of fits are shown in Fig. \ref{fig:fit_examples}. Moreover, the dependence
of $R_{\mu}^{\mathsf{max}}$ on the travelled distance $d$ was successfully
parametrised as well. It turned out that the fit parameters $a_{E_{\mu}}$
and $b_{E_{\mu}}$ from the $E_{\mu}$-dependent fit changed linearly
(in appropriate scales) with distance, as shown in Fig. \ref{fig:fitting_the_fit}.
The corresponding fitting function for the $a_{E_{\mu}}(d)$ fit was

\begin{equation}
a_{E_{\mu}}(d)=a_{\mathsf{dist}}\cdot\log_{10}(d)+b_{\mathsf{dist}},\label{eq:fitting_function-a_mu}
\end{equation}

and for $b_{E_{\mu}}(d)$ :

\begin{equation}
b_{E_{\mu}}(d)=a_{\mathsf{dist}}\cdot d+b_{\mathsf{dist}}.\label{eq:fitting_function-b_mu}
\end{equation}

Inserting Eq. \ref{eq:fitting_function-a_mu} and \ref{eq:fitting_function-b_mu}
and the fit parameters from Fig. \ref{fig:fitting_the_fit} into Eq.
\ref{eq:fitting_function} allowed for a swift computation of the
maximal expected deflection for each individual muon as:

\begin{equation}
R_{\mu}^{\mathsf{max}}\left(E_{\mu},d_{\mathsf{max}}\right)=10^{\left[\log_{10}\left(\frac{d_{\mathsf{max}}}{\mathsf{m}}\right)\cdot0.67971202-3.07154976\right]\cdot\log_{10}(\frac{E_{\mu}}{\mathsf{GeV}})-753778.571\cdot\frac{d_{\mathsf{max}}}{\mathsf{m}}+3.97088084}\,\mathsf{m},\label{eq:max_lateral_deflection_parametrisation}
\end{equation}

where $d_{\mathsf{max}}=\frac{\mathsf{site\,depth}}{\cos\theta}$
. It was confirmed that simulating more muons at highest energies
(and hence maximal travelled distances) caused more $a_{E_{\mu}}$
and $b_{E_{\mu}}$ points to align along the distance-dependent fit.
However, propagation of such energetic muons was computationally intensive
and further extension of the fit range was not needed. After concluding
the study, Eq. \ref{eq:max_lateral_deflection_parametrisation} was
implemented in muon propagation routine of gSeaGen, which was used
in the CORSIKA MC productions used within this thesis.

\subsection{Computation of DistaMax\label{sec:Computation-of-DistaMax}}

\begin{figure}[H]
\begin{centering}
\includegraphics[width=16cm]{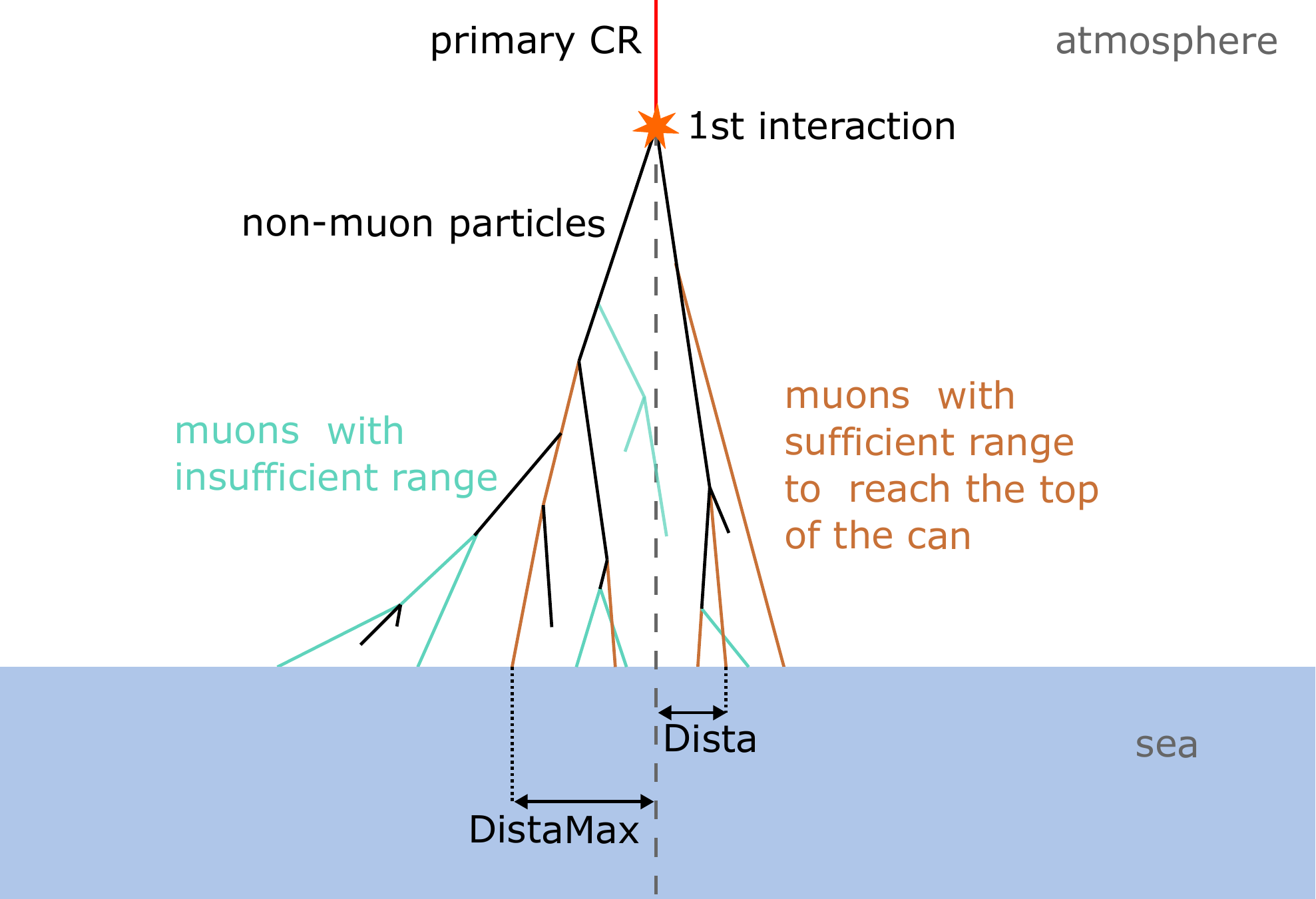}
\par\end{centering}
\centering{}\caption{Sketch demonstrating the meaning of Dista and DistaMax. A vertical
shower has been drawn for simplicity. \label{fig:DistaMax_sketch}}
\end{figure}

DistaMax is a measure of the lateral size of the CORSIKA shower used
in gSeaGen. It is the maximal lateral distance of a muon in the shower
to the primary trajectory line, evaluated at the sea level, before
the muon propagation through water (see Fig. \ref{fig:DistaMax_sketch}).
DistaMax is computed at sea level, because it is used in the propagation
routine itself. The lateral distance Dista for each muon is simply
the shortest distance from the muon position to the primary trajectory
and is computed from Eq. \ref{eq:shortest-point-line-distance}. To
take into account the fact that muons scatter on their way to the
can, Dista of each muon is increased by an estimate of how far could
it laterally deflect, given its energy and the distance it has to
travel (Eq. \ref{eq:max_lateral_deflection_parametrisation}). It
is important to note that only muons that have sufficient range (energy)
to reach the can are considered in the evaluation of DistaMax, to
avoid an overestimation. The muons below the range threshold are not
propagated at all.

\subsection{Reorientation of CORSIKA showers \label{sec:Rotation-of-CORSIKA}}

The way for gSeaGen to ensure that an event will hit the detector
is by moving it around, such that its trajectory (inferred from the
direction before propagation) will cross the detector volume. As a
vital part of this work, a fundamental rework of CORSIKA shower re-orientation
in gSeaGen was performed. Previously, the showers were only horizontally
shifted in the $x$-$y$ plane, assuming a flat geometry. Here, a
more accurate approach was introduced, taking into account the curvature
of the Earth. It was identified to play a significant role for the
horizontal showers, where the differences between the flat and curved
scenarios could reach hundredths of meters. 

\begin{figure}[H]
\begin{centering}
\includegraphics[width=16cm]{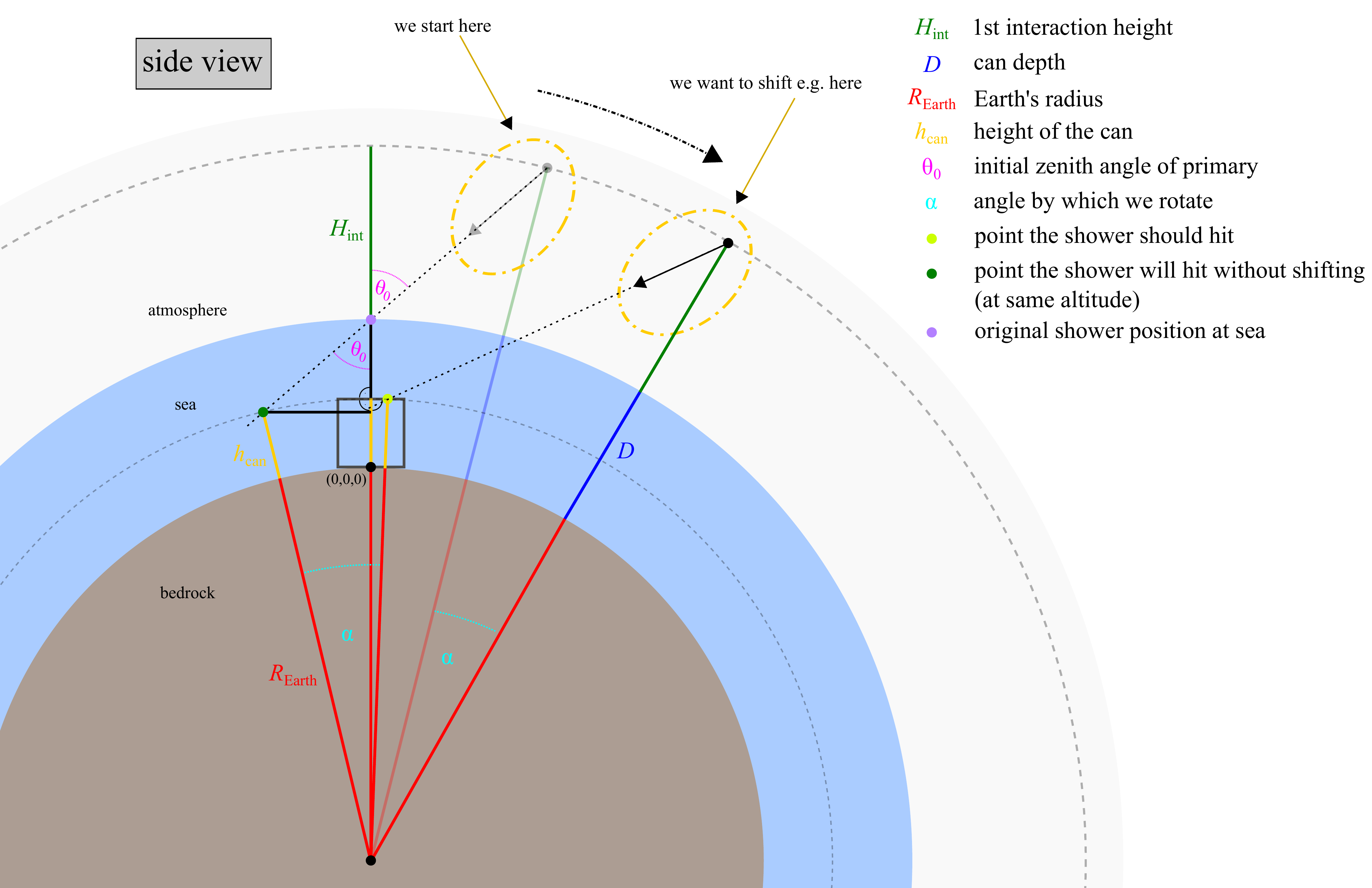}
\par\end{centering}
\centering{}\caption{Side view of the geometry of rotating the shower. It could be either
$x-y$ or $x-z$ plane, since the problem was assumed to be completely
symmetric around the z axis. \label{fig:rotation-side-view}}
\end{figure}

The geometry of the problem was assumed to be symmetric around the
$z$ axis, going through the centre of the Earth and the centre of
the can (see Fig. \ref{fig:rotation-side-view}). The Earth's shape
was approximated as a perfect sphere with radius ${\color{red}R_{\mathsf{Earth}}}=6371.315\,$km,
matching the CORSIKA simulation \cite{CORSIKA-Userguide}. This allowed
to arbitrarily rotate showers around the Earth without changing the
physics of an event (keeping the same altitude). The only requirement
for the rotation was that the shower trajectory, evaluated from the
initial position and direction of the primary, should intersect the
can. When processing a CORSIKA output file, the initial shower position
is always at the sea level at $\left(0,0,{\color{blue}{\color{blue}D}}\right)$
(\textcolor{violet}{$\bullet$ violet point} in Fig. \ref{fig:rotation-side-view}),
where \textcolor{blue}{$D$} is the depth at which the detector is
located. The origin of the coordinate system is located at the centre
of the base of the can (see Sec. \ref{sec:can}). The shower in its
original orientation will not intersect the can unless it happens
to be vertical, hence it has to be rotated. In addition, including
such a random rotation is necessary to avoid a directional bias in
the simulation.

\subsubsection{Computation of the required rotation \label{subsec:Computing-the-needed-rotation}}

\begin{figure}[H]
\begin{centering}
\includegraphics[width=10cm]{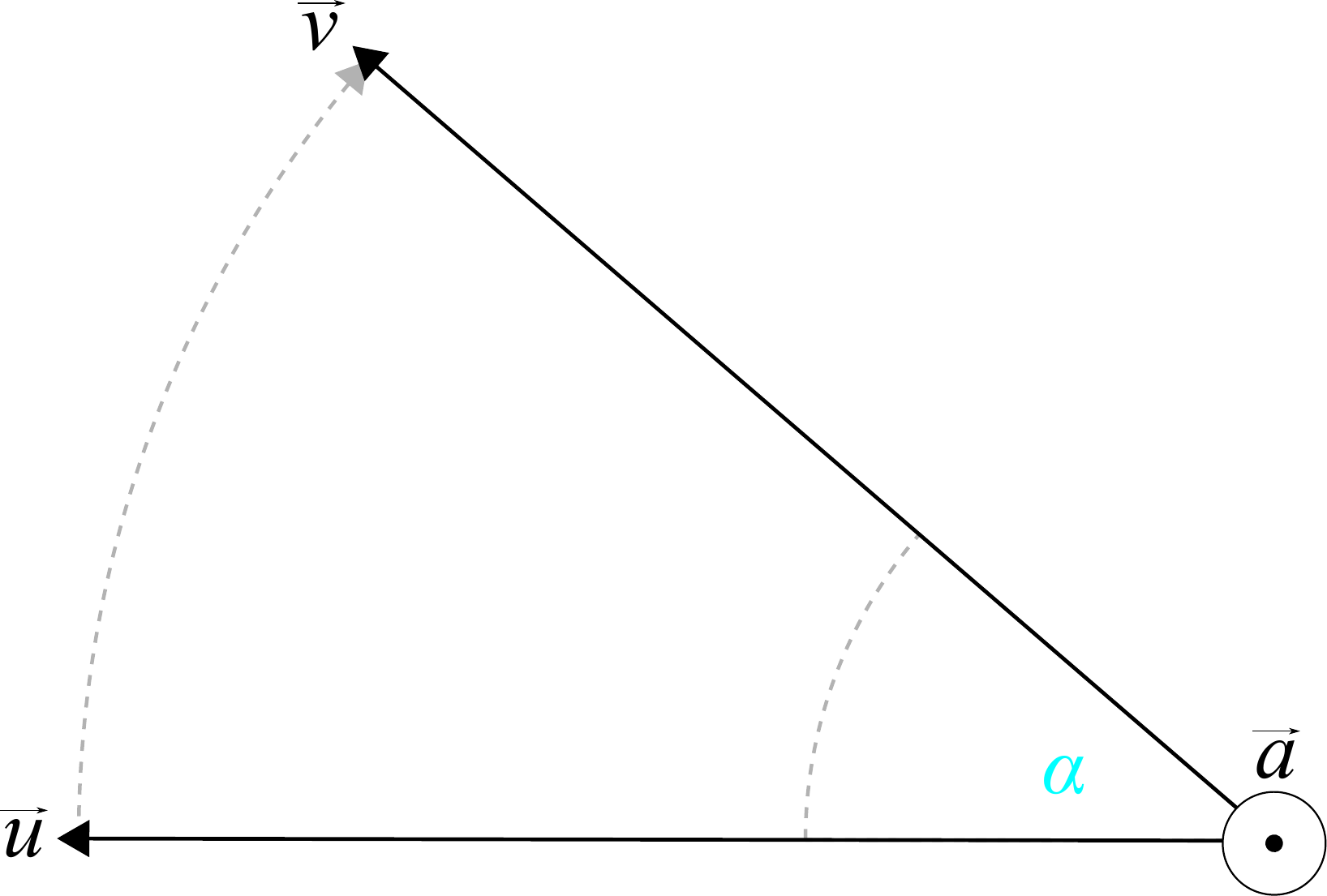}
\par\end{centering}
\centering{}\caption{Sketch of the vector $\vec{u}$ rotated by an angle ${\color{cyan}\alpha}$
around an axis of rotation, described by an orthogonal vector $\vec{a}$.
\label{fig:axis-of-rotation}}
\end{figure}

To evaluate the rotation matrix $M$, required for a shower trajectory
to intersect a given point $\textcolor{electriclime}{\bullet}$ $\left(x_{\mathsf{can}},y_{\mathsf{can}},z_{\mathsf{can}}\right)$,
one must follow the steps:
\begin{enumerate}
\item The point $\textcolor{darkgreen}{\ensuremath{\bullet}}$ $\left(x_{\mathsf{p}},y_{\mathsf{p}},z_{\mathsf{p}}\right)$,
where the initial primary trajectory intersects with the sphere of
radius $r={\color{red}R_{\mathsf{Earth}}}+z_{\mathsf{can}}$, centred
on the centre of the Earth (see Fig. \ref{fig:rotation-side-view})
must be found. It is sufficient to find the distance to that point,
since the direction $\left(c_{x,}^{\mathsf{p}},c_{y}^{\mathsf{p}},c_{z}^{\mathsf{p}}\right)=\left(c_{x,}^{\mathsf{sea}},c_{y}^{\mathsf{sea}},c_{z}^{\mathsf{sea}}\right)$
is already known (here the shorthand notation for directional cosines
from Sec. \ref{sec:The-shortest-distance} is used). Using the result
from Sec. \ref{sec:line-sphere-intersection}, the sought distance
$d_{\mathsf{intersect}}$ is the solution to the quadratic equation:\\
\[
0=a\cdot d_{\mathsf{intersect}}^{2}+b\cdot d_{\mathsf{intersect}}+c
\]
with the coefficients:\\
\begin{equation}
a=1,\label{eq:a}
\end{equation}
\begin{equation}
b=2\cdot\left(c_{x,}^{\mathsf{p}}x_{1}+c_{y}^{\mathsf{p}}y_{1}+c_{z}^{\mathsf{p}}z_{1}\right),\label{eq:b}
\end{equation}
\begin{equation}
c=x_{1}^{2}+y_{1}^{2}+z_{1}^{2}-r^{2},\label{eq:c}
\end{equation}
where $\left(x_{1},y_{1},z_{1}\right)$ can be set to \textcolor{violet}{$\bullet$}
$\left(x_{\mathsf{sea}},y_{\mathsf{sea}},z_{\mathsf{sea}}\right)$,
i.e. the original shower position at sea. There are in general two
solutions for $d_{\mathsf{intersect}}$:
\begin{equation}
d_{\mathsf{intersect}}=\frac{-b\pm\sqrt{b^{2}-4ac}}{2a},\label{eq:solutions}
\end{equation}
however only the smaller distance is of interest, as the larger one
is to the intersection point on the other side of the sphere approximating
the Earth. Thus, after inserting Eq. \ref{eq:a}, \ref{eq:b}, \ref{eq:c},
and $r={\color{red}R_{\mathsf{Earth}}}+z_{\mathsf{can}}$ into Eq.
\ref{eq:solutions}, the solution is:{\footnotesize{}
\begin{equation}
d_{\mathsf{intersect}}=-\left(c_{x}^{\mathsf{p}}x_{\mathsf{sea}}+c_{y}^{\mathsf{p}}y_{\mathsf{sea}}+c_{z}^{\mathsf{p}}z_{\mathsf{sea}}\right)\pm\sqrt{\left(c_{x}^{\mathsf{p}}x_{\mathsf{sea}}+c_{y}^{\mathsf{p}}y_{\mathsf{sea}}+c_{z}^{\mathsf{p}}z_{\mathsf{sea}}\right)^{2}-\left[x_{\mathsf{sea}}^{2}+y_{\mathsf{sea}}^{2}+z_{\mathsf{sea}}^{2}-\left({\color{red}R_{\mathsf{Earth}}}+z_{\mathsf{can}}\right)^{2}\right]}.\label{eq:solution}
\end{equation}
}Using the obtained distance from $\left(x_{\mathsf{sea}},y_{\mathsf{sea}},z_{\mathsf{sea}}\right)$,
the intersection point can be computed as: 
\begin{equation}
\left(\begin{array}{c}
x_{\mathsf{p}}\\
y_{\mathsf{p}}\\
z_{\mathsf{p}}
\end{array}\right)=\left(\begin{array}{c}
x_{\mathsf{sea}}+c_{x}^{\mathsf{p}}\cdot d_{\mathsf{intersect}}\\
y_{\mathsf{sea}}+c_{y}^{\mathsf{p}}\cdot d_{\mathsf{intersect}}\\
z_{\mathsf{sea}}+c_{z}^{\mathsf{p}}\cdot d_{\mathsf{intersect}}
\end{array}\right)
\end{equation}
\item After the $\textcolor{darkgreen}{\ensuremath{\bullet}}$ $\left(x_{\mathsf{p}},y_{\mathsf{p}},z_{\mathsf{p}}\right)$
was found, the next step is to determine a rotation necessary to move
it to $\textcolor{electriclime}{\bullet}$ $\left(x_{\mathsf{can}},y_{\mathsf{can}},z_{\mathsf{can}}\right)$.
\begin{enumerate}
\item First, the axis of rotation described by a vector $\vec{a}$ must
be identified. If the vectors pointing from the origin of the coordinate
system (centre of the Earth) to $\left(x_{\mathsf{p}},y_{\mathsf{p}},z_{\mathsf{p}}\right)$
and $\left(x_{\mathsf{can}},y_{\mathsf{can}},z_{\mathsf{can}}\right)$
will be denoted as $\vec{u}$ and $\vec{v}$ respectively, then $\vec{a}$
will be a vector orthogonal to them (see Fig. \ref{fig:axis-of-rotation}).
The orthogonal vector can be found from the cross-product: 
\[
\vec{a}=\vec{u}\times\vec{v}=\left[\begin{array}{c}
y_{\mathsf{p}}z_{\mathsf{can}}-z_{\mathsf{p}}y_{\mathsf{can}}\\
z_{\mathsf{p}}x_{\mathsf{can}}-x_{\mathsf{p}}z_{\mathsf{can}}\\
x_{\mathsf{p}}y_{\mathsf{can}}-y_{\mathsf{p}}x_{\mathsf{can}}
\end{array}\right].
\]
For the rotation, the versor (unit vector) $\hat{a}$ must be computed:{\scriptsize{}
\[
\hat{a}=\frac{\vec{a}}{\left\Vert \vec{a}\right\Vert }=\left[\begin{array}{c}
a_{x}\\
a_{y}\\
a_{z}
\end{array}\right]=\frac{1}{\sqrt{\left(y_{\mathsf{p}}z_{\mathsf{can}}-z_{\mathsf{p}}y_{\mathsf{can}}\right)^{2}+\left(z_{\mathsf{p}}x_{\mathsf{can}}-x_{\mathsf{p}}z_{\mathsf{can}}\right)^{2}+\left(x_{\mathsf{p}}y_{\mathsf{can}}-y_{\mathsf{p}}x_{\mathsf{can}}\right)^{2}}}\left[\begin{array}{c}
y_{\mathsf{p}}z_{\mathsf{can}}-z_{\mathsf{p}}y_{\mathsf{can}}\\
z_{\mathsf{p}}x_{\mathsf{can}}-x_{\mathsf{p}}z_{\mathsf{can}}\\
x_{\mathsf{p}}y_{\mathsf{can}}-y_{\mathsf{p}}x_{\mathsf{can}}
\end{array}\right].
\]
}{\scriptsize\par}
\item The second step is finding $\sin{\color{cyan}\alpha}$ and $\cos{\color{cyan}\alpha}$
by using:
\[
\sin{\color{cyan}\alpha}=\frac{\left\Vert \vec{u}\times\vec{v}\right\Vert }{\left\Vert \vec{u}\right\Vert \cdot\left\Vert \vec{v}\right\Vert }=\frac{\left\Vert \vec{a}\right\Vert }{\left\Vert \vec{u}\right\Vert \cdot\left\Vert \vec{v}\right\Vert }
\]
and
\[
\cos{\color{cyan}\alpha}=\frac{\left\Vert \vec{u}\cdot\vec{v}\right\Vert }{\left\Vert \vec{u}\right\Vert \cdot\left\Vert \vec{v}\right\Vert }.
\]
\item Finally, the rotation matrix $M$ can be computed, as all the elements
are known:{\footnotesize{}
\begin{equation}
M=\left[\begin{array}{ccc}
\cos{\color{cyan}\alpha}+a_{x}^{2}(1-\cos{\color{cyan}\alpha}) & a_{x}a_{y}(1-\cos{\color{cyan}\alpha})-a_{z}\sin{\color{cyan}\alpha} & a_{x}a_{z}(1-\cos{\color{cyan}\alpha})+a_{y}\sin{\color{cyan}\alpha}\\
a_{y}a_{x}(1-\cos{\color{cyan}\alpha})+a_{z}\sin{\color{cyan}\alpha} & \cos{\color{cyan}\alpha}+a_{y}^{2}(1-\cos{\color{cyan}\alpha}) & a_{y}a_{z}(1-\cos{\color{cyan}\alpha})-a_{x}\sin{\color{cyan}\alpha}\\
a_{z}a_{x}(1-\cos{\color{cyan}\alpha})-a_{y}\sin{\color{cyan}\alpha} & a_{z}a_{y}(1-\cos{\color{cyan}\alpha})+a_{x}\sin{\color{cyan}\alpha} & \cos{\color{cyan}\alpha}+a_{z}^{2}(1-\cos{\color{cyan}\alpha})
\end{array}\right].\label{eq:Rotation-matrix}
\end{equation}
}Derivation of Eq. \ref{eq:Rotation-matrix} may be found in Sec.
9.2 of \cite{RotationMatrixDerivation}. Matrix $M$ can be applied
both to position and direction vectors:
\[
\left[\begin{array}{c}
x\\
y\\
z
\end{array}\right]_{\mathsf{rotated}}=M\left[\begin{array}{c}
x\\
y\\
z
\end{array}\right],
\]
\[
\left[\begin{array}{c}
c_{x}\\
c_{y}\\
c_{z}
\end{array}\right]_{\mathsf{rotated}}=M\left[\begin{array}{c}
c_{x}\\
c_{y}\\
c_{z}
\end{array}\right].
\]
\end{enumerate}
\end{enumerate}

\subsubsection{Rotation of showers around the Earth}

The procedure of reorienting the CORSIKA showers is multi-staged:
\begin{enumerate}
\item The primary direction rotated such that the trajectory intersects
the middle of the can: $\left(0,0,\frac{{\color{orange}h_{\mathsf{can}}}}{2}\right)$
is computed: 
\[
\left[\begin{array}{c}
c_{x,}^{\mathsf{mid}}\\
c_{y}^{\mathsf{mid}}\\
c_{x}^{\mathsf{mid}}
\end{array}\right]=M_{\mathsf{mid}}\cdot\left[\begin{array}{c}
c_{x,}^{\mathsf{sea}}\\
c_{y}^{\mathsf{sea}}\\
c_{x}^{\mathsf{sea}}
\end{array}\right],
\]
where \textcolor{orange}{$h_{\mathsf{can}}$} is the height of the
can and $M_{\mathsf{mid}}$ is computed according to Sec. \ref{subsec:Computing-the-needed-rotation}.
\item The surface of the can is increased by DistaMax (see Sec. \ref{sec:Computation-of-DistaMax})
by adding it in projection to the can radius and height to account
for lateral spread of the muons due to the scattering. 
\item A point on the enlarged surface of the can is randomly picked. The
code selects, whether the point will be on the top cap, or on the
side of the can. The probability to land on either of the two is weighted
by the top and side areas of the can (extended by DistaMax) $A_{\mathsf{top}}$
and $A_{\mathsf{side}}$, projected onto the plane perpendicular to
$\left[c_{x,}^{\mathsf{mid}},c_{y}^{\mathsf{mid}},c_{x}^{\mathsf{mid}}\right]$.
\begin{enumerate}
\item If the top area is selected: a random point is drawn from the circle
of radius $r_{\mathsf{can}}+\mathsf{DistaMax\cdot c_{x}^{\mathsf{mid}}}$.
\item If the side area is selected: a random point is drawn from half the
cylinder side of radius $r_{\mathsf{can}}+\mathsf{DistaMax\cdot c_{x}^{\mathsf{mid}}}$
and height $h_{\mathsf{can}}+\mathsf{DistaMax\cdot\sqrt{1-\left(c_{x}^{\mathsf{mid}}\right)^{2}}}$.
Only one half of the side is used because the shower does not `see'
the whole can side. This half is subsequently rotated to face the
incoming primary.
\end{enumerate}
\item When the point on the increased can surface $\left(x_{\mathsf{can}},y_{\mathsf{can}},z_{\mathsf{can}}\right)$
is selected, the rotation matrix $M$ is computed for that point,
following Sec. \ref{subsec:Computing-the-needed-rotation}.
\item The rotation by $M$ is applied to both positions and directions of
all tracks in an event: primary, muons (which will be propagated),
and their parent particle tracks, containing additional information
on the muon history \cite{CORSIKA-EHISTORY}. The final effect is
a coherent rotation of the entire shower around the Earth.
\end{enumerate}

\section{Performance of JMuon reconstruction\label{sec:Performance-of-JMuon}}

Supplementary plots showing the results of JMuon reconstruction are
gathered here.

\subsection{Zenith reconstruction}

In this section, the performance of JMuon in the zenith reconstruction
is briefly shown. Fig. \ref{fig:Performance-of-JMuon} shows how accurate
the reconstruction is and in Fig. \ref{fig:Performance-of-JMuon-zenith-1D}
one can see that the shape of the true distribution is mostly reconstructed
well, but only for cosines above 0.5. The events reconstructed as
upgoing ($\cos\left(\theta_{\mathrm{zenith}}^{\mu\,\mathrm{bundle}}\right)<0$)
are true downgoing events with their direction erroneously flipped
by JMuon. The performance of the reconstruction improves dramatically
if a quality cut on the JMuon likelihood $\mathcal{L}$ is applied,
as demonstrated in Fig. \ref{fig:Performance-of-JMuon-with-L-cut}.
\begin{center}
\begin{figure}[H]
\begin{centering}
\subfloat[ARCA115.]{\centering{}\includegraphics[width=8cm]{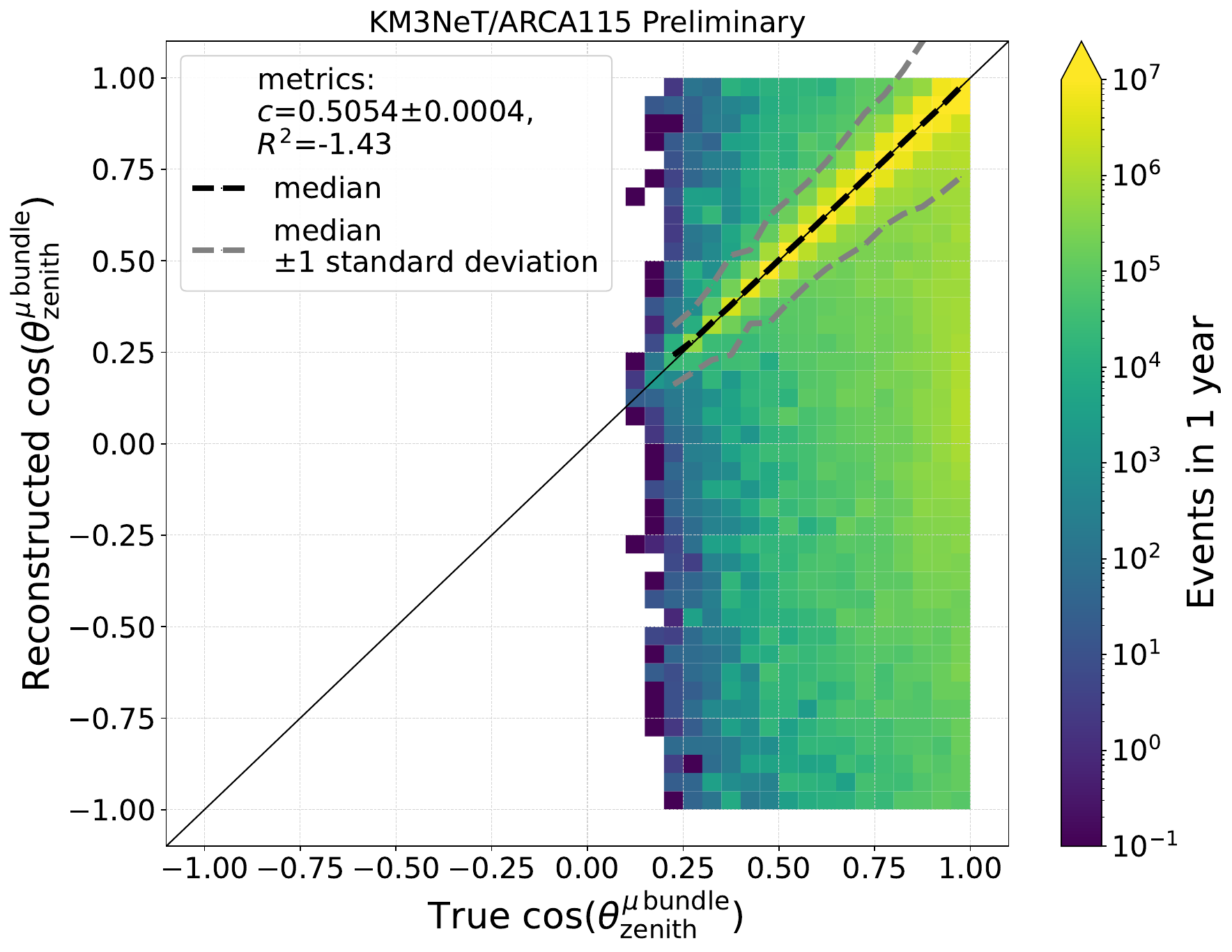}}\subfloat[ARCA6.]{\centering{}\includegraphics[width=8cm]{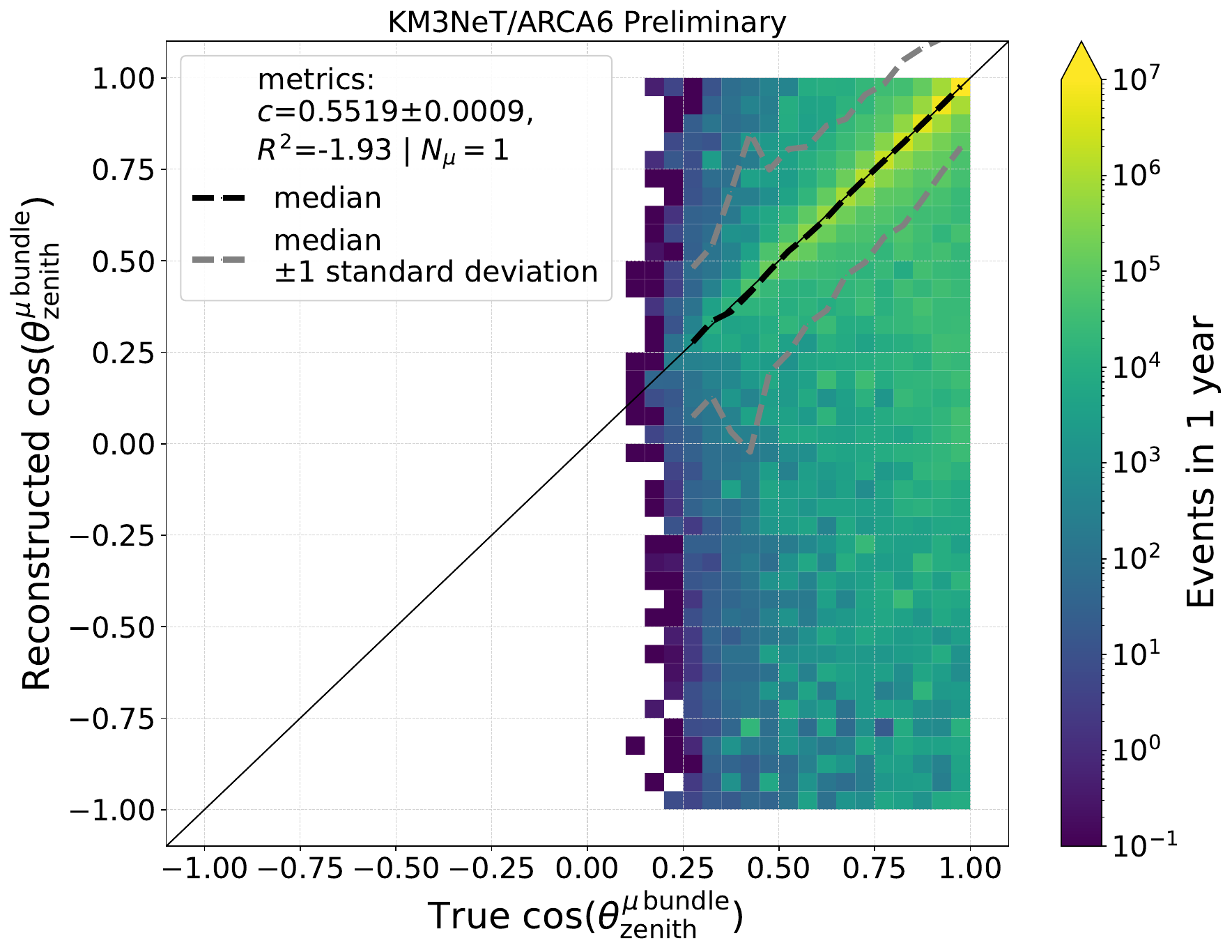}}
\par\end{centering}
\centering{}\subfloat[ORCA115.]{\centering{}\includegraphics[width=8cm]{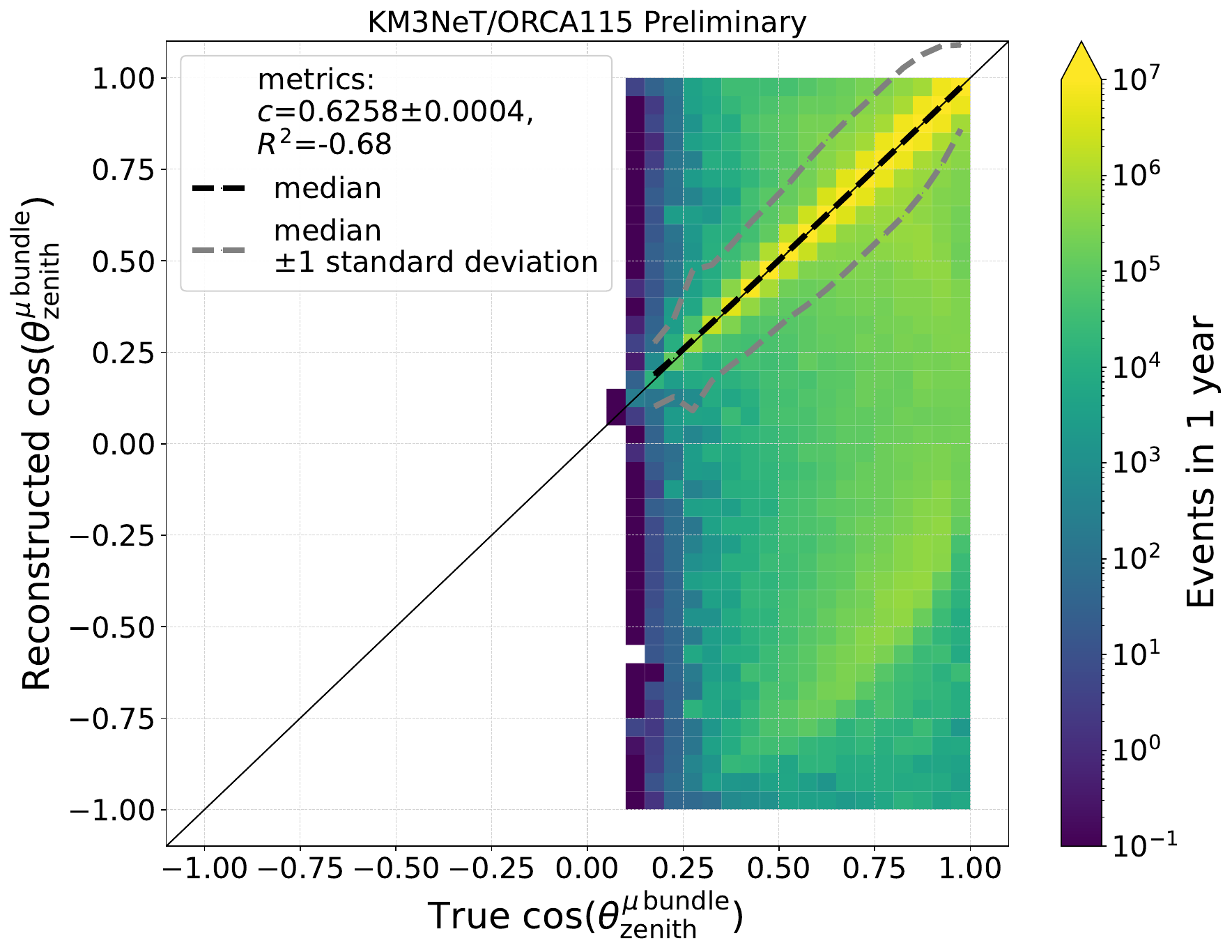}}\subfloat[ORCA6.]{\centering{}\includegraphics[width=8cm]{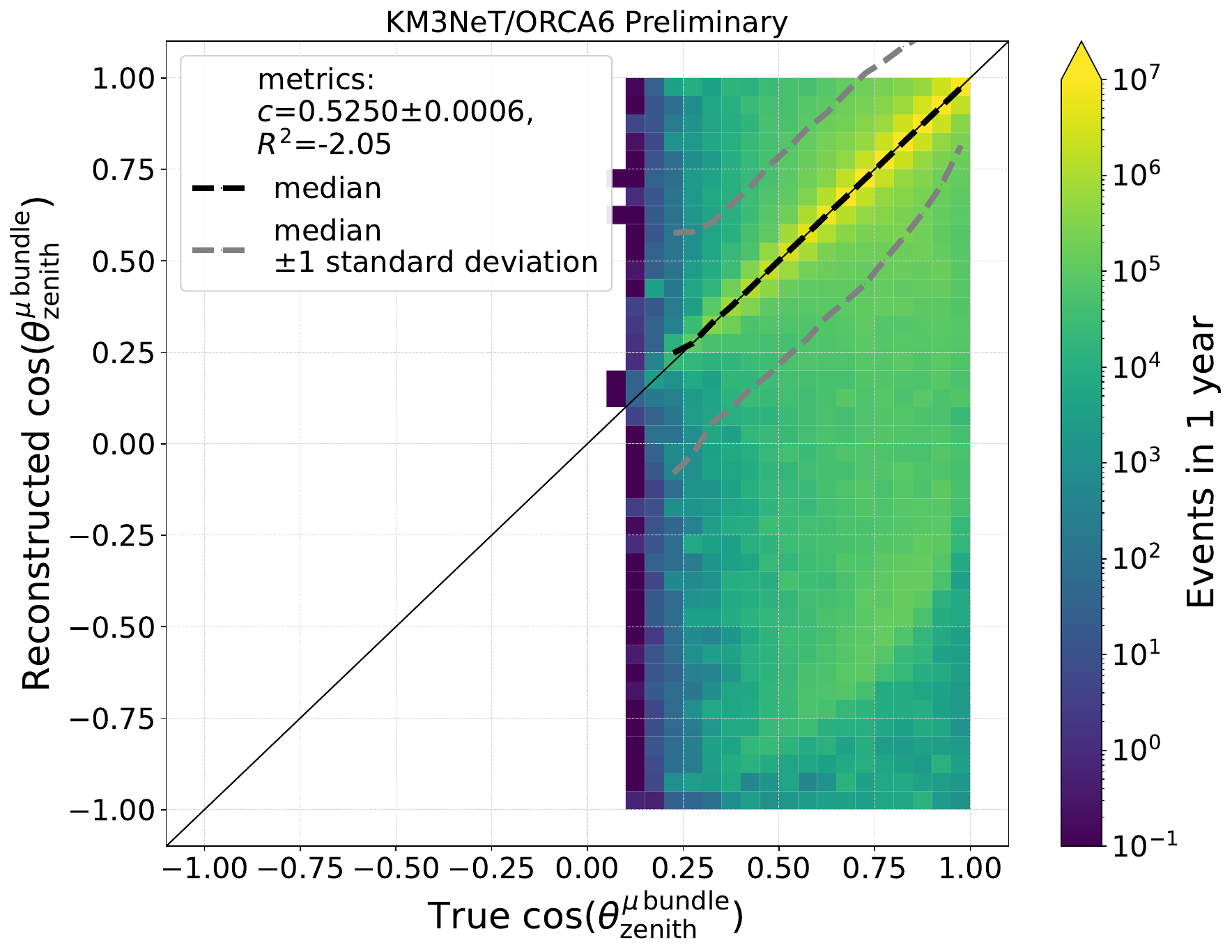}}\caption{Performance of JMuon for different detector configurations, shown
in terms of true vs reconstructed cosines of the zenith angles of
muon bundles simulated with CORSIKA. \label{fig:Performance-of-JMuon}}
\end{figure}
\par\end{center}

\begin{center}
\begin{figure}[H]
\begin{centering}
\subfloat[ARCA115.]{\centering{}\includegraphics[width=8cm]{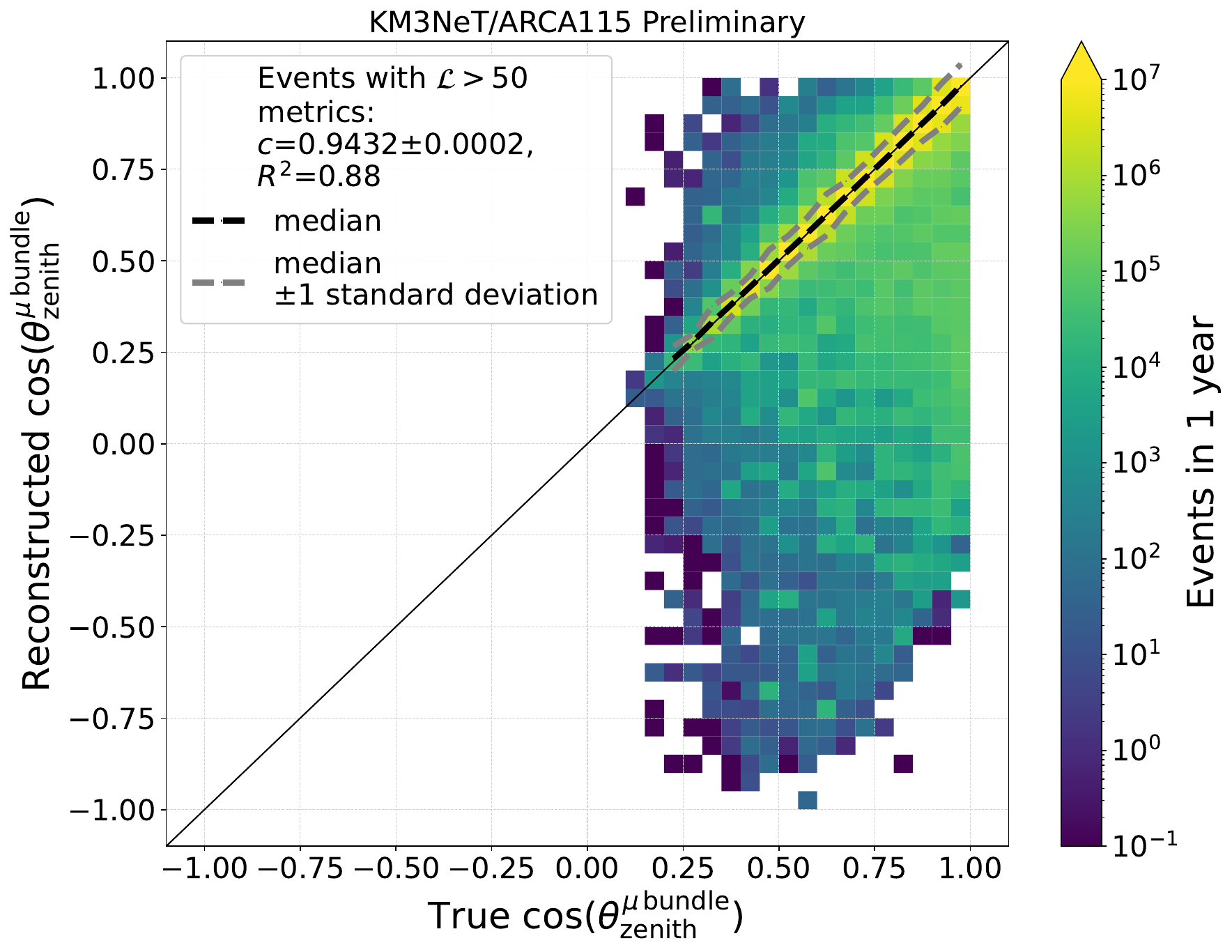}}\subfloat[ARCA6.]{\centering{}\includegraphics[width=8cm]{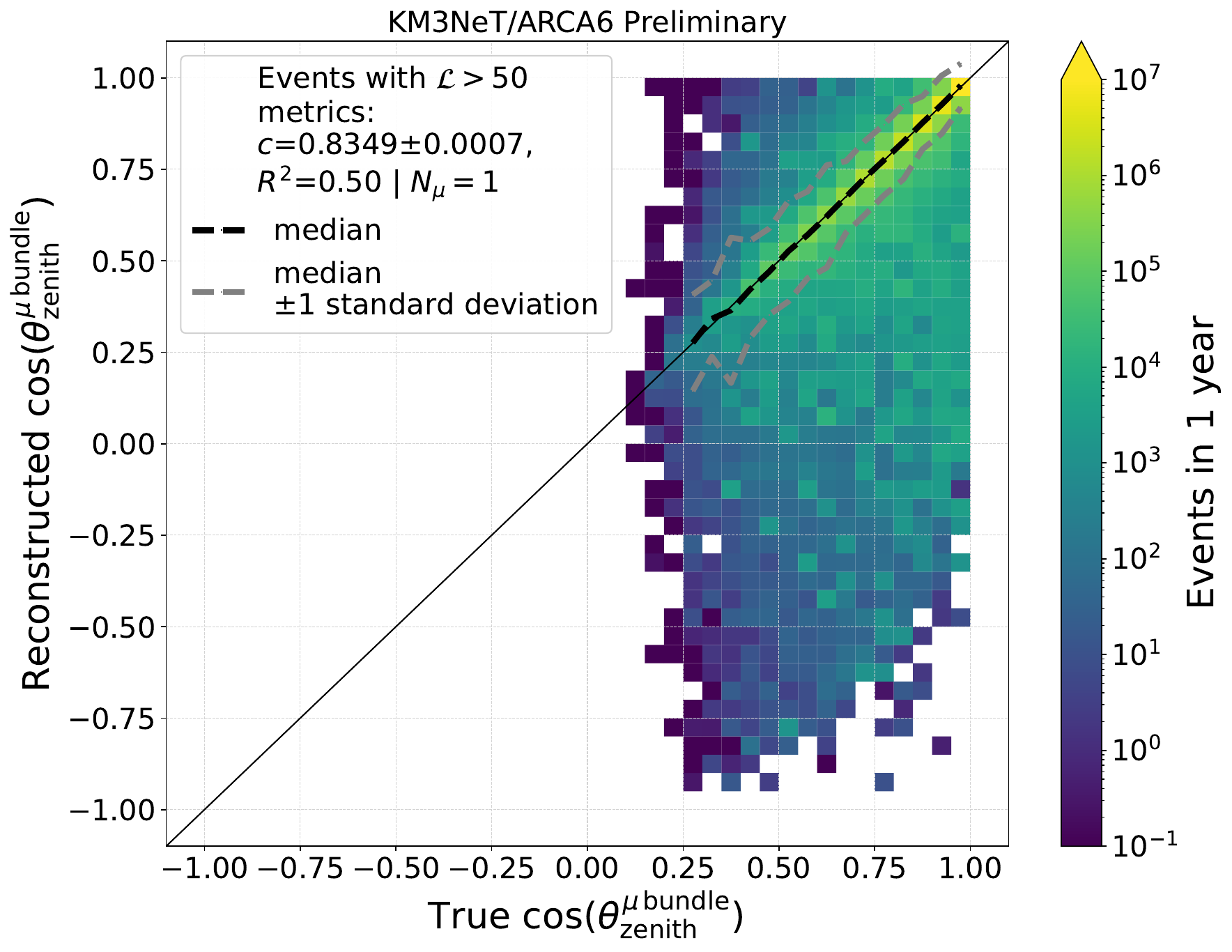}}
\par\end{centering}
\centering{}\subfloat[ORCA115.]{\centering{}\includegraphics[width=8cm]{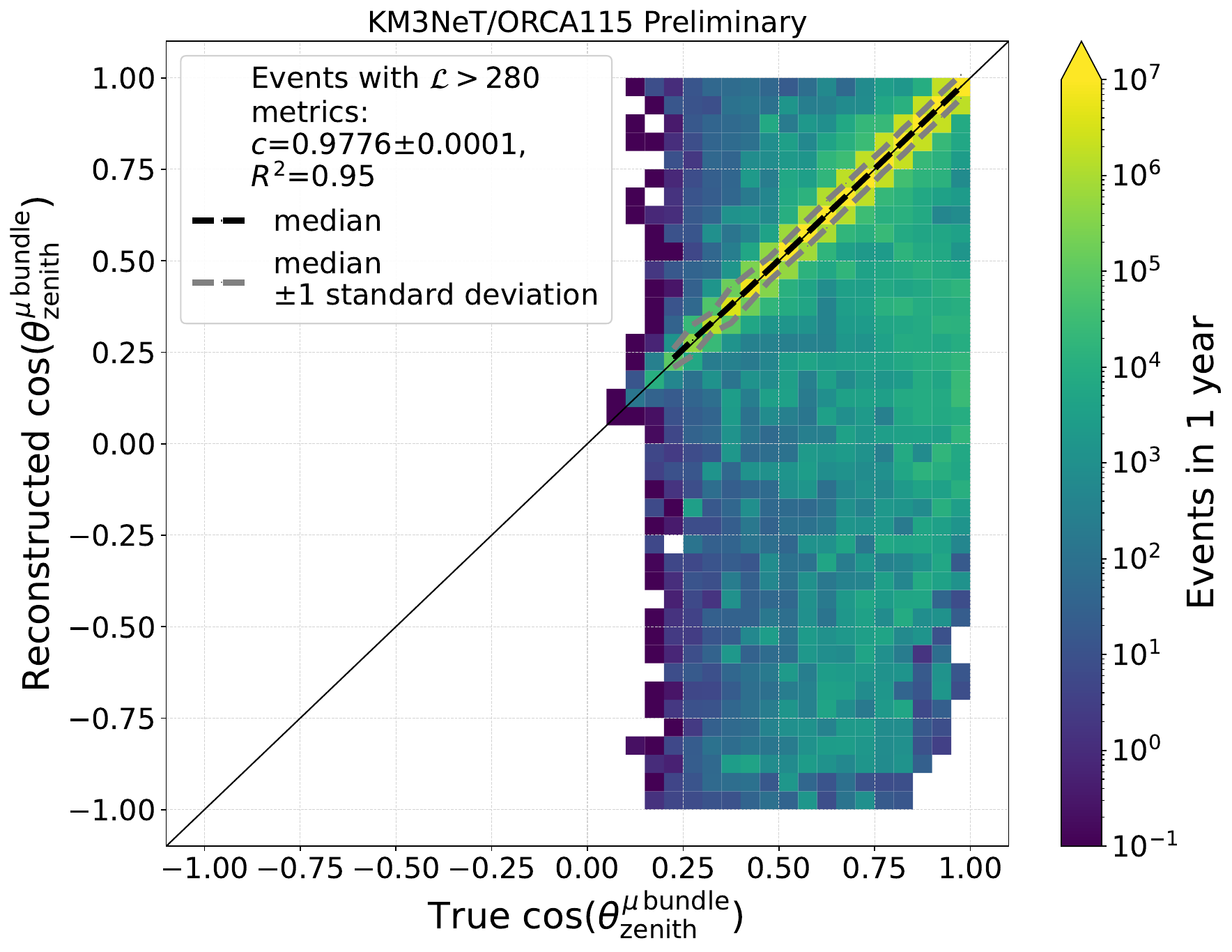}}\subfloat[ORCA6.]{\centering{}\includegraphics[width=8cm]{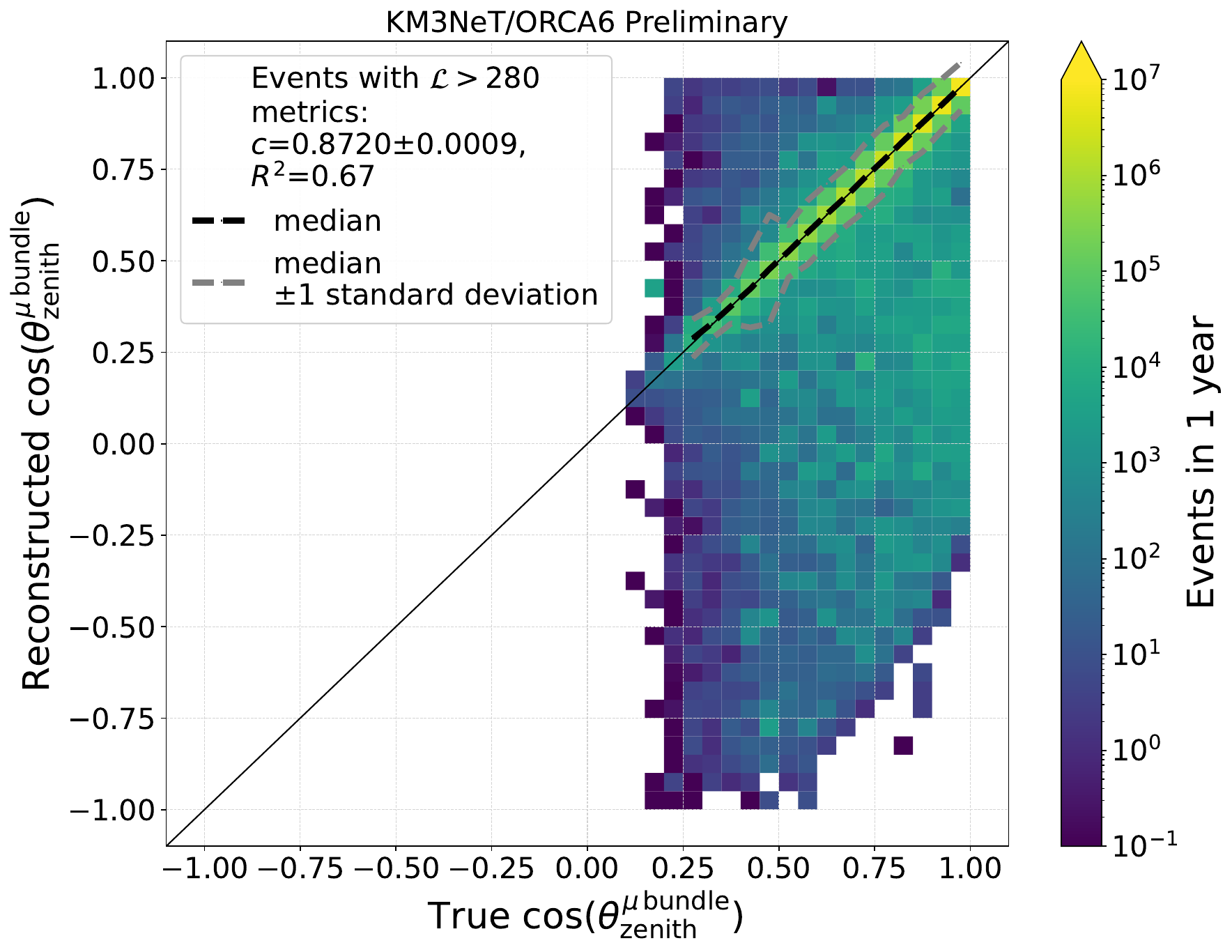}}\caption{Performance of JMuon for different detector configurations, shown
in terms of true vs reconstructed cosines of the zenith angles of
muon bundles simulated with CORSIKA. The plots show the performance
on the selected high-quality events (by cutting of the JMuon likelihood
$\mathcal{L}$). \label{fig:Performance-of-JMuon-with-L-cut}}
\end{figure}
\par\end{center}

\begin{center}
\begin{figure}[H]
\begin{centering}
\subfloat[ARCA115.]{\centering{}\includegraphics[width=8cm]{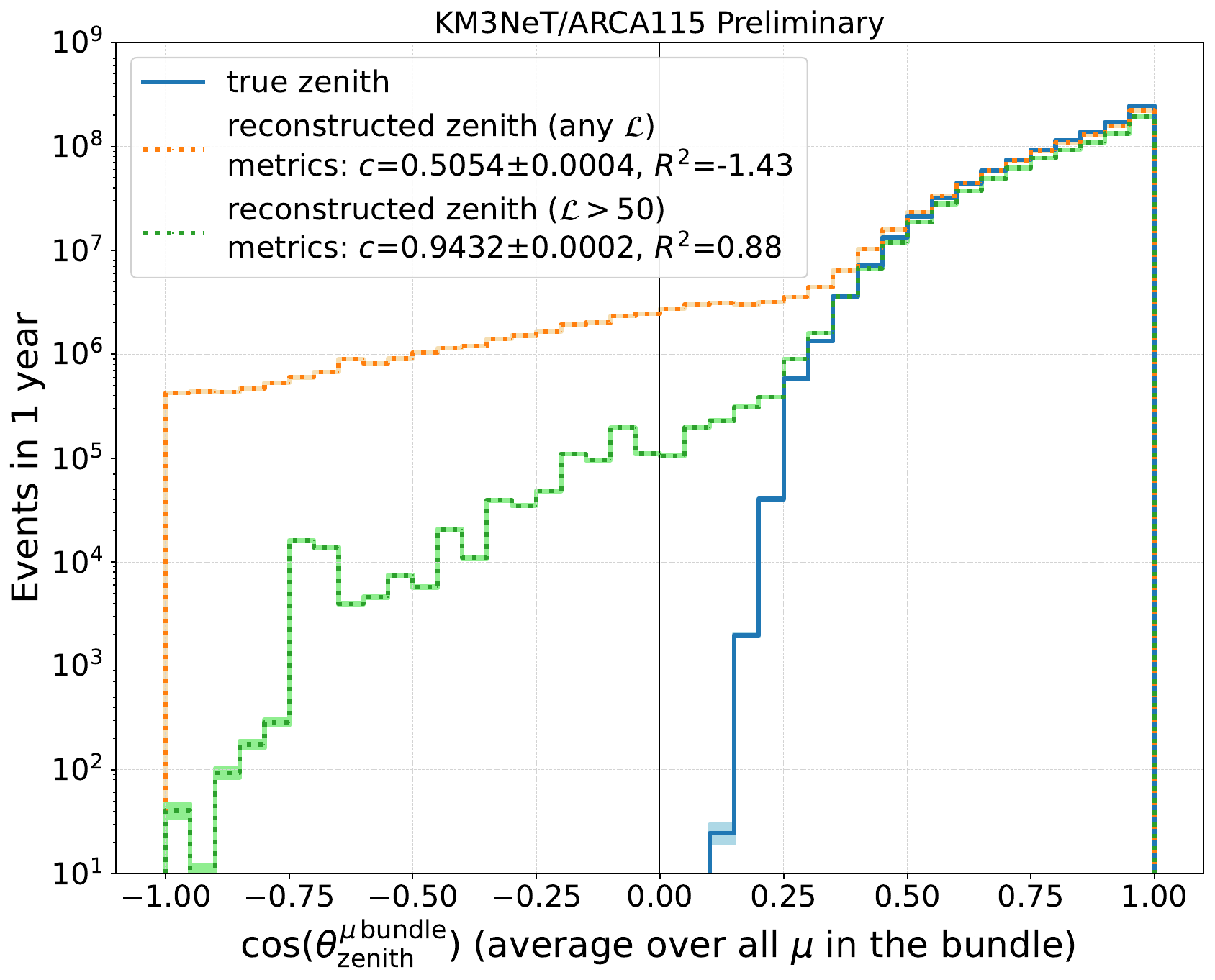}}\subfloat[ARCA6.]{\centering{}\includegraphics[width=8cm]{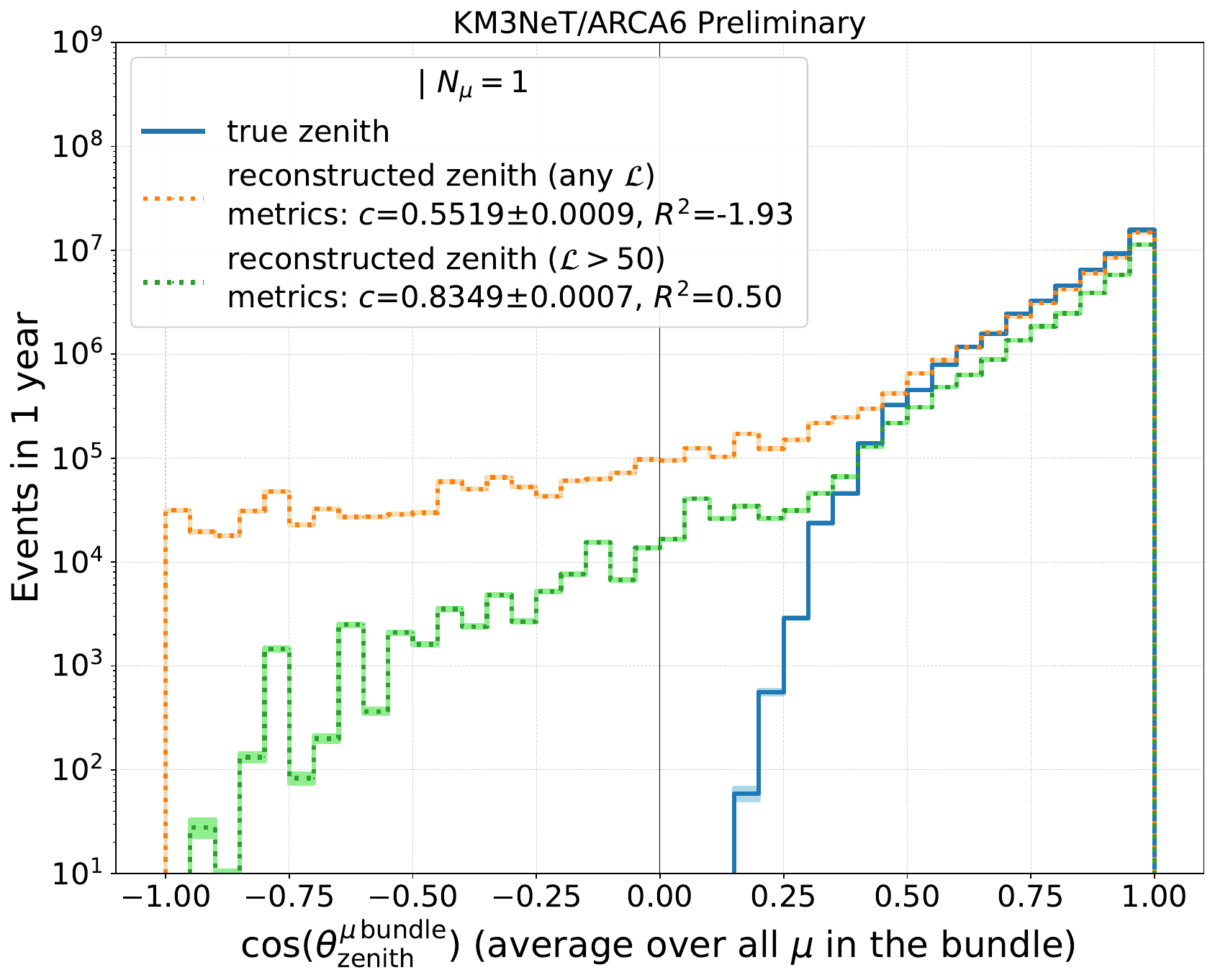}}
\par\end{centering}
\centering{}\subfloat[ORCA115.]{\centering{}\includegraphics[width=8cm]{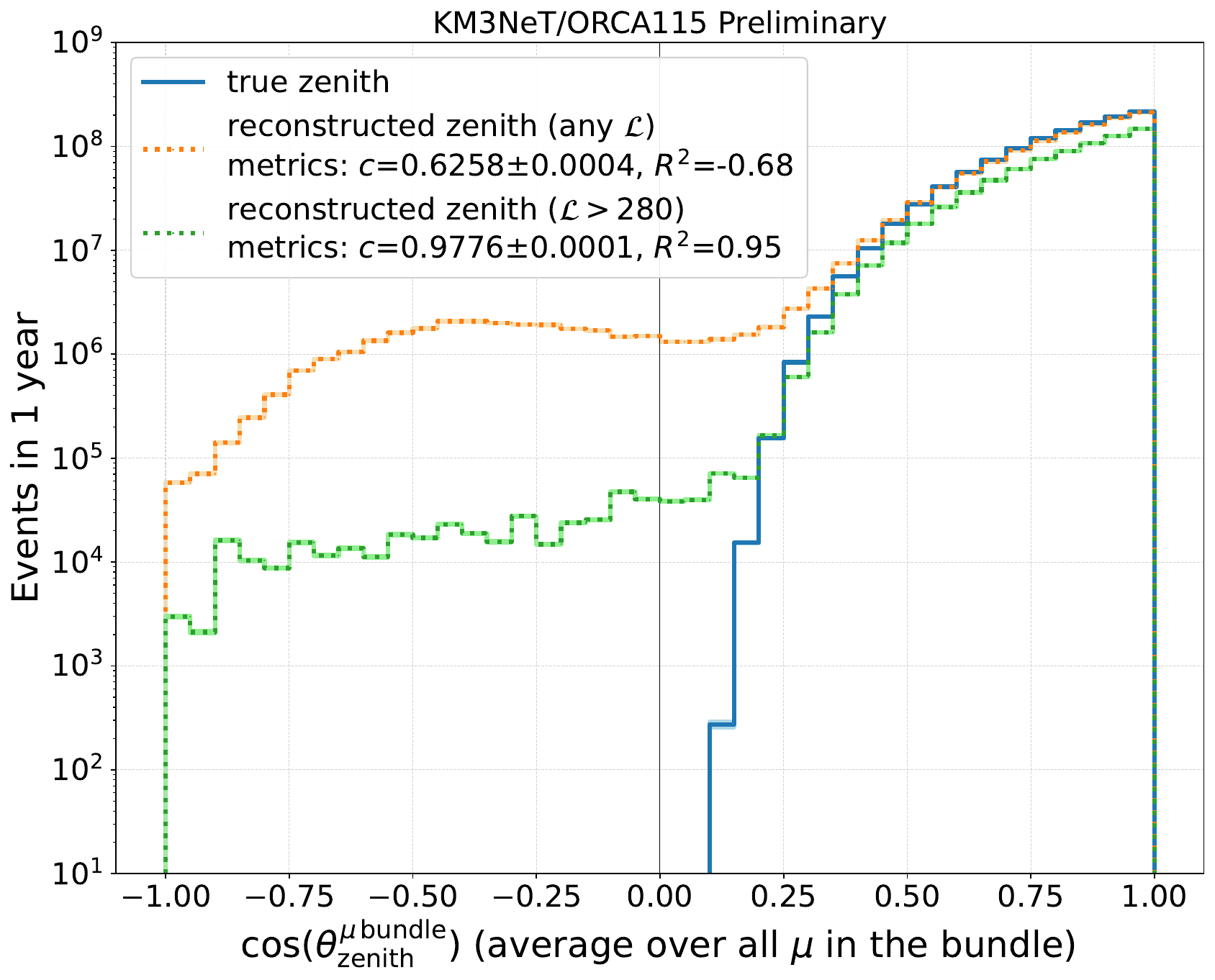}}\subfloat[ORCA6.]{\centering{}\includegraphics[width=8cm]{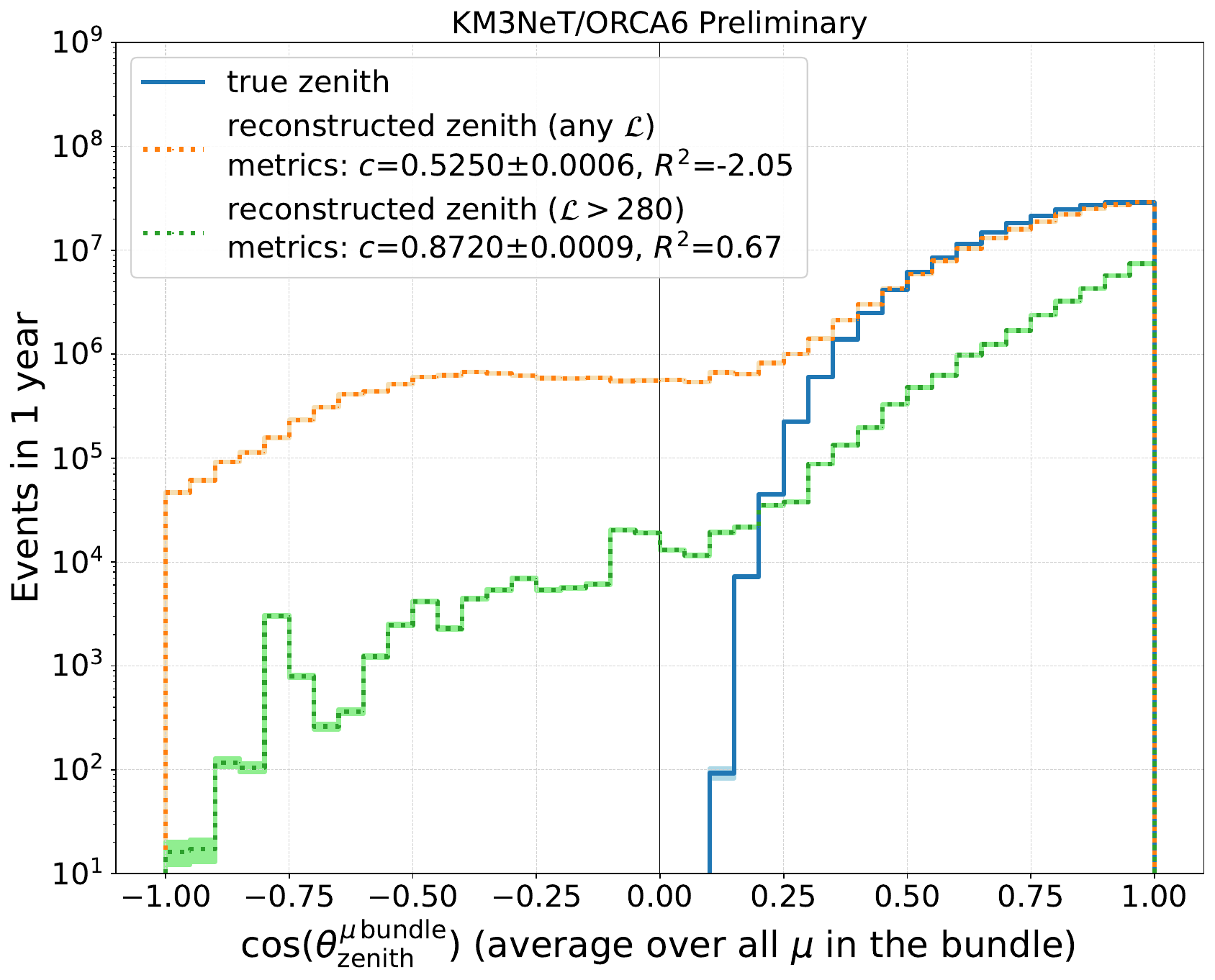}}\caption{Performance of JMuon shown in terms 1D distributions of cosines of
the zenith angles of muon bundles simulated with CORSIKA. The green
histogram is obtained by cutting of the JMuon likelihood $\mathcal{L}$.
\label{fig:Performance-of-JMuon-zenith-1D}}
\end{figure}
\par\end{center}

\subsection{Energy reconstruction}

\begin{figure}[H]
\centering{}\subfloat[All multiplicities. \label{fig:official_E_reco-A6_all}]{\centering{}\includegraphics[width=8cm]{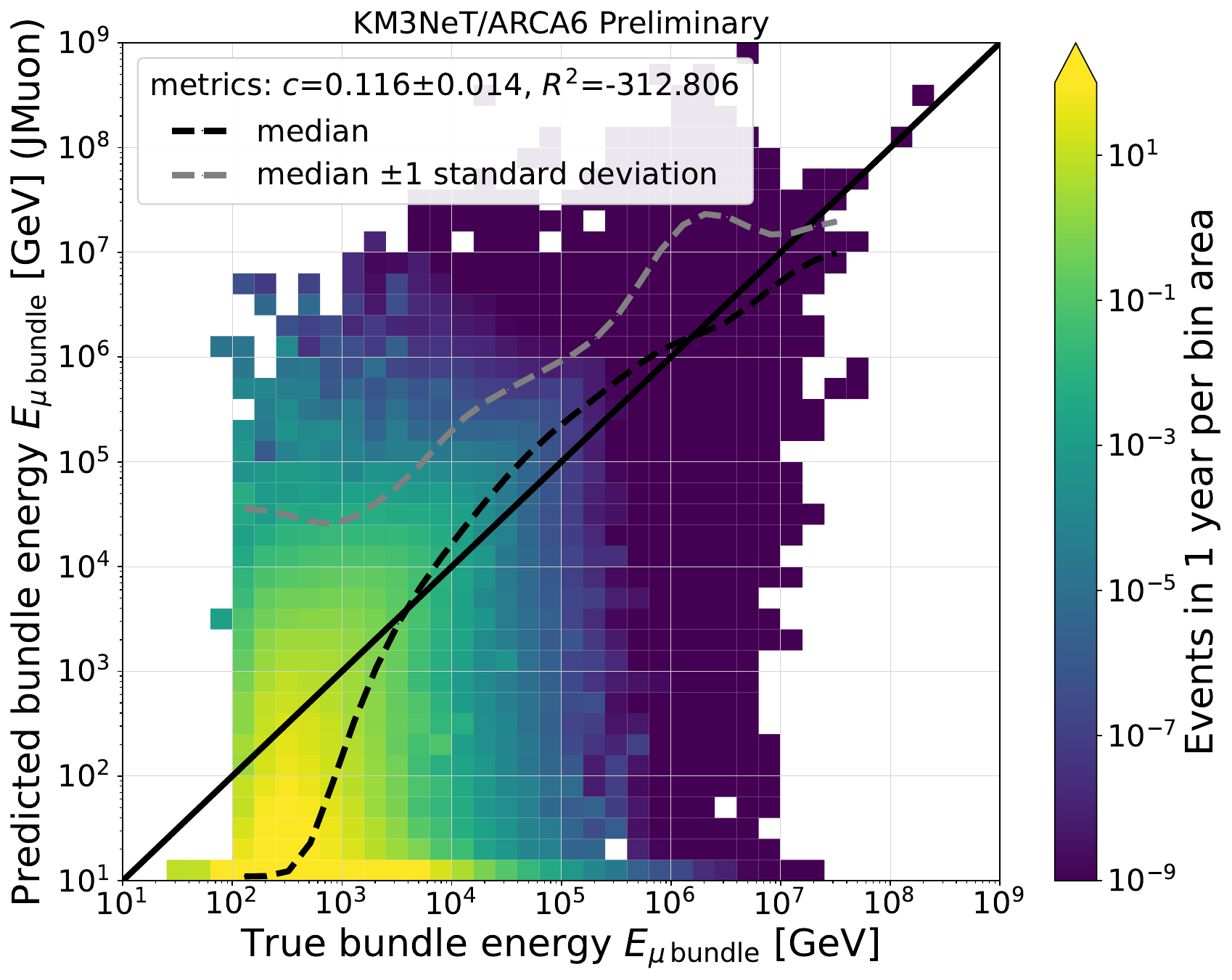}}\subfloat[Only single muon events (multiplicity 1). \label{fig:official_E_reco-A6_single_mu}]{\centering{}\includegraphics[width=8cm]{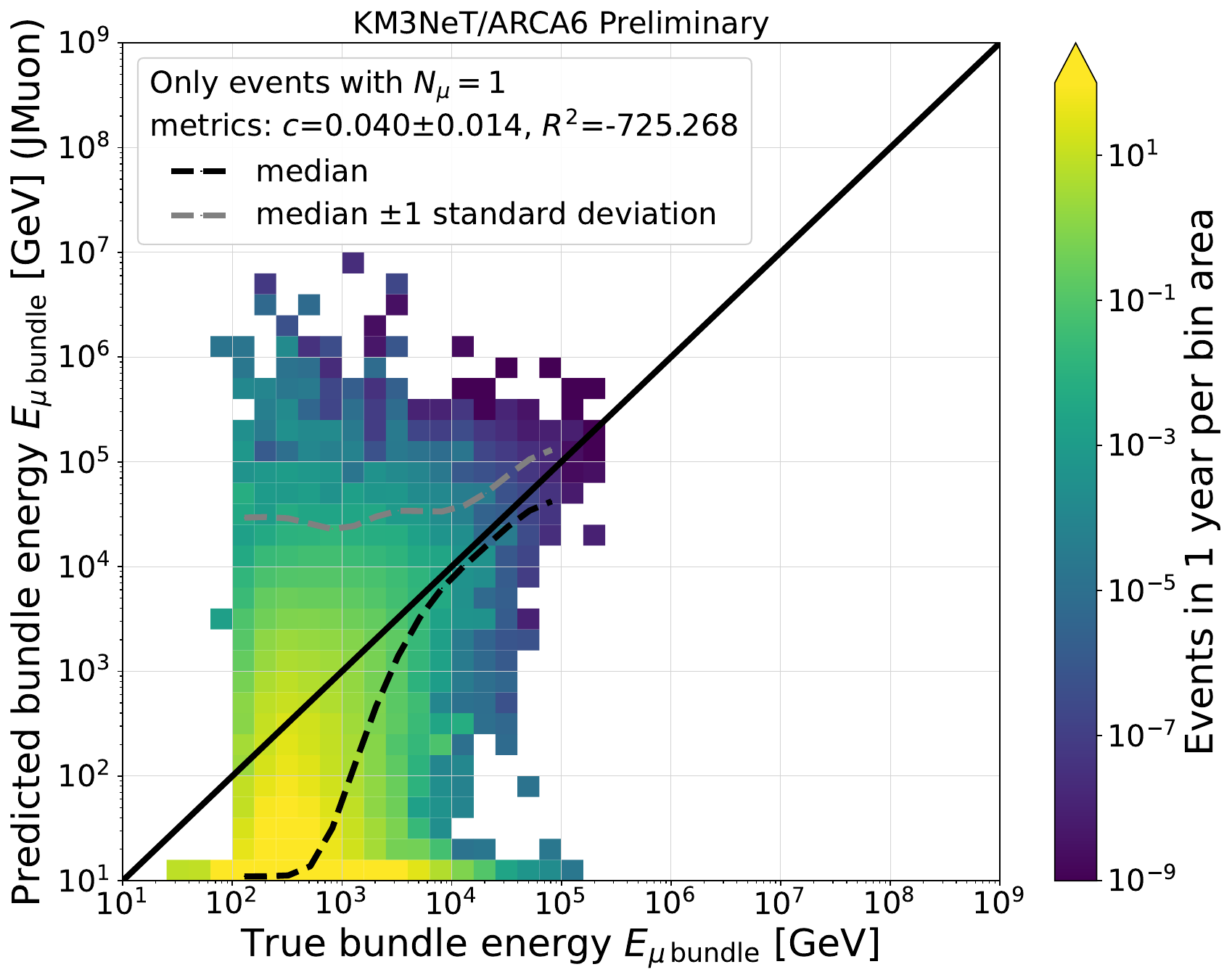}}\caption{Comparison of predicted and true bundle energy for the JMuon reconstruction.
The reconstruction was applied on CORSIKA MC for ARCA6. \label{fig:official_E_reco-A6}}
\end{figure}

\begin{figure}[H]
\centering{}\subfloat[All multiplicities. \label{fig:official_E_reco-O115_all}]{\centering{}\includegraphics[width=8cm]{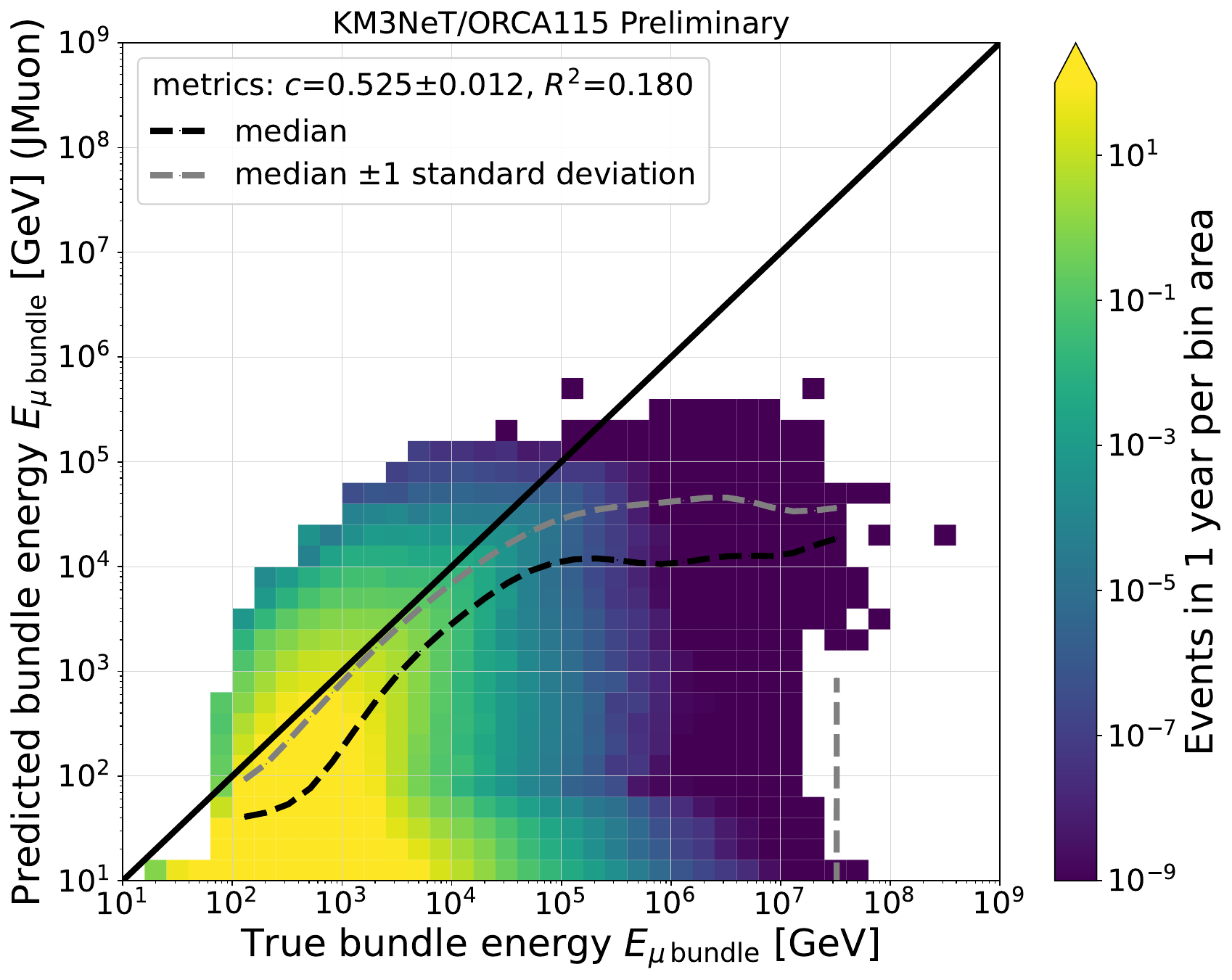}}\subfloat[Only single muon events (multiplicity 1). \label{fig:official_E_reco-O115_single_mu}]{\centering{}\includegraphics[width=8cm]{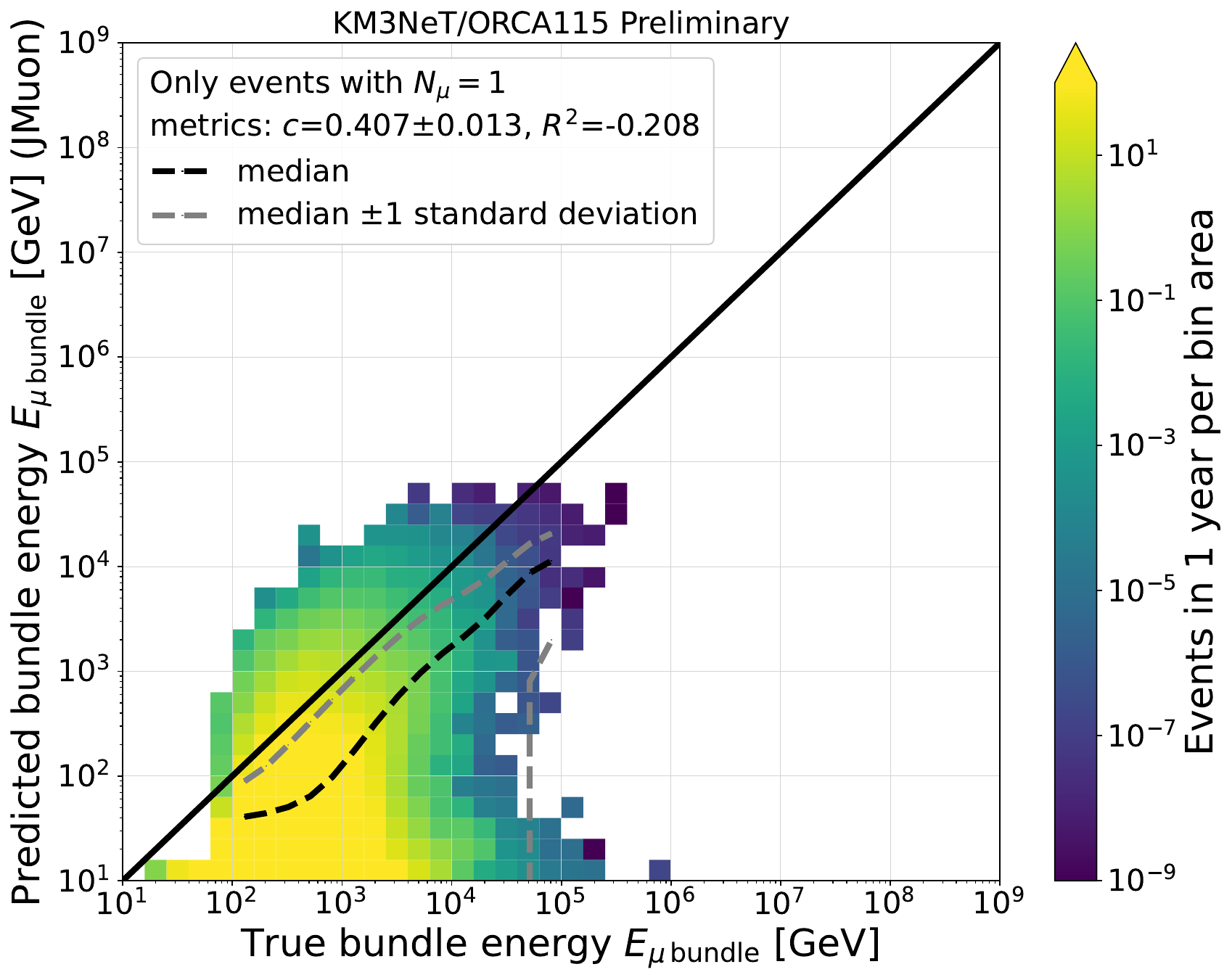}}\caption{Comparison of predicted and true bundle energy for the JMuon reconstruction.
The reconstruction was applied on CORSIKA MC for ORCA115. \label{fig:official_E_reco-O115}}
\end{figure}

\begin{figure}[H]
\centering{}\subfloat[All multiplicities. \label{fig:official_E_reco-O6_all}]{\centering{}\includegraphics[width=8cm]{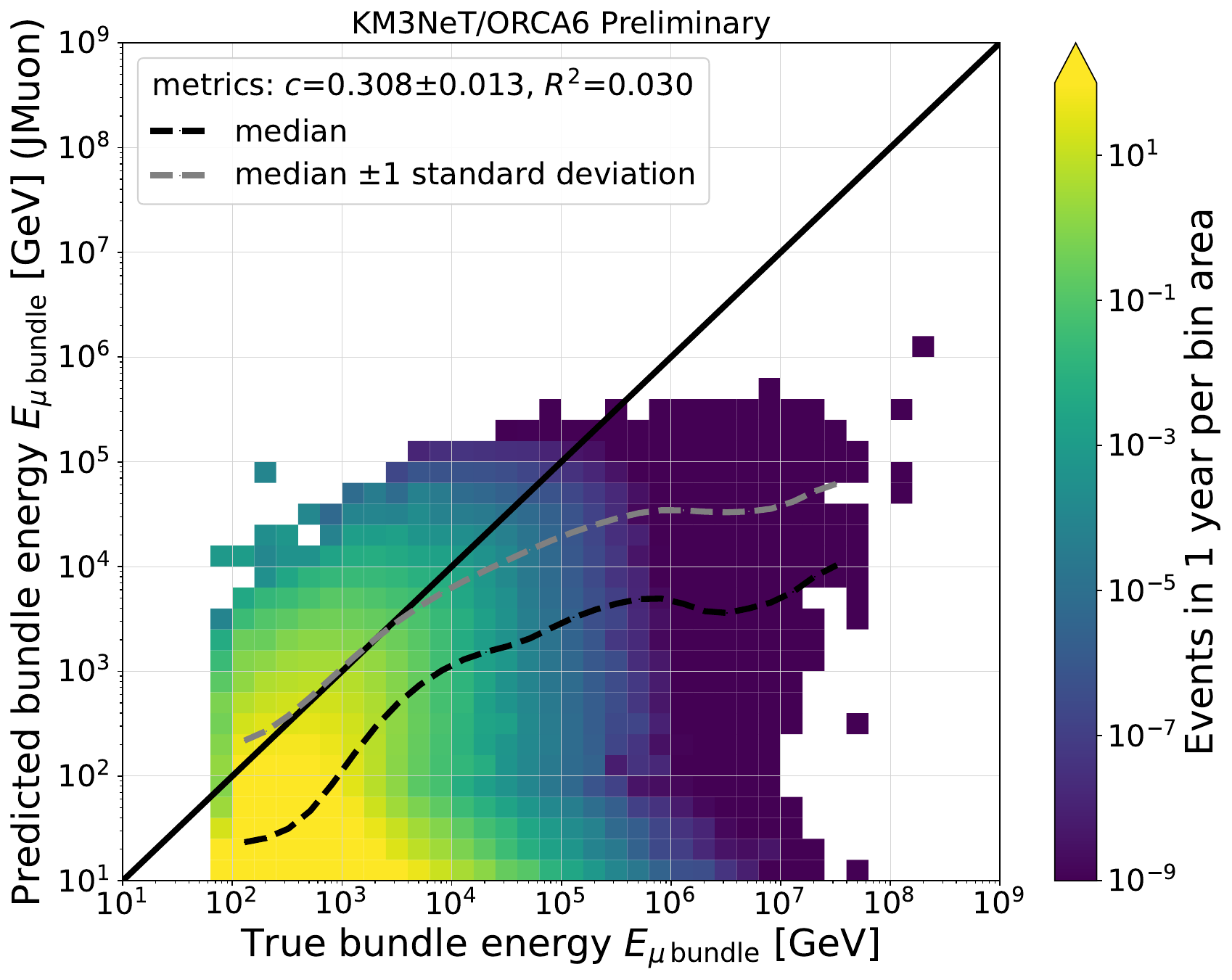}}\subfloat[Only single muon events (multiplicity 1). \label{fig:official_E_reco-O6_single_mu}]{\centering{}\includegraphics[width=8cm]{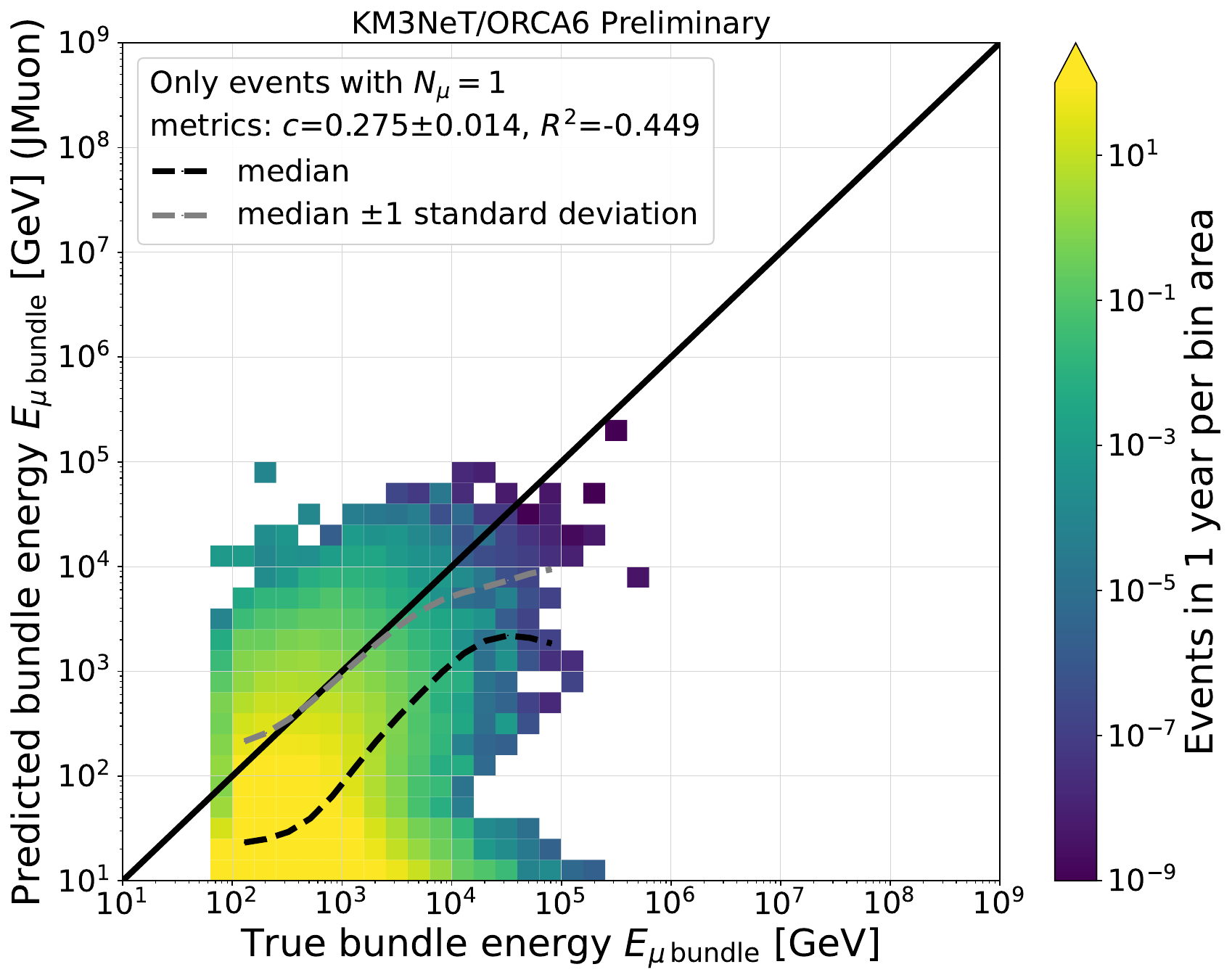}}\caption{Comparison of predicted and true bundle energy for the JMuon reconstruction.
The reconstruction was applied on CORSIKA MC for ORCA6. \label{fig:official_E_reco-O6}}
\end{figure}

Here, the results of applying JMuon to the muon bundle energy reconstruction
task are presented. Since they have been partially shown in Sec. \ref{subsec:Bundle-energy},
only the 2D plots for ARCA6, ORCA115 and ORCA6 are displayed in Fig.
\ref{fig:official_E_reco-A6}, \ref{fig:official_E_reco-O115}, and
\ref{fig:official_E_reco-O6} respectively.

\section{Muon bundle reconstruction \label{sec:Muon-bundle-reconstruction-supplement}}

The supplementary material for Chap. \ref{chap:muon-bundle-reco}
is gathered in this section.

\subsection{List of used features \label{subsec:Features-description}}

Here, each of the features used in Chap. \ref{chap:muon-bundle-reco}
is briefly described, using the following abbreviations and nomenclature:
\begin{itemize}
\item ToT – time over threshold. Time over which a PMT voltage exceeded
a certain threshold value (differs between the PMTs). The motivation
for using a threshold is noise reduction.
\item Hit – a PMT is said to have a hit when its voltage exceeds the set
threshold (equivalently, when ToT$>0$). Each hit has a position and
a direction, which correspond to the location of the PMT and the direction,
which it is facing. The hit time is the time at which the threshold
was exceeded.
\item 3DMUON and 3DSHOWER are the dedicated track and shower triggers, as
mentioned in Sec. \ref{sec:trigger}. There can also be events, where
both triggers were activated. In such a case they will be marked as
3DMUON\_3DSHOWER.
\item $c_{x}$, $c_{y}$, $c_{z}$ are the directional cosines along $x$,
$y$ and $z$ axes respectively. Here, the movement of the optical
modules in water is not taken into account.
\item Avg – average value.
\item Std – standard deviation.
\item Edge of the detector – an edge of the smallest enclosing cylinder
around the installed DUs.
\end{itemize}
\newpage{}
\begin{center}
\begin{table}[H]
\centering{}\caption{Summary of all available features for the reconstruction of muon bundle
properties. \label{tab:Summary-of-all-features}}
\end{table}
\par\end{center}

\begin{center}
\begin{longtable}[c]{|c|c|}
\hline 
{\footnotesize{}Feature} & {\footnotesize{}Description}\tabularnewline
\hline 
\hline 
{\footnotesize{}3DSHOWER\_trig\_hit\_amplitude\_sum} & {\footnotesize{}sum of 3DSHOWER hit amplitudes}\tabularnewline
\hline 
{\footnotesize{}3DSHOWER\_trig\_hit\_amplitude\_avg} & {\footnotesize{}avg 3DSHOWER hit amplitude}\tabularnewline
\hline 
{\footnotesize{}3DSHOWER\_trig\_hit\_amplitude\_std} & {\footnotesize{}std of 3DSHOWER hit amplitudes}\tabularnewline
\hline 
{\footnotesize{}3DMUON\_trig\_hit\_amplitude\_sum} & {\footnotesize{}sum of 3DMUON hit amplitudes}\tabularnewline
\hline 
{\footnotesize{}3DMUON\_trig\_hit\_amplitude\_avg} & {\footnotesize{}avg 3DMUON hit amplitude}\tabularnewline
\hline 
{\footnotesize{}3DMUON\_trig\_hit\_amplitude\_std} & {\footnotesize{}std of 3DMUON hit amplitudes}\tabularnewline
\hline 
{\footnotesize{}3DMUON\_3DSHOWER\_trig\_hit\_amplitude\_sum} & {\footnotesize{}sum of 3DMUON\_3DSHOWER hit amplitudes}\tabularnewline
\hline 
{\footnotesize{}3DMUON\_3DSHOWER\_trig\_hit\_amplitude\_avg} & {\footnotesize{}avg 3DMUON\_3DSHOWER hit amplitude}\tabularnewline
\hline 
{\footnotesize{}3DMUON\_3DSHOWER\_trig\_hit\_amplitude\_std} & {\footnotesize{}std of 3DMUON\_3DSHOWER hit amplitudes}\tabularnewline
\hline 
\multirow{2}{*}{{\footnotesize{}distance\_last\_3DSHOWER\_trig\_hit\_to\_det\_edge}} & {\footnotesize{}distance between the last 3DSHOWER}\tabularnewline
 & {\footnotesize{}hit and the edge of the detector}\tabularnewline
\hline 
\multirow{2}{*}{{\footnotesize{}distance\_first\_3DSHOWER\_trig\_hit\_to\_det\_edge}} & {\footnotesize{}distance between the first 3DSHOWER}\tabularnewline
 & {\footnotesize{}hit and the edge of the detector}\tabularnewline
\hline 
{\footnotesize{}first\_3DSHOWER\_trig\_hit\_pmt\_dir\_x} & {\footnotesize{}$c_{x}$ of the first 3DSHOWER hit}\tabularnewline
\hline 
{\footnotesize{}first\_3DSHOWER\_trig\_hit\_pmt\_dir\_y} & {\footnotesize{}$c_{y}$ of the first 3DSHOWER hit}\tabularnewline
\hline 
{\footnotesize{}first\_3DSHOWER\_trig\_hit\_pmt\_dir\_z} & {\footnotesize{}$c_{z}$ of the first 3DSHOWER hit}\tabularnewline
\hline 
{\footnotesize{}last\_3DSHOWER\_trig\_hit\_pmt\_dir\_x} & {\footnotesize{}$c_{x}$ of the last 3DSHOWER hit}\tabularnewline
\hline 
{\footnotesize{}last\_3DSHOWER\_trig\_hit\_pmt\_dir\_y} & {\footnotesize{}$c_{y}$ of the last 3DSHOWER hit}\tabularnewline
\hline 
{\footnotesize{}last\_3DSHOWER\_trig\_hit\_pmt\_dir\_z} & {\footnotesize{}$c_{z}$ of the last 3DSHOWER hit}\tabularnewline
\hline 
\multirow{2}{*}{{\footnotesize{}3DSHOWER\_trig\_hits\_duration}} & {\footnotesize{}time difference between the last}\tabularnewline
 & {\footnotesize{}and first 3DSHOWER hit}\tabularnewline
\hline 
\multirow{2}{*}{{\footnotesize{}distance\_last\_3DMUON\_trig\_hit\_to\_det\_edge}} & {\footnotesize{}distance between the last 3DMUON}\tabularnewline
 & {\footnotesize{}hit and the edge of the detector}\tabularnewline
\hline 
\multirow{2}{*}{{\footnotesize{}distance\_first\_3DMUON\_trig\_hit\_to\_det\_edge}} & {\footnotesize{}distance between the last 3DMUON}\tabularnewline
 & {\footnotesize{}hit and the edge of the detector}\tabularnewline
\hline 
{\footnotesize{}first\_3DMUON\_trig\_hit\_pmt\_dir\_x} & {\footnotesize{}$c_{x}$ of the first 3DMUON hit}\tabularnewline
\hline 
{\footnotesize{}first\_3DMUON\_trig\_hit\_pmt\_dir\_y} & {\footnotesize{}$c_{y}$ of the first 3DMUON hit}\tabularnewline
\hline 
{\footnotesize{}first\_3DMUON\_trig\_hit\_pmt\_dir\_z} & {\footnotesize{}$c_{z}$ of the first 3DMUON hit}\tabularnewline
\hline 
{\footnotesize{}last\_3DMUON\_trig\_hit\_pmt\_dir\_x} & {\footnotesize{}$c_{x}$ of the last 3DMUON hit}\tabularnewline
\hline 
{\footnotesize{}last\_3DMUON\_trig\_hit\_pmt\_dir\_y} & {\footnotesize{}$c_{y}$ of the last 3DMUON hit}\tabularnewline
\hline 
{\footnotesize{}last\_3DMUON\_trig\_hit\_pmt\_dir\_z} & {\footnotesize{}$c_{z}$ of the last 3DMUON hit}\tabularnewline
\hline 
\multirow{2}{*}{{\footnotesize{}3DMUON\_trig\_hits\_duration}} & {\footnotesize{}time difference between the last}\tabularnewline
 & {\footnotesize{}and first 3DMUON hit}\tabularnewline
\hline 
\multirow{2}{*}{{\footnotesize{}distance\_last\_3DMUON\_3DSHOWER\_trig\_hit\_to\_det\_edge}} & {\footnotesize{}distance between the last 3DMUON\_3DSHOWER}\tabularnewline
 & {\footnotesize{}hit and the edge of the detector}\tabularnewline
\hline 
\multirow{2}{*}{{\footnotesize{}distance\_first\_3DMUON\_3DSHOWER\_trig\_hit\_to\_det\_edge}} & {\footnotesize{}distance between the first 3DMUON\_3DSHOWER}\tabularnewline
 & {\footnotesize{}hit and the edge of the detector}\tabularnewline
\hline 
{\footnotesize{}first\_3DMUON\_3DSHOWER\_trig\_hit\_pmt\_dir\_x} & {\footnotesize{}$c_{x}$ of the first 3DMUON\_3DSHOWER hit}\tabularnewline
\hline 
{\footnotesize{}first\_3DMUON\_3DSHOWER\_trig\_hit\_pmt\_dir\_y} & {\footnotesize{}$c_{y}$ of the first 3DMUON\_3DSHOWER hit}\tabularnewline
\hline 
{\footnotesize{}first\_3DMUON\_3DSHOWER\_trig\_hit\_pmt\_dir\_z} & {\footnotesize{}$c_{z}$ of the first 3DMUON\_3DSHOWER hit}\tabularnewline
\hline 
{\footnotesize{}last\_3DMUON\_3DSHOWER\_trig\_hit\_pmt\_dir\_x} & {\footnotesize{}$c_{x}$ of the last 3DMUON\_3DSHOWER hit}\tabularnewline
\hline 
{\footnotesize{}last\_3DMUON\_3DSHOWER\_trig\_hit\_pmt\_dir\_y} & {\footnotesize{}$c_{y}$ of the last 3DMUON\_3DSHOWER hit}\tabularnewline
\hline 
{\footnotesize{}last\_3DMUON\_3DSHOWER\_trig\_hit\_pmt\_dir\_z} & {\footnotesize{}$c_{z}$ of the last 3DMUON\_3DSHOWER hit}\tabularnewline
\hline 
\multirow{2}{*}{{\footnotesize{}3DMUON\_3DSHOWER\_trig\_hits\_duration}} & {\footnotesize{}time difference between the last}\tabularnewline
 & {\footnotesize{}and first 3DMUON\_3DSHOWER hit}\tabularnewline
\hline 
{\footnotesize{}3DSHOWER\_trig\_hits} & {\footnotesize{}total number of 3DSHOWER hits}\tabularnewline
\hline 
{\footnotesize{}3DMUON\_trig\_hits} & {\footnotesize{}total number of 3DMUON hits.}\tabularnewline
\hline 
{\footnotesize{}3DMUON\_3DSHOWER\_trig\_hits} & {\footnotesize{}total number of 3DMUON\_3DSHOWER hits}\tabularnewline
\hline 
\multirow{2}{*}{{\footnotesize{}vertical\_span\_3DSHOWER\_trig\_hits}} & {\footnotesize{}the largest vertical distance}\tabularnewline
 & {\footnotesize{}between two 3DSHOWER hits}\tabularnewline
\hline 
\multirow{2}{*}{{\footnotesize{}horizontal\_span\_3DSHOWER\_trig\_hits}} & {\footnotesize{}the largest horizontal distance}\tabularnewline
 & {\footnotesize{}between two 3DSHOWER hits}\tabularnewline
\hline 
\multirow{2}{*}{{\footnotesize{}vertical\_span\_3DMUON\_trig\_hits}} & {\footnotesize{}the largest vertical distance}\tabularnewline
 & {\footnotesize{}between two 3DMUON hits}\tabularnewline
\hline 
\multirow{2}{*}{{\footnotesize{}horizontal\_span\_3DMUON\_trig\_hits}} & {\footnotesize{}the largest horizontal distance}\tabularnewline
 & {\footnotesize{}between two 3DMUON hits}\tabularnewline
\hline 
\multirow{2}{*}{{\footnotesize{}vertical\_span\_3DMUON\_3DSHOWER\_trig\_hits}} & {\footnotesize{}the largest vertical distance}\tabularnewline
 & {\footnotesize{}between two 3DMUON\_3DSHOWER hits}\tabularnewline
\hline 
\multirow{2}{*}{{\footnotesize{}horizontal\_span\_3DMUON\_3DSHOWER\_trig\_hits}} & {\footnotesize{}the largest horizontal distance}\tabularnewline
 & {\footnotesize{}between two 3DMUON\_3DSHOWER hits}\tabularnewline
\hline 
{\footnotesize{}overlays} & {\footnotesize{}number of overlaid triggers in an event}\tabularnewline
\hline 
\end{longtable}
\par\end{center}

\subsection{Correlation matrices and dendrograms\label{subsec:Correlation-matrices-and-dendrograms}}

The correlation matrices for ARCA6, ORCA115 and ORCA6 are shown in
Fig. \ref{fig:corr_matrix_A6}, \ref{fig:corr_matrix_O115}, and \ref{fig:corr_matrix_O6}.
The corresponding dendrograms are plotted in Fig. \ref{fig:dendro-A6},
\ref{fig:dendro-O115}, and \ref{fig:dendro-O6}.
\begin{center}
\begin{figure}[H]
\centering{}\includegraphics[width=16cm]{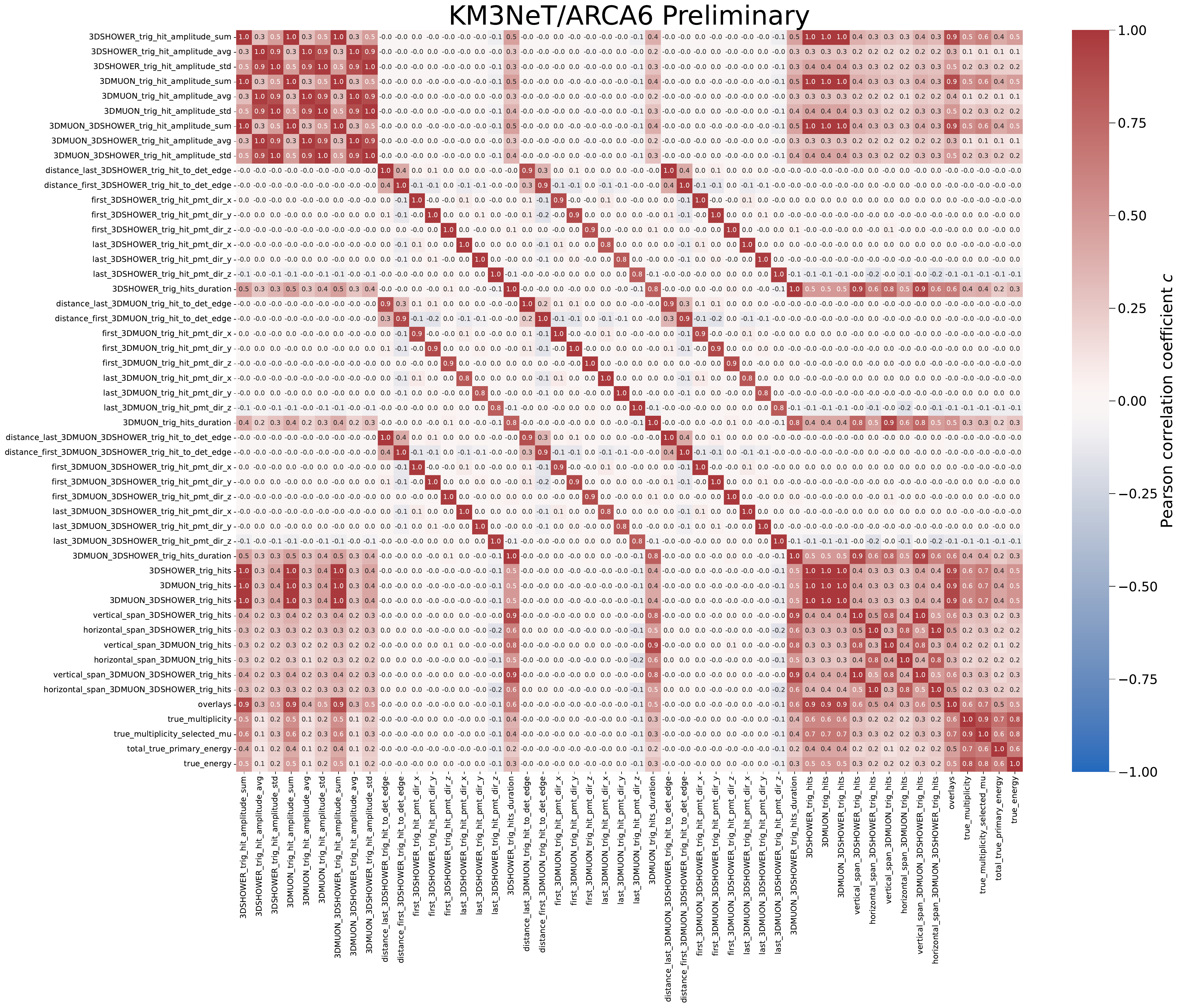}\caption{Correlation matrix with all features and potential targets for ARCA6.
\label{fig:corr_matrix_A6}}
\end{figure}
\par\end{center}

\begin{center}
\begin{figure}[H]
\centering{}\includegraphics[width=16cm]{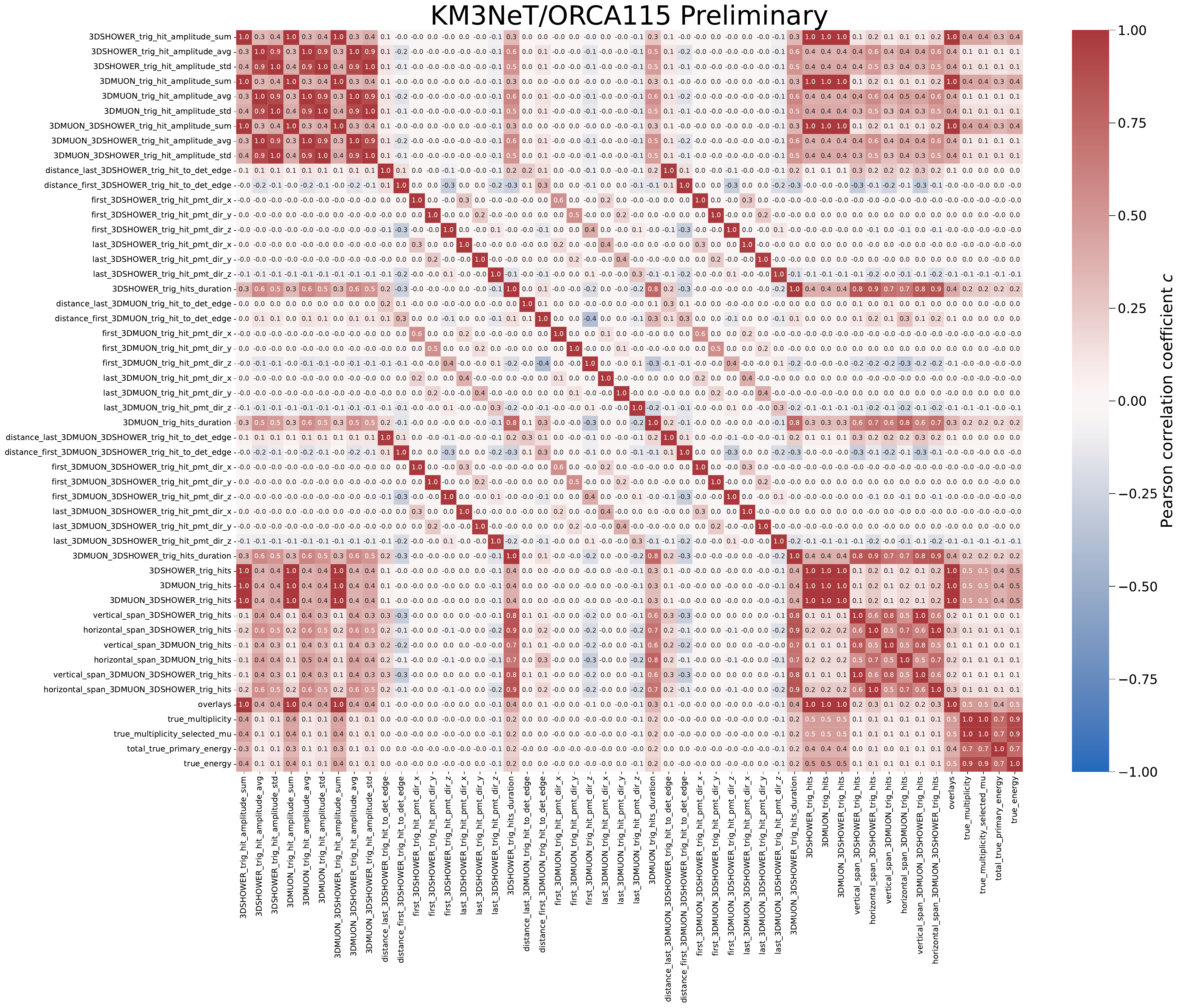}\caption{Correlation matrix with all features and potential targets for ORCA115.
\label{fig:corr_matrix_O115}}
\end{figure}
\par\end{center}

\begin{center}
\begin{figure}[H]
\centering{}\includegraphics[width=16cm]{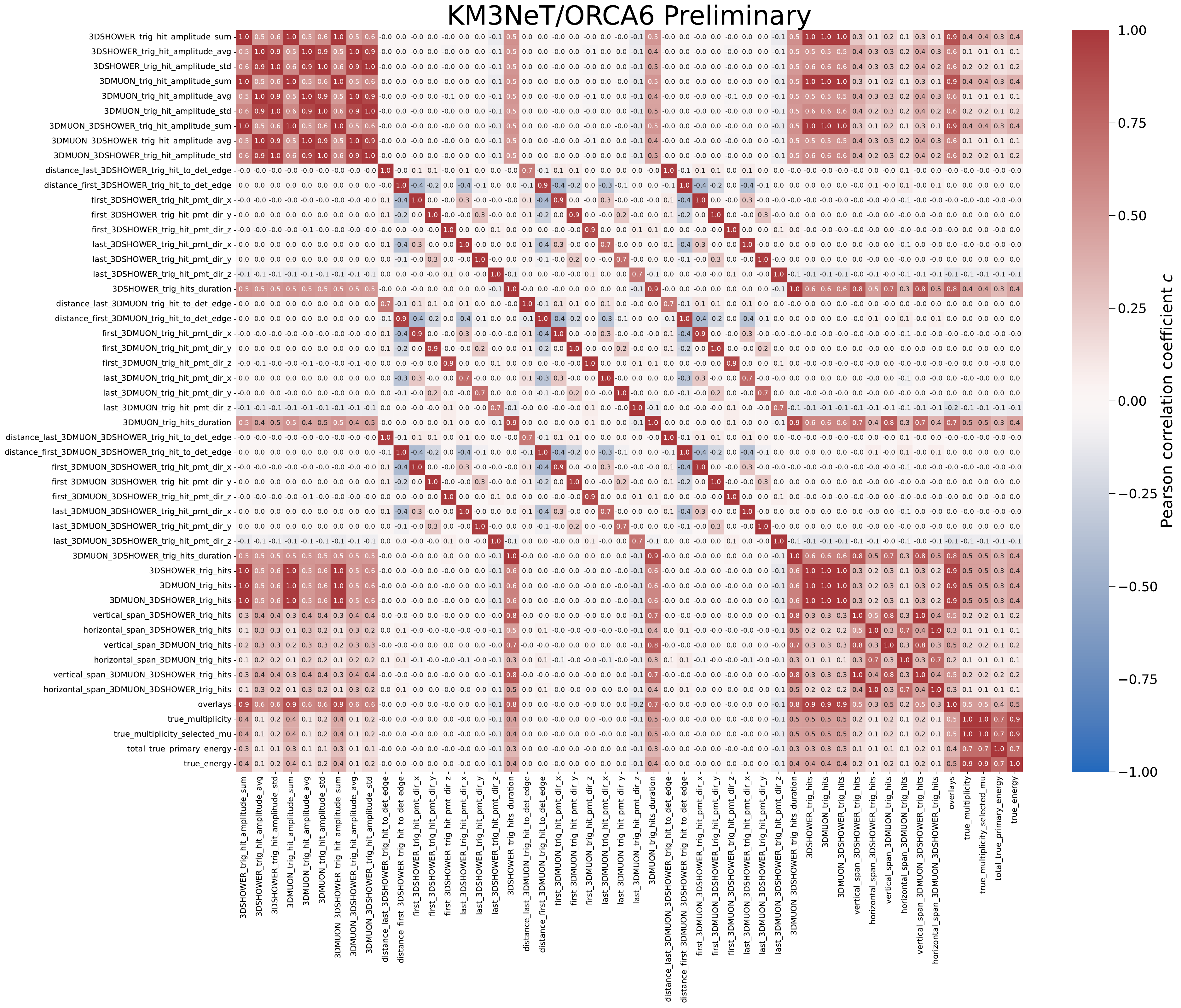}\caption{Correlation matrix with all features and potential targets for ORCA6.
\label{fig:corr_matrix_O6}}
\end{figure}
\par\end{center}

\begin{center}
\begin{figure}[H]
\centering{}\includegraphics[width=16cm]{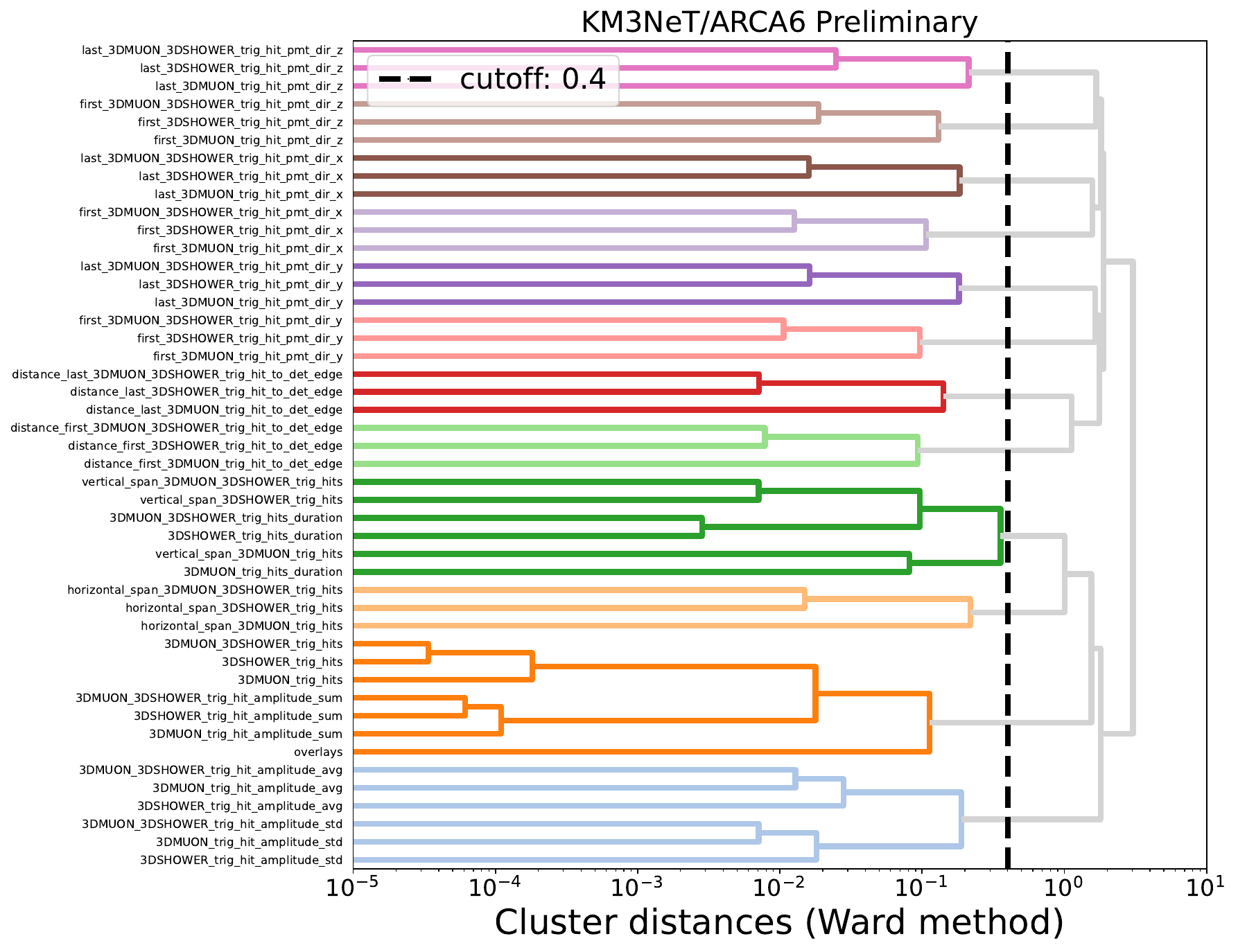}\caption{Dendrogram showing the clustering of features for ARCA6. \label{fig:dendro-A6}}
\end{figure}
\par\end{center}

\begin{center}
\begin{figure}[H]
\centering{}\includegraphics[width=16cm]{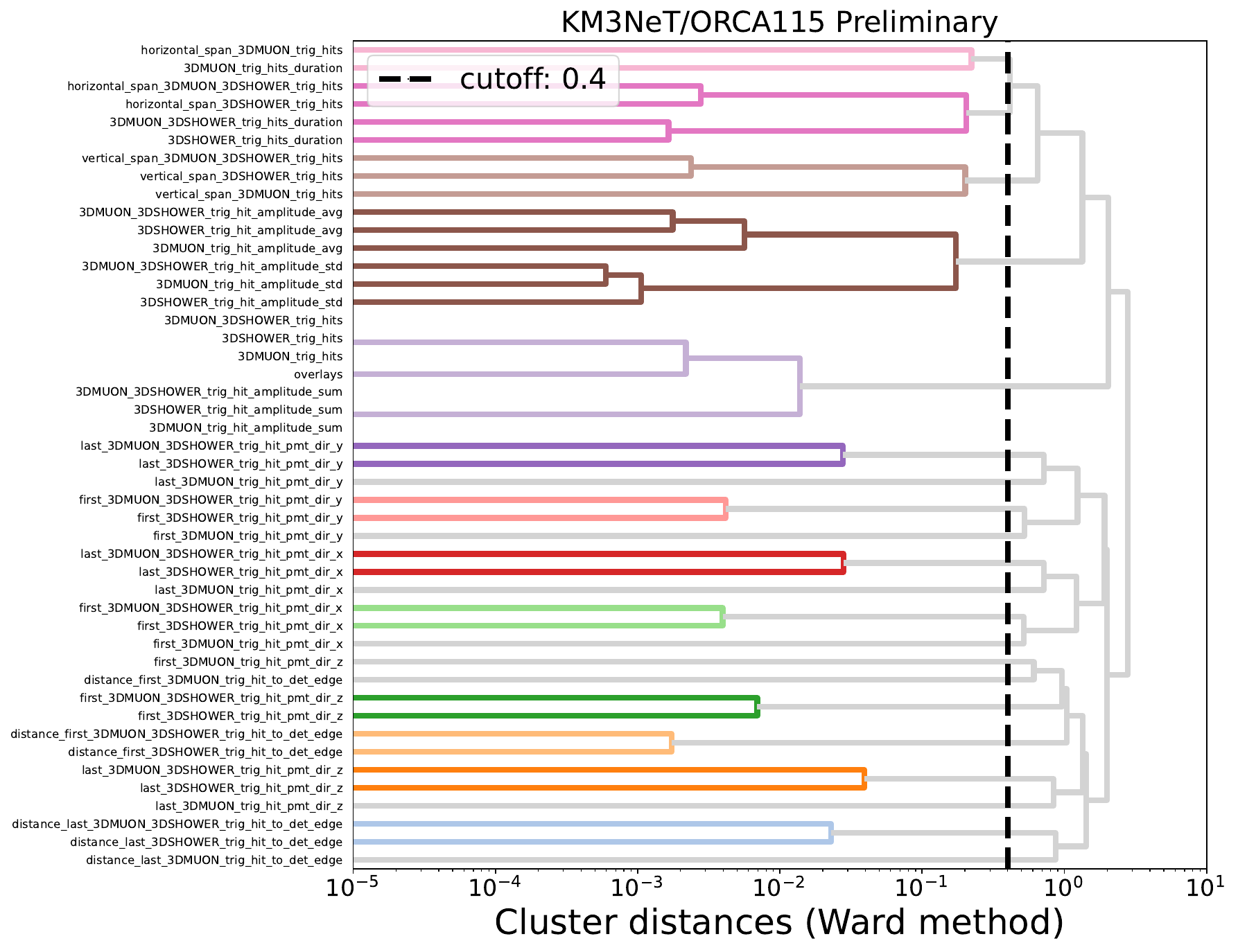}\caption{Dendrogram showing the clustering of features for ORCA115. \label{fig:dendro-O115}}
\end{figure}
\par\end{center}

\begin{center}
\begin{figure}[H]
\centering{}\includegraphics[width=16cm]{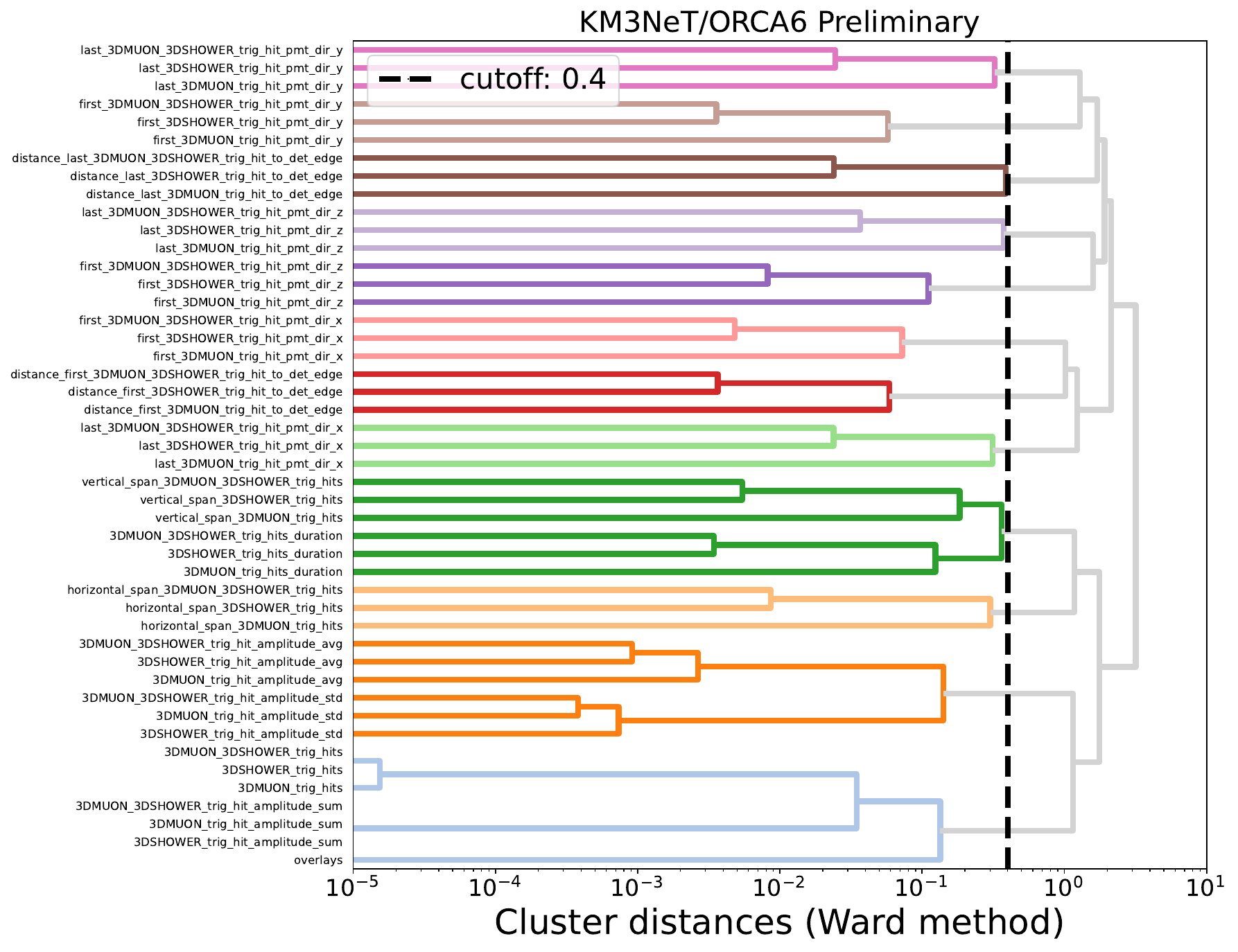}\caption{Dendrogram showing the clustering of features for ORCA6. \label{fig:dendro-O6}}
\end{figure}
\par\end{center}

\subsection{Feature importances\label{subsec:Feature-importances}}

Feature importances for energy and multiplicity reconstructions are
compiled in this section. For each of the detector configurations
(ARCA6, ORCA115, ORCA6), the first one is for the energy, and the
second one is for the multiplicity reconstruction. All the importances
were computed on untuned LightGBM.
\begin{center}
\begin{figure}[H]
\centering{}\includegraphics[width=15cm]{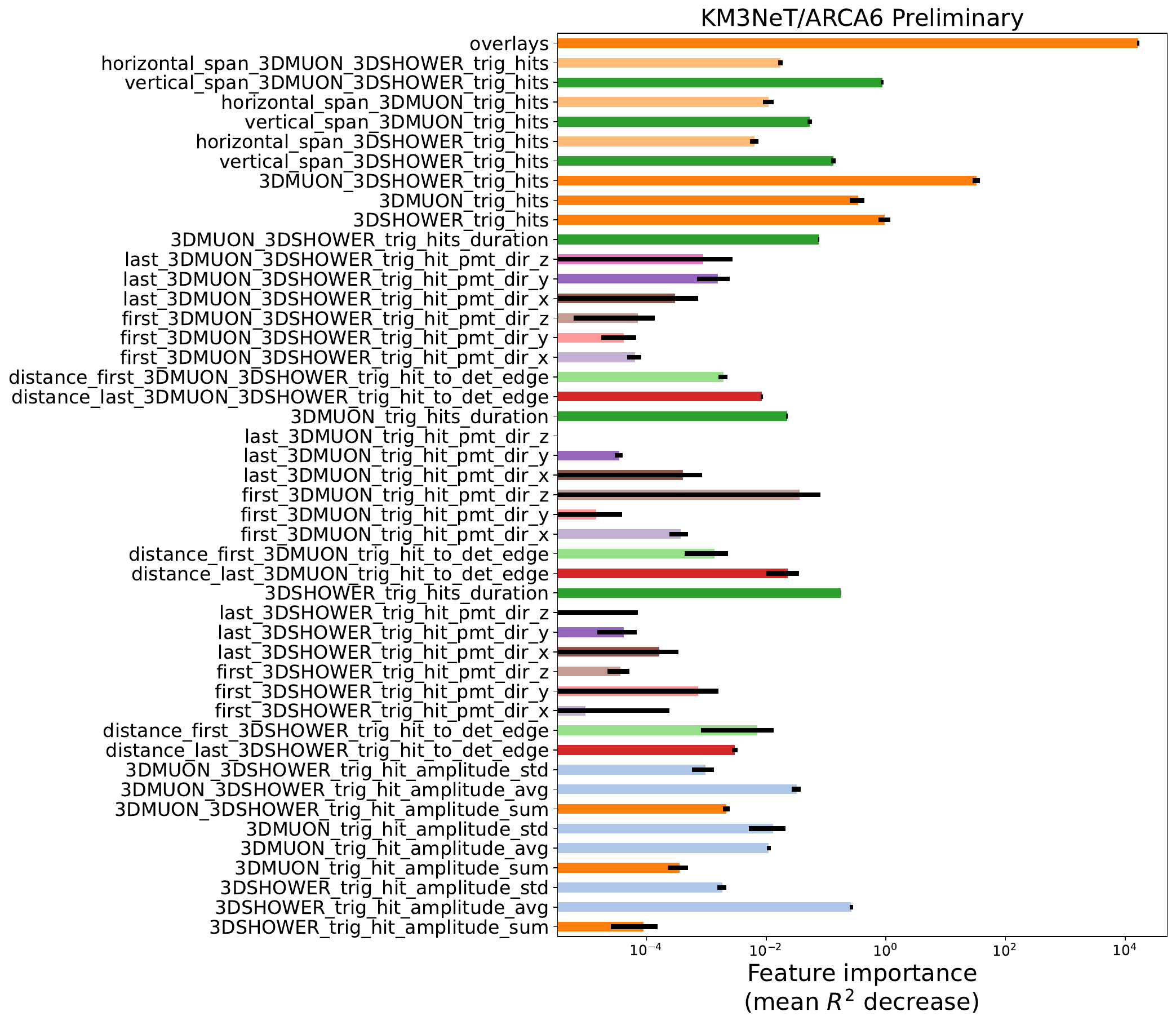}\caption{Feature importance for ARCA6 for the energy reconstruction. \label{fig:feature-importances-Ebundle-A6}}
\end{figure}
\par\end{center}

\begin{center}
\begin{figure}[H]
\centering{}\includegraphics[width=15cm]{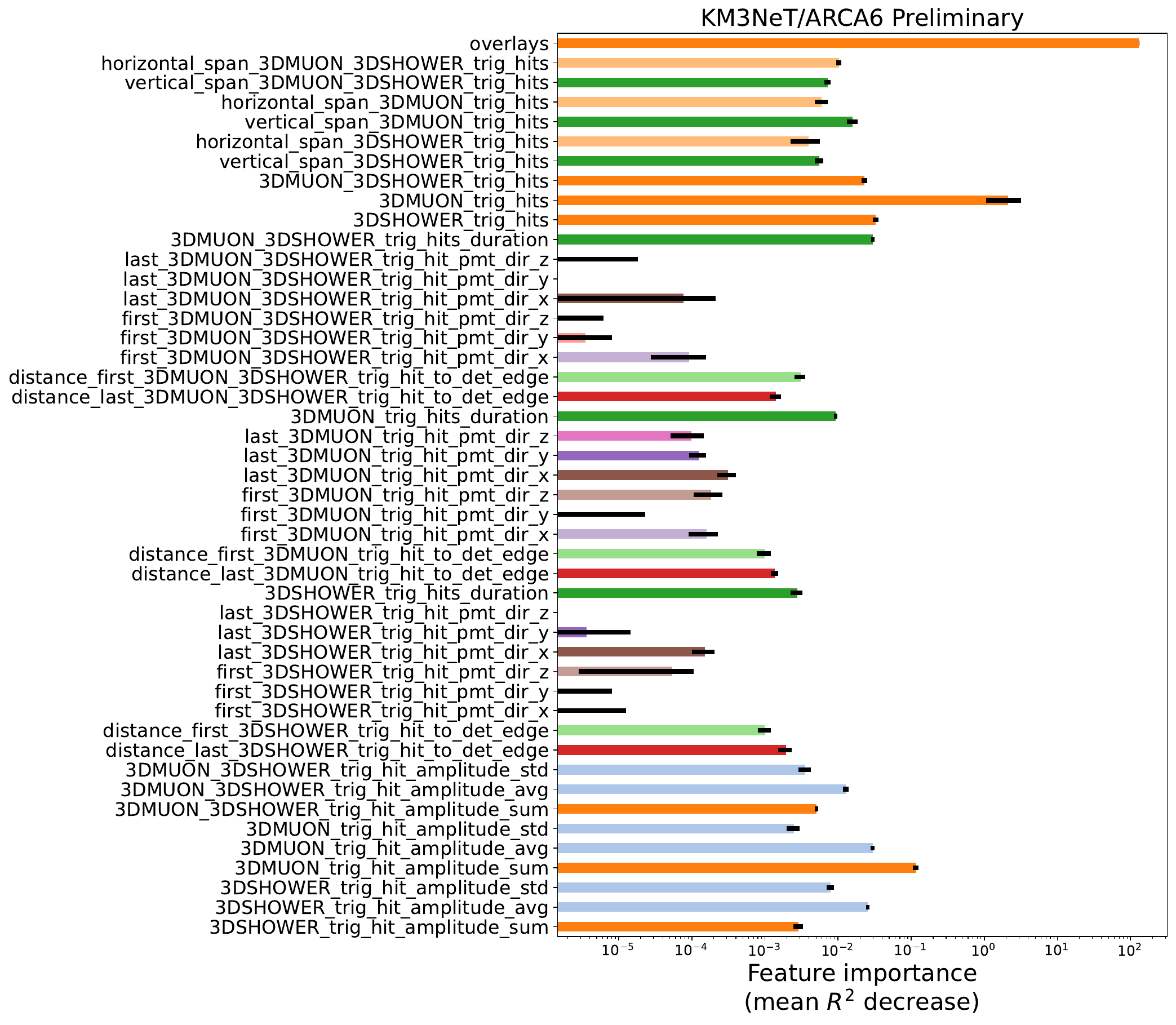}\caption{Feature importance for ARCA6 for the multiplicity reconstruction.
\label{fig:feature-importances-Nmu-A6}}
\end{figure}
\par\end{center}

\begin{center}
\begin{figure}[H]
\centering{}\includegraphics[width=15cm]{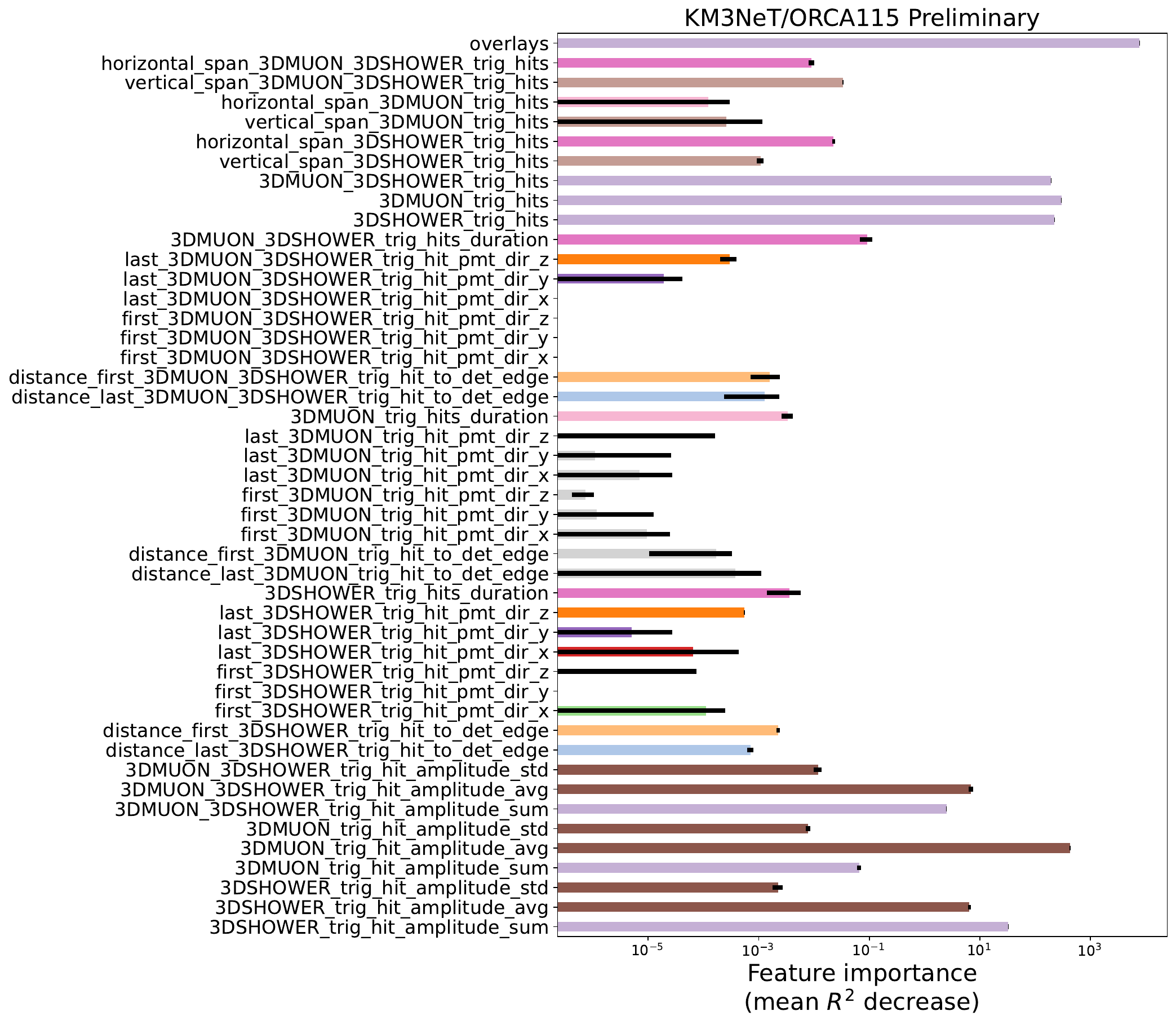}\caption{Feature importance for ORCA115 for the energy reconstruction. \label{fig:feature-importances-Ebundle-O115}}
\end{figure}
\par\end{center}

\begin{center}
\begin{figure}[H]
\centering{}\includegraphics[width=15cm]{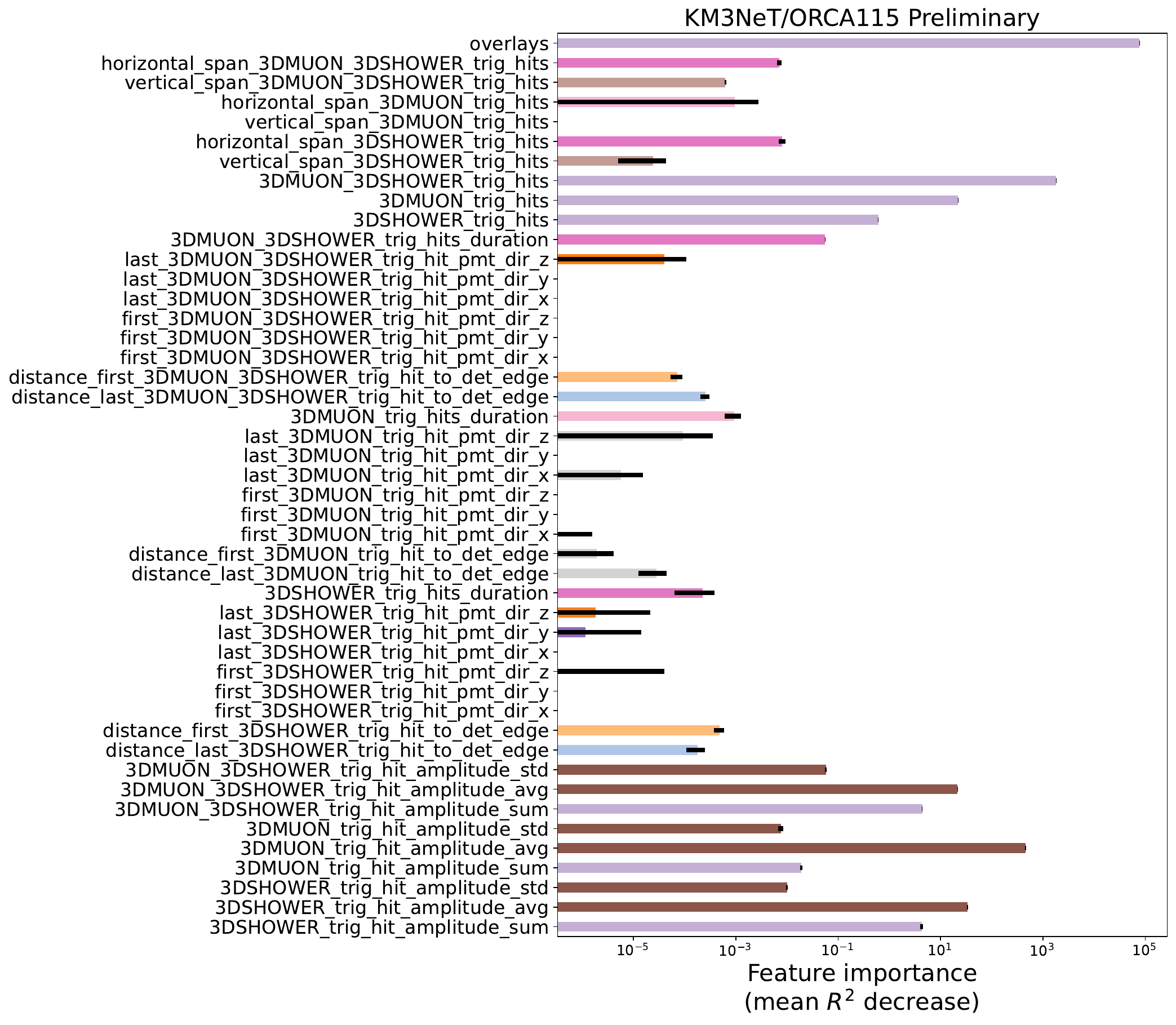}\caption{Feature importance for ORCA115 for the multiplicity reconstruction.
\label{fig:feature-importances-Nmu-O115}}
\end{figure}
\par\end{center}

\begin{center}
\begin{figure}[H]
\centering{}\includegraphics[width=15cm]{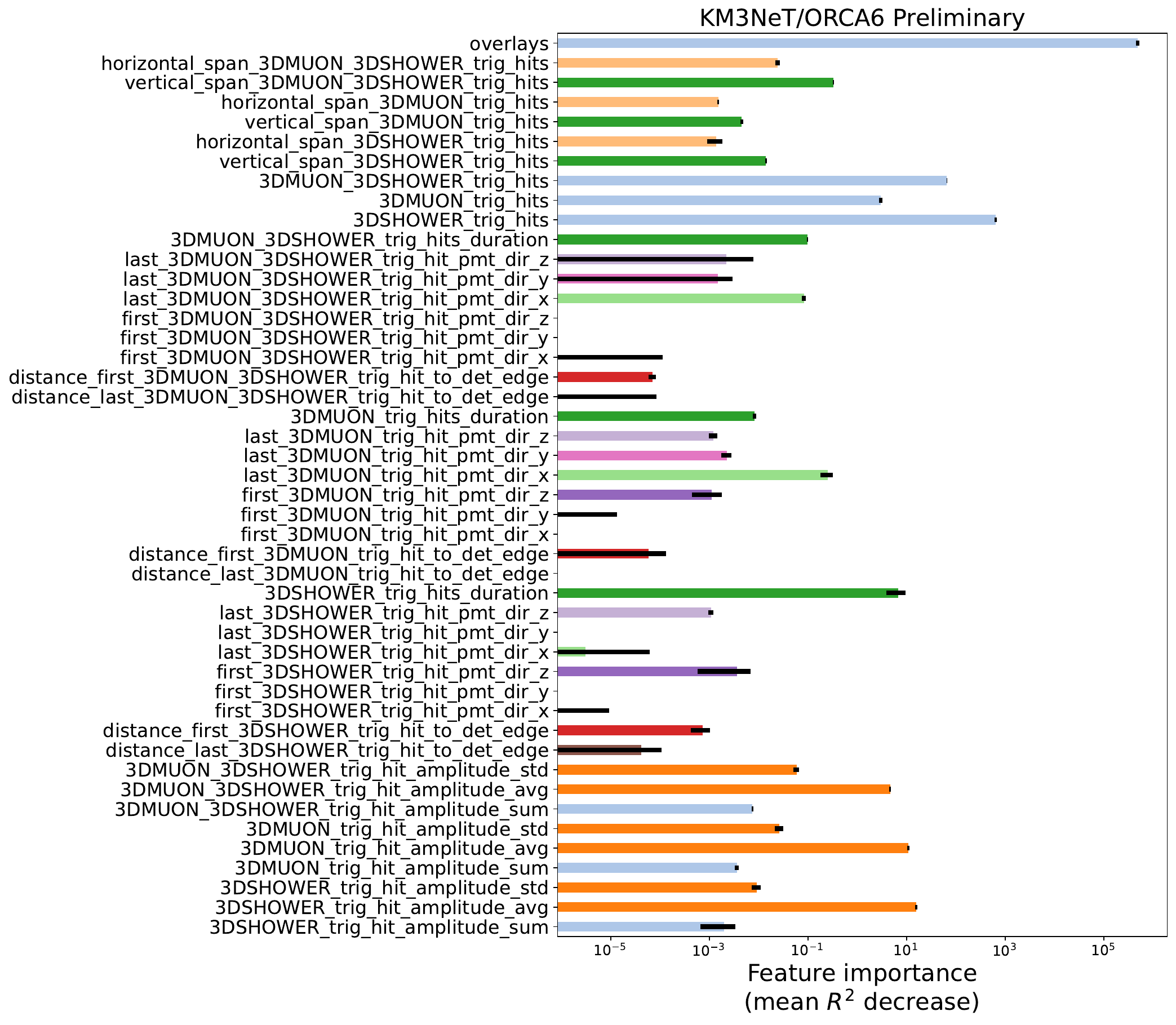}\caption{Feature importance for ORCA6 for the energy reconstruction. \label{fig:feature-importances-Ebundle-O6}}
\end{figure}
\par\end{center}

\begin{center}
\begin{figure}[H]
\centering{}\includegraphics[width=15cm]{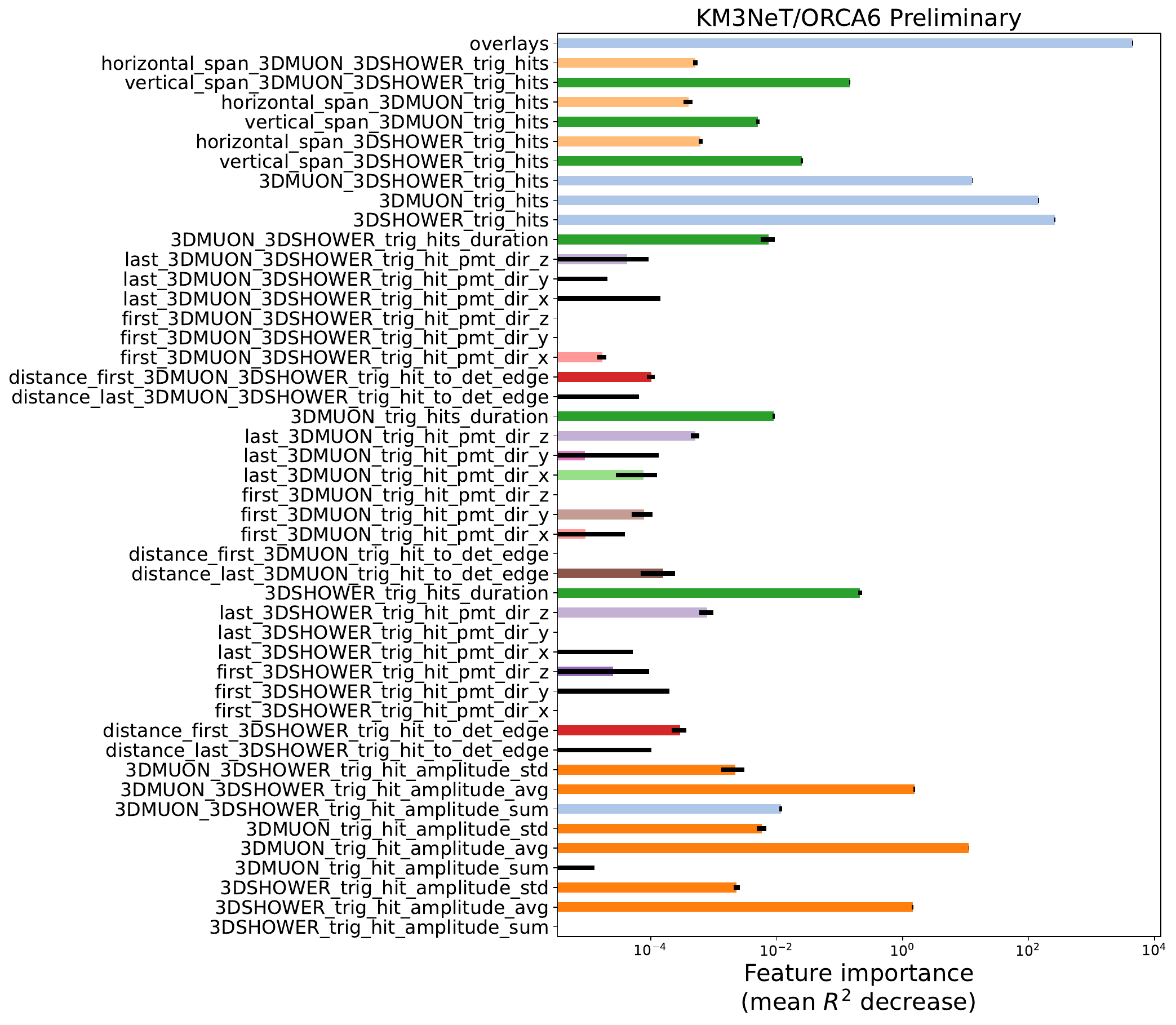}\caption{Feature importance for ORCA6 for the multiplicity reconstruction.
\label{fig:feature-importances-Nmu-O6}}
\end{figure}
\par\end{center}

\subsection{Hyperparameter tuning \label{subsec:Hyperparameter-tuning}}

The tuning of hyperparameters has been carried out using the Optuna
framework, which offers easy parallelisation options, smart parameter
space searching and pruning of unpromising trials \textcite{Optuna}.
All these traits were crucial, since scanning a multi-dimensional
hyperparameter space can be a very costly endeavour. The resulting
best-fit parameters were summarised in Tab. \ref{tab:LightGBM-tuned-hyperparameters-summary}.

\begin{table}[H]
\begin{centering}
\caption{Optimal LightGBM hyperparameters obtained with Optuna by scanning
over 120 sampled sets of parameters. A separate hyperparameter scan
has been performed for energy and multiplicity and for each of the
4 used detector configurations. \label{tab:LightGBM-tuned-hyperparameters-summary}}
\par\end{centering}
\centering{}%
\begin{tabular}{|c|c|c|c|c|c|c|c|c|}
\hline 
\multirow{2}{*}{{\footnotesize{}Parameter}} & \multicolumn{2}{c|}{{\footnotesize{}ARCA115}} & \multicolumn{2}{c|}{{\footnotesize{}ARCA6}} & \multicolumn{2}{c|}{{\footnotesize{}ORCA115}} & \multicolumn{2}{c|}{{\footnotesize{}ORCA6}}\tabularnewline
\cline{2-9} \cline{3-9} \cline{4-9} \cline{5-9} \cline{6-9} \cline{7-9} \cline{8-9} \cline{9-9} 
 & {\footnotesize{}$E$} & {\footnotesize{}$N_{\mu}$} & {\footnotesize{}$E$} & {\footnotesize{}$N_{\mu}$} & {\footnotesize{}$E$} & {\footnotesize{}$N_{\mu}$} & {\footnotesize{}$E$} & {\footnotesize{}$N_{\mu}$}\tabularnewline
\hline 
\hline 
{\footnotesize{}learning\_rate} & {\footnotesize{}0.06} & {\footnotesize{}0.08} & {\footnotesize{}0.02} & {\footnotesize{}0.05} & {\footnotesize{}0.06} & {\footnotesize{}0.05} & {\footnotesize{}0.02} & {\footnotesize{}0.06}\tabularnewline
\hline 
{\footnotesize{}num\_iterations} & {\footnotesize{}910} & {\footnotesize{}1000} & {\footnotesize{}494} & {\footnotesize{}759} & {\footnotesize{}578} & {\footnotesize{}972} & {\footnotesize{}879} & {\footnotesize{}904}\tabularnewline
\hline 
{\footnotesize{}colsample\_bytree} & {\footnotesize{}0.89} & {\footnotesize{}0.60} & {\footnotesize{}0.72} & {\footnotesize{}0.97} & {\footnotesize{}0.65} & {\footnotesize{}0.98} & {\footnotesize{}0.87} & {\footnotesize{}0.72}\tabularnewline
\hline 
{\footnotesize{}min\_child\_samples} & {\footnotesize{}194} & {\footnotesize{}462} & {\footnotesize{}541} & {\footnotesize{}158} & {\footnotesize{}21} & {\footnotesize{}515} & {\footnotesize{}695} & {\footnotesize{}918}\tabularnewline
\hline 
{\footnotesize{}min\_child\_weight} & {\footnotesize{}$2.70\cdot10^{-5}$} & {\footnotesize{}0.59} & {\footnotesize{}$1.40\cdot10^{-3}$} & {\footnotesize{}$1.62\cdot10^{-3}$} & {\footnotesize{}$5.12\cdot10^{-4}$} & {\footnotesize{}0.19} & {\footnotesize{}0.13} & {\footnotesize{}$1.51\cdot10^{-4}$}\tabularnewline
\hline 
{\footnotesize{}n\_estimators} & {\footnotesize{}675} & {\footnotesize{}492} & {\footnotesize{}568} & {\footnotesize{}783} & {\footnotesize{}474} & {\footnotesize{}259} & {\footnotesize{}19} & {\footnotesize{}87}\tabularnewline
\hline 
{\footnotesize{}num\_leaves} & {\footnotesize{}473} & {\footnotesize{}95} & {\footnotesize{}288} & {\footnotesize{}25} & {\footnotesize{}416} & {\footnotesize{}337} & {\footnotesize{}429} & {\footnotesize{}170}\tabularnewline
\hline 
{\footnotesize{}reg\_alpha} & {\footnotesize{}74.08} & {\footnotesize{}48.74} & {\footnotesize{}83.24} & {\footnotesize{}4.63} & {\footnotesize{}18.95} & {\footnotesize{}55.32} & {\footnotesize{}132.17} & {\footnotesize{}39.83}\tabularnewline
\hline 
{\footnotesize{}reg\_lambda} & {\footnotesize{}169.07} & {\footnotesize{}79.11} & {\footnotesize{}0.19} & {\footnotesize{}138.49} & {\footnotesize{}183.73} & {\footnotesize{}80.26} & {\footnotesize{}5.41} & {\footnotesize{}28.98}\tabularnewline
\hline 
{\footnotesize{}subsample} & {\footnotesize{}0.97} & {\footnotesize{}0.72} & {\footnotesize{}0.84} & {\footnotesize{}0.84} & {\footnotesize{}0.73} & {\footnotesize{}0.81} & {\footnotesize{}0.86} & {\footnotesize{}0.94}\tabularnewline
\hline 
{\footnotesize{}subsample\_for\_bin} & {\footnotesize{}10926} & {\footnotesize{}82921} & {\footnotesize{}40448} & {\footnotesize{}10488} & {\footnotesize{}58336} & {\footnotesize{}81283} & {\footnotesize{}32516} & {\footnotesize{}60497}\tabularnewline
\hline 
\end{tabular}
\end{table}

\section{Prompt muon analysis \label{sec:Prompt-muon-analysis-supplement}}

Here, all the supplementary materials for Chap. \ref{chap:prompt_ana}
were gathered.

\subsection{Hadrons implemented in CORSIKA}

The following tables list the hadrons handled by CORSIKA and the ones,
which were not implemented.

\begin{table}[H]
\begin{centering}
\caption{Hadrons implemented in CORSIKA. It is not differentiated here between
the charged and neutral species, or between particles and antiparticles
for the sake of simplicity. C, S, B indicate, whether the hadron contains
heavier quarks: $c$, $s$, or $b$ respectively. \label{tab:Hadrons-implemented-in-CORSIKA}}
\par\end{centering}
\centering{}%
\begin{tabular}{|c|c|c|c|}
\hline 
Meson & C & S & B\tabularnewline
\hline 
$\pi$ &  &  & \tabularnewline
\hline 
$K$ &  & $\times$ & \tabularnewline
\hline 
$\rho$ &  &  & \tabularnewline
\hline 
$\omega$ &  &  & \tabularnewline
\hline 
$\phi$ &  & $\times$ & \tabularnewline
\hline 
$\eta$ &  & $\times$ & \tabularnewline
\hline 
$\eta_{c}$ & $\times$ &  & \tabularnewline
\hline 
$D$ & $\times$ &  & \tabularnewline
\hline 
$D_{s}$ & $\times$ & $\times$ & \tabularnewline
\hline 
$J/\psi$ & $\times$ &  & \tabularnewline
\hline 
\end{tabular}\hspace{1.5cm}%
\begin{tabular}{|c|c|c|c|}
\hline 
Baryon & C & S & B\tabularnewline
\hline 
$p$ &  &  & \tabularnewline
\hline 
$n$ &  &  & \tabularnewline
\hline 
$\Lambda$ &  & $\times$ & \tabularnewline
\hline 
$\Lambda_{b}$ &  &  & $\times$\tabularnewline
\hline 
$\Lambda_{c}$ & $\times$ &  & \tabularnewline
\hline 
$\Sigma$ &  & $\times$ & \tabularnewline
\hline 
$\Sigma_{c}$ & $\times$ &  & \tabularnewline
\hline 
$\Sigma_{b}$ &  &  & $\times$\tabularnewline
\hline 
$\Delta$ &  &  & \tabularnewline
\hline 
$\Xi$ &  & $\times$ & \tabularnewline
\hline 
$\Xi_{c}$ & $\times$ & $\times$ & \tabularnewline
\hline 
$\Xi_{b}$ &  & $\times$ & $\times$\tabularnewline
\hline 
$\Omega$ &  & $\times$ & \tabularnewline
\hline 
$\Omega_{c}$ & $\times$ & $\times$ & \tabularnewline
\hline 
$\Omega_{b}$ &  & $\times$ & $\times$\tabularnewline
\hline 
\end{tabular}
\end{table}

\begin{table}[H]
\begin{centering}
\caption{Hadrons so far not implemented in CORSIKA. C, S, B indicate, whether
the hadron contains heavier quarks: $c$, $s$, or $b$ respectively.
\label{tab:Hadrons-not-implemented-in-CORSIKA}}
\par\end{centering}
\centering{}%
\begin{tabular}{|c|c|c|c|}
\hline 
Meson & C & S & B\tabularnewline
\hline 
$\eta_{b}$ &  &  & $\times$\tabularnewline
\hline 
$B$ &  &  & $\times$\tabularnewline
\hline 
$B_{c}$ & $\times$ &  & $\times$\tabularnewline
\hline 
$B_{s}$ &  & $\times$ & $\times$\tabularnewline
\hline 
$\Upsilon$ &  &  & $\times$\tabularnewline
\hline 
\end{tabular}\hspace{1.5cm}%
\begin{tabular}{|c|c|c|c|}
\hline 
Baryon & C & S & B\tabularnewline
\hline 
$\Xi_{cc}$ & $\times$ &  & \tabularnewline
\hline 
\end{tabular}
\end{table}

In addition to the particles listed in Tab. \ref{tab:Hadrons-not-implemented-in-CORSIKA},
some subdominant processes are not implemented in CORSIKA as well: 
\begin{itemize}
\item nuclear $\mu$ interactions,
\item heavy pair production:
\begin{itemize}
\item $\gamma\longrightarrow\mu^{+}\mu^{-}$ (cross-section suppressed by
a factor $\sim10^{-4}$ w.r.t. $\gamma\longrightarrow e^{+}e^{-}$),
\item $\gamma\longrightarrow\tau^{+}\tau^{-}$ (cross-section suppressed
by a factor $\sim10^{-10}$ w.r.t. $\gamma\longrightarrow e^{+}e^{-}$).
\end{itemize}
\end{itemize}

\subsection{Classification of parent particles into conventional and prompt by
lifetime\label{sec:Classification-of-parent-particles-by-lifetime}}

In general, a probability for a particle to decay is described by:
\begin{equation}
P\left(t\right)=e^{-\frac{t}{\gamma'\tau}},\label{eq:Poisson-decay-probability}
\end{equation}

where $\gamma=\frac{1}{\sqrt{1-\left(\frac{v}{c}\right)^{2}}}$ is
the gamma factor of the particle, $v$ is its velocity and $c$ is
the speed of light in vacuum, and $t$ is the time.

The basic criterion of classification of muons was the lifetime $\tau$
of their parent particles, as in \cite{conv-prompt-calculation}:
\begin{itemize}
\item \textcolor{red}{prompt: }if parent particle has $\tau<\tau_{K_{\mathsf{S}}^{0}}=8.95\cdot10^{-5}\,$μs,
\item \textcolor{blue}{conv: }if parent $\tau\geq\tau_{K_{\mathsf{S}}^{0}}$.
\end{itemize}
The table below compiles all parent particle types that are produced
in the CORSIKA simulation as of version v7.7410.

{\small{}}
\begin{table}[H]
{\small{}\caption{List of all parent particles found in the CORSIKA MC with their lifetimes
and classification as contributing either to \foreignlanguage{english}{\textcolor{blue}{conventional}}
or to the \foreignlanguage{english}{\textcolor{red}{prompt}} muon
flux. The masses and lifetimes are rounded values, obtained with \cite{scikit-hep_particle}.
\label{tab:List-of-all-particles-in-CORSIKA}}
}
\end{table}
{\small\par}
\begin{center}
{\small{}}%
\begin{longtable}[c]{|c|c|c|c|c|c|c|}
\hline 
\# & PDG ID & Symbol & Rest mass $m_{0}$ {[}$\frac{\mathsf{MeV}}{c^{2}}${]} & Lifetime $\tau$ {[}μs{]} & Category & Comment\tabularnewline
\endhead
\hline 
\textcolor{black}{1} & - & Cherenkov $\gamma$ & - & - & \textcolor{blue}{conv} & CORSIKA-specific\tabularnewline
\hline 
\textcolor{black}{2} & - & decaying $\mu^{\pm}$ & - & - & \textcolor{blue}{conv} & CORSIKA-specific\tabularnewline
\hline 
\textcolor{black}{3} & - & $\eta\rightarrow2\gamma$ & - & - & \textcolor{blue}{conv} & CORSIKA-specific\tabularnewline
\hline 
\textcolor{black}{4} & - & $\eta\rightarrow3\pi^{0}$ & - & - & \textcolor{blue}{conv} & CORSIKA-specific\tabularnewline
\hline 
\textcolor{black}{5} & - & $\eta\rightarrow\pi^{+}\pi^{-}\pi^{0}$ & - & - & \textcolor{blue}{conv} & CORSIKA-specific\tabularnewline
\hline 
\textcolor{black}{6} & - & $\eta\rightarrow\pi^{+}\pi^{-}\gamma$ & - & - & \textcolor{blue}{conv} & CORSIKA-specific\tabularnewline
\hline 
\textcolor{black}{7} & - & $\mu^{\pm}$ add. info & - & - & \textcolor{blue}{conv} & CORSIKA-specific\tabularnewline
\hline 
\textcolor{black}{8} & 13 & $\mu^{+}$ & 105.66 & 2.20 & \textcolor{blue}{conv} & \tabularnewline
\hline 
\textcolor{black}{9} & -13 & $\mu^{-}$ & 105.66 & 2.20 & \textcolor{blue}{conv} & \tabularnewline
\hline 
\textcolor{black}{10} & 22 & $\gamma$ & - & - & \textcolor{blue}{conv} & stable\tabularnewline
\hline 
\textcolor{black}{11} & 111 & $\pi^{0}$ & 134.98 & 8.43$\cdot10^{-11}$ & \textcolor{red}{prompt} & \tabularnewline
\hline 
\textcolor{black}{12} & 113 & $\rho^{0}$ & 775.26 & 4.41$\cdot10^{-18}$ & \textcolor{red}{prompt} & \tabularnewline
\hline 
\textcolor{black}{13} & 130 & $K_{L}^{0}$ & 497.61 & 5.11$\cdot10^{-2}$ & \textcolor{blue}{conv} & \tabularnewline
\hline 
14 & -211 & $\pi^{+}$ & 139.57 & 2.60$\cdot10^{-2}$ & \textcolor{blue}{conv} & \tabularnewline
\hline 
15 & 211 & $\pi^{-}$ & 139.57 & 2.60$\cdot10^{-2}$ & \textcolor{blue}{conv} & \tabularnewline
\hline 
16 & -213 & $\rho^{+}$ & 775.26 & 4.41$\cdot10^{-18}$ & \textcolor{red}{prompt} & \tabularnewline
\hline 
17 & 213 & $\rho^{-}$ & 775.26 & 4.41$\cdot10^{-18}$ & \textcolor{red}{prompt} & \tabularnewline
\hline 
18 & 221 & $\eta$ & 547.86 & 5.02$\cdot10^{-13}$ & \textcolor{red}{prompt} & \tabularnewline
\hline 
19 & 223 & $\omega$ & 782.65 & 7.75$\cdot10^{-17}$ & \textcolor{red}{prompt} & \tabularnewline
\hline 
20 & 310 & $K_{S}^{0}$ & 497.61 & 8.95$\cdot10^{-5}$ & \textcolor{blue}{conv} & \tabularnewline
\hline 
21 & -313 & $K^{*0}$ & 895.55 & 1.39$\cdot10^{-17}$ & \textcolor{red}{prompt} & \tabularnewline
\hline 
22 & 313 & $\bar{K}^{*0}$ & 895.55 & 1.39$\cdot10^{-17}$ & \textcolor{red}{prompt} & \tabularnewline
\hline 
23 & -321 & $K^{+}$ & 493.68 & 1.24$\cdot10^{-2}$ & \textcolor{blue}{conv} & \tabularnewline
\hline 
24 & 321 & $K^{-}$ & 493.68 & 1.24$\cdot10^{-2}$ & \textcolor{blue}{conv} & \tabularnewline
\hline 
25 & -323 & $K^{*+}$ & 891.66 & 1.30$\cdot10^{-17}$ & \textcolor{red}{prompt} & \tabularnewline
\hline 
26 & 323 & $\bar{K}^{*-}$ & 891.66 & 1.30$\cdot10^{-17}$ & \textcolor{red}{prompt} & \tabularnewline
\hline 
27 & 333 & $\phi$ & 1019.46 & 1.54$\cdot10^{-16}$ & \textcolor{red}{prompt} & \tabularnewline
\hline 
28 & -411 & $D^{+}$ & 1869.65 & 1.03$\cdot10^{-6}$ & \textcolor{red}{prompt} & \tabularnewline
\hline 
29 & 411 & $\bar{D}^{-}$ & 1869.65 & 1.03$\cdot10^{-6}$ & \textcolor{red}{prompt} & \tabularnewline
\hline 
30 & -413 & $D^{*+}$ & 2010.26 & 7.89$\cdot10^{-15}$ & \textcolor{red}{prompt} & \tabularnewline
\hline 
31 & 413 & $\bar{D}^{*-}$ & 2010.26 & 7.89$\cdot10^{-15}$ & \textcolor{red}{prompt} & \tabularnewline
\hline 
32 & -421 & $D^{0}$ & 1864.83 & 4.10$\cdot10^{-7}$ & \textcolor{red}{prompt} & \tabularnewline
\hline 
33 & 421 & $\bar{D}^{0}$ & 1864.83 & 4.10$\cdot10^{-7}$ & \textcolor{red}{prompt} & \tabularnewline
\hline 
34 & -423 & $D^{*0}$ & 2006.85 & 3.13$\cdot10^{-16}$ & \textcolor{red}{prompt} & \tabularnewline
\hline 
35 & 423 & $\bar{D}^{*0}$ & 2006.85 & 3.13$\cdot10^{-16}$ & \textcolor{red}{prompt} & \tabularnewline
\hline 
36 & -431 & $D_{s}^{+}$ & 1968.34 & 5.04$\cdot10^{-7}$ & \textcolor{red}{prompt} & \tabularnewline
\hline 
37 & 431 & $\bar{D}_{s}^{-}$ & 1968.34 & 5.04$\cdot10^{-7}$ & \textcolor{red}{prompt} & \tabularnewline
\hline 
38 & -433 & $\bar{D}_{s}^{*-}$ & 2112.20 & unknown & \textcolor{red}{prompt} & $\tau$ probably short\tabularnewline
\hline 
39 & 433 & $D_{s}^{*-}$ & 2112.20 & unknown & \textcolor{red}{prompt} & $\tau$ probably short\tabularnewline
\hline 
40 & 441 & $\eta_{c}$ & 2983.90 & 2.06$\cdot10^{-17}$ & \textcolor{red}{prompt} & \tabularnewline
\hline 
41 & 443 & $J/\psi$ & 3096.90 & 7.08$\cdot10^{-15}$ & \textcolor{red}{prompt} & \tabularnewline
\hline 
42 & -1114 & $\bar{\Delta}^{+}$ & 1232.00 & 5.62$\cdot10^{-18}$ & \textcolor{red}{prompt} & \tabularnewline
\hline 
43 & 1114 & $\Delta^{-}$ & 1232.00 & 5.62$\cdot10^{-18}$ & \textcolor{red}{prompt} & \tabularnewline
\hline 
44 & -2112 & $\bar{n}$ & 939.57 & 8.79$\cdot10^{8}$ & \textcolor{blue}{conv} & \tabularnewline
\hline 
45 & 2112 & $n$ & 939.57 & 8.79$\cdot10^{8}$ & \textcolor{blue}{conv} & \tabularnewline
\hline 
46 & -2114 & $\bar{\Delta}^{0}$ & 1232.00 & 5.62$\cdot10^{-18}$ & \textcolor{red}{prompt} & \tabularnewline
\hline 
47 & 2114 & $\Delta^{0}$ & 1232.00 & 5.62$\cdot10^{-18}$ & \textcolor{red}{prompt} & \tabularnewline
\hline 
48 & -2212 & $\bar{p}$ & 938.27 & - & \textcolor{blue}{conv} & stable\tabularnewline
\hline 
49 & 2212 & $p$ & 938.27 & - & \textcolor{blue}{conv} & stable\tabularnewline
\hline 
50 & -2214 & $\Delta^{+}$ & 1232.00 & 5.62$\cdot10^{-18}$ & \textcolor{red}{prompt} & \tabularnewline
\hline 
51 & 2214 & $\bar{\Delta}^{-}$ & 1232.00 & 5.62$\cdot10^{-18}$ & \textcolor{red}{prompt} & \tabularnewline
\hline 
52 & -2224 & $\bar{\Delta}^{--}$ & 1232.00 & 5.62$\cdot10^{-18}$ & \textcolor{red}{prompt} & \tabularnewline
\hline 
53 & 2224 & $\Delta^{++}$ & 1232.00 & 5.62$\cdot10^{-18}$ & \textcolor{red}{prompt} & \tabularnewline
\hline 
54 & -3112 & $\Sigma^{-}$ & 1197.45 & 1.48$\cdot10^{-4}$ & \textcolor{blue}{conv} & \tabularnewline
\hline 
55 & 3112 & $\bar{\Sigma}^{+}$ & 1197.45 & 1.48$\cdot10^{-4}$ & \textcolor{blue}{conv} & \tabularnewline
\hline 
56 & 3122 & $\Lambda$ & 1115.68 & 2.63$\cdot10^{-4}$ & \textcolor{blue}{conv} & \tabularnewline
\hline 
57 & -3122 & $\bar{\Lambda}$ & 1115.68 & 2.63$\cdot10^{-4}$ & \textcolor{blue}{conv} & \tabularnewline
\hline 
58 & 3212 & $\Sigma^{0}$ & 1192.64 & 7.39$\cdot10^{-14}$ & \textcolor{red}{prompt} & \tabularnewline
\hline 
59 & -3212 & $\bar{\Sigma}^{0}$ & 1192.64 & 7.39$\cdot10^{-14}$ & \textcolor{red}{prompt} & \tabularnewline
\hline 
60 & -3222 & $\Sigma^{+}$ & 1189.37 & 8.01$\cdot10^{-5}$ & \textcolor{blue}{conv} & \tabularnewline
\hline 
61 & 3222 & $\bar{\Sigma}^{-}$ & 1189.37 & 8.01$\cdot10^{-5}$ & \textcolor{blue}{conv} & \tabularnewline
\hline 
62 & -3312 & $\bar{\Xi}^{+}$ & 1321.71 & 1.64$\cdot10^{-4}$ & \textcolor{blue}{conv} & \tabularnewline
\hline 
63 & 3312 & $\Xi^{-}$ & 1321.71 & 1.64$\cdot10^{-4}$ & \textcolor{blue}{conv} & \tabularnewline
\hline 
64 & 3322 & $\Xi^{0}$ & 1314.86 & 2.90$\cdot10^{-4}$ & \textcolor{blue}{conv} & \tabularnewline
\hline 
65 & -3322 & $\bar{\Xi}^{0}$ & 1314.86 & 2.90$\cdot10^{-4}$ & \textcolor{blue}{conv} & \tabularnewline
\hline 
66 & -3334 & $\bar{\Omega}^{+}$ & 1672.45 & 8.15$\cdot10^{-5}$ & \textcolor{blue}{conv} & \tabularnewline
\hline 
67 & 3334 & $\Omega^{-}$ & 1672.45 & 8.15$\cdot10^{-5}$ & \textcolor{blue}{conv} & \tabularnewline
\hline 
68 & 4112 & $\Sigma_{c}^{0}$ & 2453.75 & 3.59$\cdot10^{-16}$ & \textcolor{red}{prompt} & \tabularnewline
\hline 
69 & -4112 & $\bar{\Sigma}_{c}^{0}$ & 2453.75 & 3.59$\cdot10^{-16}$ & \textcolor{red}{prompt} & \tabularnewline
\hline 
70 & 4114 & $\Sigma_{c}^{*0}$ & 2518.48 & 4.30$\cdot10^{-17}$ & \textcolor{red}{prompt} & \tabularnewline
\hline 
71 & -4114 & $\bar{\Sigma}_{c}^{*0}$ & 2518.48 & 4.30$\cdot10^{-17}$ & \textcolor{red}{prompt} & \tabularnewline
\hline 
72 & -4122 & $\bar{\Lambda}_{c}^{-}$ & 2286.46 & 1.99$\cdot10^{-7}$ & \textcolor{red}{prompt} & \tabularnewline
\hline 
73 & 4122 & $\Lambda_{c}^{+}$ & 2286.46 & 1.99$\cdot10^{-7}$ & \textcolor{red}{prompt} & \tabularnewline
\hline 
74 & 4132 & $\Xi_{c}^{0}$ & 2470.91 & 1.11$\cdot10^{-7}$ & \textcolor{red}{prompt} & \tabularnewline
\hline 
75 & -4132 & $\bar{\Xi}_{c}^{0}$ & 2470.91 & 1.11$\cdot10^{-7}$ & \textcolor{red}{prompt} & \tabularnewline
\hline 
76 & -4212 & $\Sigma_{c}^{+}$ & 2452.90 & 1.43$\cdot10^{-16}$ & \textcolor{red}{prompt} & \tabularnewline
\hline 
77 & 4212 & $\bar{\Sigma}_{c}^{-}$ & 2452.90 & 1.43$\cdot10^{-16}$ & \textcolor{red}{prompt} & \tabularnewline
\hline 
78 & -4214 & $\bar{\Sigma}_{c}^{*-}$ & 2517.50 & 3.87$\cdot10^{-17}$ & \textcolor{red}{prompt} & \tabularnewline
\hline 
79 & -4222 & $\Sigma_{c}^{++}$ & 2453.97 & 3.48$\cdot10^{-16}$ & \textcolor{red}{prompt} & \tabularnewline
\hline 
80 & 4222 & $\bar{\Sigma}_{c}^{--}$ & 2453.97 & 3.48$\cdot10^{-16}$ & \textcolor{red}{prompt} & \tabularnewline
\hline 
81 & -4224 & $\bar{\Sigma}_{c}^{*--}$ & 2518.41 & 4.45$\cdot10^{-17}$ & \textcolor{red}{prompt} & \tabularnewline
\hline 
82 & 4224 & $\Sigma_{c}^{*++}$ & 2518.41 & 4.45$\cdot10^{-17}$ & \textcolor{red}{prompt} & \tabularnewline
\hline 
83 & 4232 & $\Xi_{c}^{+}$ & 2467.93 & 4.41$\cdot10^{-7}$ & \textcolor{red}{prompt} & \tabularnewline
\hline 
84 & -4232 & $\bar{\Xi}_{c}^{-}$ & 2467.93 & 4.41$\cdot10^{-7}$ & \textcolor{red}{prompt} & \tabularnewline
\hline 
85 & -4332 & $\bar{\Omega}_{c}^{0}$ & 2695.20 & 2.67$\cdot10^{-7}$ & \textcolor{red}{prompt} & \tabularnewline
\hline 
86 & 4332 & $\Omega_{c}^{0}$ & 2695.20 & 2.67$\cdot10^{-7}$ & \textcolor{red}{prompt} & \tabularnewline
\hline 
\end{longtable}{\small\par}
\par\end{center}

\subsection{Verification of the prompt muon definition\label{subsec:Cross-check-of-the-prompt-def}}

All the combinations from Tab. \ref{tab:Prompt-muon-possible-combinations}
were explicitly verified with $d_{\mu\,\mathrm{prod}}$ plots. The
test was done on ORCA115 MC, since ORCA is located shallower and thus,
a greater muon statistics was available. The muons at sea level were
plotted, which means that if a shower reaches ORCA115, even muons
not reaching the can are included. The following abbreviations were
used:
\begin{itemize}
\item gmom: grandmother,
\item mom: mother,
\item nucl: nucleus,
\item prim: primary.
\end{itemize}
\begin{center}
\begin{figure}[H]
\centering{}\includegraphics[width=13.5cm]{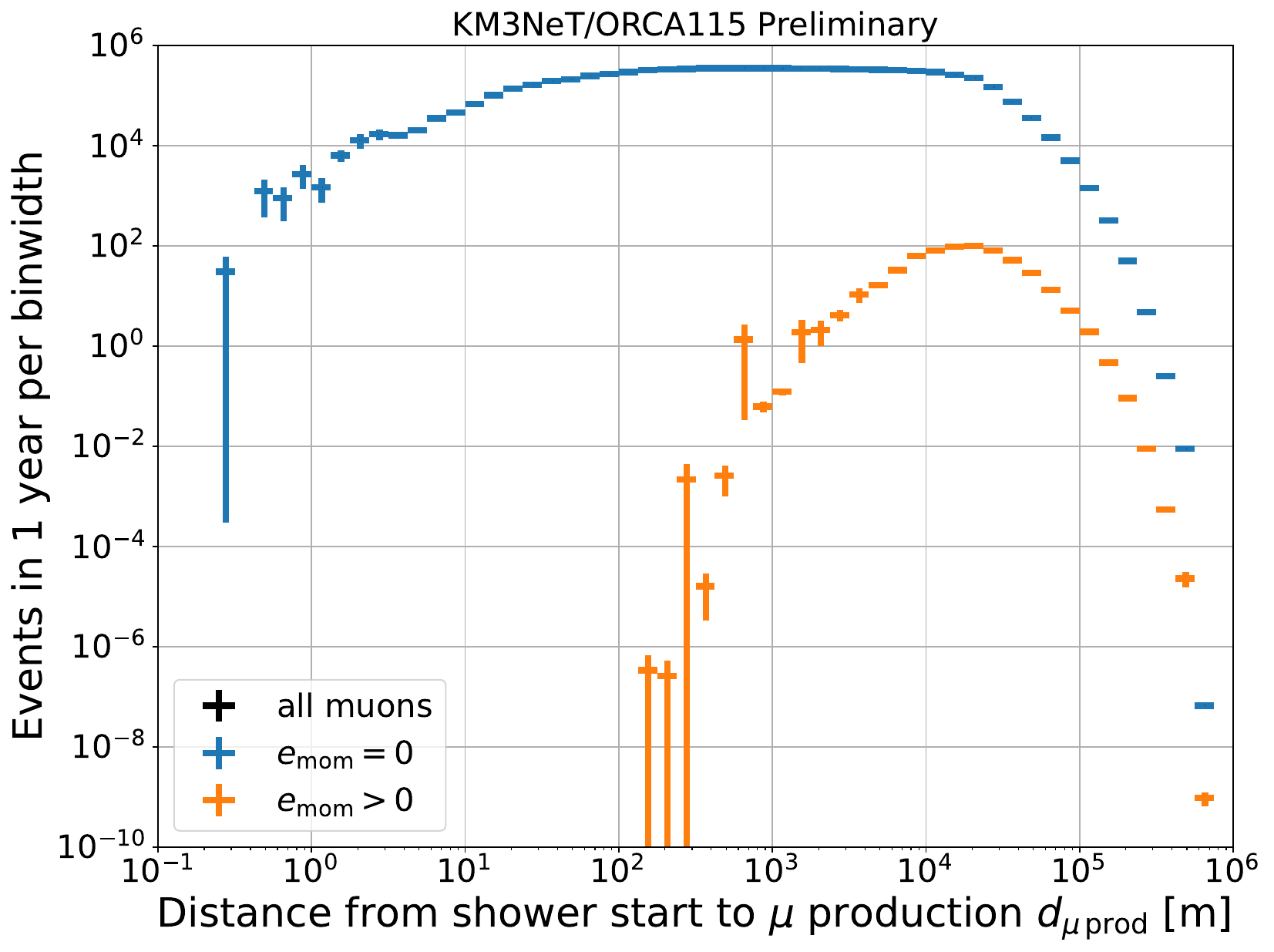}\caption{Distributions of $d_{\mu\,\mathrm{prod}}$ for $\mu$ with different
mother EM counter $e_{\mathsf{mom}}$ values. The distribution for
$e_{\mu}$ is not plotted, since it is always equal to zero. The distribution
for all muons is not visible, because it is covered by the one for
$e_{\mathsf{mom}}=0$.}
\end{figure}
\par\end{center}

\begin{center}
\begin{figure}[H]
\centering{}\includegraphics[width=13.5cm]{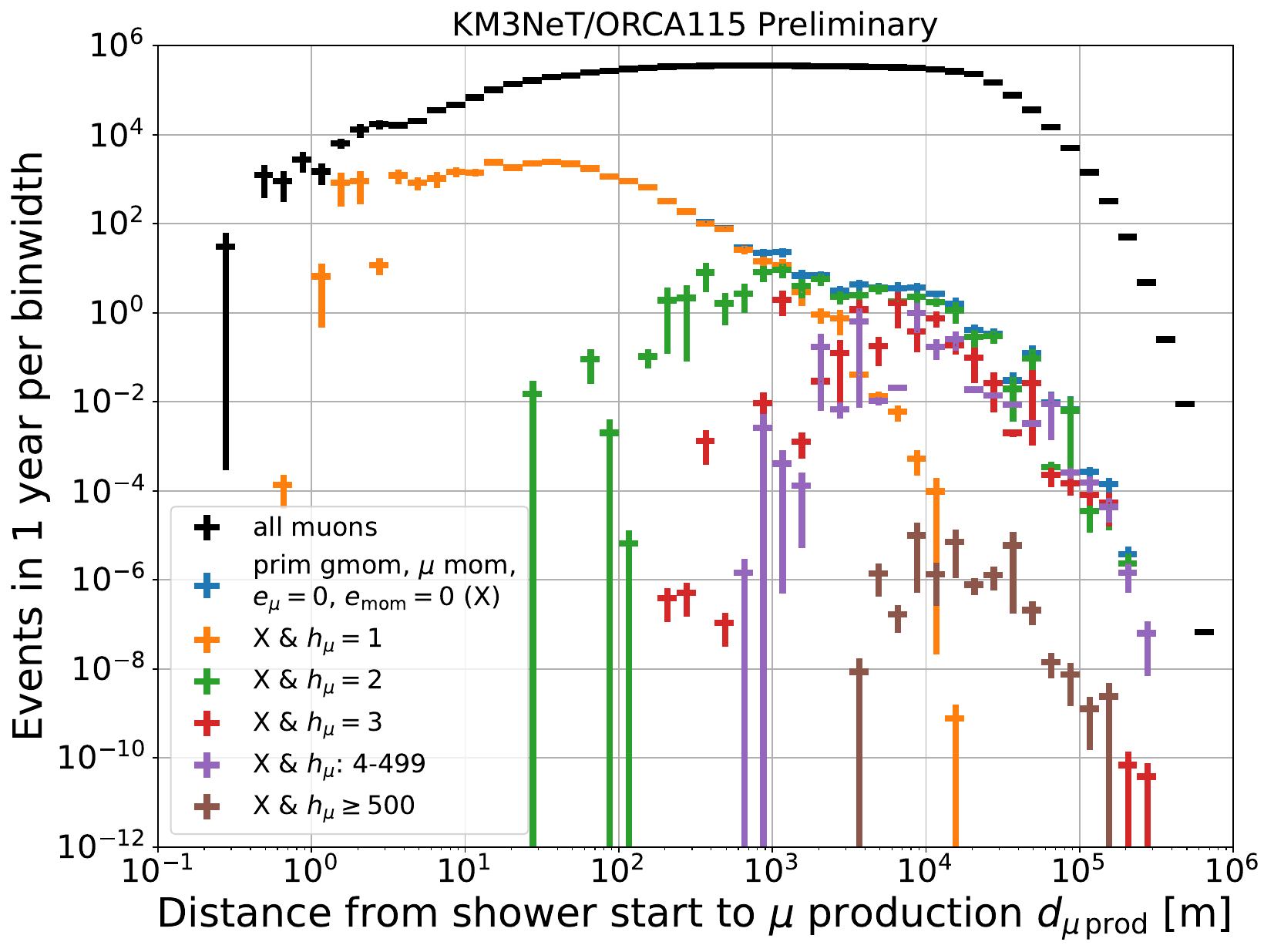}\caption{Distributions of $d_{\mu\,\mathrm{prod}}$ for $\mu$ with prim gmom
and $\mu$ mom, but different $h_{\mu}$ counter values.}
\end{figure}
\par\end{center}

\begin{center}
\begin{figure}[H]
\centering{}\includegraphics[width=13.5cm]{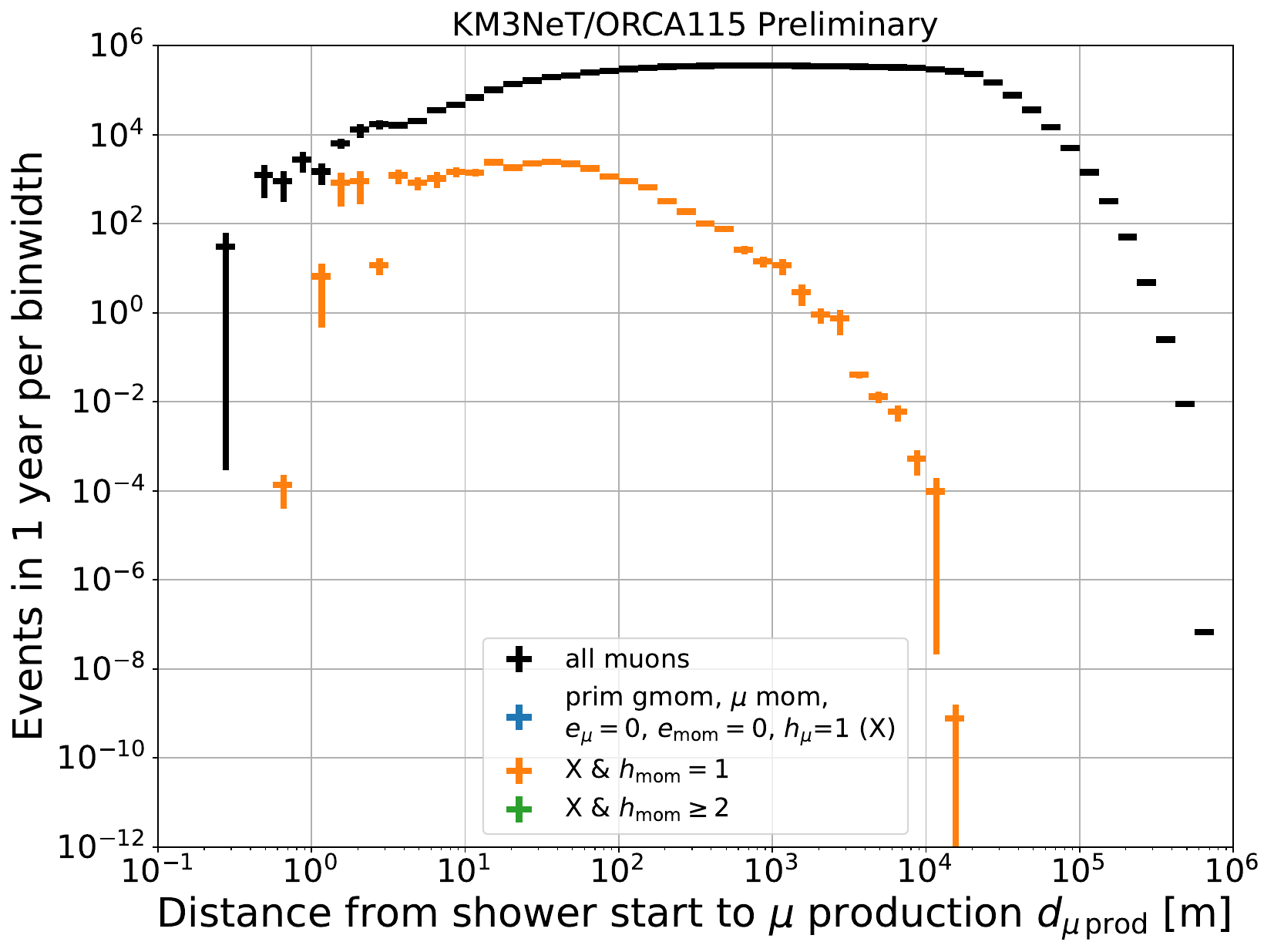}\caption{Distributions of $d_{\mu\,\mathrm{prod}}$ for $\mu$ with prim gmom,
$\mu$ mom, and $h_{\mu}=1$. The plot demonstrates that in fact for
$h_{\mu}=1$, the only contribution comes from $h_{\mathsf{mom}}=1$.}
\end{figure}
\par\end{center}

\begin{center}
\begin{figure}[H]
\centering{}\includegraphics[width=13.5cm]{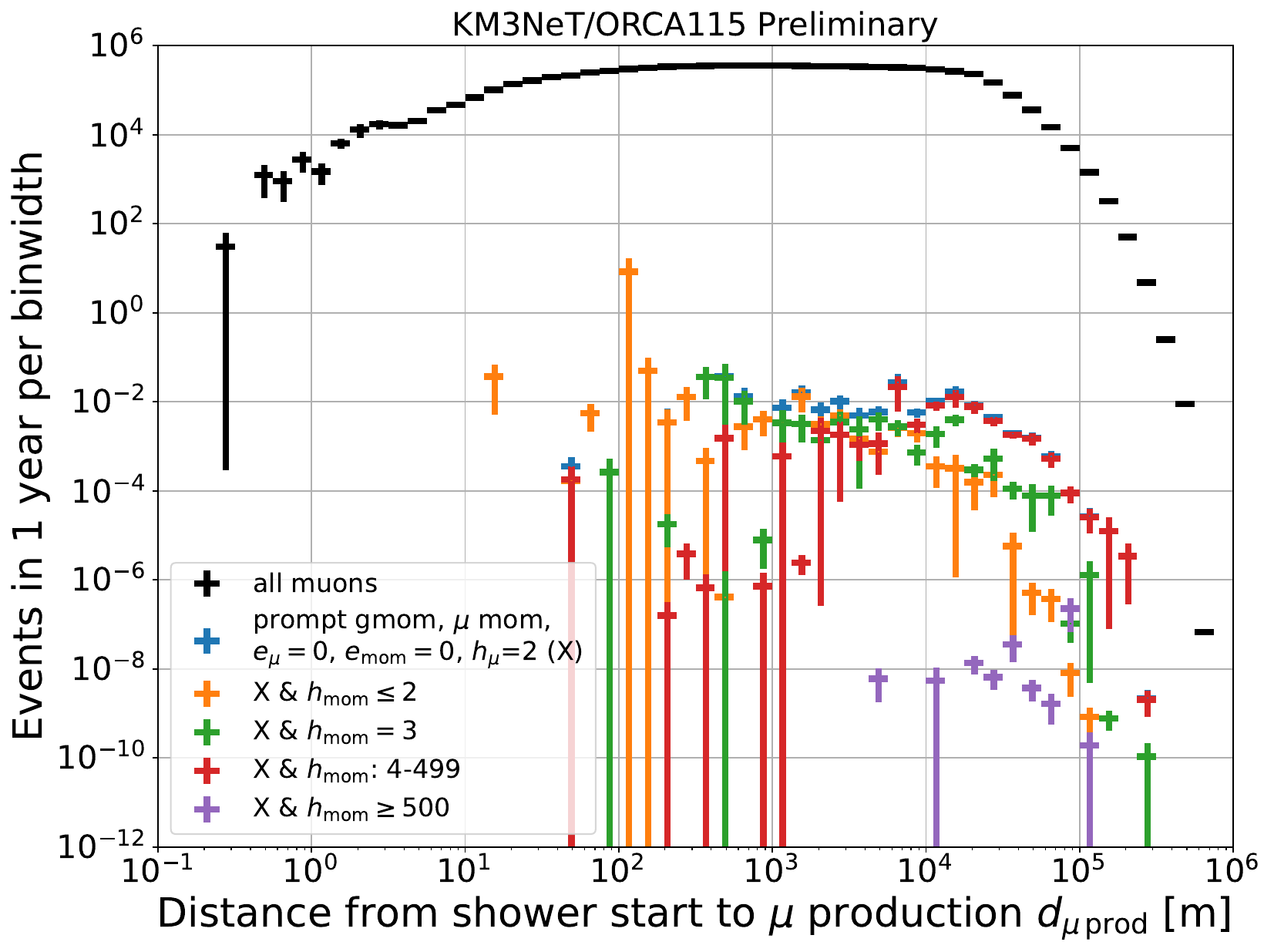}\caption{Distributions of $d_{\mu\,\mathrm{prod}}$ for $\mu$ with prompt
gmom, $\mu$ mom, and different $h_{\mu}$ values. \label{fig:d_mu_prod_prompt_mu}}
\end{figure}
\par\end{center}

\begin{center}
\begin{figure}[H]
\centering{}\includegraphics[width=13.5cm]{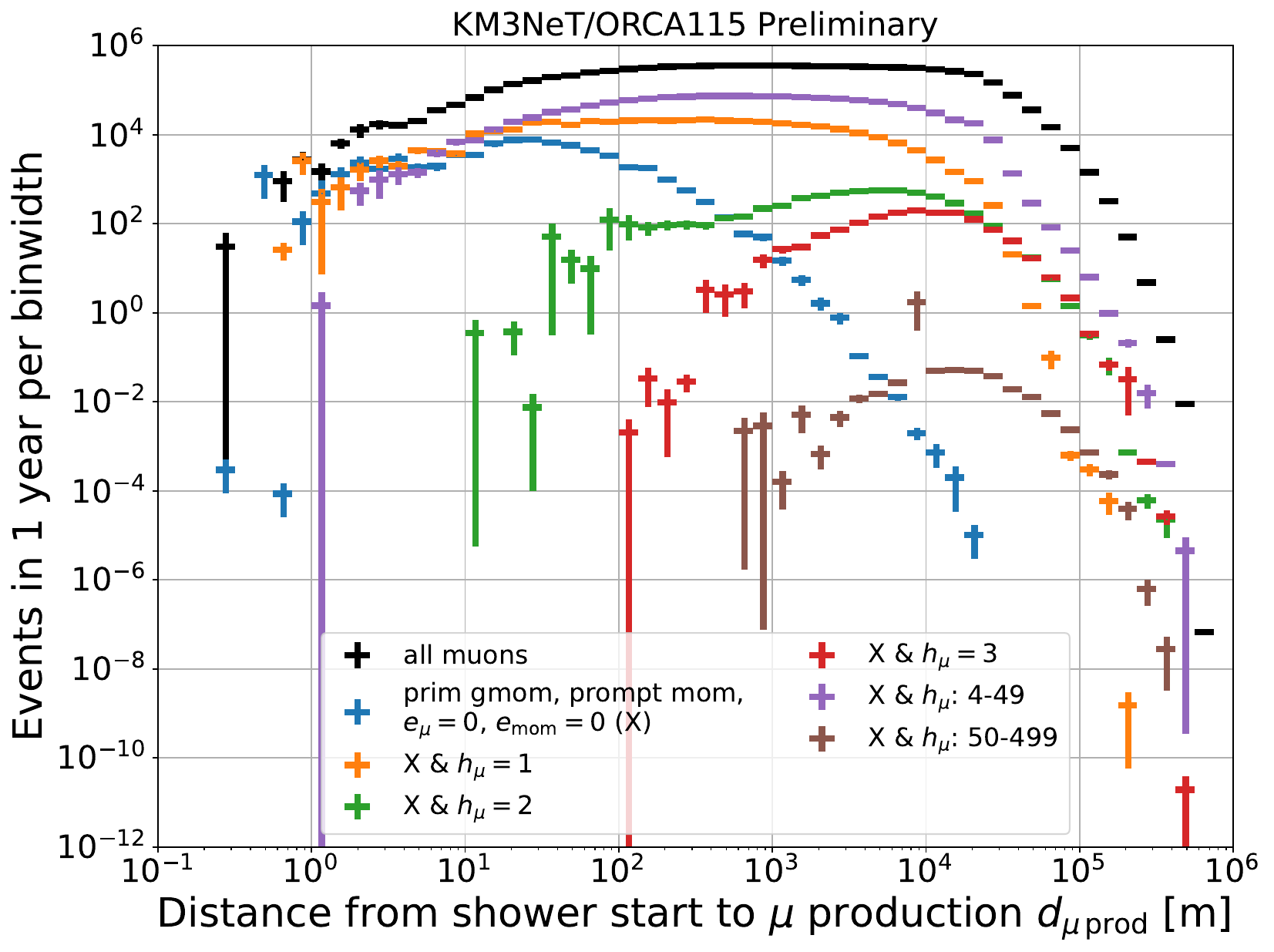}\caption{Distributions of $d_{\mu\,\mathrm{prod}}$ for $\mu$ with prim gmom
and prompt mom.}
\end{figure}
\par\end{center}

\begin{center}
\begin{figure}[H]
\centering{}\includegraphics[width=13.5cm]{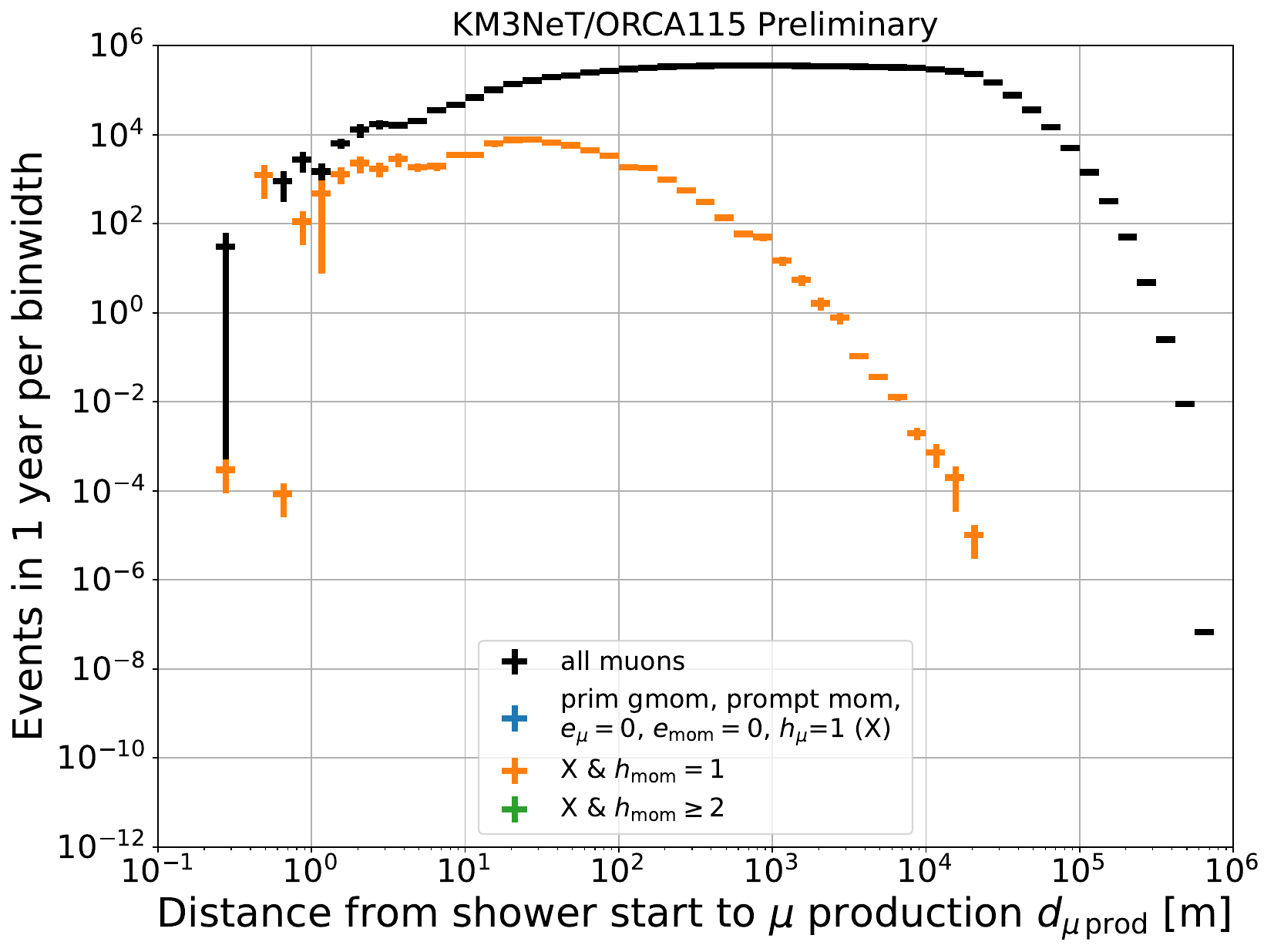}\caption{Distributions of $d_{\mu\,\mathrm{prod}}$ for $\mu$ with prim gmom,
prompt mom, and $h_{\mu}=1$.}
\end{figure}
\par\end{center}

\begin{center}
\begin{figure}[H]
\centering{}\includegraphics[width=13.5cm]{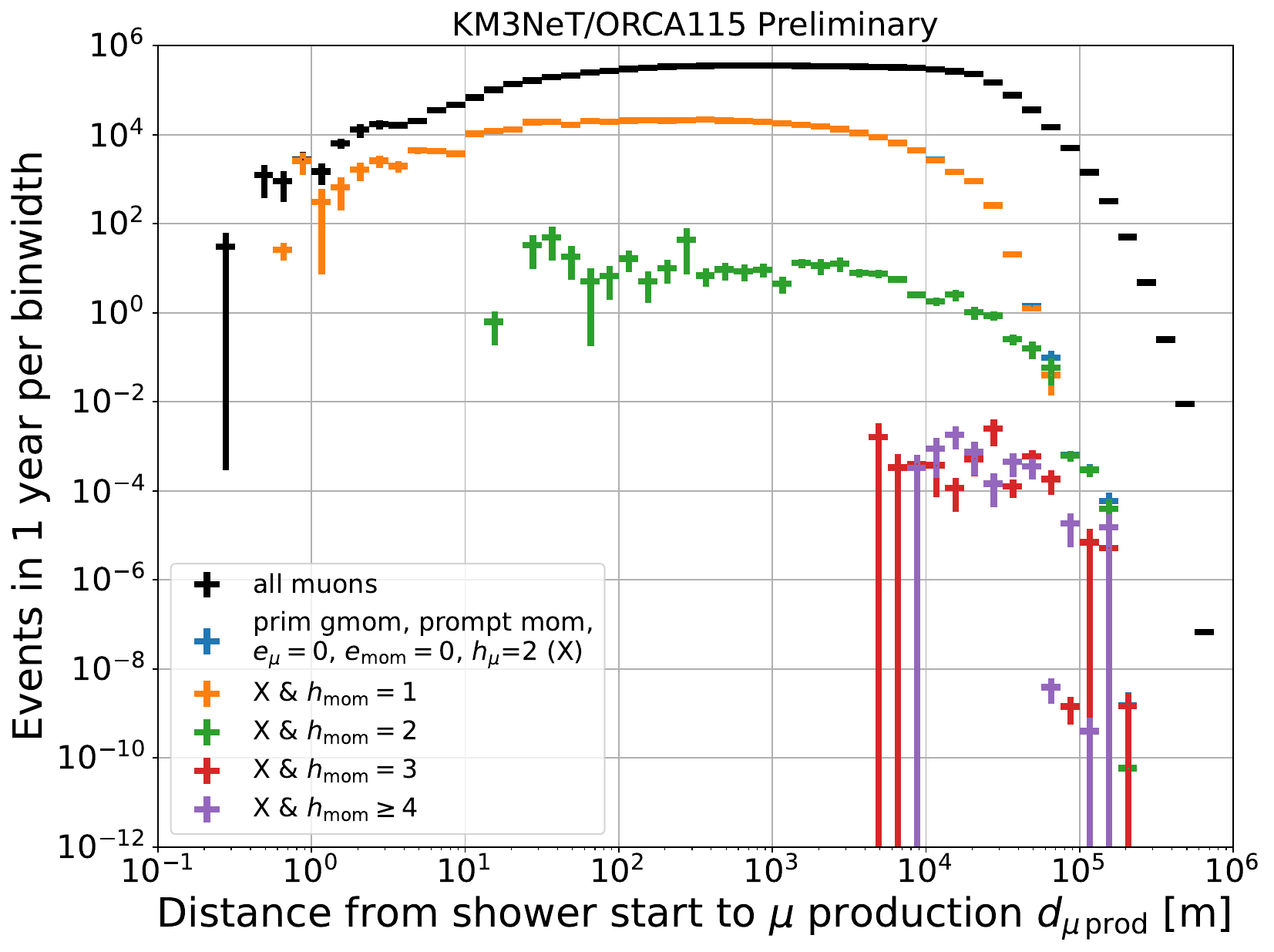}\caption{Distributions of $d_{\mu\,\mathrm{prod}}$ for $\mu$ with prim gmom,
prompt mom, and $h_{\mu}=2$.}
\end{figure}
\par\end{center}

\begin{center}
\begin{figure}[H]
\centering{}\includegraphics[width=13.5cm]{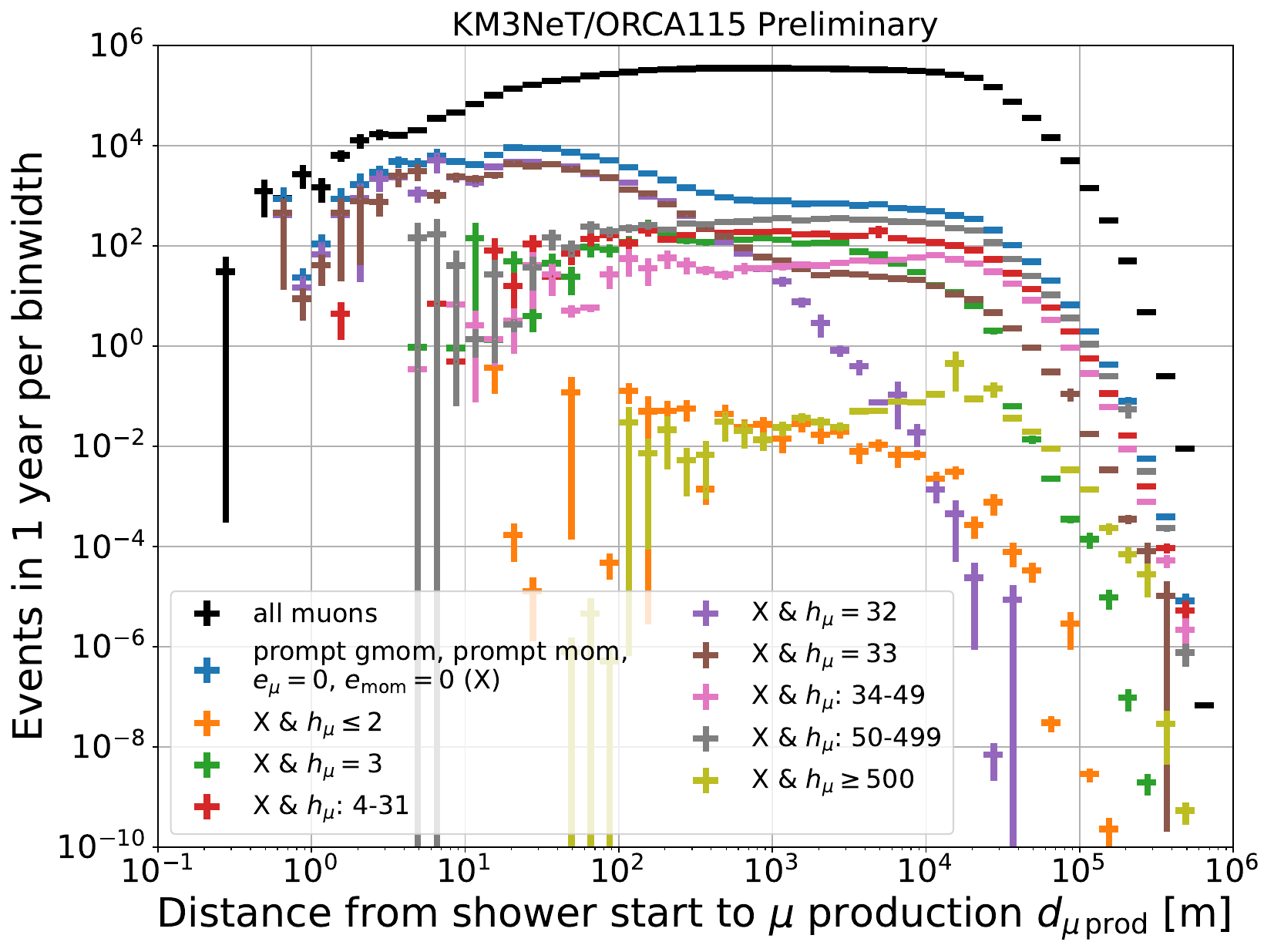}\caption{Distributions of $d_{\mu\,\mathrm{prod}}$ for $\mu$ with prompt
gmom and prompt mom.}
\end{figure}
\par\end{center}

\begin{center}
\begin{figure}[H]
\centering{}\includegraphics[width=13.5cm]{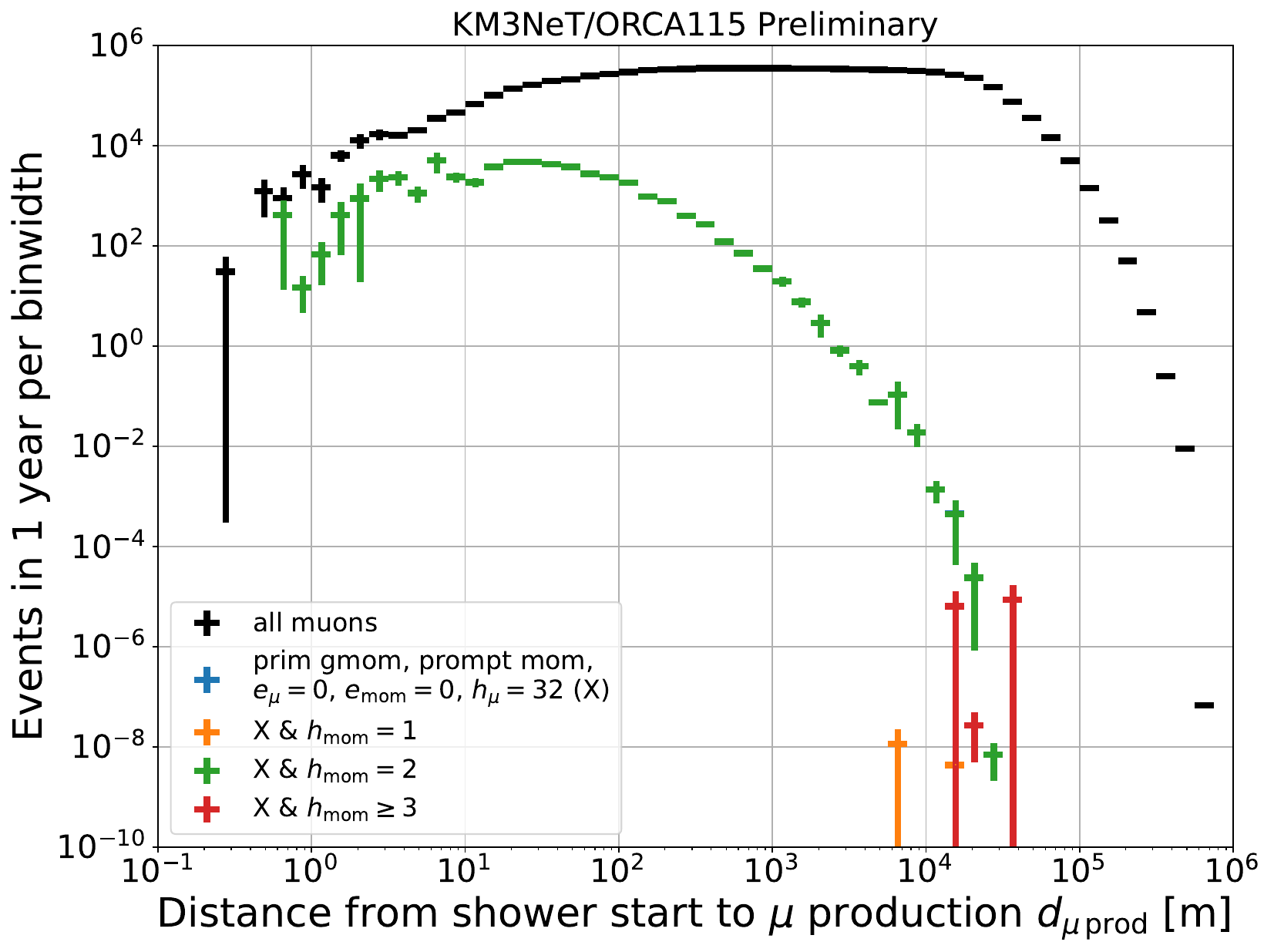}\caption{Distributions of $d_{\mu\,\mathrm{prod}}$ for $\mu$ with prompt
gmom, prompt mom, and $h_{\mu}=32$.}
\end{figure}
\par\end{center}

\begin{center}
\begin{figure}[H]
\centering{}\includegraphics[width=13.5cm]{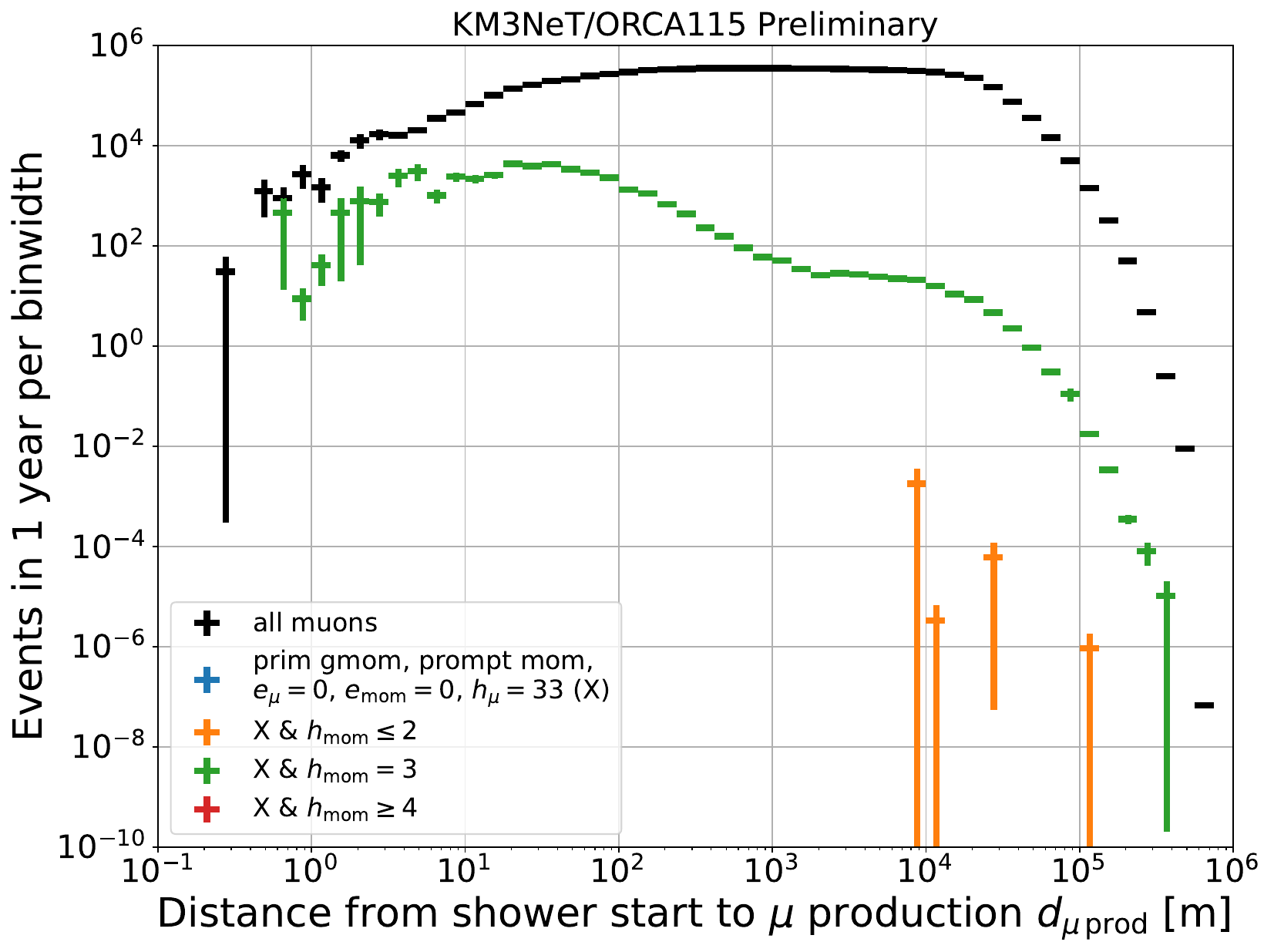}\caption{Distributions of $d_{\mu\,\mathrm{prod}}$ for $\mu$ with prompt
gmom, prompt mom, and $h_{\mu}=33$.}
\end{figure}
\par\end{center}

\chapter*{Acknowledgements\label{chap:Acknowledgements}}

Here, I would like to express my gratitude towards people and institutions,
who contributed to the successful completion of this thesis.

\section*{Supervision}

I would like to thank in the first line my thesis supervisors:
\begin{itemize}
\item \textbf{Prof. dr hab. Ewa Rondio}, for many discussions with deep
insights, which often drove me to rethink my approach and improve
my analysis. I am grateful for her generous sharing of rich experimental
experience, while making sure I do not deviate from the track too
far. Her devotion in supporting my work, despite being involved in
key roles at NCBJ and in the T2K and HK Collaborations was truly remarkable. 
\item \textbf{Dr Piotr Mijakowski}, for introducing me to the wonderful
community of the KM3NeT experiment, believing in me, and encouraging
to pursue even most ambitious goals. I owe the success of my efforts
to his guidance and expertise, in particular in the field of machine
learning.
\end{itemize}

\section*{Funding}

This work was accomplished under the financial support of the \textbf{National
Science Centre}:
\begin{itemize}
\item NCN Sonata Bis 2015/18/E/ST2/00758, `\emph{Indirect search for dark
matter with water neutrino detectors}', PI: dr Piotr Mijakowski
\item NCN Preludium 2021/41/N/ST2/01177, `\emph{Measurement of muon flux
with KM3NeT-ARCA and KM3NeT-ORCA detectors}', PI: Piotr Kalaczyński
\end{itemize}
\begin{center}
\includegraphics[height=0.7cm]{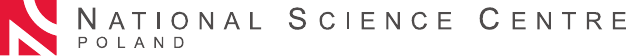}
\par\end{center}

Piotr Kalaczyński is supported by the grant `\emph{AstroCeNT: Particle
Astrophysics Science and Technology Centre}', carried out within
the International Research Agendas programme of the Foundation for
Polish Science financed by the European Union under the European Regional
Development Fund.
\begin{center}
\includegraphics[height=2.6cm]{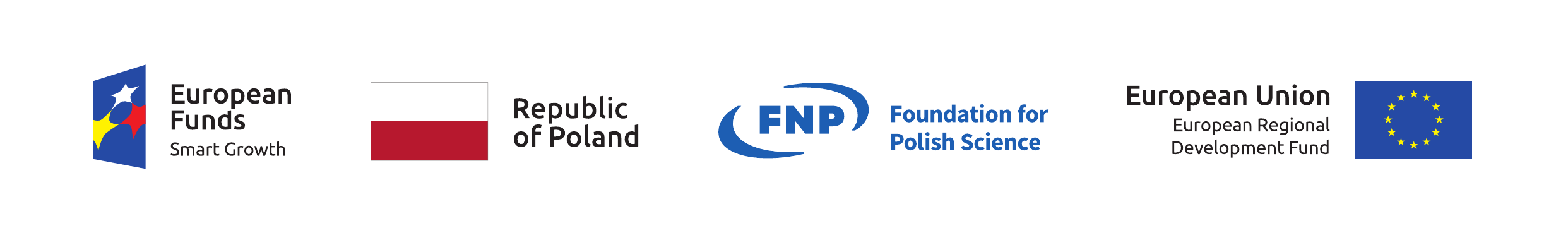}
\par\end{center}

\section*{Computing\label{sec:Computing}}

The services provided by the Świerk Computing Centre (CIŚ) were crucial
for the computationally-intensive MC simulations, which were the backbone
of this thesis. Here, I would like to extend my gratitude in particular
to \textbf{Henryk Giemza}, who was my main contact person at CIŚ and
has always been very responsive and helpful with all the trouble that
needed solving and requests that I had.
\begin{center}
\includegraphics[height=2.5cm]{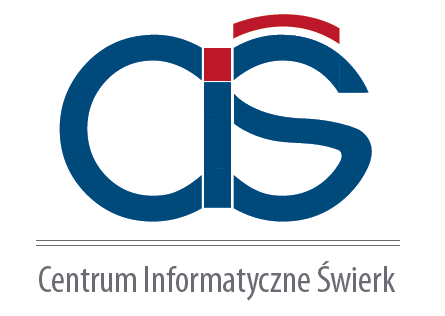}
\par\end{center}

I am also grateful to the IN2P3’s Computing Centre in Lyon, France
for delivering a smoothly running computing service, with a dedicated
KM3NeT ecosystem, which I made frequent use of.
\begin{center}
\includegraphics[height=2.5cm]{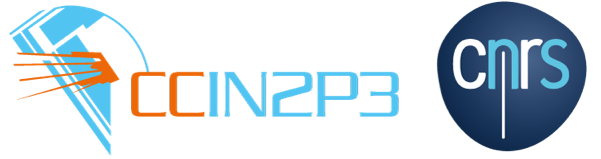}
\par\end{center}

\section*{Software}

The CORSIKA code \cite{CORSIKA} has been the main software, without
which none of the results of this thesis would have been possible.
I am not only grateful to the authors for providing the CORSIKA package
to the astroparticle physics community, but also for the organisation
of very useful CORSIKA Workshops, and for many informative discussions.
Here, I want to especially mention \textbf{dr Tanguy Pierog}, \textbf{dr
Felix Riehn}, and \textbf{dr Sergey Ostapchenko}.

I acknowledge the service provided in form of the pymsis package by
the University of Colorado Space Weather Technology, Research and
Education Center (SWx TREC), which allowed me to conveniently access
the NRLMSIS-2.0 \cite{NRLMSIS-2.0} and NRLMSISE-00 \cite{NRLMSISE-00}
model predictions .
\begin{center}
\includegraphics[height=2cm]{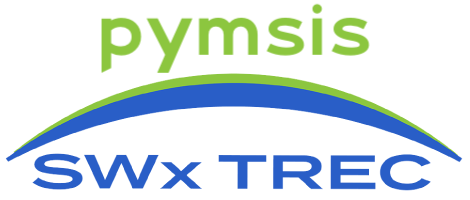}
\par\end{center}

The crflux \cite{combined_HillasGaisser_and_GaisserHonda_also_crfluxmodels_reference}
and MCEq codes \cite{MCEq}, developed by \textbf{dr Anatoli Feydnitch},
were of great help for this work. I thank Anatoli for fruitful discussions
on potential applications and extensions of his software.

\section*{Others}

I am also deeply grateful to all my professors and colleagues from
NCBJ Warsaw and KM3NeT Collaboration. They made my time as a PhD student
a memorable experience and have helped me to accumulate knowledge
and skills necessary to complete this thesis.

My PhD experience would not be the same without the great people from
the Warsaw Neutrino Group: \textbf{dr Magdalena Posiadala-Zezula},
\textbf{dr Joanna Zalipska}, \textbf{dr Katarzyna Kowalik}, \textbf{Yashwanth
S. Prabhu}, \textbf{Maitrayee Mandal}, \textbf{dr Lakshmi S. Mohan},
\textbf{dr Tomas Nosek}, \textbf{dr Katarzyna Frankiewicz}, \textbf{dr
Jerzy Mańczak},\textbf{ dr Rafał Wojaczyński}, \textbf{dr Meghna Kunhikannan
Kannichankandy}, \textbf{dr Katarzyna Grzelak} and notably: \textbf{dr
Kamil Skwarczyński}, with whom I not only shared the passion for neutrinos
\& Co., but also for climbing and exquisite memes, \textbf{dr Grzegorz
Żarnecki}, who has inspired me with his determination and dedication,\textbf{
dr Justyna Łagoda}, who was extremely helpful both in technical discussions,
and in formal issues. A honorary mention goes to \textbf{prof. Maria
„Hula” Szeptycka}$\dagger$, who has been extremely active member
of the group and showed curiosity in all the discussions. She regrettably
passed away on 23.01.2022, unable to see the conclusion of this work.

The environment within the KM3NeT Collaboration was no less welcoming
than at NCBJ. I am most thankful to have met \textbf{Tamás Gal}.\textbf{
}His expertise and devotion to KM3NeT are truly amazing. I always
wondered, how he still managed to always find the time to help me
understand the nuances of KM3NeT-specific software and programming
in general. The only person I can think of, who can match his zeal,
if undoubtedly \textbf{prof. Maarten de Jong}, who basically breathes
KM3NeT and devotes all his energy to the success of the collaboration.
I deeply valued my technical discussions with him, always displaying
sharp understanding of various complex aspects of the experiment,
and straight to the point. I very much appreciate the supportive and
encouraging attitude of \textbf{prof. Annarita Margiotta}, who worked
tirelessly to make sure that all the publications within the collaboration,
including mine conformed to best scientific practices. My introduction
to CORSIKA simulations was a smooth process thanks to \textbf{dr Konstantinos
Pikounis} and his readiness to address any of my questions or doubts.
I have many things to thank \textbf{dr Vladimir Kulikovskiy} for:
he was the best analysis reviewer I could have wished for, he introduced
me to rock climbing, and he was just a cheerful, friendly colleague,
who I could rely on. I am no less appreciative of my second reviewer,
\textbf{dr Luigi Antonio Fusco}, for pointing out things in the spirit
of improving the analysis together. The whole work was coordinated
by the Cosmic Ray Working Group leader, \textbf{dr Ronald Brujin},
who always kept track of the bigger picture. My early MC work within
KM3NeT was certainly heavily influenced by discussions with \textbf{dr
Rosa Coniglione}, a true expert on the matter. Developing the gSeaGen
code was a group effort from the very beginning. I was all the time
in tight cooperation with its original author, \textbf{dr Carla Distefano},
and main co-developer, \textbf{dr Alfonso Garcia Andres Soto}. I definitely
enjoyed the work on the gSeaGen team and hope to be able to do some
more in the future! My machine learning work has been marked by numerous
fruitful conversations and exchanges of ideas with \textbf{dr Stefan
Reck}, who worked on a similar problem in his thesis \cite{StefanReckThesis}.
It is not an exaggeration to state that I would need at least one
more year to finish the CORSIKA MC production work, if it was not
for \textbf{dr Andrey Romanov's} assistance. He not only took a significant
part of the job onto himself, but also actively worked with me to
further improve the simulation. He also contributed significantly
to my work on processing the CORSIKA files with gSeaGen and to the
evaluation of the systematic uncertainties.

I also want to mention here other people, I enjoyed collaborating
with on various subjects within KM3NeT: \textbf{prof. Antonio Capone,
dr Thomas Eberl, dr Alba Domi}, \textbf{dr Brian Ó Fearraigh}, \textbf{Louis
Bailly-Salins}, and \textbf{dr Jannik Hofestädt}.

\printbibliography 

\printnomenclature[2.5cm]{}

\addcontentsline{toc}{chapter}{Nomenclature}

\printindex

\end{document}